\documentclass[12pt]{article}
\pdfoutput=1
\usepackage{latexsym,amsmath,amssymb, graphicx, subfigure}
\usepackage[linktocpage]{hyperref}
\renewcommand{\thefootnote}{\fnsymbol{footnote}}

\usepackage{mathrsfs }
\usepackage{amsmath}
\usepackage{amsfonts}

\numberwithin{equation}{section}

\def\doubleset#1#2{\bgroup%
\def\doit#1#2{%
\setbox\dblsetbox=\hbox{$\cstyle #1$}%
\raise#2\ht\dblsetbox\copy\dblsetbox%
\hskip-\wd\dblsetbox%
\raise-#2\ht\dblsetbox\box\dblsetbox}%
\mathchoice%
{\def\cstyle{\displaystyle}\doit#1#2}%
{\def\cstyle{\textstyle}\doit#1#2}%
{\def\cstyle{\scriptstyle}\doit#1#2}%
{\def\cstyle{\scriptscriptstyle}\doit#1#2}\egroup}
\def\underarrow#1{\vbox{\ialign{##\crcr$\hfil\displaystyle
 {#1}\hfil$\crcr\noalign{\kern1pt\nointerlineskip}$\longrightarrow$\crcr}}}

\newbox\dblsetbox

\newcommand{\bR}{\mathbb{R}}

\newcommand{\bL}{\boldsymbol\lambda}

\newlength{\extraspace}
\newlength{\extraspaces}
\newcommand{\be}{\begin{equation}
\addtolength{\abovedisplayskip}{\extraspaces}
\addtolength{\belowdisplayskip}{\extraspaces}
\addtolength{\abovedisplayshortskip}{\extraspace}
\addtolength{\belowdisplayshortskip}{\extraspace}}
\newcommand{\ee}{\end{equation}}

\newcommand{\ba}{\begin{eqnarray}
\addtolength{\abovedisplayskip}{\extraspaces}
\addtolength{\belowdisplayskip}{\extraspaces}
\addtolength{\abovedisplayshortskip}{\extraspace}
\addtolength{\belowdisplayshortskip}{\extraspace}}
\newcommand{\ea}{\end{eqnarray}}

\newcommand{\bd}{\begin{displaymath}
\addtolength{\abovedisplayskip}{\extraspaces}
\addtolength{\belowdisplayskip}{\extraspaces}
\addtolength{\abovedisplayshortskip}{\extraspace}
\addtolength{\belowdisplayshortskip}{\extraspace}}
\newcommand{\ed}{\end{displaymath}}

\newcounter{saveeqn}

\newcommand{\newsection}[1]{
\vspace{12mm} \pagebreak[3] \addtocounter{section}{1}
\setcounter{equation}{0} \setcounter{subsection}{0}
\noindent{\bf \thesection. #1} \nopagebreak
\medskip
\nopagebreak
\addcontentsline{toc}{section}{\thesection. #1}}

\newcommand{\newsubsection}[1]{
\vspace{0.8cm} \pagebreak[3] \addtocounter{subsection}{1}
\setcounter{subsubsection}{0}
\noindent{ \it \thesubsection. #1} \nopagebreak \vspace{2mm}
\nopagebreak
\addcontentsline{toc}{subsection}{\thesubsection. #1}}

\begin{document}
\addtolength{\baselineskip}{1.5mm}

\thispagestyle{empty}

\vbox{} \vspace{-0.0cm}

\begin{center}
\centerline{\Large{\bf M-Theoretic Derivations of 4d--2d Dualities: From a}}
\medskip
\centerline{\Large{\bf Geometric Langlands Duality for Surfaces, to the}}
\medskip
\centerline{\Large{\bf AGT Correspondence, to Integrable Systems}}

\vspace{0.7cm}

{\bf{Meng-Chwan~Tan}}
\\[0mm]
{\it Department of Physics,
National University of Singapore}\\[0 mm]
mctan@nus.edu.sg
\end{center}

\vspace{0.7 cm}

\centerline{\bf Abstract}\smallskip

In Part I, we extend our analysis in [arXiv:0807.1107], and show that a mathematically conjectured geometric Langlands duality for complex surfaces in~\cite{BF}, and its generalizations -- which relate some cohomology of the moduli space of certain (``ramified'') $G$-instantons to the integrable representations of the\emph{ Langlands dual} of certain affine (sub) $G$-algebras, where $G$ is \emph{any }compact Lie group -- can be derived, purely physically, from the principle that the spacetime BPS spectra of \emph{string-dual} M-theory compactifications ought to be equivalent.

In Part II,  to the setup in Part I, we introduce Omega-deformation via fluxbranes and add half-BPS boundary defects via M9-branes, and show that  the celebrated AGT correspondence in~\cite{AGT, irr}, and its generalizations -- which essentially relate, among other things, some equivariant cohomology of the moduli space of certain (``ramified'') $G$-instantons to the integrable representations of the \emph{Langlands dual} of certain affine $\cal W$-algebras -- can likewise be derived from the principle that the spacetime BPS spectra of \emph{string-dual} M-theory compactifications ought to be equivalent.

In Part III, we consider various limits of our setup in Part II, and connect our story to chiral fermions and integrable systems. Among other things, we derive the Nekrasov-Okounkov conjecture in~\cite{NO} -- which relates the topological string limit of the dual Nekrasov partition function for pure $G$ to the integrable representations of the \emph{Langlands dual} of an affine $G$-algebra -- and also demonstrate that the Nekrasov-Shatashvili  limit of the ``fully-ramified'' Nekrasov instanton partition function for pure $G$ is a simultaneous eigenfunction of the quantum Toda Hamiltonians associated with the \emph{Langlands dual} of an affine $G$-algebra. Via the case with matter, we also make contact with Hitchin systems and the ``ramified'' geometric Langlands correspondence for curves.

\newpage

\renewcommand{\thefootnote}{\arabic{footnote}}
\setcounter{footnote}{0}

\tableofcontents

\newsection{Introduction, Summary and Acknowledgements}

The correspondence between 4d gauge theories and 2d CFT's have long been observed in the physical and mathematical literature. In a  mathematical work~\cite{nakajima} that dates back as early as 1994, Nakajima showed that the middle-dimensional cohomology of the moduli space of $U(N)$-instantons on a resolved ALE space of  $A_{k-1}$-type can be related to the integrable representations of an affine $SU(k)$-algebra of level $N$. Physicists then attempted to seek a physical derivation of this beautiful 4d-2d relation; in particular,  Vafa and Witten quickly realized~\cite{vafa-witten} that one needs string theory to ``see'' Nakajima's result, whence in 1995, Vafa presented evidence~\cite{vafa on nakajima proof} that the correct framework to derive Nakajima's result is in the context of heterotic-type IIA string duality, following which in 1996, Harvey and Moore argued~\cite{Harvey-Moore on nakajima proof} that it is the equivalence of the algebra of BPS states in heterotic/IIA dual pairs which is relevant. That said, a direct physical derivation  -- in the sense of an equivalence between generating functions of the middle-dimensional cohomology of the moduli space of $U(N)$-instantons  on a resolved ALE space of  $A_{k-1}$-type and the integrable representations of an affine $SU(k)$-algebra of level $N$ -- was still lacking.  

Six years later in 2002, a similar development took place in the physical literature, where it was conjectured by Nekrasov in~\cite{NN} that the equivariant cohomology of the moduli space of $SU(N)$-instantons on a (resolved) ALE space of $ADE$-type should be related to $ADE$ WZW models on the SW curve underlying the associated 4d ${\cal N} = 2$ pure  $SU(N)$ theory. Shortly thereafter in 2003, the seminal result in~\cite{NN} -- regarding the exact evaluation of the SW prepotential via the Nekrasov partition function -- was made mathematically rigorously by Nekrasov and Okounkov in~\cite{NO}, where a more refined and far-reaching 4d-2d conjecture was also proposed; they asserted that the topological string limit of the dual Nekrasov partition function of a 4d ${\cal N} = 2$ pure $G$ theory should be related to the integrable representations of the \emph{Langlands dual} of an affine $G$-algebra, where $G$ is\emph{ any }Lie group.   

Then in 2007, Dijkgraaf, Hollands, Sulkowski and Vafa finally gave a direct physical derivation in~\cite{Vafa et al} of Nakajima's result; the aforementioned generating functions were partition functions of BPS states in two different but dual frames in string/M-theory which could then be equated to each other. Right at about the same time, in an attempt to generalize the geometric Langlands duality for Lie groups~\cite{Mirkovic-Vilonen} to affine Kac-Moody groups, mathematicians Braverman and Finkelberg were also led to formulate a conjecture in~\cite{BF}, which asserts that the intersection cohomology of the moduli space of $G$-instantons on $\mathbb R^4 / \mathbb Z_k$ should be related to the integrable representations of the \emph{Langlands dual} of an affine $G$-algebra. This conjecture was henceforth known as a geometric Langlands duality for \emph{surfaces}, since it involves $G$-bundles over a complex surface as opposed to a complex curve (which is the underlying ingredient in Beilinson and Drinfeld's formulation in \cite{BD} of a geometric Langlands duality for Lie groups).   Witten, in a series of lectures delivered at the IAS in 2008~\cite{Lectures by Witten}, argued, somewhat abstractly, that a geometric Langlands duality for surfaces can be understood as an invariance of the BPS spectrum of the mysterious 6d ${\cal N} = (2,0)$ SCFT under different compactifications down to 5d.\footnote{A written account of these lectures can also be found in~\cite{GL from 6d}.}  Combining the insights from Witten's lectures and the work of  Dijkgraaf, Hollands, Sulkowski and Vafa, the author was able to give a concrete M-theoretic derivation in~\cite{ATMP} of this geometric Langlands duality for surfaces; he showed that  for the $A$--$D$ groups, the duality can be derived from the principle that the spacetime BPS spectra of string-dual M-theory compactifications ought to be equivalent. 

Next came a mini revolution in 2009, when Alday, Gaiotto and Tachikawa, motivated by the insights from Gaiotto's work in~\cite{N=2}, verified in~\cite{AGT} that the Nekrasov instanton partition function of a 4d ${\cal N} = 2$ conformal $SU(2)$ quiver theory is equivalent to a conformal block of a 2d CFT with ${\cal W}_2$-algebra symmetry that is  Liouville theory.  This celebrated 4d-2d correspondence, better known since as the AGT correspondence, was anticipated to hold for other gauge theories as well. In particular, it was soon proposed and checked to some extent in~\cite{irr}, that the correspondence should hold for 4d ${\cal N} = 2$  asymptotically-free $SU(2)$ theories;  it was also proposed and checked to some extent in~\cite{AGT-matter}, that the correspondence should hold for  a 4d ${\cal N} = 2$ conformal $SU(N)$ quiver theory whereby the corresponding 2d CFT is an $A_{N-1}$ conformal Toda field theory which has ${\cal W}_N$-algebra symmetry; and last but not least, the correspondence for a 4d ${\cal N} = 2$ pure \emph{arbitrary} $G$ theory was also proposed and checked to hold up to the first instanton level in~\cite{abcdefg}.  The basis for the AGT correspondence for $SU(N)$ -- as first pointed out by Alday and Tachikawa in~\cite{Alday-Tachikawa} -- is a conjectured relation between the equivariant cohomology of the moduli space of $SU(N)$-instantons and the integrable representations of an affine ${\cal W}_N$-algebra. This conjectured relation was first proved mathematically for finite ${\cal W}_N$-algebras in~\cite{AGT-math}, and later proved mathematically for affine ${\cal W}_N$-algebras in~\cite{Vasserot, Maulik-Okounkov}. An original effort to furnish a fundamental physical derivation of the AGT correspondence from the viewpoint of 6d ${\cal N} = (2,0)$ SCFT was also undertaken by Yagi in~\cite{junya, junya 2}, although certain assumptions made in \emph{loc.cit.} require further investigation. Also, in the Nekrasov-Shatashvili limit, the AGT correspondence in~\cite{AGT} has also been derived via a certain bispectral duality between two integrable systems in~\cite{Koroteev}.

``Ramified''  generalizations of the AGT correspondence for pure $SU(N)$ to include surface operators were also proposed and checked to some extent in~\cite{AGT-ram, Kanno-Tachikawa}, although the correspondence for pure arbitrary $G$ with a full surface operator had already been proved mathematically in 2004 by Braverman in~\cite{J-function} (as made known to physicists in~\cite{Alday-Tachikawa}). Nonetheless, based on peripheral physical evidence, it was later conjectured by Chacaltana, Distler and Tachikawa in~\cite{TD}, that the AGT correspondence should hold for pure arbitrary $G$ with not just a full but with \emph{any} surface operator, where on the 2d CFT side, one has a most general affine $\cal W$-algebra.

The AGT correspondence for $SU(N)$ was further proposed in~\cite{AGT-ALE, AGT-ALE-1} to hold on $\mathbb R^4 / \mathbb Z_m$, where on the 2d CFT side, one has an $m$-th para-${\cal W}_N$-algebra; this proposal was checked to be true for $N= m = 2$ in~\cite{AGT-ALE, AGT-ALE-checked}. Ideas for this proposal were based on physical evidence presented in~\cite{NT}, where it was also conjectured that the AGT correspondence on $\mathbb R^4 / \mathbb Z_m$  should hold not just for $SU(N)$  but for \emph{any} $ADE$ group, where on the 2d CFT side,  one has an $m$-th para-${\cal W}$-algebra derived from the affine $ADE$-algebra. 

As 2d CFT's can often be associated with integrable systems, the AGT correspondence also implies certain relations between the Nekrasov instanton partition function and integrable systems. An example which actually predates the AGT correspondence would be Nekrasov's conjecture in~\cite{NN}, which asserts that the Nekrasov instanton partition function should be related to a tau-function of Toda lattice hierarchy. A more recent example that arose from the AGT correspondence would be Alday and Tachikawa's conjecture in~\cite{Alday-Tachikawa}, which asserts that the ``fully-ramified'' Nekrasov instanton partition function should be related to Hitchin's integrable system.

Our main aim is to furnish in a pedagogical manner, a fundamental M-theoretic derivation of all the above 4d-2d relations, and more.  Let us now give a brief plan and summary of the paper.

\bigskip\noindent{\it A Brief Plan and Summary of the Paper}

In $\S$2, we will employ a chain of string dualities to physically relate distinct compactifications of M-theory down to six-dimensions, where around the five compactified directions, there can be (i) coincident M5-branes; (ii) coincident M5-branes and an orientifold fiveplane; (iii) coincident M5-branes, an orientifold fiveplane, and a worldvolume defect of the kind studied in~\cite{TD} which can be realized in M-theory by an orbifold in the transverse directions.  The relation under string dualities between multi-Taub-NUT space and D6-branes and NS5-branes, and  the relation under string dualities between Sen's four-manifold and D6-branes/O6-planes and NS5-branes/ON5-planes, play a central role in our arguments; they are described in detail in Appendix A.

In $\S$3, we will show that the Braverman-Finkelberg (BF) conjecture~\cite{BF} of a geometric Langlands duality for surfaces, can, for the $A$, $B$, $C$, $D$ and $G$ groups, be derived, purely physically, from the principle that the spacetime BPS spectra  of the \emph{string-dual} M-theory compactifications obtained in $\S$2 ought to be equivalent. As an offshoot, we would be able to also demonstrate (i) an identity of the dimension of the intersection cohomology of the moduli space of $A$-, $D$- and $G$-instantons on singular ALE spaces; (ii) a Langlands duality of the dimension of the intersection cohomology of the moduli space of $B$- and $C$-instantons on singular ALE spaces. Likewise for the $E$ and $F$ groups, we will show that the Langlands duality can be derived, purely physically, from the principle that the spacetime BPS spectra of \emph{string-dual} compactifications of M-theory and type IIB theory on singular K3 manifolds ought to be equivalent. Furthermore, for the simply-laced $A$ and $D$ groups, we would be able to also derive (1) a McKay-type correspondence of the intersection cohomologies of the moduli spaces of instantons, which serves as a generalization of Proudfoot's conjecture in~\cite{Proudfoot} to \emph{completely blown-down} ALE spaces; (2) a level-rank duality of chiral WZW models; (3) a  4d-2d Nakajima-type duality involving \emph{completely blown-down} ALE spaces. In particular, for the $A$ groups, (2), (3), and our main derivation of a geometric Langlands duality for surfaces, physically realize the commutative diagram in~\cite[$\S$1]{branching}; and for the $D$ groups, (1), (2), (3), and our main derivation of a geometric Langlands duality for surfaces, physically realize a \emph{$D$-type ALE space generalization} thereof. 

In $\S$4, we will derive a non-singular and quasi-singular generalization of the geometric Langlands duality for surfaces for the $A$ and $B$ groups. In turn, this would allow us to make contact with and generalize a closely-related field-theoretic result obtained earlier by Witten~\cite{Lectures by Witten}, and reproduce, purely physically, Nakajima's celebrated result in~\cite{nakajima}. Via the string-dual M-theory compactifications with worldvolume defects obtained in $\S$2,  we will also derive a ``ramified'' version of the geometric Langlands duality for surfaces for the $A$, $B$, $C$, $D$ and $G$ groups.

In $\S$5, to the setup in $\S$3, we will introduce Omega-deformation via the fluxbrane background studied in~\cite{susanne, orlando}, add half-BPS boundary defects realized by M9-branes~\cite{M9-branes}, and go on to show that the pure AGT correspondence for the $A$, $B$, $C$, $D$ and $G$ groups, can likewise be derived from the principle that the spacetime BPS spectra  of \emph{string-dual} M-theory compactifications ought to be equivalent. Our derivation physically reproduces the mathematical conjecture by Braverman et al. in~\cite{AGT-math}, that the Nekrasov instanton  partition function for pure $G$ is given by the norm of a coherent state in the Verma module over the \emph{Langlands dual} affine $\cal W$-algebra. Furthermore, the underlying Seiberg-Witten curve -- interpreted as an  $N$- or $2N$-fold cover of the two-punctured Gaiotto curve $\cal C$~\cite{N=2, irr} -- also arises naturally in our picture. A crucial ingredient in our derivation is the realization by a gauged WZW model of affine $\cal W$-algebras obtained from a quantum Drinfeld-Sokolov reduction procedure; this realization is described in detail in Appendix B. 

In $\S$6, we will first add worldvolume defects to our setup in $\S$3, and derive a ``ramified'' generalization of the pure AGT correspondence for the $A$, $B$, $C$, $D$ and $G$ groups. Our derivation reproduces the conjecture by  Chacaltana-Distler-Tachikawa in~\cite{TD}, that the ``ramified'' Nekrasov instanton  partition function for pure $G$ is given by the norm of a coherent state in the Verma module over the \emph{Langlands dual} affine $\cal W$-algebra associated with an \emph{arbitrary} embedding of $\frak {su}(2)$ in the underlying Lie algebra. In anticipation of a connection to integrable systems, we then specialize our formulas to the case of a full worldvolume defect. In so doing, we would be able to reproduce exactly the mathematical result by Braverman in~\cite{J-function}, that relates the ``fully-ramified'' Nekrasov instanton partition function for pure $G$ to the norm of a coherent state in the Verma module over the \emph{Langlands dual} of an affine $G$-algebra. Second, based on our setup in $\S$4.1 which underlies our earlier derivation of a non-singular generalization of the geometric Langlands duality for surfaces, we will derive a smooth $A$-type ALE generalization of the pure AGT correspondence for the  $A$, $B$, $C$, $D$ and $G$ groups. Our derivation reproduces \emph{and }generalizes to \emph{nonsimply-laced} gauge groups the conjecture by Nishioka-Tachikawa in~\cite{NT}, that the Nekrasov instanton  partition function for pure simply-laced $\cal G$ on an $A_{m-1}$-type ALE space is given by the norm of a coherent state in a Verma module over the sum of a parafermionic coset affine algebra ${\rm RCFT}[A_{m-1}, \cal G]$ and the $m$-th para-$\cal W$-algebra derived from the affine $\cal G$-algebra. In particular, our derivation furnishes us with a concrete definition of ${\rm RCFT}[A_{m-1}, G]$ even when $G \neq A$ -- see eqns.~(\ref{coset CFT-AGT-AB-ALE-paratoda})--(\ref{coset-para-AB}) and eqns.~(\ref{coset CFT-AGT-CDG-ALE-paratoda})--(\ref{coset-para-CDG}). Last but not least, via building blocks defined by M-theory compactifications with M9-brane boundaries that are in one-to-one correspondence with the three-punctured sphere and cylinder of Gaiotto's construction in~\cite{N=2},  we will derive the AGT correspondence with matter. For brevity, we will consider just conformal linear and necklace quiver theories with $n$ $SU(N)$ gauge groups, although our arguments can be straightforwardly generalized to other Gaiotto-type theories as well. Once again, the underlying Seiberg-Witten curve -- this time interpreted as an $N$-fold cover of the generically multi-punctured Gaiotto curve ${\cal C}_{\rm eff}$ that is a sphere and a torus, respectively~\cite{N=2} -- arises naturally in our picture.

And finally in $\S$7, via our results in $\S$5 and $\S$6, we will make contact with chiral fermions, integrable systems, and the ``ramified'' geometric Langlands correspondence for curves. First, by considering the topological string limit in our derivation of the AGT correspondence for a conformal necklace quiver with $n$ $SU(N)$ gauge groups, we will reproduce \emph{and} generalize a purely field-theoretic result by Nekrasov-Okounkov in~\cite{NO}, that relates the corresponding Nekrasov instanton partition function of the ${\cal N} = 2^\ast$ $SU(N)$ theory to the theory of $N$ chiral fermions on a torus. Second,  by considering the topological string limit in our derivation of the pure AGT correspondence for $G$, we will reproduce the conjecture by Nekrasov-Okounkov in~\cite{NO}, which implies that the corresponding Nekrasov instanton partition function for pure $G$ is equal to the norm of a coherent state in the integrable highest weight module over the \emph{Langlands dual} of an affine $G$-algebra of level 1. Moreover, if $G = SU(N)$, we find that the corresponding Nekrasov instanton partition function for pure $SU(N)$ is a tau-function of Toda lattice hierarchy; this also coincides with Nekrasov's conjecture in~\cite{NN}. Third, by considering the Nekrasov-Shatashvilli limit in our derivation of the ``fully-ramified'' pure AGT correspondence for $G$,  we will show that the corresponding ``fully-ramified'' Nekrasov instanton partition function for pure $G$ is a simultaneous eigenfunction of the quantum Toda Hamiltonians associated with the \emph{Langlands dual} of an affine $G$-algebra.  And last, guided by the relation between the elliptic Calogero-Moser system and the ``tamely-ramified'' Hitchin system on a single-punctured torus, we will show that  in the Nekrasov-Shatashvili limit, the corresponding ``fully-ramified'' Nekrasov instanton partition function of a conformal linear and necklace quiver theory of $n$ $SU(N)$ gauge groups is also a $\cal D$-module in the ``tamely-ramified'' geometric Langlands correspondence for $SU(N)$ at genus zero and one, respectively. In turn, this confirms the conjecture by Alday-Tachikawa in~\cite{Alday-Tachikawa}, that the aforementioned Nekrasov instanton partition function is a simultaneous eigenfunction of the quantum Hitchin Hamiltonians for $SU(N)$.

\bigskip\noindent{\it Shorter Routes Through This Paper}

As indicated in the contents page, this paper can actually be broken up into four parts. Part I, or $\S$2--$\S$4, discusses the geometric Langlands duality for surfaces and its various generalizations. Part II, or $\S$5--$\S$6, discusses the AGT correspondence and its various generalizations. Part III, or $\S$7, discusses the relation of the AGT correspondence to chiral fermions, integrable systems and the ``ramified'' geometric Langlands correspondence for curves. Part IV, or the Appendix, contains materials in support of our discussions in $\S$2 and $\S$5.  

Readers who are interested in the physical derivation of a geometric Langlands duality for surfaces, should read $\S$2.1--$\S$2.2 and $\S$3.1--$\S$3.3.  Readers who are interested in the physical derivation of a non-singular or quasi-singular generalization of the geometric Langlands duality for surfaces, should read $\S$2.1, $\S$3.1, and $\S$4.1 or $\S$4.2, respectively. Readers who are interested in the physical derivation of the ``ramified'' geometric Langlands duality for surfaces, should read $\S$2.3, $\S$3.1--$\S$3.2, and $\S$4.3. Readers who are interested in the physical derivation of (i) a McKay-type correspondence of the intersection cohomologies of the moduli spaces of instantons, (ii) a level-rank duality of chiral WZW models, and (iii) a 4d-2d Nakajima-type duality involving singular ALE spaces, should read $\S$2.1--$\S$2.2, $\S$3.1--$\S$3.2, and $\S$3.4.

Readers who are interested in the physical derivation of the pure AGT correspondence, should read $\S$2.1--$\S$2.2, $\S$3.1--$\S$3.2, and $\S$5.1--$\S$5.3. Readers who are interested in the physical derivation of a ``ramified'' generalization of the pure AGT correspondence, should read $\S$2.3, $\S$4.3, $\S$5.1--$\S$5.3, and $\S$6.1. Readers who are interested in the physical derivation of an $A$-type ALE generalization of the pure AGT correspondence, should read $\S$2.1--$\S$2.2, $\S$3.1--$\S$3.2, $\S$4.1, and $\S$6.2.  Readers who are interested in the physical derivation of the AGT correspondence with matter, should read $\S$2.1, $\S$3.1, $\S$5.1--$\S$5.2, and $\S$6.3. 

 Readers who are interested in the relation of the AGT correspondence to chiral fermions, should read  $\S$2.1, $\S$3.1, $\S$5.1--$\S$5.2, $\S$6.3, and $\S$7.1.  Readers who are interested in the relation of the AGT correspondence to the Nekrasov-Okounkov conjecture in~\cite{NO} and the tau-function of Toda lattice hierarchy, should read $\S$2.1--$\S$2.2, $\S$3.1--$\S$3.2, $\S$5.1--$\S$5.3, and $\S$7.2. Readers who are interested in the relation of the AGT correspondence to quantum affine Toda systems, should read $\S$2.1--$\S$2.2, $\S$3.1--$\S$3.2, $\S$5.1--$\S$5.3, $\S$6.1, and $\S$7.3. Readers who are interested in the relation of the AGT correspondence to the ``ramified'' geometric Langlands correspondence for curves and the Alday-Tachikawa conjecture in~\cite{Alday-Tachikawa}, should read $\S$2.1, $\S$3.1, $\S$5.1--$\S$5.2, $\S$6.1, $\S$6.3, $\S$7.3, and $\S$7.4.

\bigskip\noindent{\it Acknowledgements} 

I would like to thank V.~Balaji, P.~Bouwknegt, A.~Braverman, S.~Cherkis, M.~Douglas, K.~Maruyoshi, N.~Nekrasov, S.~Reffert, A.~Sen and S.~Wu, for illuminating exchanges. 

I would especially like to thank H.~Nakajima for his patient explanation of various related mathematical works and concepts; J.-J.~Ma and C.-B.~Zhu for their generous expertise on nilpotent orbits; D.~Orlando for his assistance with an important formula; and Y.~Tachikawa for his ever prompt and detailed reply to all my queries on the AGT correspondence, and more. 

Last but not least, I would like to thank O.~Foda for spotting some imprecisions in the previous version of this paper.

This work is supported in part by the NUS Startup Grant.

\newpage

\part{\Large A Geometric Langlands Duality for Surfaces}
 
\newsection{Dual Compactifications of M-theory with M5-Branes, OM5-Planes and 4d Worldvolume Defects}

\vspace{-0.0cm}\newsubsection{Dual Compactifications of M-theory with M5-Branes}

Consider a six-dimensional compactification of M-theory on the five-manifold ${{\bf S}^1_n} \times \mathbb R^4 / {\mathbb Z_k}$. Here, $\mathbb R^4/ \mathbb Z_k$ is a singular ALE manifold of type $A_{k-1}$; ${{\bf S}^1_n}$ is a circle of radius $R_{s}$; and the subscript `$n$'  means that we perform, in the sense of~\cite{vafa}, a  ``$\mathbb Z_n$-twist'' of the theory as we go around the circle -- that is, we evoke a $\mathbb Z_n$-outer-automorphism of $\mathbb R^4/ \mathbb Z_k$ (and of the geometrically-trivial six-dimensional spacetime) as we go around the circle and identify the circle under an order $n$ translation. Wrap on this five-manifold a stack of $N$ coincident M5-branes, such that its worldvolume, in Euclidean signature,\footnote{The six-dimensional $(2,0)$ theory that lives on the worldvolume of the stack of coincident M5-branes is a unitary, physically sensible quantum field theory with positive energy. Thus, it is possible to formulate the ``same'' theory on a space of Lorentzian or Euclidean signature via analytic continuation. For our purpose, it will be more useful to adopt a Euclidean signature.} will be given by $\mathbb R_t \times {{\bf S}^1_n} \times \mathbb R^4 / {\mathbb Z_k}$, where $\mathbb R_t$ is the ``time'' direction. In other words, let us consider the following M-theory configuration:
\be
\textrm{M-theory}: \quad \mathbb R^{5}  \times  \underbrace{\mathbb R_t \times {{\bf S}^1_n} \times \mathbb R^4 / \mathbb Z_k}_{\textrm{$N$ M5-branes}}.
\label{M-theory 1}
\ee
Taking the ``eleventh circle'' to be one of the decompactified directions along the $\mathbb R^{5}$ subspace, we see that (\ref{M-theory 1}) actually corresponds to the following ten-dimensional type IIA background with $N$ coincident NS5-branes wrapping $\mathbb R_t \times {{\bf S}^1_n} \times \mathbb R^4 / \mathbb Z_k$, where the IIA string coupling $g^A_s$ and string length $l_s$ are such that $g^A_sl_s \to \infty$:
\be
\textrm{IIA}: \quad \mathbb R^{4}  \times  \underbrace{\mathbb R_t \times  {{\bf S}^1_n}  \times \mathbb R^4 / \mathbb Z_k}_{\textrm{$N$ NS5-branes}}.
\label{IIA 2}
\ee

 Let us now T-dualize along the $\mathbb R_t$ direction of the worldvolume of the stack of NS5-branes. From $\S$A.3, we learn that T-dualizing along any one of the worldvolume directions of an NS5-brane (where the background solution is trivial), will bring us back to an NS5-brane. This means that we will arrive at the following type IIB configuration with IIB string coupling $g^B_s \sim 1$ (since $g^B_s = g^A_s l_s /r$, and $r \to \infty$, where $r$ is the radius of ${\mathbb R}_t$):
\be
\textrm{IIB}: \quad \mathbb R^{4}  \times  \underbrace{{\mathbb S}^1_{t;  r \to 0} \times {{\bf S}^1_n} \times \mathbb R^4 / \mathbb Z_k}_{\textrm{$N$ NS5-branes}}.
\label{IIB 3}
\ee

Next, let us T-dualize along a direction that is transverse to the stack of NS5-branes. As explained in $\S$A.3, since the NS5-branes are coincident, one will end up having a multi-Taub-NUT manifold $TN_N$ with an $A_{N-1}$ singularity at the origin, with no NS5-branes. Thus, as one can view one of the $\mathbb R$'s in $\mathbb R^{4}$ to be a circle of infinite radius, in doing a T-duality along this circle, we will arrive at the following type IIA background:
\be
\textrm{IIA}: \quad  TN_N^{R\to 0}  \times  {\mathbb S}^1_{t; r\to 0} \times {{\bf S}^1_n} \times {\mathbb R^4 / \mathbb Z_k},
\label{IIA 4}
\ee
where $TN_N^{R \to 0}$ is a multi-Taub-NUT manifold with an $A_{N-1}$ singularity at the origin and asymptotic radius $R \to 0$. (As explained in $\S$A.3, $R \to 0$ because we are T-dualizing along a trivially-fibered circle of infinite radius.)  At this stage, one also finds that $g^A_s \to 0$. Consequently, this can be interpreted as the following M-theory background with a very small ``eleventh circle'' ${S}^1_{11}$:
\be
\textrm{M-theory}: \quad TN_N^{R\to 0} \times  {\mathbb S}^1_{t; r \to 0} \times {{\bf S}^1_n} \times S^1_{11; r \to 0} \times {\mathbb R^4 / \mathbb Z_k}.
\label{M-theory 5}
\ee

From $\S$A.1, we learn that the singular ALE space $\mathbb R^4 / \mathbb Z_k$ is simply $TN_k$ with an $A_{k-1}$ singularity at the origin whose asymptotic radius $R \to \infty$. Note also from $\S$A.2 that M-theory on such a space is equivalent upon compactification along its circle fiber to type IIA string theory with $k$ coincident D6-branes filling out the directions transverse to the space. In other words, starting from (\ref{M-theory 5}), one can descend back to the following type IIA background:\footnote{In the following background, there is a $\mathbb Z_n$-automorphism on the D6-branes  (that descends from the $\mathbb Z_n$-automorphism on the $\mathbb R^4 / \mathbb Z_k$ in (\ref{M-theory 5}) which underlies the D6-branes solution) that permutes them $n$ times as one goes around the ${\bf S}^1_n$ circle. This permutation does not alter their description as a stack of $k$ coincident D6-branes, and is also consistent with the $\mathbb Z_n$-automorphism of their worldvolume which arises due to the $\mathbb Z_n$-automorphism of $\mathbb R^5 \times \mathbb R_t$ in (\ref{M-theory 1}).}
\be
\textrm{IIA}: \quad \underbrace{ TN_N^{R\to 0}  \times {\mathbb S}^1_{t; r \to 0} \times   {{\bf S}^1_n} \times {S}^1_{11, r \to 0}}_{\textrm{$k$ D6-branes}}  \times  {\mathbb R^3}.
\label{IIA 5}
\ee
 Note however, that we now have a type IIA theory that is strongly-coupled, since the effective type IIA string coupling from a compactification along the circle fiber is proportional to the asymptotic radius which is large. (See $\S$A.2, again.)

Let us proceed to do a T-duality along ${S}^1_{11}$, which will serve to decompactify the circle, as well as convert the D6-branes to D5-branes in a type IIB theory. By coupling this step with a type IIB S-duality that will convert the D5-branes into NS5-branes, we will arrive at the following type IIB configuration at weak-coupling:
\be
\textrm{IIB}: \quad \underbrace{ TN_N^{R\to 0}  \times {\mathbb S}^1_{t; r \to 0} \times {{\bf S}^1_n}}_{\textrm{$k$ NS5-branes}}  \times  {\mathbb R^{4}}.
\label{IIB 6}
\ee

Finally, let us do a T-duality along $\mathbb S^1_{t; r\to 0}$, which will bring us back to a type IIA background with NS5-branes and $g^A_s \to \infty$.\footnote{Recall the T-duality relation $g^A_s = g^B_s l_s / r$. Therefore, because $g^B_sl_s$, though small, is still nonzero, having $r \to 0$ would mean that $g^A_s \to \infty$.}  Lifting this IIA background back up to M-theory, we will arrive at the following configuration:
\be
\textrm{M-theory}: \quad \underbrace{TN_N^{R\to 0}  \times  {{\bf S}^1_n} \times \mathbb R_t}_{\textrm{$k$ M5-branes}}  \times  {\mathbb R^{5}},
\label{M-theory 7}
\ee
where there is a nontrivial $\mathbb Z_n$-outer-automorphism of $TN_N^{R\to 0}$ as we go around the ${\bf S}^1_n$ circle.

 Hence, from the chain of dualities described above, we conclude that the six-dimensional M-theory compactifications with $N$ and $k$ \emph{coincident} M5-branes wrapping the five compactified directions along the manifolds ${{\bf S}^1_n} \times \mathbb R^4 / \mathbb Z_k$ and ${\bf S}^1_n \times TN_N^{R\to 0}$ as shown in (\ref{M-theory 1}) and (\ref{M-theory 7}), respectively, ought to be \emph{physically dual} to each other.

\newsubsection{Dual Compactifications of M-theory with M5-Branes and OM5-Planes}

To the stack of coincident M5-branes in (\ref{M-theory 1}), one can add a fiveplane that is intrinsic to M-theory known as the OM5-plane~\cite{hanany}. Then, we would have the following M-theory configuration:  
\be
\textrm{M-theory}: \quad \mathbb R^{5}  \times  \underbrace{\mathbb R_t \times {{\bf S}^1_n} \times  \mathbb R^4 / \mathbb Z_k}_{\textrm{$N$ M5-branes/OM5-plane}},
\label{OM-theory 1}
\ee
where as before, we evoke a $\mathbb Z_n$-outer-automorphism of $\mathbb R^4 / \mathbb Z_k$ (and of the geometrically-trivial $\mathbb R^5 \times \mathbb R_t$ spacetime) as we go around the ${\bf S}^1_n$ circle and identify the circle under an order $n$ translation. 

Unlike the usual Op-planes, the OM5-plane has no (discrete torsion) variants and is thus unique. Its presence will serve to identify opposite points in the spatial directions transverse to its $\mathbb R_t \times {{\bf S}^1_n} \times \mathbb R^4 / \mathbb Z_k$ worldvolume. Consequently, the gauge symmetries associated with the stack of M5-branes will be modified, much in the same way how Op-planes modify the effective worldvolume gauge symmetry on a stack of Dp-branes by identifying open-string states with exchanged Chan-Paton indices that connect between the Dp-branes. An essential fact to note at this point is that the OM5-plane can be interpreted as a $\it{unique}$ ON$5^-_A$-plane in type IIA string theory under a compactification along an ``eleventh circle'' that is transverse to its worldvolume~\cite{hanany}; here, the `-' superscript just indicates that its presence will result in an orthogonal gauge symmetry in the type IIA theory, while the `N' just denotes that it can only be associated with NS5-branes. This means that the presence of an OM5-plane will serve to convert an existing gauge symmetry (in a certain regime) of the worldvolume theory on the stack of coincident M5-branes to that of an \emph{orthogonal} (and not symplectic) type. This fact will be important later.

Let us now take the ``eleventh circle'' to be one of the decompactified directions along the $\mathbb R^{5}$ subspace. We then see that (\ref{OM-theory 1}) actually corresponds to the following ten-dimensional type IIA background with $N$ coincident NS5-branes wrapping $ \mathbb R_t \times  {{\bf S}^1_n} \times \mathbb R^4 / \mathbb Z_k$ on top of an ON$5^-_A$-plane, where $g^A_sl_s \to \infty$:
\be
\textrm{IIA}: \quad \mathbb R^{4}  \times  \underbrace{\mathbb R_t \times  {{\bf S}^1_n} \times \mathbb R^4 / \mathbb Z_k}_{\textrm{$N$ NS5-branes/ON$5^-_A$-plane}}.
\label{OIIA 2}
\ee

Let us next T-dualize along the $\mathbb R_t$ direction of the NS5-branes/ON$5^-_A$-plane configuration. From $\S$A.6, we learn that T-dualizing along any one of the worldvolume directions of an NS5-brane/ON$5^-$-plane configuration (where the background solution is trivial), will bring us back to an NS5-brane/ON$5^-$-plane configuration. This means that we will arrive at the following type IIB configuration where $g^B_s \sim 1$:
\be
\textrm{IIB}: \quad \mathbb R^{4}  \times  \underbrace{{\mathbb S}^1_{t;  r \to 0} \times  {{\bf S}^1_n} \times \mathbb R^4 / \mathbb Z_k}_{\textrm{$N$ NS5-branes/ON$5^-_B$-plane}}.
\label{OIIB 3}
\ee
Here, the ON$5^-_B$-plane is the T-dual counterpart of the ON$5^-_A$-plane. It is also the S-dual counterpart of the usual O$5^-$-plane in type IIB theory~\cite{hanany}.

Now, let us T-dualize along a direction that is transverse to the stack of NS5-branes/ON$5^-_B$-plane. As explained in $\S$A.6, one will end up with Sen's four-manifold $SN_N$ with a $D_N$ singularity at the origin~\cite{sen} (which one can roughly regard as $TN_N$ with a $\mathbb Z_2$-identification of its ${\bf S}^1$ fiber and $\mathbb R^3$ base), with no NS5-branes and no ON$5^-_B$-plane. Thus, as one can view one of the $\mathbb R$'s in $\mathbb R^{4}$ to be a circle of infinite radius, in doing a T-duality along this circle, we will arrive at the following type IIA background:
\be
\textrm{IIA}: \quad  SN_N^{R\to 0}  \times {\mathbb S}^1_{t; r\to 0}  \times  {{\bf S}^1_n} \times  {\mathbb R^4 / \mathbb Z_k},
\label{0IIA 4}
\ee
where $SN_N^{R \to 0}$ is Sen's four-manifold with a $D_{N}$ singularity at the origin and asymptotic radius $R \to 0$. (As explained in $\S$A.6, $R \to 0$ because we are T-dualizing along a trivially-fibered circle of infinite radius.) This is consistent with the fact that a T-duality along a direction transverse to the ON$5^-_B$-plane gives rise to a solution that can be identified with a unique OM6-plane in M-theory~\cite{hanany}, which, in turn, implies the $\mathbb Z_2$-symmetry that is inherent in Sen's four-manifold~\cite{sen}. At this stage, one also finds that $g^A_s \to 0$. In other words, (\ref{0IIA 4}) can also be interpreted as the following M-theory background with a very small ``eleventh circle'' ${S}^1_{11}$:
\be
\textrm{M-theory}: \quad  SN_N^{R\to 0} \times {\mathbb S}^1_{t; r \to 0} \times {{\bf S}^1_n} \times S^1_{11; r \to 0} \times {\mathbb R^4 / \mathbb Z_k}.
\label{OM-theory 5}
\ee

From $\S$A.1, we learn that the singular ALE space $\mathbb R^4 / \mathbb Z_k$ is simply $TN_k$ with an $A_{k-1}$ singularity at the origin whose asymptotic radius $R \to \infty$. Also from $\S$A.2, we learn that M-theory on such a space is equivalent upon compactification along its circle fiber to type IIA string theory with $k$ coincident D6-branes filling out the directions transverse to this space. In other words, starting from (\ref{OM-theory 5}), one can descend back to the following type IIA background:\footnote{In the following background, there is a $\mathbb Z_n$-automorphism on the D6-branes  (that descends from the $\mathbb Z_n$-automorphism on the $\mathbb R^4 / \mathbb Z_k$ in (\ref{OM-theory 5}) which underlies the D6-branes solution) that permutes them $n$ times as one goes around the ${\bf S}^1_n$ circle. This permutation does not alter their description as a stack of $k$ coincident D6-branes, and is also consistent with the $\mathbb Z_n$-automorphism of their worldvolume which arises due to the $\mathbb Z_n$-automorphism of $\mathbb R^5 \times \mathbb R_t$ in (\ref{OM-theory 1}).}
\be
\textrm{IIA}: \quad \underbrace{SN_N^{R\to 0}  \times {\mathbb S}^1_{t; r \to 0} \times {{\bf S}^1_n} \times  {S}^1_{11, r \to 0}}_{\textrm{$k$ D6-branes}}  \times  {\mathbb R^3}.
\label{OIIA 6}
\ee
Note however, that we now have a type IIA theory that is strongly-coupled, since the effective type IIA string coupling from a compactification along the circle fiber is proportional to the asymptotic radius which is large. (See $\S$A.2, again.)

Let us proceed to do a T-duality along ${S}^1_{11}$, which will serve to decompactify the circle, as well as convert the D6-branes to D5-branes in a type IIB theory. By coupling this step with a type IIB S-duality that will convert the D5-branes into NS5-branes, we will arrive at the following type IIB configuration at weak-coupling:
\be
\textrm{IIB}: \quad \underbrace{SN_N^{R\to 0} \times {\mathbb S}^1_{t; r \to 0} \times  {{\bf S}^1_n}}_{\textrm{$k$ NS5-branes}}  \times  {\mathbb R^{4}}.
\label{OIIB 7}
\ee

Finally, let us do a T-duality along $\mathbb S^1_{t; r \to 0}$, which will bring us back to a type IIA background with NS5-branes and $g^A_s \to \infty$.\footnote{Recall the T-duality relation $g^A_s = g^B_s l_s / r$. Therefore, because $g^B_sl_s$, though small, is still nonzero, having $r \to 0$ would mean that $g^A_s \to \infty$.} Lifting this IIA background back up to M-theory, we will arrive at the following configuration:
\be
\textrm{M-theory}: \quad \underbrace{SN_N^{R\to 0}  \times  {{\bf S}^1_n} \times \mathbb R_t}_{\textrm{$k$ M5-branes}}  \times  {\mathbb R^{5}},
\label{OM-theory 8}
\ee
where there is a nontrivial $\mathbb Z_n$-outer-automorphism of $SN_N^{R\to 0}$ as we go around the ${\bf S}^1_n$ circle.

 Thus, from the chain of dualities described above, we conclude that the six-dimensional M-theory compactifications with $N$ and $k$ \emph{coincident} M5-branes wrapping the five compactified directions along the manifolds ${{\bf S}^1_n} \times \mathbb R^4 / \mathbb Z_k$ (in the presence of an OM5-plane) and ${\bf S}^1_n \times SN_N^{R\to 0}$ as shown in (\ref{OM-theory 1}) and (\ref{OM-theory 8}), respectively, ought to be\emph{ physically dual} to each other:

\newsubsection{Dual Compactifications of M-theory with M5-Branes, OM5-Planes and 4d Worldvolume Defects}

To the stack of $N$ coincident M5-branes in (\ref{M-theory 1}), one can add a 4d \emph{worldvolume} defect of the kind studied in~\cite{TD} which can be realized in M-theory  by a  $\mathbb Z_M$-orbifold in the transverse directions (see~\cite[$\S$2.2]{Kanno-Tachikawa}). For definiteness, let us consider the following M-theory configuration:
\be
\begin{array}{l|cc|cc|cc|cc|ccccc}
&0&1&2&3&4&5&6&7&8&9&10 \\
\hline
\hbox{$N$ M5's} & - & - & -& -& - & - &&&&  \\
\hbox{defect} & - & - & -& -& \times & \times & \times & \times &  &  \\
\end{array} \label{table}
\ee
Here, the `$-$' sign in the column labeled by $j$ means that the particular brane or worldvolume defect extends along the $j^{\rm th}$ direction with coordinate $x_j$; similarly, the `$\times$' sign in the column labeled by $l$ means that the $\mathbb Z_M$-orbifold realizing the worldvolume defect extends along the $l^{\rm th}$ direction with coordinate $x_l$.  We take $x_0$ and $x_1$ to be the coordinates on $\mathbb R_t$ and ${\bf S}^1_n$, respectively, so that $(x_2, x_3, x_4, x_5)$ would be the coordinates on $ \mathbb R^4 / \mathbb Z_k \simeq \mathbb C^2 /  \mathbb Z_k$. Then, if $z = x_2 + i x_3$ and $w = x_4 + i x_5$, the singular ALE manifold $ {\mathbb C}^2 /  \mathbb Z_k$ can be viewed as a complex surface $\mathbb C^2$ whose coordinates $(z,w)$ are identified under the $\mathbb Z_k$-action $(z, w) \to (\zeta z, \zeta^{-1} w)$, where $\zeta = e^{2 \pi i / k}$. According to (\ref{table}), the 4d worldvolume defect then wraps $\mathbb R_t \times {\bf S}^1_n$ and the $z$-plane. Consequently, the presence of the 4d worldvolume defect (i) modifies the theory living on $\mathbb R_t \times {\bf S}^1_n$; (ii) introduces -- when observation scales are much larger than the radius of ${\bf S}^1_n$ --  a surface worldvolume defect  (which we will describe below) in the 4d ${\cal N} = 4$ SYM theory living on the ``constant-time'' hypersurface $ \mathbb C^2 /  \mathbb Z_k$, at $w =0$. Such a 4d worldvolume defect was first considered in~\cite{Alday-Tachikawa}.

\bigskip\noindent{\it Characterization of the 4d Worldvolume Defect by a Partition of $N$}

This 4d worldvolume defect can be labeled by a partition of $N$ when $n=1$, as follows. First, set $k = 1$ for ease of illustration. (The same arguments will apply for $k > 1$, except that one must further take into account the above-mentioned identification under the $\mathbb Z_k$-action.) As usual, freeze the center-of-mass degrees of freedom of the stack of $N$ coincident M5-branes; then, along the 2345-directions, we have an ${\cal N} = 4$, $G=SU(N)$ theory on $\mathbb C^2$ with a Gukov-Witten surface operator~\cite{GW} along the $z$-plane. 

Second, note that this surface operator introduces a singularity in the gauge field $A_\mu$: if $(r,\theta)$ are the polar coordinates of the transverse  $w$-plane in $\mathbb C^2$, i.e., $w=re^{i\theta}$, the gauge field diverges as
\be
A_\mu dx^\mu \sim {\rm diag}(\alpha_1, \alpha_2, \cdots, \alpha_N)~i {d \theta},
\label{singularity}
\ee
near the surface operator. By a gauge transformation, one can assume that $1 > \alpha_i \geq \alpha_{i+1} \geq 0$.

Third, note that the commutant of $\vec{\alpha} = i\, {\rm diag}(\alpha_1, \alpha_2, \cdots, \alpha_N)$ is a subgroup $\mathbb{L} \subset G$ which is called the Levi subgroup; in other words, the gauge group $G$ reduces to $\mathbb{L}$ along the surface defined by the $z$-plane. The structure of $\vec\alpha$ can take the general form
\be
\vec\alpha= i \, {\rm diag}(\underbrace{\alpha_{(1)},\ldots,\alpha_{(1)}}_{\text{$n_1$ times}},
\underbrace{\alpha_{(2)},\ldots,\alpha_{(2)}}_{\text{$n_2$ times}},
\ldots,
\underbrace{\alpha_{(M)},\ldots,\alpha_{(M)}}_{\text{$n_M$ times}} ),
\ee
where $n_{i} > n_{i+1}$.\footnote{One can also have $n_{i+1} > n_{i}$, but we will not consider such a situation in this paper.} This means that the Levi subgroup is
\be
\mathbb L=S[U(n_1)\times U(n_2)\times \cdots \times U(n_M)], 
\ee
where $N = n_1 + n_2 + \dots + n_M$. It is in this sense that the underlying 4d worldvolume defect can be characterized by the partition $[n_1, n_2, \dots, n_M]$ of $N$, and be called one of type $\mathbb L$. 

\bigskip\noindent{\it Reduction of Gauge Group and Parabolic Subgroups}

It will be useful for later to also discuss the connection between  (i) the reduction, along the surface, of the gauge group $G$ to its Levi subgroup $\mathbb L$, and (ii) parabolic subgroups of $G_{\mathbb C}$ (the complexification of $G$). 

To this end, let  $\mathfrak{p}$ be a subalgebra of $\frak g_\mathbb C$ (the Lie algebra of $G_\mathbb C$) spanned by elements $x$ satisfying 
\be
[\vec\alpha,x] = i\lambda x, \qquad \lambda \leq 0.
\label{para-define}
\ee
Then, $\frak{p}$ is called a parabolic subalgebra, and the corresponding subgroup ${\cal P}\subset G_\mathbb C$ is called a parabolic subgroup. 

Note that since $\mathbb L$ is the commutant of $\vec \alpha$, (\ref{para-define}) means that there ought to exist a correspondence between $\mathbb L$ and $\cal P$. For example, consider $G  = SU(4)$ and $\mathbb L=S[U(2)\times U(1)^2]$; according to our above discussion, $\mathbb L$ is associated to the partition $[n_I] = [2,1,1]$ and  $\vec\alpha=i \, {\rm diag}(\alpha_{(1)},\alpha_{(1)},\alpha_{(2)},\alpha_{(3)})$; in this case, the corresponding parabolic subgroup is ${\cal P} = {\cal P}_{[2,1,1]}$, and its elements take the form 
\be
\label{matrix}
\left(\begin{array}{cc|c|c}
* & * & 0 & 0 \\
* & * & 0 & 0 \\ 
\hline
* & * & * & 0 \\
\hline
* & * & * & *
\end{array}\right),
\ee
where the sign `$*$' denotes some complex number such that the determinant of the matrix is one. 

As a second example, consider $G = SU(4)$ and $\mathbb L= U(1)^3$; according to our above discussion, $\mathbb L$ is associated to the partition $[n_I] = [1,1,1,1]$ and $\vec\alpha=i \, {\rm diag}(\alpha_{(1)},\alpha_{(2)},\alpha_{(3)},\alpha_{(4)})$; in this case, the corresponding parabolic subgroup is ${\cal P}_{[1,1,1,1]}$, and its elements can be any complex semi-lower triangular $4 \times 4$ matrix of determinant one. In general, when the Levi subgroup is $\mathbb L=U(1)^{N-1}$, the corresponding parabolic subgroup ${\cal P}_{[1, \dots,1]}$ is just the Borel subgroup $\cal B$.

As a final example, consider $G = SU(N)$ and $\mathbb L = SU(N) = G$; according to our above discussion, $\mathbb L$ is associated to the partition $[n_I] = [N]$ and $\vec\alpha = 0$;\footnote {For the partition $[N]$, we have $\vec\alpha=i \, {\rm diag}(\alpha_{(1)},\alpha_{(1)},\dots,\alpha_{(1)})$. Since  $G= SU(N)$ is the group of \emph{traceless} unitary $N \times N$ matrices of determinant one, we must also have $\sum \alpha_{(1)} = 0$. Altogether, this means that $\vec\alpha = 0$.} in other words, there is \emph{no }defect. In this case,  the corresponding parabolic subgroup ${\cal P}_{[N]}$ would be spanned by all complex $N \times N$ matrices of determinant one, i.e., ${\cal P}_{[N]} = SL(N, \mathbb C) = G_\mathbb C$.

Note that one can also understand the above correspondence between $\mathbb L$ and ${\cal P}_{[n_I]}$ to be a consequence of the fact that  $G/ \mathbb L \simeq G_\mathbb C / {\cal P}_{[n_I]}$ as Riemannian manifolds. This isomorphism also means that we can describe the reduction of the gauge group along the surface in terms of parabolic subgroups: the $SU(N)$ gauge group is reduced along the surface by an amount $SU(N) / \mathbb L$, and from the preceding observations, this is the same as ${\cal P}_{[N]} / {\cal P}_{[n_I]}$.

\bigskip\noindent{\it Dual Compactifications with M5-Branes and 4d Worldvolume Defects}

Now consider the M-theory configuration given in (\ref{table}):
\be
\textrm{M-theory}: \quad \mathbb R^{5}  \times  \underbrace{\mathbb R_t \times {{\bf S}^1_n} \times \mathbb R^4 / \mathbb Z_k}_{\textrm{$N$ M5-branes with a 4d defect}}   \equiv  \ \  \mathbb R^{3}  \times \mathbb C_{w'}/ \mathbb Z_M \times   \underbrace{\mathbb R_t \times {{\bf S}^1_n} \times  \mathbb C_z / \mathbb Z_k \times \mathbb C_w / (\mathbb Z_k  \times \mathbb Z_M)}_{\textrm{$N$ M5-branes}},
\label{M-theory defect}
\ee
where the coordinates are $(x_{10}, x_9, x_8; w'; x_0; x_1; z; w)$, with $w' = x_6 + i x_7$. Here, (i) we evoke a $\mathbb Z_n$-outer-automorphism of the transverse ten-dimensional space as we go around the ${\bf S}^1_n$ circle and identify the circle under an order $n$ translation; (ii) the 4d worldvolume defect wraps $\mathbb R_t \times {{\bf S}^1_n}$ and the $z$-plane in $\mathbb R^4 / \mathbb Z_k \simeq \mathbb C_z / \mathbb Z_k \times \mathbb C_w / \mathbb Z_k$;  (iii) $ \mathbb C_{w'}/ \mathbb Z_M$ can be regarded as the $w'$-plane identified under the $\mathbb Z_M$-action $w' \to \gamma^{-1}w'$, where $\gamma = e^{2 \pi i / M}$; (iv) $\mathbb C_z / \mathbb Z_k$ can be regarded as the $z$-plane  identified under the $\mathbb Z_k$-action $z \to \zeta z$, where $\zeta =  e^{2 \pi i / k}$; and (v) $\mathbb C_w / (\mathbb Z_k  \times \mathbb Z_M)$ can be regarded as the $w$-plane identified under the $(\mathbb Z_k  \times \mathbb Z_M)$-action $w \to \zeta^{-1}\gamma w$. The $\mathbb Z_M$-action, in addition to acting geometrically, also acts representation-theoretically: at a low-energy scale much larger than the radius of ${\bf S}^1_n$ with $n=1$, the $N$-dimensional representation of the $U(N)$ gauge group of the 4d theory living on the ``constant-time'' hypersurface $ \mathbb C_z / \mathbb Z_k \times \mathbb C_w / (\mathbb Z_k  \times \mathbb Z_M)$ gets multiplied, under the $\mathbb Z_M$-action, by (c.f.~\cite{Mehta, Biswas})    
\be
(\underbrace{\gamma,\ldots,\gamma}_{\text{$n_1$ times}},
\underbrace{\gamma^2,\ldots,\gamma^2}_{\text{$n_2$ times}},
\ldots, \underbrace{\gamma^M,\ldots,\gamma^M}_{\text{$n_M$ times}} ).
\ee

Taking the ``eleventh circle'' to be the decompactified $x_{10}$-direction along the $\mathbb R^{3}$ subspace, we see that (\ref{M-theory defect}) actually corresponds to the following ten-dimensional type IIA background with $N$ coincident NS5-branes wrapping $\mathbb R_t \times {{\bf S}^1_n} \times \mathbb C_z / \mathbb Z_k \times \mathbb C_w / (\mathbb Z_k  \times \mathbb Z_M)$, where the IIA string coupling $g^A_s$ and string length $l_s$ are such that $g^A_sl_s \to \infty$:
\be
\textrm{IIA}: \quad \mathbb R^{2}  \times \mathbb C_{w'}/ \mathbb Z_M \times   \underbrace{\mathbb R_t \times {{\bf S}^1_n} \times  \mathbb C_z / \mathbb Z_k \times \mathbb C_w / (\mathbb Z_k  \times \mathbb Z_M)}_{\textrm{$N$ NS5-branes}}.
\label{IIA 2 defect}
\ee

 Let us now T-dualize along the $\mathbb R_t$ direction of the worldvolume of the stack of NS5-branes. From $\S$A.3, we learn that T-dualizing along any one of the worldvolume directions of an NS5-brane (where the background solution is trivial), will bring us back to an NS5-brane. Therefore, we will arrive at the following type IIB configuration with IIB string coupling $g^B_s \sim 1$ (since $g^B_s = g^A_s l_s /r$, and $r \to \infty$, where $r$ is the radius of ${\mathbb R}_t$):
\be
\textrm{IIB}: \quad \mathbb R^{2}  \times \mathbb C_{w'}/ \mathbb Z_M \times   \underbrace{{\mathbb S}^1_{t;  r \to 0} \times {{\bf S}^1_n} \times  \mathbb C_z / \mathbb Z_k \times \mathbb C_w / (\mathbb Z_k  \times \mathbb Z_M)}_{\textrm{$N$ NS5-branes}}.
\label{IIB 3 defect}
\ee

Next, let us T-dualize along the $x_9$-direction transverse to the stack of NS5-branes. As explained in $\S$A.3, since the NS5-branes are coincident, one will end up having a multi-Taub-NUT manifold $TN_N$ with an $A_{N-1}$ singularity at the origin, with no NS5-branes. To this end, note that one can view $\mathbb R$ along the $x_9$-direction to be a circle of infinite radius. In doing a T-duality along this circle, we arrive at the following type IIA background:
\be
\textrm{IIA}: \quad  TN_N^{R\to 0}\vert_{\mathbb C_{w'}/ \mathbb Z_M}  \times  {\mathbb S}^1_{t; r\to 0} \times {{\bf S}^1_n} \times  \mathbb C_z / \mathbb Z_k \times \mathbb C_w / (\mathbb Z_k  \times \mathbb Z_M).
\label{IIA 4 defect}
\ee
Here, $TN_N^{R \to 0}\vert_{\mathbb C_{w'}/ \mathbb Z_M}$ is a multi-Taub-NUT manifold with an $A_{N-1}$ singularity at the origin whose $w'$-plane (spanning the $x_6$-$x_7$ directions of its $\mathbb R^3$ base in the $x_6$-$x_7$-$x_8$ directions that supports a nontrivial ${\bf S}^1$-fibration in the $x_9$-direction) is further identified under the $\mathbb Z_M$-action $w' \to \gamma^{-1} w'$, and whose asymptotic radius $R \to 0$. (As explained in $\S$A.3, $R \to 0$ because we are T-dualizing along a trivially-fibered circle of infinite radius.) At this stage, one also finds that $g^A_s \to 0$. Consequently, this can be interpreted as the following M-theory background with a very small ``eleventh circle'' ${S}^1_{11}$:
\be
\textrm{M-theory}: \quad TN_N^{R\to 0}\vert_{\mathbb C_{w'}/ \mathbb Z_M}  \times  {\mathbb S}^1_{t; r\to 0} \times {{\bf S}^1_n} \times S^1_{11; r \to 0} \times TN_k^{R\to \infty}\vert_{\mathbb C_{w}/ \mathbb Z_M}.
\label{M-theory 5 defect}
\ee
To arrive at this configuration, we have noted that from $\S$A.1, the singular ALE space $\mathbb C_z / \mathbb Z_k \times \mathbb C_w / \mathbb Z_k$ is simply $TN_k$ with an $A_{k-1}$ singularity at the origin whose asymptotic radius $R \to \infty$. Here, $TN_k^{R\to \infty}\vert_{\mathbb C_{w}/ \mathbb Z_M}$  is a multi-Taub-NUT manifold with an $A_{k-1}$ singularity at the origin whose $w$-plane (spanning the $x_4$-$x_5$ directions of its $\mathbb R^3$ base in the $x_3$-$x_4$-$x_5$ directions which supports a nontrivial ${\bf S}^1$-fibration in the $x_2$-direction) is further identified under the $\mathbb Z_M$-action $w \to \gamma w$, and whose asymptotic radius $R \to \infty$.

From $\S$A.2, we learn that M-theory on the space $TN_k^{R\to \infty}$ is equivalent upon compactification along its circle fiber to type IIA string theory with $k$ coincident D6-branes filling out the directions transverse to the space. In other words, starting from (\ref{M-theory 5 defect}), one can descend back to the following type IIA background:\footnote{In the following background, there is a $\mathbb Z_n$-automorphism on the D6-branes  (that descends from the $\mathbb Z_n$-automorphism on the $TN_k^{R\to \infty}$ in (\ref{M-theory 5 defect}) which underlies the D6-branes solution) that permutes them $n$ times as one goes around the ${\bf S}^1_n$ circle. This permutation does not alter their description as a stack of $k$ coincident D6-branes, and is also consistent with the $\mathbb Z_n$-automorphism of their worldvolume which arises due to the $\mathbb Z_n$-automorphism of $\mathbb R^{3}  \times \mathbb C_{w'}/ \mathbb Z_M \times \mathbb R_t$ in (\ref{M-theory defect}).}
\be
\textrm{IIA}: \quad \underbrace{ TN_N^{R\to 0}\vert_{\mathbb C_{w'}/ \mathbb Z_M}  \times  {\mathbb S}^1_{t; r\to 0} \times {{\bf S}^1_n} \times S^1_{11; r \to 0}}_{\textrm{$k$ D6-branes}}  \times {\mathbb R} \times  {\mathbb C_w / \mathbb Z_M}.
\label{IIA 5 defect}
\ee
Note however, that we now have a type IIA theory that is strongly-coupled, since the effective type IIA string coupling from a compactification along the circle fiber is proportional to the asymptotic radius which is large. (See $\S$A.2, again.)

Let us proceed to do a T-duality along ${S}^1_{11}$, which will serve to decompactify the circle, as well as convert the D6-branes to D5-branes in a type IIB theory. By coupling this step with a type IIB S-duality that will convert the D5-branes into NS5-branes, we will arrive at the following type IIB configuration at weak-coupling:
\be
\textrm{IIB}: \quad \underbrace{ TN_N^{R\to 0}\vert_{\mathbb C_{w'}/ \mathbb Z_M}  \times  {\mathbb S}^1_{t; r\to 0} \times {{\bf S}^1_n}}_{\textrm{$k$ NS5-branes}}  \times \mathbb R^2 \times  {\mathbb C_w / \mathbb Z_M}.
\label{IIB 6 defect}
\ee

Finally, let us do a T-duality along $\mathbb S^1_{t; r\to 0}$, which will bring us back to a type IIA background with NS5-branes and $g^A_s \to \infty$.\footnote{Recall the T-duality relation $g^A_s = g^B_s l_s / r$. Therefore, because $g^B_sl_s$, though small, is still nonzero, having $r \to 0$ would mean that $g^A_s \to \infty$.}  Lifting this IIA background back up to M-theory, we will arrive at the following configuration:
\be
\textrm{M-theory}: \quad \underbrace{TN_N^{R\to 0}\vert_{\mathbb C_{w'}/ \mathbb Z_M} \times {{\bf S}^1_n} \times \mathbb R_t}_{\textrm{$k$ M5-branes}}  \times  {\mathbb C_w / \mathbb Z_M}  \times  \mathbb R^3 \ \  \equiv \  \underbrace{TN_N^{R\to 0}  \times  {{\bf S}^1_n} \times \mathbb R_t}_{\textrm{$k$ M5-branes with a 4d defect}}  \times  {\mathbb R^{5}},
\label{M-theory defect dual}
\ee
where the 4d worldvolume defect wraps ${\bf S}^1_n \times \mathbb R_t$ and the $x_8$-$x_9$ directions in $TN_N^{R\to 0}$. (Recall that the $x_9$-direction is spanned by the ${\bf S}^1$-fiber of $TN_N^{R\to 0}$, while the $x_6$-$x_7$-$x_8$-directions are spanned by its $\mathbb R^3$ base.) Also, there is a nontrivial $\mathbb Z_n$-outer-automorphism of the ten-dimensional transverse space as we go around the ${\bf S}^1_n$ circle.

Note that the $\mathbb Z_M$-action, in addition to acting geometrically, also acts representation-theoretically: when $n=1$, the $k$-dimensional representation of the $U(k)$ gauge group of the 4d theory along $TN_N^{R\to 0}\vert_{\mathbb C_{w'}/ \mathbb Z_M}$ gets multiplied, under the $\mathbb Z_M$-action, by (c.f.~\cite{Mehta, Biswas})      
\be
(\underbrace{\gamma,\ldots,\gamma}_{\text{$n'_1$ times}},
\underbrace{\gamma^2,\ldots,\gamma^2}_{\text{$n'_2$ times}},
\ldots, \underbrace{\gamma^M,\ldots,\gamma^M}_{\text{$n'_M$ times}} ),
\ee 
where $k = n'_1 + n'_2 + \dots + n'_M$. Note that the partition $[n'_1, n'_2, \dots, n'_M]$ of $k$ depends on the partition $[n_1, n_2, \dots, n_M]$ of $N$, as one would expect. We shall elaborate on this in $\S$4.3.

Assuming that the center-of-mass degrees of freedom of the stack of $k$ coincident M5-branes are frozen, the presence of the 4d worldvolume defect means that  at a low-energy scale much larger than the radius of ${\bf S}^1_n$ with $n=1$, the $SU(k)$ gauge group of the 4d ${\cal N} = 4$ theory living on the ``constant-time'' hypersurface $TN_N^{R\to 0}$ is broken to a Levi subgroup $\mathbb L' \subset SU(k)$  along the $x_8$-$x_9$ directions that is the commutant of 
\be
\vec{\alpha}' = (\underbrace{\alpha_{(1)},\ldots,\alpha_{(1)}}_{\text{$n'_1$ times}},
\underbrace{\alpha_{(2)},\ldots,\alpha_{(2)}}_{\text{$n'_2$ times}},
\ldots,
\underbrace{\alpha_{(M)},\ldots,\alpha_{(M)}}_{\text{$n'_M$ times}} ).
\ee

At any rate, from the chain of dualities described above, we conclude that the six-dimensional M-theory compactifications with $N$ and $k$ \emph{coincident} M5-branes wrapping the five compactified directions along the manifolds ${{\bf S}^1_n} \times \mathbb R^4 / \mathbb Z_k$ and ${\bf S}^1_n \times TN_N^{R\to 0}$ in the presence of 4d worldvolume defects as shown in (\ref{M-theory defect}) and (\ref{M-theory defect dual}), respectively, ought to be \emph{physically dual} to each other.

\bigskip\noindent{\it Dual Compactifications with M5-Branes, OM5-Planes and 4d Worldvolume Defects}

To the stack of coincident M5-branes with a 4d worldvolume defect in (\ref{M-theory defect}), one can, as was done in $\S$2.2, add an OM5-plane~\cite{hanany}. Then, we would have the following six-dimensional M-theory compactification:  
\be
\textrm{M-theory}: \quad \mathbb R^{5}  \times  \underbrace{\mathbb R_t \times {{\bf S}^1_n} \times  \mathbb R^4 / \mathbb Z_k}_{\textrm{$N$ M5 + OM5 + 4d defect}} \  \equiv \ \ \mathbb R^{3}  \times \mathbb C_{w'}/ \mathbb Z_M \times   \underbrace{\mathbb R_t \times {{\bf S}^1_n} \times  \mathbb C_z / \mathbb Z_k \times \mathbb C_w / (\mathbb Z_k  \times \mathbb Z_M)}_{\textrm{$N$ M5 + OM5}},
\label{OM-theory defect}
\ee
where the coordinates are $(x_{10}, x_9, x_8; w'; x_0; x_1; z; w)$, with $w' = x_6 + i x_7$. Here, (i) we evoke a $\mathbb Z_n$-outer-automorphism of the transverse ten-dimensional space as we go around the ${\bf S}^1_n$ circle and identify the circle under an order $n$ translation; (ii) the 4d worldvolume defect wraps $\mathbb R_t \times {{\bf S}^1_n}$ and the $z$-plane in $\mathbb R^4 / \mathbb Z_k \simeq \mathbb C_z / \mathbb Z_k \times \mathbb C_w / \mathbb Z_k$;  (iii) $ \mathbb C_{w'}/ \mathbb Z_M$ can be regarded as the $w'$-plane identified under the $\mathbb Z_M$-action $w' \to \gamma^{-1}w'$, where $\gamma = e^{2 \pi i / M}$; (iv) $\mathbb C_z / \mathbb Z_k$ can be regarded as the $z$-plane  identified under the $\mathbb Z_k$-action $z \to \zeta z$, where $\zeta = e^{2 \pi i / k}$; and (v) $\mathbb C_w / (\mathbb Z_k  \times \mathbb Z_M)$ can be regarded as the $w$-plane identified under the $(\mathbb Z_k  \times \mathbb Z_M)$-action $w \to \zeta^{-1}\gamma w$.

Combining our arguments behind (\ref{M-theory defect})--(\ref{M-theory defect dual}) with those behind (\ref{OM-theory 1})--(\ref{OM-theory 8}), we arrive at the following \emph{physically dual} six-dimensional M-theory compactification: 
\be
\textrm{M-theory}: \quad \underbrace{SN_N^{R\to 0}\vert_{\mathbb C_{w'}/ \mathbb Z_M} \times {{\bf S}^1_n} \times \mathbb R_t}_{\textrm{$k$ M5-branes}}  \times  {\mathbb C_w / \mathbb Z_M}  \times  \mathbb R^3 \ \  \equiv \  \underbrace{SN_N^{R\to 0}  \times  {{\bf S}^1_n} \times \mathbb R_t}_{\textrm{$k$ M5 + 4d defect}}  \times  {\mathbb R^{5}},
\label{OM-theory 8 defect}
\ee
where the 4d worldvolume defect wraps ${\bf S}^1_n \times \mathbb R_t$ and the $x_8$-$x_9$ directions in $SN_N^{R\to 0}$, Sen's four-manifold with a $D_N$ singularity at the origin whose asymptotic radius $R \to 0$. (Note that the $x_9$-direction is spanned by the ${\bf S}^1$-fiber of $SN_N^{R\to 0}$, while the $x_6$-$x_7$-$x_8$-directions are spanned by its $\mathbb R^3$ base. See $\S$A.4 for further details, if desired.) Also, there is a nontrivial $\mathbb Z_n$-outer-automorphism of the transverse ten-dimensional space as we go around the ${\bf S}^1_n$ circle.

\newsection{An M-Theoretic Derivation of a Geometric Langlands Duality for Surfaces}

\def\cN{{\cal N}}

\newsubsection{An Equivalence of  Spacetime BPS Spectra and a Geometric Langlands Duality for Surfaces for the $A$--$B$ Groups}

We shall now derive, purely physically, a geometric Langlands duality for surfaces for the $A$--$B$ groups. As a start, note that in $\S$2.1, we showed that the following six-dimensional M-theory compactification on the five-manifold $X_5 = \mathbb R^4 / \mathbb Z_k  \times {{\bf S}^1_n}$  with $N$ coincident M5-branes around it,
\be
 \textrm{M-theory}:\quad   \underbrace{\mathbb R^4 / \mathbb Z_k  \times {{\bf S}^1_n} \times \mathbb R_t}_{\textrm{$N$ M5-branes}}\times \mathbb R^{5},
 \label{M-theory 1 discussion}
 \ee
where we evoke a $\mathbb Z_n$-outer-automorphism of $\mathbb R^4 / \mathbb Z_k$ (and of the geometrically-trivial $\mathbb R^5 \times \mathbb R_t$ spacetime) as we go around the ${\bf S}^1_n$ circle and identify the circle under an order $n$ translation, is\emph{ physically dual} to the following six-dimensional M-theory compactification on  the five-manifold $\tilde X_5 = {\bf S}^1_n  \times TN_N^{R\to 0}$ with $k$ coincident M5-branes around it,
\be
\textrm{M-theory}:\quad  {\mathbb R^{5}} \times \underbrace{\mathbb R_t \times {{\bf S}^1_n}  \times TN_N^{R\to 0}  }_{\textrm{$k$ M5-branes}},
 \label{M-theory 7 discussion}
 \ee
where there is a nontrivial $\mathbb Z_n$-outer-automorphism of $TN_N^{R \to 0}$ as we go around the ${\bf S}^1_n$ circle of radius $R_s$.

Notice that because $\mathbb R^4 / \mathbb Z_k$ and $TN_N^{R\to 0}$ are hyperk\"ahler four-manifolds which break half of the thirty-two supersymmetries in M-theory,  the resulting six-dimensional spacetime theories along $ \mathbb R_t \times \mathbb R^5$ in (\ref{M-theory 1 discussion}) and (\ref{M-theory 7 discussion}), respectively, will both have 6d ${\cN} = (1,1)$ supersymmetry. As usual, there are spacetime BPS states which are annihilated by a subset of the sixteen supersymmetry generators of the 6d $\cN = (1,1)$ supersymmetry algebra; in particular, a generic (half) BPS state in six dimensions would be annihilated by eight supercharges~\cite{DVV}.  Since the supersymmetries of the worldvolume theory of the stack of M5-branes are represented by the spacetime supersymmetries which are unbroken across the brane-spacetime barrier -- in this instance, only half of the sixteen spacetime supersymmetries are unbroken across the brane-spacetime barrier because the M5-branes are half-BPS objects  --  a generic spacetime BPS state would correspond to a worldvolume ground state that is annihilated by all eight worldvolume supercharges.\footnote{By a worldvolume ground state, we really mean a state that is annihilated (in Lorentz signature) by the positive semi-definite operator $H-P$ of the worldvolume supersymmetry algebra $\{Q_\alpha, Q_\beta\} = H-P$, i.e., a minimal energy state that saturates the bound $H \geq P$, where $H$ is the Hamiltonian operator which generates translations along $\mathbb R_t$; $P$ is the momentum operator around ${\bf S}^1_n$;  and the $Q_\alpha$'s and $Q_\beta$'s -- where $\alpha, \beta = 1, \dots, 8$ -- are the eight worldvolume supercharges.\label{worldvolume ground state}}  For example, in a six-dimensional compactification of M-theory with an M5-brane wrapping $K3 \times {\bf S}^1$, where $K3$ is a hyperk\"aher four-manifold,  the sixteen spacetime BPS states which furnish the massless representations of the 6d $\cN = (1,1)$ spacetime supersymmetry algebra correspond to the ground states of the worldvolume theory of the M5-brane~\cite{DVV}. 

For our immediate purpose of deriving purely physically a geometric Langlands duality for surfaces, it suffices to ascertain the spectrum of such spacetime BPS states in the M-theory compactifications (\ref{M-theory 1 discussion}) and (\ref{M-theory 7 discussion}). To do so, we would first need to describe the quantum worldvolume theory of the stack of M5-branes.

\bigskip\noindent{\it{Quantum Worldvolume Theory of the Stack of M5-branes}}

In ten dimensions or less, the fundamental string, and in particular its magnetically-dual NS5-brane, have their origins in the M2- and M5-branes of eleven-dimensional M-theory, respectively. From this fact, it is clear that the fivebranes must be as fundamental as the strings themselves. Moreover, one can also expect that upon quantizing the worldvolume theory of the fivebranes, we would get a spectrum spanned by a tower of excited states, much like when we quantize the worldsheet theory of a fundamental string. 

Indeed, the quantum worldvolume theory of $l$ coincident M5-branes is described by tensionless self-dual strings which live in the six-dimensional worldvolume itself~\cite{RD ref}. In the low-energy limit, the theory of these strings reduces to a non-gravitational 6d $\cN = (2,0)$ $A_{l-1}$ superconformal field theory of $l-1$ massless tensor multiplets.\footnote{Actually, there are, to begin with, $l$ such tensor multiplets from the $l$ M5-branes. However, a single tensor multiplet has been omitted, as we have implicitly frozen its scalars that describe the (transverse) center-of-mass degrees of freedom of the $l$ M5-branes.}
 Each of these $l-1$ multiplets consists of a chiral two-form $Y$ (i.e., with self-dual field strength $dY = \ast dY$), an $Sp(4)$ symplectic Majorana-Weyl fermion $\psi$, and  an $SO(5)$ vector $\phi^A$ of scalars that parameterize the five transverse positions of the M5-branes in eleven dimensions. ($Sp(4) \simeq SO(5)$ is the $R$-symmetry of the $\cN = (2,0)$ superconformal algebra.)  

Alternatively, one can also describe the quantum worldvolume theory via a sigma-model on instanton moduli space~\cite{RD ref,RD}; in particular, if the worldvolume is given by $M \times {\bf S}^1_n \times \mathbb R_t$, where $M$ is a generic hyperk\"ahler four-manifold, one can, in an appropriate gauge, compute the spectrum of ground states of the quantum worldvolume theory (that are annihilated by all of its supercharges), as the spectrum of  physical observables in the topological sector of a two-dimensional ${\cal N} = (4,4)$ sigma-model on ${\bf S}^1_n \times \mathbb R_t$ with target the hyperk\"ahler moduli space ${\cal M}_{G}(M)$ of $G$-instantons on $M$. On the side of (\ref{M-theory 1 discussion}) where $l = N$, we have $G=SU(N)$ if $n=1$, and $G = SO(N+1)$ if $n = 2$ and $N$ is\emph{ even}. 

To arrive at the above claim that the spectrum of ground states of the quantum worldvolume theory is captured by the spectrum of physical observables in the topological sector of the sigma-model, note that (i) the eight supercharges of the  ${\cal N} = (4,4)$ sigma-model on ${\bf S}^1_n \times \mathbb R_t$ represent the eight supersymmetries of the 6d $\cN = (2,0)$ quantum worldvolume theory which are left unbroken on $M \times {\bf S}^1_n \times \mathbb R_t$; (ii) the physical observables of any two-dimensional supersymmetric sigma-model that are annihilated by all of its supercharges necessarily span its topological sector.

To arrive at the above claim about $G$, $n$ and $N$, first note that at a low-energy scale much larger than $R_s$ whence the 6d $\cN = (2,0)$ $A_{N-1}$ SCFT is effectively compactified on  ${\bf S}^1_n$, we get 5d maximally supersymmetric $SU(N)$ theory on $M \times \mathbb R_t$. Next, notice that a $\mathbb Z_n$-outer-automorphism of $M$ would also result in a $\mathbb Z_n$-outer-automorphism of  the $SU(N)$ gauge group (since it is associated with a principal $SU(N)$-bundle over $M \times \mathbb R_t$); as such, the gauge group is effectively $G = SU(N)$ or $SO(N+1)$, depending on the aforementioned values of $n$ and $N$~\cite{Yuji 2 yrs}. Since instantons on $M$ originate from static particle-like BPS configurations of the 5d gauge theory on $M \times \mathbb R_t$, our claim follows. 

The existence of such static particle-like BPS configurations on  $M \times \mathbb R_t$ which manifest as $G$-instantons on $M$, can be understood as follows. Upon compactifying on ${\bf S}^1_n$ (which one can always regard as the ``eleventh circle''), M5-branes which wrap $M \times {\bf S}^1_n \times \mathbb R_t$ reduce to D4-branes in type IIA string theory which wrap $M \times \mathbb R_t$. In type IIA string theory, one can have (half-BPS) D0-branes within the $M \times \mathbb R_t$ worldvolume of the D4-branes~\cite{Douglas}. These D0-branes correspond to the static particle-like BPS configurations in question. 

Being D0-branes, they are charged under a one-form RR gauge field which arises from the Kaluza-Klein (KK) reduction on ${\bf S}^1_n$. Consequently, a single D0-brane of unit RR charge has momentum $1 / R_s$ along ${\bf S}^1_n$. In the case where $n=1$, we do not ``twist'' the theory as we go around ${\bf S}^1_n$; the scalar fields $\varphi$ of the sigma-model are therefore periodic around this circle: if $\sigma$ parameterizes the (compact) spatial direction of the sigma-model worldsheet, then $\varphi(\sigma + 2 \pi) = \varphi(\sigma)$. Hence, the operator $e^{2 \pi i R_s p_s}$ which effects the translation $\sigma \to \sigma + 2 \pi$, where $p_s$ is the momentum along ${\bf S}^1_n$, is such that $e^{2 \pi i R_s p_s} = 1$, i.e., $p_s = m /R_s$, where $m \in \mathbb Z_{\geq 0}$.\footnote{The case of $m$ being negative is \emph{a priori} possible, but its correspondence to D0-brane charge means that we have to  restrict to non-negative values of $m$ only.} Hence, the KK mode, or the D0-brane charge, is $m$. This is the usual story for KK reduction on a circle, where there can be bound states of $m$ D0-branes that manifest as instantons on $M$ with instanton numbers $m$.   

In the case where $n > 1$, we must ``$\mathbb Z_n$-twist'' the theory as we go around ${\bf S}^1_n$; in particular, this circle will be identified under an order $n$ translation.  As such, we must now include a twisted sector in the sigma-model.  In the twisted sector, the scalar fields $\varphi$ of the sigma-model are periodic only up to a $\mathbb Z_n$-factor around the circle, i.e., $\varphi(\sigma + 2 \pi) = e^{2 \pi i r_j / n} \varphi(\sigma)$, where $r_j = 1, 2, \dots, n-1$. In other words,  $e^{2 \pi i R_s p_s} = e^{2 \pi i r_j / n}$, or $p_s = {m \over R_s} +  {r_j \over n R_s}$, where $m \in \mathbb Z_{\geq 0}$. Together with the untwisted sector, we then have $p_s ={m / R_s},  ( m + {1 \over n}) /R_s, ( m + {2 \over n}) /R_s, \dots, \left( m + {n-1 \over n}\right)/R_s$. Therefore, the KK modes, or the D0-brane charges, are $m, m + {1\over n}, m+ {2 \over n}, \dots, m + {n-1\over n}$.  Hence, there can be bound states of $m, m + {1\over n}, m+ {2 \over n}, \dots, \, {\rm and} \ m + {n-1\over n}$ D0-branes that manifest as instantons on $M$, giving rise to the instanton numbers $m, m + {1\over n}, m+ {2 \over n}, \dots, \,  {\rm and} \ m + {n-1\over n}$, respectively.\footnote{Fractional branes such as these which give rise to fractional instanton numbers in this instance, have also appeared elsewhere in string theory -- see~\cite[$\S$13.2]{CJ} and references therein.}  

Altogether from the last two paragraphs, it would mean that instantons on the \emph{spin} manifold $M$ (modulo the noncompactness of $M$) have instanton numbers that take values in $\mathbb Z_{\geq 0} / n$. In particular, for $n=1$ whence we have $SU(N)$-instantons on the side of (\ref{M-theory 1 discussion}), the instanton numbers take only non-negative integer values, as is well-known. For $n=2$ and even $N$ whence we have $SO(N +1)$-instantons, the instanton numbers take values in $\mathbb Z_{\geq 0}/2$; this is consistent with the fact that for nonsimply-connected groups such as $SO(N+1)$, the instanton numbers may not always be integral. Indeed, our results agree with~\cite{vafa-witten}, $\S$3.2, first paragraph, where for $SO(3)$, it was shown that the instanton number takes values in $\mathbb Z_{\geq 0}/2$; moreover, our results also agree with~\cite[Appendix B]{Siye Wu}, where for all other $SO(N+1)$, it was shown that the instanton numbers take only non-negative integer values. In the latter case of all other $SO(N+1)$ where one only has integral instanton numbers, there necessarily has to be further binding of pairs of bound states consisting of full and one-half-fractional D0-branes whose respective charges take the forms $m+ {1 \over 2}$ and $(m +1) - {1 \over 2}$, such that we effectively have an integral number of D0-branes only.

\bigskip\noindent{\it{Spacetime BPS States from the ${\cal N} = (4,4)$ Sigma-Model on ${\bf S}^1_n \times \mathbb R_t$}}

As explained earlier, the spectrum of spacetime BPS states would correspond to the spectrum of ground states of the quantum worldvolume theory of the M5-branes; in turn, as claimed and justified thereafter, this is captured by the spectrum of physical observables in the topological sector of the ${\cal N} = (4,4)$ sigma-model on ${\bf S}^1_n \times \mathbb R_t$ that are annihilated by all of its eight supercharges. As such, the spacetime BPS states would correspond to  differential forms on the target space ${\cal M}_{G}(M)$. These differential forms are necessarily (i) harmonic, as all eight supercharges -- which have a well-known sigma-model interpretation~\cite{tsm} as de Rham differentials and their adjoints on ${\cal M}_{G}(M)$ -- annihilate them; (ii) square-integrable, as they are expected to be well-defined even on a noncompact space like  ${\cal M}_{G}(M)$. Therefore, the spacetime BPS states  would correspond to ${\bf L}^2$-harmonic forms which span the ${\bf L}^2$-cohomology of (some natural compactification of) ${\cal M}_{G}(M)$.\footnote{The good ultraviolet behavior of any string theory -- in this case, one described by a sigma-model with $\cN = (4,4)$ supersymmetry -- would lead to a natural compactification of ${\cal M}_{G}(M)$. \label{uv behavior}}$^,$\footnote{It is a theorem that on any complete manifold which is therefore compact, an ${\bf L}^2$-harmonic form represents a class in the ${\bf L}^2$-cohomology~\cite{Hitchin}.}

\bigskip\noindent{\it The Gradings on ${\cal M}_G(M)$}

In order to determine in detail the spectrum of spacetime BPS states in the M-theory compactification (\ref{M-theory 1 discussion}), we must first and foremost ascertain how ${\cal M}_{G}(M)$, where $M = \mathbb R^4 / \mathbb Z_k$,  is graded. Firstly, it is clear that ${\cal M}_{G}(M)$ has got to be graded by the instanton number $a$. 

Secondly, note that a $G$-bundle on a generic four-manifold $X$ is topologically classified by $p_2 \in H^2(X,  \pi_1(G))$. As such, it would appear that ${\cal M}_{G}(M)$ ought to also be graded by $p_2$. However, because $M$ is a complete blowdown of the fully-resolved ALE space $\widetilde{ \mathbb R^4 / \mathbb Z_k}$, we have $H_2(M, \pi_1(G))  = 0$; that is, $p_2$ is effectively zero.

Thirdly, since the theory is supposed to be physically consistent,  the instanton action ought to be finite in an integration over $M$. As $M = \mathbb R^4/\mathbb Z_k$ is noncompact, this implies that only flat connections survive at infinity; in other words, we have, at infinity, a choice of conjugacy classes of the homomorphism $\rho_\infty:\pi_1(M) \to G$, where $\pi_1(M) = \mathbb Z_k$. Note also that a $G$-bundle on $\mathbb R^4/\mathbb Z_k$ is the same as a $\mathbb Z_k$-equivariant $G$-bundle on $\mathbb R^4$; since the origin 0 is a fixed point of the $\mathbb Z_k$-action, it follows that the $\mathbb Z_k$-action acts in the fiber of the bundle at 0. Such an action is given by a conjugacy class of the homomorphism $\rho_0: \mathbb Z_k \to G$. In short, this means that in addition to $a$, ${\cal M}_{G}(M)$ has  also got to be graded by the conjugacy classes of the homomorphisms $\rho_0$ and $\rho_\infty$ one is allowed to pick at the origin and infinity of $M = \mathbb R^4/ \mathbb Z_k$, respectively. Therefore, ${\cal M}_{G}(M)$ consists of components labeled by $(a, \rho_0, \rho_\infty)$;  that is, one can write
\be
{\cal M}_{G}(M) = \bigoplus_{a, \rho_0, \rho_\infty} {\cal M}^{\rho_0,  a}_{G, \rho_\infty}(M).
\ee
Note that $a$ is not really independent of $\rho_0$ and $\rho_\infty$, as we shall now explain. 

\bigskip\noindent{\it More About the Instanton Number}

Notice that $M = \mathbb R^4/ \mathbb Z_k$ is defined by imposing an order $k$ cyclic identification of $\mathbb R^4$. This means that the total number of D0-branes ought to be $kd$, where $d$ is the number of D0-branes in a fundamental region of $M$. Moreover, according to our earlier explanations, for $G = SU(N)$, $d$ must take values in $\mathbb Z_{\geq 0}$; for $G = SO(3)$, $d$ must take values in $\mathbb Z_{\geq 0}/2$; and for all other $G = SO(N+1)$ with even $N$, $d$ must again take values in $\mathbb Z_{\geq 0}$. In all, this means that we can write the instanton number as $a = kd = k n' (i - j)$, where for $G = SU(N)$, $SO(3)$ and $SO(N+1)$, $n' =1$, $1$ and 2, while $i,j$ are certain integers divided by $1$, $2$ and $2$. In all cases, $i \geq j$, as $d$ must be non-negative. Here, one can interpret $(i-j)$ as the contribution from the bound states of D0-branes, and $n' \neq 1$ if bound states consisting of fractional D0-branes necessarily need to be paired to form bound states of full D0-branes. 

That said, since $M$ is noncompact, the total instanton number must actually be shifted by an amount which depends on the conjugacy class of $\rho_\infty$. (See \cite[$\S$4.4]{vafa-witten}.) In our language, this means that we have to omit D0-branes at infinity -- which are necessarily associated with flat gauge fields that consequently have zero instanton number and are thus topologically trivial -- when counting the total instanton number. Since a conjugacy class of $\rho_\infty$ can be interpreted as a dominant coweight $\bar\mu$ of $G$,\footnote{To understand this claim, note that conjugacy classes of a homomorphism $\rho: U(1) \to G$ are classified by highest weights of the Langlands dual group $G^\vee$. Furthermore, these highest weights are associated with irreducible representations of $G^\vee$. In turn, this means that they ought to be dominant. (See~\cite[$\S$13.2]{CFT text}.) Thus, since weights of $G^\vee$ are also coweights of $G$, we have our claim.} and since the instanton number is a scalar, it would mean that we can actually write the shifted instanton number as $a = kn'(i - j)  - b(\bar \mu, \bar \mu)$, where $b$ is some positive real constant, and $(~,~)$ is just the scalar product in coweight space. 

Last but not least, note that in our counting of the total instanton number performed hitherto, we have implicitly overlooked the D0-branes at the origin of $M$: in writing $a = k n' (i-j)$ in the paragraph before last, we have accounted for the D0-branes away from the origin which have $k$ mirror partners under the order $k$ cyclic identification, but \emph{not} the D0-branes at the origin which do not have any mirror partners (since the origin is a fixed-point of the identification).  Thus, just like how we can exclude the D0-branes at infinity by subtracting $b(\bar \mu, \bar \mu)$ from the total instanton number, we can include the D0-branes at the origin by adding $\tilde b(\bar \lambda, \bar \lambda)$ to the total instanton number, where $\tilde b$ is some positive real constant, and $\bar \lambda$ is a dominant coweight of $G$ which corresponds to a conjugacy class of $\rho_0$.  In short, we can write the instanton number as
\be
a = k n'(i-j) + \tilde b(\bar \lambda, \bar \lambda) - b(\bar \mu, \bar \mu),
\label{a}
\ee
where for $G = SU(N)$, $SO(3)$ and $SO(N+1)$, $n' =1$, $1$ and $2$, while $i,j$ --  whereby $i \geq j$ -- are certain integers divided by $1$, $2$ and $2$, respectively. Hence, as mentioned earlier, we find that $a$ is not really independent of  $\rho_0$ (or $\bar \lambda$) and $\rho_\infty$ (or $\bar \mu$).

For $n=1$ whence we have $G = SU(N)$ with $n'=1$ and $i,j$ being certain integers, expression (\ref{a}) is indeed consistent with results from the mathematical literature (which only addresses the case of simply-connected groups like $SU(N)$): eqn.~(\ref{a}) coincides with~\cite[eqn.~(4.3)]{BF} after we set $\tilde b = b = 1/2$ and identify $(i, j)$ with $(l, m)$ of \emph{loc.~cit..}

\bigskip\noindent{\it The Spectrum of Spacetime BPS States in the M-Theory Compactification (\ref{M-theory 1 discussion})}

We are now ready to state the generic Hilbert space ${\cal H}_{\rm BPS}$ of spacetime BPS states in the M-theory compactification (\ref{M-theory 1 discussion}). To this end, let us first denote by ${\textrm H}^\ast_{{\bf L}^2}{\cal U}({\cal M}^{\lambda}_{G, \mu}(\mathbb R^4/ \mathbb Z_k))$, the ${\bf L}^2$-cohomology of the Uhlenbeck compactification ${\cal U}({\cal M}^{\lambda}_{G, \mu}(\mathbb R^4/ \mathbb Z_k))$ of the component ${\cal M}^{\lambda}_{G, \mu}(\mathbb R^4/ \mathbb Z_k)$ of the highly singular moduli space ${\cal M}_{G}(\mathbb R^4/ \mathbb Z_k)$ labeled by the triples $\lambda = (k, \bar \lambda, i)$ and $\mu = (k, \bar \mu, j)$ (where $a$ is correspondingly given by (\ref{a})).\footnote{To define a cohomology on a space, one first needs to compactify the space; see footnote~\ref{uv behavior} as to why string theory ought to lead to a natural compactification of the target space. Then, according to~\cite{BF}, a suitable compactification in this case would be given by the Uhlenbeck compactification.\label{compactification}}$^,$\footnote{Although the ${\cal M}_{G}(\mathbb R^4/ \mathbb Z_k)$ target space of the string described by the sigma-model is highly singular, it is well-known that the physics remains well-behaved. \label{well-behaved}} Then, since one can express ${\textrm H}^\ast_{{\bf L}^2}{\cal U}({\cal M}^{\lambda}_{G, \mu}(\mathbb R^4/ \mathbb Z_k))$  as the intersection cohomology ${\rm IH}^\ast{\cal U}({\cal M}^{\lambda}_{G, \mu}(\mathbb R^4/ \mathbb Z_k))$~\cite{Goresky}, we can write
\be
{\cal H}_{\rm BPS} = \bigoplus_{\lambda, \mu} {\cal H}^{\lambda, \mu}_{\rm BPS}  =  \bigoplus_{\lambda, \mu} ~{\rm IH}^\ast {\cal U}({\cal M}^{\lambda}_{G, \mu}(\mathbb R^4/ \mathbb Z_k)).
\label{BPS-M}
\ee
Notice that because we cannot have a negative number of D0-branes, we must have $a \geq 0$. In turn, this implies, via (\ref{a}) and the  condition $i \geq j$, that 
\be
\lambda \geq \mu.
\label{weight condition}
\ee
 As $k \in \mathbb Z_+$ and $\bar \lambda$ and $\bar \mu$ are dominant coweights of $G$, the triples $\lambda$ and $\mu$ can be regarded as dominant coweights of the corresponding affine Kac-Moody group $G_{\rm aff}$ of level $k$. (See~\cite[$\S$14.3.1]{CFT text}.) Thus, we find that (\ref{weight condition}) is also consistent with~\cite[Theorem 4.8]{BF}, which implies that ${\cal H}^{\lambda, \mu}_{\rm BPS}$ is empty unless $\lambda \geq \mu$. 
 
 \bigskip\noindent{\it The Partition Function of Spacetime BPS States for $G = SU(N)$}

 One can of course go on to state the partition function of spacetime BPS states. The partition function, which counts (with weights) the total number of  states, can be obtained by taking a trace in the Hilbert space of states. Note at this point that taking such a trace is geometrically equivalent to identifying the two ends of the sigma-model worldsheet ${\bf S}^1_n \times \mathbb R_t$ to form a torus. Let the modulus of this torus be $\tau = \tau_1 + i \tau_2$; then, if $n=1$, the partition function for simply-connected $G = SU(N)$ can be written as (c.f.~\cite{MS})
 \be
 Z_{SU(N)}^{\rm BPS} = {\rm Tr}_{{\cal H}_{\rm BPS}} \, q^P,
 \label{Zbps}
 \ee
where $q= e^{2 \pi i \tau}$, and $P$ is the momentum operator along ${\bf S}^1_{n}$. 

Since $P$ measures the number of D0-branes (as each D0-brane has unit momentum along ${\bf S}^1_n$), according to our analysis leading up to (\ref{a}), we have $P = k(i-j) + {1 \over 2}(\bar \lambda, \bar \lambda)$. Together with (\ref{BPS-M}), we can therefore write (\ref{Zbps}) as
\be
 Z_{SU(N)}^{\rm BPS}   =  \sum_{\lambda} \, q^{m_\lambda}  \, \sum_{\bar\mu}  \sum_{m \geq 0}  \, {\rm dim} \, {\rm IH}^\ast {\cal U}({\cal M}^{\lambda, m}_{SU(N), \bar\mu}(\mathbb R^4/ \mathbb Z_k)) \,  q^m.
 \label{Z_SU(N)} 
\ee
Here,
\be
m_\lambda = h_{ \lambda} - {c_{ \lambda}  \over 24};
\label{m}
\ee
$m = k (i-j)$ is a non-negative integer, as  $i, j$ are integers such that $(i-j) \in \mathbb Z_{\geq 0}$; the non-negative number 
\be
h_{ \lambda} = {(\bar \lambda, \bar\lambda + 2 \rho^\vee) \over 2(k + h)},
\label{h}
\ee 
where $\rho^\vee$ and $h$ are the Weyl vector and dual Coxeter number of the\emph{ Langlands dual} group ${SU(N)}^\vee$, respectively; and the number
\be
{c_{ \lambda}}  =  -24 \tilde b {(\bar \lambda, \bar \lambda)}  +  {12 (\bar \lambda, \bar \lambda + 2 \rho^\vee) \over {(k + h)}},
\label{c}
\ee
where $\tilde b = 1/2$ in this $SU(N)$ case.

In this instance, $\lambda = (k, \bar \lambda, i)$ and $\mu = (k, \bar \mu, j)$ can also be regarded as \emph{dominant weights} of the corresponding \emph{Langlands dual} affine Kac-Moody group ${SU(N)}^\vee_{\rm aff}$ of level $k$.

 \bigskip\noindent{\it The Partition Function of Spacetime BPS States for $G = SO(N+1)$}

Now, let $n=2$ whence the theory is ``$\mathbb Z_2$-twisted'' as we go around ${\bf S}^1_n$. In this case, the total partition function of spacetime BPS states for nonsimply-connected $G = SO(N+1)$ (where $N \geq 2$ is even), can be written as 
\be
Z^{\rm BPS}_{SO(N+1)} = {\rm Tr}_{{\cal H}^0_{\rm BPS}} \, {\mathscr P}_2 \, q^{P_0} + {\rm Tr}_{{\cal H}^{1}_{\rm BPS}} \, {\mathscr P}_2 \, q^{P_{1}},
\label{pf underlying}
\ee
where $\mathscr P_2$ is a projection onto $\mathbb Z_2$-invariant states, and the super(sub)script `$0$' or `$1$' indicates that the operator or space in question is that of the untwisted or twisted sector, respectively.\footnote{The reason why one has to add a twisted sector whilst projecting onto $\mathbb Z_2$-invariant states -- like in any consistent 2d CFT with a cyclic identification along its spatial direction -- is because the spacetime BPS states come from the topological sector of the sigma-model which is therefore conformal.\label{twisted sector}} 

The meaning of $\mathscr P_2$ in the trace over ${{\cal H}^0_{\rm BPS}}$ can be understood explicitly as follows. First, note that in the \emph{untwisted} sector, we have the dominant coweights $\lambda_0 = (k, {\bar \lambda}_0, i_0)$ and $\mu_0 = (k, {\bar \mu}_0, j_0)$ of $SO(N+1)_{\rm aff}$ of level $k$, where  $\lambda_0 \geq \mu_0$; according to our earlier discussion, $i_0$ and $j_0$ are integers whereby $(i_0-j_0) \in \mathbb Z_{\geq 0}$, and to satisfy this condition unequivocally, one ought to have $i_0 \in \mathbb Z_{\geq 0}$ and $-j_0 \in \mathbb Z_{\geq 0}$;  that is,  $\lambda_0$ and $\mu_0$ are dominant coweights with integer grading. Second, note that the intersection cohomology ${\rm IH}^\ast {\cal U}({\cal M}^{\lambda_0}_{SO(N+1), \mu_0}(\mathbb R^4/ \mathbb Z_k))$ which represents ${\cal H}^{\lambda_0, \mu_0}_{\rm BPS} \subset {\cal H}^0_{\rm BPS} \subset {\cal H}_{\rm BPS}$ (see (\ref{BPS-M})), corresponds to the space of physical observables of the $\cN=(4,4)$ sigma-model that take the form ${\cal O}_0 = f _{c  \dots e; \bar c \dots \bar e}(\varphi^d_0, \varphi^{\bar d}_0) \eta^c_0 \dots \eta^e_0 \eta^{\bar c}_0 \dots \eta^{\bar e}_0$, where the $\varphi_0$'s and $\eta_0$'s are untwisted bose and fermi fields of the sigma-model which are periodic and antiperiodic around ${\bf S}^1_n$, respectively,\footnote{Unlike the commuting bose fields, the anti-commutativity of the fermi fields means that whenever a fermi field passes another in a correlation function as it is being translated around  ${\bf S}^1_n$, the correlation function picks up a minus sign. As such, the fermi fields are effectively antiperiodic around ${\bf S}^1_n$.\label{anti-commutativity}} i.e., 
\be
\varphi^{c, \bar d}_0(\sigma + 2 \pi) = \varphi^{c, \bar d}_0(\sigma) \qquad {\rm and} \qquad \eta^{c, \bar d}_0 (\sigma + 2 \pi) = - \eta^{c, \bar d}_0(\sigma).
\label{field twist 1}
\ee
Here, the indices run as $c, \bar d = 1, 2, \dots, {\rm dim}_{\mathbb C} \, {\cal U}({\cal M}^{\lambda_0, m_0}_{SO(N+1), {\bar\mu}_0}(\mathbb R^4/ \mathbb Z_k))$, where $m_0 =  kn'(i_0 - j_0)$ -- the eigenvalue of $ P_0 - (\bar \lambda_0, \bar \lambda_0) /2$ --  is a non-negative integer, and $n' =2$ or $1$ if $N > 2$ or $N = 2$, respectively.   The insertion of $\mathscr P_2$ then means that in computing the trace over ${\cal H}^0_{\rm BPS}$, one must consider only ${\cal O}_0$'s which are invariant under the $\mathbb Z_2$-transformations $\varphi \to -\varphi$  and $\eta \to - \eta$. For later convenience, let us denote the space of such $\mathbb Z_2$-invariant  ${\cal O}_0$'s by  $\overline{{\rm IH}^\ast {\cal U}}({\cal M}^{\lambda_0, m_0}_{SO(N+1), \bar\mu_0}(\mathbb R^4/ \mathbb Z_k)) \subset  {\rm IH}^\ast {\cal U}({\cal M}^{\lambda_0, m_0}_{SO(N+1), \bar \mu_0}(\mathbb R^4/ \mathbb Z_k))$.

Similarly, the meaning of $\mathscr P_2$ in the trace over ${{\cal H}^1_{\rm BPS}}$ can be understood explicitly as follows. First, note that in the \emph{twisted} sector, we have the dominant coweights $\lambda_1 = (k, {\bar \lambda}_1, i_1)$ and $\mu_1 = (k, {\bar \mu}_1, j_1)$ of $SO(N+1)_{\rm aff}$ of level $k$, where  $\lambda_1 \geq \mu_1$; according to our earlier discussion, $i_1$ and $j_1$ are integers divided by 2 such that $(i_1-j_1) \in \mathbb Z_{\geq 0} + {1 \over 2}$, and to satisfy this condition unequivocally, one ought to have $i_1 \in \mathbb Z_{\geq 0}$ and $-j_1 \in \mathbb Z_{\geq 0} + {1 \over 2}$; in other words, $\lambda_1$ and $\mu_1$ ought to be dominant coweights with integer and half-integer grading, respectively. Second, note that the intersection cohomology ${\rm IH}^\ast {\cal U}({\cal M}^{\lambda_1}_{SO(N+1), \mu_1}(\mathbb R^4/ \mathbb Z_k))$ which represents ${\cal H}^{\lambda_1, \mu_1}_{\rm BPS} \subset {\cal H}^{1}_{\rm BPS} \subset H_{\rm BPS}$ (see (\ref{BPS-M})), corresponds to the space of physical observables of the $\cN=(4,4)$ sigma-model that take the form ${\cal O}_{1} = f _{c  \dots e; \bar c \dots \bar e}(\varphi^d_{1}, \varphi^{\bar d}_{1}) \eta^c_{1} \dots \eta^e_{1} \eta^{\bar c}_{1} \dots \eta^{\bar e}_{1}$. Here, the $\varphi_{1}$'s and $\eta_{1}$'s are \emph{twisted} bose and fermi fields of the sigma-model which are thus antiperiodic and periodic around ${\bf S}^1_n$, respectively; specifically, we have 
\be
\varphi^{c}_1(\sigma + 2 \pi) = e^{2 \pi i \nu \over n}\varphi^{c}_1(\sigma) = - \varphi^{c}_1(\sigma),    \qquad \eta^{c}_1(\sigma + 2 \pi) = - e^{2 \pi i \nu \over n}\eta^{c}_1(\sigma) = \eta^{c}_1(\sigma), 
\label{field twist 2}
\ee
and
\be
\varphi^{\bar d}_1(\sigma + 2 \pi) = e^{-{2 \pi i \nu \over n}}\varphi^{\bar d}_1(\sigma) = - \varphi^{\bar d}_1(\sigma),    \qquad \eta^{\bar d}_1 (\sigma + 2 \pi) = - e^{-{2 \pi i \nu \over n}}\eta^{\bar d}_1(\sigma) = \eta^{\bar d}_1(\sigma),
\label{field twist 3}
\ee
as $n =2$ and the twist parameter $\nu = 1$. Also, $c, \bar d = 1, 2, \dots, {\rm dim}_{\mathbb C} \, {\cal U}({\cal M}^{\lambda_1, m_1}_{SO(N+1), \bar\mu_1}(\mathbb R^4/ \mathbb Z_k))$, where $m_1 = k n' (i_1 - j_1)$ -- the eigenvalue of  $P_1 - (\bar \lambda_1, \bar \lambda_1) /2$ -- is a non-negative integer divided 2,  and $n' =2$ or $1$ if $N > 2$ or $N = 2$, respectively.  The insertion of $\mathscr P_2$ then means that in computing the trace over ${\cal H}^{1}_{\rm BPS}$, one must consider only ${\cal O}_{1}$'s which are invariant under the $\mathbb Z_2$-transformations $\varphi \to -\varphi$  and $\eta \to - \eta$. Let us denote the space of such $\mathbb Z_2$-invariant  ${\cal O}_{1}$'s by  $\overline {{\rm IH}^\ast {\cal U}}({\cal M}^{\lambda_1, m_1}_{SO(N+1), \bar\mu_1}(\mathbb R^4/ \mathbb Z_k)) \subset  {\rm IH}^\ast {\cal U}({\cal M}^{\lambda_1, m_1}_{SO(N+1), \bar \mu_1}(\mathbb R^4/ \mathbb Z_k))$. Then, together with what was said in the previous paragraph, and by relabeling the integer-graded coweights $\lambda_0$ and $\lambda_1$ as $\lambda$, we can write 
\be
Z^{\rm BPS}_{SO(N+1)} =  Z^{{\rm BPS}, 0}_{SO(N+1)} + Z^{{\rm BPS}, 1}_{SO(N+1)},
\label{Z-SO(N+1)-1}
\ee
where 
\be
Z^{\rm BPS, 0}_{SO(N+1)} = \sum_{\lambda, \bar \mu_0}  q^{m_{\lambda}}  \sum_{m_0 \geq 0}   {\rm dim} \, \overline {{\rm IH}^\ast {\cal U}}({\cal M}^{\lambda, m_0}_{SO(N+1), \bar\mu_0}(\mathbb R^4/ \mathbb Z_k)) \, q^{m_0}, 
\label{Z-SO(N+1)-2}
\ee
and 
\be
Z^{\rm BPS, 1}_{SO(N+1)} = \sum_{\lambda, \bar \mu_1}  q^{m_{\lambda}}  \sum_{m_1 \geq 0}   {\rm dim} \, \overline {{\rm IH}^\ast {\cal U}}({\cal M}^{\lambda, m_1}_{SO(N+1), \bar\mu_1}(\mathbb R^4/ \mathbb Z_k)) \, q^{m_1}.
\label{Z-SO(N+1)-3}
\ee
The phase factor $m_{\lambda}$ takes the form in (\ref{m}). 
 
 In this instance, the dominant coweights  $\lambda = (k, \bar \lambda, i)$ and $\mu_{0,1} = (k, \bar \mu_{0,1}, j_{0,1})$ of ${SO(N+1)}_{\rm aff}$ are also  (un)twisted dominant weights of the $\mathbb Z_2$-twisted affine Kac-Moody group ${SU(N)}^{(2)}_{\rm aff}$; furthermore, ${SU(N)}^{(2)}_{\rm aff}$ is equal to $SO(N+1)^\vee_{\rm aff}$. In other words, $\lambda$ and $\mu_{0,1}$ can also be regarded as \emph{dominant weights} of the \emph{Langlands dual} affine Kac-Moody group ${SO(N+1)}^\vee_{\rm aff}$ of level $k$.

Additionally, notice that (\ref{Z-SO(N+1)-1})--(\ref{Z-SO(N+1)-3}) imply that the \emph{effective} Hilbert space ${{\cal H}}^{\rm eff}_{\rm BPS}$ of spacetime BPS states (which one obtains after taking into account the projection $\mathscr P_2$ in the trace over all underlying states in (\ref{pf underlying})) ought to be given by
\be
\label{HBPS-eff}
{{\cal H}}^{\rm eff}_{\rm BPS} = \bigoplus_{\lambda} \bigoplus_{\nu =0,1}  \bigoplus_{\mu_\nu}  \, {\overline{\cal H}}^{\lambda, \mu_\nu}_{\rm BPS}   =   \bigoplus_{\lambda} \bigoplus_{\nu =0,1}  \bigoplus_{\mu_\nu}  \, \overline {{\rm IH}^\ast {\cal U}}({\cal M}^{\lambda}_{SO(N+1), \mu_\nu}(\mathbb R^4/ \mathbb Z_k)),
\ee
where $\nu =0$ or $1$ if the sector is untwisted or twisted, respectively.

\bigskip\noindent{\it The Spectrum of Spacetime BPS States in the M-Theory Compactification (\ref{M-theory 7 discussion})}

Let us now turn our attention to the\emph{ physically dual} M-theory compactification (\ref{M-theory 7 discussion}) with $k$ coincident M5-branes. One can proceed as before to ascertain the spacetime BPS states by computing the ground states of the M5-brane quantum worldvolume theory over $\mathbb R_t \times {\bf S}^1_n \times TN^{R\to 0}_N$. As explained early on in this subsection, one can, if $n=1$ for example, interpret the spacetime BPS states as the physical observables in the topological sector of the sigma-model on  ${\bf S}^1_n \times \mathbb R_t$ with target the moduli space of $U(k)$-instantons on $TN^{R\to 0}_N$.\footnote{The reason why we have instantons of $U(k)$ (and not $SU(k)$) is because in duality step (\ref{IIA 5}), the center-of-mass degrees of freedom of the $k$ D6-branes are not frozen. \label{not frozen 1}} 

That said, since we would like to make contact with a geometric Langlands duality for surfaces, we shall seek a different description of these spacetime BPS states, i.e, worldvolume ground states. To this end, recall  that the low-energy limit of the worldvolume theory (minus the center-of-mass degrees of freedom) is a 6d  $\cN = (2,0)$ $A_{k-1}$ \emph{superconformal }field theory of massless tensor multiplets. Hence, where the ground states are concerned, one can regard the worldvolume theory to be conformally-invariant. Since it is conformally-invariant, one can rescale the worldvolume to bring the region near infinity to a finite distance close to the origin without altering the theory.  Thus, one can, for the purpose of computing ground states, simply analyze the physics near infinity.

Near infinity, the ${\bf S}^1_R$ circle fiber of $TN^{R \to 0}_N$ has radius $R \to 0$. To make sense of this limit, notice that a compactification along the circle fiber would take us down to a type IIA theory whereby the stack of $k$ coincident M5-branes would now correspond to a stack of $k$ coincident D4-branes. In addition, as explained in $\S$A.2, since the circle fibration is a monopole bundle over an ${\bf S}^2$ at infinity of charge $N$, we would also have $N$ D6-branes spanning the directions transverse to its $\mathbb R^3$ base; since $TN^{R \to 0}_N$ has an $A_{N-1}$ singularity at the origin, these $N$ D6-branes will be coincident. In other words, we have, in the limit $R \to 0$, the following type IIA configuration:
\be
\textrm{IIA}: \quad \underbrace{ {\mathbb R}^5 \times  {{\bf S}^1_n} \times {\mathbb R}_t \times {\mathbb R}^3}_{\textrm{I-brane on ${{\bf S}^1_n} \times {\mathbb R}_t = N \textrm{D6} \cap k\textrm{D4}$}}.
\label{equivalent IIA system 1}
\ee
Here, we have  a stack of $N$ coincident D6-branes whose worldvolume is given by ${\mathbb R}^5 \times {{\bf S}^1_n} \times {\mathbb R}_t$, and a stack of $k$ coincident D4-branes whose worldvolume is given by $ {{\bf S}^1_n} \times {\mathbb R}_t \times \mathbb R^3$. Generically, there ought to be, on the stack of D4- and D6-branes, a $U(k)$ and $U(N)$ gauge group, respectively. Notice also that the two stacks intersect along ${{\bf S}^1_n} \times {\mathbb R}_t$ to form a D4-D6 I-brane system. A set of D4- and D6-branes that intersect along two flat directions is a supersymmetric configuration. In our case, we have 2d $\cN = (8,0)$ supersymmetry on the I-brane. 

We now argue that the sought-after spectrum of M5-brane worldvolume ground states can be computed solely from the I-brane theory along ${\bf S}^1_n \times \mathbb R_t$.  Firstly, notice that the 4-6 open strings which stretch between the D4- and D6-branes descend from open M2-branes whose topology is a disc with an ${\bf S}^1_R$ boundary that ends on the M5-branes. Secondly, the interval filling the disc and thus, the tension of these open M2-branes, goes to zero as the type IIA open strings approach the I-brane and become massless. This means that the massless type IIA open strings which live along the I-brane descend from tensionless self-dual closed strings of topology ${\bf S}^1_R$ that live in the M5-brane worldvolume. Thirdly, the $R \to 0$ limit can be viewed as a low-energy limit of these tensionless self-dual closed strings whence their corresponding spectrum would be spanned by the M5-brane worldvolume ground states that we are after. Altogether, these three points mean that the spectrum of M5-branes worldvolume ground states can be computed solely from the field theory associated with the massless 4-6 strings that live along the I-brane.\footnote{Apart from the fact that in the sigma-model computation of these ground states, we also consider a 2d theory along ${\bf S}^1_n \times \mathbb R_t$, one can also see that this claim is physically consistent as the I-brane theory has $\cN = (8,0)$ supersymmetry which it inherits from the ambient spacetime, while the ground states are supposed to be invariant under eight spacetime supersymmetries  which are also inherited from the ambient spacetime.} Therefore, let us henceforth focus on the I-brane theory.

The massless modes of the 4-6 open strings reside entirely in the Ramond sector. However, in the NS sector, there are massive modes. Note at this point from $\S$A.2 that the asymptotic radius $R$ is given by $g^A_s l_s$, where $g^A_s$ and $l_s$ are the type IIA string coupling and string length, respectively. Since we are really analyzing the system at fixed coupling $g^A_s$, the $R \to 0$ limit can be interpreted as the $l_s \to 0$ low-energy limit, consistent with the regime that the aforementioned tensionless strings with topology ${\bf S}^1_R$ are in. In this limit, all the massive modes decouple; one is then left with the massless modes only.  

The massless modes are well-known to be chiral fermions on the two-dimensional I-brane~\cite{Vafa ref, Vafa ref 2}. If we have $k$ D4-branes and $N$ D6-branes, the $kN$ complex chiral fermions
\be
\psi_{i, \bar a}(z), \ \psi^{\dagger}_{\bar i, a}(z), \qquad  i = 1, \dots, k, \quad a = 1, \dots, N,
\label{4.9}
\ee
will transform in the bifundamental representations $(k, \bar N)$ and $(\bar k, N)$ of $U(k) \times U(N)$, where `$z$' is the complex coordinate on the 2d I-brane worldsheet.  Being massless, the chiral fermions are necessarily free. Their action is then given (modulo an overall coupling constant) by
\be
I = \int d^2z \ \psi^{\dagger} \bar\partial_{A + A'} \psi,
\label{L}
\ee
where $A$ and $A'$ are the restrictions to the I-brane worldsheet ${\bf S}^1_n \times \mathbb R_t$ of the $U(k)$ and $U(N)$ gauge fields associated with the D4-branes and D6-branes, respectively. In fact, the fermions couple (up to certain discrete identifications under the $\mathbb Z_k$ and $\mathbb Z_N$ centers of $U(k)$ and $U(N)$) to the gauge group
\be
U(1) \times SU(k) \times SU(N),
\label{coupling}
\ee
where the $U(1)$ is the anti-diagonal. This point will be relevant shortly. 

At any rate, note that the chiral fermions on the I-brane are actually gauge-anomalous. Nevertheless, by repeating the arguments in~\cite[eqn.~(4.12)--(4.24)]{ATMP} whilst noting that ${\bf S}^1_n$ is topologically equivalent to an ordinary circle, we find that the overall system consisting of the chiral fermions on the I-brane and the gauge fields in the bulk of the D-branes, is gauge-invariant and therefore physically consistent, as expected.

The system of $kN$ $\it{complex}$ free fermions has central charge $kN$ and gives a direct realization of $\widehat{u}(kN)^{(n)}_1$, the integrable module over the $\mathbb Z_n$-twisted affine Lie algebra $\frak{u}(kN)^{(n)}_{\textrm{aff},1}$ of level 1.\footnote{To understand this claim, see~\cite[$\S$15.5.6]{CFT text}, and note that (i) the identification under an order $n$ translation of the circle ${\bf S}^1_n$ results in a $\mathbb Z_n$-twist of the underlying affine Lie algebra; (ii) a twisted version of an affine Lie algebra has the same central charge and level as its untwisted version (c.f.~\cite[$\S$3]{abcdefg}). \label{central charge}} Moreover, there exists the following twisted affine embedding which preserves conformal invariance:
\be
\frak{u}(1)^{(n)}_{{\rm aff}, kN} \otimes \frak{su}(k)^{(n)}_{{\rm aff},N} \otimes \frak{su}(N)^{(n)}_{{\rm aff},k} \subset \frak{u}(kN)^{(n)}_{{\rm aff},1},
\label{affine embedding}
\ee
where this can be viewed as an affine analog of the gauge symmetry in (\ref{coupling}).\footnote{Conventionally, affine Lie algebra embeddings are expressed in the additive notation. Nevertheless, the multiplicative notation will be used here and henceforth so that the  the analogy with the underlying gauge symmetries would make sense.\label{analogy}}  As such, the total Fock space ${\cal F}^{\otimes kN}$ of the $kN$ complex free fermions can be expressed as
\be
{\cal F}^{\otimes kN} = \textrm{WZW}_{\widehat{u}(1)^{(n)}_{kN}} \otimes \textrm{WZW}_{\widehat{su}(k)^{(n)}_N} \otimes \textrm{WZW}_{\widehat{su}(N)^{(n)}_k},
\label{Fock}
\ee
where  $\textrm{WZW}_{\widehat{u}(1)^{(n)}_{kN}}$,  $\textrm{WZW}_{\widehat{su}(k)^{(n)}_N}$ and $\textrm{WZW}_{\widehat{su}(N)^{(n)}_k}$ are the spectra of states furnished by $\widehat{u}(1)^{(n)}_{kN}$, $\widehat{su}(k)^{(n)}_N$ and $\widehat{su}(N)^{(n)}_k$, respectively, which can be realized in the relevant $\it{chiral}$ WZW models.  Consequently, the partition function of the I-brane theory will be expressed in terms of the \emph{chiral} characters of  $\widehat{u}(1)^{(n)}_{kN}$, $\widehat{su}(k)^{(n)}_N$ and $\widehat{su}(N)^{(n)}_k$.

Note that ${\cal F}^{\otimes kN}$ is the Fock space of the $kN$ complex free fermions which have $\it{not}$ yet been coupled to $A$ and $A'$. Upon coupling to the gauge fields, the characters that appear in the overall partition function of the I-brane theory will be reduced. In a generic situation, the free fermions will couple to the gauge group $U(1) \times SU(k) \times SU(N)$ (see  (\ref{coupling})). However, in our case, only the $U(k)$ gauge field associated with the D4-branes is dynamical; the $U(N)$ gauge field associated with the D6-branes should $\it{not}$ be dynamical as the geometry of $TN^{R \to 0}_N$ is fixed in our description -- the center-of-mass degrees of freedom of the $N$ NS5-branes which give rise to the $TN^{R \to 0}_N$ geometry via steps (\ref{IIB 3}) and (\ref{IIA 4}), are frozen. Also, it has been argued in \cite{Vafa et al} that for a multi-Taub-NUT space whose ${\bf S}^1$ fiber has a finite radius at infinity, there can be additional topological configurations of the gauge field (in the form of monopoles that go around the ${\bf S}^1$ fiber at infinity) which render the $U(1)$ gauge field non-dynamical; nonetheless, it is clear that one cannot have such configurations when the radius of the ${\bf S}^1$ fiber at infinity is either infinitely large or zero. Therefore, the free fermions will, in our case, couple dynamically to the gauge group $U(1) \times SU(k)$. Schematically, this means that we are dealing with the following partially gauged CFT
\be
\frak{u}(kN)^{(n)}_{{\rm aff},1} / [\frak{u}(1)^{(n)}_{{\rm aff},kN} \otimes \frak{su}(k)^{(n)}_{{\rm aff},N}].
\label{partially gauged CFT - AB}
\ee
In particular, the $\frak{u}(1)^{(n)}_{{\rm aff},kN}$ and $\frak{su}(k)^{(n)}_{{\rm aff},N}$ chiral WZW models will be replaced by the corresponding topological $G/G$ models. As a result, all chiral characters except those of  $\widehat{su}(N)^{(n)}_k$ which appear in the overall partition function of the uncoupled free fermions system on the I-brane, will reduce to constant complex factors after coupling to the dynamical $SU(k)$ and $U(1)$ gauge fields. Thus, modulo these constant complex factors which serve only to shift the energy levels of the ground states by numbers dependent on the highest affine weights of  $\widehat{u}(1)^{(n)}_{kN}$ and $\widehat{su}(k)^{(n)}_N$, the $\it{effective}$ overall partition function of the I-brane theory will be expressed solely in terms of the chiral characters of  $\widehat{su}(N)^{(n)}_k$. 

In summary, the sought-after spectrum of spacetime BPS states in the M-theory compactification (\ref{M-theory 7 discussion}) would be realized by $\textrm{WZW}_{\widehat{su}(N)^{(n)}_k}$. This observation is indeed physically consistent because according to  footnote~\ref{worldvolume ground state}, the spacetime BPS states satisfy $H=P$ -- here, $H$ and $P$ are the Hamiltonian and momentum operators which generate translations along $\mathbb R_t$ and ${\bf S}^1_n$, respectively -- while a chiral WZW model on ${\bf S}^1_n \times \mathbb R_t$, having no right-moving excitations, has a spectrum whereby $H=P$.  

\bigskip\noindent{\it A Geometric Langlands Duality for Surfaces for the $A_{N-1}$ Groups} 

Let us now consider $n=1$ whence there is no ``twist'' at all, i.e., $\widehat{su}(N)^{(n)}_k$ is simply $\widehat{su}(N)_k$, the integrable module over the untwisted affine Lie algebra ${\frak su}(N)_{{\rm aff}, k}$ of level $k$.  Then, unitarity of any WZW model requires that $\textrm{WZW}_{\widehat{su}(N)_k}$ be generated by dominant highest weight modules over $\frak{su}(N)_{{\rm aff},k}$, such that a generic state in any one such module can be expressed as~\cite{CFT text}
\be
|{\tilde \mu}'\rangle = E^{- \tilde\alpha}_{-n} \dots E^{-\tilde\beta}_{-m} |\tilde \lambda\rangle, \qquad \forall~~~n,m \geq 0 ~~ \textrm{and} ~~ \tilde\alpha, \tilde\beta > 0.
\label{generic WZW state}
\ee
Here, the $E^{-\tilde\gamma}_{-l}$'s are lowering operators that correspond to the respective modes of the currents of $\frak{su}(N)_{{\rm aff},k}$ (in a Cartan-Weyl basis) which are associated with the complement of the Cartan subalgebra; $|\tilde \lambda \rangle$ is a highest weight state associated with a dominant highest affine weight $\tilde \lambda$; $\tilde \mu' = \tilde \lambda -\tilde \alpha \dots -\tilde\beta$ is an affine weight in the weight system ${\widehat\Omega}_{\tilde\lambda}$ of $\widehat{su}(N)^{\tilde\lambda}_k$ -- the module of dominant highest weight $\tilde\lambda$ of level $k$ -- which is not necessarily dominant; and $\tilde\alpha,\tilde\beta$ are positive affine roots.  

Note that each module labeled by a dominant highest affine weight $\tilde\lambda$ can be decomposed into a direct sum of finite-dimensional subspaces each spanned by states of the form $|\tilde\mu'\rangle$ for $\it{all}$ possible positive affine roots $\tilde\alpha, \dots, \tilde\beta$. These finite-dimensional subspaces of states are the $\tilde\mu'$-weight spaces $\widehat{su}(N){}^{\tilde\lambda}_{k,\tilde\mu'} \subset \widehat{su}(N)^{\tilde\lambda}_k$. Note at this point that there is a Weyl group symmetry on these weight spaces that maps $\tilde\mu'$ to a dominant weight $\tilde\mu$ in ${\widehat\Omega}_{\tilde\lambda}$ which also leaves the chiral character of $\widehat{su}(N)^{\tilde\lambda}_k$ and thus, the partition function of the chiral WZW model, invariant.\footnote{See~\cite[eqns.~(14.143), (14.145), (14.165), (14.166) and (15.119)]{CFT text}, noting that $z_j$ in \emph{loc.~cit.} corresponds to the Coulomb moduli in our story which must therefore be set to zero since the $N$ D6-branes are coincident.} As such, one can also express the spectrum of states of the chiral WZW model as
 \be
 \textrm{WZW}_{\widehat{su}(N)_k} = \bigoplus_{\tilde\lambda, \tilde\mu} \, \textrm{WZW}_{\widehat{su}(N)^{\tilde \lambda}_{k, \tilde\mu}}.
 \label{WZW space-SU(N)}
 \ee

Now, the physical duality of the M-theory compactifications (\ref{M-theory 1 discussion}) and (\ref{M-theory 7 discussion}) means that their respective spacetime BPS spectra ought to be equivalent, i.e., $ \textrm{WZW}_{\widehat{su}(N)_k}$ ought to be equal to $\cal H_{\rm BPS}$ of (\ref{BPS-M}). Indeed, since $\frak{su}(N)_{\textrm{aff}}$ is isomorphic to its Langlands dual counterpart  $\frak{su}(N)^\vee_{\textrm{aff}}$, $\tilde\lambda$ and $\tilde\mu$ are also \emph{dominant weights} of the \emph{Langlands dual} affine Kac-Moody group $SU(N)^\vee_{\rm aff}$ whence we can identify them with $\lambda$ and $\mu$ of~(\ref{BPS-M}), respectively;  moreover, both $\cal H_{\rm BPS}$ and $ \textrm{WZW}_{\widehat{su}(N)_k}$ are labeled by $k$. Specializing to the $\{\lambda, \mu\}$-sector of the spectra of spacetime BPS states, we can thus write 
\be
{\cal H^{\lambda, \mu}_{\rm BPS}} = \textrm{WZW}_{\widehat{su}(N)^{ \lambda}_{k, \mu}}.
\label{H=WZW}
\ee
As $\textrm{WZW}_{\widehat{su}(N)^\lambda_{k, \mu}}$ is furnished by ${\widehat{su}(N)^\lambda_{k, \mu}}$, and since $\frak{su}(N)_{\textrm{aff}} \simeq \frak{su}(N)^\vee_{\textrm{aff}} $ whence $\widehat{su}(N)^\lambda_{k, \mu}$ is isomorphic to the submodule  ${^L\widehat{su}(N)^\lambda_{k, \mu}}$ over $\frak{su}(N)^\vee_{\textrm{aff}}$, via (\ref{BPS-M}), we can also express (\ref{H=WZW}) as 
\be
\boxed{{\rm IH}^\ast {\cal U}({\cal M}^{\lambda}_{SU(N), \mu}(\mathbb R^4/ \mathbb Z_k)) = {^L\widehat{su}(N)^\lambda_{k, \mu}}}
\label{GL-relation A}
\ee
Note that this is exactly~\cite[Conjecture 4.14(3)]{BF} for simply-connected $G=SU(N)$! This completes our purely physical derivation of a geometric Langlands duality for surfaces for the $SU(N) = A_{N-1}$ groups. 

\bigskip\noindent{\it An Identity of the Dimension of the Intersection Cohomology of the Moduli space of $A_{N-1}$-Instantons on $\mathbb R^4 / \mathbb Z_k$}

Let us now revisit the partition function (\ref{Z_SU(N)}). For simplicity, let us focus on a particular $\lambda'$-sector, where $\lambda' = (k, \bar\lambda', 0)$; that is, consider
\be
 Z_{SU(N), \lambda'}^{\rm BPS} (q)  =   q^{m_{\lambda'}}  \, \sum_{\bar\mu'}  \sum_{m \geq 0}  \, {\rm dim} \, {\rm IH}^\ast {\cal U}({\cal M}^{\lambda', m}_{SU(N), \bar\mu'}(\mathbb R^4/ \mathbb Z_k)) \,  q^m,
 \label{Z_SU(N), revisit}
 \ee 
where $q= e^{2 \pi i \tau}$, and $m_{\lambda'}$ is as given in (\ref{m}). From (\ref{GL-relation A}), we have\footnote{Here, we recall that for any $\lambda = (k, \bar \lambda, i)$ and $\mu = (k, \bar \mu, j)$, we have $m = k(i-j)$ whereby $(i-j) \in  \mathbb Z_{\geq 0}$. Thus, for $\lambda' = (k, \bar{\lambda'}, 0)$, we have $\mu' = (k, \bar\mu', j')$ such that the integer $j' = - {m \over k} \leq 0$, where $-j'$ is known as the grade of the $\mu'$-string in the mathematical literature, or the energy level of the $| \mu' \rangle$ state in the physical context. \label{i-j-A}} 
\be
{\rm dim} \, {\rm IH}^\ast {\cal U}({\cal M}^{\lambda', m}_{SU(N), \bar\mu'}(\mathbb R^4/ \mathbb Z_k)) = {\rm mult}_{\lambda'} (\bar\mu')\vert_{ {m'}}, 
\ee
where ${\rm mult}_{\lambda'} (\bar\mu')\vert_{{m'}}$ is the multiplicity of the $|\mu' \rangle$ state of non-negative energy level $m' = {m / k}$ in $^L\widehat{su}(N)^{\lambda'}_{k, \mu'}$. Consequently, via (\ref{m})--(\ref{c}), we can write
\begin{eqnarray}
 Z_{SU(N), \lambda'}^{\rm BPS} (\tilde q)   = {\tilde q}^{m'_{\lambda'} - c/24} \, \sum_{\bar\mu'}  \sum_{m' \geq 0}  \, {\rm mult}_{\lambda'} (\bar\mu')\vert_{m'} \,  {\tilde q}^{m'}  =   \chi_{^L{{\widehat {su}}(N)}_k}^{\lambda'} (\tilde q),
 \label{chiral WZW SU(N)}
\end{eqnarray}
where $\tilde q = e^{2 \pi i \tilde \tau}$ and $\tilde \tau = k \tau$.  Here
\be
 \chi_{^L{{\widehat {su}}(N)}_k}^{\lambda'} (\tilde q) =  {\rm Tr}_{\lambda'} \, {\tilde q}^{L_0 + m'_{\lambda'} - c /24}, 
 \label{chi-SU(N)}
 \ee
 and
 \be
 m'_{\lambda'} = h'_{ \lambda'} -{ (c'_{\lambda'} - c) \over 24},
\ee
while
\be
c= kN, \quad h'_{ \lambda'} = {h_{ \lambda'} \over k}, \quad {\rm and} \quad c'_{\lambda'} = {c_{ \lambda'} \over k},
\label{c-SU(N)}
\ee
where $h_{ \lambda'}$ and $c_{ \lambda'}$ are as given in (\ref{h}) and (\ref{c}), respectively. Also, $L_0$ -- whose  eigenvalue is $m' \in \mathbb Z_{\geq 0}$ -- can be interpreted as the Hamiltonian operator of a 2d theory that is effectively defined on a torus of modulus $\tilde\tau$. Hence, it is clear from (\ref{chiral WZW SU(N)})--(\ref{c-SU(N)}) that $ Z_{SU(N), \lambda'}^{\rm BPS}$ is equal to the $\lambda'$-sector of the partition function of a chiral ${\frak {su}(N)}^\vee_{\rm aff}$ WZW model on ${\bf S}^1_n \times \mathbb R_t$ with (i) central charge $kN$; (ii) ground state energy level shifted by a number $m'_{\lambda'}$. This observation is consistent with our earlier conclusion about the I-brane partition function, as expected. 

Let us now consider the modified partition function 
\be
{\tilde Z}_{SU(N), \lambda'}^{\rm BPS} (\tilde q) = {\tilde q}^{\tilde m_{\lambda'}}  Z_{SU(N), \lambda'}^{\rm BPS} (\tilde q).
\label{mod pf}
\ee
where
 \be
\tilde m_{\lambda'} = {(k-1) h'_{\lambda'} + {(c'_{\lambda'} - c'_{SU(N)^\vee, k}) \over 24 }} \quad {\rm while} \quad c'_{SU(N)^\vee, k} =  {k \, {\rm dim} \, \frak{su}(N)^\vee  \over{ (k +h)}}.
 \ee
Notice that ${\tilde Z}_{SU(N), \lambda'}^{\rm BPS}$ is just $Z_{SU(N), \lambda'}^{\rm BPS}$ but with instanton number shifted by $\tilde m_{\lambda'} $. In the dual 2d theory picture, this is tantamount to a trivial redefinition of the ground state energy level. Hence, ${\tilde Z}_{SU(N), \lambda'}^{\rm BPS}$ and $Z_{SU(N), \lambda'}^{\rm BPS}$ can be thought to define the ``same'' physical theory.

From (\ref{chi-SU(N)}), one can see that $ {\tilde \chi}_{^L{{\widehat {su}}(N)}_k}^{\lambda'} =  {\tilde q}^{\tilde m_{\lambda'}} \chi_{^L{{\widehat {su}}(N)}_k}^{\lambda'}$ is a character of $^L\widehat{su}(N)^{\lambda'}_{k}$, where ${\tilde m_{\lambda'}} +  h'_{\lambda'}  - c'_{\lambda'} /24$ is the corresponding modular anomaly. As such, (\ref{chiral WZW SU(N)}) would mean that the partition function ${\tilde Z}_{SU(N), \lambda'}^{\rm BPS} $ ought to transform as a representation of the modular group $SL(2, \mathbb Z)$; specifically, we have (c.f.~\cite[eqn.~(14.235)]{CFT text}) the relation
\be
{ \tilde Z}_{SU(N), \lambda'}^{\rm BPS} (- 1 / \tilde \tau) = \sum_{\lambda} {\cal S}_{\lambda' \lambda} \, { \tilde Z}_{SU(N), \lambda}^{\rm BPS} (\tilde \tau), 
\label{3.33}
\ee
where ${\cal S}$ is a $\tilde \tau$-independent unitary matrix (given by~\cite[ eqn.~(14.217)]{CFT text}) associated with the \emph{Langlands dual} affine Lie algebra $\frak {su}(N)^\vee_{\rm aff}$, which represents the $SL(2,\mathbb Z)$ transformation $S: \tilde\tau \to - 1 / \tilde\tau$ in the space of $\lambda$-vector-valued partition functions ${ \tilde Z}_{SU(N), \lambda}^{\rm BPS}$.  

Via (\ref{mod pf}) and (\ref{Z_SU(N), revisit}), the relation (\ref{3.33}) implies, in the limit of large $k$, the following identity involving the intersection cohomology of the moduli space of $SU(N)$-instantons on $\mathbb R^4/ \mathbb Z_k$: 
\be
\boxed{\sum_{\bar\mu'}  \sum_{m \geq 0}  \, {\rm dim} \, {\rm IH}^\ast {\cal U}({\cal M}^{\lambda', m}_{SU(N), \bar\mu'}(\mathbb R^4/ \mathbb Z_k))  =    \sum_{\lambda}  \sum_{\bar\mu}  \sum_{m \geq 0} {\cal S}^m_{\lambda' \lambda}  \,  {\rm dim} \, {\rm IH}^\ast {\cal U}({\cal M}^{\lambda, m}_{SU(N), \bar\mu}(\mathbb R^4/ \mathbb Z_k))}
\ee
where the components ${\cal S}^m_{\lambda' \lambda}$ are given by
\be
\boxed{{\cal S}^m_{\lambda' \lambda}  = {\hat q}^{\tilde m_{\lambda'}\vert_{k \gg 1}} {\tilde q}^{m + \tilde m_{\lambda}+ m_{\lambda}}{\cal S}_{\lambda' \lambda}}
\ee
Here, $\hat q = e^{2 \pi i / \tilde \tau}$. (See also footnote~\ref{i-j-A}.)

In other words, in the limit of large $k$, the total dimension of the intersection cohomology of the component of the moduli space of $A_{N-1}$-instantons on $\mathbb R^4/ \mathbb Z_k$ labeled by a highest weight $\lambda'$ or $\lambda$  -- and therefore, the dimension of the corresponding sector of the Hilbert space of spacetime BPS states -- is found to be intimately related to one another via $\frak {su}(N)^\vee_{\rm aff}$-dependent unitary modular transformations!

\bigskip\noindent{\it A Geometric Langlands Duality for Surfaces for the $B_{N/2}$ Groups}

Let us now restrict ourselves to \emph{even} $N$, and consider $n=2$ whence there is a ``$\mathbb Z_2$-twist'', i.e., the relevant module is $\widehat{su}(N)^{(2)}_k$, the integrable module over the $\mathbb Z_2$-twisted affine Lie algebra ${\frak su}(N)^{(2)}_{{\rm aff}, k}$ of level $k$.  Then, unitarity of any WZW model requires that $\textrm{WZW}_{\widehat{su}(N)^{(2)}_k}$ be generated by dominant highest weight modules over $\frak{su}(N)^{(2)}_{{\rm aff},k}$. By repeating the arguments that led us to write (\ref{generic WZW state})--(\ref{GL-relation A}) in the untwisted case, whilst noting that the Weyl group symmetry mentioned therein persists in this case to map non-dominant affine weights $\tilde \mu'_\nu$ to dominant ones $\tilde \mu_\nu$ even though the grading of $\tilde \mu'_\nu$ (captured by its last index $\tilde j'_\nu$)  may not be integral, we find that we can express the spectrum of states of the corresponding chiral WZW model as
 \be
 \textrm{WZW}_{\widehat{su}(N)^{(2)}_k} =  \bigoplus_{\tilde\lambda} \bigoplus_{\nu =0,1} \, \bigoplus_{\tilde\mu_\nu} \, \overline{\textrm{WZW}}_{\widehat{su}(N)^{(2), \tilde \lambda}_{k, \tilde\mu_\nu}}.
 \label{WZW space-SO(N+1)}
 \ee
 Here, the overhead bar means that we project onto $\mathbb Z_2$-invariant states (as required of twisted CFT's); $\nu =0$ or $1$ indicates that the sector is untwisted or twisted, respectively; $\tilde\lambda$ and $\tilde\mu_\nu$ are the (un)twisted  dominant affine weights of the $\mathbb Z_2$-twisted affine Kac-Moody group $SU(N)^{(2)}_{\rm aff}$ of level $k$; the space $\widehat{su}(N)^{(2), \tilde \lambda}_{k, \tilde\mu_\nu}$ is the  $\tilde\mu_\nu$-weight space of $\widehat{su}(N)^{(2), \tilde \lambda}_{k}$, the module over ${\frak su}(N)^{(2)}_{{\rm aff}, k}$ of dominant highest weight $\tilde\lambda$ of level $k$.

Now, the physical duality of the M-theory compactifications (\ref{M-theory 1 discussion}) and (\ref{M-theory 7 discussion}) means that their respective spacetime BPS spectra ought to be equivalent, i.e., $ \textrm{WZW}_{\widehat{su}(N)^{(2)}_k}$ ought to be equal to $\cal H^{\rm eff}_{\rm BPS}$ of (\ref{HBPS-eff}). Indeed, since $\frak{su}(N)^{(2)}_{\textrm{aff}}$ is isomorphic to $\frak{so}(N+1)^\vee_{\textrm{aff}}$, it would mean that $\tilde\lambda$ and $\tilde\mu_\nu$ are also \emph{dominant weights} of the \emph{Langlands dual} affine Kac-Moody group $SO(N+1)^\vee_{\rm aff}$ whence we can identify them with $\lambda$ and $\mu_\nu$ of~(\ref{HBPS-eff}), respectively;  moreover, both $\cal H^{\rm eff}_{\rm BPS}$ and $ \textrm{WZW}_{\widehat{su}(N)^{(2)}_k}$ are labeled by $k$. Specializing to the $\{\lambda, \mu_\nu \}$-sector of the spectra of spacetime BPS states, we can therefore write 
\be
{\overline{\cal H}}^{\lambda, \mu_\nu}_{\rm BPS} = \overline{\textrm{WZW}}_{\widehat{su}(N)^{(2), \lambda}_{k, \mu_\nu}}.
\label{H=WZW-SO(N+1)}
\ee
As $\overline{\textrm{WZW}}_{\widehat{su}(N)^{(2), \lambda}_{k, \mu_\nu}}$ is furnished by the $\mathbb Z_2$-invariant projection ${\widehat{su}(N)^{(2),  \lambda}_{k, \mu_\nu}}\vert_{\mathscr P_2}$ of $\widehat{su}(N)^{(2),  \lambda}_{k, \mu_\nu}$, and since $\frak{su}(N)^{(2)}_{\textrm{aff}} \simeq \frak{so}(N +1)^\vee_{\textrm{aff}} $ whence ${\widehat{su}(N)^{(2),  \lambda}_{k, \mu_\nu}}\vert_{\mathscr P_2}$ is isomorphic to the submodule ${^L\widehat{so}(N+1)^{\lambda}_{k, \mu_\nu}}$ over $\frak{so}(N+1)^\vee_{\textrm{aff}}$, via (\ref{HBPS-eff}), we can also express (\ref{H=WZW-SO(N+1)}) as 
\be
\boxed{\overline{{\rm IH}^\ast {\cal U}}({\cal M}^{\lambda}_{SO(N+1), \mu_\nu}(\mathbb R^4/ \mathbb Z_k)) =  {^L\widehat{so}}(N+1)^{\lambda}_{k, \mu_\nu}}
\label{GL-relation B}
\ee
for $\nu = 0$ and $1$. Thus, we have arrived at a $G=SO(N+1)$ generalization of~\cite[Conjecture 4.14(3)]{BF}! This completes our purely physical derivation of a geometric Langlands duality for surfaces for the $SO(N+1) = B_{N/2}$ groups.

\bigskip\noindent{\it A Langlands Duality of the Dimension of the Intersection Cohomology of the Moduli Space of $B_{N/2}$-Instantons on A-Type ALE Spaces}

Let us now revisit the partition function given by (\ref{Z-SO(N+1)-1})--(\ref{Z-SO(N+1)-3}). For simplicity, let us focus on a particular $\lambda'$-sector, where $\lambda' = (k, \bar\lambda', 0)$; that is, consider
\be
Z^{\rm BPS}_{SO(N+1), \lambda'} (q) = q^{m_{\lambda'}} \sum_{\nu =0,1} \sum_{\bar \mu'_\nu}    \sum_{m_\nu \geq 0}   {\rm dim} \, \overline {{\rm IH}^\ast {\cal U}}({\cal M}^{\lambda', m_\nu}_{SO(N+1), \bar\mu'_\nu}(\mathbb R^4/ \mathbb Z_k)) \, q^{m_\nu}, 
\label{Z-SO(N+1)-revisit}
\ee
where $q= e^{2 \pi i \tau}$, and $m_{\lambda'}$ takes the form given in (\ref{m}). From our discussion leading up to (\ref{GL-relation B}), we have\footnote{Here, we recall that for any $\lambda = (k, \bar \lambda, i)$ and $\mu_\nu = (k, \bar \mu_\nu, j_\nu)$, we have $m_\nu = kn'(i-j_\nu)$ whereby $(i-j_\nu) \in \mathbb Z_{\geq 0} + {\nu \over 2}$ and $n' =2$ or $1$ if $N > 2$ or $N = 2$, respectively. Thus, for $\lambda' = (k, \bar{\lambda'}, 0)$, we have $\mu'_\nu = (k, \bar\mu'_\nu, j'_\nu)$ such that $j'_\nu = - {m_\nu \over {kn'}} \leq 0$, where $-j'_\nu$ is known as the grade of the $\mu'_\nu$-string in the mathematical literature, or the energy level of the $| \mu'_\nu \rangle$ state in the physical context. \label{i-j-B}} 
\be
{\rm dim} \, \overline{{\rm IH}^\ast {\cal U}}({\cal M}^{\lambda', m_\nu}_{SO(N+1), \bar\mu'_\nu}(\mathbb R^4/ \mathbb Z_k)) = \overline{{\rm mult}}_{\lambda'} (\bar\mu'_\nu)\vert_{ {m'_\nu}}, 
\ee
where $\overline{\rm mult}_{\lambda'} (\bar\mu'_\nu)\vert_{{m'_\nu}}$ is the multiplicity of the $|\mu'_\nu \rangle$ state of non-negative energy level $m'_\nu = {m_\nu / kn'}$ in ${\widehat{su}(N)^{(2),  \lambda'}_{k, \mu'_\nu}}\vert_{\mathscr P_2}$, and $n' =2$ or $1$ if $N > 2$ or $N = 2$, respectively. Consequently, via (\ref{m})--(\ref{c}), we can write
\begin{eqnarray}
 Z_{SO(N+1), \lambda'}^{\rm BPS} (\tilde q)   = {\tilde q}^{m'_{\lambda'} - c/24} \, \sum_{\nu =0,1}\sum_{\bar\mu'_\nu}  \sum_{m'_\nu \geq 0}  \, \overline{\rm mult}_{\lambda'} (\bar\mu'_\nu)\vert_{m'_\nu} \,  {\tilde q}^{m'_\nu}  =  \sum_{\nu =0,1} \, \chi_{{{\widehat {su}}(N)}^{(2)}_{k}}^{\lambda',\nu} (\tilde q),
 \label{chiral WZW SO(N+1)}
\end{eqnarray}
where $\tilde q = e^{2 \pi i \tilde \tau}$ and $\tilde \tau = kn' \tau$.  Here
\be
\chi_{{{\widehat {su}}(N)}^{(2)}_{k}}^{\lambda',\nu} (\tilde q) =  {\rm Tr}_{\lambda'} \, \mathscr P_2 \, {\tilde q}^{L_{0, \nu} + m'_{\lambda'} - c /24}, 
 \label{chi-SO(N+1)}
 \ee
 where as before, $\mathscr P_2$ singles out the $\mathbb Z_2$-invariant states, and
 \be
 m'_{\lambda'} = h'_{ \lambda'} -{ (c'_{\lambda', \tilde b} - c) \over 24}.
\ee
The constants are
\be
c= kN, \quad h'_{ \lambda'} = {(\bar \lambda', \bar\lambda' + 2 \rho) \over 2kn'(k + h^\vee)}, \quad c'_{\lambda', \tilde b} = {-{24 \tilde b {(\bar \lambda', \bar \lambda')} \over kn'}  +  {12 (\bar \lambda', \bar \lambda' + 2 \rho) \over {kn'(k + h^\vee)}}},
\label{c-SO(N+1)}
\ee
such that $\tilde b$ is some positive real constant (first introduced in (\ref{a})), and $\rho$ and $h^\vee$ are the Weyl vector and dual Coxeter number associated with ${\frak {su}(N)}^{(2)}_{\rm aff}$, respectively.  Also, $L_{0, \nu}$ -- whose  eigenvalue is $m'_\nu \in \mathbb Z_{\geq 0} + {\nu \over 2}$ -- can be interpreted as the Hamiltonian operator of a 2d theory that is effectively defined on a torus of modulus $\tilde \tau$. Hence, it is clear from (\ref{chiral WZW SO(N+1)})--(\ref{c-SO(N+1)}) that $ Z_{SO(N+1), \lambda'}^{\rm BPS}$ is equal to the $\lambda'$-sector of the partition function of a chiral ${\frak {su}}(N)^{(2)}_{\rm aff}$ WZW model on ${\bf S}^1_n \times \mathbb R_t$ with (i) central charge $kN$; (ii) ground state energy level shifted by a number $m'_{\lambda'}$.  This observation is consistent with our earlier conclusion about the I-brane partition function, as expected. 

Let us now consider the modified partition function 
\be
{\tilde Z}_{SO(N+1), \lambda'}^{\rm BPS} (\tilde q) = {\tilde q}^{\tilde m_{\lambda'}}  Z_{SO(N+1), \lambda'}^{\rm BPS} (\tilde q),
\label{mod pf-SO(N+1)}
\ee
 where (c.f.~\cite[eqns.~(12.7.5) and (13.11.5)]{Kac})
 \be
\tilde m_{\lambda'} = -h'_{\lambda'} + {c'_{\lambda', \tilde b} \over 24} +{ {|\bar\lambda' + \rho|^2} \over 2(k+ h^\vee)} - {{\rm dim} \, {\frak{su}(N)} \over 48}.
\label{tilde m-B}
 \ee
Notice that ${\tilde Z}_{SO(N+1), \lambda'}^{\rm BPS}$ is just $Z_{SO(N+1), \lambda'}^{\rm BPS}$ but with instanton number shifted by $\tilde m_{\lambda'} $. In the dual 2d theory picture, this is tantamount to a trivial redefinition of the ground state energy level. Hence, ${\tilde Z}_{SO(N+1), \lambda'}^{\rm BPS}$ and $Z_{SO(N+1), \lambda'}^{\rm BPS}$ can be thought to define the ``same'' physical theory.

From (\ref{chi-SO(N+1)}), one can see that ${\tilde \chi}_{{{\widehat {su}}(N)}^{(2)}_{k}}^{\lambda',\nu}  = {\tilde q}^{\tilde m_{\lambda'}} \chi_{{{\widehat {su}}(N)}^{(2)}_{k}}^{\lambda',\nu}$ is a ($\mathbb Z_2$-invariant) character of the $\nu$-sector of $\widehat{su}(N)^{{(2)}, \lambda'}_{k}$, where $\tilde m_{\lambda'} + h'_{\lambda'} - {c'_{\lambda', \tilde b} / 24}$ is the corresponding modular anomaly. As such, (\ref{mod pf-SO(N+1)}), (\ref{chiral WZW SO(N+1)}) and~\cite[ Theorem 13.9]{Kac} mean that the partition function ${\tilde Z}_{SO(N+1), \lambda'}^{\rm BPS} $ ought to transform under $S: \tilde \tau \to - 1 / \tilde\tau$ as follows: 
\be
{ \tilde Z}_{SO(N+1), \lambda'}^{\rm BPS} (- 1 / \tilde \tau)  = \sum_{\xi} {\cal S}_{\lambda' \xi} \, {\tilde \chi}_{{{\widehat {so}}(N+2)}^{(2)}_{k}}^{\xi}  (\tilde \tau / 2).
\label{3.33-SO(N+1)}
\ee
Here, ${\cal S}$ is a $\tilde \tau$-independent matrix (given in~\cite[Theorem 13.9]{Kac});  $\xi$ is a dominant  highest weight of the $\mathbb Z_2$-twisted affine Kac-Moody group $SO(N+2)^{(2)}_{\rm aff}$ of level $k$; ${\tilde \chi}_{{{\widehat {so}}(N +2)}^{(2)}_{k}}^{\xi}  = {\tilde q}^{\tilde m_{\xi}} \chi_{{{\widehat {so}}(N+2)}^{(2)}_{k}}^{\xi}$, where ${\tilde m_{\xi}}$ is as in  (\ref{tilde m-B}) but with ${\frak{su}(N)}$ replaced by $\frak {so}(N+2)$.  Notice that the group type on the LHS and RHS of (\ref{3.33-SO(N+1)}) are not the same; nevertheless, the characters on the RHS of (\ref{3.33-SO(N+1)}) will be given by the partition function ${ \tilde Z}_{USp(N), \xi}^{\rm BPS} (\tilde \tau / 2)$ associated with $USp(N)$-instantons on $\mathbb R^4 / \mathbb Z_{k}$ (see discussion leading up to (\ref{GL-relation C})); that is,  
\be
{ \tilde Z}_{SO(N+1), \lambda'}^{\rm BPS} (- 1 / \tilde \tau)  = \sum_{\xi} {\cal S}_{\lambda' \xi} \, { \tilde Z}_{USp(N), \xi}^{\rm BPS} (\tilde \tau / 2).
\label{3.33-SO(N+1) dual}
\ee

Via (\ref{mod pf-SO(N+1)}), (\ref{Z-SO(N+1)-revisit}) and (\ref{Z-USp(2N-2)-revisit}), the relation (\ref{3.33-SO(N+1) dual}) implies, in the limit of large $k$, the following identity involving the intersection cohomology of the moduli space of instantons: 
\be
\label{IC-SO-USp}
\hspace{-0.0cm}\boxed{\sum_{\nu =0,1} \sum_{\bar \mu'_\nu}    \sum_{m_\nu \geq 0}   {\rm dim} \, \overline {{\rm IH}^\ast {\cal U}}({\cal M}^{\lambda', m_\nu}_{SO(N+1), \bar\mu'_\nu}(\mathbb R^4/ \mathbb Z_k)) 
  =  \sum_{\xi} \sum_{\delta =0,1} \sum_{\bar\zeta_\delta}  \sum_{m_\delta \geq 0}   {\cal S}^{m_\delta}_{\lambda' \xi}  \,  {\rm dim} \, \overline{{\rm IH}^\ast {\cal U}}({\cal M}^{\xi, m_\delta}_{USp(N), \bar\zeta_\delta}(\mathbb R^4/ \mathbb Z_{k}))}
\ee
where the components ${\cal S}^{m_\delta}_{\lambda' \xi}$ are given by
\be
\boxed{{\cal S}^{m_\delta}_{\lambda' \xi}  = {\hat q}^{\tilde m_{\lambda'}\vert_ {k \gg 1}}{\tilde q}^{(m_\delta + \tilde m_{\xi}+ m_{\xi})/2}{\cal S}_{\lambda' \xi}}
\ee
Here, ${\hat q} = e^{2 \pi i / \tilde \tau}$; $\lambda' = (k, \bar\lambda', 0)$ and $\mu'_\nu = (k, \bar\mu'_\nu, j_\nu)$ are dominant coweights of the affine Kac-Moody group $SO(N+1)_{\rm aff}$ of level $k$, where $\bar\lambda'$ and $\bar \mu'_\nu$ are the corresponding dominant coweights of $SO(N+1)$, and for $N > 2$, ${m_\nu \over 2k} = - j_\nu \in \mathbb Z_{\geq 0} + {\nu \over 2}$ (see footnote~\ref{i-j-B}); $\xi = (k, \bar \xi, 0)$ and $\zeta_\delta = (k, \bar\zeta_\delta, j_\delta)$  are dominant coweights of the affine Kac-Moody group $USp(N)_{\rm aff}$ of level $k$, where $\bar \xi$ and $\bar \zeta_\delta$ are the corresponding dominant coweights of $USp(N)$, and ${m_\delta \over 2k} = - j_\delta \in \mathbb Z_{\geq 0} + {\delta \over 2}$.

At any rate, it is clear from (\ref{IC-SO-USp}) that in the limit of large $k$, the total dimension of the intersection cohomology of the moduli space of $G$-instantons on $\mathbb R^4/ \mathbb Z_k$ in the $\lambda'$-sector, can be expressed in terms of the dimensions of the intersection cohomology of the various components of the moduli space of $G^\vee$-instantons on $\mathbb R^4/ \mathbb Z_{k}$, where $G = SO(N+1)$ with even $N$. In other words, we have a \emph{Langlands duality }of the dimension of the intersection cohomology of the moduli space of $B_{N/2}$-instantons on A-type ALE spaces!

\newsubsection{An Equivalence of Spacetime BPS Spectra and a Geometric Langlands Duality for Surfaces for the $C$--$D$--$G$ Groups}

We shall now derive, purely physically, a geometric Langlands duality for surfaces for the $C$--$D$--$G$ groups. For a start, note that in $\S$2.2, we showed that the following six-dimensional M-theory compactification on the five-manifold $X_5 = \mathbb R^4 / \mathbb Z_k  \times {{\bf S}^1_n}$  with $N$ coincident M5-branes and an OM5-plane around it,
\be
 \textrm{M-theory}:\quad    \underbrace{\mathbb R^4 / \mathbb Z_k  \times {{\bf S}^1_n} \times \mathbb R_t}_{\textrm{$N$ M5-branes/OM5-plane}} \times \mathbb R^{5},
 \label{M-theory 1 OM5 discussion}
 \ee
where we evoke a $\mathbb Z_n$-outer-automorphism of $\mathbb R^4 / \mathbb Z_k$ (and of the geometrically-trivial $\mathbb R^5 \times \mathbb R_t$ spacetime) as we go around the ${\bf S}^1_n$ circle and identify the circle under an order $n$ translation, is\emph{ physically dual} to the following six-dimensional M-theory compactification on  the five-manifold $\tilde X_5 = {\bf S}^1_n  \times SN_N^{R\to 0}$ with $k$ coincident M5-branes around it,
\be
\textrm{M-theory}:\quad   {\mathbb R^{5}} \times \underbrace{\mathbb R_t \times {{\bf S}^1_n}  \times SN_N^{R\to 0}  }_{\textrm{$k$ M5-branes}},
 \label{M-theory 7 OM5 discussion}
 \ee
where there is a nontrivial $\mathbb Z_n$-outer-automorphism of $SN_N^{R \to 0}$ as we go around the ${\bf S}^1_n$ circle of radius $R_s$. 

The case at hand is almost the same as that in the previous subsection except that we now have (i) an extra OM5-plane in the former compactification with $N$ M5-branes; (ii) the hyperk\"ahler manifold $SN_N^{R \to 0}$ in the latter compactification with $k$ M5-branes. Consequently, as in the previous subsection, the resulting six-dimensional spacetime theories along $ \mathbb R_t \times \mathbb R^5$ in (\ref{M-theory 1 OM5 discussion}) and (\ref{M-theory 7 OM5 discussion}) will both have 6d ${\cN} = (1,1)$ supersymmetry; moreover, the sought-after spacetime BPS states which are annihilated by eight supersymmetry generators of the 6d ${\cN} = (1,1)$ supersymmetry algebra, would be furnished by the ground states of the worldvolume theory of the ($N$ M5)/OM5 stack. To derive purely physically in this case a geometric Langlands duality for surfaces, it suffices to ascertain the spectra of such spacetime BPS states in the M-theory compactifications (\ref{M-theory 1 OM5 discussion}) and (\ref{M-theory 7 OM5 discussion}). To this end, let us now describe the quantum worldvolume theory of the ($N$ M5)/OM5 stack in (\ref{M-theory 1 OM5 discussion}).  (The worldvolume theory of the $k$ M5-branes in (\ref{M-theory 7 OM5 discussion}) has already been described in detail in the previous subsection -- one just has  to replace $TN_N^{R \to 0}$ with $SN_N^{R \to 0}$ in the description therein.)

\bigskip\noindent{\it{Quantum Worldvolume Theory of the  ($N$ M5)/OM5 Stack in (\ref{M-theory 1 OM5 discussion})}}

The presence of the OM5-plane in (\ref{M-theory 1 OM5 discussion}) modifies the worldvolume theory on the stack of $N$ coincident M5-branes discussed earlier, in two ways. First and foremost, instead of an $SU(N)$ symmetry group, we now have an $SO(2N)$ symmetry group; in particular, the low-energy limit of the quantum worldvolume theory is now a non-gravitational 6d $\cN = (2,0)$ $D_{N}$ superconformal field theory of $N$ massless tensor multiplets~\cite{hanany}. As such, one can, in an appropriate gauge, compute the spectrum of ground states of the quantum worldvolume theory, as the spectrum of physical observables in the topological sector of a two-dimensional ${\cal N} = (4,4)$ sigma-model on ${\bf S}^1_n \times \mathbb R_t$ with target the hyperk\"ahler moduli space ${\cal M}_{G}(M)$ of $G$-instantons on $M = \mathbb R^4 / \mathbb Z_k$. Here, we have $G = SO(2N)$ or $USp(2N -2)$ if $n=1$ or $2$, respectively;  $G = G_2$ if $n=3$ and $N = 4$~\cite{Yuji 2 yrs}.

Second, upon compactification along ${\bf S}^1_n$, one can have D0-branes in the worldvolume of the corresponding ($N$ D4)/${\rm O4}^-$ stack that correspond to $G$-instantons.\footnote{The `-' superscript in ${\rm Op}^-$ for any p, just means that it is associated with an orthogonal (as opposed to symplectic) gauge symmetry.} In the case where $n=1$, we do not ``twist'' the theory as we go around ${\bf S}^1_n$; the KK mode, or the D0-brane charge, is then $m$, where $m \in \mathbb Z_{\geq 0}$.  In the case where $n > 1$, we must ``$\mathbb Z_n$-twist'' the theory as we go around ${\bf S}^1_n$;  the KK modes, or the D0-brane charges, are then $m, m + {1\over n}, m+ {2 \over n}, \dots, m + {n-1\over n}$; in other words, we can have bound states of full/fractional D0-branes with the aforementioned charges.

The implications of this $\mathbb Z_n$-twist are as follows. Consider $n=1$ whence $G=SO(2N)$. Since there are no fractional branes, the total number of D0-branes in the entire configuration would be given by a non-negative integer; that is, the instanton number is $d \in \mathbb Z_{\geq 0}$. This is consistent with the fact that although $SO(2N)$ is a nonsimply-connected Lie group, because $M$ is spin, the instanton number is nonetheless integral.\footnote{This can be deduced from a generalization of the analysis in~\cite{Siye Wu}. I would like to thank Siye Wu for his expertise on this matter.}    

For $n=2$ whence $G = USp(2N-2)$, the one-half-fractional D0-branes that result from the ``$\mathbb Z_2$ twist'' can form bound states with full D0-branes. Pairs of such bound states whose corresponding charges are of the forms $m+ {1 \over 2}$ and  $(m +1) - {1 \over 2}$, can further bind together such that the total number $d$ of D0-branes is effectively an integer; in other words, the instanton number is $d \in \mathbb Z_{\geq 0}$. This is consistent with the fact that $USp(2N-2)$ is a simply-connected Lie group whence the instanton number is expected to be integral. 

For $n=3$ and $N=4$ whence $G=G_2$, the one-third-fractional D0-branes that result from the ``$\mathbb Z_3$ twist'' can form bound states with full D0-branes. Pairs of such bound states whose corresponding charges are of the forms $m+ {1 \over 3}$ and  $(m +1) - {1 \over 3}$, can further bind together such that the total number $d$ of D0-branes is effectively an integer; in other words, the instanton number is again $d \in \mathbb Z_{\geq 0}$. This is consistent with the fact that $G_2$ is also a simply-connected Lie group whence the instanton number is again expected to be integral.

\bigskip\noindent{\it{Spacetime BPS States from the ${\cal N} = (4,4)$ Sigma-Model on ${\bf S}^1_n \times \mathbb R_t$}}

According to what we have said above, the spectrum of spacetime BPS states would correspond to the spectrum of physical observables in the topological sector of the ${\cal N} = (4,4)$ sigma-model on ${\bf S}^1_n \times \mathbb R_t$. As explained in the previous subsection, the spacetime BPS states would then correspond to ${\bf L}^2$-harmonic forms which span the ${\bf L}^2$-cohomology of (some natural compactification of) ${\cal M}_{G}(M)$, where 
\be
{\cal M}_{G}(M) = \bigoplus_{a, \rho_0, \rho_\infty} {\cal M}^{\rho_0,  a}_{G, \rho_\infty}(M).
\ee
Here, $a$ is the instanton number; $\rho_{\infty}: \pi_1(M) \to G$ is a homomorphism associated with flat gauge fields at infinity, where $\pi_1(M) = \pi_1(\mathbb R^4 / \mathbb Z_k) = \mathbb Z_k$; and $\rho_{0}:  \mathbb Z_k \to G$ is a homomorphism associated with the $\mathbb Z_k$-action in the fiber of the $G$-bundle at the origin.

\bigskip\noindent{\it More About the Instanton Number}

Notice that $M = \mathbb R^4/ \mathbb Z_k$ is defined by imposing an order $k$ cyclic identification of $\mathbb R^4$; therefore, the total number of D0-branes ought to be given by $kd$, where $d$ is the effective number of D0-branes in each fundamental region of $M$. 

According to our explanations five, four and three paragraphs earlier, for $G = SO(2N)$, $d$ must take values in $\mathbb Z_{\geq 0}$; for $G = USp(2N-2)$, $d$ must take values in $\mathbb Z_{\geq 0}$ because of further binding of bound states that contain one-half-fractional D0-branes; and for $G = G_2$, $d$ must also take values in $\mathbb Z_{\geq 0}$ because of further binding of bound states that contain one-third-fractional D0-branes. In all, this means that we can write the instanton number as $a = kd = k n' (i - j)$, where for $G = SO(2N)$, $USp(2N -2)$ and $G_2$, $n' =1$, 2 and 3, while $i,j$ are certain integers divided by $1$, $2$ and $3$. In all cases, $i \geq j$, since $d$ must be non-negative. Here, one can interpret $(i-j)$ as the contribution from the bound states of D0-branes, and $n' \neq 1$ if there exists bound states consisting of fractional D0-branes which necessarily need to be paired to form bound states of full D0-branes.

That said, since $M$ is noncompact, the total instanton number must actually be shifted by an amount which depends on the conjugacy class of $\rho_\infty$. According to our explanations in the previous subsection, it would mean that we can actually write the shifted instanton number as $a = kn'(i - j)   - b(\bar \mu, \bar \mu)$, where $b$ is some positive real constant, $\bar \mu$ is a dominant coweight of $G$ which corresponds to a conjugacy class of $\rho_\infty$, and $(~,~)$ is just the scalar product in coweight space. 

Last but not least, note that in our counting of the total instanton number performed hitherto, we have implicitly overlooked the D0-branes at the origin of $M$: in writing $a = k n' (i-j)$ in the paragraph before last, we have accounted for the D0-branes away from the origin which have $k$ mirror partners under the order $k$ cyclic identification, but \emph{not} the D0-branes at the origin which do not have any mirror partners (since the origin is a fixed-point of the identification).  According to our explanations in the previous subsection, we can include the D0-branes at the origin by adding $\tilde b(\bar \lambda, \bar \lambda)$ to the total instanton number, where $\tilde b$ is some positive real constant, and $\bar \lambda$ is a dominant coweight of $G$ which corresponds to a conjugacy class of $\rho_0$.  In short, we can write the instanton number as
\be
a = kn'(i-j) +  \tilde b(\bar \lambda, \bar \lambda) - b(\bar \mu, \bar \mu),
\label{a-OM5}
\ee
where for $G = SO(2N)$, $USp(2N-2)$ and $G_2$, $n' =1$, 2 and 3, while $i \geq j$ are certain integers divided by $1$, $2$ and $3$.

For $n=2$ whence we have $G = USp(2N -2)$ with $n'=2$ and $i,j$ being certain integers divided by 2, expression (\ref{a-OM5}) is indeed consistent with results from the mathematical literature (which only addresses the case of simply-connected groups like $USp(2N-2)$): eqn.~(\ref{a-OM5}) coincides with~\cite[eqn.~(4.3)]{BF} after we set $\tilde b = b = 1/2$ and identify the \emph{integer} $n' (i-j)$ with the integer $(l-m)$ of \emph{loc.~cit..}

Likewise, for $n=3$ and $N=4$ whence we have $G = G_2$ with $n'=3$ and $i,j$ being certain integers divided by 3,  expression (\ref{a-OM5}) is also consistent with results from the mathematical literature (which only addresses the case of simply-connected groups like $G_2$): eqn.~(\ref{a-OM5}) coincides with~\cite[eqn.~(4.3)]{BF} after we set $\tilde b = b = 1/2$ and identify the \emph{integer} $n' (i-j)$ with the integer $(l-m)$ of \emph{loc.~cit..}

\bigskip\noindent{\it The Spectrum of Spacetime BPS States in the M-Theory Compactification (\ref{M-theory 1 OM5 discussion})}

We are now ready to state the generic Hilbert space ${\cal H}_{\rm BPS}$ of spacetime BPS states in the M-theory compactification (\ref{M-theory 1 OM5 discussion}). To this end, let us first denote by ${\textrm H}^\ast_{{\bf L}^2}{\cal U}({\cal M}^{\lambda}_{G, \mu}(\mathbb R^4/ \mathbb Z_k))$, the ${\bf L}^2$-cohomology of the Uhlenbeck compactification ${\cal U}({\cal M}^{\lambda}_{G, \mu}(\mathbb R^4/ \mathbb Z_k))$ of the component ${\cal M}^{\lambda}_{G, \mu}(\mathbb R^4/ \mathbb Z_k)$ of the highly singular moduli space ${\cal M}_{G}(\mathbb R^4/ \mathbb Z_k)$ labeled by the triples $\lambda = (k, \bar \lambda, i)$ and $\mu = (k, \bar \mu, j)$ (where $a$ is correspondingly given by (\ref{a-OM5})).\footnote{See also footnote~\ref{compactification} and~\ref{well-behaved} as to why (i) we need to compactify the moduli space; (ii) the physical theory is well-behaved despite the highly singular nature of the moduli space.} Then, since one can express ${\textrm H}^\ast_{{\bf L}^2}{\cal U}({\cal M}^{\lambda}_{G, \mu}(\mathbb R^4/ \mathbb Z_k))$  as the intersection cohomology ${\rm IH}^\ast{\cal U}({\cal M}^{\lambda}_{G, \mu}(\mathbb R^4/ \mathbb Z_k))$~\cite{Goresky}, we can write
\be
{\cal H}_{\rm BPS} = \bigoplus_{\lambda, \mu} {\cal H}^{\lambda, \mu}_{\rm BPS}  =  \bigoplus_{\lambda, \mu} ~{\rm IH}^\ast {\cal U}({\cal M}^{\lambda}_{G, \mu}(\mathbb R^4/ \mathbb Z_k)).
\label{BPS-M-OM5}
\ee
Notice that because we cannot have a negative number of D0-branes, we must have $a \geq 0$. In turn, this implies, via (\ref{a-OM5}) and the  condition $i \geq j$, that
\be
\lambda \geq \mu.
\label{weight condition-OM5}
\ee
 As $k \in \mathbb Z_+$ and $\bar \lambda$ and $\bar \mu$ are dominant coweights of $G$, the triples $\lambda$ and $\mu$ can be regarded as dominant coweights of the corresponding affine Kac-Moody group $G_{\rm aff}$ of level $k$.

 \bigskip\noindent{\it The Partition Function of Spacetime BPS States for $G = SO(2N)$}

One can of course go on to state the partition function of spacetime BPS states. The partition function, which counts (with weights) the total number of  states, can be obtained by taking a trace in the Hilbert space of states. Note at this point that taking such a trace is geometrically equivalent to identifying the two ends of the sigma-model worldsheet ${\bf S}^1_n \times \mathbb R_t$ to form a torus. Let the modulus of this torus be $\tau = \tau_1 + i \tau_2$; then, if $n=1$, the partition function for nonsimply-connected $G = SO(2N)$ can (according to our explanations in the previous subsection) be written as
 \be
 Z_{SO(2N)}^{\rm BPS} = {\rm Tr}_{{\cal H}_{\rm BPS}} \, q^P,
 \label{Zbps-OM5}
 \ee
where $q= e^{2 \pi i \tau}$, and $P$ is the momentum operator along ${\bf S}^1_{n}$.

Since $P$ measures the number of D0-branes (as each D0-brane has unit momentum along ${\bf S}^1_n$), according to our analysis leading up to (\ref{a-OM5}), we have $P = k(i-j) + {\tilde b}(\bar \lambda, \bar \lambda)$. Together with (\ref{BPS-M-OM5}), we can therefore write (\ref{Zbps-OM5}) as
\be
 Z_{SO(2N)}^{\rm BPS}  =  \sum_{\lambda} \, q^{m_\lambda}  \, \sum_{\bar\mu}  \sum_{m \geq 0}  \, {\rm dim} \, {\rm IH}^\ast {\cal U}({\cal M}^{\lambda, m}_{SO(2N), \bar\mu}(\mathbb R^4/ \mathbb Z_k)) \,  q^m.
 \label{Z_SO(2N)} 
\ee
Here,
\be
m_\lambda = h_{ \lambda} - {c_{ \lambda}  \over 24};
\label{m-OM5}
\ee
$m = k (i-j) \in \mathbb Z_{\geq 0}$, as  $i, j$ are integers such that $(i-j) \in \mathbb Z_{\geq 0}$; the non-negative number 
\be
h_{ \lambda} = {(\bar \lambda, \bar\lambda + 2 \rho^\vee) \over 2(k + h)},
\label{h-OM5}
\ee 
where $\rho^\vee$ and $h$ are the Weyl vector and dual Coxeter number of the\emph{ Langlands dual} group ${SO(2N)}^\vee$, respectively; and the number
\be
{c_{ \lambda}}  =  -24 \tilde b {(\bar \lambda, \bar \lambda)}  +  {12 (\bar \lambda, \bar \lambda + 2 \rho^\vee) \over {(k + h)}}.
\label{c-OM5}
\ee

In this instance, $\lambda = (k, \bar \lambda, i)$ and $\mu = (k, \bar \mu, j)$ can also be regarded as \emph{dominant weights} of the corresponding \emph{Langlands dual} affine Kac-Moody group ${SO(2N)}^\vee_{\rm aff}$ of level $k$.

 \bigskip\noindent{\it The Partition Function of Spacetime BPS States for $G =USp(2N-2)$}

Now, let $n=2$ whence the theory is ``$\mathbb Z_2$-twisted'' as we go around ${\bf S}^1_n$. In this case, the total partition function of spacetime BPS states for simply-connected $G = USp(2N - 2)$ can be written as 
\be
Z^{\rm BPS}_{USp(2N-2)} = {\rm Tr}_{{\cal H}^0_{\rm BPS}} \, {\mathscr P}_2 \, q^{P_0} + {\rm Tr}_{{\cal H}^{1}_{\rm BPS}} \, {\mathscr P}_2 \, q^{P_{1}},
\label{pf underlying-USp(2n-2)}
\ee
where $\mathscr P_2$ is a projection onto $\mathbb Z_2$-invariant states, and the super(sub)script `$0$' or `$1$' indicates that the operator or space in question is that of the untwisted or twisted sector, respectively (see footnote~\ref{twisted sector}).

The meaning of $\mathscr P_2$ in the trace over ${{\cal H}^0_{\rm BPS}}$ can be understood explicitly as follows. First, note that in the \emph{untwisted} sector, we have the dominant coweights $\lambda_0 = (k, {\bar \lambda}_0, i_0)$ and $\mu_0 = (k, {\bar \mu}_0, j_0)$ of $USp(2N-2)_{\rm aff}$ of level $k$, where  $\lambda_0 \geq \mu_0$; according to our discussions hitherto, $i_0$ and $j_0$ are integers whereby $(i_0-j_0) \in \mathbb Z_{\geq 0}$, and to satisfy this condition unequivocally, one ought to have $i_0 \in \mathbb Z_{\geq 0}$ and $-j_0 \in \mathbb Z_{\geq 0}$; that is,  $\lambda_0$ and $\mu_0$ are dominant coweights with integer grading. Second, note that the intersection cohomology ${\rm IH}^\ast {\cal U}({\cal M}^{\lambda_0}_{USp(2N-2), \mu_0}(\mathbb R^4/ \mathbb Z_k))$ which represents ${\cal H}^{\lambda_0, \mu_0}_{\rm BPS} \subset {\cal H}^0_{\rm BPS} \subset {\cal H}_{\rm BPS}$ (see (\ref{BPS-M-OM5})), corresponds to the space of physical observables of the $\cN=(4,4)$ sigma-model that take the form ${\cal O}_0 = f _{c  \dots e; \bar c \dots \bar e}(\varphi^d_0, \varphi^{\bar d}_0) \eta^c_0 \dots \eta^e_0 \eta^{\bar c}_0 \dots \eta^{\bar e}_0$, where the $\varphi_0$'s and $\eta_0$'s are untwisted bose and fermi fields of the sigma-model which are periodic and antiperiodic around ${\bf S}^1_n$, respectively (see footnote~\ref{anti-commutativity}), i.e., 
\be
\varphi^{c, \bar d}_0(\sigma + 2 \pi) = \varphi^{c, \bar d}_0(\sigma) \qquad {\rm and} \qquad \eta^{c, \bar d}_0 (\sigma + 2 \pi) = - \eta^{c, \bar d}_0(\sigma).
\ee
Here, the indices run as $c, \bar d = 1, 2, \dots, {\rm dim}_{\mathbb C} \, {\cal U}({\cal M}^{\lambda_0, m_0}_{USp(2N-2), {\bar\mu}_0}(\mathbb R^4/ \mathbb Z_k))$, where $m_0 =  2k(i_0 - j_0)$ -- the eigenvalue of $ P_0 - \tilde b (\bar \lambda_0, \bar \lambda_0)$ -- is a  non-negative integer. The insertion of $\mathscr P_2$ then means that in computing the trace over ${\cal H}^0_{\rm BPS}$, one must consider only ${\cal O}_0$'s which are invariant under the $\mathbb Z_2$-transformations $\varphi \to -\varphi$  and $\eta \to - \eta$. For later convenience, let us denote the space of such $\mathbb Z_2$-invariant  ${\cal O}_0$'s by  $\overline{{\rm IH}^\ast {\cal U}}({\cal M}^{\lambda_0, m_0}_{USp(2N-2), \bar\mu_0}(\mathbb R^4/ \mathbb Z_k)) \subset  {\rm IH}^\ast {\cal U}({\cal M}^{\lambda_0, m_0}_{USp(2N-2), \bar \mu_0}(\mathbb R^4/ \mathbb Z_k))$.

Similarly, the meaning of $\mathscr P_2$ in the trace over ${{\cal H}^1_{\rm BPS}}$ can be understood explicitly as follows. First, note that in the \emph{twisted} sector, we have the dominant coweights $\lambda_1 = (k, {\bar \lambda}_1, i_1)$ and $\mu_1 = (k, {\bar \mu}_1, j_1)$ of $USp(2N-2)_{\rm aff}$ of level $k$, where  $\lambda_1 \geq \mu_1$; according to our discussions hitherto, $i_1$ and $j_1$ are integers divided by 2 such that $(i_1-j_1) \in \mathbb Z_{\geq 0} + {1 \over 2}$, and to satisfy this condition unequivocally, one ought to have $i_1 \in \mathbb Z_{\geq 0}$ and $ - j_1 \in \mathbb Z_{\geq 0} + {1 \over 2}$; in other words, $\lambda_1$ and $\mu_1$ ought to be dominant coweights with integer and half-integer grading, respectively. Second, note that the intersection cohomology ${\rm IH}^\ast {\cal U}({\cal M}^{\lambda_1}_{USp(2N-2), \mu_1}(\mathbb R^4/ \mathbb Z_k))$ which represents ${\cal H}^{\lambda_1, \mu_1}_{\rm BPS} \subset {\cal H}^{1}_{\rm BPS} \subset H_{\rm BPS}$ (see (\ref{BPS-M-OM5})), corresponds to the space of physical observables of the $\cN=(4,4)$ sigma-model that take the form ${\cal O}_{1} = f _{c  \dots e; \bar c \dots \bar e}(\varphi^d_{1}, \varphi^{\bar d}_{1}) \eta^c_{1} \dots \eta^e_{1} \eta^{\bar c}_{1} \dots \eta^{\bar e}_{1}$. Here, the $\varphi_{1}$'s and $\eta_{1}$'s are \emph{twisted} bose and fermi fields of the sigma-model which are thus antiperiodic and periodic around ${\bf S}^1_n$, respectively; specifically, we have 
\be
\varphi^{c}_1(\sigma + 2 \pi) = e^{2 \pi i \nu \over n}\varphi^{c}_1(\sigma) = - \varphi^{c}_1(\sigma),    \qquad \eta^{c}_1(\sigma + 2 \pi) = - e^{2 \pi i \nu \over n}\eta^{c}_1(\sigma) = \eta^{c}_1(\sigma), 
\ee
and
\be
\varphi^{\bar d}_1(\sigma + 2 \pi) = e^{-{2 \pi i \nu \over n}}\varphi^{\bar d}_1(\sigma) = - \varphi^{\bar d}_1(\sigma),    \qquad \eta^{\bar d}_1 (\sigma + 2 \pi) = - e^{-{2 \pi i \nu \over n}}\eta^{\bar d}_1(\sigma) = \eta^{\bar d}_1(\sigma),
\ee
as $n =2$ and the twist parameter $\nu = 1$. Also, $c, \bar d = 1, 2, \dots, {\rm dim}_{\mathbb C} \, {\cal U}({\cal M}^{\lambda_1, m_1}_{USp(2N-2), \bar\mu_1}(\mathbb R^4/ \mathbb Z_k))$, where $m_1 = 2k(i_1 - j_1)$ -- the eigenvalue of  $P_1 - \tilde b(\bar \lambda_1, \bar \lambda_1)$ -- is a non-negative integer.  The insertion of $\mathscr P_2$ then means that in computing the trace over ${\cal H}^{1}_{\rm BPS}$, one must consider only ${\cal O}_{1}$'s which are invariant under the $\mathbb Z_2$-transformations $\varphi \to -\varphi$  and $\eta \to - \eta$. Let us denote the space of such $\mathbb Z_2$-invariant  ${\cal O}_{1}$'s by  $\overline {{\rm IH}^\ast {\cal U}}({\cal M}^{\lambda_1, m_1}_{USp(2N-2), \bar\mu_1}(\mathbb R^4/ \mathbb Z_k)) \subset  {\rm IH}^\ast {\cal U}({\cal M}^{\lambda_1, m_1}_{USp(2N-2), \bar \mu_1}(\mathbb R^4/ \mathbb Z_k))$. Then, together with what was said in the previous paragraph, and by relabeling the integer-graded coweights $\lambda_0$ and $\lambda_1$ as $\lambda$, we can write 
\be
Z^{\rm BPS}_{USp(2N-2)} =  Z^{{\rm BPS}, 0}_{USp(2N-2)} + Z^{{\rm BPS}, 1}_{USp(2N-2)},
\label{Z-USp(2N-2)-1}
\ee
where 
\be
Z^{\rm BPS, 0}_{USp(2N-2)} = \sum_{\lambda, \bar \mu_0}  q^{m_{\lambda}}  \sum_{m_0 \geq 0}   {\rm dim} \, \overline {{\rm IH}^\ast {\cal U}}({\cal M}^{\lambda, m_0}_{USp(2N-2), \bar\mu_0}(\mathbb R^4/ \mathbb Z_k)) \, q^{m_0}, 
\label{Z-USp(2N-2)-2}
\ee
and 
\be
Z^{\rm BPS, 1}_{USp(2N-2)} = \sum_{\lambda, \bar \mu_1}  q^{m_{\lambda}}  \sum_{m_1 \geq 0}   {\rm dim} \, \overline {{\rm IH}^\ast {\cal U}}({\cal M}^{\lambda, m_1}_{USp(2N-2), \bar\mu_1}(\mathbb R^4/ \mathbb Z_k)) \, q^{m_1}.
\label{Z-USp(2N-2)-3}
\ee
The phase factor $m_{\lambda}$ takes the form in (\ref{m-OM5}) with $\tilde b = 1/2$.
 
 In this instance, the dominant coweights  $\lambda = (k, \bar \lambda, i)$ and $\mu_{0,1} = (k, \bar \mu_{0,1}, j_{0,1})$ of ${USp(2N-2)}_{\rm aff}$ are also  (un)twisted dominant weights of the $\mathbb Z_2$-twisted affine Kac-Moody group ${SO(2N)}^{(2)}_{\rm aff}$; furthermore, ${SO(2N)}^{(2)}_{\rm aff}$ is equal to $USp(2N-2)^\vee_{\rm aff}$. In other words, $\lambda$ and $\mu_{0,1}$ can also be regarded as \emph{dominant weights} of the \emph{Langlands dual} affine Kac-Moody group ${USp(2N-2)}^\vee_{\rm aff}$ of level $k$.

Additionally, notice that (\ref{Z-USp(2N-2)-1})--(\ref{Z-USp(2N-2)-3}) imply that the \emph{effective} Hilbert space ${{\cal H}}^{\rm eff}_{\rm BPS}$ of spacetime BPS states  (which one obtains after taking into account the projection $\mathscr P_2$ in the trace over all underlying states in (\ref{pf underlying-USp(2n-2)})) ought to be given by
\be
\label{HBPS-eff-USp(2N-2)}
{{\cal H}}^{\rm eff}_{\rm BPS} = \bigoplus_{\lambda} \bigoplus_{\nu =0,1}  \bigoplus_{\mu_\nu} \, {\overline{\cal H}}^{\lambda, \mu_\nu}_{\rm BPS}   =   \bigoplus_{\lambda} \bigoplus_{\nu =0,1}  \bigoplus_{\mu_\nu} \, \overline {{\rm IH}^\ast {\cal U}}({\cal M}^{\lambda}_{USp(2N-2), \mu_\nu}(\mathbb R^4/ \mathbb Z_k)),
\ee
where $\nu =0$ or $1$ if the sector is untwisted or twisted, respectively.

 \bigskip\noindent{\it The Partition Function of Spacetime BPS States for $G =G_2$}

Now, let $n=3$ whence the theory is ``$\mathbb Z_3$-twisted'' as we go around ${\bf S}^1_n$. In the case where $N=4$, the total partition function of spacetime BPS states for simply-connected $G = G_2$ can be written as 
\be
Z^{\rm BPS}_{G_2} = {\rm Tr}_{{\cal H}^0_{\rm BPS}} \, {\mathscr P}_3 \, q^{P_0} + {\rm Tr}_{{\cal H}^{1}_{\rm BPS}} \, {\mathscr P}_3 \, q^{P_{1}} + {\rm Tr}_{{\cal H}^{2}_{\rm BPS}} \, {\mathscr P}_3 \, q^{P_{2}},
\label{pf underlying-G_2}
\ee
where $\mathscr P_3$ is a projection onto $\mathbb Z_3$-invariant states, and the super(sub)scripts `$0$' and `$1$'/`$2$' indicate that the operator or space in question is that of the untwisted and twisted sectors, respectively. (See footnote~\ref{twisted sector}, whose explanations also hold for the $\mathbb Z_3$ case at hand.)

The meaning of $\mathscr P_3$ in the trace over ${{\cal H}^0_{\rm BPS}}$ can be understood explicitly as follows. First, note that in the \emph{untwisted} sector, we have the dominant coweights $\lambda_0 = (k, {\bar \lambda}_0, i_0)$ and $\mu_0 = (k, {\bar \mu}_0, j_0)$ of ${G_2}_{\rm aff}$ of level $k$, where  $\lambda_0 \geq \mu_0$; according to our discussions hitherto, $i_0$ and $j_0$ are integers whereby $(i_0-j_0) \in \mathbb Z_{\geq 0}$, and to satisfy this condition unequivocally, one ought to have $i_0 \in \mathbb Z_{\geq 0}$ and $-j_0 \in \mathbb Z_{\geq 0}$;  that is,  $\lambda_0$ and $\mu_0$ are dominant coweights with integer grading. Second, note that the intersection cohomology ${\rm IH}^\ast {\cal U}({\cal M}^{\lambda_0}_{G_2, \mu_0}(\mathbb R^4/ \mathbb Z_k))$ which represents ${\cal H}^{\lambda_0, \mu_0}_{\rm BPS} \subset {\cal H}^0_{\rm BPS} \subset {\cal H}_{\rm BPS}$ (see (\ref{BPS-M-OM5})), corresponds to the space of physical observables of the $\cN=(4,4)$ sigma-model that take the form ${\cal O}_0 = f _{c  \dots e; \bar c \dots \bar e}(\varphi^d_0, \varphi^{\bar d}_0) \eta^c_0 \dots \eta^e_0 \eta^{\bar c}_0 \dots \eta^{\bar e}_0$, where the $\varphi_0$'s and $\eta_0$'s are untwisted bose and fermi fields of the sigma-model which are periodic and antiperiodic around ${\bf S}^1_n$, respectively (see footnote~\ref{anti-commutativity}), i.e., 
\be
\varphi^{c, \bar d}_0(\sigma + 2 \pi) = \varphi^{c, \bar d}_0(\sigma) \qquad {\rm and} \qquad \eta^{c, \bar d}_0 (\sigma + 2 \pi) = - \eta^{c, \bar d}_0(\sigma).
\ee
Here, the indices run as $c, \bar d = 1, 2, \dots, {\rm dim}_{\mathbb C} \, {\cal U}({\cal M}^{\lambda_0, m_0}_{G_2, {\bar\mu}_0}(\mathbb R^4/ \mathbb Z_k))$, where $m_0 =  3k(i_0 - j_0)$ -- the eigenvalue of $ P_0 - \tilde b (\bar \lambda_0, \bar \lambda_0)$ -- is a  non-negative integer. The insertion of $\mathscr P_3$ then means that in computing the trace over ${\cal H}^0_{\rm BPS}$, one must consider only ${\cal O}_0$'s which are invariant under the $\mathbb Z_3$-transformations:
\be
\varphi^c  \to e^{2\pi i \theta_j} \varphi^c,  \qquad  \varphi^{\bar d} \to e^{-2\pi i \theta_j} \varphi^{\bar d}, \qquad \eta^c \to e^{2\pi i \theta_j} \eta^c,  \qquad   \eta^{\bar d} \to e^{-2\pi i \theta_j} \eta^{\bar d},
\label{Z3-G2}
\ee
where $\theta_j = {m_j / 3}$, and $m_j = 1, 2$. For later convenience, let us denote the space of such $\mathbb Z_3$-invariant  ${\cal O}_0$'s by  $\overline{{\rm IH}^\ast {\cal U}}({\cal M}^{\lambda_0, m_0}_{G_2, \bar\mu_0}(\mathbb R^4/ \mathbb Z_k)) \subset  {\rm IH}^\ast {\cal U}({\cal M}^{\lambda_0, m_0}_{G_2, \bar \mu_0}(\mathbb R^4/ \mathbb Z_k))$.

The meaning of $\mathscr P_3$ in the trace over ${{\cal H}^\gamma_{\rm BPS}}$, where $\gamma = 1$ or 2, can also be understood explicitly as follows. First, note that in the `$\gamma$'-\emph{twisted} sector, we have the dominant coweights $\lambda_\gamma = (k, {\bar \lambda}_\gamma, i_\gamma)$ and $\mu_\gamma = (k, {\bar \mu}_\gamma, j_\gamma)$ of ${G_2}_{\rm aff}$ of level $k$, where  $\lambda_\gamma \geq \mu_\gamma$; according to our discussions hitherto, $i_\gamma$ and $j_\gamma$ are integers divided by 3 such that $(i_\gamma-j_\gamma) \in \mathbb Z_{\geq 0} + {\gamma \over 3}$, and to satisfy this condition unequivocally, one ought to have $i_\gamma \in \mathbb Z_{\geq 0}$ and $ - j_\gamma \in \mathbb Z_{\geq 0} + {\gamma \over 3}$; in other words, $\lambda_\gamma$ and $\mu_\gamma$ ought to be dominant coweights with integer and one-third-integer grading, respectively. Second, note that the intersection cohomology ${\rm IH}^\ast {\cal U}({\cal M}^{\lambda_\gamma}_{G_2, \mu_\gamma}(\mathbb R^4/ \mathbb Z_k))$ which represents ${\cal H}^{\lambda_\gamma, \mu_\gamma}_{\rm BPS} \subset {\cal H}^{\gamma}_{\rm BPS} \subset H_{\rm BPS}$ (see (\ref{BPS-M-OM5})), corresponds to the space of physical observables of the $\cN=(4,4)$ sigma-model that take the form ${\cal O}_{\gamma} = f _{c  \dots e; \bar c \dots \bar e}(\varphi^d_{\gamma}, \varphi^{\bar d}_{\gamma}) \eta^c_{\gamma} \dots \eta^e_{\gamma} \eta^{\bar c}_{\gamma} \dots \eta^{\bar e}_{\gamma}$. Here, the $\varphi_{\gamma}$'s and $\eta_{\gamma}$'s are \emph{twisted} bose and fermi fields of the sigma-model which have the following boundary conditions around ${\bf S}^1_n$:
\be
\varphi^{c}_\gamma(\sigma + 2 \pi) = e^{2 \pi i \gamma \over 3}\varphi^{c}_\gamma(\sigma),    \qquad \eta^{c}_\gamma(\sigma + 2 \pi) = - e^{2 \pi i \gamma \over 3}\eta^{c}_1(\sigma), 
\ee
and
\be
\varphi^{\bar d}_\gamma(\sigma + 2 \pi) = e^{-{2 \pi i \gamma \over 3}}\varphi^{\bar d}_\gamma(\sigma),    \qquad \eta^{\bar d}_\gamma (\sigma + 2 \pi) = - e^{-{2 \pi i \gamma \over 3}}\eta^{\bar d}_\gamma(\sigma),
\ee
as $n =3$ and the twist parameter is $\gamma$. Also, $c, \bar d = 1, 2, \dots, {\rm dim}_{\mathbb C} \, {\cal U}({\cal M}^{\lambda_\gamma, m_\gamma}_{G_2, \bar\mu_\gamma}(\mathbb R^4/ \mathbb Z_k))$, where $m_\gamma = 3k(i_\gamma - j_\gamma)$ -- the eigenvalue of  $P_\gamma - \tilde b(\bar \lambda_\gamma, \bar \lambda_\gamma)$ -- is a non-negative integer.  The insertion of $\mathscr P_3$ then means that in computing the trace over ${\cal H}^{\gamma}_{\rm BPS}$, one must consider only ${\cal O}_{\gamma}$'s which are invariant under the $\mathbb Z_3$-transformations in (\ref{Z3-G2}). Let us denote the space of such $\mathbb Z_3$-invariant  ${\cal O}_{\gamma}$'s by  $\overline {{\rm IH}^\ast {\cal U}}({\cal M}^{\lambda_\gamma, m_\gamma}_{G_2, \bar\mu_\gamma}(\mathbb R^4/ \mathbb Z_k)) \subset  {\rm IH}^\ast {\cal U}({\cal M}^{\lambda_\gamma, m_\gamma}_{G_2, \bar \mu_\gamma}(\mathbb R^4/ \mathbb Z_k))$. Then, together with what was said in the previous paragraph, and by relabeling the integer-graded coweights $\lambda_0$ and $\lambda_{\gamma}$ as $\lambda$, we can write 
\be
Z^{\rm BPS}_{G_2} =   \bigoplus_{\nu =0}^2 Z^{{\rm BPS}, \nu}_{G_2},
\label{Z-G_2-1}
\ee
where 
\be
Z^{\rm BPS, \nu}_{G_2} = \sum_{\lambda, \bar \mu_\nu}  q^{m_{\lambda}}  \sum_{m_\nu \geq 0}   {\rm dim} \, \overline {{\rm IH}^\ast {\cal U}}({\cal M}^{\lambda, m_\nu}_{G_2, \bar\mu_\nu}(\mathbb R^4/ \mathbb Z_k)) \, q^{m_\nu}. 
\label{Z-G_2-2}
\ee
The phase factor $m_{\lambda}$ takes the form in (\ref{m-OM5}) with $\tilde b = 1/2$. 
 
 In this instance, the dominant coweights  $\lambda = (k, \bar \lambda, i)$ and $\mu_{\nu} = (k, \bar \mu_{\nu}, j_{\nu})$ of ${G_2}_{\rm aff}$ are also  (un)twisted dominant weights of the $\mathbb Z_3$-twisted affine Kac-Moody group ${SO(8)}^{(3)}_{\rm aff}$; furthermore, ${SO(8)}^{(3)}_{\rm aff}$ is equal to ${G_2}^\vee_{\rm aff}$. In other words, $\lambda$ and $\mu_{\nu}$ can also be regarded as \emph{dominant weights} of the \emph{Langlands dual} affine Kac-Moody group ${G_2}^\vee_{\rm aff}$ of level $k$.

Additionally, notice that (\ref{Z-G_2-1})--(\ref{Z-G_2-2}) imply that the \emph{effective} Hilbert space ${{\cal H}}^{\rm eff}_{\rm BPS}$ of spacetime BPS states (which one obtains after taking into account the projection $\mathscr P_3$ in the trace over all underlying states in (\ref{pf underlying-G_2})) ought to be given by
\be
\label{HBPS-eff-G_2}
{{\cal H}}^{\rm eff}_{\rm BPS} = \bigoplus_{\lambda} \bigoplus_{\nu =0}^2  \bigoplus_{\mu_\nu} \, {\overline{\cal H}}^{\lambda, \mu_\nu}_{\rm BPS}   =   \bigoplus_{\lambda} \bigoplus_{\nu =0}^2  \bigoplus_{\mu_\nu} \, \overline {{\rm IH}^\ast {\cal U}}({\cal M}^{\lambda}_{G_2, \mu_\nu}(\mathbb R^4/ \mathbb Z_k)),
\ee
where $\nu \neq 0$ if the sector is twisted.

\bigskip\noindent{\it The Spectrum of Spacetime BPS States in the M-Theory Compactification (\ref{M-theory 7 OM5 discussion})}

Let us now turn our attention to the\emph{ physically dual} M-theory compactification (\ref{M-theory 7 OM5 discussion}) with $k$ coincident M5-branes. One can proceed as before to ascertain the spacetime BPS states by computing the ground states of the M5-brane quantum worldvolume theory over $\mathbb R_t \times {\bf S}^1_n \times SN^{R\to 0}_N$. According to our earlier explanations, one can, if $n=1$ for example, interpret the spacetime BPS states as the physical observables in the topological sector of the sigma-model on  ${\bf S}^1_n \times \mathbb R_t$ with target the moduli space of $U(k)$-instantons on $SN^{R\to 0}_N$.\footnote{The reason why we have instantons of $U(k)$ (and not $SU(k)$) is because in duality step (\ref{OIIA 6}), the center-of-mass degrees of freedom of the $k$ D6-branes are not frozen. \label{not frozen 2}} 

That said, since we would like to make contact with a geometric Langlands duality for surfaces, we shall seek a different description of these spacetime BPS states, i.e, worldvolume ground states. To this end, recall  that the low-energy limit of the worldvolume theory is a 6d  $\cN = (2,0)$ $D_{k}$ \emph{superconformal }field theory of massless tensor multiplets. Hence, where the ground states are concerned, one can regard the worldvolume theory to be conformally-invariant. Since it is conformally-invariant, one can rescale the worldvolume to bring the region near infinity to a finite distance close to the origin without altering the theory.  Thus, one can, for the purpose of computing ground states, simply analyze the physics near infinity.

Near infinity, the ${\bf S}^1_R$ circle fiber of $SN^{R \to 0}_N$ has radius $R \to 0$. To make sense of this limit, notice that a compactification along the circle fiber would take us down to a type IIA theory whereby the stack of $k$ coincident M5-branes would now correspond to a stack of $k$ coincident D4-branes. In addition, as explained in $\S$A.5, we will also have $N$ D6-branes and an O$6^-$-plane spanning the directions transverse to its $\mathbb R^3/ {\cal I}_3$ base, where ${\cal I}_3$ acts as ${\vec r} \to -{\vec r}$ in $\mathbb R^3$; moreover, since $SN^{R \to 0}_N$ has a $D_{N}$ singularity at the origin, the D6-branes will be coincident. In other words, we have, in the limit $R \to 0$, the following type IIA configuration:
\be
\textrm{IIA}: \quad \underbrace{ {\mathbb R}^5 \times {{\bf S}^1_n} \times {\mathbb R}_t  \times {\mathbb R}^3/{\cal I}_3}_{\textrm{I-brane on ${{{\bf S}^1_n} \times {\mathbb R}_t} = N \textrm{D6}/\textrm{O6}^- \cap k\textrm{D4}$}}.
\label{equivalent IIA system 2}
\ee
Here, we have  a stack of $N$ coincident D6-branes on top of an O$6^-$-plane whose worldvolume is given by ${\mathbb R}^5 \times {{\bf S}^1_n} \times {\mathbb R}_t$, and a stack of $k$ coincident D4-branes whose worldvolume is given by ${{\bf S}^1_n} \times {\mathbb R}_t \times \mathbb R^3/{\cal I}_3$; these two stacks intersect along ${{\bf S}^1_n} \times {\mathbb R}_t$ to form a D4-D6/O$6^-$ I-brane system.

The proceeding analysis of this system is identical to the one for the system (\ref{equivalent IIA system 1}).  In particular, the sought-after worldvolume ground states will correspond to the states of the I-brane theory on ${\bf S}^1_n \times \mathbb R_t$ defined by the massless modes of the 4-6 open strings of the D4-D6/O$6^-$ system. Furthermore, this I-brane theory is a theory of massless free chiral fermions, and as in the case of (\ref{equivalent IIA system 1}), the chiral fermions will couple to certain gauge fields. In order to determine what these gauge fields are, let us now discuss what gauge groups should appear in the D4-D6/O$6^-$ I-brane system.

By a T-duality along three directions, we can get to a D1-D9/O$9^-$ system, where O$9^-$ is a spacetime-filling orientifold. One can compare this to an analogous D5-D9/O$9^{\pm}$ system studied in \cite{Gimon}, where the gauge groups are of different types on the D5- and D9-branes; they are either orthogonal on the D5-branes and symplectic on the D9-branes or vice-versa, depending on the sign in O$9^{\pm}$. This is due to the fact that there are four possible mixed Neumann-Dirichlet boundary conditions for the 5-9 open strings which stretch between the corresponding D-branes. On the other hand, there are eight possible mixed Neumann-Dirichlet boundary conditions for the 1-9 open strings stretched between D-branes in the D1-D9/O$9^-$ system; in other words, orthogonal gauge groups appear on \emph{both} the D1- and D9-branes. By T-dualizing back to a D4-D6/O$6^-$ system, one can conclude that generically, there ought to be, in the presence of the O$6^-$-plane, an $SO(\alpha)$ and $SO(2N)$ gauge group on the $k$ D4- and $N$ D6-branes, respectively, where $\alpha$ depends on $k$. 

To ascertain what $\alpha$ is, note that according to~\cite{nissan}, the total central charge of the \emph{real} chiral fermions   should not change as we move the D4- and D6-branes around; in particular, it should not change as we move the stack of coincident D4- and D6-branes away from the O$6^-$-plane. When we move the stack of coincident D4- and D6-branes away from the O$6^-$-plane, we effectively have the $U(k) \times U(N)$ theory described by (\ref{4.9})--(\ref{L}). Thus, $\alpha$ must be such that the total central charge of the real chiral fermions is $kN$. 

Since a single real chiral fermion will contribute one-half to the central charge, we ought to have a total of $2kN$ real chiral fermions. As the $2kN$ real chiral fermions are furnished by the massless modes of the 4-6 open strings, they necessarily transform in the bifundamental representation of $SO(\alpha) \times SO(2N)$; this would mean that $\alpha = k$. Hence, along the I-brane with complex coordinate `$z$', the $2kN$ real chiral fermions ought to be given by
\be
\psi_{i, a}(z), \quad {\rm where} \quad  i = 1, \dots, k, \quad {\rm and} \quad a = 1, \dots, 2N.
\ee
And, as their indices imply, they ought to transform in the bifundamental representation $(k, 2N)$ of $SO(k) \times SO(2N)$.  Their action is given (modulo an overall coupling constant) by
\be
I = \int d^2z \ \psi \bar\partial_{{\cal A} + {\cal A}'} \psi,
\label{L'}
\ee
where $\cal A$ and ${\cal A}'$ are the restrictions to the I-brane worldsheet ${\bf S}^1_n \times \mathbb R_t$ of the $SO(k)$ and $SO(2N)$ gauge fields associated with the $k$ D4- and $N$ D6-branes. In other words, the fermions couple to the gauge group
\be
SO(k) \times SO(2N).
\label{coupling'}
\ee
The I-brane theory is anomalous under the corresponding gauge transformations, but like in the earlier case of (\ref{equivalent IIA system 1}), one can show that the overall D4-D6/O$6^-$ system is anomaly-free and thus physically consistent.

The system of $2kN$ real free chiral fermions of central charge $kN$ gives a direct realization of $\widehat{so}(2kN)^{(n)}_1$, the integrable module over the $\mathbb Z_n$-twisted affine Lie algebra $\frak{so}(2kN)^{(n)}_{\textrm{aff}, 1}$ of level 1.\footnote{To understand this claim, see~\cite[$\S$15.5.2]{CFT text}, and note that (i) the identification under an order $n$ translation of the circle ${\bf S}^1_n$ results in a $\mathbb Z_n$-twist of the underlying affine Lie algebra; (ii) a twisted version of an affine Lie algebra has the same central charge and level as its untwisted version (c.f.~\cite[$\S$3]{abcdefg}).\label{central charge SO(2N)}}  Moreover, there exists the following twisted affine embedding which preserves conformal invariance~\cite{Hasegawa}:
\be
\label{D-embedding}
\frak{so}(k)^{(n)}_{{\rm aff}, 2N} \otimes \frak{so}(2N)^{(n)}_{{\rm aff}, k} \subset \frak{so}(2kN)^{(n)}_{{\rm aff}, 1},
\ee
where this can be viewed as an affine analog of the gauge symmetry in (\ref{coupling'}) (see footnote~\ref{analogy}). As such, the total Fock space ${\cal F}^{\otimes 2kN}$ of the $2kN$ real free fermions can be expressed as
\be
{\cal F}^{\otimes 2kN} = \textrm{WZW}_{\widehat{so}(k)^{(n)}_{2N}} \otimes \textrm{WZW}_{\widehat{so}(2N)^{(n)}_{k}},
\label{Fock SO(2N)}
\ee
where $\textrm{WZW}_{\widehat{so}(k)^{(n)}_{2N}}$ and $\textrm{WZW}_{\widehat{so}(2N)^{(n)}_{k}}$ are the spectra of states furnished by $\widehat{so}(k)^{(n)}_{2N}$ and $\widehat{so}(2N)^{(n)}_{k}$, respectively, which can be realized in the relevant $\it{chiral}$ WZW models. Consequently, the partition function of the I-brane theory will be expressed in terms of the \emph{chiral} characters of $\widehat{so}(k)^{(n)}_{2N}$ and $\widehat{so}(2N)^{(n)}_{k}$.

Note that ${\cal F}^{\otimes 2kN}$ is the Fock space of the $2kN$ real free fermions which have $\it{not}$ yet been coupled to $\cal A$ and ${\cal A}'$. Upon coupling to the gauge fields, the characters that appear in the overall partition function of the I-brane theory will be reduced. In a generic situation, the free fermions will couple to the gauge group $SO(k) \times SO(2N)$ (see  (\ref{coupling'})). However, in this case, only the $SO(k)$ gauge field associated with the D4-branes is dynamical; the $SO(2N)$ gauge field associated with the D6-branes/O$6^-$-plane should $\it{not}$ be dynamical as the geometry of $SN^{R \to 0}_N$ is fixed in our description -- the center-of-mass degrees of freedom of the $N$ NS5-branes/ON$5^-_B$-plane which give rise to the $SN^{R \to 0}_N$ geometry via steps (\ref{OIIB 3}) and (\ref{0IIA 4}), are frozen. Therefore, the free fermions will, in this case, couple dynamically to the gauge group $SO(k)$ only. Schematically, this means that we are dealing with the following partially gauged CFT
\be
\frak{so}(2kN)^{(n)}_{{\rm aff},1} / \frak{so}(k)^{(n)}_{{\rm aff}, 2N}.
\label{partially gauged CFT - CDG}
\ee
In particular, the $\frak{so}(k)^{(n)}_{{\rm aff}, 2N}$ chiral WZW model will be replaced by the corresponding topological $G/G$ model. As a result, the chiral characters of $\widehat{so}(k)^{(n)}_{2N}$ which appear in the overall partition function of the uncoupled free fermions system on the I-brane, will reduce to constant complex factors after coupling to the dynamical $SO(k)$ gauge field. Hence, modulo these constant complex factors which serve only to shift the energy levels of the ground states by numbers dependent on the highest affine weights of $\widehat{so}(k)^{(n)}_{2N}$, the $\it{effective}$ overall partition function of the I-brane theory will be expressed solely in terms of the chiral characters of $\widehat{so}(2N)^{(n)}_{k}$. 

In summary, the sought-after spectrum of spacetime BPS states in the M-theory compactification (\ref{M-theory 7 OM5 discussion}) would be realized by $\textrm{WZW}_{\widehat{so}(2N)^{(n)}_{k}}$. This observation is indeed physically consistent because according to  footnote~\ref{worldvolume ground state}, the spacetime BPS states satisfy $H=P$ -- here, $H$ and $P$ are the Hamiltonian and momentum operators which generate translations along $\mathbb R_t$ and ${\bf S}^1_n$, respectively -- while a chiral WZW model on ${\bf S}^1_n \times \mathbb R_t$, having no right-moving excitations, has a spectrum whereby $H=P$.  

\bigskip\noindent{\it A Geometric Langlands Duality for Surfaces for the $D_{N}$ Groups} 

Let us now consider $n=1$ whence there is no ``twist'' at all, i.e., $\widehat{so}(2N)^{(n)}_{k}$ is simply $\widehat{so}(2N)_{k}$, the integrable module over the untwisted affine Lie algebra ${\frak so}(2N)_{{\rm aff}, k}$ of level $k$.  Then, unitarity of any WZW model requires that $\textrm{WZW}_{\widehat{so}(2N)_{k}}$ be generated by dominant highest weight modules over $\frak{so}(2N)_{{\rm aff},k}$, such that a generic state in any one such module can be expressed as~\cite{CFT text}
\be
|{\tilde \mu}'\rangle = E^{- \tilde\alpha}_{-n} \dots E^{-\tilde\beta}_{-m} |\tilde \lambda\rangle, \qquad \forall~~~n,m \geq 0 ~~ \textrm{and} ~~ \tilde\alpha, \tilde\beta > 0.
\label{generic WZW state OM5}
\ee
Here, the $E^{-\tilde\gamma}_{-l}$'s are lowering operators that correspond to the respective modes of the currents of $\frak{so}(2N)_{{\rm aff},k}$ (in a Cartan-Weyl basis) which are associated with the complement of the Cartan subalgebra; $|\tilde \lambda \rangle$ is a highest weight state associated with a dominant highest affine weight $\tilde \lambda$; $\tilde \mu' = \tilde \lambda -\tilde \alpha \dots -\tilde\beta$ is an affine weight in the weight system ${\widehat\Omega}_{\tilde\lambda}$ of $\widehat{so}(2N)^{\tilde\lambda}_{k}$ -- the module of dominant highest weight $\tilde\lambda$ of level $k$ -- which is not necessarily dominant; and $\tilde\alpha,\tilde\beta$ are positive affine roots.  

Note that each module labeled by a dominant highest affine weight $\tilde\lambda$ can be decomposed into a direct sum of finite-dimensional subspaces each spanned by states of the form $|\tilde\mu'\rangle$ for $\it{all}$ possible positive affine roots $\tilde\alpha, \dots, \tilde\beta$. These finite-dimensional subspaces of states are the $\tilde\mu'$-weight spaces $\widehat{so}(2N){}^{\tilde\lambda}_{k,\tilde\mu'} \subset \widehat{so}(2N)^{\tilde\lambda}_{k}$. Note at this point that there is a Weyl group symmetry on these weight spaces that maps $\tilde\mu'$ to a dominant weight $\tilde\mu$ in ${\widehat\Omega}_{\tilde\lambda}$ which also leaves the chiral character of $\widehat{so}(2N)^{\tilde\lambda}_{k}$ and thus, the partition function of the chiral WZW model, invariant.\footnote{See~\cite[eqns.~(14.143), (14.145), (14.165), (14.166) and (15.119)]{CFT text}, noting that $z_j$ in \emph{loc.~cit.} corresponds to the Coulomb moduli in our story which must therefore be set to zero since the $N$ D6-branes are coincident.} As such, one can also express the spectrum of states of the chiral WZW model as
 \be
 \textrm{WZW}_{\widehat{so}(2N)_{k}} = \bigoplus_{\tilde\lambda, \tilde\mu} \, \textrm{WZW}_{\widehat{so}(2N)^{\tilde \lambda}_{k, \tilde\mu}}.
 \label{WZW space-SO(2N)}
 \ee

Now, the physical duality of the M-theory compactifications (\ref{M-theory 1 OM5 discussion}) and (\ref{M-theory 7 OM5 discussion}) means that their respective spacetime BPS spectra ought to be equivalent, i.e., $ \textrm{WZW}_{\widehat{so}(2N)_{k}}$ ought to be equal to $\cal H_{\rm BPS}$ of (\ref{BPS-M-OM5}). Indeed, since $\frak{so}(2N)_{\textrm{aff}}$ is isomorphic to its Langlands dual counterpart  $\frak{so}(2N)^\vee_{\textrm{aff}}$, $\tilde\lambda$ and $\tilde\mu$ are also \emph{dominant weights} of the \emph{Langlands dual} affine Kac-Moody group $SO(2N)^\vee_{\rm aff}$ whence we can identify them with $\lambda$ and $\mu$ of~(\ref{BPS-M-OM5}), respectively;  moreover, both $\cal H_{\rm BPS}$ and $ \textrm{WZW}_{\widehat{so}(2N)_{k}}$ are labeled by $k$. Specializing to the $\{\lambda, \mu\}$-sector of the spectra of spacetime BPS states, we can thus write 
\be
{\cal H^{\lambda, \mu}_{\rm BPS}} = \textrm{WZW}_{\widehat{so}(2N)^{ \lambda}_{k, \mu}}.
\label{H=WZW-SO(2N)}
\ee
As $\textrm{WZW}_{\widehat{so}(2N)^\lambda_{k, \mu}}$ is furnished by $\widehat{so}(2N)^\lambda_{k, \mu}$, and since $\frak{so}(2N)_{\rm aff} \simeq \frak{so}(2N)^\vee_{\textrm{aff}} $ whence ${\widehat{so}(2N)^\lambda_{k, \mu}}$ is isomorphic to the submodule ${^L\widehat{so}(2N)^\lambda_{k, \mu}}$ over $\frak{so}(2N)^\vee_{\textrm{aff}}$, via (\ref{BPS-M-OM5}), we can also express (\ref{H=WZW-SO(2N)}) as 
\be
\boxed{{\rm IH}^\ast {\cal U}({\cal M}^{\lambda}_{SO(2N), \mu}(\mathbb R^4/ \mathbb Z_k)) = {^L\widehat{so}(2N)^\lambda_{k, \mu}}}
\label{GL-relation D}
\ee
Thus, we have arrived at a $G=SO(2N)$ generalization of~\cite[Conjecture 4.14(3)]{BF}! This completes our purely physical derivation of a geometric Langlands duality for surfaces for the $SO(2N) = D_{N}$ groups.

\bigskip\noindent{\it An Identity of the Dimension of the Intersection Cohomology of the Moduli space of $D_{N}$-Instantons on $\mathbb R^4 / \mathbb Z_k$}

Let us now revisit the partition function (\ref{Z_SO(2N)}). For simplicity, let us focus on a particular $\lambda'$-sector, where $\lambda' = (k, \bar\lambda', 0)$; that is, consider
\be
 Z_{SO(2N), \lambda'}^{\rm BPS} (q)  =   q^{m_{\lambda'}}  \, \sum_{\bar\mu'}  \sum_{m \geq 0}  \, {\rm dim} \, {\rm IH}^\ast {\cal U}({\cal M}^{\lambda', m}_{SO(2N), \bar\mu'}(\mathbb R^4/ \mathbb Z_k)) \,  q^m,
 \label{Z_SO(2N), revisit}
 \ee 
where $q= e^{2 \pi i \tau}$, and $m_{\lambda'}$ is as given in (\ref{m-OM5}). From (\ref{GL-relation D}), we have\footnote{Here, we recall that for any $\lambda = (k, \bar \lambda, i)$ and $\mu = (k, \bar \mu, j)$, we have $m = k(i-j)$ whereby $(i-j) \in  \mathbb Z_{\geq 0}$. Thus, for $\lambda' = (k, \bar{\lambda'}, 0)$, we have $\mu' = (k, \bar\mu', j')$ such that the integer $j' = - {m \over k} \leq 0$, where $-j'$ is known as the grade of the $\mu'$-string in the mathematical literature, or the energy level of the $| \mu' \rangle$ state in the physical context.\label{i-j-D}} 
\be
{\rm dim} \, {\rm IH}^\ast {\cal U}({\cal M}^{\lambda', m}_{SO(2N), \bar\mu'}(\mathbb R^4/ \mathbb Z_k)) = {\rm mult}_{\lambda'} (\bar\mu')\vert_{ {m'}}, 
\ee
where ${\rm mult}_{\lambda'} (\bar\mu')\vert_{{m'}}$ is the multiplicity of the $|\mu' \rangle$ state of non-negative energy level $m' = {m / k}$ in $^L\widehat{so}(2N)^{\lambda'}_{k, \mu'}$. Consequently, via (\ref{m-OM5})--(\ref{c-OM5}), we can write
\begin{eqnarray}
 Z_{SO(2N), \lambda'}^{\rm BPS} (\tilde q)   = {\tilde q}^{m'_{\lambda'} - c/24} \, \sum_{\bar\mu'}  \sum_{m' \geq 0}  \, {\rm mult}_{\lambda'} (\bar\mu')\vert_{m'} \,  {\tilde q}^{m'}  =   \chi_{^L{{\widehat {so}}(2N)}_{k}}^{\lambda'} (\tilde q),
 \label{chiral WZW SO(2N)}
\end{eqnarray}
where $\tilde q = e^{2 \pi i \tilde \tau}$ and $\tilde \tau = k \tau$.  Here
\be
 \chi_{^L{{\widehat {so}}(2N)}_{k}}^{\lambda'} (\tilde q) =  {\rm Tr}_{\lambda'} \, {\tilde q}^{L_0 + m'_{\lambda'} - c /24}, 
 \label{chi-SO(2N)}
 \ee
 and
 \be
 m'_{\lambda'} = h'_{ \lambda'} -{ (c'_{\lambda'} - c) \over 24},
\ee
while
\be
c= kN, \quad h'_{ \lambda'} = {h_{ \lambda'} \over k}, \quad {\rm and} \quad c'_{\lambda'} = {c_{ \lambda'} \over k},
\label{c-SO(2N)}
\ee
where $h_{ \lambda'}$ and $c_{ \lambda'}$ are as given in (\ref{h-OM5}) and (\ref{c-OM5}), respectively. Also, $L_0$ -- whose  eigenvalue is $m' \in \mathbb Z_{\geq 0}$ -- can be interpreted as the Hamiltonian operator of a 2d theory that is effectively defined on a torus of modulus $\tilde \tau$. Hence, it is clear from (\ref{chiral WZW SO(2N)})--(\ref{c-SO(2N)}) that $ Z_{SO(2N), \lambda'}^{\rm BPS}$ is equal to the $\lambda'$-sector of the partition function of a chiral ${\frak {so}(2N)}^\vee_{\rm aff}$ WZW model on ${\bf S}^1_n \times \mathbb R_t$ with (i) central charge $kN$; (ii) ground state energy level shifted by a number $m'_{\lambda'}$. This observation is consistent with our earlier conclusion about the I-brane partition function, as expected. 

Let us now consider the modified partition function 
\be
{\tilde Z}_{SO(2N), \lambda'}^{\rm BPS} (\tilde q) = {\tilde q}^{\tilde m_{\lambda'}}  Z_{SO(2N), \lambda'}^{\rm BPS} (\tilde q),
\label{mod pf-SO(2N)}
\ee
where
 \be
\tilde m_{\lambda'} = {(k-1) h'_{\lambda'} + {(c'_{\lambda'} - c'_{SO(2N)^\vee, k}) \over 24 }} \quad {\rm while} \quad c'_{SO(2N)^\vee, k} =  {k \, {\rm dim} \, \frak{so}(2N)^\vee  \over{ (k +h)}}.
 \ee
Notice that ${\tilde Z}_{SO(2N), \lambda'}^{\rm BPS}$ is just $Z_{SO(2N), \lambda'}^{\rm BPS}$ but with instanton number shifted by $\tilde m_{\lambda'} $. In the dual 2d theory picture, this is tantamount to a trivial redefinition of the ground state energy level. Hence, ${\tilde Z}_{SO(2N), \lambda'}^{\rm BPS}$ and $Z_{SO(2N), \lambda'}^{\rm BPS}$ can be thought to define the ``same'' physical theory.

From (\ref{chi-SO(2N)}), one can see that $ {\tilde \chi}_{^L{{\widehat {so}}(2N)}_{k}}^{\lambda'} = {\tilde q}^{\tilde m_{\lambda'}}  \chi_{^L{{\widehat {so}}(2N)}_{k}}^{\lambda'}$ is a character of $^L\widehat{so}(2N)^{\lambda'}_{k}$,  where ${\tilde m_{\lambda'}} +  h'_{\lambda'}  - c'_{\lambda'} /24$ is the corresponding modular anomaly. As such, (\ref{chiral WZW SO(2N)}) would mean that the partition function ${\tilde Z}_{SO(2N), \lambda'}^{\rm BPS} $ ought to transform as a representation of the modular group $SL(2, \mathbb Z)$; specifically, we have (c.f.~\cite[eqn.~(14.235)]{CFT text}) the relation
\be
{ \tilde Z}_{SO(2N), \lambda'}^{\rm BPS} (- 1 / \tilde \tau) = \sum_{\lambda} {\cal S}_{\lambda' \lambda} \, { \tilde Z}_{SO(2N), \lambda}^{\rm BPS} (\tilde \tau), 
\label{3.33-OM5}
\ee
where ${\cal S}$ is a $\tilde \tau$-independent unitary matrix (given by~\cite[eqn.~(14.217)]{CFT text}) associated with the \emph{Langlands dual} affine Lie algebra $\frak {so}(2N)^\vee_{\rm aff}$, which represents the $SL(2,\mathbb Z)$ transformation $S: \tilde\tau \to - 1 / \tilde\tau$ in the space of $\lambda$-vector-valued partition functions ${ \tilde Z}_{SO(2N), \lambda}^{\rm BPS}$.  

Via (\ref{mod pf-SO(2N)}) and (\ref{Z_SO(2N), revisit}), the relation (\ref{3.33-OM5}) implies, in the limit of large $k$, the following identity involving the intersection cohomology of the moduli space of $SO(2N)$-instantons on $\mathbb R^4/ \mathbb Z_k$: 
\be
\boxed{\sum_{\bar\mu'}  \sum_{m \geq 0}  \, {\rm dim} \, {\rm IH}^\ast {\cal U}({\cal M}^{\lambda', m}_{SO(2N), \bar\mu'}(\mathbb R^4/ \mathbb Z_k))  =    \sum_{\lambda}  \sum_{\bar\mu}  \sum_{m \geq 0} {\cal S}^m_{\lambda' \lambda}  \,  {\rm dim} \, {\rm IH}^\ast {\cal U}({\cal M}^{\lambda, m}_{SO(2N), \bar\mu}(\mathbb R^4/ \mathbb Z_k))}
\ee
where the components ${\cal S}^m_{\lambda' \lambda}$ are given by
\be
\boxed{{\cal S}^m_{\lambda' \lambda}  = {\hat q}^{{\tilde m_{\lambda'}}\vert_{k \gg 1}}{\tilde q}^{m + \tilde m_{\lambda}+ m_{\lambda}}{\cal S}_{\lambda' \lambda}}
\ee
Here, ${\hat q} = e^{2 \pi i / \tilde \tau}$. (See also footnote~\ref{i-j-D}.)

In other words, in the limit of large $k$, the total dimension of the intersection cohomology of the component of the moduli space of $D_{N}$-instantons on $\mathbb R^4/ \mathbb Z_k$ labeled by a highest weight $\lambda'$ or $\lambda$  -- and therefore, the dimension of the corresponding sector of the Hilbert space of spacetime BPS states -- is found to be intimately related to one another via $\frak {so}(2N)^\vee_{\rm aff}$-dependent unitary modular transformations!

\bigskip\noindent{\it A Geometric Langlands Duality for Surfaces for the $C_{N-1}$ Groups}

Let us now consider $n=2$ whence there is a ``$\mathbb Z_2$-twist'', i.e., the relevant module is $\widehat{so}(2N)^{(2)}_{k}$, the integrable module over the $\mathbb Z_2$-twisted affine Lie algebra ${\frak so}(2N)^{(2)}_{{\rm aff}, k}$ of level $k$.  Then, unitarity of any WZW model requires that $\textrm{WZW}_{\widehat{so}(2N)^{(2)}_{k}}$ be generated by dominant highest weight modules over $\frak{so}(2N)^{(2)}_{\textrm{aff},k}$. By repeating the arguments that led us to write (\ref{generic WZW state OM5})--(\ref{GL-relation D}) in the untwisted case, whilst noting that the Weyl group symmetry mentioned therein persists in this case to map non-dominant affine weights $\tilde \mu'_\nu$ to dominant ones $\tilde \mu_\nu$ even though the grading of $\tilde \mu'_\nu$ (captured by its last index $\tilde j'_\nu$)  may not be integral, we find that we can express the spectrum of states of the corresponding chiral WZW model as
 \be
 \textrm{WZW}_{\widehat{so}(2N)^{(2)}_{k}} =  \bigoplus_{\tilde\lambda} \bigoplus_{\nu =0,1} \, \bigoplus_{\tilde\mu_\nu} \, \overline{\textrm{WZW}}_{\widehat{so}(2N)^{(2), \tilde \lambda}_{k, \tilde\mu_\nu}}.
 \label{WZW space-USp(2N-2)}
 \ee
 Here, the overhead bar means that we project onto $\mathbb Z_2$-invariant states (as required of twisted CFT's); $\nu =0$ or $1$ indicates that the sector is untwisted or twisted, respectively; $\tilde\lambda$ and $\tilde\mu_\nu$ are the (un)twisted  dominant affine weights of the $\mathbb Z_2$-twisted affine Kac-Moody group $SO(2N)^{(2)}_{\rm aff}$ of level $k$; the space $\widehat{so}(2N)^{(2), \tilde \lambda}_{k, \tilde\mu_\nu}$ is the  $\tilde\mu_\nu$-weight space of $\widehat{so}(2N)^{(2), \tilde \lambda}_{k}$, the module over ${\frak so}(2N)^{(2)}_{{\rm aff}, k}$ of dominant highest weight $\tilde\lambda$ of level $k$.

Now, the physical duality of the M-theory compactifications (\ref{M-theory 1 OM5 discussion}) and (\ref{M-theory 7 OM5 discussion}) means that their respective spacetime BPS spectra ought to be equivalent, i.e., $ \textrm{WZW}_{\widehat{so}(2N)^{(2)}_{k}}$ ought to be equal to $\cal H^{\rm eff}_{\rm BPS}$ of (\ref{HBPS-eff-USp(2N-2)}). Indeed, since $\frak{so}(2N)^{(2)}_{\textrm{aff}}$ is isomorphic to $\frak{usp}(2N - 2)^\vee_{\textrm{aff}}$, it would mean that $\tilde\lambda$ and $\tilde\mu_\nu$ are also \emph{dominant weights} of the \emph{Langlands dual} affine Kac-Moody group $USp(2N-2)^\vee_{\rm aff}$ whence we can identify them with $\lambda$ and $\mu_\nu$ of~(\ref{HBPS-eff-USp(2N-2)}), respectively;  moreover, both $\cal H^{\rm eff}_{\rm BPS}$ and $ \textrm{WZW}_{\widehat{so}(2N)^{(2)}_{k}}$ are labeled by $k$. Specializing to the $\{\lambda, \mu_\nu \}$-sector of the spectra of spacetime BPS states, we can therefore write 
\be
{\overline{\cal H}}^{\lambda, \mu_\nu}_{\rm BPS} = \overline{\textrm{WZW}}_{\widehat{so}(2N)^{(2), \lambda}_{k, \mu_\nu}}.
\label{H=WZW-USp(2N-2)}
\ee
As $\overline{\textrm{WZW}}_{\widehat{so}(2N)^{(2), \lambda}_{k, \mu_\nu}}$ is furnished by the $\mathbb Z_2$-invariant projection ${\widehat{so}(2N)^{(2),  \lambda}_{k, \mu_\nu}}\vert_{\mathscr P_2}$ of $\widehat{so}(2N)^{(2),  \lambda}_{k, \mu_\nu}$, and since $\frak{so}(2N)^{(2)}_{\textrm{aff}} \simeq \frak{usp}(2N -2)^\vee_{\textrm{aff}} $ whence ${\widehat{so}(2N)^{(2),  \lambda}_{k, \mu_\nu}}\vert_{\mathscr P_2}$ is isomorphic to the submodule ${^L\widehat{usp}(2N-2)^{\lambda}_{k, \mu_\nu}}$ over $\frak{usp}(2N-2)^\vee_{\textrm{aff}}$, via (\ref{HBPS-eff-USp(2N-2)}), we can also express (\ref{H=WZW-USp(2N-2)}) as 
\be
\boxed{\overline{{\rm IH}^\ast {\cal U}}({\cal M}^{\lambda}_{USp(2N-2), \mu_\nu}(\mathbb R^4/ \mathbb Z_k)) =  {^L\widehat{usp}}(2N-2)^{\lambda}_{k, \mu_\nu}}
\label{GL-relation C}
\ee
for $\nu = 0$ and $1$. Note that (\ref{GL-relation C}) is~\cite[Conjecture 4.14(3)]{BF} for simply-connected $G = USp(2n-2)$! This completes our purely physical derivation of a geometric Langlands duality for surfaces for the $USp(2N-2) = C_{N - 1}$ groups.

\bigskip\noindent{\it A Langlands Duality of the Dimension of the Intersection Cohomology of the Moduli Space of $C_{N-1}$-Instantons on A-Type ALE Spaces}

Let us now revisit the partition function given by (\ref{Z-USp(2N-2)-1})--(\ref{Z-USp(2N-2)-3}). For simplicity, let us focus on a particular $\lambda'$-sector, where $\lambda' = (k, \bar\lambda', 0)$; that is, consider
\be
Z^{\rm BPS}_{USp(2N-2), \lambda'} (q) = q^{m_{\lambda'}} \sum_{\nu =0,1} \sum_{\bar \mu'_\nu}    \sum_{m_\nu \geq 0}   {\rm dim} \, \overline {{\rm IH}^\ast {\cal U}}({\cal M}^{\lambda', m_\nu}_{USp(2N-2), \bar\mu'_\nu}(\mathbb R^4/ \mathbb Z_k)) \, q^{m_\nu}, 
\label{Z-USp(2N-2)-revisit}
\ee
where $q= e^{2 \pi i \tau}$, and $m_{\lambda'}$ takes the form given in (\ref{m-OM5}) with $\tilde b = 1/2$. From our discussion leading up to (\ref{GL-relation C}), we have\footnote{Here, we recall that for any $\lambda = (k, \bar \lambda, i)$ and $\mu_\nu = (k, \bar \mu_\nu, j_\nu)$, we have $m_\nu = 2k(i-j_\nu)$ whereby $(i-j_\nu) \in \mathbb Z_{\geq 0} + {\nu \over 2}$. Thus, for $\lambda' = (k, \bar{\lambda'}, 0)$, we have $\mu'_\nu = (k, \bar\mu'_\nu, j'_\nu)$ such that $j'_\nu = - {m_\nu \over {2k}} \leq 0$, where $-j'_\nu$ is known as the grade of the $\mu'_\nu$-string in the mathematical literature, or the energy level of the $| \mu'_\nu \rangle$ state in the physical context. \label{i-j-C}} 
\be
{\rm dim} \, \overline{{\rm IH}^\ast {\cal U}}({\cal M}^{\lambda', m_\nu}_{USp(2N-2), \bar\mu'_\nu}(\mathbb R^4/ \mathbb Z_k)) = \overline{{\rm mult}}_{\lambda'} (\bar\mu'_\nu)\vert_{ {m'_\nu}}, 
\ee
where $\overline{\rm mult}_{\lambda'} (\bar\mu'_\nu)\vert_{{m'_\nu}}$ is the multiplicity of the $|\mu'_\nu \rangle$ state of non-negative energy level $m'_\nu = {m_\nu / 2k}$ in ${\widehat{so}(2N)^{(2),  \lambda'}_{k, \mu'_\nu}}\vert_{\mathscr P_2}$. Consequently, via (\ref{m-OM5})--(\ref{c-OM5}), we can write
\begin{eqnarray}
\hspace{-0.5cm} Z_{USp(2N-2), \lambda'}^{\rm BPS} (\tilde q)   = {\tilde q}^{m'_{\lambda'} - c/24} \, \sum_{\nu =0,1}\sum_{\bar\mu'_\nu}  \sum_{m'_\nu \geq 0}  \, \overline{\rm mult}_{\lambda'} (\bar\mu'_\nu)\vert_{m'_\nu} \,  {\tilde q}^{m'_\nu}  =  \sum_{\nu =0,1} \, \chi_{{{\widehat {so}}(2N)}^{(2)}_{k}}^{\lambda',\nu} (\tilde q),
 \label{chiral WZW USp(2N-2)}
\end{eqnarray}
where $\tilde q = e^{2 \pi i \tilde \tau}$ and $\tilde \tau = 2k \tau$.  Here
\be
\chi_{{{\widehat {so}}(2N)}^{(2)}_{k}}^{\lambda',\nu} (\tilde q) =  {\rm Tr}_{\lambda'} \, \mathscr P_2 \, {\tilde q}^{L_{0, \nu} + m'_{\lambda'} - c /24}, 
 \label{chi-USp(2N-2)}
 \ee
 where as before, $\mathscr P_2$ singles out the $\mathbb Z_2$-invariant states, and
 \be
 m'_{\lambda'} = h'_{ \lambda'} -{ (c'_{\lambda', \tilde b} - c) \over 24}.
\ee
The constants are
\be
c= kN, \quad h'_{ \lambda'} = {(\bar \lambda', \bar\lambda' + 2 \rho) \over 4k(k + h^\vee)}, \quad c'_{\lambda', \tilde b} = {-{24 \tilde b {(\bar \lambda', \bar \lambda')} \over 2k}  +  {12 (\bar \lambda', \bar \lambda' + 2 \rho) \over {2k(k + h^\vee)}}},
\label{c-USp(2N-2)}
\ee
such that $\tilde b$ is some positive real constant (first introduced in (\ref{a-OM5})), and $\rho$ and $h^\vee$ are the Weyl vector and dual Coxeter number associated with ${\frak {so}(2N)}^{(2)}_{\rm aff}$, respectively.  Also, $L_{0, \nu}$ -- whose  eigenvalue is $m'_\nu \in \mathbb Z_{\geq 0} + {\nu \over 2}$ -- can be interpreted as the Hamiltonian operator of a 2d theory that is effectively defined on a torus of modulus $\tilde \tau$. Hence, it is clear from (\ref{chiral WZW USp(2N-2)})--(\ref{c-USp(2N-2)}) that $ Z_{USp(2N-2), \lambda'}^{\rm BPS}$ is equal to the $\lambda'$-sector of the partition function of a chiral ${\frak {so}}(2N)^{(2)}_{\rm aff}$ WZW model on ${\bf S}^1_n \times \mathbb R_t$ with (i) central charge $kN$; (ii) ground state energy level shifted by a number $m'_{\lambda'}$.  This observation is consistent with our earlier conclusion about the I-brane partition function, as expected. 

Let us now consider the modified partition function 
\be
{\tilde Z}_{USp(2N-2), \lambda'}^{\rm BPS} (\tilde q) = {\tilde q}^{\tilde m_{\lambda'}}  Z_{USp(2N-2), \lambda'}^{\rm BPS} (\tilde q),
\label{mod pf-USp(2N-2)}
\ee
where (c.f.~\cite[eqns.~(12.7.5) and (13.11.5)]{Kac})
 \be
\tilde m_{\lambda'} = -h'_{\lambda'} + {c'_{\lambda', \tilde b} \over 24} +{ {|\bar\lambda' + \rho|^2} \over 2(k+ h^\vee)} - {{\rm dim} \, {\frak{so}(2N)} \over 48}.
\label{tilde m-USp}
 \ee
Notice that ${\tilde Z}_{USp(2N-2), \lambda'}^{\rm BPS}$ is just $Z_{USp(2N-2), \lambda'}^{\rm BPS}$ but with instanton number shifted by $\tilde m_{\lambda'} $. In the dual 2d theory picture, this is tantamount to a trivial redefinition of the ground state energy level. Hence, ${\tilde Z}_{USp(2N-2), \lambda'}^{\rm BPS}$ and $Z_{USp(2N-2), \lambda'}^{\rm BPS}$ can be thought to define the ``same'' physical theory.

From (\ref{chi-USp(2N-2)}), one can see that ${\tilde \chi}_{{{\widehat {so}}(2N)}^{(2)}_{k}}^{\lambda',\nu}  = {\tilde q}^{\tilde m_{\lambda'}} \chi_{{{\widehat {so}}(2N)}^{(2)}_{k}}^{\lambda',\nu}$ is a ($\mathbb Z_2$-invariant) character of the $\nu$-sector of $\widehat{so}(2N)^{{(2)}, \lambda'}_{k}$, where $\tilde m_{\lambda'} + h'_{\lambda'} - {c'_{\lambda', \tilde b} / 24}$ is the corresponding modular anomaly. As such, (\ref{mod pf-USp(2N-2)}), (\ref{chiral WZW USp(2N-2)}) and~\cite[Theorem 13.9]{Kac} mean that the partition function ${\tilde Z}_{USp(2N-2), \lambda'}^{\rm BPS} $ ought to transform under $S: \tilde \tau \to - 1 / \tilde\tau$ as follows: 
\be
{ \tilde Z}_{USp(2N-2), \lambda'}^{\rm BPS} (- 1 / \tilde \tau)  = \sum_{\xi} {\cal S}_{\lambda' \xi} \, {\tilde \chi}_{{{\widehat {su}}(2N - 2)}^{(2)}_{k}}^{\xi}  (\tilde \tau / 2). 
\label{3.33-USp(2N-2)}
\ee
Here, ${\cal S}$ is a $\tilde \tau$-independent matrix (given in~\cite[Theorem 13.9]{Kac}); $\xi$ is a dominant highest weight of the $\mathbb Z_2$-twisted affine Kac-Moody group $SU(2N - 2)^{(2)}_{\rm aff}$ of level $k$; ${\tilde \chi}_{{{\widehat {su}}(2N - 2)}^{(2)}_{k}}^{\xi} = {\tilde q}^{\tilde m_{\xi}} {\chi}_{{{\widehat {su}}(2N - 2)}^{(2)}_{k}}^{\xi}$, where $\tilde m_{\xi}$ is as in (\ref{tilde m-USp}) but with ${\frak{so}(2N)}$ replaced by ${\frak{su}(2N-2)}$.  Notice that the group type on the LHS and RHS of (\ref{3.33-USp(2N-2)}) are not the same; nevertheless, from the main result of the previous subsection, it is clear that the characters on the RHS of (\ref{3.33-USp(2N-2)}) will be given by the partition function ${ \tilde Z}_{SO(2N-1), \xi}^{\rm BPS} (\tilde \tau / 2)$ associated with $SO(2N -1)$-instantons on $\mathbb R^4 / \mathbb Z_{k}$, i.e.,  
\be
{ \tilde Z}_{USp(2N-2), \lambda'}^{\rm BPS} (- 1 / \tilde \tau)  = \sum_{\xi} {\cal S}_{\lambda' \xi} \, { \tilde Z}_{SO(2N -1), \xi}^{\rm BPS} (\tilde \tau / 2).
\label{3.33-USp(2N-2) dual}
\ee

Via (\ref{mod pf-USp(2N-2)}), (\ref{Z-USp(2N-2)-revisit}) and (\ref{Z-SO(N+1)-revisit}), the relation (\ref{3.33-USp(2N-2) dual}) implies, in the limit of large $k$, the following identity involving the intersection cohomology of the moduli space of instantons: 
\be
\label{IC-SO-USp-OM5}
\hspace{-0.0cm}\boxed{\sum_{\nu =0,1} \sum_{\bar \mu'_\nu}    \sum_{m_\nu \geq 0}   {\rm dim} \, \overline {{\rm IH}^\ast {\cal U}}({\cal M}^{\lambda', m_\nu}_{USp(2N-2), \bar\mu'_\nu}(\mathbb R^4/ \mathbb Z_k)) 
  =  \sum_{\xi} \sum_{\delta =0,1} \sum_{\bar\zeta_\delta}  \sum_{m_\delta \geq 0}   {\cal S}^{m_\delta}_{\lambda' \xi}  \,  {\rm dim} \, \overline{{\rm IH}^\ast {\cal U}}({\cal M}^{\xi, m_\delta}_{SO(2N-1), \bar\zeta_\delta}(\mathbb R^4/ \mathbb Z_{k}))}
\ee
where the components ${\cal S}^{m_\delta}_{\lambda' \xi}$ are given by
\be
\boxed{{\cal S}^{m_\delta}_{\lambda' \xi}  = {\hat q^{\tilde m_{\lambda'}\vert_{k \gg 1}}} {\tilde q}^{(m_\delta + \tilde m_{\xi}+ m_{\xi})/2}{\cal S}_{\lambda' \xi}}
\ee
Here, ${\hat q} = e^{2 \pi i / \tilde \tau}$; $\tilde m_{\lambda'}\vert_ {k \gg 1} = -{{{\rm dim} \, \frak{so} (2N)} / 48}$; $\lambda' = (k, \bar \lambda', 0)$ and $\mu'_\nu = (k, \bar\mu'_\nu, j_\nu)$ are dominant coweights of the affine Kac-Moody group $USp(2N-2)_{\rm aff}$ of level $k$, where $\bar \lambda'$ and $\bar \mu'$ are the corresponding dominant coweights of $USp(2N-2)$, and ${m_\nu \over 2k} = - j_\nu \in  \mathbb Z_{\geq 0} + {\nu \over 2}$ (see footnote~\ref{i-j-C}); $\xi = (k, \bar \xi, 0)$ and $\zeta_\delta = (k, \bar\zeta_\delta, j_\delta)$  are dominant coweights of the affine Kac-Moody group $SO(2N-1)_{\rm aff}$ of level $k$, where $\bar \xi$ and $\bar \zeta_\delta$ are the corresponding dominant coweights of $SO(2N-1)$, and for $N > 2$, ${m_\delta \over 2k} = - j_\delta \in \mathbb Z_{\geq 0} + {\delta \over 2}$.

At any rate, it is clear from (\ref{IC-SO-USp-OM5}) that in the limit of large $k$, the total dimension of the intersection cohomology of the moduli space of $G$-instantons on $\mathbb R^4/ \mathbb Z_k$ in the $\lambda'$-sector, can be expressed in terms of the dimensions of the intersection cohomology of the various components of the moduli space of $G^\vee$-instantons on $\mathbb R^4/ \mathbb Z_{k}$, where $G = USp(2N-2)$. In other words, we have a \emph{Langlands duality }of the dimension of the intersection cohomology of the moduli space of $C_{N-1}$-instantons on A-type ALE spaces! 

\bigskip\noindent{\it A Comparison With (\ref{IC-SO-USp})}

Recall that we also obtained a Langlands duality of the dimension of the intersection cohomology of the moduli space of $B_{N-1}$-instantons on A-type ALE spaces in (\ref{IC-SO-USp}) (after we relabel $N/2$ therein as $N-1$) which tells us that
\be
\label{IC-SO-USp-OM5-compare}
\hspace{-0.2cm}\sum_{\delta =0,1} \sum_{\bar \zeta'_\delta}    \sum_{m_\delta \geq 0}   {\rm dim} \, \overline {{\rm IH}^\ast {\cal U}}({\cal M}^{\xi', m_\delta}_{SO(2N -1), \bar\zeta'_\delta}(\mathbb R^4/ \mathbb Z_{k})) 
  =  \sum_{\lambda} \sum_{\nu =0,1} \sum_{\bar\mu_\nu}  \sum_{m_\nu \geq 0}   {\cal S}^{m_\nu}_{\xi' \lambda}  \,  {\rm dim} \, \overline{{\rm IH}^\ast {\cal U}}({\cal M}^{\lambda, m_\nu}_{USp(2N - 2), \bar\mu_\nu}(\mathbb R^4/ \mathbb Z_{k})),
\ee
where  the components ${\cal S}^{m_\nu}_{\xi' \lambda}$ are given by
\be
{\cal S}^{m_\nu}_{\xi' \lambda}  = {\hat q}^{\tilde m_{\xi'}\vert_ {k \gg 1}}{\tilde q}^{(m_\nu + \tilde m_{\lambda}+ m_{\lambda})/2}{\cal S}_{\xi' \lambda}.  
\ee
Here, $\tilde m_{\xi'}\vert_ {k \gg 1} = -{{{\rm dim} \, \frak{su} (2N-2)} / 48}$; $\xi' = (k, \bar \xi', 0)$ and $\zeta'_\delta = (k, \bar \zeta'_\delta, j_\delta)$ are dominant coweights of the affine Kac-Moody group $SO(2N-1)_{\rm aff}$ of level $k$, where $\bar \xi'$ and $\bar \zeta'_\delta$ are the corresponding dominant coweights of $SO(2N - 1)$, and for $N > 2$, ${m_\delta \over 2k} = - j_\delta \in \mathbb Z_{\geq 0} + {\delta \over 2}$; $\lambda = (k, \bar \lambda, 0)$ and $\mu_\nu = (k, \bar\mu_\nu, j_\nu)$  are dominant coweights of the affine Kac-Moody group $USp(2N - 2)_{\rm aff}$ of level $k$, where $\bar \lambda$ and $\bar \mu_\nu$ are the corresponding dominant coweights of $USp(2N - 2)$, and ${m_\nu \over 2k} = - j_\nu \in \mathbb Z_{\geq 0} + {\nu \over 2}$. 

Assuming that $N > 2$, notice that the relations (\ref{IC-SO-USp-OM5}) and (\ref{IC-SO-USp-OM5-compare}) map into each other when we exchange $SO(2N-1) \leftrightarrow USp(2N - 2)$ and thus $\frak{su} (2N-2) \leftrightarrow \frak {so}(2N)$,\footnote{To understand this, recall that $\frak {so} (2N -1)^\vee_{\rm aff} \simeq \frak{su} (2N-2)^{(2)}_{\rm aff}$ and $ \frak{usp} (2N-2)^\vee_{\rm aff} \simeq \frak {so}(2N)^{(2)}_{\rm aff}$; hence, the exchange $SO (2N -1) \leftrightarrow USp (2N-2)$ would imply the exchange $\frak{su} (2N-2) \leftrightarrow \frak {so}(2N)$.} $(\xi', \zeta'_\delta, \xi, \zeta_\delta) \leftrightarrow (\lambda', \mu'_\nu, \lambda, \mu_\nu)$, and $m_\delta \leftrightarrow m_\nu$. This is expected, since the groups $B_{N-1}$ and $C_{N-1}$ are themselves Langlands dual to each other.

\bigskip\noindent{\it A Geometric Langlands Duality for Surfaces for the $G_{2}$ Group}

Let us now consider  $N=4$ and $n=3$ whence there is a ``$\mathbb Z_3$-twist'', i.e., the relevant module is $\widehat{so}(8)^{(3)}_{k}$, the integrable module over the $\mathbb Z_3$-twisted affine Lie algebra ${\frak so}(8)^{(3)}_{{\rm aff}, k}$ of level $k$.  Then, unitarity of any WZW model requires that $\textrm{WZW}_{\widehat{so}(8)^{(3)}_{k}}$ be generated by dominant highest weight modules over $\frak{so}(8)^{(3)}_{\textrm{aff},k}$. By repeating the arguments that led us to write (\ref{generic WZW state OM5})--(\ref{GL-relation D}) in the untwisted case, whilst noting that the Weyl group symmetry mentioned therein persists in this case to map non-dominant affine weights $\tilde \mu'_\nu$ to dominant ones $\tilde \mu_\nu$ even though the grading of $\tilde \mu'_\nu$ (captured by its last index $\tilde j'_\nu$)  may not be integral, we find that we can express the spectrum of states of the corresponding chiral WZW model as
 \be
 \textrm{WZW}_{\widehat{so}(8)^{(3)}_{k}} =  \bigoplus_{\tilde\lambda} \bigoplus_{\nu =0,1,2} \, \bigoplus_{\tilde\mu_\nu} \, \overline{\textrm{WZW}}_{\widehat{so}(8)^{(3), \tilde \lambda}_{k, \tilde\mu_\nu}}.
 \label{WZW space-G_2}
 \ee
 Here, the overhead bar means that we project onto $\mathbb Z_3$-invariant states (as required of twisted CFT's); $\nu \neq 0$ indicates that the sector is twisted; $\tilde\lambda$ and $\tilde\mu_\nu$ are the (un)twisted  dominant affine weights of the $\mathbb Z_3$-twisted affine Kac-Moody group $SO(8)^{(3)}_{\rm aff}$ of level $k$; the space $\widehat{so}(8)^{(3), \tilde \lambda}_{k, \tilde\mu_\nu}$ is the  $\tilde\mu_\nu$-weight space of $\widehat{so}(8)^{(3), \tilde \lambda}_{k}$, the module over ${\frak so}(8)^{(8)}_{{\rm aff}, k}$ of dominant highest weight $\tilde\lambda$ of level $k$.

Now, the physical duality of the M-theory compactifications (\ref{M-theory 1 OM5 discussion}) and (\ref{M-theory 7 OM5 discussion}) means that their respective spacetime BPS spectra ought to be equivalent, i.e., $ \textrm{WZW}_{\widehat{so}(8)^{(3)}_{k}}$ ought to be equal to $\cal H^{\rm eff}_{\rm BPS}$ of (\ref{HBPS-eff-G_2}). Indeed, since $\frak{so}(8)^{(3)}_{\textrm{aff}}$ is isomorphic to ${{\frak g}^\vee_2}_{\textrm{aff}}$, it would mean that $\tilde\lambda$ and $\tilde\mu_\nu$ are also \emph{dominant weights} of the \emph{Langlands dual} affine Kac-Moody group ${G^\vee_2}_{\textrm{aff}}$ whence we can identify them with $\lambda$ and $\mu_\nu$ of~(\ref{HBPS-eff-G_2}), respectively;  moreover, both $\cal H^{\rm eff}_{\rm BPS}$ and $ \textrm{WZW}_{\widehat{so}(8)^{(3)}_{k}}$ are labeled by $k$. Specializing to the $\{\lambda, \mu_\nu \}$-sector of the spectra of spacetime BPS states, we can therefore write 
\be
{\overline{\cal H}}^{\lambda, \mu_\nu}_{\rm BPS} = \overline{\textrm{WZW}}_{\widehat{so}(8)^{(3), \lambda}_{k, \mu_\nu}}.
\label{H=WZW-G_2}
\ee
As $\overline{\textrm{WZW}}_{\widehat{so}(8)^{(3), \lambda}_{k, \mu_\nu}}$ is furnished by the $\mathbb Z_3$-invariant projection ${\widehat{so}(8)^{(3),  \lambda}_{k, \mu_\nu}}\vert_{\mathscr P_3}$ of $\widehat{so}(8)^{(3),  \lambda}_{k, \mu_\nu}$, and since $\frak{so}(8)^{(3)}_{\textrm{aff}} \simeq {{\frak g}^\vee_2}_{\textrm{aff}}$ whence ${\widehat{so}(8)^{(3),  \lambda}_{k, \mu_\nu}}\vert_{\mathscr P_3}$ is isomorphic to the submodule $({^L\widehat{g_2}})^\lambda_{k, \mu_\nu}$ over ${{\frak g}^\vee_2}_{\textrm{aff}}$, via (\ref{HBPS-eff-G_2}), we can also express (\ref{H=WZW-G_2}) as 
\be
\boxed{\overline{{\rm IH}^\ast {\cal U}}({\cal M}^{\lambda}_{G_2, \mu_\nu}(\mathbb R^4/ \mathbb Z_k)) = ({^L\widehat{g_2}})^\lambda_{k, \mu_\nu}}
\label{GL-relation G}
\ee
for $\nu = 0, 1$ and 2. Note that (\ref{GL-relation G}) is~\cite[ Conjecture 4.14(3)]{BF} for simply-connected $G = G_2$! This completes our purely physical derivation of a geometric Langlands duality for surfaces for the $G_2$ group.

\bigskip\noindent{\it An Identity of the Dimension of the Intersection Cohomology of the Moduli Space of $G_2$-Instantons on $\mathbb R^4 / \mathbb Z_k$}

Let us now revisit the partition function given by (\ref{Z-G_2-1})--(\ref{Z-G_2-2}). For simplicity, let us focus on a particular $\lambda'$-sector, where $\lambda' = (k, \bar\lambda', 0)$; that is, consider
\be
Z^{\rm BPS}_{G_2, \lambda'} (q) = q^{m_{\lambda'}} \sum_{\nu =0}^2 \sum_{\bar \mu'_\nu}    \sum_{m_\nu \geq 0}   {\rm dim} \, \overline {{\rm IH}^\ast {\cal U}}({\cal M}^{\lambda', m_\nu}_{G_2, \bar\mu'_\nu}(\mathbb R^4/ \mathbb Z_k)) \, q^{m_\nu}, 
\label{Z-G_2-revisit}
\ee
where $q= e^{2 \pi i \tau}$, and $m_{\lambda'}$ takes the form given in (\ref{m-OM5}) with $\tilde b = 1/2$. From our discussion leading up to (\ref{GL-relation G}), we have\footnote{Here, we recall that for any $\lambda = (k, \bar \lambda, i)$ and $\mu_\nu = (k, \bar \mu_\nu, j_\nu)$, we have $m_\nu = 3k(i-j_\nu)$ whereby $(i-j_\nu) \in \mathbb Z_{\geq 0} + {\nu \over 3}$. Thus, for $\lambda' = (k, \bar{\lambda'}, 0)$, we have $\mu'_\nu = (k, \bar\mu'_\nu, j'_\nu)$ such that $j'_\nu = - {m_\nu \over {3k}} \leq 0$, where $-j'_\nu$ is known as the grade of the $\mu'_\nu$-string in the mathematical literature, or the energy level of the $| \mu'_\nu \rangle$ state in the physical context. \label{i-j-G}} 
\be
{\rm dim} \, \overline{{\rm IH}^\ast {\cal U}}({\cal M}^{\lambda', m_\nu}_{G_2, \bar\mu'_\nu}(\mathbb R^4/ \mathbb Z_k)) = \overline{{\rm mult}}_{\lambda'} (\bar\mu'_\nu)\vert_{ {m'_\nu}}, 
\ee
where $\overline{\rm mult}_{\lambda'} (\bar\mu'_\nu)\vert_{{m'_\nu}}$ is the multiplicity of the $|\mu'_\nu \rangle$ state of non-negative energy level $m'_\nu = {m_\nu / 3k}$ in ${\widehat{so}(8)^{(3),  \lambda'}_{k, \mu'_\nu}}\vert_{\mathscr P_3}$. Consequently, via (\ref{m-OM5})--(\ref{c-OM5}), we can write
\begin{eqnarray}
\hspace{-0.0cm} Z_{G_2, \lambda'}^{\rm BPS} (\tilde q)   = {\tilde q}^{m'_{\lambda'} - c/24} \, \sum^2_{\nu =0}\sum_{\bar\mu'_\nu}  \sum_{m'_\nu \geq 0}  \, \overline{\rm mult}_{\lambda'} (\bar\mu'_\nu)\vert_{m'_\nu} \,  {\tilde q}^{m'_\nu}  =  \sum^2_{\nu =0} \, \chi_{{{\widehat {so}}(8)}^{(3)}_{k}}^{\lambda',\nu} (\tilde q),
 \label{chiral WZW G_2}
\end{eqnarray}
where $\tilde q = e^{2 \pi i \tilde \tau}$ and $\tilde \tau = 3k \tau$.  Here
\be
\chi_{{{\widehat {so}}(8)}^{(3)}_{k}}^{\lambda',\nu} (\tilde q) =  {\rm Tr}_{\lambda'} \, \mathscr P_3 \, {\tilde q}^{L_{0, \nu} + m'_{\lambda'} - c /24}, 
 \label{chi-G_2}
 \ee
 where as before, $\mathscr P_3$ singles out the $\mathbb Z_3$-invariant states, and
 \be
 m'_{\lambda'} = h'_{ \lambda'} -{ (c'_{\lambda', \tilde b} - c) \over 24}.
\ee
The constants are
\be
c= kN, \quad h'_{ \lambda'} = {(\bar \lambda', \bar\lambda' + 2 \rho) \over 6k(k + h^\vee)}, \quad c'_{\lambda', \tilde b} = {-{24 \tilde b {(\bar \lambda', \bar \lambda')} \over 3k}  +  {12 (\bar \lambda', \bar \lambda' + 2 \rho) \over {3k(k + h^\vee)}}},
\label{c-G_2}
\ee
such that $\tilde b$ is some positive real constant (first introduced in (\ref{a-OM5})), and $\rho$ and $h^\vee$ are the Weyl vector and dual Coxeter number associated with ${\frak {so}(8)}^{(3)}_{\rm aff}$, respectively.  Also, $L_{0, \nu}$ -- whose  eigenvalue is $m'_\nu \in \mathbb Z_{\geq 0} + {\nu \over 3}$ -- can be interpreted as the Hamiltonian operator of a 2d theory that is effectively defined on a torus of modulus $\tilde \tau$. Hence, it is clear from (\ref{chiral WZW G_2})--(\ref{c-G_2}) that $ Z_{G_2, \lambda'}^{\rm BPS}$ is equal to the $\lambda'$-sector of the partition function of a chiral ${\frak {so}}(8)^{(3)}_{\rm aff}$ WZW model on ${\bf S}^1_n \times \mathbb R_t$ with (i) central charge $kN$; (ii) ground state energy level shifted by a number $m'_{\lambda'}$.  This observation is consistent with our earlier conclusion about the I-brane partition function, as expected. 

Let us now consider the modified partition function 
\be
{\tilde Z}_{G_2, \lambda'}^{\rm BPS} (\tilde q) = {\tilde q}^{\tilde m_{\lambda'}}  Z_{G_2, \lambda'}^{\rm BPS} (\tilde q),
\label{mod pf-G_2}
\ee
where (c.f.~\cite[eqns.~(12.7.5) and (13.11.5)]{Kac})
 \be
\tilde m_{\lambda'} = -h'_{\lambda'} + {c'_{\lambda', \tilde b} \over 24} +{ {|\bar\lambda' + \rho|^2} \over 2(k+ h^\vee)} - {{\rm dim} \, {\frak{so}(8)} \over 72}.
\label{tilde m-G_2}
 \ee
Notice that ${\tilde Z}_{G_2, \lambda'}^{\rm BPS}$ is just $Z_{G_2, \lambda'}^{\rm BPS}$ but with instanton number shifted by $\tilde m_{\lambda'} $. In the dual 2d theory picture, this is tantamount to a trivial redefinition of the ground state energy level. Hence, ${\tilde Z}_{G_2, \lambda'}^{\rm BPS}$ and $Z_{G_2, \lambda'}^{\rm BPS}$ can be thought to define the ``same'' physical theory.

From (\ref{chi-G_2}), one can see that ${\tilde \chi}_{{{\widehat {so}}(8)}^{(3)}_{k}}^{\lambda',\nu}  = {\tilde q}^{\tilde m_{\lambda'}} \chi_{{{\widehat {so}}(8)}^{(3)}_{k}}^{\lambda',\nu}$ is a ($\mathbb Z_3$-invariant) character of the $\nu$-sector of $\widehat{so}(8)^{{(3)}, \lambda'}_{k}$, where $\tilde m_{\lambda'} + h'_{\lambda'} - {c'_{\lambda', \tilde b} / 24}$ is the corresponding modular anomaly. As such, (\ref{mod pf-G_2}), (\ref{chiral WZW G_2}) and~\cite[ Theorem 13.9]{Kac} mean that the partition function ${\tilde Z}_{G_2, \lambda'}^{\rm BPS} $ ought to transform under $S: \tilde \tau \to - 1 / \tilde\tau$ as follows: 
\be
{ \tilde Z}_{G_2, \lambda'}^{\rm BPS} (- 1 / \tilde \tau)  = \sum_{\lambda} {\cal S}_{\lambda' \lambda} \, {\tilde \chi}_{{{\widehat {so}}(8)}^{(3)}_{k}}^{\lambda}  (\tilde \tau / 3) =  \sum_{\lambda} {\cal S}_{\lambda' \lambda} \, { \tilde Z}_{G_2, \lambda}^{\rm BPS} (\tilde \tau / 3). 
\label{3.33-G_2}
\ee
Here, ${\cal S}$ is a $\tilde \tau$-independent matrix (given in~\cite[Theorem 13.9]{Kac}) associated with the \emph{twisted} affine Lie algebra $\frak {so}(8)^{(3)}_{{\rm aff}}$, and $\lambda$ is a dominant highest coweight of the affine Kac-Moody group ${G_2}_{\rm aff}$ of level $k$.

Via (\ref{mod pf-G_2}) and (\ref{Z-G_2-revisit}), the relation (\ref{3.33-G_2}) implies, in the large $k$ limit, the following identity involving the intersection cohomology of the moduli space of instantons: 
\be
\label{IC-SO-G_2-OM5}
\hspace{-0.0cm}\boxed{\sum_{\nu =0}^2 \sum_{\bar \mu'_\nu}    \sum_{m_\nu \geq 0}   {\rm dim} \, \overline {{\rm IH}^\ast {\cal U}}({\cal M}^{\lambda', m_\nu}_{G_2, \bar\mu'_\nu}(\mathbb R^4/ \mathbb Z_k)) 
  =  \sum_{\lambda} \sum_{\nu =0}^2 \sum_{\bar\mu_\nu}  \sum_{m_\nu \geq 0}   {\cal S}^{m_\nu}_{\lambda' \lambda}  \,  {\rm dim} \, \overline{{\rm IH}^\ast {\cal U}}({\cal M}^{\lambda, m_\nu}_{G_2, \bar\mu_\nu}(\mathbb R^4/ \mathbb Z_{k}))}
\ee
where the components ${\cal S}^{m_\nu}_{\lambda' \lambda}$ are given by
\be
\boxed{{\cal S}^{m_\nu}_{\lambda' \lambda}  = {\hat q}^{\tilde m_{\lambda'}} {\tilde q}^{(m_\nu + \tilde m_{\lambda}+ m_{\lambda})/3}{\cal S}_{\lambda' \lambda}}
\ee
Here, $\hat q = e^{2 \pi i / \tilde \tau}$. (See also footnote~\ref{i-j-G}.)

In other words, in the limit of large $k$, the total dimension of the intersection cohomology of the component of the moduli space of $G_2$-instantons on $\mathbb R^4/ \mathbb Z_k$ labeled by a highest weight $\lambda'$ or $\lambda$  -- and therefore, the dimension of the corresponding sector of the Hilbert space of spacetime BPS states -- is found to be intimately related to one another via $\frak {so}(8)^{(3)}_{\rm aff}$-dependent modular transformations!

\newsubsection{An Equivalence of Spacetime BPS Spectra and a Geometric Langlands Duality for Surfaces for the $E$--$F$ Groups}

We shall now derive, purely physically, a geometric Langlands duality for surfaces for the $E$--$F$ groups. Let us start with the $E_6$ case. (The derivation for the $E_{7,8}$ case is similar, and we shall skip it for brevity.)  

\bigskip\noindent{\it A Geometric Langlands Duality for Surfaces for the $E_6$ Group}

To this end, first consider type IIA theory on a circle ${\bf S}^1_r$ with radius $r$; this is T-dual to type IIB theory on a circle ${\bf S}^1_{1/r}$ with radius $1/r$. Next, further compactify both theories on a singular K3 manifold with an $E_6$-singularity. Then, let the remaining noncompact directions be spanned by $\mathbb R_t \times \mathbb R^4/\mathbb Z_k$. Lastly, lift the IIA configuration to M-theory via an ``eleventh circle'' of radius $r' \to 0$. In all, this means that we have the following \emph{physically dual} compactifications: 
\begin{eqnarray}
\label{M-theory-IIB duality}
{\textrm {M-theory}}: & {\bf S}^1_{11; r' \to 0} \times {\rm K3}_{E_6}  \times {\bf S}^1_r \times \mathbb R_t \times \mathbb R^4/\mathbb Z_k  \nonumber \\
&\Updownarrow \\
{\textrm {Type IIB}}: & {\rm K3}_{E_6} \times {\bf S}^1_{1/r} \times \mathbb R_t \times \mathbb R^4/\mathbb Z_k \nonumber.
\end{eqnarray}
Let us choose $r \approx 1$ so that the compact four-manifold ${\rm K3}_{E_6}$ and the ``eleventh circle''  ${\bf S}^1_{11; r' \to 0}$ are much smaller than the noncompact spaces ${\bf S}^1_{r} \times \mathbb R^4/\mathbb Z_k$ and ${\bf S}^1_{1/r} \times \mathbb R^4/\mathbb Z_k$; then, we can view (\ref{M-theory-IIB duality}) as a duality of six-dimensional string compactifications on  ${\bf S}^1_{11; r' \to 0} \times {\rm K3}_{E_6}$ and ${\rm K3}_{E_6}$, whereby the corresponding spacetime is ${\bf S}^1_{r} \times \mathbb R_t  \times \mathbb R^4/\mathbb Z_k$ and ${\bf S}^1_{1/r} \times \mathbb R_t  \times \mathbb R^4/\mathbb Z_k$ on the M-theory and IIB side, respectively. 

In the low-energy limit, the six-dimensional spacetime theory on the IIB side is the  $\cN = (2,0)$ $E_6$ theory on ${\bf S}^1_{1/r} \times \mathbb R_t  \times \mathbb R^4/\mathbb Z_k$. Note that for an $\cN = (2,0)$ theory on ${\bf S}^1 \times \mathbb R_t  \times M_4$, where $M_4$ is any hyperk\"ahler four-manifold, the theory is topological along $M_4$ (and conformal along ${\bf S}^1 \times \mathbb R_t$)~\cite{junya}.\footnote{In \emph{loc.~cit.}, it was shown that one can twist the theory such that there are two topological scalar supercharges on a generic four-manifold $M_4$. However, when $M_4$ is hyperk\"ahler, there will be an enhancement to eight supersymmetries on $M_4$ whence the untwisted and twisted theories are one and the same thing; hence our claim.\label{junya's twist}} In particular, this means that the BPS spectrum of minimal energy states of the  $\cN = (2,0)$ $E_6$ theory on ${\bf S}^1_{1/r} \times \mathbb R_t  \times \mathbb R^4/\mathbb Z_k$ -- which are states annihilated by all eight unbroken supercharges whence they satisfy $H=P$, where $H$ and $P$ are the Hamiltonian and momentum operators which generate translations along $\mathbb R_t$ and ${\bf S}^1_{1/r}$, respectively\footnote{In the context of our derivation of the duality for the $A$--$B$--$C$--$D$--$G$ groups  in $\S$3.1--3.2, these minimal energy states correspond to the ground states of the M5-brane worldvolume $\cN = (2,0)$ theory (described in footnote~\ref{worldvolume ground state}) which are similarly annihilated by all eight unbroken worldvolume supercharges whence they satisfy $H=P$.\label{minimal energy state}} -- is invariant under topological deformations of $\mathbb R^4/\mathbb Z_k$.  

Let us ascertain this BPS spectrum of minimal energy states in the case where $\mathbb R^4/\mathbb Z_k$ has yet to be topologically deformed. According to footnote~\ref{minimal energy state}, and our explanations in $\S$3.1, the Hilbert space of such BPS states would be given by
\be
{\cal H}^{E_6}_{{\rm BPS}} = \bigoplus_{\lambda, \mu} ~{\rm IH}^\ast {\cal U}({\cal M}^{\lambda}_{E_6, \mu}(\mathbb R^4/ \mathbb Z_k)).
\label{BPS-E6}
\ee
Here, ${\rm IH}^\ast {\cal U}({\cal M}^{\lambda}_{E_6, \mu}(\mathbb R^4/ \mathbb Z_k))$ is the intersection cohomology of the Uhlenbeck compactification ${\cal U}({\cal M}^{\lambda}_{E_6, \mu}(\mathbb R^4/ \mathbb Z_k))$ of the moduli space ${\cal M}^{\lambda}_{E_6, \mu}(\mathbb R^4/ \mathbb Z_k)$ of $E_6$-instantons on $\mathbb R^4/ \mathbb Z_k$ in the $\{\lambda, \mu\}$-sector; $\lambda$ and $\mu$ can be regarded as\emph{ dominant weights} of the corresponding \emph{Langlands dual} affine Kac-Moody group $E^\vee_{{6 \, \rm aff}}$ of level $k$; and
\be
\lambda \geq \mu.
\label{weight condition-E6}
\ee

Let us now topologically deform $\mathbb R^4/\mathbb Z_k$ and scale it down to zero size. Then, the 6d $\cN = (2,0)$ $E_6$ theory on ${\bf S}^1_{1/r} \times \mathbb R_t  \times \mathbb R^4/\mathbb Z_k$ will essentially reduce to a 2d theory along ${\bf S}^1_{1/r} \times \mathbb R_t$ with $\cN = (8,0)$ supersymmetry. Since the BPS spectrum ought to be invariant under such a topological deformation of  $\mathbb R^4/\mathbb Z_k$,  the BPS states that span ${\cal H}^{E_6}_{{\rm BPS}}$ in (\ref{BPS-E6}) should be given by the minimal energy states of this 2d $\cN = (8,0)$ theory along ${\bf S}^1_{1/r} \times \mathbb R_t$ which satisfy $H=P$.

In order to better understand this 2d $\cN = (8,0)$ theory along ${\bf S}^1_{1/r} \times \mathbb R_t$, we can appeal to the physically dual M-theory compactification in (\ref{M-theory-IIB duality}) -- in the limit that  $\mathbb R^4/\mathbb Z_k$ goes to zero size, the aforementioned 2d theory on the IIB side would be given by the 2d theory along  ${\bf S}^1_{r} \times \mathbb R_t$ on the M-theory side. That said, before we proceed any further, recall that $\mathbb R^4/\mathbb Z_k \simeq TN^{R \to \infty}_k$,  where $TN^R_K$ is the singular $k$-centered Taub-NUT manifold, and $R$ is the asymptotic radius of its circle fiber. Notice also that we are free to effect the topological deformation of $TN^{R \to \infty}_k$ to zero size in two steps: first, by shrinking its circle fiber, then, by shrinking its remaining $\mathbb R^3$ base. 
 
When we shrink the circle fiber of $TN^{R \to \infty}_k$ completely, we have, on the M-theory side, a reduction to the following type IIA background: 
\be
\textrm{IIA}: \quad \underbrace{ {\bf S}^1_{11; r' \to 0} \times {\rm K3}_{E_6}  \times {\bf S}^1_r \times \mathbb R_t}_{\textrm{$k$ D6-branes}}  \times  {\mathbb R^3}.
\label{IIA-E6}
\ee
According to the discussion in~\cite[$\S$1]{Vafa ref 2}, and recalling that ${\bf S}^1_{11; r' \to 0} \times {\rm K3}_{E_6}$ is much smaller than $ {\bf S}^1_r $ such that we effectively have a 5d $E_6$ Yang-Mills theory along $ {\bf S}^1_r \times \mathbb R_t \times \mathbb R^3$ in the low-energy long distance limit, we find that due to the presence of the $k$ D6-branes, there would be an additional term in the 5d Yang-Mills Lagrangian of the form
\be
I_{CS} = \int_{{\bf S}^1_r \times \mathbb R_t \times \mathbb R^3} {\tilde H}_2 \wedge CS(A),
\label{add term}
\ee  
where $\tilde H_2$ is a RR two-form field strength that is magnetically dual to the RR eight-form field strength sourced by the D6-branes, and $CS(A)$ is the usual Chern-Simons three-form associated with the $E_6$ gauge field $A$, i.e.,
\be
CS(A) = {\rm Tr} (A \wedge dA + {2 \over 3} A \wedge A \wedge A).
\ee
In addition, we have the following equation of motion for the $\tilde H_2$ field:
\be
d\tilde H_2 = k \cdot \delta_3(\cal B),
\label{EOM H-field}
\ee
where $\delta_3(\cal B)$ is a Poincar\'e-dual delta three-form that is supported at the intersection $\cal B$ of the D6-branes and $ {\bf S}^1_r \times \mathbb R_t \times \mathbb R^3$, i.e., ${\cal B} = {\bf S}^1_r \times \mathbb R_t$.

Under a gauge transformation of the $A$ field
\be
\delta A = D\epsilon, 
\label{gauge tx-A}
\ee
where $\epsilon$ is a position-dependent gauge parameter, we have
\be
CS(A) \to CS(A) + d{\rm Tr} (\epsilon dA),
\ee
and because of the equation of motion (\ref{EOM H-field}) for the $\tilde H_2$ field, the additional $I_{CS}$ term in the 5d Yang-Mills theory along ${\bf S}^1_r \times \mathbb R_t \times \mathbb R^3$ gets shifted by 
\be
\delta I_{CS} = - k \int_{{\bf S}^1_r \times \mathbb R_t} {\rm Tr} (\epsilon dA). 
\label{gauge anomaly}
\ee
In other words, the Lagrangian of the 5d Yang-Mills theory is not invariant under gauge transformations; it has a gauge anomaly given by (\ref{gauge anomaly}). 

Nevertheless, as explained in~\cite[$\S$1]{Vafa ref 2}, there ought to be an anomaly originating from the D6-branes that exactly cancels $\delta I_{CS}$ so that the whole system is anomaly-free. In particular, since the worldvolume of the D6-branes is effectively two-dimensional from a  compactification on ${\bf S}^1_{11; r' \to 0} \times {\rm K3}_{E_6}$, the aforementioned anomaly should come from the part of the D6-branes that wraps ${\bf S}^1_r \times \mathbb R_t$ -- that is, under the gauge transformation (\ref{gauge tx-A}), the Lagrangian of the 2d theory along ${\bf S}^1_r \times \mathbb R_t$ ought to be shifted by
\be
-\delta I_{CS} = k \int_{{\bf S}^1_r \times \mathbb R_t} {\rm Tr} (\epsilon dA).
\label{cancel}
\ee

Note that a \emph{chiral} $E_6$ WZW model at level $k$  exhibits exactly the anomaly (\ref{cancel}) under gauge transformations~\cite{gauged WZW}; one can thus conclude that the 2d theory along ${\bf S}^1_r \times \mathbb R_t$ must support such a chiral WZW model. Indeed, as the worldvolume of the D6-branes is effectively two-dimensional, according to the discussion in~\cite[$\S$1]{Vafa ref 2}, the 2d theory along ${\bf S}^1_r \times \mathbb R_t$ can support gauged chiral fermions; via the process of chiral bosonization~\cite{Ketov}, these gauged chiral fermions can be expressed in terms of chiral bosons embedded in a theory of non-chiral bosons (at the free fermion radius)  gauged to $A$; in turn, this system can be related to a chiral WZW model. 

Let us now shrink the $\mathbb R^3$ base in (\ref{IIA-E6}). According to the duality with the type IIB compactification (which is topological along $\mathbb R^4 / \mathbb Z_k$ in the long distance limit of interest), this step should not modify the remaining 2d theory along ${\bf S}^1_r \times \mathbb R_t$. As such, the equivalent 2d $\cN = (8,0)$ theory along ${\bf S}^1_{1/r} \times \mathbb R_t$ on the type IIB side, can be understood to support a chiral $E_6$ WZW model at level $k$. Moreover, since $H=P$ in any chiral WZW model (as it has no right-moving excitations), one can conclude that the minimal energy states of the 2d $\cN = (8,0)$ theory along ${\bf S}^1_{1/r} \times \mathbb R_t$ which correspond to the BPS states in (\ref{BPS-E6}), ought to be furnished by the spectrum of the chiral $E_6$ WZW model at level $k$. Thus, since we have an isomorphism of affine Lie algebras $\frak {e}_{6 \, \rm aff} \simeq \frak {e}^\vee_{6 \, \rm aff}$ whence we have an isomorphism of the corresponding integrable modules $[\widehat{e_6}]_k \simeq [^L\widehat{e_6}]_k$ of level $k$, where ${\frak g}^\vee_{\rm aff}$ is the \emph{Langlands dual} affine Lie algebra and $[^L\widehat{g}]_m$ is the integrable module over it of level $m$, we can write 
\be
\label{E-duality}
\boxed{{\rm IH}^\ast {\cal U}({\cal M}^{\lambda}_{E_6, \mu}(\mathbb R^4/ \mathbb Z_k)) = [^L\widehat{e_6}]^\lambda_{k, \mu}}
\ee
where $[^L\widehat{g}]^\alpha_{m, \beta}$ is a submodule over ${\frak g}^\vee_{\rm aff}$ of level $m$ labeled by a highest dominant weight $\alpha$ and a dominant weight $\beta$.  Note that (\ref{E-duality}) is exactly~\cite[Conjecture 4.14(3)]{BF} for simply-connected $G=E_6$! This completes our purely physical derivation of a geometric Langlands duality for surfaces for the $E_6$ group.

\bigskip\noindent{\it A Geometric Langlands Duality for Surfaces for the $F_4$ Group}

Let us now proceed to discuss the $F_4$ case. To this end, let us effect a  ``$\mathbb Z_2$-twist'' of the six-dimensional spacetime theories of the dual compactifications in (\ref{M-theory-IIB duality}) as we go around the ${\bf S}^1$ circles, i.e., we evoke a $\mathbb Z_2$-outer-automorphism of $\mathbb R_t \times \mathbb R^4/ \mathbb Z_k$ therein as we go around the ${\bf S}^1$ circles and identify the circles under an order $2$ translation. In other words, we now have the following \emph{physically dual} compactifications:
\begin{eqnarray}
\label{M-theory-IIB duality-F4}
{\textrm {M-theory}}: & {\bf S}^1_{11; r' \to 0} \times {\rm K3}_{E_6}  \times {\bf S}^1_{r; \, 2} \times \mathbb R_t \times \mathbb R^4/\mathbb Z_k\vert_2  \nonumber \\
&\Updownarrow \\
{\textrm {Type IIB}}: & {\rm K3}_{E_6} \times {\bf S}^1_{1/r; \, 2} \times \mathbb R_t \times \mathbb R^4/\mathbb Z_k\vert_2 \nonumber,
\end{eqnarray}
where $r \approx 1$; ${\rm K3}_{E_6}$ and ${\bf S}^1_{11; r' \to 0}$ are much smaller than  ${\bf S}^1_{r; \, 2} \times \mathbb R_t \times \mathbb R^4/\mathbb Z_k\vert_2$ and ${\bf S}^1_{1/r; \, 2} \times \mathbb R_t \times \mathbb R^4/\mathbb Z_k\vert_2$; and the subscript `$2$' denotes the above-described $\mathbb Z_2$-action (which is trivial on $\mathbb R_t$) along the indicated manifold.

In the low-energy limit, the six-dimensional spacetime theory on the IIB side is a ``$\mathbb Z_2$-twisted'' $\cN = (2,0)$ $E_6$ theory on ${\bf S}^1_{1/r; \, 2} \times \mathbb R_t  \times \mathbb R^4/\mathbb Z_k\vert_2$. This theory is topological along $\mathbb R^4/\mathbb Z_k\vert_2$ (and conformal along ${\bf S}^1_{1/r; \, 2} \times \mathbb R_t$) (c.f.~\cite{junya} and footnote~\ref{junya's twist}). In particular, this means that the BPS spectrum of minimal energy states of the ``$\mathbb Z_2$-twisted'' $\cN = (2,0)$ $E_6$ theory on ${\bf S}^1_{1/r; \, 2} \times \mathbb R_t  \times \mathbb R^4/\mathbb Z_k\vert_2$ -- which are states annihilated by all eight unbroken supercharges whence they satisfy $H=P$, where $H$ and $P$ are the Hamiltonian and momentum operators which generate translations along $\mathbb R_t$ and ${\bf S}^1_{1/r; \, 2}$, respectively -- is invariant under topological deformations of $\mathbb R^4/\mathbb Z_k\vert_2$.

Let us ascertain this BPS spectrum of minimal energy states in the case where $\mathbb R^4/\mathbb Z_k\vert_2$ has yet to be topologically deformed. According to (i) footnote~\ref{minimal energy state}; (ii) our explanations in $\S$3.1; (iii) the fact that a $\mathbb Z_2$-outer-automorphism of $\mathbb R^4/\mathbb Z_k$ would also result in a $\mathbb Z_2$-outer-automorphism of   a principal $E_6$-bundle over $\mathbb R^4/\mathbb Z_k \times \mathbb R_t$ such that at long distances, the gauge group of the 5d maximally supersymmetric Yang-Mills theory on $\mathbb R_t \times \mathbb R^4/\mathbb Z_k\vert_2$ is effectively $F_4$~\cite{Yuji 2 yrs}; we find that the Hilbert space of such BPS states would be given by
\be
{\cal H}^{F_4}_{{\rm BPS}} =   \bigoplus_{\lambda} \bigoplus_{\nu =0,1}  \bigoplus_{\mu_\nu}  \, \overline {{\rm IH}^\ast {\cal U}}({\cal M}^{\lambda}_{F_4, \mu_\nu}(\mathbb R^4/ \mathbb Z_k)).
\label{BPS-F4}
\ee
Here, $\overline{{\rm IH}^\ast {\cal U}}({\cal M}^{\lambda}_{F_4, \mu_\nu}(\mathbb R^4/ \mathbb Z_k))$ is the $\mathbb Z_2$-invariant intersection cohomology of the Uhlenbeck compactification ${\cal U}({\cal M}^{\lambda}_{F_4, \mu_\nu}(\mathbb R^4/ \mathbb Z_k))$ of the moduli space ${\cal M}^{\lambda}_{F_4, \mu_\nu}(\mathbb R^4/ \mathbb Z_k)$ of $F_4$-instantons on $\mathbb R^4/ \mathbb Z_k$ in the $\{\lambda, \mu_\nu\}$-sector (as described around (\ref{field twist 1})--(\ref{field twist 3}), where $SO(N+1)$ therein is replaced by $F_4$); $\nu = 0$ or 1 in the untwisted or twisted sector, respectively; $\lambda$ and $\mu_\nu$ can be regarded as\emph{ dominant weights} of the corresponding \emph{Langlands dual} affine Kac-Moody group $F^\vee_{{4 \, \rm aff}}$ of level $k$; and
\be
\lambda \geq \mu_\nu.
\label{weight condition-F4}
\ee

Let us now topologically deform $\mathbb R^4/\mathbb Z_k\vert_2$ and scale it down to zero size. Then, the 6d ``$\mathbb Z_2$-twisted''  $\cN = (2,0)$ $E_6$ theory on ${\bf S}^1_{1/r; \, 2} \times \mathbb R_t  \times \mathbb R^4/\mathbb Z_k\vert_2$ will essentially reduce to a 2d ``$\mathbb Z_2$-twisted'' theory along ${\bf S}^1_{1/r; \, 2} \times \mathbb R_t$ with $\cN = (8,0)$ supersymmetry. Since the BPS spectrum ought to be invariant under such a topological deformation of  $\mathbb R^4/\mathbb Z_k\vert_2$,  the BPS states that span ${\cal H}^{F_4}_{{\rm BPS}}$ in (\ref{BPS-F4}) should be given by the minimal energy states of this 2d ``$\mathbb Z_2$-twisted'' $\cN = (8,0)$ theory along ${\bf S}^1_{1/r; \, 2} \times \mathbb R_t$ which satisfy $H=P$. 

Repeating the arguments from (\ref{IIA-E6})--(\ref{E-duality}), whilst bearing in mind that the $\mathbb Z_2$-outer-automorphism is trivial on a flat space such as the $\mathbb R_t \times \mathbb R^3$ manifold in (\ref{IIA-E6}), we find that the 2d ``$\mathbb Z_2$-twisted'' $\cN = (8,0)$ theory along ${\bf S}^1_{1/r; \, 2} \times \mathbb R_t$ ought to support a $\mathbb Z_2$-twisted chiral $E_6$ WZW model at level $k$.  Moreover, since $H=P$ in any chiral WZW model, twisted or not, one can conclude that the minimal energy states of the 2d ``$\mathbb Z_2$-twisted'' $\cN = (8,0)$ theory along ${\bf S}^1_{1/r; \, 2} \times \mathbb R_t$ which correspond to the BPS states in (\ref{BPS-F4}), ought to be furnished by the spectrum of the $\mathbb Z_2$-twisted chiral $E_6$ WZW model at level $k$. Thus, since we have an isomorphism of affine Lie algebras $\frak {e}^{(2)}_{6 \, \rm aff} \simeq \frak {f}^\vee_{4 \, \rm aff}$ whence we have an isomorphism of the corresponding integrable modules $[{\widehat{e_6}}^{(2)}]_k \simeq [^L\widehat{f_4}]_k$ of level $k$, where ${\frak g}^{(2)}_{\rm aff}$ is a $\mathbb Z_2$-twisted affine Lie algebra and $[{\widehat{g}}^{(2)}]_m$ is the integrable module over it of level $m$, we can write 
\be
\label{F-duality}
\boxed{{\rm IH}^\ast {\cal U}({\cal M}^{\lambda}_{F_4, \mu_\nu}(\mathbb R^4/ \mathbb Z_k)) = [^L\widehat{f_4}]^\lambda_{k, \mu_\nu}}
\ee
for $\nu = 0$ and 1. Thus, we have arrived at a $G = F_4$ generalization of~\cite[Conjecture 4.14(3)]{BF}! This completes our purely physical derivation of a geometric Langlands duality for surfaces for the $F_4$ group.

\newsubsection{A McKay-Type Correspondence of Instantons, a Level-Rank Duality of Chiral WZW Models, and a 4d-2d Nakajima-Type Duality}

We shall now derive, purely physically for the simply-laced $A$--$D$ groups, a McKay-type correspondence of the intersection cohomology of the moduli spaces of instantons, a level-rank duality of chiral WZW models, and a 4d-2d Nakajima-type duality involving completely blown-down ALE spaces. To this end, recall from $\S$3.1--$\S$3.2 that we have the dual M-theory compactifications 
\be
\underbrace{\mathbb R^4 / \mathbb Z_k  \times {{\bf S}^1} \times \mathbb R_t}_{\textrm{$N$ M5-branes}}\times \mathbb R^{5} \quad \Longleftrightarrow  \quad {\mathbb R^{5}} \times \underbrace{\mathbb R_t \times {{\bf S}^1}  \times TN_N^{R\to 0}  }_{\textrm{$k$ M5-branes}},
 \label{PF dual A}
 \ee
and
\be
\underbrace{\mathbb R^4 / \mathbb Z_k  \times {{\bf S}^1} \times \mathbb R_t}_{\textrm{$N$ M5-branes/OM5-plane}}\times \mathbb R^{5} \quad  \Longleftrightarrow \quad {\mathbb R^{5}} \times \underbrace{\mathbb R_t \times {{\bf S}^1}  \times SN_N^{R\to 0}  }_{\textrm{$k$ M5-branes}}.
 \label{PF dual D}
\ee

\bigskip\noindent{\it The $A$ Groups}

According to our discussions in $\S$3.1, the Hlibert space ${\cal H}_{SU(N)}$ of spacetime BPS states associated with the LHS of (\ref{PF dual A}) is given by 
\be
{\cal H}_{SU(N)} = \bigoplus_{\lambda, \mu} {\cal H}^{\lambda, \mu}_{SU(N)}  =  \bigoplus_{\lambda, \mu} ~{\rm IH}^\ast {\cal U}({\cal M}^{\lambda}_{SU(N), \mu}(\mathbb R^4/ \mathbb Z_k)),
\label{BPS-M-PF-A-1}
\ee
where $\lambda \geq \mu$,  and $\lambda$ and $\mu$ can be regarded as dominant coweights of the corresponding affine Kac-Moody group $SU(N)_{\rm aff}$ of level $k$.

On the other hand, note that where the spectrum of ground states of the worldvolume theory of a stack of M5-branes wrapping $M_4 \times {\bf S}^1 \times \mathbb R_t$ is concerned, one can -- if $M_4$ is a hyperk\"ahler four-manifold -- regard the theory to be topological along $M_4$ (and conformal along ${\bf S}^1 \times \mathbb R_t$) (c.f.~\cite{junya} and footnote~\ref{junya's twist}); in other words, where computing the spacetime BPS states is concerned, one can replace on the RHS of (\ref{PF dual A}), the singular multi-Taub-NUT space $TN_N^{R\to 0}$ (whose circle fiber has radius $R \to 0$ at infinity) with ${TN}^{R \to \infty}_N \simeq {\mathbb R^4 / \mathbb Z_N}$.\footnote{As explained in $\S$3.1, there is a technical subtlety associated with monopoles that go around the finite-sized circle fiber at infinity. However, since our discussion is restricted to the limits $R \to \{0, \infty \}$, we can ignore this technical subtlety whence our claim is consistent.} Hence, assuming that the geometry of $\mathbb R^4/\mathbb Z_k$ is frozen, according to our discussions in $\S$3.1, the Hlibert space ${\cal H}_{SU(k)}$ of spacetime BPS states associated with the RHS of (\ref{PF dual A}) would be given by 
\be
{\cal H}_{SU(k)} = \bigoplus_{\hat\lambda, \hat\mu} {\cal H}^{\hat\lambda, \hat\mu}_{SU(k)}  =  \bigoplus_{\hat\lambda, \hat\mu} ~{\rm IH}^\ast {\cal U}({\cal M}^{\hat\lambda}_{SU(k), \hat\mu}(\mathbb R^4/ \mathbb Z_N)),
\label{BPS-M-PF-A-2}
\ee
where $\hat\lambda \geq \hat\mu$,  and $\hat\lambda$ and $\hat\mu$ can be regarded as dominant coweights of the corresponding affine Kac-Moody group $SU(k)_{\rm aff}$ of level $N$.

The duality of the compactifications in (\ref{PF dual A}) then means that ${\cal H}_{SU(N)}  = {\cal H}_{SU(k)}$, i.e., 
\be
\boxed{\bigoplus_{\lambda, \mu} ~{\rm IH}^\ast {\cal U}({\cal M}^{\lambda}_{SU(N), \mu}(\mathbb R^4/ \mathbb Z_k)) = \bigoplus_{\hat\lambda, \hat\mu} ~{\rm IH}^\ast {\cal U}({\cal M}^{\hat\lambda}_{SU(k), \hat\mu}(\mathbb R^4/ \mathbb Z_N))}
\label{McKay IC}
\ee
Note at this point that the McKay correspondence~\cite{McKay} relates a finite subgroup $\Gamma \subset SU(2)$ to the Lie algebra of the $A$--$D$--$E$ groups; in particular, it relates the subgroup $\Gamma = \mathbb Z_r$ to the Lie algebra of the $A_{r-1}$ group. Since (\ref{McKay IC}) relates the moduli space of $A_{N-1}$-instantons on a $\mathbb Z_k$-orbifold to the moduli space of $A_{k-1}$-instantons on a $\mathbb Z_N$-orbifold,  one can regard (\ref{McKay IC}) as a McKay-type correspondence of the intersection cohomology of the (Uhlenbeck compactification of the) moduli spaces of $A$-instantons! This is a generalization of Proudfoot's conjecture in~\cite{Proudfoot} to \emph{completely blown-down} ALE spaces. 

Via our discussions leading up to (\ref{GL-relation A}), we find that (\ref{McKay IC}) also implies that we have an equivalence of chiral WZW models
\be
\boxed{{\rm WZW}_{\widehat{su}(N)_{k}} = {\rm WZW}_{\widehat{su}(k)_{N}}}
\label{level-rank-A}
\ee
Thus, we have a level-rank duality of chiral WZW models for the $A$ groups!

Moreover, (\ref{McKay IC}), and the discussion leading up to (\ref{GL-relation A}), also mean that 
\be
\boxed{\bigoplus_{\hat\lambda, \hat\mu} ~{\rm IH}^\ast {\cal U}({\cal M}^{\hat\lambda}_{SU(k), \hat\mu}(\mathbb R^4/ \mathbb Z_N)) = \bigoplus_{\tilde\lambda, \tilde\mu}~{\widehat{su}(N)^{\tilde\lambda}_{k, \tilde\mu}}}
\label{nakajima-type-A}
\ee
where $\tilde \lambda$ and $\tilde \mu$ are dominant affine weights such that $\tilde \lambda \geq \tilde \mu$. Notice that in (\ref{nakajima-type-A}), the $\mathbb Z_N$-singularity on the LHS is related to the (affine) $A_{N-1}$ Lie algebra on the RHS in the sense of a McKay correspondence; furthermore, the rank $k$ of the gauge group on the LHS equals the level $k$ of the affine Lie algebra on the RHS; in other words, we have a 4d-2d Nakajima-type duality involving \emph{completely blown-down} $A$-type ALE spaces!

Notice also that if we start with (\ref{nakajima-type-A}) and apply (\ref{level-rank-A}), we would get  the same result as (\ref{GL-relation A}) (with $(N, k)$ therein relabeled as $(k, N)$);  in other words, we have a physical realization of the commutative diagram in~\cite[$\S$1]{branching}!

\bigskip\noindent{\it The $D$ Groups}

According to our discussions in $\S$3.2, the Hlibert space ${\cal H}_{SO(2N)}$ of spacetime BPS states associated with the LHS of (\ref{PF dual D}) is given by 
\be
{\cal H}_{SO(2N)} = \bigoplus_{n, \rho_0, \rho_\infty} {\cal H}^{n, \rho_0, \rho_\infty}_{SO(2N)}  =  \bigoplus_{n, \rho_0, \rho_\infty} ~{\rm IH}^\ast {\cal U}({\cal M}^{n, \rho_0, \rho_\infty}_{SO(2N)}(\mathbb R^4/ \mathbb Z_k)),
\label{BPS-M-PF-D-1}
\ee
where $n$ is the $SO(2N)$-instanton number; $\rho_0: \mathbb Z_k \to SO(2N)$ is the homomorphism associated with the $\mathbb Z_k$-action in the fiber of the $SO(2N)$-bundle at the origin; $\rho_\infty: \mathbb Z_k \to SO(2N)$ is the homomorphism associated with a choice of flat $SO(2N)$-connection at infinity. 

On the other hand, since Sen's singular manifold  $SN_N^{R\to 0}$ is also hyperk\"ahler, according to our explanations above, where computing the spacetime BPS states on the RHS of (\ref{PF dual D}) is concerned, one can replace $SN_N^{R\to 0}$ (whose circle fiber has radius $R \to 0$ at infinity) with ${SN}^{R \to \infty}_N \simeq {\mathbb R^4 / \mathbb D_N}$, where ${\mathbb D}_N$ is the binary dihedral group of order $2N$. (See $\S$A.4.) Hence, assuming that the geometry of $\mathbb R^4/\mathbb Z_k$ is frozen, according to our discussions in $\S$3.2, the Hlibert space ${\cal H}_{\overline{SU(k)}}$ of spacetime BPS states associated with the RHS of (\ref{PF dual D}) would be given by  
\be
{\cal H}_{\overline {SU(k)}} = \bigoplus_{\hat n, \hat \rho_0, \hat \rho_\infty} {\cal H}^{\hat n, \hat \rho_0, \hat \rho_\infty}_{\overline {SU(k)}}  =  \bigoplus_{\hat n, \hat \rho_0, \hat \rho_\infty} ~{\rm IH}^\ast {\cal U}({\cal M}^{\hat n, \hat \rho_0, \hat \rho_\infty}_{SU(k)}(\mathbb R^4/ \mathbb D_N)),
\label{BPS-M-PF-D-2}
\ee
where $\hat n$ is the $SU(k)$-instanton number; $\hat\rho_0: \mathbb D_N \to SU(k)$ is the homomorphism associated with the $\mathbb D_N$-action in the fiber of the $SU(k)$-bundle at the origin; $\hat\rho_\infty: \mathbb D_N \to SU(k)$ is the homomorphism associated with a choice of flat $SU(k)$-connection at infinity. 

The duality of the compactifications in (\ref{PF dual D}) then means that ${\cal H}_{SO(2N)}  = {\cal H}_{\overline{SU(k)}}$, i.e., 
\be
\boxed{ \bigoplus_{n, \rho_0, \rho_\infty} ~{\rm IH}^\ast {\cal U}({\cal M}^{n, \rho_0, \rho_\infty}_{SO(2N)}(\mathbb R^4/ \mathbb Z_k)) =  \bigoplus_{\hat n, \hat \rho_0, \hat \rho_\infty} ~{\rm IH}^\ast {\cal U}({\cal M}^{\hat n, \hat \rho_0, \hat \rho_\infty}_{SU(k)}(\mathbb R^4/ \mathbb D_N))}
\label{McKay IC-D}
\ee
Note at this point that the McKay correspondence also relates the subgroup ${\mathbb D}_r \subset SU(2)$ to the Lie algebra of the group $SO(2r)$. Hence, since (\ref{McKay IC-D}) relates the moduli space of $D_{N}$-instantons on a $\mathbb Z_k$-orbifold to the moduli space of $A_{k-1}$-instantons on a $\mathbb D_N$-orbifold,  one can regard (\ref{McKay IC-D}) as a McKay-type correspondence of the intersection cohomology of the (Uhlenbeck compactification of the)  moduli spaces of $A$--$D$ instantons! This is another generalization of Proudfoot's conjecture in~\cite{Proudfoot} to \emph{completely blown-down} ALE spaces. 

Via our discussions leading up to (\ref{GL-relation D}), we find that the LHS of (\ref{McKay IC-D}) is equal to $\bigoplus_{\alpha, \beta}{\widehat{so}(2N)^\alpha_{k, \beta}}$. Here, $\alpha \geq \beta$,  and $\alpha$ and $\beta$ can be regarded as dominant weights of the corresponding affine Kac-Moody group $SO(2N)_{\rm aff}$ of level $k$.

By reversing the arguments employed in going from (\ref{OM-theory 1}) to (\ref{OM-theory 8}) whilst replacing ${SN}^{R \to 0}_N$ in (\ref{OM-theory 8}) with ${SN}^{R \to \infty}_N$, we have the following duality relation
\be
 \underbrace{\mathbb R_t \times {{\bf S}^1}  \times {SN}^{R \to \infty}_N}_{\textrm{$k$ M5-branes}} \times {\mathbb R^{5}} \quad  \Longleftrightarrow \quad \mathbb R^{5} \times \underbrace{TN^{R \to 0}_k  \times {{\bf S}^1} \times \mathbb R_t}_{\textrm{$N$ M5-branes/OM5-plane}}.
 \label{PF dual D-SN}
\ee
Applying to this duality relation the analysis in $\S$3.1--$\S$3.2, whilst bearing in mind that (i) a D6-D4/O$4^-$ I-brane system is T-dual to a D4-D6/O$6^-$ I-brane system (studied in $\S$3.2) whence the gauge groups on the D6- and D4-branes are \emph{both} of orthogonal type; (ii) ${SN}^{R \to \infty}_N \simeq {\mathbb R^4 / \mathbb D_N}$; we find that the RHS of  (\ref{McKay IC-D})  is equal to $\bigoplus_{\hat \alpha, \hat \beta} {\widehat{so}(k)^{\hat\alpha}_{2N, \hat\beta}}$. Here, $\hat\alpha \geq \hat\beta$,  and $\hat\alpha$ and $\hat\beta$ can be regarded as dominant weights of the corresponding affine Kac-Moody group $SO(k)_{\rm aff}$ of level $2N$. 

Therefore, from the preceding two paragraphs,  we find that (\ref{McKay IC-D}) also implies that we have an equivalence of chiral WZW models
\be
\boxed{{\rm WZW}_{\widehat{so}(2N)_{k}} = {\rm WZW}_{\widehat{so}(k)_{2N}}}
\label{level-rank-D}
\ee
Thus, we have a level-rank duality of chiral WZW models for the $D$ groups! 

In turn, (\ref{level-rank-D}), and the discussion leading up to (\ref{GL-relation D}), would mean that (\ref{McKay IC-D}) can also be written as
\be
\boxed{ \bigoplus_{n', \rho'_0, \rho'_\infty} ~{\rm IH}^\ast {\cal U}({\cal M}^{n', \rho'_0, \rho'_\infty}_{SO(k)}(\mathbb R^4/ \mathbb Z_{2N})) =  \bigoplus_{\hat n, \hat \rho_0, \hat \rho_\infty} ~{\rm IH}^\ast {\cal U}({\cal M}^{\hat n, \hat \rho_0, \hat \rho_\infty}_{SU(k)}(\mathbb R^4/ \mathbb D_N))}
\label{Swop IC-D}
\ee
where $n'$ is the $SO(k)$-instanton number; $\rho'_0: \mathbb Z_{2N} \to SO(k)$ is the homomorphism associated with the $\mathbb Z_{2N}$-action in the fiber of the $SO(k)$-bundle at the origin; $\rho'_\infty: \mathbb Z_{2N} \to SO(k)$ is the homomorphism associated with a choice of flat $SO(k)$-connection at infinity.

Moreover, (\ref{McKay IC-D}), and the discussion leading up to (\ref{GL-relation D}), also mean that 
\be
\boxed{\bigoplus_{\hat n, \hat \rho_0, \hat \rho_\infty} ~{\rm IH}^\ast {\cal U}({\cal M}^{\hat n, \hat \rho_0, \hat \rho_\infty}_{SU(k)}(\mathbb R^4/ \mathbb D_N)) = \bigoplus_{\tilde\lambda, \tilde\mu}~{\widehat{so}(2N)^{\tilde\lambda}_{k, \tilde\mu}}}
\label{nakajima-type-D}
\ee
where $\tilde \lambda$ and $\tilde \mu$ are dominant affine weights such that $\tilde \lambda \geq \tilde \mu$.  Notice that in (\ref{nakajima-type-D}), the $\mathbb D_N$-singularity on the LHS is related to the (affine) $D_{N}$ Lie algebra on the RHS in the sense of a McKay correspondence; furthermore, the rank of the gauge group on the LHS is equal to the level of the affine Lie algebra on the RHS which equals to $k$; in other words, we have a 4d-2d Nakajima-type duality involving \emph{completely blown-down} $D$-type ALE spaces! 

Notice also that if we start with (\ref{nakajima-type-D}) and apply (\ref{level-rank-D}), we would get  the same result as if we started with (\ref{Swop IC-D}) and applied (\ref{GL-relation D}) (with $(2N, k)$ therein relabeled as $(k, 2N)$);  in other words, we have a physical realization of a \emph{$D$-type ALE space generalization} of the commutative diagram in~\cite[$\S$1]{branching}!

\newsection{Generalizations of the Geometric Langlands Duality for Surfaces}

\newsubsection{A Non-Singular Generalization of the Geometric Langlands Duality for Surfaces}

Let us now derive a non-singular generalization of the geometric Langlands duality for surfaces for the $A$--$B$ groups. To this end, let us replace $\mathbb R^4/\mathbb Z_k$ in (\ref{M-theory 1}) with its fully-resolved $\it{smooth}$ counterpart $\widetilde{\mathbb R^4 / \mathbb Z_k}$ which has $k$ centers being \emph{completely} separated. By repeating the arguments behind (\ref{M-theory 1})--(\ref{M-theory 7}), we find that the following six-dimensional M-theory compactification
\be
\textrm{M-theory}: \quad \mathbb R^{5}  \times  \underbrace{\mathbb R_t \times {\bf S}^1_n \times \widetilde{\mathbb R^4 / \mathbb Z_k}}_{\textrm{$N$ M5-branes}},
\label{M-theory 1 Witten discussion}
\ee
where we evoke a $\mathbb Z_n$-outer-automorphism of $\widetilde{\mathbb R^4 / \mathbb Z_k}$ (and of the geometrically-trivial $\mathbb R^5 \times \mathbb R_t$ spacetime) as we go around the ${\bf S}^1_n$ circle and identify the circle under an order $n$ translation, is \emph{physically dual} to the following six-dimensional M-theory compactification
\be
\textrm{M-theory}: \quad \underbrace{TN_N^{R\to 0}  \times  {{\bf S}^1_n}\times \mathbb R_t}_{\textrm{$k$ non-coincident M5-branes}}  \times  {\mathbb R^{5}},
\label{M-theory 7 Witten discussion}
\ee
where there is a nontrivial $\mathbb Z_n$-outer-automorphism of the singular multi-Taub-NUT space $TN_N^{R\to 0}$ (whose circle fiber at infinity approaches zero radius) as we go around the ${\bf S}^1_n$ circle. Notice that in contrast to the $\mathbb R^4/\mathbb Z_k$ case, due to the fully separated $k$ centers of $ \widetilde{\mathbb R^4 / \mathbb Z_k}$, the $k$ M5-branes will be $\textrm{\it{non-coincident}}$. 

\bigskip\noindent{\it The Spectrum of Spacetime BPS States in the M-Theory Compactification (\ref{M-theory 1 Witten discussion})}

In order to describe the Hilbert space of spacetime BPS states furnished by the ground states of the quantum worldvolume theory of the $N$ coincident M5-branes in (\ref{M-theory 1 Witten discussion}), first note that because $\widetilde{\mathbb R^4 / \mathbb Z_k}$ is a hyperk\"ahler manifold like $\mathbb R^4/\mathbb Z_k$,  the Gieseker compactification ${\cal G}({\cal M}_{G}(\widetilde{\mathbb R^4 / \mathbb Z_k})$ of the moduli space ${\cal M}_{G}(\widetilde{\mathbb R^4 / \mathbb Z_k})$ of $G$-instantons on $\widetilde{\mathbb R^4 / \mathbb Z_k}$ -- where  $G= SU(N)$ if $n=1$, and $G = SO(N+1)$ if  $n=2$ and $N$ is even -- will also inherit a hyperk\"ahler structure, consistent with the ${\cal N} =(4,4)$ supersymmetry of the corresponding sigma-model which describes the quantum worldvolume theory of the M5-branes. The worldvolume ground states, being annihilated by all eight supercharges of the sigma-model, will span its topological sector, and as explained in the $\mathbb R^4/\mathbb Z_k$ case, the ground states and therefore the spacetime BPS states, will thus correspond to harmonic forms in the ${\bf L}^2$-cohomology of ${\cal G}({\cal M}_{G}(\widetilde{\mathbb R^4 / \mathbb Z_k})$. Moreover, since the hyperk\"ahler structure of ${\cal G}({\cal M}_{G}(\widetilde{\mathbb R^4 / \mathbb Z_k})$ is smooth, its ${\bf L}^2$-cohomology will coincide with its middle-dimensional cohomology~\cite{Hitchin}.

Second, note that for the instanton action to be finite in an integration over noncompact $\widetilde{\mathbb R^4 / \mathbb Z_k}$, we need to have flat albeit nontrivial connections far away from the origin of $\widetilde{\mathbb R^4 / \mathbb Z_k}$. The hyperk\"ahler metrics on  $\widetilde{\mathbb R^4 / \mathbb Z_k}$ are asymptotic at infinity to $\mathbb  R^4/ \mathbb Z_k$; because gauge-inequivalent classes of flat connections far away from the origin correspond to conjugacy classes of homomorphisms $\rho_{\infty}$ from the fundamental group at infinity to $G$, and that moreover, as explained in $\S$3.1, conjugacy classes of the homomorphism $\rho: Z_l \to G$ are in one-to-one correspondence with dominant coweights of the affine Kac-Moody group $G_{\textrm{aff}}$ of level $l$, we find that distinct choices of flat connections far away from the origin will correspond to distinct dominant coweights $\mu = (k, \bar \mu, j)$ of $G_{\textrm{aff}}$ of level $k$, where $j$ is a number. 

Third, recall that in the case of $\mathbb R^4 / \mathbb Z_k$, the $k$ centers coincide with multiplicity $k$ at the origin such that a $\mathbb Z_k$-type singularity develops whence we have a $\mathbb Z_k$-action in the fiber of the $G$-bundle at 0. On the other hand, in the case of $\widetilde{\mathbb R^4 / \mathbb Z_k}$,  we have instead $k$ non-coincident centers of multiplicity 1 each -- in other words, we have instead a $\mathbb Z_1$-action in the fiber of the $G$-bundle over each of the $k$ positions ${\vec p}_m$ of the non-coincident centers. Since this action is given by a conjugacy class of the homomorphism $\rho: Z_1 \to G$, we can associate $k$ distinct dominant coweights $\lambda^{(m)} = (1, \bar \lambda^{(m)}, i^{(m)})$ of $G_{\textrm{aff}}$ of level $1$  with the $k$ non-coincident centers, where the $i^{(m)}$'s are numbers. Nonetheless, consistency with the results of $\S$3.1 (where all $k$ centers coincide) constrains the $i^{(m)}$'s to be zero.

Fourth, according to our analysis leading up to (\ref{a}), and the fact that the $\lambda^{(m)}$'s  ought to be linearly-independent of one another, we find that the $G$-instantons -- which again correspond to D0-branes within the D4-brane worldvolume in the type IIA picture -- are such that the associated \emph{non-negative} instanton numbers are
\be
a = -kn'j + \tilde b(\bar {\boldsymbol \lambda}, \bar {\boldsymbol \lambda}) - b(\bar \mu, \bar \mu),
\label{a-TNk}
\ee 
where for $G = SU(N)$, $SO(3)$ and $SO(N+1)$, $n' =1$, $1$ and $2$ while $j$ is a non-positive integer divided by $1$, $2$ and $2$, respectively. Also, $\bar  {\boldsymbol \lambda} = \sum_{i=1}^k \bar \lambda^{(i)}$; $\tilde b$ and $b$ are some positive real constants; and $( \, , )$ is the scalar product in finite coweight space. For $n=1$ whence we have $G = SU(N)$ with $n'=1$ and $j$ being a non-positive integer, expression (\ref{a-TNk}) is indeed consistent with results from the mathematical literature (which only addresses the case of simply-connected groups like $SU(N)$): eqn.~(\ref{a-TNk}) coincides with~\cite[below Conjecture 3.2]{BF2} after we set $\tilde b = b = 1/2$ and identify $a/k$ with $d/k$ of \emph{loc.~cit..}\footnote{The instanton number of \emph{loc.~cit.} is defined as $a/k$ and not $a$ because they consider the effective instanton number counted only by D-branes in the fundamental region of $\widetilde{\mathbb R^4 / \mathbb Z_k}$. \label{fraction}}

Last but not least, recall that a $G$-bundle on $\widetilde{\mathbb R^4 / \mathbb Z_k}$ is topologically classified by $p_2 \in H^2(\widetilde{\mathbb R^4 / \mathbb Z_k}, \pi_1(G))$. In particular, $p_2$ vanishes for simply-connected $G = SU(N)$, but \emph{not }for nonsimply-connected $G = SO(N+1)$; in fact, since $\pi_1(SO(N+1)) = \mathbb Z_2$, we find that $SO(N+1)$-bundles on $\widetilde{\mathbb R^4 / \mathbb Z_k}$ are topologically classified by the second Stiefel-Whitney class $w_2 \in H^2(\widetilde{\mathbb R^4 / \mathbb Z_k}, \mathbb Z_2)$.  Note also that $w_2$ can be interpreted as a linear sum of $\mathbb Z_2$-valued non-abelian magnetic fluxes that pass through the $k-1$ two-spheres in $\widetilde{\mathbb R^4 / \mathbb Z_k}$~\cite{`t Hooft, vafa-witten}; this observation will be important shortly.

Thus, from the five points above, it is clear that ${\cal G}({\cal M}_{G}(\widetilde{\mathbb R^4 / \mathbb Z_k}))$ ought to be graded by $\{\lambda^{(1)}, \dots, \lambda^{(k)} \}$, $\mu$, and $w_2$ (where $a$ is correspondingly given by (\ref{a-TNk})). We are now ready to state the generic Hilbert space ${\cal H}_{\rm BPS}$ of spacetime BPS states in the M-theory compactification (\ref{M-theory 1 Witten discussion}). Let us denote by ${\textrm H}^\ast_{\rm mid}{\cal G}({\cal M}^{w_2, \bL}_{G, \mu}(\widetilde{\mathbb R^4 / \mathbb Z_k}))$, the middle-dimensional cohomology of the Gieseker compactification ${\cal G}({\cal M}^{w_2, \bL}_{G, \mu}(\widetilde{\mathbb R^4 / \mathbb Z_k}))$ of the component ${\cal M}^{w_2, \bL}_{G, \mu}(\widetilde{\mathbb R^4 / \mathbb Z_k})$ of the moduli space ${\cal M}_{G}(\widetilde{\mathbb R^4 / \mathbb Z_k})$ labeled by $\boldsymbol\lambda = \sum_{i=1}^k \lambda^{(i)} = (k, \bar {\bL}, 0)$, $\mu = (k, \bar\mu, j)$ and $w_2$; then, we can write
\be
{\cal H}_{\rm BPS} = \bigoplus_{ w_2, \bL, \mu} {\cal H}^{w_2, \bL, \mu}_{\rm BPS}  =  \bigoplus_{w_2, \bL, \mu} {\textrm H}^\ast_{\rm mid}{\cal G}({\cal M}^{w_2, \bL}_{G, \mu}(\widetilde{\mathbb R^4 / \mathbb Z_k})),
\label{BPS-TNk}
\ee
where $w_2 = 0$ if $n=1$, and $\bL \geq \mu$ (since $a$ is non-negative).

\bigskip\noindent{\it The Partition Function of Spacetime BPS States in (\ref{M-theory 1 Witten discussion}) for  $G = SU(N)$}

Consider the $n=1$ case whence we have $G = SU(N)$, $n'=1$, and $- j \in \mathbb Z_{\geq 0}$. By repeating the arguments that led us to (\ref{Z_SU(N)}), and by noting that $P$ in (\ref{Zbps}) is now equal to $-kj + {1 \over 2}(\bar {\boldsymbol \lambda}, \bar {\boldsymbol \lambda})$ while $w_2 = 0$ for $SU(N)$-instantons, we can write the corresponding partition function of spacetime BPS states in any $\bL$-sector as
\be
 Z_{SU(N), \bL}^{\rm BPS}  =  q^{m_{\bL}}  \, \sum_{\bar \mu} \sum_{m \geq 0}  \, {\rm dim} \, {\textrm H}^\ast_{\rm mid}{\cal G}({\cal M}^{0, \bL, m}_{SU(N), \bar \mu}(\widetilde{\mathbb R^4 / \mathbb Z_k})) \,  q^m, 
\label{partition function Witten SU(N)}
\ee
where $m = -kj$ is a non-negative integer; $q = e^{2 \pi i \tau}$; and $\tau = \tau_1 + i \tau_2$ is the modulus of the torus formed by identifying the two ends of of the ${\bf S}^1_n \times \mathbb R_t$ worldsheet of the sigma-model. Here,
\be
{m_{\bL}} = h_{\bL} - {c_{\bL}  \over 24};
\label{m-Witten}
\ee
the non-negative number 
\be
h_{\bL} = {({\bar {\bL}}, {\bar {\bL}} + 2 \rho^\vee) \over 2(k + h)},
\label{h-Witten}
\ee 
where $\rho^\vee$ and $h$ are the Weyl vector and dual Coxeter number of the\emph{ Langlands dual} group ${SU(N)}^\vee$, respectively; and the number
\be
{c_{ \bL}}  =  -24 \tilde b{(\bar \bL, \bar \bL)}  +  {12 (\bar \bL, \bar \bL + 2 \rho^\vee) \over {(k + h)}},
\label{c-Witten}
\ee
where $\tilde b = 1/2$ in this $SU(N)$ case.

In this instance, $\bL$  and $\mu$ can also be regarded as \emph{dominant weights} of the corresponding \emph{Langlands dual} affine Kac-Moody group ${SU(N)}^\vee_{\rm aff}$ of level $k$. 

\bigskip\noindent{\it The Partition Function of Spacetime BPS States in (\ref{M-theory 1 Witten discussion}) for $G = SO(N+1)$}

Now consider the $n=2$ case with even $N$ whence the theory is ``$\mathbb Z_2$-twisted'' as we go around ${\bf S}^1_n$ and $G = SO(N+1)$; as usual, we would have an untwisted and twisted sector labeled by $\nu = 0$ and $1$, respectively. By repeating the arguments that led us to (\ref{Z-SO(N+1)-1})--(\ref{Z-SO(N+1)-3}), and by noting that  $w_2 \neq 0$ for $SO(N+1)$-instantons, we can write the  corresponding partition function of spacetime BPS states in any $\bL$-sector as
\be
Z^{\rm BPS}_{SO(N+1), \bL}  =   q^{m_{\bL}} \sum_{w_2} \sum_{\nu = 0, 1}  \sum_{\bar\mu_\nu}  \sum_{m_\nu \geq 0}   {\rm dim} \, \overline {{\textrm H}^{\ast, \nu}_{\rm mid}{\cal G}}({\cal M}^{w_2, \bL, m_\nu}_{SO(N+1), \bar\mu_\nu}(\widetilde{\mathbb R^4 / \mathbb Z_k})) \, q^{m_\nu}.
\label{Z-SO(N+1)-1-Witten}
\ee
Here,  $\overline{{\textrm H}^{\ast, \nu}_{\rm mid}{\cal G}}(\cal M)$ is generated by physical observables in the fields $\varphi_\nu$ and $\eta_\nu$  which obey (\ref{field twist 1})--(\ref{field twist 3}), that are also invariant under the $\mathbb Z_2$ transformations $\varphi_\nu \to - \varphi_\nu$ and $\eta_\nu \to -\eta_\nu$; the non-negative number $m_\nu = -kn' j_\nu$, where $n' = 1$ or $2$ if $N = 2$ or $N > 2$, respectively; $j_\nu \in \mathbb Z_{\geq 0} + {\nu \over 2}$; and $\mu_\nu = (k, \bar \mu_\nu, j_\nu)$. The phase factor $m_{\bL}$ takes the form in (\ref{m-Witten}). 
 
 In this instance,  $\bL$ and $\mu_\nu$ can also be regarded as (un)twisted dominant weights of the  $\mathbb Z_2$-twisted affine Kac-Moody group ${SU(N)}^{(2)}_{\rm aff}$; furthermore, ${SU(N)}^{(2)}_{\rm aff}$ is equal to $SO(N+1)^\vee_{\rm aff}$. In other words, $\bL$ and $\mu_\nu$ can also be regarded as \emph{dominant weights} of the \emph{Langlands dual} affine Kac-Moody group ${SO(N+1)}^\vee_{\rm aff}$ of level $k$.

Additionally, notice that (\ref{Z-SO(N+1)-1-Witten}) also implies that the \emph{effective} Hilbert space ${{\cal H}}^{\rm eff}_{\rm BPS}$ of spacetime BPS states ought to be given by
\be
\label{HBPS-eff-Witten}
{{\cal H}}^{\rm eff}_{\rm BPS} =  \bigoplus_{w_2}  \bigoplus_{\bL} \bigoplus_{\nu =0,1}  \bigoplus_{\mu_\nu}  \, {\overline{\cal H}}^{w_2, \bL, \mu_\nu}_{\rm BPS}   =     \bigoplus_{w_2}  \bigoplus_{\bL} \bigoplus_{\nu =0,1}  \bigoplus_{\mu_\nu} \, \overline{{\textrm H}^{\ast, \nu}_{\rm mid}{\cal G}}({\cal M}^{w_2, \bL}_{SO(N+1), \mu_\nu}(\widetilde{\mathbb R^4 / \mathbb Z_k})),
\ee
where $\nu =0$ or $1$ if the sector is untwisted or twisted, respectively.

\bigskip\noindent{\it The Spectrum of Spacetime BPS States in the M-Theory Compactification (\ref{M-theory 7 Witten discussion})}

We shall now describe the spacetime BPS states given by the ground states of the quantum worldvolume theory of the M5-branes in the physically dual compactification (\ref{M-theory 7 Witten discussion}). Repeating the arguments in $\S$3.1, we find that the spacetime BPS states will be given by the states of the I-brane theory in the following type IIA configuration:
\be
\textrm{IIA}: \quad \underbrace{ {\mathbb R}^3 \times  {{\bf S}^1_n} \times {\mathbb R}_t \times {\mathbb R}^5}_{\textrm{I-brane on ${{\bf S}^1} \times {\mathbb R}_t = k \ \textrm{non-coincident D4}\cap N \textrm{D6}$}}.
\label{equivalent IIA system Witten}
\ee
Here, we have a stack of $k$ $\textrm{\it{non-coincident}}$  D4-branes whose worldvolume is given by $\mathbb R^3 \times {{\bf S}^1_n} \times {\mathbb R}_t$, and a stack of $N$ coincident D6-branes whose worldvolume is given by ${{\bf S}^1_n} \times {\mathbb R}_t \times {\mathbb R}^5$; the two stacks intersect along ${{\bf S}^1_n} \times {\mathbb R}_t$ to form a D4-D6 I-brane system.

It is useful to note at this point that the analysis surrounding (\ref{4.9})--(\ref{Fock}) has also been carried out for a T-dual D5-D5 I-brane system in \cite{nissan}. In particular, one can also understand the embedding (\ref{affine embedding}) as a splitting into the factors $\frak{u}(1)^{(n)}_{\textrm {aff},kN} \times \frak{su}(k)^{(n)}_{\textrm {aff},N} \times \frak{su}(N)^{(n)}_{\textrm {aff},k}$ of the free fermion bilinear currents which nevertheless preserves the total central charge.\footnote{Recall from footnote~\ref{central charge} that the $\mathbb Z_n$-twist does not modify the central charge.} According to the T-dual analysis in \cite{nissan} of an I-brane that results from stacks of intersecting D5-branes which are separated, the free fermion bilinear currents along the I-brane in (\ref{equivalent IIA system Witten}) ought to split into the factors $\frak{u}(1)^{(n)}_{\textrm {aff},kN} \times (\frak{u}(1)^{(n)}_{\textrm {aff},N})^{k-1} \times \frak{su}(N)^{(n)}_{\textrm {aff},k} \times [\frak{su}(k)^{(n)}_{\textrm {aff},N} / (\frak{u}(1)^{(n)}_{\textrm {aff},N})^{k-1}]$. As such, the system of $kN$ complex free fermions with central charge $kN$ will, in this case, give a realization of the total integrable module over the affine Lie algebra
\be
\frak{u}(1)^{(n)}_{\textrm {aff},kN} \otimes [\frak{u}(1)^{(n)}_{\textrm {aff},N}]^{k-1} \otimes \frak{su}(N)^{(n)}_{\textrm {aff},k} \otimes \left [\frak{su}(k)^{(n)}_{\textrm {aff},N} / [\frak{u}(1)^{(n)}_{\textrm {aff},N}]^{k-1} \right ].
\label{affine embedding Witten}
\ee
The total central charge is still $kN$ -- as argued in \emph{loc.~cit.}, the central charge does not change as we move along the Coulomb branch to separate the D-branes. Indeed, it is also invariant under the exchange $k \leftrightarrow N$.

Note at this juncture that we also have the following (conformal) equivalence of coset realizations (c.f.~\cite{CFT text}):
\be
{\frak{su}(k)^{(n)}_{{\rm aff}, N} \over{[\frak{u}(1)^{(n)}_{{\rm aff}, N}}]^{k-1}}  = {[\frak{su}(N)^{(n)}_{{\rm aff},1}]^{k} \over {\frak{su}(N)^{(n)}_{{\rm aff}, k}}}.
\label{coset}
\ee
Substituting this in (\ref{affine embedding Witten}), we find that we effectively have the following total integrable module over the affine Lie algebra
\be
\frak{u}(1)^{(n)}_{{\rm aff}, kN} \otimes [\frak{u}(1)^{(n)}_{{\rm aff}, N}]^{k-1} \otimes [\frak{su}(N)^{(n)}_{{\rm aff}, 1}]^k
\label{effective affine embedding Witten}
\ee
of central charge $kN$. This means that the total Fock space ${\cal F}^{kN}$ of the $\it{uncoupled}$ $kN$ complex free fermions can be realized as
\be
{\cal F}^{\otimes kN} = \textrm{WZW}_{\widehat{u}(1)^{(n)}_{kN}} \otimes [\textrm{WZW}_{\widehat{u}(1)^{(n)}_{N}}]^{k-1} \otimes [\textrm{WZW}_{\widehat{su}(N)^{(n)}_1}]^k,
\label{Fock Witten}
\ee
where $\textrm{WZW}_{\widehat{u}(1)^{(n)}_{kN}}$, $[\textrm{WZW}_{\widehat{u}(1)^{(n)}_{N}}]^{k-1}$, and $[\textrm{WZW}_{\widehat{su}(N)^{(n)}_1}]^k$ are the irreducible integrable modules $\widehat{u}(1)^{(n)}_{kN}$, $[\widehat{u}(1)^{(n)}_{N}]^{k-1}$ and $[\widehat{su}(N)^{(n)}_1]^k$ over the corresponding affine Lie algebras that can be realized by the spectra of states of the corresponding $\it{chiral}$ WZW models. Consequently, the partition function of the uncoupled I-brane theory will be expressed in terms of the (product of) chiral characters of $\widehat{u}(1)^{(n)}_{kN}$, $\widehat{u}(1)^{(n)}_{N}$  and $\widehat{su}(N)^{(n)}_1$.

Next, we must couple the free fermions to the gauge fields which are dynamical. Since the $k$ D4-branes are non-coincident, the free fermions will generically couple to the gauge group $U(1) \times U(1)^{k-1} \times SU(N)$, where the $U(1)^{k-1}$ factor is the Cartan tori of $SU(k)$. As explained in $\S$3.1, since the radius of the circle fiber of $TN^{R \to 0}_N$ goes to zero at infinity, the free fermions will couple dynamically to the $U(1)$ gauge field. In addition, because the geometry of $TN^{R \to 0}_N$ is fixed in our setup (recall that the center-of-mass degrees of freedom of the $N$ NS5-branes which give rise to the $TN^{R \to 0}_N$ geometry via steps (\ref{IIB 3}) and (\ref{IIA 4}), are frozen), in contrast to the gauge field on the D4-branes, the $SU(N)$ gauge field on the $N$ D6-branes should $\it{not}$ be dynamical. Hence, we conclude that the free fermions couple dynamically only to the gauge group $U(1) \times U(1)^{k-1}$. Schematically, this means that we are dealing with the following partially gauged CFT
\be
\label{effective CFT-fully-resolved-A}
{{\frak{u}(1)^{(n)}_{{\rm aff}, kN} \over \frak{u}(1)^{(n)}_{{\rm aff}, kN} }  \otimes {\frak{u}(1)^{(n)}_{{\rm aff}, N}]^{k-1} \over [\frak{u}(1)^{(n)}_{{\rm aff}, N}]^{k-1} } \otimes [\frak{su}(N)^{(n)}_{{\rm aff}, 1}]^k}.
\ee
In particular, the $\frak{u}(1)^{(n)}_{{\rm aff}, kN}$ WZW model and the $k-1$ number of $\frak{u}(1)^{(n)}_{{\rm aff}, N}$ WZW models will be replaced by the corresponding topological $G/G$ models. Consequently, all chiral characters except those of  $\widehat{su}(N)^{(n)}_1$ which appear in the overall partition function of the uncoupled free fermions system on the I-brane, will reduce to constant complex factors after coupling to the dynamical $U(1)$ and $U(1)^{k-1}$ gauge fields. As such, modulo these constant complex factors which serve only to shift the energy levels of the ground states by numbers dependent on the highest affine weights of  $\widehat{u}(1)^{(n)}_{kN}$ and $\widehat{u}(1)^{(n)}_N$, the $\it{effective}$ overall partition function of the I-brane theory will be expressed solely in terms of the product of $k$ chiral characters of  $\widehat{su}(N)^{(n)}_1$;  that is, the sought-after spectrum of spacetime BPS states in the M-theory compactification (\ref{M-theory 7 Witten discussion}) would be realized by $[\textrm{WZW}_{\widehat{su}(N)^{(n)}_1}]^k$. 

\bigskip\noindent{\it A Geometric Langlands Duality for $\widetilde{\mathbb R^4 / \mathbb Z_k}$ for the $A_{N-1}$ Groups} 

Let us now consider $n=1$ whence there is no ``twist'' at all, i.e., $\widehat{su}(N)^{(n)}_1$ is simply $\widehat{su}(N)_1$, the integrable module over the untwisted affine Lie algebra ${\frak su}(N)_{{\rm aff}, 1}$ of level $1$. Then, the physical duality of the M-theory compactifications (\ref{M-theory 1 Witten discussion}) and (\ref{M-theory 7 Witten discussion}) means that their respective spacetime BPS spectra ought to be equivalent, i.e., $[\textrm{WZW}_{\widehat{su}(N)_1}]^k$ ought to be equal to $\cal H_{\rm BPS}$ of (\ref{BPS-TNk}). Indeed,  both $\cal H_{\rm BPS}$ and $[\textrm{WZW}_{\widehat{su}(N)_1}]^k$ are labeled by $k$; moreover, $\frak{su}(N)_{\textrm{aff}} \cong \frak{su}(N)^\vee_{\textrm{aff}}$ whence we can identify $\widehat{su}(N)_1$ with the module ${^L\widehat{su}}(N)_1$ over  $\frak{su}(N)^\vee_{\textrm{aff}}$; such a module -- associated with the $l^{\rm th}$ WZW model -- is labeled by a dominant highest weight $\tilde\lambda^{(l)}$ of $SU(N)^\vee_{\rm aff}$ of level 1, which we can naturally identify as $\lambda^{(l)}$ in $\bL = \sum_{i=1}^k \lambda^{(i)}$ of (\ref{BPS-TNk}). Thus, in any $\{\bL, \mu\}$-sector of the spectra of spacetime BPS states, we can write 
\be
 {\cal H}^{0, \bL, \mu}_{\rm BPS}= \left[\bigotimes_{i=1}^k \textrm{WZW}_{^L\widehat{su}(N)^{\lambda^{(i)}}_{1}} \right]_\mu,
\label{H=WZW-Witten}
\ee
where the subscript `$\mu$' just refers to the $\mu$-weight space of the spectrum of states of the total WZW model. 

As $\textrm{WZW}_{^L\widehat{su}(N)^{\lambda^{(i)}}_{1}}$ is furnished by ${^L\widehat{su}(N)^{\lambda^{(i)}}_{1}}$, via (\ref{BPS-TNk}), we can also express (\ref{H=WZW-Witten}) as 
\be
\boxed{{\textrm H}^\ast_{\rm mid}{\cal G}({\cal M}^{0, \bL}_{SU(N), \mu}(\widetilde{\mathbb R^4 / \mathbb Z_k})) = \left[\bigotimes_{i = 1}^{k} \, {^L\widehat{su}(N)^{\lambda^{(i)}}_{1, \, {{\vec p}_i}}} \right]_\mu}
\label{GL-relation A-Witten}
\ee
where the label ${\vec p}_i$ can be interpreted as the position of the $i^{\rm th}$ center of $\widetilde{\mathbb R^4 / \mathbb Z_k}$ that the module is associated with. This is an  $\widetilde{\mathbb R^4 / \mathbb Z_k}$ non-singular generalization of~\cite[Conjecture 4.14(3)]{BF} for the simply-connected $SU(N) = A_{N-1}$ groups!

\bigskip\noindent{\it{A More General Statement and Witten's Field-Theoretic Result}}

Note that the partition function of the spacetime BPS states realized by  $[\textrm{WZW}_{^L\widehat{su}(N)_1}]^k \simeq [\textrm{WZW}_{\widehat{su}(N)_1}]^k$ can be written in any $\bL$-sector  as~\cite{CFT text}
\begin{eqnarray}
Z_{\bL}  =  q^{\delta}  \left[\bigotimes_{i=1}^k \, \textrm{Tr}_{\lambda^{(i)}}  \, \,  e^{-2\pi i \sum_l u_l J^l_0} q^{L_0 - c'/24} \right]
 =  q^{\delta}  \left [\bigotimes_{i=1}^k\, {\Theta^{\textrm{level 1}}_{\lambda^{(i)}}(0, q) \over {\eta(q)^{N-1}}} \right].
\label{partition function Witten 2}
\end{eqnarray}
Here, $\delta$ represents the overall shift in the ground state energy level due to the aforementioned $G/G$ topological models; $L_0 \in \mathbb Z_{\geq 0}$ is the general state energy level; $\eta(q)$ is the usual Dedekind eta-function; $\Theta^{\textrm{level 1}}_{\lambda^{(i)}}(\xi, q)$ is the generalized theta-function associated with the highest weight module over $\frak{su}(N)_{\textrm{aff},1}$ labeled by $\lambda^{(i)}$ with central charge $c'=N-1$; and $\xi = \sum_{l} u_l J^l_0 = 0$, because the Coulomb moduli $u_l$ of the $N$ $\it{coincident}$ D6-branes ought to vanish, as the corresponding $SU(N)$ gauge group is not broken down to its Cartan tori associated with the bilinear currents $J^l_0$. At any rate, note that the reason why $Z_{\bL}$ can be expressed in terms of modular forms even though our chiral WZW model is defined on a cylinder ${\bf S}^1_n \times \mathbb R_t$ and not an elliptic curve, is because in taking the trace as indicated in (\ref{partition function Witten 2}), we are effectively gluing the two ends of the cylinder together whence it becomes an elliptic curve.  

Once again, the equivalence of the spacetime BPS spectra of the compactifications (\ref{M-theory 1 Witten discussion}) and (\ref{M-theory 7 Witten discussion}) implies that $Z_{SU(N), \bL}^{\rm BPS}$ of (\ref{partition function Witten SU(N)}) ought to be equal to $Z_{\bL}$ of (\ref{partition function Witten 2}), i.e., 
\be
\boxed{ \bigotimes_{i=1}^k\, {\Theta^{\textrm{level 1}}_{\lambda^{(i)}}(0, q) \over {\eta(q)^{N-1}}}  = \sum_{\bar \mu} \sum_{m' \geq 0}  \, {\rm dim} \, {\textrm H}^\ast_{\rm mid}{\cal G}({\cal M}^{0, \bL, m'}_{G, \bar \mu}(\widetilde{\mathbb R^4 / \mathbb Z_k})) \,  q^{m'   - kc' / 24}}
\label{Witten's relation}
\ee
where $G = A_{N-1}$ type whence ${\rm rank} (G) = N-1$; ${\Theta^{\textrm{level 1}}_{\lambda^{(i)}}(0, q) / {\eta(q)^{\textrm{rank}(G)}}}$ is the character of the integrable representation (associated with $\lambda^{(i)}$) of the loop group ${\cal L}G$ at level 1; $m' = m + {m_{\bL}}$, where $m/k \in \mathbb Z_{\geq 0}$ and $m_{\bL}$ is as given in (\ref{m-Witten}); and $kc' / 24 = \delta$.\footnote{To understand this equality, first note that $\delta = k(h_\alpha -1/24)$ as it originates from the $k$ topological $U(1)/U(1)$ models; here, $h_\alpha$ is the conformal dimension of the ground state of the dominant highest weight module of a chiral $\frak{u}(1)_{\rm aff}$ WZW model with dominant highest affine weight $\alpha$. Next, note that the spectrum of this WZW model can be described by the spectrum of a free chiral boson on the I-brane ${\bf S}^1_n \times \mathbb R_t$; as such, $h_\alpha = {1\over 2} (nr + mr/2)^2$, where $m,n \in \mathbb Z_{\geq 0}$ and $r$ is the radius of ${\bf S}^1_n$~\cite{CFT text}. Therefore, since the radius $r$ can be arbitrary, one can always find a solution to $h_{\alpha} = N/24$ for some  $n$ and $m$ -- that is, we can set $\delta = kc' / 24$, as claimed. \label{Heisenberg}}

Incidentally, a $\widetilde {TN}_k$ specialization of (\ref{Witten's relation}) (where there ought to be, on the LHS, an additional contribution from the Fock space of a free chiral boson because of monopoles that go around the finite-sized circle fiber at infinity), has also been derived by Witten in \cite{Lectures by Witten} via purely field-theoretic considerations  (see also~\cite[eqn.~(5.17)]{GL from 6d}); in particular,  he understood the $\widetilde {TN}_k$ specialization of (\ref{Witten's relation}) to be a consequence of an invariance in the BPS spectrum of a 6d $(2,0)$ $A_{N-1}$ superconformal field theory on $\widetilde {TN}_k \times {\bf S}^1 \times \mathbb R_t$ under different limits of a compactification down to five dimensions. Witten's derivation in~\cite{Lectures by Witten} thus serves as a non string-theoretic corroboration of (\ref{Witten's relation}) for $\widetilde {TN}_k$  that is rooted in six-dimensional superconformal field theory.

That said, one cannot, within the purely field-theoretic framework of~\cite{Lectures by Witten}, derive (\ref{Witten's relation}) for $\widetilde{\mathbb R^4 / \mathbb Z_k}$  -- see~\cite[Remark 5.3]{GL from 6d}.  On the other hand, the purely field-theoretic analysis in \cite{Lectures by Witten} shows that (\ref{Witten's relation}) for $\widetilde {TN}_k$ ought to also hold for the other simply-laced $D_N$ and $E_{6,7,8}$ groups. In our M-theoretic setup with M5-branes, there is no direct way to realize an $E_{6,7,8}$ type symmetry in their worldvolume theory. However, as explained in $\S$3.2, one can realize a $D_N$ type symmetry by adding an OM5-plane to the stack of M5-branes. For brevity, we shall not work out the $D_N$ case; rather, we shall -- after the following excursion to reproduce purely physically a closely-related and celebrated mathematical result -- continue our analysis for the nonsimply-laced $B_{N/2}$ groups.

\bigskip\noindent{\it Reproducing Nakajima's Celebrated Result} 

As mentioned in the last section, where the spectrum of ground states of the worldvolume theory of a stack of M5-branes wrapping $M_4 \times {\bf S}^1 \times \mathbb R_t$ is concerned, one can -- if $M_4$ is a hyperk\"ahler four-manifold -- regard the theory to be topological along $M_4$ (and conformal along ${\bf S}^1 \times \mathbb R_t$) (c.f.~\cite{junya} and footnote~\ref{junya's twist}).  Moreover, if the gauge group is $SU(N)$, there are no non-abelian magnetic fluxes that pass through the $k-1$ two-spheres in $\widetilde{\mathbb R^4 / \mathbb Z_k}$.  Altogether, this means that $\sum_{\bL} \, Z_{SU(N), \bL}^{\rm BPS}$ of (\ref{partition function Witten SU(N)}) ought to be equal to $ Z_{SU(N)}^{\rm BPS}$ of (\ref{Z_SU(N)}) which, via (\ref{GL-relation A}), is equal to the partition function of the chiral WZW model whose spectrum is ${\rm WZW}_{\widehat{su}(N)_{k}}$; then, by the level-rank duality of chiral WZW models for the $A$ groups in (\ref{level-rank-A}), we finally find that $\sum_{\bL} \, Z_{SU(N), \bL}^{\rm BPS}$ ought to be equal to the partition function of the chiral WZW model whose spectrum is ${\rm WZW}_{\widehat{su}(k)_{N}}$; in other words, we can (up to some modular anomaly) write
\be
\boxed{\sum_{m' \geq 0}  \, {\rm dim} \, {\textrm H}^\ast_{\rm mid}{\cal G}({\cal M}^{\Lambda, m'}_{SU(N)}(\widetilde{\mathbb R^4 / \mathbb Z_k})) \,  q^{m' - c/24} =  \sum_\gamma c^\Lambda_\gamma \, \Theta^{{\rm level} \, N}_{\gamma,  \frak {su}(k)}(q)}
\label{nakajima-A}
\ee
where $c$, $c^\Lambda_\gamma$ and $ \Theta^{{\rm level} \, N}_{\gamma, \frak {su}(k)}$ are the central charge, string-functions and theta-functions associated with the integrable module over $\frak{su}(k)_{\textrm{aff}, N}$ of dominant highest weight $\Lambda$; $\gamma$ are weights of $\frak{su}(k)_{\textrm{aff}, N}$; and $m' = m + m_{\Lambda}$, where $m$ is a non-negative integer, while $m_{\Lambda}$ is a number which depends on $\Lambda$. 

To arrive at (\ref{nakajima-A}) and the accompanying statements, we have made use of the fact that (i) the McKay correspondence implies that $\bar \mu$ in the earlier formulas -- which represents a conjugacy class of the homomorphism $\phi_\infty: \mathbb Z_k \to SU(N)$ at infinity -- can be mapped to $\Lambda$ (see~\cite[$\S$4.4]{vafa-witten}); (ii) the level-rank duality in (\ref{level-rank-A}) implies that the dominant highest weight ${\bL}$ of the integrable module over $\frak{su}(N)_{\textrm{aff}, k}$, can likewise be mapped to $ \Lambda$.

In short, we have obtained in (\ref{nakajima-A}) Nakajima's celebrated result in~\cite{nakajima} for $SU(N)$!

\bigskip\noindent{\it A Geometric Langlands Duality for $\widetilde{\mathbb R^4 / \mathbb Z_k}$ for the $B_{N/2}$ Groups}

Let us now restrict ourselves to \emph{even} $N$, and consider $n=2$ whence there is a ``$\mathbb Z_2$-twist'', i.e., the relevant module is $\widehat{su}(N)^{(2)}_1$, the integrable module over the $\mathbb Z_2$-twisted affine Lie algebra ${\frak su}(N)^{(2)}_{{\rm aff}, 1}$ of level $1$. Let  $\{ \lambda^{(1)'}, \dots, \lambda^{(k)'} \}$ be a set of dominant highest weights of this module; $\bL' = \sum_{i=1}^k \lambda^{(i)'} = (k, \bar {\bL}', 0)$; and $\bar {\bL}' = \sum_{i=1}^k \bar{\lambda}^{(i)'}$. Then, by repeating the arguments that led us to (\ref{WZW space-SO(N+1)}), bearing in mind that we now have $ [\textrm{WZW}_{\widehat{su}(N)^{(2)}_1}]^k $ instead of $\textrm{WZW}_{\widehat{su}(N)^{(2)}_k}$, we find that we can write
 \be
 [\textrm{WZW}_{\widehat{su}(N)^{(2)}_1}]^k =  \bigoplus_{w'_2} \bigoplus_{\bL'}  \bigoplus_{\nu =0,1} \, \left[\bigotimes_{i=1}^k\overline{\textrm{WZW}}_{\widehat{su}(N)^{(2), \, \lambda^{(i)'}}_{1, \nu}}\right]_{w'_2},
 \label{WZW space-SO(N+1)-Witten}
 \ee
 where
\be
w'_2 = \sum_{a=1}^{k-1} v_a {\bar\alpha}_a.
\label{xi-SO(N+1)-Witten}
\ee
Here, the $k-1$ numbers $v_a$ correspond to the nonvanishing Coulomb moduli of the $k$ \emph{fully separated} D4-branes (with center-of-mass locked in the first $U(1)$ factor of (\ref{affine embedding Witten})) whose magnitudes correlate with the sizes of the $k-1$ two-spheres in $\widetilde{\mathbb R^4 / \mathbb Z_k}$; $\bar\alpha_a = \bar \alpha$, where $ \bar \alpha$ is the finite part of the dominant highest affine weight $\alpha$ that labels a dominant highest weight module of a chiral $\frak{u}(1)_{\rm aff}$ WZW model; the overhead bar means that we project onto $\mathbb Z_2$-invariant states (as required of twisted CFT's); $\nu =0$ or $1$ indicates that the sector is untwisted or twisted, respectively; $\widehat{su}(N)^{(2), \, \lambda^{(i)'}}_{1, \nu}$ is a dominant highest weight module labeled by the dominant highest weight $\lambda^{(i)'}$, and whose general state energy level is $h_\nu \in \mathbb Z_{\geq 0} + {\nu \over 2}$; and the subscript `$w'_2$' means that the \emph{overall} ground state energy level is further shifted by  $ w'_2 / \tau$. 

Looking at the RHS of (\ref{xi-SO(N+1)-Witten}), we see that we can interpret $w'_2$ as a linear sum of $\mathbb Z_2$-valued non-abelian magnetic fluxes through the $k-1$ two-spheres in $\widetilde{\mathbb R^4 / \mathbb Z_k}$: starting at the origin, the D4-branes can move either in the positive or negative direction whence the $v_r$'s can take either positive or negative values, and by a natural identification of the $\bar \alpha$'s as the standard area of the $k-1$ two-spheres which define a basis of $H_2(\widetilde{\mathbb R^4 / \mathbb Z_k}, \mathbb Z)$, we have our claim. A somewhat related analysis has also been carried out in~\cite[$\S$2.4]{Vafa et al}, where it was shown that $w'_2$ can indeed be associated with fluxes through the $k-1$ two-spheres of  $\widetilde{\mathbb R^4 / \mathbb Z_k}$. Thus, let us henceforth identify $w'_2$ as $w_2$ of (\ref{BPS-TNk}).

Now the physical duality of the M-theory compactifications (\ref{M-theory 1 Witten discussion}) and (\ref{M-theory 7 Witten discussion}) means that their respective spacetime BPS spectra ought to be equivalent, i.e., $[\textrm{WZW}_{\widehat{su}(N)^{(2)}_1}]^k$ ought to be equal to $\cal H^{\rm eff}_{\rm BPS}$ of (\ref{HBPS-eff-Witten}). Indeed,  both $\cal H^{\rm eff}_{\rm BPS}$ and $[\textrm{WZW}_{\widehat{su}(N)^{(2)}_1}]^k$ are labeled by $k$; moreover, it is clear that one can identify $\bL'$ of (\ref{WZW space-SO(N+1)-Witten}) with $\bL$ of (\ref{HBPS-eff-Witten}); it is also clear that as in $\S$3.1, one can identify $h_\nu$ (which is implicit in (\ref{WZW space-SO(N+1)-Witten}))  with $j_\nu$ (which is implicit in (\ref{HBPS-eff-Witten})).  Hence, in any $(w_2, \bL, \nu, \mu_\nu)$-sector of the spectra of spacetime BPS states, we can  write 
\be
 {\overline{\cal H}}^{w_2, \bL, \mu_\nu}_{\rm BPS}   =  \left[\bigotimes_{i=1}^k\overline{\textrm{WZW}}_{\widehat{su}(N)^{(2), \, \lambda^{(i)}}_{1, \nu}}\right]_{w_2, \, \mu_\nu},
\label{H=WZW-SO(N+1)-Witten}
\ee
where the subscript `$\mu_\nu$' just refers to the $\mu_\nu$-weight space of the spectrum of states of the total WZW model in the $\nu$-sector. As $\overline{\textrm{WZW}}_{\widehat{su}(N)^{(2), \, \lambda^{(i)}}_{1, \nu}}$ is furnished by the $\mathbb Z_2$-invariant projection ${\widehat{su}(N)^{(2), \, \lambda^{(i)}}_{1, \nu}}\vert_{\mathscr P_2}$ of $\widehat{su}(N)^{(2), \, \lambda^{(i)}}_{1, \nu}$, and since $\frak{su}(N)^{(2)}_{\textrm{aff}} \simeq \frak{so}(N +1)^\vee_{\textrm{aff}} $ whence ${\widehat{su}(N)^{(2),  \, \lambda^{(i)}}_{1, \nu}}\vert_{\mathscr P_2}$ is isomorphic to the submodule ${^L\widehat{so}(N+1)^{\lambda^{(i)}}_{1, \nu}}$ over $\frak{so}(N+1)^\vee_{\textrm{aff}}$, via (\ref{HBPS-eff-Witten}), we can also express (\ref{H=WZW-SO(N+1)-Witten}) as 
\be
\boxed{\overline{{\textrm H}^{\ast, \nu}_{\rm mid}{\cal G}}({\cal M}^{w_2, \bL}_{SO(N+1),  \mu_\nu}(\widetilde{\mathbb R^4 / \mathbb Z_k})) =   \left[ \bigotimes_{i=1}^k \, {^L\widehat{so}}(N+1)^{\lambda^{(i)}}_{1, \, \nu, \, {{\vec p}_i}} \right]_{w_2, \, \mu_\nu}}
\label{GL-relation B-Witten}
\ee
for $\nu = 0$ and $1$, where the label ${\vec p}_i$ can be interpreted as the position of the $i^{\rm th}$ center of $\widetilde{\mathbb R^4 / \mathbb Z_k}$ that the module is associated with. This is an $\widetilde{\mathbb R^4 / \mathbb Z_k}$ non-singular generalization of~\cite[Conjecture 4.14(3)]{BF} for the nonsimply-connected $SO(N +1) = B_{N/2}$ groups!

\bigskip\noindent{\it A Nonsimply-Laced Generalization of Witten's Field-Theoretic Result}

Note that the partition function of the spacetime BPS states realized by  $[\overline{\textrm{WZW}}_{\widehat{su}(N)^{(2)}_1}]^k \simeq [{\textrm{WZW}}_{^L\widehat{so}(N+1)_1}]^k$ can be written in any $(w_2, \bL, \nu)$-sector as
\begin{eqnarray}
\hspace{-0.5cm} Z_{w_2, \bL, \nu}  =  q^{\delta + \tilde w_2}    \left[\bigotimes_{i=1}^k \, \textrm{Tr}_{\lambda^{(i)}}  \,  q^{L_{0, \nu} - c'/24} \right]
 = q^{\delta + \tilde w_2}  \left [\bigotimes_{i=1}^k\, \left( \sum_{\gamma} c^{\lambda^{(i)}}_{\gamma, \nu} \,  {^L\Theta}^{\textrm {level 1}}_{\gamma, \nu}(q) \right) \right].
\label{partition function Witten 3}
\end{eqnarray}
Here, $\delta + {\tilde w}_2$ represents the overall shift in the ground state energy level due to the aforementioned $G/G$ topological models, and it is equal to  $kc' / 24 + {w_2 / \tau}$ (see footnote~\ref{Heisenberg}); $L_{0, \nu} \in \mathbb Z_{\geq 0} + {\nu \over 2}$ is the general state energy level; $c^{\lambda^{(i)}}_{\gamma, \nu}$ and $^L\Theta^{\textrm{level 1}}_{\gamma, \nu}$ are string-functions and theta-functions associated with the $\nu$-sector of the underlying dominant highest weight module over $\frak{so}(N +1)^\vee_{\textrm{aff}, 1}$ of central charge $c'=N-1$; and $\gamma$ is a weight of $\frak{so}(N +1)^\vee_{\textrm{aff}, 1}$.

Once again, the equivalence of the spacetime BPS spectra of the compactifications (\ref{M-theory 1 Witten discussion}) and (\ref{M-theory 7 Witten discussion}) implies that $Z_{SO(N+1), \bL}^{\rm BPS}$ of (\ref{Z-SO(N+1)-1-Witten}) in the $(w_2, \nu)$-sector ought to be equal to $ Z_{w_2, \bL, \nu}$ of (\ref{partition function Witten 3}), i.e., 
\be
 \boxed{ \bigotimes_{i=1}^k\, \left( \sum_{\gamma} c^{\lambda^{(i)}}_{\gamma, \nu} \,  {^L\Theta}^{\textrm {level 1}}_{\gamma, \nu}(q) \right) =    \sum_{\bar\mu_\nu}  \sum_{m'_\nu \geq 0}   {\rm dim} \, \overline {{\textrm H}^{\ast, \nu}_{\rm mid}{\cal G}}({\cal M}^{\tilde w_2, \bL, m'_\nu}_{SO(N+1), \bar\mu_\nu}(\widetilde{\mathbb R^4 / \mathbb Z_k})) 
  q^{m'_\nu - {\tilde w}_2  - kc' / 24}}
\label{Witten's relation SO(N+1)}
\ee
where $G = B_{N/2}$ type; $\sum_{\gamma} c^{\lambda^{(i)}}_{\gamma, \nu} \,  {^L\Theta}^{\textrm {level 1}}_{\gamma, \nu}(q)$ is the character in the $\nu$-sector of the integrable representation (associated with ${\lambda^{(i)}}$) of the \emph{Langlands dual} loop group ${\cal L}G^\vee$ at level 1; $m'_\nu = m_\nu + m_{\bL}$, where $m_\nu/k n' \in  \mathbb Z_{\geq 0} + {\nu \over 2}$ and $m_{\bL}$ is as given in (\ref{m-Witten});  and $n' = 1$ or $2$ if $N = 2$ or $N > 2$, respectively. 

Last but not least, notice that a $\widetilde {TN}_k$ specialization of (\ref{Witten's relation SO(N+1)}) (where there ought to be, on the LHS, an additional contribution from the Fock space of a $\mathbb Z_2$-twisted free chiral boson because of monopoles that go around the finite-sized circle fiber at infinity), would just serve as a nonsimply-laced $B_{N/2}$ group generalization of Witten's field-theoretic result in~\cite{Lectures by Witten}.

\newsubsection{A Quasi-Singular Generalization of the Geometric Langlands Duality for Surfaces}

We shall now continue to derive a quasi-singular generalization of the geometric Langlands duality for surfaces for the $A$--$B$ groups. To this end, let us replace $\mathbb R^4/\mathbb Z_k$ in (\ref{M-theory 1}) with a \emph{partially-resolved} $k$-centered ALE manifold $\widetilde {\mathbb R^4 / \mathbb Z_{k-l, l}}$, where $k-l$ and $l$ centers are coincident and fully-separated, respectively. By repeating the arguments behind (\ref{M-theory 1})--(\ref{M-theory 7}), we find that the following six-dimensional M-theory compactification
\be
\textrm{M-theory}: \quad \mathbb R^{5}  \times  \underbrace{ \mathbb R_t \times {{\bf S}^1_n} \times \widetilde {\mathbb R^4 / \mathbb Z_{k-l, l}}}_{\textrm{$N$ M5-branes}},
\label{M-theory 1 Witten discussion-partial}
\ee
where we evoke a $\mathbb Z_n$-outer-automorphism of $\widetilde {\mathbb R^4 / \mathbb Z_{k-l, l}}$ (and of the geometrically-trivial $\mathbb R^5 \times \mathbb R_t$ spacetime) as we go around the ${\bf S}^1_n$ circle and identify the circle under an order $n$ translation, is \emph{physically dual} to the following six-dimensional M-theory compactification
\be
\textrm{M-theory}: \quad \underbrace{TN_N^{R\to 0}  \times  {{\bf S}^1_n}\times \mathbb R_t}_{\textrm{$l$ out of $k$ M5-branes are non-coincident}}  \times  {\mathbb R^{5}},
\label{M-theory 7 Witten discussion-partial}
\ee
where there is a nontrivial $\mathbb Z_n$-outer-automorphism of the singular multi-Taub-NUT space $TN_N^{R\to 0}$ (whose circle fiber at infinity approaches zero radius)  as we go around the ${\bf S}^1_n$ circle. In contrast to the $\widetilde {\mathbb R^4 / \mathbb Z_k}$ case of the previous subsection, only $l$ out of $k$ centers are (fully) separated in $ \widetilde {\mathbb R^4 / \mathbb Z_{k-l, l}}$; as such, only $l$ out of $k$ M5-branes will be non-coincident in (\ref{M-theory 7 Witten discussion-partial}); the rest of the $k-l$ M5-branes remain coincident.

\bigskip\noindent{\it The Spectrum of Spacetime BPS States in the M-Theory Compactification (\ref{M-theory 1 Witten discussion-partial})}

In order to describe the Hilbert space of spacetime BPS states furnished by the ground states of the quantum worldvolume theory of the M5-branes in (\ref{M-theory 1 Witten discussion-partial}), first note that because $ \widetilde {\mathbb R^4 / \mathbb Z_{k-l, l}}$, like $\widetilde{\mathbb R^4 / \mathbb Z_k}$, is also a hyperk\"ahler manifold, we can repeat our arguments in the previous subsection and conclude that the spacetime BPS states are given by the ${\bf L}^2$-cohomology of some compactification of the moduli space ${\cal M}_{G}( \widetilde {\mathbb R^4 / \mathbb Z_{k-l, l}})$ of $G$-instantons on $ \widetilde {\mathbb R^4 / \mathbb Z_{k-l, l}}$, where  $G= SU(N)$ if $n=1$, and $G = SO(N+1)$ if  $n=2$ and $N$ is even. Since $ \widetilde {\mathbb R^4 / \mathbb Z_{k-l, l}}$ is only a partial resolution of $\mathbb R^4 / \mathbb Z_k$, it is (quasi-)singular; thus, like in the $\mathbb R^4 / \mathbb Z_k$ case,  the spacetime BPS states would be given by the intersection cohomology ${\rm IH}^\ast{\cal U}({\cal M}_{G}( \widetilde {\mathbb R^4 / \mathbb Z_{k-l, l}}))$ of the Uhlenbeck compactification ${\cal U}({\cal M}_{G}( \widetilde {\mathbb R^4 / \mathbb Z_{k-l, l}}))$.

Second, note that for the instanton action to be finite in an integration over noncompact $ \widetilde {\mathbb R^4 / \mathbb Z_{k-l, l}}$, we need to have flat albeit nontrivial connections far away from the origin of $ \widetilde {\mathbb R^4 / \mathbb Z_{k-l, l}}$. Since $ \widetilde {\mathbb R^4 / \mathbb Z_{k-l, l}}$ is topologically equivalent to $\mathbb  R^4/ \mathbb Z_k$ at infinity, according to our discussion in the previous subsection, distinct choices of such flat connections will correspond to distinct dominant coweights $\mu = (k, \bar \mu, j)$ of $G_{\textrm{aff}}$ of level $k$, where $j$ is a number. 

Third, recall that in the case of $\mathbb R^4 / \mathbb Z_k$, the $k$ centers coincide with multiplicity $k$ at the origin such that a $\mathbb Z_k$-type singularity develops whence we have a $\mathbb Z_k$-action in the fiber of the $G$-bundle at 0. On the other hand, in the case of $ \widetilde {\mathbb R^4 / \mathbb Z_{k-l, l}}$,  we have instead (i) $k-l$ centers that coincide at position ${\vec p}_{c}$ with multiplicity $k-l$; and (ii) $l$ non-coincident centers at positions ${\vec p}_{1}, \dots, {\vec p}_l$ with multiplicity 1 each. In other words, we have instead   (i) a $\mathbb Z_{k-l}$-action in the fiber of the $G$-bundle over  ${\vec p}_{c}$; and (ii) a $\mathbb Z_1$-action in the fiber of the $G$-bundle over ${\vec p}_1, \dots, {\vec p}_l$. Since the $\mathbb Z_r$-action is given by a conjugacy class of the homomorphism $\rho: Z_r \to G$, one can  (i) associate a dominant coweight $\lambda_c = (k-l, \bar \lambda_c, i_c)$ of $G_{\textrm{aff}}$ of level $k-l$ with the centers at ${\vec p}_c$, where $i_c$ is a number; and (ii) associate $l$ distinct dominant coweights $\lambda^{(m)} = (1, \bar \lambda^{(m)}, i^{(m)})$ of $G_{\textrm{aff}}$ of level $1$  with the non-coincident centers at ${\vec p}_{1}, \dots, {\vec p}_l$, where the $i^{(m)}$'s are numbers. Nonetheless, consistency with the results of $\S$3.1 (where all $k$ centers coincide) constrains $i_c$ and the $i^{(m)}$'s to be zero.

Fourth, according to our analysis leading up to (\ref{a}), and the fact that $\lambda_c$ and the $\lambda^{(m)}$'s ought to be linearly-independent of one another, we find that the $G$-instantons -- which again correspond to D0-branes within the D4-brane worldvolume in the type IIA picture -- are such that the associated \emph{non-negative} instanton numbers are
\be
a = -kn'j + \tilde b(\bar {\boldsymbol \lambda}, \bar {\boldsymbol \lambda}) - b(\bar \mu, \bar \mu),
\label{a-TNk-partial}
\ee 
where for $G = SU(N)$, $SO(3)$ and $SO(N+1)$,  $n' =1$, $1$ and $2$ while $j$ is a non-positive integer divided by $1$, $2$ and $2$, respectively. Also, $\bar  {\boldsymbol \lambda} = \bar{\lambda}_c +  \sum_{r=1}^l \bar \lambda^{(r)}$; $\tilde b$ and $b$ are some positive real constants; and $( \, , )$ is the scalar product in finite coweight space. For $n=1$ whence we have $G = SU(N)$ with $n'=1$ and $j$ being a non-positive integer, expression (\ref{a-TNk-partial}) is indeed consistent with results from the mathematical literature (which only addresses the case of simply-connected groups like $SU(N)$): eqn.~(\ref{a-TNk-partial}) coincides with~\cite[below Conjecture 3.2]{BF2}  after we set $\tilde b = b = 1/2$ and identify $a/k$ with $d/k$ of \emph{loc.~cit..} (see also footnote~\ref{fraction}).

Last but not least, recall that a $G$-bundle on $ \widetilde {\mathbb R^4 / \mathbb Z_{k-l, l}}$ is topologically classified by $p_2 \in H^2( \widetilde {\mathbb R^4 / \mathbb Z_{k-l, l}}, \pi_1(G))$. In particular, $p_2$ vanishes for simply-connected $G = SU(N)$, but \emph{not }for nonsimply-connected $G = SO(N+1)$; in fact, since $\pi_1(SO(N+1)) = \mathbb Z_2$, we find that $SO(N+1)$-bundles on $ \widetilde {\mathbb R^4 / \mathbb Z_{k-l, l}}$ are topologically classified by the class $w_2 \in H^2( \widetilde {\mathbb R^4 / \mathbb Z_{k-l, l}}, \mathbb Z_2)$.  Note also that $w_2$ can be interpreted as a linear sum of $\mathbb Z_2$-valued non-abelian magnetic fluxes that pass through the $l$ two-spheres in $ \widetilde {\mathbb R^4 / \mathbb Z_{k-l, l}}$~\cite{`t Hooft, vafa-witten}; this observation will be important shortly.

Thus, from the five points above, it is clear that ${\cal U}({\cal M}_{G}( \widetilde {\mathbb R^4 / \mathbb Z_{k-l, l}}))$ ought to be graded by $\{ \lambda_c, \lambda^{(1)}, \dots, \lambda^{(l)} \}$, $\mu$ and $w_2$ (where $a$ is correspondingly given by (\ref{a-TNk-partial})). We are now ready to state the generic Hilbert space ${\cal H}_{\rm BPS}$ of spacetime BPS states in the M-theory compactification (\ref{M-theory 1 Witten discussion-partial}). Let us denote by ${\textrm {IH}}^\ast{\cal U}({\cal M}^{w_2, \bL}_{G, \mu}( \widetilde {\mathbb R^4 / \mathbb Z_{k-l, l}}))$, the intersection cohomology of the Uhlenbeck compactification ${\cal U}({\cal M}^{w_2, \bL}_{G, \mu}( \widetilde {\mathbb R^4 / \mathbb Z_{k-l, l}}))$ of the component ${\cal M}^{w_2, \bL}_{G, \mu}( \widetilde {\mathbb R^4 / \mathbb Z_{k-l, l}})$ of the moduli space ${\cal M}_{G}( \widetilde {\mathbb R^4 / \mathbb Z_{k-l, l}})$ labeled by $\bL = \lambda_c + \sum_{r=1}^l \lambda^{(r)} = (k, \bar{\bL}, 0)$, $\mu = (k, \bar\mu, j)$ and $w_2$; then, we can write
\be
{\cal H}_{\rm BPS} = \bigoplus_{ w_2, \bL, \mu} {\cal H}^{w_2, \bL, \mu}_{\rm BPS}  =  \bigoplus_{w_2, \bL, \mu} {\rm IH}^\ast{\cal U}({\cal M}^{w_2, \bL}_{G, \mu}( \widetilde {\mathbb R^4 / \mathbb Z_{k-l, l}})),
\label{BPS-TNk-partial}
\ee
where $w_2 = 0$ if $n=1$, and $\bL \geq \mu$ (since $a$ is non-negative).

\bigskip\noindent{\it The Partition Function of Spacetime BPS States in (\ref{M-theory 1 Witten discussion-partial}) for  $G = SU(N)$}

Consider the $n=1$ case whence we have $G = SU(N)$, $n'=1$, and $- j \in \mathbb Z_{\geq 0}$. By repeating the arguments that led us to (\ref{Z_SU(N)}), and by noting that $P$ in (\ref{Zbps}) is now equal to $-kj + {1 \over 2}(\bar {\boldsymbol \lambda}, \bar {\boldsymbol \lambda})$ while $w_2 = 0$ for $SU(N)$-instantons, we can write the corresponding partition function of spacetime BPS states in any $\bL$-sector as
\be
 Z_{SU(N), \bL}^{\rm BPS}  =  q^{m_{\bL}}  \, \sum_{\bar \mu} \sum_{m \geq 0}  \, {\rm dim} \, {\rm IH}^\ast{\cal U}({\cal M}^{0, \bL, m}_{SU(N), \bar \mu}(\widetilde {\mathbb R^4 / \mathbb Z_{k-l, l}})) \,  q^m, 
\label{partition function Witten SU(N)-partial}
\ee
where $m = -kj$ is a non-negative integer; $q = e^{2 \pi i \tau}$; and $\tau = \tau_1 + i \tau_2$ is the modulus of the torus formed by identifying the two ends of of the ${\bf S}^1_n \times \mathbb R_t$ worldsheet of the sigma-model. Here,
\be
{m_{\bL}} = h_{\bL} - {c_{\bL}  \over 24};
\label{m-Witten-partial}
\ee
the non-negative number 
\be
h_{\bL} = {({\bar {\bL}}, {\bar {\bL}} + 2 \rho^\vee) \over 2(k + h)},
\label{h-Witten-partial}
\ee 
where $\rho^\vee$ and $h$ are the Weyl vector and dual Coxeter number of the\emph{ Langlands dual} group ${SU(N)}^\vee$, respectively; and the number
\be
{c_{ \bL}}  =  -24 \tilde b{(\bar \bL, \bar \bL)}  +  {12 (\bar \bL, \bar \bL + 2 \rho^\vee) \over {(k + h)}},
\label{c-Witten-partial}
\ee
where $\tilde b = 1/2$ in this $SU(N)$ case.

In this instance, $\bL$  and $\mu$ can also be regarded as \emph{dominant weights} of the corresponding \emph{Langlands dual} affine Kac-Moody group ${SU(N)}^\vee_{\rm aff}$ of level $k$.

\bigskip\noindent{\it The Partition Function of Spacetime BPS States in (\ref{M-theory 1 Witten discussion-partial}) for $G = SO(N+1)$}

Now consider the $n=2$ case with even $N$ whence the theory is ``$\mathbb Z_2$-twisted'' as we go around ${\bf S}^1_n$ and $G = SO(N+1)$; as usual, we would have an untwisted and twisted sector labeled by $\nu = 0$ and $1$, respectively. By repeating the arguments that led us to (\ref{Z-SO(N+1)-1})--(\ref{Z-SO(N+1)-3}), and by noting that  $w_2 \neq 0$ for $SO(N+1)$-instantons, we can write the  corresponding partition function of spacetime BPS states in any $\bL$-sector as
\be
Z^{\rm BPS}_{SO(N+1), \bL}  =   q^{m_{\bL}} \sum_{w_2} \sum_{\nu = 0, 1}  \sum_{\bar\mu_\nu}  \sum_{m_\nu \geq 0}   {\rm dim} \, \overline {{\rm IH}^{\ast, \nu}{\cal U}}({\cal M}^{w_2, \bL, m_\nu}_{SO(N+1), \bar\mu_\nu}(\widetilde {\mathbb R^4 / \mathbb Z_{k-l, l}})) \, q^{m_\nu}.
\label{Z-SO(N+1)-1-Witten-partial}
\ee
Here,  $\overline{{\rm IH}^{\ast, \nu}{\cal U}}(\cal M)$ is generated by physical observables in the fields $\varphi_\nu$ and $\eta_\nu$  which obey (\ref{field twist 1})--(\ref{field twist 3}), that are also invariant under the $\mathbb Z_2$ transformations $\varphi_\nu \to - \varphi_\nu$ and $\eta_\nu \to -\eta_\nu$; the non-negative number $m_\nu = -kn' j_\nu$, where $n' = 1$ or $2$ if $N = 2$ or $N > 2$, respectively; $j_\nu \in \mathbb Z_{\geq 0} + {\nu \over 2}$; and $\mu_\nu = (k, \bar \mu_\nu, j_\nu)$. The phase factor $m_{\bL}$ takes the form in (\ref{m-Witten-partial}). 
 
 In this instance,  $\bL$ and $\mu_\nu$ can also be regarded as (un)twisted dominant weights of the  $\mathbb Z_2$-twisted affine Kac-Moody group ${SU(N)}^{(2)}_{\rm aff}$; furthermore, ${SU(N)}^{(2)}_{\rm aff}$ is equal to $SO(N+1)^\vee_{\rm aff}$. In other words, $\bL$ and $\mu_\nu$ can also be regarded as \emph{dominant weights} of the \emph{Langlands dual} affine Kac-Moody group ${SO(N+1)}^\vee_{\rm aff}$ of level $k$.

Additionally, notice that (\ref{Z-SO(N+1)-1-Witten-partial}) also implies that the \emph{effective} Hilbert space ${{\cal H}}^{\rm eff}_{\rm BPS}$ of spacetime BPS states ought to be given by
\be
\label{HBPS-eff-Witten-partial}
{{\cal H}}^{\rm eff}_{\rm BPS} =  \bigoplus_{w_2}  \bigoplus_{\bL} \bigoplus_{\nu =0,1}  \bigoplus_{\mu_\nu}  \, {\overline{\cal H}}^{w_2, \bL, \mu_\nu}_{\rm BPS}   =     \bigoplus_{w_2}  \bigoplus_{\bL} \bigoplus_{\nu =0,1}  \bigoplus_{\mu_\nu} \, \overline{{\rm IH}^{\ast, \nu}{\cal U}}({\cal M}^{w_2, \bL}_{SO(N+1), \mu_\nu}(\widetilde {\mathbb R^4 / \mathbb Z_{k-l, l}})),
\ee
where $\nu =0$ or $1$ if the sector is untwisted or twisted, respectively.

\bigskip\noindent{\it The Spectrum of Spacetime BPS States in the M-Theory Compactification (\ref{M-theory 7 Witten discussion-partial})}

We shall now describe the spacetime BPS states given by the ground states of the quantum worldvolume theory of the M5-branes in the physically dual compactification (\ref{M-theory 7 Witten discussion-partial}). Repeating the arguments in $\S$3.1, we find that the spacetime BPS states will be given by the states of the I-brane theory in the following type IIA configuration:
\be
\textrm{IIA}: \quad \underbrace{ {\mathbb R}^3 \times  {{\bf S}^1_n} \times {\mathbb R}_t \times {\mathbb R}^5}_{\textrm{I-brane on ${{\bf S}^1} \times {\mathbb R}_t = \ \textrm{($l$)$k-l$ (non-)coincident D4}\cap N \textrm{D6}$}}.
\label{equivalent IIA system Witten-partial}
\ee
Here, we have a stack of $l$ non-coincident and $k-l$ coincident D4-branes whose worldvolume is given by $\mathbb R^3 \times {{\bf S}^1_n} \times {\mathbb R}_t$, and a stack of $N$ coincident D6-branes whose worldvolume is given by ${{\bf S}^1_n} \times {\mathbb R}_t \times {\mathbb R}^5$; the two stacks intersect along ${{\bf S}^1_n} \times {\mathbb R}_t$ to form a D4-D6 I-brane system.

According to our analysis in the previous subsection, the free fermion bilinear currents along the I-brane in (\ref{equivalent IIA system Witten-partial}) ought to split into the factors $\frak{u}(1)^{(n)}_{\textrm {aff},kN} \times (\frak{u}(1)^{(n)}_{\textrm {aff},N})^{l-1} \times \frak{su}(k-l)^{(n)}_{\textrm {aff},N}  \times \frak{su}(N)^{(n)}_{\textrm {aff},k} \times \{ \frak{su}(k)^{(n)}_{\textrm {aff},N} / [(\frak{u}(1)^{(n)}_{\textrm {aff},N})^{l-1} \times \frak{su}(k-l)^{(n)}_{\textrm {aff},N}] \}$. As such, the system of $kN$ complex free fermions with central charge $kN$ will, in this case, give a realization of the total integrable module over the affine Lie algebra
\be
\frak{u}(1)^{(n)}_{\textrm {aff},kN} \otimes [\frak{u}(1)^{(n)}_{\textrm {aff},N}]^{l-1} \otimes \frak{su}(k-l)^{(n)}_{\textrm {aff},N}  \otimes \frak{su}(N)^{(n)}_{\textrm {aff},k} \otimes \left({ \frak{su}(k)^{(n)}_{\textrm {aff},N} \over [\frak{u}(1)^{(n)}_{\textrm {aff},N}]^{l-1} \otimes \frak{su}(k-l)^{(n)}_{\textrm {aff},N} }\right).
\label{affine embedding Witten-partial}
\ee
The total central charge is still $kN$ -- as argued in \emph{loc.~cit.}, the central charge does not change as we move along the Coulomb branch to separate the D-branes. Indeed, it is also invariant under the exchange $k \leftrightarrow N$.

Note at this juncture that from (\ref{coset}), we also have the following (conformal) equivalence of coset realizations:
\be
{\frak{su}(k)^{(n)}_{{\rm aff}, N}  \over{[\frak{u}(1)^{(n)}_{{\rm aff}, N}}]^{l-1}}  = {[\frak{su}(N)^{(n)}_{{\rm aff},1}]^{k} \otimes [\frak{u}(1)^{(n)}_{{\rm aff}, N}]^{k-l} \over {\frak{su}(N)^{(n)}_{{\rm aff}, k}}} ,
\ee
and
\be
{[\frak{su}(N)^{(n)}_{{\rm aff},1}]^{k-l} \over \frak{su}(k-l)^{(n)}_{{\rm aff}, N}} = {\frak{su}(N)^{(n)}_{{\rm aff}, k-l} \over [\frak{u}(1)^{(n)}_{\textrm {aff},N}]^{k-l-1} }.
\ee
Substituting this in (\ref{affine embedding Witten-partial}), we find that we effectively have the following total integrable module over the affine Lie algebra
\be
\frak{u}(1)^{(n)}_{\textrm {aff},kN} \otimes [\frak{u}(1)^{(n)}_{\textrm {aff},N}]^{l} \otimes \frak{su}(k-l)^{(n)}_{\textrm {aff},N}  \otimes \left([\frak{su}(N)^{(n)}_{\textrm {aff},1}]^l \otimes \frak{su}(N)^{(n)}_{{\rm aff}, k-l}\right)
\label{effective affine embedding Witten-partial}
\ee
of central charge $kN$. This means that the total Fock space ${\cal F}^{kN}$ of the $\it{uncoupled}$ $kN$ complex free fermions can be realized as
\be
{\cal F}^{\otimes kN} = \textrm{WZW}_{\widehat{u}(1)^{(n)}_{kN}} \otimes [\textrm{WZW}_{\widehat{u}(1)^{(n)}_{N}}]^{l} \otimes \textrm {WZW}_{\widehat{su}(k-l)^{(n)}_{N}}  \otimes \left( [\textrm{WZW}_{\widehat{su}(N)^{(n)}_1}]^l \otimes \textrm{WZW}_{\widehat{su}(N)^{(n)}_{k-l}} \right),
\label{Fock Witten-partial}
\ee
where $\textrm{WZW}_{\widehat{u}(1)^{(n)}_{kN}}$, $[\textrm{WZW}_{\widehat{u}(1)^{(n)}_{N}}]^{l}$, $\textrm {WZW}_{\widehat{su}(k-l)^{(n)}_{N}}$, $ [\textrm{WZW}_{\widehat{su}(N)^{(n)}_1}]^l$ and $\textrm{WZW}_{\widehat{su}(N)^{(n)}_{k-l}}$ are the irreducible integrable modules $\widehat{u}(1)^{(n)}_{kN}$, $[\widehat{u}(1)^{(n)}_{N}]^{l}$, $\widehat{su}(k-l)^{(n)}_{N}$, $[\widehat{su}(N)^{(n)}_1]^l$ and $\widehat{su}(N)^{(n)}_{k-l}$ over the corresponding affine Lie algebras that can be realized by the spectra of states of the corresponding $\it{chiral}$ WZW models. Consequently, the partition function of the uncoupled I-brane theory will be expressed in terms of the (product of) chiral characters of $\widehat{u}(1)^{(n)}_{kN}$, $\widehat{u}(1)^{(n)}_{N}$, $\widehat{su}(k-l)^{(n)}_{N}$, $\widehat{su}(N)^{(n)}_1$ and $\widehat{su}(N)^{(n)}_{k-l}$.

Next, we must couple the free fermions to the gauge fields which are dynamical. Since only $l$ out of the $k$ D4-branes are non-coincident, the free fermions will generically couple to the gauge group $U(1) \times U(1)^{l} \times SU(k-l)  \times SU(N)$, where the $U(1)^{l} \times SU(k-l)$ factor is associated with the $k$ D4-branes which are distributed as described. As explained in $\S$3.1, since the radius of the circle fiber of $TN^{R \to 0}_N$ goes to zero at infinity, the free fermions will couple dynamically to the $U(1)$ gauge field. In addition, because the geometry of $TN^{R \to 0}_N$ is fixed in our setup (recall that the center-of-mass degrees of freedom of the $N$ NS5-branes which give rise to the $TN^{R \to 0}_N$ geometry via steps (\ref{IIB 3}) and (\ref{IIA 4}), are frozen), in contrast to the gauge fields on the D4-branes, the $SU(N)$ gauge field on the $N$ D6-branes should $\it{not}$ be dynamical. Hence, we conclude that the free fermions couple dynamically only to the gauge group $U(1) \times U(1)^{l} \times SU(k-l)$. Schematically, this means that we are dealing with the following partially gauged CFT
\be
{\frak{u}(1)^{(n)}_{\textrm {aff},kN} \over \frak{u}(1)^{(n)}_{\textrm {aff},kN}} \otimes {[\frak{u}(1)^{(n)}_{\textrm {aff},N}]^{l} \over  [\frak{u}(1)^{(n)}_{\textrm {aff},N}]^{l}} \otimes {\frak{su}(k-l)^{(n)}_{\textrm {aff},N} \over {\frak{su}(k-l)^{(n)}_{\textrm {aff},N}}}  \otimes [\frak{su}(N)^{(n)}_{\textrm {aff},1}]^l \otimes \frak{su}(N)^{(n)}_{{\rm aff}, k-l}.
\ee
In particular, the $\frak{u}(1)^{(n)}_{{\rm aff}, kN}$ WZW model, the $l$ number of $\frak{u}(1)^{(n)}_{{\rm aff}, N}$ WZW models, and the  $\frak{su}(k-l)^{(n)}_{\textrm {aff},N}$ WZW model, will be replaced by the corresponding topological $G/G$ models. Consequently, all chiral characters except those of  $\widehat{su}(N)^{(n)}_1$ and $\widehat{su}(N)^{(n)}_{k-l}$ which appear in the overall partition function of the uncoupled free fermions system on the I-brane, will reduce to constant complex factors after coupling to the dynamical $U(1)$ and $U(1)^{l} \times SU(k-l)$ gauge fields. As such, modulo these constant complex factors which serve only to shift the energy levels of the ground states by numbers dependent on the highest affine weights of  $\widehat{u}(1)^{(n)}_{kN}$, $\widehat{u}(1)^{(n)}_N$ and $\widehat{su}(k-l)^{(n)}_{N}$, the $\it{effective}$ overall partition function of the I-brane theory will be expressed solely in terms of the product of $l$ chiral characters of  $\widehat{su}(N)^{(n)}_1$ and the chiral characters of $\widehat{su}(N)^{(n)}_{k-l}$;  that is, the sought-after spectrum of spacetime BPS states in the M-theory compactification (\ref{M-theory 7 Witten discussion-partial}) would be realized by $\textrm{WZW}_{\widehat{su}(N)^{(n)}_{k-l}} \otimes [\textrm{WZW}_{\widehat{su}(N)^{(n)}_1}]^l$.

\bigskip\noindent{\it A Geometric Langlands Duality for $\widetilde {\mathbb R^4 / \mathbb Z_{k-l, l}}$ for the $A_{N-1}$ Groups} 

Let us now consider $n=1$ whence there is no ``twist'' at all, i.e., $\widehat{su}(N)^{(n)}_{k_r}$ is simply $\widehat{su}(N)_{k_r}$, the integrable module over the untwisted affine Lie algebra ${\frak su}(N)_{{\rm aff}, k_r}$ of level $k_r$. Then, the physical duality of the M-theory compactifications (\ref{M-theory 1 Witten discussion-partial}) and (\ref{M-theory 7 Witten discussion-partial}) means that their respective spacetime BPS spectra ought to be equivalent, i.e., $\textrm{WZW}_{\widehat{su}(N)^{(n)}_{k-l}} \otimes [\textrm{WZW}_{\widehat{su}(N)^{(n)}_1}]^l$ ought to be equal to $\cal H_{\rm BPS}$ of (\ref{BPS-TNk-partial}). Indeed,  both $\cal H_{\rm BPS}$ and $\textrm{WZW}_{\widehat{su}(N)^{(n)}_{k-l}} \otimes [\textrm{WZW}_{\widehat{su}(N)^{(n)}_1}]^l$ are labeled by $k-l$ and $l$; moreover, $\frak{su}(N)_{\textrm{aff}} \cong \frak{su}(N)^\vee_{\textrm{aff}}$ whence we can identify $\widehat{su}(N)_{k_r}$ with the module ${^L\widehat{su}}(N)_{k_r}$ over  $\frak{su}(N)^\vee_{\textrm{aff}}$. The module $\textrm{WZW}_{^L\widehat{su}(N)_{k-l}}$ associated with the unique WZW model, is labeled by a dominant highest weight $\tilde\lambda_c$ of $SU(N)^\vee_{\rm aff}$ of level $k-l$ which we can naturally identify as $\lambda_c$ in $\bL = \lambda_c + \sum_{r =1}^l \lambda^{(r)}$ of (\ref{BPS-TNk-partial}).  The module $\textrm{WZW}_{^L\widehat{su}(N)_1}$ associated with one of the $l$ WZW models, is labeled by a dominant highest weight $\tilde\lambda^{(r)}$ of $SU(N)^\vee_{\rm aff}$ of level 1 which we can naturally identify as $\lambda^{(r)}$ in  $\bL = \lambda_c + \sum_{r =1}^l \lambda^{(r)}$ of (\ref{BPS-TNk-partial}). Thus, in any $\{\bL, \mu\}$-sector of the spectra of spacetime BPS states, we can write 
\be
 {\cal H}^{0, \bL, \mu}_{\rm BPS}= \left[\bigotimes_{r=0}^l \textrm{WZW}_{^L\widehat{su}(N)^{\lambda^{(r)}}_{k_r}} \right]_\mu,
\label{H=WZW-Witten-partial}
\ee
where $\lambda^{(0)} = \lambda_c$; $k_0 = k-l$; $k_r = 1$ for $r \geq 1$; and the subscript `$\mu$' just refers to the $\mu$-weight space of the spectrum of states of the total WZW model. 

As $\textrm{WZW}_{^L\widehat{su}(N)^{\lambda^{(r)}}_{k_r}}$ is furnished by ${^L\widehat{su}(N)^{\lambda^{(r)}}_{k_r}}$, via (\ref{BPS-TNk-partial}), we can also express (\ref{H=WZW-Witten-partial}) as 
\be
\boxed{{\rm IH}^\ast{\cal U}({\cal M}^{0, \bL}_{SU(N), \mu}(\widetilde {\mathbb R^4 / \mathbb Z_{k-l, l}})) = \left[\bigotimes_{r = 0}^{l} \, {^L\widehat{su}(N)^{\lambda^{(r)}}_{k_r, \, {{\vec p}_r}}} \right]_\mu}
\label{GL-relation A-Witten-partial}
\ee
where the label ${\vec p}_r$ can be interpreted as the position of the $r^{\rm th}$ center of $\widetilde {\mathbb R^4 / \mathbb Z_{k-l, l}}$ that the module is associated with, and $\vec p_0 = \vec p_c$. This is an  $\widetilde {\mathbb R^4 / \mathbb Z_{k-l, l}}$ quasi-singular generalization of~\cite[Conjecture 4.14(3)]{BF} for the simply-connected $SU(N) = A_{N-1}$ groups!

\bigskip\noindent{\it{A More General Statement}}

Note that the partition function of the spacetime BPS states realized by  $\textrm{WZW}_{^L\widehat{su}(N)_{k-l}} \otimes [\textrm{WZW}_{^L\widehat{su}(N)_1}]^l \simeq \textrm{WZW}_{\widehat{su}(N)_{k-l}} \otimes [\textrm{WZW}_{\widehat{su}(N)_1}]^l$ can be written in any $\bL$-sector  as~\cite{CFT text}
\begin{eqnarray}
\hspace{-1.0cm} Z_{\bL}  =  q^{\delta}    \left[\bigotimes_{r=0}^l \, \textrm{Tr}_{\lambda^{(r)}}  \, q^{L_0 - c'_r/24} \right]
 =  q^{\delta}  \left [\bigotimes_{r=0}^l\, \left(\sum_{\gamma} c^{\lambda^{(r)}}_{\gamma} \,  {\Theta}^{\textrm {level $k_r$}}_{\gamma}(q) \right) \right].
\label{partition function Witten 2-partial}
\end{eqnarray}
Here, $\delta$ represents the overall shift in the ground state energy level due to the aforementioned $G/G$ topological models; $L_{0} \in \mathbb Z_{\geq 0}$ is the general state energy level; $c^{\lambda^{(r)}}_{\gamma}$ and $\Theta^{\textrm{level $k_r$}}_{\gamma}$ are string-functions and theta-functions associated with the underlying dominant highest weight module over $\frak{su}(N)_{\textrm{aff}, k_r}$ of central charge $c'_r$; $\gamma$ is a  weight of $\frak{su}(N)_{\textrm{aff}, k_r}$; and in writing the first equality, we have set $\xi = \sum_{a} u_a J^a_0 = 0$ because the Coulomb moduli $u_a$ of the $N$ $\it{coincident}$ D6-branes ought to vanish as the corresponding $SU(N)$ gauge group is not broken down to its Cartan tori associated with the bilinear currents $J^a_0$.

Once again, the equivalence of the spacetime BPS spectra of the compactifications (\ref{M-theory 1 Witten discussion-partial}) and (\ref{M-theory 7 Witten discussion-partial}) implies that $Z_{SU(N), \bL}^{\rm BPS}$ of (\ref{partition function Witten SU(N)-partial}) ought to be equal to $Z_{\bL}$ of (\ref{partition function Witten 2-partial}), i.e., 
\be
\boxed{\bigotimes_{r=0}^l\, \left(\sum_{\gamma} c^{\lambda^{(r)}}_{\gamma} \,  {\Theta}^{\textrm {level $k_r$}}_{\gamma}(q) \right)  = \sum_{\bar \mu} \sum_{m' \geq 0}  \, {\rm dim} \, {\rm IH}^\ast{\cal U}({\cal M}^{0, \bL, m'}_{G, \bar \mu}(\widetilde {\mathbb R^4 / \mathbb Z_{k-l, l}})) \,  q^{m'   - c_\delta/ 24}}
\label{Witten's relation-partial}
\ee
where $G = A_{N-1}$ type; $\sum_{\gamma} c^{\lambda^{(r)}}_{\gamma} \,  {\Theta}^{\textrm {level $k_r$}}_{\gamma}(q) $ is the character of the integrable representation (associated with $\lambda^{(r)}$) of the loop group ${\cal L}G$ at level $k_r$; $m' = m + {m_{\bL}}$, where $m/k \in \mathbb Z_{\geq 0}$ and $m_{\bL}$ is as given in (\ref{m-Witten-partial}); and $c_\delta/  24 = (\sum^l_{r=0} c'_r) /24 =   \delta$.

\bigskip\noindent{\it A Geometric Langlands Duality for $\widetilde {\mathbb R^4 / \mathbb Z_{k-l, l}}$ for the $B_{N/2}$ Groups}

Let us now restrict ourselves to \emph{even} $N$, and consider $n=2$ whence there is a ``$\mathbb Z_2$-twist'', i.e., the relevant module is $\widehat{su}(N)^{(2)}_{k_r}$, the integrable module over the $\mathbb Z_2$-twisted affine Lie algebra ${\frak su}(N)^{(2)}_{{\rm aff}, 1}$ of level $k_r$. Let  $\{ \lambda'_c, \lambda^{(1)'}, \dots, \lambda^{(l)'} \}$ be a set of dominant highest weights of this module; $\bL' = \lambda'_c + \sum_{r=1}^l \lambda^{(r)'} = (k, \bar {\bL}', 0)$; and $\bar {\bL}' =  {\bar\lambda}'_c +  \sum_{r=1}^l \bar{\lambda}^{(l)'}$. Then, by repeating the arguments that led us to (\ref{WZW space-SO(N+1)}), bearing in mind that we now have $\textrm{WZW}_{\widehat{su}(N)^{(2)}_{k-l}} \otimes [\textrm{WZW}_{\widehat{su}(N)^{(2)}_1}]^l$ instead of $\textrm{WZW}_{\widehat{su}(N)^{(2)}_k}$, we find that we can write
 \be
\textrm{WZW}_{\widehat{su}(N)^{(2)}_{k-l}} \otimes [\textrm{WZW}_{\widehat{su}(N)^{(2)}_1}]^l =  \bigoplus_{w'_2} \bigoplus_{\bL'}  \bigoplus_{\nu =0,1} \, \left[\bigotimes_{r=0}^l\overline{\textrm{WZW}}_{\widehat{su}(N)^{(2), \, \lambda^{(r)'}}_{k_r, \nu}}\right]_{w'_2},
 \label{WZW space-SO(N+1)-Witten-partial}
 \ee
 where $\lambda^{(0)'} = \lambda'_c$, $k_0 = k-l$, $k_r = 1$ for $r \geq 1$, and
\be
w'_2 = \sum_{b=1}^{l} v_b {\bar\alpha}_b.
\label{xi-SO(N+1)-Witten-partial}
\ee
Here, the $l$ numbers $v_b$ correspond to the nonvanishing Coulomb moduli of the $l$ \emph{fully separated} D4-branes  whose magnitudes correlate with the sizes of the $l$ two-spheres in $\widetilde {\mathbb R^4 / \mathbb Z_{k-l, l}}$; $\bar\alpha_b = \bar \alpha$, where $ \bar \alpha$ is the finite part of the dominant highest affine weight $\alpha$ that labels a dominant highest weight module of a chiral $\frak{u}(1)_{\rm aff}$ WZW model; the overhead bar means that we project onto $\mathbb Z_2$-invariant states (as required of twisted CFT's); $\nu =0$ or $1$ indicates that the sector is untwisted or twisted, respectively; $\widehat{su}(N)^{(2), \, \lambda^{(r)'}}_{k_r, \nu}$ is a dominant highest weight module labeled by the dominant highest weight $\lambda^{(r)'}$, and whose general state energy level is $h_\nu \in \mathbb Z_{\geq 0} + {\nu \over 2}$; and the subscript `$w'_2$' means that the \emph{overall} ground state energy level is further shifted by  $ w'_2 / \tau$. 

Looking at the RHS of (\ref{xi-SO(N+1)-Witten-partial}), we see that we can interpret $w'_2$ as a linear sum of $\mathbb Z_2$-valued non-abelian magnetic fluxes through the $l$ two-spheres in $\widetilde {\mathbb R^4 / \mathbb Z_{k-l, l}}$: starting at the origin, the D4-branes can move either in the positive or negative direction whence the $v_i$'s can take either positive or negative values, and by a natural identification of the $\bar \alpha$'s as the standard area of the $l$ two-spheres which define a basis of $H_2(\widetilde {\mathbb R^4 / \mathbb Z_{k-l, l}}, \mathbb Z)$, we have our claim. A somewhat related analysis has also been carried out in~\cite[$\S$2.4]{Vafa et al}, where it was also shown that $w'_2$ can indeed be associated with fluxes through the $l$ two-spheres of  $\widetilde {\mathbb R^4 / \mathbb Z_{k-l, l}}$. Thus, let us henceforth identify $w'_2$ as $w_2$ of (\ref{BPS-TNk-partial}).

Now the physical duality of the M-theory compactifications (\ref{M-theory 1 Witten discussion-partial}) and (\ref{M-theory 7 Witten discussion-partial}) means that their respective spacetime BPS spectra ought to be equivalent, i.e., $\textrm{WZW}_{\widehat{su}(N)^{(2)}_{k-l}} \otimes [\textrm{WZW}_{\widehat{su}(N)^{(2)}_1}]^l$  ought to be equal to $\cal H^{\rm eff}_{\rm BPS}$ of (\ref{HBPS-eff-Witten-partial}). Indeed,  both $\cal H^{\rm eff}_{\rm BPS}$ and $\textrm{WZW}_{\widehat{su}(N)^{(2)}_{k-l}} \otimes [\textrm{WZW}_{\widehat{su}(N)^{(2)}_1}]^l$  are labeled by $k-l$ and $l$; moreover, it is clear that one can identify $\bL'$ of (\ref{WZW space-SO(N+1)-Witten-partial}) with $\bL$ of (\ref{HBPS-eff-Witten-partial}); it is also clear that as in $\S$3.1, one can identify $h_\nu$ (which is implicit in (\ref{WZW space-SO(N+1)-Witten-partial}))  with $j_\nu$ (which is implicit in (\ref{HBPS-eff-Witten-partial})).  Hence, in any $(w_2, \bL, \nu, \mu_\nu)$-sector of the spectra of spacetime BPS states, we can  write 
\be
 {\overline{\cal H}}^{w_2, \bL, \mu_\nu}_{\rm BPS}   =  \left[\bigotimes_{r=0}^l\overline{\textrm{WZW}}_{\widehat{su}(N)^{(2), \, \lambda^{(r)}}_{k_r, \nu}}\right]_{w_2, \, \mu_\nu},
\label{H=WZW-SO(N+1)-Witten-partial}
\ee
where $\lambda^{(0)} = \lambda_c$, $k_0 = k-l$, $k_r = 1$ for $r \geq 1$, and the subscript `$\mu_\nu$' just refers to the $\mu_\nu$-weight space of the spectrum of states of the total WZW model in the $\nu$-sector. As $\overline{\textrm{WZW}}_{\widehat{su}(N)^{(2), \, \lambda^{(r)}}_{k_r, \nu}}$ is furnished by the $\mathbb Z_2$-invariant projection ${\widehat{su}(N)^{(2), \, \lambda^{(r)}}_{k_r, \nu}}\vert_{\mathscr P_2}$ of $\widehat{su}(N)^{(2), \, \lambda^{(r)}}_{k_r, \nu}$, and since $\frak{su}(N)^{(2)}_{\textrm{aff}} \simeq \frak{so}(N +1)^\vee_{\textrm{aff}} $ whence ${\widehat{su}(N)^{(2),  \, \lambda^{(r)}}_{k_r, \nu}}\vert_{\mathscr P_2}$ is isomorphic to the submodule ${^L\widehat{so}(N+1)^{\lambda^{(r)}}_{k_r, \nu}}$ over $\frak{so}(N+1)^\vee_{\textrm{aff}}$, via (\ref{HBPS-eff-Witten-partial}), we can also express (\ref{H=WZW-SO(N+1)-Witten-partial}) as 
\be
\boxed{\overline{{\rm IH}^{\ast, \nu}{\cal U}}({\cal M}^{w_2, \bL}_{SO(N+1),  \mu_\nu}(\widetilde {\mathbb R^4 / \mathbb Z_{k-l, l}})) =   \left[ \bigotimes_{r=0}^l \, {^L\widehat{so}}(N+1)^{\lambda^{(r)}}_{k_r, \, \nu, \, {{\vec p}_r}} \right]_{w_2, \, \mu_\nu}}
\label{GL-relation B-Witten-partial}
\ee
for $\nu = 0$ and $1$, where the label ${\vec p}_r$ can be interpreted as the position of the $r^{\rm th}$ center of $\widetilde {\mathbb R^4 / \mathbb Z_{k-l, l}}$ that the module is associated with, and $\vec p_0 = \vec p_c$. This is an $\widetilde {\mathbb R^4 / \mathbb Z_{k-l, l}}$ non-singular generalization of~\cite[Conjecture 4.14(3)]{BF} for the nonsimply-connected $SO(N +1) = B_{N/2}$ groups!

\bigskip\noindent{\it A More General Statement}

Note that the partition function of the spacetime BPS states realized by  $\overline{\textrm{WZW}}_{\widehat{su}(N)^{(2)}_{k-l}} \otimes [\overline{\textrm{WZW}}_{\widehat{su}(N)^{(2)}_1}]^l \simeq {\textrm{WZW}}_{^L\widehat{so}(N+1)_{k-l}} \otimes[{\textrm{WZW}}_{^L\widehat{so}(N+1)_1}]^l$ can be written in any $(w_2, \bL, \nu)$-sector as
\begin{eqnarray}
\hspace{-0.7cm} Z_{w_2, \bL, \nu}  =  q^{\delta + \tilde w_2}    \left[\bigotimes_{r=0}^l \, \textrm{Tr}_{\lambda^{(r)}}  \,  q^{L_{0, \nu} - c'_r/24} \right]
 = q^{\delta + \tilde w_2}  \left [\bigotimes_{r=0}^l\, \left( \sum_{\gamma} c^{\lambda^{(r)}}_{\gamma, \nu} \,  {^L\Theta}^{\textrm {level $k_r$}}_{\gamma, \nu}(q) \right) \right].
\label{partition function Witten 3-partial}
\end{eqnarray}
Here, $\delta + {\tilde w}_2$ represents the overall shift in the ground state energy level due to the aforementioned $G/G$ topological models, where $\tilde w_2 = {w_2 / \tau}$; $L_{0, \nu} \in \mathbb Z_{\geq 0} + {\nu \over 2}$ is the general state energy level; $c^{\lambda^{(r)}}_{\gamma, \nu}$ and $^L\Theta^{\textrm{level $k_r$}}_{\gamma, \nu}$ are string-functions and theta-functions associated with the $\nu$-sector of the underlying dominant highest weight module over $\frak{so}(N +1)^\vee_{\textrm{aff}, k_r}$ of central charge $c'_r$; and $\gamma$ is a  weight of $\frak{so}(N +1)^\vee_{\textrm{aff}, k_r}$.

Once again, the equivalence of the spacetime BPS spectra of the compactifications (\ref{M-theory 1 Witten discussion-partial}) and (\ref{M-theory 7 Witten discussion-partial}) implies that $Z_{SO(N+1), \bL}^{\rm BPS}$ of (\ref{Z-SO(N+1)-1-Witten-partial}) in the $(w_2, \nu)$-sector ought to be equal to $ Z_{w_2, \bL, \nu}$ of (\ref{partition function Witten 3-partial}), i.e., 
\be
\boxed{ \bigotimes_{r=0}^l\, \left( \sum_{\gamma} c^{\lambda^{(r)}}_{\gamma, \nu} \,  {^L\Theta}^{\textrm {level $k_r$}}_{\gamma, \nu}(q) \right)  =    \sum_{\bar\mu_\nu}  \sum_{m'_\nu \geq 0}   {\rm dim} \, \overline {{\rm IH}^{\ast, \nu}{\cal U}}({\cal M}^{\tilde w_2, \bL, m'_\nu}_{SO(N+1), \bar\mu_\nu}(\widetilde {\mathbb R^4 / \mathbb Z_{k-l, l}})) 
  q^{m'_\nu - {\tilde w}_2  - c_\delta/ 24}}
\label{Witten's relation SO(N+1)-partial}
\ee
where $G = B_{N/2}$ type; $\sum_{\gamma} c^{\lambda^{(r)}}_{\gamma, \nu} \,  {^L\Theta}^{\textrm {level $k_r$}}_{\gamma, \nu}(q)$ is the character in the $\nu$-sector of the integrable representation (associated with ${\lambda^{(r)}}$) of the \emph{Langlands dual} loop group ${\cal L}G^\vee$ at level $k_r$; $m'_\nu = m_\nu + m_{\bL}$, where $m_\nu/k n' \in  \mathbb Z_{\geq 0} + {\nu \over 2}$ and $m_{\bL}$ is as given in (\ref{m-Witten-partial}); $n' = 1$ or $2$ if $N = 2$ or $N > 2$, respectively; $c_\delta/  24 = (\sum^l_{r=0} c'_r) /24 =   \delta$.

\bigskip\noindent{\it Blowing Down to the Fully-Singular Case of $\S$3.1}

Now let $l = 0$ so that all $k$ centers are coincident at $\vec p_c = \vec p_0$ whence $\widetilde {\mathbb R^4 / \mathbb Z_{k-l, l}}$ blows down to the fully-singular $\mathbb R^4 / \mathbb Z_k$ manifold considered in $\S$3.1. In this instance, $k_0 = k$, and  there are no two-spheres for the non-abelian magnetic fluxes to pass through, i.e., $w_2 = 0$. As before, to a flat connection at infinity, we can associate a dominant coweight $\mu = (k, \bar \mu, j)$ of $G_{\textrm{aff}}$ of level $k$, where $j$ is a number. Also, since all $k$ centers are coincident at a single point $\vec p_0$, we have a $\mathbb Z_{k}$-action in the fiber of the $G$-bundle only over  ${\vec p}_{0}$; in other words, associated to this sole $\mathbb Z_{k}$-action is $\bL = \lambda =  (k, \bar \lambda, 0)$, a dominant highest coweight of $G_{\textrm{aff}}$ of level $k$.   It is then clear that all of our above formulas for $\widetilde {\mathbb R^4 / \mathbb Z_{k-l, l}}$ indeed reduce to their $\mathbb R^4 / \mathbb Z_k$-counterpart  in $\S$3.1, as expected.  In particular, (\ref{GL-relation A-Witten-partial}), (\ref{Witten's relation-partial}), (\ref{GL-relation B-Witten-partial}) and (\ref{Witten's relation SO(N+1)-partial}), will reduce to (\ref{GL-relation A}), (\ref{chiral WZW SU(N)}), (\ref{GL-relation B}) and (\ref{chiral WZW SO(N+1)}), respectively. This serves as a consistency check of our results herein.

\bigskip\noindent{\it Blowing Up to the Non-Singular Case of $\S$4.1}

Now let $l = k-1$ so that all $k$ centers are non-coincident at positions $\vec p_0, \dots, \vec p_{k-1}$ whence $\widetilde {\mathbb R^4 / \mathbb Z_{k-l, l}}$ blows up to the non-singular $\widetilde {\mathbb R^4 / \mathbb Z_k}$ manifold considered in $\S$4.1. In this instance, $k_r = 1$, where $ 0 \leq r \leq k-1$; also, there are $k-1$ two-spheres for the non-abelian magnetic fluxes to pass through, i.e., $w_2 \neq 0$,  and it is as given in (\ref{xi-SO(N+1)-Witten-partial}). As before, to a flat connection at infinity, we can associate a dominant coweight $\mu = (k, \bar \mu, j)$ of $G_{\textrm{aff}}$ of level $k$, where $j$ is a number. In addition, since all $k$ centers are non-coincident with multiplicity 1 each, we have a $\mathbb Z_{1}$-action in the fiber of the $G$-bundle over the positions $\vec p_0, \dots, \vec p_{k-1}$; in other words,  associated to these $\mathbb Z_{1}$-actions is $\bL = \sum_{r=0}^{k-1} \, \lambda^{(r)} =  (k, \bar {\bL}, 0)$, a dominant highest coweight of $G_{\textrm{aff}}$ of level $k$, where $\lambda^{(r)} =  (1, \bar \lambda^{(r)}, 0)$ is a dominant highest coweight of $G_{\textrm{aff}}$ of level $1$ (associated with the underlying $\mathbb Z_{1}$-action), and $\bar {\bL} = \sum_{r=0}^{k-1} \, \bar {\lambda}^{(r)}$. Last but not least, since $\widetilde {\mathbb R^4 / \mathbb Z_k}$ is a non-singular manifold, intersection cohomology ought to be replaced by middle-dimensional cohomology throughout. It is then clear -- after noting that (i) the index `$i$' in $\S$4.1 is such that $ i = r+1$; (ii) there is only \emph{one} string-function $c^{\lambda^{(r)}}_{\lambda^{(r)}} = \eta(q)^{-N+1}$ which can be associated with $\widehat{su}(N)_1$ -- that all of our above formulas for $\widetilde {\mathbb R^4 / \mathbb Z_{k-l, l}}$ indeed reduce to their $\widetilde {\mathbb R^4 / \mathbb Z_k}$-counterpart  in $\S$4.1, as expected.  In particular, (\ref{GL-relation A-Witten-partial})), (\ref{Witten's relation-partial}), (\ref{GL-relation B-Witten-partial}) and (\ref{Witten's relation SO(N+1)-partial}), will reduce to  (\ref{GL-relation A-Witten}), (\ref{Witten's relation}), (\ref{GL-relation B-Witten}) and (\ref{Witten's relation SO(N+1)}), respectively. This serves as a yet another consistency check of our results herein.

\newsubsection{4d Worldvolume Defects and a ``Ramified'' Geometric Langlands Duality for Surfaces}

In this final subsection, we will derive a ``ramified'' version of the geometric Langlands duality for surfaces demonstrated in $\S$3.1--$\S$3.2. The ``ramification'' in our sense, is effected by the presence of the 4d worldvolume defect of the kind studied in~\cite{TD} and described in $\S$2.3. Let us now try to understand, note, discuss and describe a few essential things before we proceed to derive our main result.

\bigskip\noindent{\it On the Appearance of the Langlands Dual Affine Lie Algebra in $\S$3}

First, let us try to understand, from a hitherto unconsidered perspective, why (the representations of) the \emph{Langlands dual} affine Lie algebras appear on the RHS of the ``unramified'' duality relations (\ref{GL-relation A}), (\ref{GL-relation B}), (\ref{GL-relation D}), (\ref{GL-relation C}) and (\ref{GL-relation G}) for the $A$, $B$, $C$, $D$ and $G$ groups.  

To this end, recall that we could express the RHS of the duality relations  for the simply-laced $A$ and $D$ groups   in terms of (the representations of) the Langlands dual affine Lie algebra $\frak g^\vee_{\rm aff}$,  because $\frak g_{\rm aff} \simeq \frak g^\vee_{\rm aff}$ for simply-laced Lie algebras $\frak g$.  Also recall that we could express the RHS of the duality relations  for the $B$, $C$ and $G$ groups  in terms of (the representations of) the Langlands dual affine Lie algebra $\frak g^\vee_{\rm aff}$,  because $\frak {su}(N)^{(2)}_{\rm aff} \simeq \frak {so}(N+1)^\vee_{\rm aff}$ (where $N$ is even),  $\frak {so}(2N)^{(2)}_{\rm aff} \simeq \frak {usp}(2N - 2)^\vee_{\rm aff}$ and $\frak {so}(8)^{(3)}_{\rm aff} \simeq \frak {g}^\vee_{2 \, \rm aff}$. 

Interestingly, the appearance of (the representations of) $\frak g^\vee_{\rm aff}$ can also be understood without appealing to the above-stated isomorphism of twisted affine Lie algebras, as follows.  Consider the dual M-theory compactification (\ref{M-theory 7 discussion}) relevant to the RHS of the duality relations for the $A$ and $B$ groups; mapping this to a strongly-coupled type IIA compactification, geometric engineering and the $\mathbb Z_n$-outer-automorphism of $TN_N^{R\to 0}$ (as we go around the ${\bf S}^1_n$ circle) then tell us that the Lie algebra underlying the enhanced gauge symmetry of the 6d supergravity theory along $\mathbb R^4 \times  \mathbb R_t  \times {\bf S}^1_n$ ought to be ${\frak {su}(N)}^\vee$ (for any $N$) or $\frak{usp}(N) = {\frak {so}(N+1)}^\vee$ (for even $N$) when $n=1$ or $2$, respectively~\cite{vafa}.  Now consider  the dual M-theory compactification (\ref{M-theory 7 OM5 discussion}) relevant to the RHS of the duality relations for the $C$, $D$ and $G$ groups; mapping this to a strongly-coupled type IIA compactification, geometric engineering and the $\mathbb Z_n$-outer-automorphism of $SN_N^{R\to 0}$ (as we go around the ${\bf S}^1_n$ circle) then tell us that the Lie algebra underlying the enhanced gauge symmetry of the 6d supergravity theory along $\mathbb R^4 \times  \mathbb R_t  \times {\bf S}^1_n$ ought to be ${\frak {so}(2N)}^\vee$, $\frak {so}(2N-1) = {\frak {usp}(2N-2)}^\vee$, or  $\frak {g}^\vee_2$ (if $N=4$), when $n=1$, $2$, or $3$, respectively~\cite{vafa}. In sum, this means that in the type IIA limit of the dual M-theory compactifications  (\ref{M-theory 7 discussion}) and  (\ref{M-theory 7 OM5 discussion}), the symmetries of the 2d theory along ${\bf S}^1_n \times \mathbb R_t$ ought to be rooted in $\frak g^\vee$; in particular, we ought to have,  along ${\bf S}^1_n \times \mathbb R_t$,  a  chiral WZW theory with $\frak g^\vee_{\rm aff}$-symmetry. In other words, the appearance of (the representations of) $\frak g^\vee_{\rm aff}$ on the RHS of the aforementioned ``unramified'' duality relations, can also be understood to be a consequence of a (dual) compactification on a ``$\mathbb Z_n$-twisted'' $TN_N^{R\to 0}$  or $SN_N^{R\to 0}$ manifold.

\bigskip\noindent{\it The Characterization of the 4d Worldvolume Defect by Nilpotent Orbits}

Second, note that the 4d worldvolume defect is characterized by a homomorphism $\rho: \frak {sl}(2)\to \frak g_{\mathbb C}$~\cite{TD}, where $ \frak g_{\mathbb C}$ is the complexification of $\frak g$. In turn, via the Jacobson-Morozov theorem -- which states that the classification of such homomorphisms $\rho$ (up to conjugacy) is equivalent to the classification of nilpotent elements $e$ in $\frak g_{\mathbb C}$ (also up to conjugacy) through the correspondence $e=\rho(\sigma^+)$, where $\rho(\sigma^+) \in \frak {sl}(2) \subset \frak g_{\mathbb C}$ -- the 4d worldvolume defect would be characterized by nilpotent orbits $O_e$ of $\frak g_\mathbb C$.\footnote{The (adjoint) orbit for an element $e \in \frak g_{\mathbb C}$ is the set of elements in $\frak g_{\mathbb C}$ that are $G_{\mathbb C}$-conjugate to $e$, i.e., are of the form, $\mathrm{ad}(g)\cdot e$ for some $g$ in $G_{\mathbb C}$. We denote the orbit containing $e$ by $O_e = G_{\mathbb C} \cdot e$. See~\cite{Mcgovern} for more details.} 

When $ \frak g_{\mathbb C} =\frak {sl}(N)$, a nilpotent orbit $O_e$ can also be labeled by a partition $p = [n_1, \dots, n_M] $ of $N$, where $N = n_1 + n_2 + \dots + n_M$. This just reflects the fact pointed out in $\S$2.3, that the 4d worldvolume defect can be (i) labeled by a partition $p$ of $N$ when $n=1$ (i.e., when $\frak g_\mathbb C = \frak{sl}(N)$); (ii) called a defect of type $\mathbb L$, where $\mathbb L$ -- which can be related to $p$ -- is the effective gauge group of the underlying $SU(N)$ theory when restricted along the defect.   

When $\frak g_\mathbb C \neq \frak {sl}(N)$, nilpotent orbits can likewise be labeled by some partition $p'$. However, it is presently unclear how one can, in this case, relate $p'$ to $\mathbb L$. That said, we expect such a relation to exist -- presumably via the concept of orbit induction -- although we would not need to appeal to it in this paper.

Also, in the theory of nilpotent orbits, there is a Spaltenstein map~\cite{Mcgovern}
\be
d:\{\textrm{nilpotent orbits of }\frak g_\mathbb C \} \to \{\textrm{nilpotent orbits of }{}{\frak g^\vee_\mathbb C}\}.
\ee
In particular, we have $d(O_e) = O_{e^\vee}$, where $O_{e^\vee}$ -- which is a nilpotent orbit of ${\frak g^\vee_\mathbb C}$ -- is labeled by a \emph{dual} partition $p^\vee$.  For $\frak g_\mathbb C = \frak {sl}(N)$, $p^\vee = p^t$, where $p^t$ is the transpose of $p$.  

 \bigskip\noindent{\it The Dual 4d Worldvolume Defect}

Third, let us discuss the properties of the \emph{dual }4d worldvolume defect in the dual M-theory compactifications (\ref{M-theory defect dual}) and (\ref{OM-theory 8 defect}).  As the defect spans $\mathbb R_t \times {{\bf S}^1_n} $ and two other dimensions in $TN_N^{R\to 0}$ and $SN_N^{R\to 0}$, it would also be ``$\mathbb Z_n$-twisted'' in the directions along $TN_N^{R\to 0}$ and $SN_N^{R\to 0}$. From the perspective of the 4d maximally supersymmetric Yang-Mills theory along $M_4$, where $M_4$ is either $TN_N^{R\to 0}$ or $SN_N^{R\to 0}$, the ``$\mathbb Z_n$-twist'' of both the 2d defect and $M_4$ would mean that the gauge group -- which is $SU(k)$ (for any $k$) or $SO(k+1)$ (for even $k$) when $n=1$ or $2$, respectively -- would reduce, along the 2d defect, to a Levi subgroup thereof. Consequently, from the viewpoint of the directions in $M_4$ which are transverse to the defect, the defect would be characterized by nilpotent orbits of $\frak {sl}(k)$.  

On the other hand, according to our earlier explanations, in the type IIA limit of the dual M-theory compactifications (\ref{M-theory defect dual}) and (\ref{OM-theory 8 defect}), it is $\frak g^\vee$ which underlies the enhanced gauge symmetry of the 6d supergravity theory along $\mathbb R^4 \times  \mathbb R_t  \times {\bf S}^1_n$. Consequently, from the viewpoint of the $\mathbb R^4$-space transverse to the defect, the defect would be characterized by \emph{dual} nilpotent orbits $O_{e^\vee}$.     

\bigskip\noindent{\it The Moduli Space of ``Ramified'' $G$-Instantons on $\mathbb R^4 / \mathbb Z_k$}

Last but not least, let us describe the moduli space of ``ramified'' $G$-instantons on $\mathbb R^4 / \mathbb Z_k$, where $G$ is the Lie group corresponding to $\frak g$. To this end, note that according to~\cite[$\S$4.4]{BF}, the moduli space ${\cal M}^{a, \rho_0}_{G, \rho_\infty}(\mathbb R^4/ \mathbb Z_k)$ of ``unramified'' $G$-instantons on  $\mathbb R^4 / \mathbb Z_k$ which appears throughout $\S$3, can also be interpreted as the $\mathbb Z_k$-invariant part ${\cal M}^{a, \rho_0}_{G, \rho_\infty}(\mathbb R^4)^{\mathbb Z_k}$ of the moduli space ${\cal M}^a_{G}(\mathbb R^4)$ of ``unramified'' $G$-instantons on  $\mathbb R^4$ determined by $\{\rho_0, \rho_\infty \}$. Here, the positive number $a$ is the instanton number; the homomorphism $\rho_\infty: \mathbb Z_k \to G$ determines -- via the $G$-action on ${\cal M}^a_{G}(\mathbb R^4)$ -- the underlying $\mathbb Z_k$-action on ${\cal M}^a_{G}(\mathbb R^4)$;  the homomorphism $\rho_0: \mathbb Z_k \to G$ determines the $\mathbb Z_k$-action in the fibers of the underlying $\mathbb Z_k$-equivariant $G$-bundles on $\mathbb R^4$ at the origin.\footnote{Recall that a $G$-bundle on $\mathbb R^4/\mathbb Z_k$ is the same as $\cal F$ -- a $\mathbb Z_k$-equivariant $G$-bundle on $\mathbb R^4$, where ${\cal F} \in {\cal M}^a_{G, \rho_\infty}(\mathbb R^4)^{\mathbb Z_k}$. Since the origin 0 is a fixed point of the $\mathbb Z_k$-action, it follows that the $\mathbb Z_k$-action acts in the fiber of $\cal F$ at $0 \in \mathbb R^4$.}

Now introduce a ``ramification'' along the $z$-plane in $\mathbb R^4 / \mathbb Z_k \simeq \mathbb C_z / \mathbb Z_k \times \mathbb C_w / \mathbb Z_k$;\footnote{The simultaneous $\mathbb Z_k$-action on the $z$- and $w$-plane is described below (\ref{table}).} in other words, let the gauge group $G$ reduce to a Levi subgroup $\mathbb L$ along the plane $\mathbb C_z$.  Then, according to the previous paragraph, the moduli space ${\cal M}^{a', \rho'_0}_{G, \mathbb L, \rho'_\infty}(\mathbb R^4/ \mathbb Z_k)$ of ``ramified'' $G$-instantons on  $\mathbb R^4 / \mathbb Z_k$, can also be interpreted as the $\mathbb Z_k$-invariant part ${\cal M}^{a', \rho'_0}_{G, \mathbb L, \rho'_\infty}(\mathbb R^4)^{\mathbb Z_k}$ of the moduli space ${\cal M}^{a'}_{G, \mathbb L}(\mathbb R^4)$ of ``ramified'' $G$-instantons on  $\mathbb R^4$ determined by $\{\rho'_0, \rho'_\infty \}$. Here, the positive number $a' = a + {\rm Tr} \, \alpha\frak m$ is the ``ramified'' instanton number, where Tr is a quadratic form on $\frak g$~\cite{GW}; $\alpha \in \frak t$ is the holonomy parameter that is the commutant of $\mathbb L$, where $\frak t$ is the Lie algebra of the Cartan subgroup $\mathbb T \subset G$; ${\frak m} \in \Lambda_{\rm cochar}$ is the ``magnetic charge'', where $ \Lambda_{\rm cochar}$ is the cocharacter lattice of $G$; the homomorphism $\rho'_\infty: \mathbb Z_k \to \mathbb L$ determines -- via the $\mathbb L$-action on ${\cal M}^{a'}_{G, \mathbb L}(\mathbb R^4)$~\cite{B-SO} -- the underlying $\mathbb Z_k$-action on ${\cal M}^{a'}_{G, \mathbb L}(\mathbb R^4)$, as well as the flat gauge fields at infinity which ensure the finiteness of the instanton action; the homomorphism $\rho'_0: \mathbb Z_k \to \mathbb L$  determines the $\mathbb Z_k$-action in the fibers of the underlying $\mathbb Z_k$-equivariant ``ramified'' $G$-bundles on $\mathbb R^4$ at the origin (where $G$ reduces to $\mathbb L$).\footnote{Note that a ``ramified'' $G$-bundle on $\mathbb R^4/\mathbb Z_k$ is the same as $\cal F'$ -- a $\mathbb Z_k$-equivariant ``ramified'' $G$-bundle on $\mathbb R^4$, where ${\cal F'} \in {\cal M}^{a'}_{G, \mathbb L, \rho'_\infty}(\mathbb R^4)^{\mathbb Z_k}$. Since the origin 0 is a fixed point of the $\mathbb Z_k$-action, it follows that the $\mathbb Z_k$-action acts in the fiber of $\cal F'$ at $0 \in \mathbb R^4$.}

\bigskip\noindent{\it  A ``Ramified'' Geometric Langlands Duality for Surfaces for the $A$--$B$ Groups}

Armed with the above facts and observations, we are now ready to derive our main result for the $A$--$B$ groups. As the follow-on analysis is largely similar to that in $\S$3.1, we shall be brief in our exposition.

Recall from (\ref{M-theory defect}) and (\ref{M-theory defect dual}) that the six-dimensional M-theory compactification 
\be
\textrm{M-theory}: \quad \mathbb R^{5}  \times  \underbrace{\mathbb R_t \times {{\bf S}^1_n} \times \mathbb R^4 / \mathbb Z_k}_{\textrm{$N$ M5-branes with a 4d defect}},
\label{M-theory defect-AB}
\ee
where the 4d worldvolume defect wraps $\mathbb R_t \times {\bf S}^1_n$ and the $z$-plane in $\mathbb R^4 / \mathbb Z_k \simeq \mathbb C_z / \mathbb Z_k \times \mathbb C_w / \mathbb Z_k$, is\emph{ physically dual} to the following six-dimensional M-theory compactification 
\be
\textrm{M-theory}: \quad \underbrace{TN_N^{R\to 0}  \times  {{\bf S}^1_n} \times \mathbb R_t}_{\textrm{$k$ M5-branes with a 4d defect}}  \times  {\mathbb R^{5}},
\label{M-theory defect dual-AB}
\ee
where the 4d worldvolume defect wraps ${\bf S}^1_n \times \mathbb R_t$, the ${\bf S}^1$-fiber of $TN_N^{R\to 0}$, and a single direction along the $\mathbb R^3$ base of $TN_N^{R\to 0}$. 

According to $\S$3.1, the spacetime BPS states in (\ref{M-theory defect-AB}) and (\ref{M-theory defect dual-AB}) -- which are annihilated by eight of the sixteen supercharges of the 6d ${\cal N} = (1, 1)$ supersymmetry algebra of the underlying Yang-Mills theories along $\mathbb R^{5}  \times  \mathbb R_t$ -- ought to have equivalent spectra. As in $\S$3.1, it is through this equivalence of spectra that we will be able to derive our main result. As such, let us first ascertain the spacetime BPS states in (\ref{M-theory defect-AB}).

Via the arguments which led us to (\ref{BPS-M}), we find that for the $SU(N) = A_{N-1}$ groups, the Hlibert space ${\cal H}^{\mathbb L_A}_{{\rm BPS}}$ of spacetime BPS states in (\ref{M-theory defect-AB}) ought to be given by
\be
{\cal H}^{{\mathbb L}_A}_{{\rm BPS}} = \bigoplus_{a', \rho'_0, \rho'_\infty}~{\rm IH}^\ast {\cal U}({\cal M}^{a', \rho'_0}_{SU(N), \mathbb L_A, \rho'_\infty}(\mathbb R^4)^{\mathbb Z_k}),
\label{BPS-M-ram-A}
\ee 
where ${\rm IH}^\ast {\cal U}(\cal M)$ is the intersection cohomology of the Uhlenbeck compactification of $\cal M$, and $\mathbb L_A \subset SU(N)$ is a Levi subgroup determined by the defect.

Via the arguments which led us to (\ref{HBPS-eff}), we find that for the $ SO(N +1) = B_{N/2}$ groups (where $N$ is even), the Hlibert space ${\cal H}^{\mathbb L_B}_{{\rm BPS}}$ of spacetime BPS states in (\ref{M-theory defect-AB}) ought to be given by
\be
{\cal H}^{\mathbb L_B}_{{\rm BPS}} = \bigoplus_{a', \rho'_0, \rho'_\infty}~\overline{{\rm IH}^\ast {\cal U}}({\cal M}^{a', \rho'_0}_{SO(N+1), \mathbb L_B, \rho'_\infty}(\mathbb R^4)^{\mathbb Z_k}),
\label{BPS-M-ram-B}
\ee 
where $\overline{{\rm IH}^\ast {\cal U}}(\cal M) \subset {\rm IH}^\ast {\cal U}(\cal M)$ is the $\mathbb Z_2$-invariant subspace of ${\rm IH}^\ast {\cal U}(\cal M)$ (as described in the paragraphs leading up to (\ref{HBPS-eff})), and $\mathbb L_B \subset SO(N+1)$ is a Levi subgroup determined by the defect. 

Let us now ascertain the spacetime BPS states in (\ref{M-theory defect dual-AB}). Via the arguments which led us to (\ref{equivalent IIA system 1}), we arrive at the following equivalent type IIA configuration to  (\ref{M-theory defect dual-AB}):
\be
\textrm{IIA}: \quad \underbrace{ {\mathbb R}^5 \times  {{\bf S}^1_n} \times {\mathbb R}_t \times {\mathbb R}^3}_{\textrm{I-brane on ${{\bf S}^1_n} \times {\mathbb R}_t = N \textrm{D6} \cap k\textrm{D4} \cap {\rm 3d \, defect}$}}.
\label{equivalent IIA system 1-ram}
\ee
Here, we have  a stack of $N$ coincident D6-branes whose worldvolume is given by ${\mathbb R}^5 \times {{\bf S}^1_n} \times {\mathbb R}_t$; a stack of $k$ coincident D4-branes whose worldvolume is given by $ {{\bf S}^1_n} \times {\mathbb R}_t \times \mathbb R^3$; and a 3d worldvolume defect which wraps ${\bf S}^1_n \times {\mathbb R}_t \times \mathbb R$,  where $\mathbb R \subset \mathbb R^3$. The two stacks of branes and the defect intersect along ${{\bf S}^1_n} \times {\mathbb R}_t$ to form a D4-D6 ``ramified'' I-brane system. Via the arguments following (\ref{equivalent IIA system 1}), we find that the spacetime BPS states in (\ref{M-theory defect dual-AB}) ought to be captured by this ``ramified'' I-brane system. 

Note at this point that according to our preparatory discussion of the (dual) defect in (\ref{M-theory defect dual-AB}), from the affine Lie algebraic perspective of the ``ramified'' I-brane theory along ${\bf S}^1_n \times {\mathbb R}_t$ where $n=1$ or 2 (with even $N$), the gauge group associated with the $k$ D4-branes is $\mathbb L_k \subset SU(k)$, where $\mathbb L_k$ is a Levi subgroup, while the gauge group associated with the $N$ D6-branes is $\mathbb L^\vee_A$ or $\mathbb L^\vee_B$, respectively, where ${\cal G}^\vee$ is the Langlands dual of the group $\cal G$.

Let us now repeat the arguments that enabled us to go from (\ref{equivalent IIA system 1}) to (\ref{partially gauged CFT - AB}). Bearing in mind the statements of the previous paragraph -- which tell us that from the affine Lie algebraic perspective of the ``ramified'' I-brane  theory along ${\bf S}^1_n \times {\mathbb R}_t$ where $n=1$ or 2 (with even $N$), the \emph{dynamical} gauge symmetry associated with the D4-branes is now $\mathbb L_k$ and not $SU(k)$, while an amount $SU(N)^\vee / \mathbb L^\vee_A$ or $SO(N+1)^\vee / \mathbb L^\vee_B$ is being ``unfrozen'' from the original \emph{non-dynamical} $SU(N)^\vee$ or $SO(N+1)^\vee$ gauge symmetry associated with the D6-branes -- we find that when  $n=1$ or 2 (with even $N$), the free chiral fermions that underlie the  ``ramified'' I-brane theory will effectively couple to the gauge group $U(1) \times \mathbb L_k \times SU(N)^\vee / \mathbb L^\vee_A$ or $U(1) \times \mathbb L_k \times SO(N+1)^\vee / \mathbb L^\vee_B$, respectively.  

Therefore, when $n=1$, the ``ramified'' I-brane theory would be given by a partially gauged CFT that can be schematically expressed as
\be
{\frak{u}(1)_{{\rm aff}, kN} \over \frak{u}(1)_{{\rm aff},kN} } \otimes { \frak{su}(k)_{{\rm aff},N} \over  \frak{l}(k)_{{\rm aff},N'}} \otimes  {\frak{su}(N)^\vee_{{\rm aff},k} \over { [\frak{su}(N)^\vee_{{\rm aff},k} /\frak {l}(N)^\vee_{{\rm aff},k'}]}}.
\label{partially gauged CFT - A - ram}
\ee
Here, $\frak{g}_{{\rm aff}, r}$ is the affine Lie algebra of level $r$ associated with the underlying Lie group; $\frak{g}^\vee_{{\rm aff}, r}$ is its \emph{Langlands dual}; $\frak l(k)$ is the Lie algebra of $\mathbb L_k$; and $\frak l(N)$ is the Lie algebra of $\mathbb L_A$. Note that the chiral fermions on the ``ramified'' I-brane are actually gauge-anomalous. Nevertheless, by repeating the arguments in~\cite[eqn.~(4.12)--(4.24)]{ATMP}, we find that the overall system consisting of the chiral fermions on the ``ramified'' I-brane and the gauge fields in the bulk of the D-branes, is gauge-invariant and therefore physically consistent, as expected. Furthermore, $N' = N$ and $k' = k$, as the simple roots of $\mathbb L_k$ and $\mathbb L^\vee_A$ form a subset of the simple roots of $SU(k)$ and $SU(N)^\vee$, respectively. (See~\cite[$\S$VI.1]{Ketov}.) 

At any rate, let $\widehat{g}_r$ and $^L\widehat{g}_r$ be the integrable modules over the affine Lie algebras $\frak{g}_{\textrm{aff},r}$ and $\frak{g}^\vee_{\textrm{aff},r}$ which can be realized as the spectra of states $\textrm{WZW}_{\widehat{g}_r}$ and $\textrm{WZW}_{^L\widehat{g}_r}$  in the corresponding $\it{chiral}$ WZW models. Then, (\ref{partially gauged CFT - A - ram}) would mean that after coupling to the gauge fields, (i) the original $\frak{u}(1)_{{\rm aff},kN}$ chiral WZW model will be replaced by the corresponding topological $G/G$ model; (ii) the original  $\frak{su}(k)_{{\rm aff},N}$ chiral WZW model will be replaced by an ${ \frak{su}(k)_{{\rm aff},N} /  \frak{l}(k)_{{\rm aff},N}}$ chiral coset model; and (iii) the original $\frak{su}(N)^\vee_{{\rm aff},k}$ chiral WZW model will be replaced by an $\frak {l}(N)^\vee_{{\rm aff},k}$ chiral WZW model. As such,  the chiral character of $\widehat{u}(1)_{kN}$ in the overall partition function of the uncoupled free fermions system on the ``ramified'' I-brane, will reduce to a constant complex factor. Modulo this constant complex factor which serves only to shift the underlying modular anomaly of the remaining chiral characters, the $\it{effective}$ overall partition function of the ``ramified'' I-brane theory would be expressed solely in terms of the chiral characters of $\widehat{su}(k)_{N} /  \widehat{l}(k)_{N}$ and  $^L\widehat{l}(N)_{k}$. Therefore, when $n=1$, the sought-after spectrum of spacetime BPS states in (\ref{M-theory defect dual-AB}) would be realized by 
\be
(\textrm{WZW}_{\widehat{su}(k)_{N}} /  \textrm{WZW}_{\widehat{l}(k)_{N}}) \otimes \textrm{WZW}_{^L\widehat{l}(N)_{k}}.
\label{WZW-ram-A}
\ee

When $n=2$ (with even $N$), the ``ramified'' I-brane theory would be given by a partially gauged CFT that can be schematically expressed as
\be
{\frak{u}(1)^{(2)}_{{\rm aff}, kN} \over \frak{u}(1)^{(2)}_{{\rm aff},kN} } \otimes { \frak{su}(k)^{(2)}_{{\rm aff},N} \over  \frak{l}(k)^{(2)}_{{\rm aff},N''}} \otimes  {\frak{so}(N+1)^\vee_{{\rm aff},k} \over { [\frak{so}(N+1)^\vee_{{\rm aff},k} /\frak {l}(N+1)^\vee_{{\rm aff},k''}]}}.
\label{partially gauged CFT - B - ram}
\ee
Here, $\frak{g}^{(2)}_{{\rm aff}, r}$ is the $\mathbb Z_2$-twisted affine Lie algebra of level $r$ associated with the underlying Lie group, and $\frak l(N+1)$ is the Lie algebra of $\mathbb L_B$. Note that the chiral fermions on the ``ramified'' I-brane are actually gauge-anomalous. Nevertheless, by repeating the arguments in~\cite[eqn.~(4.12)--(4.24)]{ATMP} whilst noting that ${\bf S}^1_2$ is topologically equivalent to an ordinary circle, we find that the overall system consisting of the chiral fermions on the ``ramified'' I-brane and the gauge fields in the bulk of the D-branes, is gauge-invariant and therefore physically consistent, as expected. Furthermore, similar to the $n=1$ case, $k'' = k$ and $N'' = N$, as the simple roots of $\mathbb L^\vee_B$ and $\mathbb L_k$ form a subset of the simple roots of $SO(N+1)^\vee$ and $SU(k)$, respectively. (See~\cite[$\S$VI.1]{Ketov} and footnote~\ref{central charge}.) 

Now note that (\ref{partially gauged CFT - B - ram}) would mean that after coupling to the gauge fields, (i) the original $\frak{u}(1)^{(2)}_{{\rm aff},kN}$ chiral WZW model will be replaced by the corresponding topological $G/G$ model; (ii) the original  $\frak{su}(k)^{(2)}_{{\rm aff},N}$ chiral WZW model will be replaced by an ${ \frak{su}(k)^{(2)}_{{\rm aff},N} /  \frak{l}(k)^{(2)}_{{\rm aff},N}}$ chiral coset model; and (iii) the original $\frak{so}(N+1)^\vee_{{\rm aff},k}$ chiral WZW model will be replaced by an $\frak {l}(N+1)^\vee_{{\rm aff},k}$ chiral WZW model. As such,  the chiral character of $\widehat{u}(1)^{(2)}_{kN}$ in the overall partition function of the uncoupled free fermions system on the ``ramified'' I-brane, will reduce to a constant complex factor. Modulo this constant complex factor which serves only to shift the underlying modular anomaly of the remaining chiral characters, the $\it{effective}$ overall partition function of the ``ramified'' I-brane theory would be expressed solely in terms of the chiral characters of $\widehat{su}(k)^{(2)}_{N} /  \widehat{l}(k)^{(2)}_{N}$ and  $^L\widehat{l}(N+1)_{k}$. Therefore, when $n=2$ (with even $N$), the sought-after spectrum of spacetime BPS states in (\ref{M-theory defect dual-AB}) would be realized by 
\be
(\textrm{WZW}_{\widehat{su}(k)^{(2)}_{N}} /  \textrm{WZW}_{\widehat{l}(k)^{(2)}_{N}}) \otimes \textrm{WZW}_{^L\widehat{l}(N+1)_{k}}.
\label{WZW-ram-B}
\ee 

We are finally ready to state our main result. When $n=1$, the equivalence of the spectra of spacetime BPS states in (\ref{M-theory defect-AB}) and (\ref{M-theory defect dual-AB}) would mean that the Hilbert space (\ref{BPS-M-ram-A}) ought to be equal to the chiral CFT spectrum in (\ref{WZW-ram-A}), i.e.,
\be
\boxed{\bigoplus_{a', \rho'_0, \rho'_\infty}~{{\rm IH}^\ast {\cal U}}({\cal M}^{a', \rho'_0}_{SU(N), \mathbb L_A, \rho'_\infty}(\mathbb R^4)^{\mathbb Z_k}) =  \textrm{WZW}_{^L\widehat{l}(N)_{k}} \otimes {\textrm{WZW}_{\widehat{su}(k)_{N}} \over \textrm{WZW}_{\widehat{l}(k)_{N}}}}
\label{GL-A-ram}
\ee
This is a ``\emph{ramified}'' generalization of the geometric Langlands duality for surfaces for the $SU(N) = A_{N-1}$ groups in (\ref{GL-relation A}).

When $n=2$ (with even $N$), the equivalence of the spectra of spacetime BPS states in (\ref{M-theory defect-AB}) and (\ref{M-theory defect dual-AB}) would mean that the Hilbert space (\ref{BPS-M-ram-B}) ought to be equal to the chiral CFT spectrum in (\ref{WZW-ram-B}), i.e.,
\be
\boxed{\bigoplus_{a', \rho'_0, \rho'_\infty}~\overline{{\rm IH}^\ast {\cal U}}({\cal M}^{a', \rho'_0}_{SO(N+1), \mathbb L_B, \rho'_\infty}(\mathbb R^4)^{\mathbb Z_k}) = \textrm{WZW}_{^L\widehat{l}(N+1)_{k}} \otimes {\textrm{WZW}_{\widehat{su}(k)^{(2)}_{N}} \over  \textrm{WZW}_{\widehat{l}(k)^{(2)}_{N}}}}
\label{GL-B-ram}
\ee
This is a ``\emph{ramified}'' generalization of the geometric Langlands duality for surfaces for the $SO(N+1) = B_{N/2}$ groups in (\ref{GL-relation B}).

Notice that for a trivial defect whence $\mathbb L_A = SU(N)$, $\mathbb L_B = SO(N+1)$ and $\mathbb L_k = SU(k)$, (\ref{GL-A-ram}) and (\ref{GL-B-ram}) would simplify to the ``unramified'' case in (\ref{GL-relation A}) and (\ref{GL-relation B}), respectively. Moreover, (\ref{GL-A-ram}) and (\ref{GL-B-ram}) also agree with and generalize the mathematical results in~\cite[$\S$16]{B-SO} (which analyzes the $k=1$ case only).

 Last but not least, note that because $\mathbb L_k$ and $\mathbb L_A$ are associated with the dual and original defect in (\ref{M-theory defect dual-AB}) and (\ref{M-theory defect-AB}), respectively, $\mathbb L_k$ ought to be ``dual'' to $\mathbb L_A$. Indeed, let us generalize the arguments behind (\ref{McKay IC}) to include a defect of the kind considered herein; one would then get an equivalence relation similar to (\ref{McKay IC}) which involves the LHS of (\ref{GL-A-ram}); in turn, via the RHS of (\ref{GL-A-ram}) and the level-rank duality (\ref{level-rank-A}), one can conclude that $\mathbb L_k$ and $\mathbb L_A$ are ``dual'' in the sense that $\widehat{l}(k)_{N} = \widehat{l}(N)_{k}$, where $\frak l(k)$ and $\frak l(N)$ define $\mathbb L_k$ and $\mathbb L_A$, respectively.

\bigskip\noindent{\it  A ``Ramified'' Geometric Langlands Duality for Surfaces for the $C$--$D$--$G$ Groups}

Let us now proceed to derive our main result for the $C$--$D$--$G$ groups. To this end, recall from (\ref{OM-theory defect}) and (\ref{OM-theory 8 defect}) that the six-dimensional M-theory compactification 
\be
\textrm{M-theory}: \quad \mathbb R^{5}  \times  \underbrace{\mathbb R_t \times {{\bf S}^1_n} \times \mathbb R^4 / \mathbb Z_k}_{\textrm{$N$ M5 + OM5 + 4d defect}},
\label{M-theory defect-CDG}
\ee
where the 4d worldvolume defect wraps $\mathbb R_t \times {\bf S}^1_n$ and the $z$-plane in $\mathbb R^4 / \mathbb Z_k \simeq \mathbb C_z / \mathbb Z_k \times \mathbb C_w / \mathbb Z_k$, is\emph{ physically dual} to the following six-dimensional M-theory compactification 
\be
\textrm{M-theory}: \quad \underbrace{SN_N^{R\to 0}  \times  {{\bf S}^1_n} \times \mathbb R_t}_{\textrm{$k$ M5 + 4d defect}}  \times  {\mathbb R^{5}},
\label{M-theory defect dual-CDG}
\ee
where the 4d worldvolume defect wraps ${\bf S}^1_n \times \mathbb R_t$, the ${\bf S}^1$-fiber of $SN_N^{R\to 0}$, and a single direction along the $\mathbb R^3$ base of $SN_N^{R\to 0}$. 

According to $\S$3.2, the spacetime BPS states in (\ref{M-theory defect-CDG}) and (\ref{M-theory defect dual-CDG}) -- which are annihilated by eight of the sixteen supercharges of the 6d ${\cal N} = (1, 1)$ supersymmetry algebra of the underlying Yang-Mills theories along $\mathbb R^{5}  \times  \mathbb R_t$ -- ought to have equivalent spectra. As in $\S$3.2, it is through this equivalence of spectra that we will be able to derive our main result. As such, let us first ascertain the spacetime BPS states in (\ref{M-theory defect-CDG}).

Via the arguments which led us to (\ref{BPS-M-OM5}), we find that for the $SO(2N) = D_N$ groups, the Hlibert space ${\cal H}^{\mathbb L_D}_{{\rm BPS}}$ of spacetime BPS states in (\ref{M-theory defect-CDG}) ought to be given by
\be
{\cal H}^{{\mathbb L}_D}_{{\rm BPS}} = \bigoplus_{a', \rho'_0, \rho'_\infty}~{\rm IH}^\ast {\cal U}({\cal M}^{a', \rho'_0}_{SO(2N), \mathbb L_D, \rho'_\infty}(\mathbb R^4)^{\mathbb Z_k}),
\label{BPS-M-ram-D}
\ee 
where ${\rm IH}^\ast {\cal U}(\cal M)$ is the intersection cohomology of the Uhlenbeck compactification of $\cal M$, and $\mathbb L_D \subset SO(2N)$ is a Levi subgroup determined by the defect.

Via the arguments which led us to (\ref{HBPS-eff-USp(2N-2)}), we find that for the $USp (2N-2) = C_{N-1}$ groups, the Hlibert space ${\cal H}^{\mathbb L_C}_{{\rm BPS}}$ of spacetime BPS states in (\ref{M-theory defect-CDG}) ought to be given by
\be
{\cal H}^{\mathbb L_C}_{{\rm BPS}} = \bigoplus_{a', \rho'_0, \rho'_\infty}~\overline{{\rm IH}^\ast {\cal U}}({\cal M}^{a', \rho'_0}_{Usp(2N - 2), \mathbb L_C, \rho'_\infty}(\mathbb R^4)^{\mathbb Z_k}),
\label{BPS-M-ram-C}
\ee 
where $\overline{{\rm IH}^\ast {\cal U}}(\cal M) \subset {\rm IH}^\ast {\cal U}(\cal M)$ is the $\mathbb Z_2$-invariant subspace of ${\rm IH}^\ast {\cal U}(\cal M)$ (as described in the paragraphs leading up to (\ref{HBPS-eff-USp(2N-2)})), and $\mathbb L_C \subset USp(2N-2)$ is a Levi subgroup determined by the defect. 

Via the arguments which led us to (\ref{HBPS-eff-G_2}), we find that for the $G_2$ group, the Hlibert space ${\cal H}^{\mathbb L_G}_{{\rm BPS}}$ of spacetime BPS states in (\ref{M-theory defect-CDG}) ought to be given by
\be
{\cal H}^{\mathbb L_G}_{{\rm BPS}} = \bigoplus_{a', \rho'_0, \rho'_\infty}~\widetilde{{\rm IH}^\ast {\cal U}}({\cal M}^{a', \rho'_0}_{G_2, \mathbb L_G, \rho'_\infty}(\mathbb R^4)^{\mathbb Z_k}),
\label{BPS-M-ram-G}
\ee 
where $\widetilde{{\rm IH}^\ast {\cal U}}(\cal M) \subset {\rm IH}^\ast {\cal U}(\cal M)$ is the $\mathbb Z_3$-invariant subspace of ${\rm IH}^\ast {\cal U}(\cal M)$ (as described in the paragraphs leading up to  (\ref{HBPS-eff-G_2})), and $\mathbb L_G \subset G_2$ is a Levi subgroup determined by the defect.

Let us now ascertain the spacetime BPS states in (\ref{M-theory defect dual-CDG}). Via the arguments which led us to (\ref{equivalent IIA system 2}), we arrive at the following equivalent type IIA configuration to  (\ref{M-theory defect dual-CDG}):
\be
\textrm{IIA}: \quad \underbrace{ {\mathbb R}^5 \times {{\bf S}^1_n} \times {\mathbb R}_t  \times {\mathbb R}^3/{\cal I}_3}_{\textrm{I-brane on ${{\bf S}^1_n} \times {\mathbb R}_t =  N \textrm{D6}/O$6$^-$} \cap  k\textrm{D4} \cap {\rm 3d \, defect}}.
\label{equivalent IIA system 2-ram}
\ee
Here, we have  a stack of $N$ coincident D6-branes on top of an O$6^-$-plane whose worldvolume is given by ${\mathbb R}^5 \times {{\bf S}^1_n} \times {\mathbb R}_t$; a stack of $k$ coincident D4-branes whose worldvolume is given by ${{\bf S}^1_n} \times {\mathbb R}_t \times \mathbb R^3/{\cal I}_3$, where ${\cal I}_3$ acts as a reflection about the origin in $\mathbb R^3$; and a 3d worldvolume defect which wraps ${\bf S}^1_n \times {\mathbb R}_t \times \mathbb R$,  where $\mathbb R \subset \mathbb R^3$. The two stacks of branes and the defect intersect along ${{\bf S}^1_n} \times {\mathbb R}_t$ to form a D4-D6/O$6^-$ ``ramified'' I-brane system. Via the arguments following (\ref{equivalent IIA system 2}), we find that the spacetime BPS states in (\ref{M-theory defect dual-CDG}) ought to be captured by this ``ramified'' I-brane system. 

Note at this point that according to our preparatory discussion of the (dual) defect in (\ref{M-theory defect dual-CDG}), from the affine Lie algebraic perspective of the ``ramified'' I-brane theory along ${\bf S}^1_n \times {\mathbb R}_t$ where $n=1$,  2 or 3 (with $N=4$), the gauge group associated with the $k$ D4-branes is $\mathbb L_{k} \subset SO(k)$, where $\mathbb L_{k}$ is a Levi subgroup, while the gauge group associated with the $N$ D6-branes is $\mathbb L^\vee_D$, $\mathbb L^\vee_C$ or $\mathbb L^\vee_G$, respectively, where ${\cal G}^\vee$ is the Langlands dual of the group $\cal G$.

Let us now repeat the arguments that enabled us to go from (\ref{equivalent IIA system 2}) to (\ref{partially gauged CFT - CDG}). Bearing in mind the statements of the previous paragraph -- which tell us that from the affine Lie algebraic perspective of the ``ramified'' I-brane  theory along ${\bf S}^1_n \times {\mathbb R}_t$ where $n=1$, 2 or 3 (with $N = 4$), the \emph{dynamical} gauge symmetry associated with the D4-branes is now $\mathbb L_{k}$ and not $SO(k)$, while an amount $SO(2N)^\vee / \mathbb L^\vee_D$, $USp(2N - 2)^\vee / \mathbb L^\vee_C$ or $G^\vee_2 / \mathbb L^\vee_G$ is being ``unfrozen'' from the original \emph{non-dynamical} $SO(2N)^\vee$, $USp(2N -2)^\vee$ or $G_2$ gauge symmetry associated with the D6-branes -- we find that when  $n=1$, 2 or 3 (with $N = 4$), the free chiral fermions that underlie the  ``ramified'' I-brane theory will effectively couple to the gauge group $\mathbb L_{k} \times SO(2N)^\vee / \mathbb L^\vee_D$, $\mathbb L_{k} \times USp(2N - 2)^\vee / \mathbb L^\vee_C$ or $\mathbb L_{k} \times G^\vee_2 / \mathbb L^\vee_G$, respectively.  

Therefore, when $n=1$, the ``ramified'' I-brane theory would be given by a partially gauged CFT that can be schematically expressed as
\be
{ \frak{so}(k)_{{\rm aff},2N} \over  \frak{l}(k)_{{\rm aff},2N'}} \otimes  {\frak{so}(2N)^\vee_{{\rm aff},k} \over { [\frak{so}(2N)^\vee_{{\rm aff},k} /\frak {l}(2N)^\vee_{{\rm aff},k'}]}}.
\label{partially gauged CFT - D - ram}
\ee
Here, $\frak{g}_{{\rm aff}, r}$ is the affine Lie algebra of level $r$ associated with the underlying Lie group; $\frak{g}^\vee_{{\rm aff}, r}$ is its \emph{Langlands dual}; $\frak l(k)$ is the Lie algebra of $\mathbb L_{k}$; and $\frak l(2N)$ is the Lie algebra of $\mathbb L_D$. Note that the chiral fermions on the ``ramified'' I-brane are actually gauge-anomalous. Nevertheless, by arguments similar to those in~\cite[eqn.~(4.12)--(4.24)]{ATMP}, we find that the overall system consisting of the chiral fermions on the ``ramified'' I-brane and the gauge fields in the bulk of the D-branes, is gauge-invariant and therefore physically consistent, as expected. Furthermore,  $2N' = 2N$ and $k' = k$, as the simple roots of $\mathbb L_{k}$ and $\mathbb L^\vee_D$ form a subset of the simple roots of $SO(k)$ and $SO(2N)^\vee$, respectively. (See~\cite[$\S$VI.1]{Ketov}.)

At any rate, let $\widehat{g}_r$ and $^L\widehat{g}_r$ be the integrable modules over the affine Lie algebras $\frak{g}_{\textrm{aff},r}$ and $\frak{g}^\vee_{\textrm{aff},r}$ which can be realized as the spectra of states $\textrm{WZW}_{\widehat{g}_r}$ and $\textrm{WZW}_{^L\widehat{g}_r}$  in the corresponding $\it{chiral}$ WZW models. Then, (\ref{partially gauged CFT - D - ram}) would mean that after coupling to the gauge fields, (i) the original  $\frak{so}(k)_{{\rm aff},2N}$ chiral WZW model will be replaced by an ${ \frak{so}(k)_{{\rm aff},2N} /  \frak{l}(k)_{{\rm aff},2N}}$ chiral coset model; and (ii) the original $\frak{so}(2N)^\vee_{{\rm aff},k}$ chiral WZW model will be replaced by an $\frak {l}(2N)^\vee_{{\rm aff},k}$ chiral WZW model. As such,  the $\it{effective}$ overall partition function of the ``ramified'' I-brane theory would be expressed in terms of the chiral characters of $\widehat{so}(k)_{2N} /  \widehat{l}(k)_{2N}$ and  $^L\widehat{l}(2N)_{k}$. Therefore, when $n=1$, the sought-after spectrum of spacetime BPS states in (\ref{M-theory defect dual-CDG}) would be realized by 
\be
(\textrm{WZW}_{\widehat{so}(k)_{2N}} /  \textrm{WZW}_{\widehat{l}(k)_{2N}}) \otimes \textrm{WZW}_{^L\widehat{l}(2N)_{k}}.
\label{WZW-ram-D}
\ee 

When $n=2$, the ``ramified'' I-brane theory would be given by a partially gauged CFT that can be schematically expressed as
\be
{ \frak{so}(k)^{(2)}_{{\rm aff},2N} \over  \frak{l}(k)^{(2)}_{{\rm aff},2N''}} \otimes  {\frak{usp}(2N-2)^\vee_{{\rm aff},k} \over { [\frak{usp}(2N -2)^\vee_{{\rm aff},k} /\frak {l}(2N-2)^\vee_{{\rm aff},k''}]}}.
\label{partially gauged CFT - C - ram}
\ee
Here, $\frak{g}^{(2)}_{{\rm aff}, r}$ is the $\mathbb Z_2$-twisted affine Lie algebra of level $r$ associated with the underlying Lie group, and $\frak l(2N-2)$ is the Lie algebra of $\mathbb L_C$.  Note that the chiral fermions on the ``ramified'' I-brane are actually gauge-anomalous. Nevertheless, by arguments similar to those in~\cite[eqn.~(4.12)--(4.24)]{ATMP} whilst noting that ${\bf S}^1_2$ is topologically equivalent to an ordinary circle, we find that the overall system consisting of the chiral fermions on the ``ramified'' I-brane and the gauge fields in the bulk of the D-branes, is gauge-invariant and therefore physically consistent, as expected. Furthermore, similar to the $n=1$ case, $k'' = k$ and $2N'' = 2N$, as the simple roots of $\mathbb L^\vee_C$ and $\mathbb L_{k}$ form a subset of the simple roots of $USp(2N-2)^\vee$ and $SO(k)$, respectively. (See~\cite[$\S$VI.1]{Ketov} and footnote~\ref{central charge SO(2N)}.)

Now note that (\ref{partially gauged CFT - C - ram}) would mean that after coupling to the gauge fields, (i) the original  $\frak{so}(k)^{(2)}_{{\rm aff},2N}$ chiral WZW model will be replaced by an ${ \frak{so}(k)^{(2)}_{{\rm aff},2N} /  \frak{l}(k)^{(2)}_{{\rm aff},2N}}$ chiral coset model; and (ii) the original $\frak{usp}(2N-2)^\vee_{{\rm aff},k}$ chiral WZW model will be replaced by an $\frak {l}(2N-2)^\vee_{{\rm aff},k}$ chiral WZW model. As such,  the $\it{effective}$ overall partition function of the ``ramified'' I-brane theory would be expressed solely in terms of the chiral characters of $\widehat{so}(k)^{(2)}_{2N} /  \widehat{l}(k)^{(2)}_{2N}$ and  $^L\widehat{l}(2N-2)_{k}$. Therefore, when $n=2$, the sought-after spectrum of spacetime BPS states in (\ref{M-theory defect dual-CDG}) would be realized by 
\be
(\textrm{WZW}_{\widehat{so}(k)^{(2)}_{2N}} /  \textrm{WZW}_{\widehat{l}(k)^{(2)}_{2N}}) \otimes \textrm{WZW}_{^L\widehat{l}(2N-2)_{k}}.
\label{WZW-ram-C}
\ee 

When $n=3$ (with $N=4$), the ``ramified'' I-brane theory would be given by a partially gauged CFT that can be schematically expressed as
\be
{ \frak{so}(k)^{(3)}_{{\rm aff},2N} \over  \frak{l}(k)^{(3)}_{{\rm aff},2N'''}} \otimes  {\frak{g}^\vee_{2 \, {\rm aff},k} \over { [\frak{g}^\vee_{2 \, {\rm aff},k} /\frak {l}(G)^\vee_{{\rm aff},k'''}]}}.
\label{partially gauged CFT - G - ram}
\ee
Here, $\frak{g}^{(3)}_{{\rm aff}, r}$ is the $\mathbb Z_3$-twisted affine Lie algebra of level $r$ associated with the underlying Lie group, and $\frak l(G)$ is the Lie algebra of $\mathbb L_G$. Note that the chiral fermions on the ``ramified'' I-brane are actually gauge-anomalous. Nevertheless, by arguments similar to those in~\cite[eqn.~(4.12)--(4.24)]{ATMP} whilst noting that ${\bf S}^1_3$ is topologically equivalent to an ordinary circle, we find that the overall system consisting of the chiral fermions on the ``ramified'' I-brane and the gauge fields in the bulk of the D-branes, is gauge-invariant and therefore physically consistent, as expected. Furthermore, similar to the $n=1$ and $2$ cases, $k''' = k$ and $2N''' = 2N$, as the simple roots of $\mathbb L^\vee_G$ and $\mathbb L_{k}$ form a subset of the simple roots of $G^\vee_2$ and $SO(k)$, respectively. (See~\cite[$\S$VI.1]{Ketov} and footnote~\ref{central charge SO(2N)}.)

Now note that (\ref{partially gauged CFT - G - ram}) would mean that after coupling to the gauge fields, (i) the original  $\frak{so}(k)^{(3)}_{{\rm aff},2N}$ chiral WZW model will be replaced by an ${ \frak{so}(k)^{(3)}_{{\rm aff},2N} /  \frak{l}(k)^{(3)}_{{\rm aff},2N}}$ chiral coset model; and (ii) the original $\frak{g}^\vee_{2 \, {\rm aff},k}$ chiral WZW model will be replaced by an $\frak {l}(G)^\vee_{{\rm aff},k}$ chiral WZW model. As such,  the $\it{effective}$ overall partition function of the ``ramified'' I-brane theory would be expressed solely in terms of the chiral characters of $\widehat{so}(k)^{(3)}_{2N} /  \widehat{l}(k)^{(3)}_{2N}$ and  $^L\widehat{l}(G)_{k}$. Therefore, when $n=3$ (with $N=4$), the sought-after spectrum of spacetime BPS states in (\ref{M-theory defect dual-CDG}) would be realized by 
\be
(\textrm{WZW}_{\widehat{so}(k)^{(3)}_{2N}} /  \textrm{WZW}_{\widehat{l}(k)^{(3)}_{2N}}) \otimes \textrm{WZW}_{^L\widehat{l}(G)_{k}}.
\label{WZW-ram-G}
\ee 

We are finally ready to state our main result. When $n=1$, the equivalence of the spectra of spacetime BPS states in (\ref{M-theory defect-CDG}) and (\ref{M-theory defect dual-CDG}) would mean that the Hilbert space (\ref{BPS-M-ram-D}) ought to be equal to the chiral CFT spectrum in (\ref{WZW-ram-D}), i.e.,
\be
\boxed{\bigoplus_{a', \rho'_0, \rho'_\infty}~{\rm IH}^\ast {\cal U}({\cal M}^{a', \rho'_0}_{SO(2N), \mathbb L_D, \rho'_\infty}(\mathbb R^4)^{\mathbb Z_k}) =  \textrm{WZW}_{^L\widehat{l}(2N)_{k}} \otimes {\textrm{WZW}_{\widehat{so}(k)_{2N}} \over  \textrm{WZW}_{\widehat{l}(k)_{2N}}}}
\label{GL-D-ram}
\ee
This is a ``\emph{ramified}'' generalization of the geometric Langlands duality for surfaces for the $SO(2N) = D_N$ groups in (\ref{GL-relation D}).		

When $n=2$, the equivalence of the spectra of spacetime BPS states in (\ref{M-theory defect-CDG}) and (\ref{M-theory defect dual-CDG}) would mean that the Hilbert space (\ref{BPS-M-ram-C}) ought to be equal to the chiral CFT spectrum in (\ref{WZW-ram-C}), i.e.,
\be
\boxed{\bigoplus_{a', \rho'_0, \rho'_\infty}~\overline{{\rm IH}^\ast {\cal U}}({\cal M}^{a', \rho'_0}_{Usp(2N - 2), \mathbb L_C, \rho'_\infty}(\mathbb R^4)^{\mathbb Z_k}) = \textrm{WZW}_{^L\widehat{l}(2N-2)_{k}} \otimes {\textrm{WZW}_{\widehat{so}(k)^{(2)}_{2N}} \over  \textrm{WZW}_{\widehat{l}(k)^{(2)}_{2N}}}}
\label{GL-C-ram}
\ee
This is a ``\emph{ramified}'' generalization of the geometric Langlands duality for surfaces for the $USp(2N-2) = C_{N-1}$ groups in (\ref{GL-relation C}).		

When $n=3$ (with $N=4$), the equivalence of the spectra of spacetime BPS states in (\ref{M-theory defect-CDG}) and (\ref{M-theory defect dual-CDG}) would mean that the Hilbert space (\ref{BPS-M-ram-G}) ought to be equal to the chiral CFT spectrum in (\ref{WZW-ram-G}), i.e.,
\be
\boxed{\bigoplus_{a', \rho'_0, \rho'_\infty}~\widetilde{{\rm IH}^\ast {\cal U}}({\cal M}^{a', \rho'_0}_{G_2, \mathbb L_G, \rho'_\infty}(\mathbb R^4)^{\mathbb Z_k})= \textrm{WZW}_{^L\widehat{l}(G)_{k}} \otimes {\textrm{WZW}_{\widehat{so}(k)^{(3)}_{2N}} \over  \textrm{WZW}_{\widehat{l}(k)^{(3)}_{2N}}}}
\label{GL-G-ram}
\ee
This is a ``\emph{ramified}'' generalization of the geometric Langlands duality for surfaces for the $G_2$ group in (\ref{GL-relation G}).

Notice that for a trivial defect whence $\mathbb L_D = SO(2N)$, $\mathbb L_C = USp(2N-2)$, $\mathbb L_G = G_2$ and $\mathbb L_{k} = SO(k)$, (\ref{GL-D-ram}), (\ref{GL-C-ram}) and (\ref{GL-G-ram}) would simplify to the ``unramified'' case in (\ref{GL-relation D}), (\ref{GL-relation C}) and (\ref{GL-relation G}), respectively. Moreover, (\ref{GL-D-ram}), (\ref{GL-C-ram}) and (\ref{GL-G-ram}) also agree with and generalize the mathematical results in~\cite[$\S$16]{B-SO} (which analyzes the $k=1$ case only).

 Last but not least, note that because $\mathbb L_{k}$ and $\mathbb L_D$ are associated with the dual and original defect in (\ref{M-theory defect dual-CDG}) and (\ref{M-theory defect-CDG}), respectively, $\mathbb L_{k}$ ought to be ``dual'' to $\mathbb L_D$. Indeed, let us generalize the arguments behind (\ref{McKay IC-D}) and (\ref{PF dual D-SN}) to include a defect of the kind considered herein; one would then get a ``ramified'' version of (\ref{McKay IC-D}) and (\ref{PF dual D-SN}), and together with (\ref{GL-D-ram}) and (\ref{level-rank-D}), one can conclude that $\mathbb L_{k}$ and $\mathbb L_D$ are ``dual'' in the sense that $\widehat{l}(k)_{2N} = \widehat{l}(2N)_{k}$, where $\frak l(k)$ and $\frak l(2N)$ define $\mathbb L_{k}$ and $\mathbb L_D$, respectively. 
 
 \newpage
 
 \part{\Large The AGT Correspondence}

\newsection{An M-Theoretic Derivation of the Pure AGT Correspondence}

\newsubsection{Turning on Omega-Deformation}

Let $k=1$ in (\ref{M-theory 1 discussion}), (\ref{M-theory 7 discussion}), (\ref{M-theory 1 OM5 discussion}) and (\ref{M-theory 7 OM5 discussion}); in other words, consider the \emph{physically dual} six-dimensional M-theory compactifications
\be
\underbrace{\mathbb R^4  \times {{\bf S}^1_n} \times \mathbb R_t}_{\textrm{$N$ M5-branes}}\times \mathbb R^{5} \quad \Longleftrightarrow  \quad {\mathbb R^{5}} \times \underbrace{\mathbb R_t \times {{\bf S}^1_n}  \times TN_N^{R\to 0}  }_{\textrm{$1$ M5-branes}},
 \label{AGT-dual pair-A}
 \ee
 and
\be
\underbrace{\mathbb R^4  \times {{\bf S}^1_n} \times \mathbb R_t}_{\textrm{$N$ M5-branes + OM5-plane}}\times \mathbb R^{5} \quad \Longleftrightarrow  \quad {\mathbb R^{5}} \times \underbrace{\mathbb R_t \times {{\bf S}^1_n}  \times SN_N^{R\to 0}  }_{\textrm{$1$ M5-branes}}.
 \label{AGT-dual pair-D} 
 \ee
 As explained in $\S$2.1 and $\S$2.2, on the LHS of (\ref{AGT-dual pair-A}) and (\ref{AGT-dual pair-D}), there is a $\mathbb Z_n$-outer-automorphism of ${\mathbb R^4}$ as we go around the ${\bf S}^1_n$ circle and identify the circle under an order $n$ translation; on the RHS of (\ref{AGT-dual pair-A}) and (\ref{AGT-dual pair-D}), there is a $\mathbb Z_n$-outer-automorphism of the singular multi-Taub-NUT space $TN_N^{R\to 0}$ and Sen's singular four-manifold $SN_N^{R\to 0}$ (whose circle fibers at infinity approach zero radius) as we go around the ${\bf S}^1_n$ circle. 

Recall from our arguments that brought us from (\ref{M-theory 1}) to (\ref{M-theory 7}), and from (\ref{OM-theory 1}) to (\ref{OM-theory 8}), that the above M5-branes and M5-branes + OM5-plane in (\ref{AGT-dual pair-A}) and (\ref{AGT-dual pair-D}) span the following directions:
\be
\begin{array}{l|c|c|cccc|cccc|c}
&0&1&2&3&4&5&6&7&8&9&10 \\
\hline
\hbox{$N$ M5's/OM5} & - & - & -& -& - & - &&&& & \\
\hbox{1 M5} & - & - & & &  &  & - &- &-& - & \\
\end{array} \label{table-AGT}
\ee
Here, the `$-$' sign in the column labeled by $j$ means that the particular brane extends along the $j^{\rm th}$ direction with coordinate $x_j$.  We take $x_0$ and $x_1$ to be the coordinates on $\mathbb R_t$ and ${\bf S}^1_n$, so that $(x_2, x_3, x_4, x_5)$ would be the coordinates on $ \mathbb R^4$; then, if $z = x_2 + i x_3$ and $w = x_4 + i x_5$, $\mathbb R^4$ can be viewed as a complex surface $\mathbb C^2$ whose coordinates are $(z,w)$. On the other hand, $(x_6, x_7, x_8, x_9)$ would be the coordinates on $TN_N^{R\to 0}$ and $SN_N^{R\to 0}$, and if $u = x_6 + i x_7$ and $v = x_8 + i x_9$, $TN_N^{R\to 0}$ and $SN_N^{R\to 0}$ can likewise be viewed as a complex surface whose singularity at the origin would be modeled by $\mathbb C^2 / \mathbb Z_N$ and $\mathbb C^2 / \mathbb D_N$, respectively, where $(u,v)$ are the coordinates on $\mathbb C^2$.

\bigskip\noindent{\it Omega-Deformation via a Fluxbrane}

Now, on the LHS of (\ref{AGT-dual pair-A}) and (\ref{AGT-dual pair-D}), turn on Omega-deformation~\cite{NN, NO} with \emph{real} parameters $\epsilon_1$ and $\epsilon_2$ along the $z$- and $w$-planes, respectively, via a fluxbrane as described in~\cite{susanne, orlando}:
\be
\begin{array}{l|c|c|cc|cc|cc|cc|c}
&0&1&2&3&4&5&6&7&8&9&10 \\
\hline
\hbox{$N$ M5's/OM5} & - & - & -& -& - & - &&&& & \\
\hbox{fluxbrane} & \times & \otimes & \hspace{0.25cm}\epsilon_1 & &  \hspace{0.25cm} \epsilon_2 &  &  \hspace{0.25cm} \epsilon_3 & & \times & \times & \circ \\ 
\end{array} \label{table-AGT-fluxbrane}
\ee
Here, the `$\times$'s denote the fluxbrane directions; `$\otimes$' denotes the ${\bf S}^1_n$ circle direction; and $\circ$ denotes the ``eleventh circle''. In addition, there is also a rotation along the $u$-plane with rotation parameter $\epsilon_3 = \epsilon_1 + \epsilon_2$, and it is tantamount to a topological twist (that involves an $R$-symmetry) which helps preserve some supersymmetry that would otherwise be completely broken by the $(\epsilon_1, \epsilon_2)$ rotations along the $(z, w)$ planes. 

In short, the LHS of  (\ref{AGT-dual pair-A}) and (\ref{AGT-dual pair-D}) in the presence of the fluxbrane denoted in (\ref{table-AGT-fluxbrane}), can be written as 
\be
\underbrace{\mathbb R^4\vert_{\epsilon_1, \epsilon_2}  \times {{\bf S}^1_n} \times \mathbb R_t}_{\textrm{$N$ M5-branes}}\times \mathbb R^{5}\vert_{\epsilon_3; \,  x_{6,7}}, \quad {\rm and} \quad \underbrace{\mathbb R^4\vert_{\epsilon_1, \epsilon_2}  \times {{\bf S}^1_n} \times \mathbb R_t}_{\textrm{$N$ M5-branes + OM5-plane}}\times \mathbb R^{5}\vert_{\epsilon_3; \,  x_{6,7}},
\label{AGT-fluxbrane}
\ee
where $\mathbb R^4 \vert_{\epsilon_1, \epsilon_2}$ is a completely Omega-deformed $\mathbb R^4$, and $\mathbb R^{5}\vert_{\epsilon_3; \,  x_{6,7}}$ is an $\mathbb R^5$ that is partially Omega-deformed along the $x_6$-$x_7$ plane with parameter $\epsilon_3$. 

 Repeating  in the presence of this fluxbrane, the chain of arguments that brought us from (\ref{M-theory 1}) to (\ref{M-theory 7}), and from (\ref{OM-theory 1}) to (\ref{OM-theory 8}) -- bearing in mind that since we do \emph{not} perform a T-duality along the ${\bf S}^1_n$ circle in the $x_1$-direction which would convert the fluxbrane into a fluxtrap, (i) the T-dualities we perform in the directions along the fluxbrane would not induce additional deformations to the B-field at any step of the duality chain; (ii) in performing steps (\ref{IIA 5}) and (\ref{OIIA 6}), the $\mathbb R^3$ base of the $TN_1$ space and the D6-brane normal to it would be deformed, purely geometrically, such that the ten-dimensional background metric will (omitting the contributions from $TN^{R \to 0}_N$ and $SN^{R \to 0}_N$) be given by\footnote{I would like to acknowledge Domenico Orlando's assistance with the following formula.}
 \begin{eqnarray}
{ {U(r)^{1/2}} \over 2 \pi} \left[ dr^2 + r^2 d \omega^2 + r^2 (\epsilon_1 \beta  dx_1 + \epsilon_2 \beta  dx_1 + d \phi)^2 {\rm sin}^2 \omega \right] \nonumber \\
   + { 1 \over {2 \pi \, U(r)^{1/2}}}\left[dx^2_{0,\dots 1} + dx^2_{8, \dots, 10} + d \rho^2_3 + \rho_3^2 (d \lambda_3 - \epsilon_3 \beta dx_1)^2 \right],
   \label{metric}
  \end{eqnarray} 
 where $(r, \phi, \omega)$ are the usual spherical coordinates on $\mathbb R^3$ spanning the $x_3$-$x_4$-$x_5$ directions, $(\rho_3, \lambda_3)$ are the radial-angular coordinates along the $u$-plane, the function $U(r)$ is the background contribution of the D6-branes, $\beta$ is the radius of the ${\bf S}^1_n$ circle, and the second term in (\ref{metric}) is the worldvolume metric of the D6-branes; and (iii) according to (\ref{metric}), the T-duality we perform  after steps  (\ref{IIA 5}) and (\ref{OIIA 6}) in the $x_{10}$-direction normal to the fluxbrane will not induce additional deformations to the B-field either -- we can, after proceeding with steps (\ref{IIB 6}) and (\ref{M-theory 7}), and steps (\ref{OIIB 7}) and (\ref{OM-theory 8}), express the dual configuration on the RHS of (\ref{AGT-dual pair-A}) and (\ref{AGT-dual pair-D}) in the presence of the now \emph{dual} fluxbrane, as
\be
\begin{array}{l|c|c|cccc|cc|cc|ccc}
&0&1&2&3&4&5&6&7&8&9&10 \\
\hline
\hbox{1 M5} & - & - & & &  &  & - &- &-& - &  \\
\hbox{dual fluxbrane} & \times & \otimes & \hspace{0.0cm} &  \hspace{0.0cm} \epsilon_1, &  \epsilon_2 &  &  \hspace{0.25cm} \epsilon_3  &  & \times & \times & \circ \\ 
\end{array} \label{table-AGT-fluxbrane-dual}
\ee
Here, ``$\epsilon_1, \epsilon_2$'' along the $x_2$-$x_3$-$x_4$-$x_5$ directions means that there are two simultaneous rotations along the $x_4$-$x_5$ plane with rotation parameters $\epsilon_1$ and $\epsilon_2$.

 In short, the RHS of the duality relations (\ref{AGT-dual pair-A}) and (\ref{AGT-dual pair-D}) in the presence of the dual fluxbrane denoted in (\ref{table-AGT-fluxbrane-dual}), can be written as 
 \be
 {\mathbb R^{5}}\vert_{\epsilon_3; \, x_{4,5}} \times \underbrace{\mathbb R_t \times {{\bf S}^1_n}  \times TN_N^{R\to 0}\vert_{\epsilon_3; \, x_{6,7}}}_{\textrm{$1$ M5-branes}} \quad {\rm and} \quad {\mathbb R^{5}}\vert_{\epsilon_3; \, x_{4,5}} \times \underbrace{\mathbb R_t \times {{\bf S}^1_n}  \times SN_N^{R\to 0}\vert_{\epsilon_3; \, x_{6,7}}}_{\textrm{$1$ M5-branes}},
 \label{AGT-fluxbrane-dual}
 \ee
 where the subscript in $TN_N^{R\to 0}\vert_{\epsilon_3; \, x_{6,7}}$ and $SN_N^{R\to 0}\vert_{\epsilon_3; \, x_{6,7}}$ indicates that they are partially Omega-deformed along the $x_6$-$x_7$ plane with parameter $\epsilon_3$.  

In (\ref{AGT-fluxbrane}), since each independent $\epsilon$ parameter breaks $1/2$ of the total number of supersymmetries~\cite{orlando}, there are, in the eleven-dimensional background, effectively $32 \times 1/2 \times 1/2 = 8$ conserved supercharges. Thus, the worldvolume theory of the $N$ M5-branes and $N$ M5-branes + OM5-plane on $\mathbb R^4 \vert_{\epsilon_1, \epsilon_2} \times {\bf S}^1_n \times \mathbb R_t$ has $ 8  \times 1/2= 4$ conserved supercharges, and the corresponding 6d spacetime theory along $\mathbb R_t \times \mathbb R^{5}\vert_{\epsilon_3; \,  x_{6,7}}$ (which spans the $x_0$-$x_6$-$x_7$-$x_8$-$x_9$-$x_{10}$ directions) has $\cN = (1,0)$ supersymmetry.

 On the other hand in (\ref{AGT-fluxbrane-dual}), because the $\epsilon_3$-rotation along the $x_6$-$x_7$ plane in hyperk\"ahler $TN^{R \to 0}_N$ and $SN^{R \to 0}_N$ does not break any additional supersymmetries (c.f.~\cite{orlando}), and because there is a rotation only of a \emph{single} plane in the $x_4$-$x_5$ directions which, consequently, breaks just $1/2$ of the existing number of supersymmetries,   there are, in the eleven-dimensional background, effectively $32 \times 1/2 \times 1/2 = 8$ conserved supercharges. Thus, the worldvolume theory of the single M5-brane on $\mathbb R_t \times {{\bf S}^1_n}  \times TN_N^{R\to 0}\vert_{\epsilon_3; \, x_{6,7}}$ and $\mathbb R_t \times {{\bf S}^1_n}  \times SN_N^{R\to 0}\vert_{\epsilon_3; \, x_{6,7}}$ has $ 8  \times 1/2= 4$ conserved supercharges, and the \emph{dual}  6d spacetime theory along $\mathbb R_t \times {\mathbb R^{5}}\vert_{\epsilon_3; \, x_{4,5}}$ (that spans the $x_0$-$x_2$-$x_3$-$x_4$-$x_5$-$x_{10}$ directions) also has $\cN = (1,0)$ supersymmetry, consistent with the duality of the 6d compactifications (\ref{AGT-fluxbrane}) and (\ref{AGT-fluxbrane-dual}). Moreover, (\ref{AGT-fluxbrane-dual}), like its dual compactification  (\ref{AGT-fluxbrane}), is invariant under the exchange $\epsilon_1 \leftrightarrow \epsilon_2$. 

\bigskip\noindent{\it Omega-Deformation and Spacetime Half-BPS States in a Variant of (\ref{AGT-fluxbrane})}

The Omega-deformation due to the fluxbrane in~(\ref{table-AGT-fluxbrane}) can also be explained in terms of the partition function of spacetime half-BPS states in a compactification that is a slight variant of (\ref{AGT-fluxbrane}). 

To this end, first recall from $\S$3.1 and $\S$3.2 that the spacetime quarter-BPS states on the LHS of (i) (\ref{AGT-dual pair-A}) and (ii) (\ref{AGT-dual pair-D}) will correspond to the quantum states of the worldvolume theory of the (i) $N$ M5-branes and (ii) $N$ M5-branes + OM5-plane given by the topological sector of an $\cN = (4,4)$ sigma-model with worldsheet $\Sigma = {\bf S}^1_n \times \mathbb R_t$ and target manifold ${\cal M}_G$ the moduli space of $G$-instantons on $\mathbb R^4$, where (i) for $n=1$ or 2 (with even $N$), $G = SU(N)$ or $SO(N+1)$, and (ii) for $n=1$, 2 or 3 (with $N = 4$), $G = SO(2N)$, $USp(2N-2)$ or $G_2$. In other words, where the spacetime quarter-BPS states on the LHS of (\ref{AGT-dual pair-A}) and (\ref{AGT-dual pair-D})  are concerned, we can regard the sigma-model to be topological whence we are free to deform $\Sigma$ into a short cylinder  ${\bf S}^1_n \times \mathbb I_t$, where $\mathbb I_t$ is an interval whose length is much smaller than $\beta$. 

Since the far past and far future are now brought to finite distances whence the eleven-dimensional fields no longer decay to zero at the beginning and end of time, one would need to specify nontrivial boundary conditions at the ten-dimensional ends of $\mathbb I_t$. Therefore, let us pick, for our purpose, a physically consistent common half-BPS boundary condition that preserves only a certain one-half of the sixteen worldvolume supersymmetries, such that the remaining eight worldvolume supersymmetries continue to define the $\cN = (4,4)$ supersymmetry of the underlying sigma-model whose worldsheet is now $\Sigma_{n,t} = {\bf S}^1_n \times \mathbb I_t$, whence the spacetime quarter-BPS states mentioned in the last paragraph -- which, at the tips of $\mathbb I_t$, are now spacetime half-BPS states due to the supersymmetry-breaking boundary condition we picked -- would again be captured by the topological sector of the sigma-model. This common half-BPS boundary condition can, for example, be effected by inserting a pair of M9-branes~\cite{M9-branes} whose worldvolumes at the tips of $\mathbb I_t$ span the ten directions along ${\bf S}^1_n \times \mathbb R^4 \times \mathbb R^5$, whence the M5-branes/OM5-plane would intersect them along ${\bf S}^1_n \times \mathbb R^4$. 

If Omega-deformation is now turned on via a fluxbrane as shown in~(\ref{table-AGT-fluxbrane}), i.e., if we consider instead of (\ref{AGT-fluxbrane}) the following compactifications
\be
\underbrace{\mathbb R^4\vert_{\epsilon_1, \epsilon_2}  \times \Sigma_{n,t}}_{\textrm{$N$ M5-branes}}\times \mathbb R^{5}\vert_{\epsilon_3; \,  x_{6,7}}, \quad {\rm and} \quad \underbrace{\mathbb R^4\vert_{\epsilon_1, \epsilon_2}  \times \Sigma_{n,t}}_{\textrm{$N$ M5-branes + OM5-plane}}\times \mathbb R^{5}\vert_{\epsilon_3; \,  x_{6,7}},
\label{AGT-fluxbrane-interval}
\ee
our discussion hitherto would mean in particular that as one traverses around the ${\bf S}^1_n$ circle, the $x_2$-$x_3$ and the $x_4$-$x_5$ planes in $\mathbb R^4\vert_{\epsilon_1, \epsilon_2}$ would be rotated by angles $\epsilon_1$ and $\epsilon_2$ together with an $SU(2)_R$-symmetry rotation of the $G$ gauge theory along $\mathbb R^4\vert_{\epsilon_1, \epsilon_2}$,\footnote{Here, the $R$-symmetry is that of a 4d $\cN =2$ supersymmetry algebra that underlies the $G$ gauge theory along $\mathbb R^4 \vert_{\epsilon_1, \epsilon_2}$ which the surviving worldvolume supercharges are supposed to be associated with.} such that at low-energy distances much larger than $\mathbb I_t$, the partition function of spacetime half-BPS states in (\ref{AGT-fluxbrane-interval}) (which \emph{a priori} is defined as a trace that is tantamount to gluing the two ends of  $\Sigma_{n,t} = {\bf S}^1_n \times \mathbb I_t$ into a two-torus ${\bf S}^1_n \times \mathbb S_t$)  would be given by the following 5d (since $\mathbb S_t \ll \beta$) worldvolume expression (c.f.~\cite[eqns.~(29) and (43)]{NN lectures})
\be
Z_{\rm BPS} (\epsilon_1, \epsilon_2, \vec a, \beta) =  \sum_m {\rm Tr}_{{\cal H}_m} \, {\rm exp} \, \beta (\epsilon_1 J_1 +  \epsilon_2 J_2 + {\vec a \cdot \vec T}),
\label{5d BPS}
\ee 
where $\vec T = (T_1 \dots, T_{{\rm rank} \, G})$ are the generators of the Cartan subgroup of $G$;  $\vec a = (a_1, \dots, a_{{\rm rank} \, G})$ are the corresponding purely imaginary Coulomb moduli of the $G$ gauge theory on $\mathbb R^4 \vert_{\epsilon_1, \epsilon_2}$; $J_{1, 2}$ are the rotation generators of the $x_2$-$x_3$ and $x_4$-$x_5$ planes, respectively, corrected with an appropriate amount of the $SU(2)_R$-symmetry to commute with the two surviving worldvolume supercharges; and ${\cal H}_m$ is the space of holomorphic functions on the moduli space ${\cal M}_{G, m}$ of $G$-instantons on $\mathbb R^4$ with instanton number $m$. 

In fact, as ${\bf g} = {\rm exp} \, \beta (\epsilon_1 J_1 +  \epsilon_2 J_2 + {\vec a \cdot \vec T})$ is a symmetry group of ${\cal M}_{G, m}$, the appearance of ${\bf g}  \in U(1) \times U(1) \times { T}$ in (\ref{5d BPS}) means that Omega-deformation also effects a $\bf g$-automorphism of ${\cal M}_{G, m}$ as we traverse around the ${\bf S}^1_n$ circle, where ${T} \subset G$ is the Cartan subgroup. This point will be important in the next two subsections.

\bigskip\noindent{\it Omega-Deformation in the Dual of (\ref{AGT-fluxbrane-interval}) and Rotations in the Type IIA Spacetime Theory}

According to the duality of the six-dimensional compactifications (\ref{AGT-fluxbrane}) and (\ref{AGT-fluxbrane-dual}), the dual of (\ref{AGT-fluxbrane-interval}) would be given by
 \be
 {\mathbb R^{5}}\vert_{\epsilon_3; \, x_{4,5}} \times \underbrace{\Sigma_{n,t}  \times TN_N^{R\to 0}\vert_{\epsilon_3; \, x_{6,7}}}_{\textrm{$1$ M5-branes}} \quad {\rm and} \quad {\mathbb R^{5}}\vert_{\epsilon_3; \, x_{4,5}} \times \underbrace{\Sigma_{n,t}  \times SN_N^{R\to 0}\vert_{\epsilon_3; \, x_{6,7}}}_{\textrm{$1$ M5-branes}},
 \label{AGT-fluxbrane-dual-interval}
 \ee
where we have the \emph{same} common half-BPS boundary condition  as in  (\ref{AGT-fluxbrane-interval}) at the tips of $ \mathbb I_t$, that is effected by a pair of M9-branes whose worldvolumes at the tips of  $\mathbb I_t$ span the ten directions transverse to it.  

Let us for a moment turn off Omega-deformation in (\ref{AGT-fluxbrane-dual-interval}), i.e., set $\epsilon_3 = 0$. Notice then that (\ref{AGT-fluxbrane-dual-interval}) is equivalent to a (strongly-coupled) type IIA compactification where geometric engineering and the $\mathbb Z_n$-outer-automorphism of the compactification four-manifolds (i) $TN_N^{R\to 0}$ and (ii) $SN_N^{R\to 0}$ (as we go around the ${\bf S}^1_n$ circle) tell us~\cite{vafa} that the Lie algebra underlying the enhanced gauge symmetry of the resulting 6d spacetime theory along $\mathbb R^4 \times  \Sigma_{n,t}$ ought to be (i) the Langlands dual Lie algebra $\frak {su}(N)^\vee$ or $\frak {so}(N+1)^\vee$ when $n =1$ or $2$ (with even $N$), and (ii) the Langlands dual Lie algebra $\frak {so}(2N)^\vee$ or $\frak {usp}(2N - 2)^\vee$ or $\frak {g}^\vee_2$  when $n =1$ or $2$ or $3$ (with $N =4$). In particular, the symmetries of the 2d theory along $\Sigma_{n,t}$ ought to be rooted in $\frak g^\vee$, where $\frak g$ is the Lie algebra of $G$.

Now turn Omega-deformation back on. Then, as one traverses around the ${\bf S}^1_n$ circle, among other things, the $x_4$-$x_5$ plane in ${\mathbb R^{4}}\vert_{\epsilon_3; \, x_{4,5}}$  would be rotated by an angle of $\epsilon_1 + \epsilon_2 = \epsilon_3$ together with an $SU(2)_R$-symmetry rotation of the gauge theory along $ {\mathbb R^{4}}\vert_{\epsilon_3; \, x_{4,5}}$.\footnote{Here, the $R$-symmetry is that of a 4d $\cN =2$ supersymmetry algebra that underlies the gauge theory along ${\mathbb R^{4}}\vert_{\epsilon_3; \, x_{4,5}} $ which the surviving worldvolume supercharges are supposed to be associated with.}  This type IIA spacetime perspective of the Omega-deformation would be relevant in the next two subsections.

\newsubsection{An Equivalence of Spacetime BPS Spectra and a Pure AGT Correspondence for the $A$--$B$ Groups}

We shall now derive, purely physically, a pure AGT correspondence for the $A$--$B$ groups. To this end, recall from (\ref{AGT-fluxbrane-interval}) and (\ref{AGT-fluxbrane-dual-interval}) that we have the following \emph{physically dual} M-theory compactifications
\be
\underbrace{\mathbb R^4\vert_{\epsilon_1, \epsilon_2}  \times \Sigma_{n,t}}_{\textrm{$N$ M5-branes}}\times \mathbb R^{5}\vert_{\epsilon_3; \,  x_{6,7}}  \quad \Longleftrightarrow  \quad   {\mathbb R^{5}}\vert_{\epsilon_3; \, x_{4,5}} \times \underbrace{{\cal C}  \times TN_N^{R\to 0}\vert_{\epsilon_3; \, x_{6,7}}}_{\textrm{$1$ M5-branes}},
\label{AGT-M-duality-AB}
\ee
where we have a common half-BPS boundary condition at the tips of $\mathbb I_t \subset \Sigma_{n,t} = {\bf S}^1_n \times \mathbb I_t$; the radius of ${\bf S}^1_n$ is $\beta$; $\mathbb I_t \ll \beta$; and $\cal C$ is \emph{a priori} the same as $\Sigma_{n,t} $. As usual, there is a $\mathbb Z_n$-outer-automorphism of ${\mathbb R^4}\vert_{\epsilon_1, \epsilon_2}$ and $ TN_N^{R\to 0}\vert_{\epsilon_3; \, x_{6,7}}$ as we go around the ${\bf S}^1_n$ circle and identify the circle under an order $n$ translation, and the $\epsilon_i$'s are parameters of the Omega-deformation along the indicated planes described in detail in the last subsection.  

\bigskip\noindent{\it The Spectrum of Spacetime BPS States on the LHS of (\ref{AGT-M-duality-AB})}

Let us first ascertain the spectrum of spacetime BPS states on the LHS of (\ref{AGT-M-duality-AB}) that define $Z_{\rm BPS} (\epsilon_1, \epsilon_2, \vec a, \beta)$ in (\ref{5d BPS}). In the absence of Omega-deformation whence $\epsilon_i = 0$, according to our discussion in the previous subsection, the spacetime BPS states would be captured by the topological sector of the $\cN = (4,4)$ sigma-model on  $\Sigma_{n,t}$ with target the moduli space ${\cal M}_{G}$ of $G$-instantons on $\mathbb R^4$, where for $n=1$ or 2 (with even $N$), $G = SU(N)$ or $SO(N+1)$, respectively. However, in the presence of Omega-deformation, recall from our discussion immediately after (\ref{5d BPS}) that as one traverses a closed loop in $\Sigma_{n,t}$, there would be a  $\bf g$-automorphism of ${\cal M}_{G}$, where ${\bf g} \in U(1) \times U(1) \times T$, and $T \subset G$ is the Cartan subgroup. Consequently, the spacetime BPS states of interest would, in the presence of Omega-deformation, be captured by the topological sector of a non-dynamically ${\bf g}$-gauged version of the aforementioned sigma-model.\footnote{The relation between $\bf g$-automorphisms of the sigma-model target space and a non-dynamical $\bf g$-gauging of its worldsheet theory, is explained in~\cite[$\S$2.4 and $\S$5]{tsm}. For self-containment, let us review the idea here. Consider a sigma-model with worldsheet $\Sigma_{n,t}$, target space ${\cal M}_G$, and bosonic scalar fields $\Phi$. In the usual case where there is no action on ${\cal M}_G$ as we traverse a closed loop in $\Sigma_{n,t}$, one would consider in the sigma-model path-integral, the space of maps $\Phi : \Sigma_{n,t} \to {\cal M}_G$, which can be viewed as the space of trivial sections of a trivial bundle $X = {\cal M}_G \times \Sigma_{n,t}$. If however there is a  $\bf g$-automorphism of ${\cal M}_G$ as we traverse a closed loop in $\Sigma_{n,t}$, $X$ would have to be a nontrivial bundle given by $ {\cal M}_G \hookrightarrow X \to \Sigma_{n,t}$; then, $\Phi(z, \bar z)$ will not represent a map $\Sigma_{n,t} \to  {\cal M}_G$, but rather, it will be a nontrivial section of $X$. Thus, since $\Phi$ is no longer a function but a nontrivial section of a nontrivial bundle, its ordinary derivatives must be replaced by covariant derivatives. As the nontrivial structure group of $X$ is now $\bf g$, replacing ordinary derivatives by covariant derivatives would mean introducing on $\Sigma_{n,t}$ gauge fields $A^a$, which, locally, can be regarded as $({\rm Lie} \, {\bf g})$-valued one-forms with the usual gauge transformation law ${A^a}' = g^{-1} A^a g + g^{-1} dg$, where $g \in {\bf g}$. This is equivalent to gauging the sigma-model non-dynamically by $\bf g$. \label{gauging worldsheet}} Hence, according to~\cite{mine-equivariant} and our arguments in $\S$3.1 which led us to (\ref{BPS-M}), we can express the Hilbert space ${\cal H}^{\Omega}_{\rm BPS}$ of spacetime BPS states on the LHS of (\ref{AGT-M-duality-AB}) as
\be
{\cal H}^\Omega_{\rm BPS} = \bigoplus_{m} {\cal H}^\Omega_{{\rm BPS}, m}  =  \bigoplus_{m} ~{\rm IH}^\ast_{U(1)^2 \times T} \, {\cal U}({\cal M}_{G, m}), 
\label{BPS-AGT-AB}
\ee
where ${\rm IH}^\ast_{U(1)^2 \times T} \, {\cal U}({\cal M}_{G, m})$ is the $\mathbb Z_n$-invariant (in the sense of (\ref{HBPS-eff}) when $n=2$) $U(1)^2 \times T$-equivariant intersection cohomology of the Uhlenbeck compactification ${\cal U}({\cal M}_{G, m})$ of the (singular) moduli space ${\cal M}_{G, m}$ of $G$-instantons on $\mathbb R^4$ with instanton number $m$.

\bigskip\noindent{\it The Spectrum of Spacetime BPS States on the RHS of (\ref{AGT-M-duality-AB})}

Let us next ascertain the corresponding spectrum of spacetime BPS states on the RHS of (\ref{AGT-M-duality-AB}). Bearing in mind footnote~\ref{junya's twist} which tells us that the underlying worldvolume theory of the single M5-brane is conformal along $TN_N^{R\to 0}\vert_{\epsilon_3; \, x_{6,7}}$  in (\ref{AGT-M-duality-AB}), by repeating our arguments in $\S$3.1 which led us to (\ref{equivalent IIA system 1}) and beyond, and from our discussion surrounding (\ref{metric}), we find that the spacetime BPS states would be furnished by the I-brane theory in the following type IIA configuration:
\be
\textrm{IIA}: \quad \underbrace{ {\mathbb R}^5\vert_{\epsilon_3; x_{4,5}} \times {\cal C} \times {\mathbb R}^3\vert_{\epsilon_3; x_{6,7}}}_{\textrm{I-brane on ${\cal C} = N \textrm{D6} \cap 1\textrm{D4}$}}.
\label{equivalent IIA system 1 - AGT}
\ee
Here, we have  a stack of $N$ coincident D6-branes whose worldvolume is given by ${\mathbb R}^5\vert_{\epsilon_3; x_{4,5}} \times {\cal C}$, and a single D4-brane whose worldvolume is given by ${\cal C} \times \mathbb R^3\vert_{\epsilon_3; x_{6,7}}$. 

Let us for a moment turn off Omega-deformation in (\ref{equivalent IIA system 1 - AGT}), i.e., let $\epsilon_3 = \epsilon_1 + \epsilon_2 = 0$. Then, by applying to (\ref{equivalent IIA system 1 - AGT}) our analysis in $\S$3.1 which eventually led us to (\ref{GL-relation A}) and (\ref{GL-relation B}) from (\ref{equivalent IIA system 1}), we learn that the spacetime BPS states would be furnished by chiral fermions on ${\cal C} $ which couple to the dynamical $U(1)$ gauge degrees of freedom on the single D4-brane that, in turn, can be effectively represented by a  chiral WZW model at level 1 on ${\cal C} $, ${\rm WZW}^{{\rm level} \, 1}_{{\frak g}^\vee_{{\rm aff}}}$, where ${\frak g}^\vee_{{\rm aff}}$ is the \emph{Langlands dual} of the affine $G$-algebra $\frak g_{{\rm aff}}$. This is consistent with our observation after (\ref{AGT-fluxbrane-dual-interval}) that  the symmetries of the 2d theory along ${\cal C}$ ought to be rooted in the Langlands dual Lie algebra $\frak g^\vee$ (and therefore $\frak g^\vee_{\rm aff}$). 

Now turn Omega-deformation back on. As indicated in (\ref{equivalent IIA system 1 - AGT}), as one traverses around a closed loop in ${\cal C}$, the $x_4$-$x_5$ plane in ${\mathbb R^{4}}\vert_{\epsilon_3; \, x_{4,5}} \subset {\mathbb R^{5}}\vert_{\epsilon_3; \, x_{4,5}}$ would be rotated by an angle of $\epsilon_3$ together with an $SU(2)_R$-symmetry rotation of the supersymmetric $SU(N)$ gauge theory along $ {\mathbb R^{4}}\vert_{\epsilon_3; \, x_{4,5}}$. According to our discussion in the last subsection which led us to (\ref{5d BPS}) and slightly beyond, we find that Omega-deformation in this instance would effect a ${\bf g}'$-automorphism of ${\cal M}_{SU(N), m}$ as we traverse around a closed loop in ${\cal C}$, where ${\cal M}_{SU(N), m}$ is the moduli space of $SU(N)$-instantons on $\mathbb R^4$ with instanton number $m$; ${\bf g}' =  {\rm exp} \, \beta (\epsilon_3 J_3 + {{\vec a} \cdot {\vec T}'})$; $J_3$ is the rotation generator of the $x_4$-$x_5$ plane corrected with an appropriate amount of $SU(2)_R$-symmetry to commute with the D6-brane worldvolume supercharges; ${\vec T}' = (T'_1 \dots, T'_{{\rm rank} \, SU(N)})$ are the generators of the Cartan subgroup $T' \subset SU(N)$;  and ${\vec a} = (a_1, \dots, a_{{\rm rank} \, SU(N)})$ are the corresponding purely imaginary Coulomb moduli of the $SU(N)$ gauge theory on $\mathbb R^4 \vert_{\epsilon_3; \, x_{4,5}}$. In fact, since ${\cal M}_{SU(N), m}$ is also the space of self-dual connections of an $SU(N)$-bundle on $\mathbb R^4$, and since these self-dual connections correspond to differential one-forms valued in the Lie algebra $\frak {su}(N)$, Omega-deformation also means that there is a ${\bf g}'$-automorphism of the space of elements of $\frak {su}(N)$ and thus $SU(N)$,  as we traverse a closed loop in ${\cal C}$.

Note at this point that in the above, $\frak g^\vee_{\rm aff}  \simeq \frak {su}(N)^{(n)}_{\rm aff}$, where $\frak {su}(N)^{(n)}_{\rm aff}$ is a $\mathbb Z_n$-twisted affine $SU(N)$-algebra. What this means is that ${\rm WZW}^{{\rm level} \, 1}_{{\frak g}^\vee_{{\rm aff}}}$ can be regarded as a (chiral half of a) $SU(N)$ WZW model at level 1 that is $\mathbb Z_n$-twisted on ${\cal C}$. Since a $\cal G$ WZW model on $\Sigma$ is a bosonic sigma-model on $\Sigma$ with target the $\cal G$-manifold, according to the last paragraph,  it would mean that Omega-deformation would effect a ${\bf g}'$-automorphism  of the target space of ${\rm WZW}^{{\rm level} \, 1}_{{\frak g}^\vee_{{\rm aff}}}$ as we traverse a closed loop in ${\cal C}$, where ${\bf g}' \in U(1) \times T'$.  In turn, according to footnote~\ref{gauging worldsheet}, it would mean that in the presence of Omega-deformation, we would have to non-dynamically gauge ${\rm WZW}^{{\rm level} \, 1}_{{\frak g}^\vee_{{\rm aff}}}$ by $ U(1) \times T'$. 

That being said, notice also from (\ref{equivalent IIA system 1 - AGT}) that as one traverses around a closed loop in ${\cal C}$, the $x_6$-$x_7$ plane in ${\mathbb R^{3}}\vert_{\epsilon_3; \, x_{6,7}}$ would be rotated by an angle of $\epsilon_3$ together with an $R$-symmetry rotation of the supersymmetric $U(1)$ gauge theory living on the single D4-brane, i.e., Omega-deformation is also being turned on along the D4-brane. Now recall from our arguments leading up to (\ref{partially gauged CFT - AB}) that because the $U(1)$ gauge field on the D4-brane --  unlike the $SU(N)$ gauge field on the D6-branes -- is dynamical,  one has to reduce away in the I-brane system the $U(1)$ WZW model associated with the D4-brane. Similarly,  the aforedescribed Omega-deformation along the D4-brane would act not to enlarge  but to \emph{reduce} the $ U(1) \times T'$ Omega-deformation factor in the previous paragraph by $R = U(1) \times \mathbb T$, where $U(1) \subset R$ is associated with the $\epsilon_3$-rotation of the $x_6$-$x_7$ plane in ${\mathbb R^{3}}\vert_{\epsilon_3; \, x_{6,7}}$, and ${\mathbb T} \subset R$ is the Cartan of the gauge group on the D4-brane, i.e., $\mathbb T = U(1)$. In short, we would in fact have to non-dynamically gauge ${\rm WZW}^{{\rm level} \, 1}_{{\frak g}^\vee_{{\rm aff}}}$ not by $ U(1) \times T'$ but by ${\cal T} \subset T'$. 

At any rate, because $SU(N)/T' \simeq SL(N, \mathbb C) / B_+$, where $B_+$ is a Borel subgroup, it would mean that $SU(N)/{\cal T} \simeq (SL(N, \mathbb C) / B_+) \times (T' / \cal T)$. Also, $T' / \cal T$ is never bigger than the Cartan subgroup $C \subset B_+ = C \times N_+$, where $N_+$ is the subgroup of strictly upper triangular matrices which are nilpotent and traceless whose Lie algebra is $\frak n_+$. Altogether, this means that our gauged WZW model which corresponds to the coset model $SU(N)/{\cal T}$, can also be studied as an $S$-gauged  $SL(N, \mathbb C)$ WZW model which corresponds to the coset model  $SL(N, \mathbb C) / S$, where $N_+ \subseteq S \subset B_+$. As physically consistent $\cal H$-gauged $\cal G$ WZW models are such that $\cal H$ is necessarily a connected subgroup of $\cal G$, it will mean that $S = N_+$. Therefore, what we ought to ultimately consider is an $N_+$-gauged $SL(N, \mathbb C)$ WZW model. 

Before we proceed any further, let us make a slight deviation to highlight an important point regarding the effective geometry of ${\cal C}$. As the simple roots of $N_+$ form a subset of the simple roots of $SL(N,\mathbb C)$, the level of the affine $N_+$-algebra ought to be the equal to the level of the affine $SL(N,\mathbb C)$-algebra~\cite{Ketov} which is 1.  However, it is clear from our discussion hitherto that the affine $N_+$-algebra, in particular its level, will depend nontrivially on the Omega-deformation parameters which may or may not take integral values; in other words, its level will \emph{not} be equal to 1. A resolution to this conundrum is as follows. A deviation of the level of the affine $N_+$-algebra from 1 would translate into a corresponding deviation of its central charge; since a central charge arises due to an introduction of a macroscopic scale in the 2d system which results from a curvature along ${\cal C}$~\cite{CFT text}, it would mean that Omega-deformation ought to deform  the \emph{a priori} flat ${\cal C}  = \Sigma_{n,t}$ into a curved Riemann surface with the same topology -- that is, a Riemann sphere with two punctures -- such that the anomalous deviation in the central charge and thus level, can be consistently ``absorbed'' in the process.\footnote{A geometrical modification of ${\cal C}$ due to Omega-deformation has also been justified in~\cite{orlando}.\label{deform C}} Thus, we effectively have $ {\cal C} =  {\bf S}^2 / \{ 0, \infty \}$, so $\cal C$ can be viewed as an ${\bf S}^1_n$ fibration of $\mathbb I_t$ whose fiber has zero radius at the two end points $z = 0$ and $z = \infty$, where `$z$' is a holomorphic coordinate on $\cal C$.

Coming back to our main discussion, it is clear that in the schematic notation of $\S$3.1, our $N_+$-gauged $SL(N, \mathbb C)$ WZW model can be expressed as the partially gauged chiral CFT
\be
\frak {sl}(N)^{(n)}_{\rm aff, 1} / {\frak{n}_+}^{(n)}_{{\rm aff}, p}
\label{AGT-AB-chiral CFT}
\ee
on $\cal C$, where the level $p$ would, according to our discussions hitherto, necessarily depend on the Omega-deformation parameters $\epsilon'_1 = \beta \epsilon_1$ and $\epsilon'_2 = \beta \epsilon_2$. ($p$, being a purely real number, should not depend on the purely imaginary parameter $\vec a' = \beta \vec a$).

In sum, the sought-after spacetime BPS states ought to be given by the states of the partially gauged chiral CFT in (\ref{AGT-AB-chiral CFT}), and via $\S$B and~\cite[eqn.~(6.67)]{review}, we find that this chiral CFT realizes  ${\cal W}({\frak g}^\vee_{\rm aff})$ -- a  $\mathbb Z_n$-twisted version of the affine $\cal W$-algebra ${\cal W}(\widehat {\frak {sl}(N)})$ obtained from $\frak {sl}(N)_{\rm aff}$ via a quantum Drinfeld-Sokolov reduction. In other words,  the states of the chiral CFT would be furnished by a Verma module  $\widehat{{\cal W}}({\frak g}^\vee_{\rm aff})$ over ${\cal W}({\frak g}^\vee_{\rm aff})$, and the Hilbert space ${\cal H}^{\Omega'}_{\rm BPS}$ of spacetime BPS states on the RHS of (\ref{AGT-M-duality-AB}) can be expressed as
\be
{\cal H}^{\Omega'}_{\rm BPS}  = \widehat{{\cal W}}({\frak g}^\vee_{\rm aff}).
\label{AGT-AB-H=W}
\ee

\bigskip\noindent{\it A Pure AGT Correspondence for the $A$--$B$ Groups}

Clearly, the physical duality of the compactifications in (\ref{AGT-M-duality-AB}) will mean that ${\cal H}^\Omega_{\rm BPS}$ in (\ref{BPS-AGT-AB}) is equivalent to ${\cal H}^{\Omega'}_{\rm BPS}$ in (\ref{AGT-AB-H=W}), i.e.,
\be
 \boxed{\bigoplus_{m} ~{\rm IH}^\ast_{U(1)^2 \times T} \, {\cal U}({\cal M}_{G, m}) = \widehat{{\cal W}}({\frak g}^\vee_{\rm aff})}
 \label{AGT-duality-A}
\ee
Thus, we have a 4d-2d duality relation in the sense of  (\ref{GL-relation A}) and (\ref{GL-relation B}). 

According to footnote~\ref{central charge} and (\ref{cc-simply-laced}) -- bearing in mind that (i) $\textrm{dim} \,\frak {sl}(N) = N^2 -1$;  (ii) ${\rm rank} \, \frak {sl}(N) = N -1$; and (iii) $h^\vee_{\frak {sl}(N)} = N$ -- the central charge of  ${\cal W}({\frak g}^\vee_{\rm aff})$ is
\be
 c_{A}  = (N-1) -  (N^3-N) \left( \alpha_+ + \alpha_- \right)^2,
\label{cc-A}
\ee
where $\alpha_+ \alpha_- = -1$; $\alpha_+ = 1/ \sqrt{k'+ N}$; and $k' \in \mathbb R$ is the effective level of the underlying affine Lie algebra $\frak {sl}(N)_{\rm aff}$. Note at this point that (\ref{AGT-AB-chiral CFT}) means that we can also write $c_A = c(\frak {sl}(N)^{(n)}_{\rm aff, 1}) - c({\frak{n}_+}^{(n)}_{{\rm aff}, p})$, and since the central charge $c(\frak {sl}(N)^{(n)}_{\rm aff, 1}) = N -1$, according to (\ref{cc-A}), we can also write $c({\frak{n}_+}^{(n)}_{{\rm aff}, p}) = (N^3-N) \left( \alpha_+ + \alpha_- \right)^2$. 

As mentioned, $p$ would depend on the Omega-deformation parameters $\epsilon'_1 = \beta \epsilon_1$ and $\epsilon'_2 = \beta \epsilon_2$; thus, so would $\alpha_{\pm}$. Because $\alpha_+ \alpha_- = -1$, it would mean that we can write $\alpha_+ = if(\epsilon'_1, \epsilon'_2)$ and $\alpha_- = i f^{-1}(\epsilon'_1, \epsilon'_2)$, where $f(\epsilon'_1, \epsilon'_2)$ is some possibly complex function. Because (\ref{AGT-M-duality-AB}) is symmetric under the exchange $\epsilon'_1 \leftrightarrow \epsilon'_2$, so must $c_A$; in particular, we ought to have $f(\epsilon'_1, \epsilon'_2)+  f^{-1}(\epsilon'_1, \epsilon'_2) = f(\epsilon'_2, \epsilon'_1)+  f^{-1}(\epsilon'_2, \epsilon'_1)$.  Because $\alpha_+ = 1/ \sqrt{k'+ N} = if(\epsilon'_1, \epsilon'_2)$ would go from positive real to negative purely complex as we vary $k'$, it would mean that $f(\epsilon'_1, \epsilon'_2)$ must also go from negative purely complex to negative real as we vary the $\epsilon'_i$'s. Because we have a geometrical ${\bf g}'' = {\rm exp} [(\epsilon'_1 + \epsilon'_2)J_3] = {\rm exp} [(\lambda \epsilon'_1 + \lambda \epsilon'_2) \lambda^{-1}J_3]$ automorphism associated with the Omega-deformation in (\ref{equivalent IIA system 1 - AGT}), and since $c_A$ is only a function of $\epsilon'_{1,2}$ and not of $J_3$, we ought to have $c_A(\epsilon'_1, \epsilon'_2) = c_A(\lambda\epsilon'_1, \lambda\epsilon'_2)$; in other words, $\alpha_+ + \alpha_-$ ought to be invariant under $\epsilon'_i \to \lambda \epsilon'_i$, where $\lambda$ is some real constant. Altogether therefore, it would mean that we can write $\alpha_+ =  - i \sqrt {\epsilon'_1 / \epsilon'_2} =  - i \sqrt {\epsilon_1 / \epsilon_2}$ and $\alpha_- =  - i \sqrt {\epsilon'_2 / \epsilon'_1} =  - i \sqrt {\epsilon_2 / \epsilon_1}$; in turn, (\ref{cc-A}) would be given by
\be
\label{c-A-epsilon}
\boxed{c_{A, \epsilon_{1,2} }  = (N-1)  + (N^3 - N) {(\epsilon_1 + \epsilon_2)^2 \over {\epsilon_1\epsilon_2}}}
\ee
where in addition, we would have
\be
\label{k'}
\boxed{k'  = - N - b^{-2} \qquad {\rm and} \qquad b = \sqrt {\epsilon_1 / \epsilon_2}}
\ee
so that one can also write
\be
c_{A, \epsilon_{1,2} } = c(\frak {su}(N)^{(n)}_{\rm aff, 1}) + c(\Omega_{\epsilon_{1,2}}),
\ee
where
\be
 c(\Omega_{\epsilon_{1,2}}) =  h^\vee_{\frak {su}(N)} {\rm dim} \, \frak {su}(N) \left(b + {1 \over b}\right)^2
\label{cc-A-Omega}
\ee
can be regarded as an Omega-deformation-induced central charge. (Notice that there is no $\beta$-dependence in the formulas (\ref{c-A-epsilon})--(\ref{cc-A-Omega}); this is consistent with the fact that these formulas are supposed to be globally-defined on $\cal C$ independent of the varying radius of ${\bf S}^1_n \subset \cal C$.)

The Verma module $\widehat{{\cal W}}({\frak g}^\vee_{\rm aff})$ is generated by the application of creation operators $W^{(s_i)}_{m < 0}$ on its $\mathbb Z_n$-twisted highest  weight state $| \Delta \rangle$, where the $W^{(s_i)}_{m < 0}$'s are the negative modes of the spin-$s_i$ fields $W^{(s_i)}(z)$ on $\cal C$ which span ${\cal W}({\frak g}^\vee_{\rm aff})$, and $m \in \mathbb Z/n$. On the other hand, $| \Delta \rangle$ is annihilated by the annihilation operators $W^{(s_i)}_{m > 0}$, where the $W^{(s_i)}_{m > 0}$'s are the positive modes of the $W^{(s_i)}(z)$ fields on $\cal C$ which also span ${\cal W}({\frak g}^\vee_{\rm aff})$. Nonetheless, we have $W^{(s_i)}_{ 0} | \Delta \rangle = \Delta^{(s_i)} | \Delta \rangle$, where the $\Delta^{(s_i)}$'s are Weyl-invariant polynomials in ${\bf a} = {\bf J}_0 + (\alpha_+ + \alpha_-) \rho$; ${\bf J}_0 = (J^1_0, \dots, J_0^{{\rm rank} \, \frak {sl}(N)})$ are the zeroth modes of the ${\rm rank} \, \frak {sl}(N)$ untwisted scalar bosonic fields in the free-field realization of ${\cal W}({\frak g}^\vee_{\rm aff})$; and $\rho$ is the Weyl vector of $\frak {sl}(N)$~\cite{review}. For example, $W^{(2)}_{ 0} | \Delta \rangle = L_0 | \Delta \rangle = \Delta^{(2)} | \Delta \rangle$, where $L_0$ is the zeroth mode of the stress tensor $T(z) = W^{(2)}(z)$, and $\Delta^{(2)} = \left({\bf a}^2 - (\alpha_+ + \alpha_-)^2 \rho^2 \right) /2 = {\bf a}^2/2  + (N^3 - N)(\epsilon_1 + \epsilon_2)^2/24 \epsilon_1 \epsilon_2$. (See~\cite[eqn.~(6.18)]{review},  and note that just like the quantities in (\ref{c-A-epsilon})--(\ref{cc-A-Omega}), $\Delta^{(2)}$ should be $\beta$-independent, as is the case.) 

Recall at this point that $L_0$ generates translations along the ${\bf S}^1_n$ fiber in $\cal C$, and since the presence of Omega-deformation means that there is a rotation of an $\mathbb R^4$ space and the gauge field over it as we go around the ${\bf S}^1_n$ (c.f.~our earlier discussion on a ${\bf g}'$-automorphism), $L_0$ should be related to the rotation parameters $(\epsilon_1, \epsilon_2, \vec a)$. Indeed, we saw in the last paragraph that $L_0$ has eigenvalues which depend on $(\epsilon_1, \epsilon_2, \bf a)$, and since $\bf a$, like $\vec a$, is a vector whose number of components even coincides with that of $\vec a$ when $G = SU(N)$, we can naturally identify $ \bf a$ with $-i\vec a$, where a factor of $-i$ is needed because $\vec a$ is purely imaginary while $\bf a$ is purely real. That said, because of (\ref{AGT-duality-A}), it would mean that the symmetries of $\Delta^{(2)}$ ought to be compatible with the symmetries of the partition function $Z_{\rm BPS} (\epsilon_1, \epsilon_2, \vec a, \beta)$ of ${\cal H}^{\Omega}_{\rm BPS}$ in (\ref{5d BPS}); in particular, since $Z_{\rm BPS} (\epsilon_1, \epsilon_2, \vec a, \beta)$ is invariant under the simultaneous rescalings $(\beta, \epsilon_1, \epsilon_2, \vec a) \to ( \zeta^{-1} \beta, \zeta \epsilon_1, \zeta \epsilon_2, \zeta \vec a)$, where $\zeta$ is some real constant, the $\beta$-independent $\Delta^{(2)}$ must be  invariant under the simultaneous rescalings $(\epsilon_1, \epsilon_2, \vec a) \to (\zeta \epsilon_1, \zeta \epsilon_2, \zeta \vec a)$. Furthermore, because (\ref{AGT-M-duality-AB}) is symmetric under the exchange $\epsilon_1 \leftrightarrow \epsilon_2$, so must $\Delta^{(2)}$. In sum, we ought to have ${\bf a} \sim -i  \vec a / \sqrt {\epsilon_1\epsilon_2}$, whence we can write
\be
\label{L_0}
\boxed{W^{(2)}_{ 0} | \Delta \rangle = \Delta^{(2)} | \Delta \rangle}
\ee 
where
\be
\label{Delta(2)}
\boxed{\Delta^{(2)} = {(N^3 - N) \over 24}{(\epsilon_1 + \epsilon_2)^2 \over \epsilon_1 \epsilon_2} - { \gamma  {\vec a}^2 \over  \epsilon_1\epsilon_2}}
\ee
for some real constant $\gamma$.
 
In the limit that $\beta \to 0$, it is well-known~\cite{abcd} that
 \be
 \label{abcd BPS relation}
 Z_{\rm BPS} (\epsilon_1, \epsilon_2, \vec a, \beta) =  \sum_m Z_{{\rm BPS}, m} (\epsilon_1, \epsilon_2, \vec a, \beta)
 \ee 
 of (\ref{5d BPS}) behaves such that $Z_{{\rm BPS}, m} (\epsilon_1, \epsilon_2, \vec a, \beta \to 0)  \sim \beta^{-2mh^\vee_{\frak g}} \, Z^{\rm 4d}_{{\rm BPS}, m} (\epsilon_1, \epsilon_2, \vec a)$, whence the Nekrasov instanton partition function $Z_{\rm inst}  (\Lambda, \epsilon_1, \epsilon_2, \vec a)= \sum_m \Lambda^{2mh^\vee_{\frak g}} \, Z^{\rm 4d}_{{\rm BPS}, m} (\epsilon_1, \epsilon_2, \vec a) $ can be written as 
\be
Z_{\rm inst}(\Lambda, \epsilon_1, \epsilon_2, \vec a) =  \sum_m \Lambda^{2mh^\vee_{\frak g}} \, Z'_{{\rm BPS}, m} (\epsilon_1, \epsilon_2, \vec a, \beta \to 0), 
\label{Z_inst-5d}
\ee
where $Z'_{{\rm BPS}, m} = l_m \beta^{2mh^\vee_{\frak g}} Z_{{\rm BPS}, m}$; $l_m$ is some constant; and $\Lambda$ can be interpreted as the inverse of the observed scale of the $\mathbb R^4\vert_{\epsilon_1, \epsilon_2}$ space on the LHS of (\ref{AGT-M-duality-AB}). 

The expression for $Z_{\rm inst}$ in (\ref{Z_inst-5d}) is indeed consistent with (a) its original definition in~\cite{NN} as a sum of weighted integrals  over ${\cal U}({\cal M}_{G,m})$ of the exponent of the Hamiltonian of a $U(1)^2 \times T$ action against the symplectic measure,\footnote{More precisely, it is the  Gieseker compactification ${\cal G}({\cal M}_{G,m})$ of ${\cal M}_{G,m}$ that is considered in~\cite{NN}, where ${\cal G}({\cal M}_{G,m})$ is just a smooth resolution of the singular Uhlenbeck compactification ${\cal U}({\cal M}_{G,m})$. However, we will continue to formulate our results in terms of ${\cal U}({\cal M}_{G,m})$ to be consistent with the earlier parts of the paper, and to also make contact with the mathematical literature~\cite{AGT-math, Vasserot} on the subject.\label{about Geiseker compactification}}  and (b) the fact that $Z'_{{\rm BPS}, m}$ counts (with weights) the states in ${\cal H}^\Omega_{{\rm BPS}, m} = {\rm IH}^\ast_{U(1)^2 \times T} \, {\cal U}({\cal M}_{G, m})$. To see this, first note that from (a), we can also write
\be
Z_{\rm inst} (\Lambda, \epsilon_1, \epsilon_2, \vec a) = \sum_m \Lambda^{2mh^\vee_{\frak g}} \, \int_{{\cal U}({\cal M}_{G,m})} {\rm exp} \, [ \omega + \mu( \epsilon_1, \epsilon_2, \vec a) ],
\label{Z_inst-NN}
\ee 
where $\omega$ is a symplectic form on ${\cal U}({\cal M}_{G,m})$, invariant under the $U(1)^2 \times T$ action, and $\mu: {\cal U}({\cal M}_{G,m}) \to \xi^\ast$ is a moment map, where $\xi = {\rm Lie}(U(1)^2 \times T) = (\epsilon_1, \epsilon_2, \vec a)$.\footnote{The expression for $Z_{\rm inst}$ was originally stated in~\cite{NN} in terms of the smooth Geiseker compactification ${\cal G}({\cal M}_{G,m})$ instead of ${\cal U}({\cal M}_{G,m})$. Nevertheless, since the equivariant cohomology ${\rm H}^\ast_{U(1)^2 \times T} \, {\cal G}({\cal M}_{G,m})$ is equal to the equivariant intersection cohomology  ${\rm IH}^\ast_{U(1)^2 \times T} \, {\cal U}({\cal M}_{G, m})$ (c.f. footnote~\ref{about Geiseker compactification} and~\cite[$\S$4]{AGT-math}), we can also state $Z_{\rm inst}$ in terms of ${\cal U}({\cal M}_{G,m})$, as was done mathematically in~\cite[$\S$6]{AGT-math}.}

Next, note that $\tilde \omega = \omega + \mu$ is a  $U(1)^2 \times T$-equivariant symplectic form on (singular) ${\cal U}({\cal M}_{G,m})$~\cite{sternberg}, and moreover, it is a class in ${\rm IH}^\ast_{U(1)^2 \times T} \, {\cal U}({\cal M}_{G, m})$; thus, by the Duistermaat-Heckmann theorem, we can write the terms on the RHS of (\ref{Z_inst-NN}) as~\cite{sternberg}
\be
\int_{{\cal U}({\cal M}_{G,m})} {\rm exp} \, [ \omega + \mu(\vec a, \epsilon_1, \epsilon_2) ] = (2 \pi)^d \sum_{{{\vec p}_m}} \, {e^{\mu_{{{\vec p}_m}}(\xi)} \over \Pi^d_{i=1} \,\alpha_{i, {{\vec p}_m}}(\xi)},
\label{rhs of Z_inst-NN}
\ee
where $d = {\rm dim}_\mathbb C \, {\cal U}({\cal M}_{G,m})$; the set $\{{{\vec p}_m} \}$ are the fixed-points of the $U(1)^2 \times T$-action on ${\cal U}({\cal M}_{G,m})$;  $\mu_{{{\vec p}_m}}$ is the restriction of $\mu$ to ${{\vec p}_m}$; and $\alpha_{i, {{\vec p}_m}}(\xi)$ are the weights of the $U(1)^2 \times T$-action on the tangent space to ${{\vec p}_m}$. 

Last but not least, note that equivariant localization~\cite{Atiyah-Bott} implies that ${\rm IH}^\ast_{U(1)^2 \times T} \, {\cal U}({\cal M}_{G, m})$ must be endowed with an orthogonal basis $\{| {{\vec p}_m} \rangle \}$  that is in one-to-one correspondence with the fixed-point set $\{ {{\vec p}_m} \}$.\footnote{See~\cite[eqn.~(3.10)]{Kanno-Tachikawa} where this fact was also exploited.}  Thus, since  according to (b), $Z'_{{\rm BPS}, m}$ is a weighted count of the states in ${\cal H}^\Omega_{{\rm BPS}, m} = {\rm IH}^\ast_{U(1)^2 \times T} \, {\cal U}({\cal M}_{G, m})$, it would mean that one can write
\be
Z'_{{\rm BPS}, m} (\epsilon_1, \epsilon_2, \vec a, \beta \to 0) =  \sum_{{{\vec p}_m}} \,  l^2_{{{\vec p}_m}} (\epsilon_1, \epsilon_2, \vec a)\langle {{\vec p}_m} | {{\vec p}_m} \rangle,
\label{rhs of Z_inst-5d}
\ee
where $l_{{{\vec p}_m}} (\epsilon_1, \epsilon_2, \vec a) \in \mathbb R$, and the dependence on $\epsilon_1$, $\epsilon_2$ and $\vec a$ arises because the energy level of each state (given by the eigenvalue of the $L_0$ operator which generates translation along ${\bf S}^1_n \subset \Sigma_{n,t}$ in (\ref{AGT-M-duality-AB}) whence there is an Omega-deformation twist of the theory along the orthogonal spaces indicated therein) ought to depend on these Omega-deformation parameters. Comparing (\ref{Z_inst-NN}) with (\ref{Z_inst-5d}), and then comparing (\ref{rhs of Z_inst-NN}) with (\ref{rhs of Z_inst-5d}), we get
\be
  { (2 \pi)^d \, e^{\mu_{{{\vec p}_m}}(\epsilon_1, \epsilon_2, \vec a)} \over \Pi^d_{i=1} \,\alpha_{i, {{\vec p}_m}}(\epsilon_1, \epsilon_2, \vec a)} = l^2_{{{\vec p}_m}} (\epsilon_1, \epsilon_2, \vec a) \langle {{\vec p}_m} | {{\vec p}_m} \rangle.
\ee 
Thus, we find our assertion that the expression for $Z_{\rm inst}$ in (\ref{Z_inst-5d}) is indeed consistent with facts (a) and (b), to be true.

Notice that (\ref{rhs of Z_inst-5d}) also means that
\be
Z'_{{\rm BPS}, m} (\epsilon_1, \epsilon_2, \vec a, \beta \to 0) =   \langle \Psi_m | \Psi_m \rangle,
\label{Z'}
\ee
where
\be
|\Psi_m \rangle = \bigoplus_{{{\vec p}_m}} \,  l_{{{\vec p}_m}}    | {{\vec p}_m} \rangle.
\ee 
Here, the state $| \Psi_m \rangle \in {\rm IH}^\ast_{U(1)^2 \times T} \, {\cal U}({\cal M}_{G, m})$, and $ \langle \cdot | \cdot \rangle$ is a Poincar\'e pairing in the sense of~\cite[$\S$2.6]{J-function}.

Now consider the state
\be
| \Psi \rangle = \bigoplus_m  \Lambda^{m h^\vee_{\frak g}} | \Psi_m \rangle.
\ee
By substituting (\ref{Z'}) in the RHS of (\ref{Z_inst-5d}), and by noting that $\langle \Psi_m | \Psi_n \rangle = \delta_{mn}$, one can immediately see that
\be
Z_{\rm inst} (\Lambda, \epsilon_1, \epsilon_2, \vec a) = \langle \Psi |  \Psi \rangle,
\label{Psi | Psi}
\ee
where $| \Psi \rangle \in \bigoplus_m {\rm IH}^\ast_{U(1)^2 \times T} \, {\cal U}({\cal M}_{G, m})$.  In turn, the duality relation (\ref{AGT-duality-A}) would mean that 
\be
\boxed{| \Psi \rangle = | q, \Delta \rangle}
\label{psi = delta}
\ee
whence  
\be
\label{q | q}
\boxed{Z_{\rm inst} (\Lambda, \epsilon_1, \epsilon_2, \vec a) = \langle q, \Delta   | q, \Delta \rangle}
\ee
where $| q, \Delta \rangle \in \widehat{{\cal W}}({\frak g}^\vee_{\rm aff})$. (The meaning of the label `$q$' will be clear shortly.) Since the RHS of (\ref{q | q}) is defined at $\beta \to 0$ (see the RHS of (\ref{Z_inst-5d})), and since we have in $\cal C$ a \emph{common} boundary condition at $z = 0$ and $z=\infty$,  $| q, \Delta \rangle$ and $\langle q, \Delta |$ ought to be a state and its dual associated with the puncture at $z = 0$ and $z=\infty$, respectively (as $z = 0, \infty$ are the points in $\cal C$ where the ${\bf S}^1_n$ fiber   has zero radius). This is depicted in fig.~1.  

\begin{figure}
  \centering
    \includegraphics[width=0.4\textwidth]{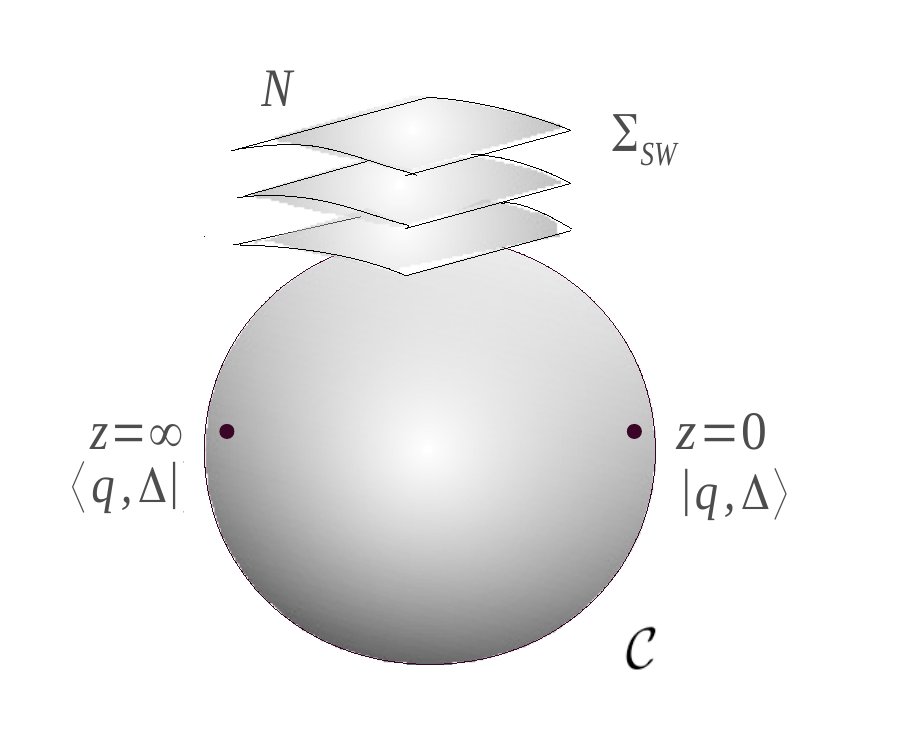}
  \caption{$\cal C$ and its $N$-fold cover $\Sigma_{SW}$ with the states $\langle q, \Delta |$ and $| q, \Delta \rangle$ at $z=0$ and $\infty$}
\end{figure}

At any rate, since we have $N$ D6-branes and $1$ D4-brane wrapping $\cal C$ (see (\ref{equivalent IIA system 1 - AGT})), we effectively have an $N \times 1 = N$-fold cover $\Sigma_{SW}$ of $\cal C$. This is also depicted in fig.~1. Incidentally, $\Sigma_{SW}$ is also the Seiberg-Witten curve which underlies $Z_{\rm inst} (\Lambda, \epsilon_1, \epsilon_2, \vec a)$! Moreover, it is by now well-established (see~\cite{Lerche} and references therein) that for $n=1$, i.e., $G = SU(N)$, $\Sigma_{SW}$ can be described in terms of the algebraic relation 
\be
\Sigma_{SW}: \lambda^N + \phi_2(z) \lambda^{N-2} + \dots + \phi_N(z) = 0, 
\ee
where $\lambda = y  dz /z$ (for some complex variable $y$) is a section of $T^\ast \cal C$; the $\phi_s(z)$'s are $(s,0)$-holomorphic differentials on $\cal C$ given by
 \be
 \label{phi(z)}
\phi_j (z)= u_j \left({dz \over z}\right)^j \quad {\rm and} \quad \phi_N(z) = \left( z + u_N + {\Lambda^N \over z}\right)\left({dz \over z}\right)^N,
\ee
where $j = 2, 3, \dots, N-1$; while for weights $\lambda_1, \dots, \lambda_N$ of the $N$-dimensional representation of $SU(N)$, and for $s=2, 3, \dots, N$,
\be
\label{u's}
u_s = (-1)^{s+1} \sum_{k_1 \neq \dots \neq k_s} e_{\lambda_{k_1}} e_{\lambda_{k_2}} \dots e_{\lambda_{k_s}} (\vec a) \quad {\rm and} \quad e_{\lambda_{r}}  = \vec a \cdot \lambda_r.  
\ee
This is consistent with our results established in $\S$B that for $G = SU(N)$, we have, on $\cal C$, the following $(s_i, 0)$-holomorphic differentials 
\be
\label{W's}
W^{(s_i)}(z) = \left(\sum_{l \in \mathbb Z} {W^{(s_i)}_l \over z^l} \right) \left( dz \over z \right)^{s_i}, \quad {\rm where} \quad s_i = e_i + 1 = 2, 3, \dots, N,
\ee
whence we can naturally identify, up to some constant factor,  $\phi_s (z)$ with  $W^{(s)}(z)$. (In fact, a $U(1)$ $R$-symmetry of the 4d theory along $\mathbb R^4\vert_{\epsilon_1, \epsilon_2}$ on the LHS of (\ref{AGT-M-duality-AB}) which underlies $Z_{\rm inst} (\Lambda, \epsilon_1, \epsilon_2, \vec a)$ and $\Sigma_{SW}$, can be identified with the rotational symmetry of ${\bf S}^1_n$; the duality relation (\ref{AGT-M-duality-AB}) then means that the corresponding $U(1)$ $R$-charge of the $\phi_s(z)$  operators that define  $\Sigma_{SW}$, ought to match, up to a constant, the conformal dimension of the $W^{(s)}(z)$ operators on $\cal C$, which is indeed the case.) 

At $z = 0$ where we have the state $|q, \Delta \rangle$, we find, after comparing (\ref{phi(z)}) with (\ref{W's}), that 
\be
\label{W-2}
\boxed{W^{(s)}_{l \geq 2} \, |q, \Delta \rangle = 0, \quad {\rm for} \quad s = 2, 3, \dots, N}
\ee
$W^{(s)}_0|q, \Delta \rangle \sim u_s|q, \Delta \rangle$, and $W^{(N)}_1 |q, \Delta \rangle = q |q, \Delta \rangle  \sim \Lambda^N |q, \Delta \rangle$. To determine the exact form of the relation involving $W^{(s)}_0$, note that as in our derivation of (\ref{Delta(2)}), i.e., the eigenvalue $\Delta^{(2)}$ of $W^{(2)}_0$,  the eigenvalues of $W^{(s)}_0$ must be invariant under the simultaneous rescalings $(\epsilon_1, \epsilon_2, \vec a) \to (\zeta \epsilon_1, \zeta \epsilon_2, \zeta \vec a)$ and the exchange $\epsilon_1 \leftrightarrow \epsilon_2$; since (\ref{u's}) tells us that $u_s$ is of order $s$ in $\vec a$, it must be that
\be
\label{Ws0}
\boxed{W^{(s)}_0|q, \Delta \rangle  = { u_s \over (\epsilon_1 \epsilon_2)^{s / 2}} \,  |q, \Delta \rangle, \quad {\rm for} \quad s = 2, 3, \dots, N}
\ee
To determine the exact form of the relation involving $W^{(N)}_1$, recall that since the underlying worldvolume theory of the $N$ M5-branes on the LHS of (\ref{AGT-M-duality-AB}) is scale-invariant, it would mean that in addition to possessing the symmetries of $Z_{\rm BPS} (\epsilon_1, \epsilon_2, \vec a, \beta)$ in (\ref{5d BPS}), the $W^{(N)}_1$-eigenvalue $q \sim \Lambda^N$ ought to also be invariant under the rescaling $(\Lambda, \beta) \to (\zeta^{-1} \Lambda, \zeta \beta)$; thus, as the rescaling $(\beta, \epsilon_1, \epsilon_2, \vec a) \to ( \zeta \beta, \zeta^{-1} \epsilon_1, \zeta^{-1} \epsilon_2, \zeta^{-1} \vec a)$ is a symmetry of $Z_{\rm BPS} (\epsilon_1, \epsilon_2, \vec a, \beta)$, the $\beta$- and $\vec a$-independent $q$ must be  invariant under the rescaling $(\Lambda, \epsilon_1, \epsilon_2) \to (\zeta^{-1} \Lambda, \zeta^{-1} \epsilon_1, \zeta^{-1} \epsilon_2)$. Furthermore, because (\ref{AGT-M-duality-AB}) is symmetric under the exchange $\epsilon_1 \leftrightarrow \epsilon_2$, so must $q$. In sum,  it must be that
\be
\label{WN1}
\boxed{W^{(N)}_1 |q, \Delta \rangle = q |q, \Delta \rangle, \quad q = {\Lambda^N \over (\epsilon_1\epsilon_2)^{N/2} }}
\ee
Recall here that the $W^{(s_i)}_l$'s generate ${\cal W}({\frak {su}(N)}^\vee_{\rm aff})$, and that on $\widehat{{\cal W}}({\frak {su}(N)}^\vee_{\rm aff})$,  the  $W^{(s_i)}_{l < 0}$'s and $W^{(s_i)}_{l > 0}$'s act as creation and annihilation operators, respectively; in particular, $W^{(N)}_1$ is an annihilation operator, so (\ref{WN1}) means that $ |q, \Delta \rangle$ is actually a\emph{ coherent state}, i.e., an eigenstate of an annihilation operator.

What about when $n=2$ (with even $N$) whence we have $G = SO(N+1)$? According to~\cite{abcdefg}, instead of (\ref{phi(z)}), we now have
\be
\label{twisted phi(z)}
\phi_s (z)= u_s \left({dz \over z}\right)^s,  \quad {\tilde \phi}_j (z)= 0, \quad {\tilde \phi}_N(z) = \left(z^{1/2} + {\Lambda^N \over z^{1/2}}\right)\left({dz \over z}\right)^N,
\ee 
where the $\tilde \phi_s(z)$'s are also $(s,0)$-holomorphic differentials on $\cal C$ with modes in $\mathbb Z$ and $\mathbb Z + 1/2$. This is again consistent with our results established in $\S$B and after (\ref{AGT-AB-chiral CFT}) that for $n=2$ (with even $N$), we have, on $\cal C$, the following $(s_i, 0)$-holomorphic differentials 
\be
\label{twisted W's}
W^{(s_i)}(z) = \left(\sum_{l \in \mathbb Z} {W^{(s_i)}_l \over z^l} \right) \left( dz \over z \right)^{s_i}, \quad {\tilde W}^{(s_i)}(z) = \left(\sum_{l \in \mathbb Z} {{\tilde W}^{(s_i)}_{l + 1/2} \over z^{l+1/2}} \right) \left( dz \over z \right)^{s_i}, \quad  s_i = 2, 3, \dots, N,
 \ee
whence we can naturally identify, up to some constant factor,  $\phi_s (z)$ with  $W^{(s)}(z)$ and  $\tilde\phi_s (z)$ with  $\tilde W^{(s)}(z)$.

At $z = 0$ where the state $|q, \Delta \rangle$ is, we find, after comparing (\ref{twisted phi(z)}) with (\ref{twisted W's}), that instead of (\ref{W-2}), we have
\be
\label{W-1}
\boxed{W^{(s)}_{l \geq 1} \, |q, \Delta \rangle = 0, \quad {\rm for} \quad s = 2, 3, \dots, N}
\ee
We also have (\ref{Ws0}), and
\be
\label{twisted W-2}
\boxed{\tilde W^{(s)}_{l \geq 3/2} \, |q, \Delta \rangle = 0, \quad {\rm for} \quad s = 2, 3, \dots, N}
\ee
and instead of (\ref{WN1}), we have $\tilde W^{(N)}_{1/2} |q, \Delta \rangle = q |q, \Delta \rangle  \sim \Lambda^N |q, \Delta \rangle$. By employing the same reasoning used to derive (\ref{WN1}), we find that  
\be
\label{tildeWN1}
\boxed{\tilde W^{(N)}_{1/2} |q, \Delta \rangle = q |q, \Delta \rangle, \quad q = {\Lambda^N \over (\epsilon_1\epsilon_2)^{N/2} }}
\ee
Recall here that the $W^{(s_i)}_l$'s and $\tilde W^{(s_i)}_{l + 1/2}$'s generate ${\cal W}({\frak {su}(N)}^{(2)}_{\rm aff}) = {\cal W}({\frak {so}(N+1)}^\vee_{\rm aff})$, and that on $\widehat{{\cal W}}({\frak {so}(N+1)}^\vee_{\rm aff})$,  the  $\{W^{(s_i)}_{l < 0}, \tilde W^{(s_i)}_{l < 0}\}$ and $\{W^{(s_i)}_{l > 0}, \tilde W^{(s_i)}_{l > 0}\}$ act as creation and annihilation operators, respectively; in particular, $\tilde W^{(N)}_{1/2}$ is an annihilation operator, so (\ref{tildeWN1}) means that $ |q, \Delta \rangle$ is again a\emph{ coherent state}.

Thus, in arriving at the above boxed relations (i) (\ref{AGT-duality-A}), (\ref{c-A-epsilon}), (\ref{k'}), (\ref{L_0}), (\ref{Delta(2)}), (\ref{psi = delta}), (\ref{q | q}), (\ref{W-2}), (\ref{Ws0}), (\ref{WN1}) and (ii) (\ref{AGT-duality-A}), (\ref{c-A-epsilon}), (\ref{k'}), (\ref{L_0}), (\ref{Delta(2)}), (\ref{psi = delta}), (\ref{q | q}), (\ref{Ws0}), (\ref{W-1}), (\ref{twisted W-2}), (\ref{tildeWN1}),  we have just furnished a fundamental physical derivation of the pure AGT correspondence for the (i) $A_{N-1}$ and (ii) $B_{N/2}$ groups!

\newsubsection{An Equivalence of Spacetime BPS Spectra and a Pure AGT Correspondence for the $C$--$D$--$G$ Groups}

We shall now derive, purely physically, a pure AGT correspondence for the $C$--$D$--$G$ groups. To this end, recall from (\ref{AGT-fluxbrane-interval}) and (\ref{AGT-fluxbrane-dual-interval}) that we have the following \emph{physically dual} M-theory compactifications
\be
\underbrace{\mathbb R^4\vert_{\epsilon_1, \epsilon_2}  \times \Sigma_{n,t}}_{\textrm{$N$ M5-branes + OM5-plane}}\times \mathbb R^{5}\vert_{\epsilon_3; \,  x_{6,7}}  \quad \Longleftrightarrow  \quad   {\mathbb R^{5}}\vert_{\epsilon_3; \, x_{4,5}} \times \underbrace{{\cal C}  \times SN_N^{R\to 0}\vert_{\epsilon_3; \, x_{6,7}}}_{\textrm{$1$ M5-branes}},
\label{AGT-M-duality-CDG}
\ee
where we have a common half-BPS boundary condition at the tips of $\mathbb I_t \subset \Sigma_{n,t} = {\bf S}^1_n \times \mathbb I_t$; the radius of ${\bf S}^1_n$ is $\beta$; $\mathbb I_t \ll \beta$; and $\cal C$ is \emph{a priori} the same as $\Sigma_{n,t} $. As usual, there is a $\mathbb Z_n$-outer-automorphism of ${\mathbb R^4}\vert_{\epsilon_1, \epsilon_2}$ and $ SN_N^{R\to 0}\vert_{\epsilon_3; \, x_{6,7}}$ as we go around the ${\bf S}^1_n$ circle and identify the circle under an order $n$ translation, and the $\epsilon_i$'s are parameters of the Omega-deformation along the indicated planes described in detail in $\S$5.1.  

\bigskip\noindent{\it The Spectrum of Spacetime BPS States on the LHS of (\ref{AGT-M-duality-CDG})}

Let us first ascertain the spectrum of spacetime BPS states on the LHS of (\ref{AGT-M-duality-CDG}) that define $Z_{\rm BPS} (\epsilon_1, \epsilon_2, \vec a, \beta)$ in (\ref{5d BPS}). In the absence of Omega-deformation whence $\epsilon_i = 0$, according to our discussion in $\S$5.1, the spacetime BPS states would be captured by the topological sector of the $\cN = (4,4)$ sigma-model on  $\Sigma_{n,t}$ with target the moduli space ${\cal M}_{G}$ of $G$-instantons on $\mathbb R^4$, where for $n=1$, 2 or 3 (with $N = 4$), $G = SO(2N)$, $USp(2N-2)$ or $G_2$, respectively. However, in the presence of Omega-deformation, recall from our discussion immediately after (\ref{5d BPS}) that as one traverses a closed loop in $\Sigma_{n,t}$, there would be a  $\bf g$-automorphism of ${\cal M}_{G}$, where ${\bf g} \in U(1) \times U(1) \times T$, and $T \subset G$ is the Cartan subgroup. Consequently, the spacetime BPS states of interest would, in the presence of Omega-deformation, be captured by the topological sector of a non-dynamically ${\bf g}$-gauged version of the aforementioned sigma-model (c.f.~footnote~\ref{gauging worldsheet}).\emph{} Hence, according to~\cite{mine-equivariant} and our arguments in $\S$3.2 which led us to (\ref{BPS-M-OM5}), we can express the Hilbert space ${\cal H}^{\Omega}_{\rm BPS}$ of spacetime BPS states on the LHS of (\ref{AGT-M-duality-CDG}) as
\be
{\cal H}^\Omega_{\rm BPS} = \bigoplus_{m} {\cal H}^\Omega_{{\rm BPS}, m}  =  \bigoplus_{m} ~{\rm IH}^\ast_{U(1)^2 \times T} \, {\cal U}({\cal M}_{G, m}), 
\label{BPS-AGT-CDG}
\ee
where ${\rm IH}^\ast_{U(1)^2 \times T} \, {\cal U}({\cal M}_{G, m})$ is the $\mathbb Z_n$-invariant (in the sense of (\ref{HBPS-eff-USp(2N-2)}) and (\ref{HBPS-eff-G_2}) when $n=2$ and $3$, respectively) $U(1)^2 \times T$-equivariant intersection cohomology of the Uhlenbeck compactification ${\cal U}({\cal M}_{G, m})$ of the (singular) moduli space ${\cal M}_{G, m}$ of $G$-instantons on $\mathbb R^4$ with instanton number $m$.

\bigskip\noindent{\it The Spectrum of Spacetime BPS States on the RHS of (\ref{AGT-M-duality-CDG})}

Let us next ascertain the corresponding spectrum of spacetime BPS states on the RHS of (\ref{AGT-M-duality-CDG}). Bearing in mind footnote~\ref{junya's twist} which tells us that the underlying worldvolume theory of the single M5-brane is conformal along $SN_N^{R\to 0}\vert_{\epsilon_3; \, x_{6,7}}$  in (\ref{AGT-M-duality-CDG}), by repeating our arguments in $\S$3.2 which led us to (\ref{equivalent IIA system 2}) and beyond, and from our discussion surrounding (\ref{metric}), we find that the spacetime BPS states would be furnished by the I-brane theory in the following type IIA configuration:
\be
\textrm{IIA}: \quad \underbrace{ {\mathbb R}^5\vert_{\epsilon_3; x_{4,5}} \times {\cal C} \times {\mathbb R}^3/{\cal I}_3\vert_{\epsilon_3; x_{6,7}}}_{\textrm{I-brane on ${\cal C} = N \textrm{D6}/\textrm{O6}^- \cap 1\textrm{D4}$}}.
\label{equivalent IIA system 2 - AGT}
\ee
Here, we have  a stack of $N$ coincident D6-branes on top  of  an O$6^-$-plane whose worldvolume is given by ${\mathbb R}^5\vert_{\epsilon_3; x_{4,5}} \times {\cal C}$, and a single D4-brane whose worldvolume is given by ${\cal C} \times \mathbb R^3/{\cal I}_3\vert_{\epsilon_3; x_{6,7}}$, where ${\cal I}_3$ acts as ${\vec r} \to -{\vec r}$ in $\mathbb R^3$.

Let us for a moment turn off Omega-deformation in (\ref{equivalent IIA system 2 - AGT}), i.e., let $\epsilon_3 = \epsilon_1 + \epsilon_2 = 0$. Then, by applying to (\ref{equivalent IIA system 2 - AGT}) our analysis in $\S$3.2 which eventually led us to (\ref{GL-relation D}), (\ref{GL-relation C}) and (\ref{GL-relation G}) from (\ref{equivalent IIA system 2}), we learn that the spacetime BPS states would be furnished by chiral fermions on ${\cal C} $ which couple to the dynamical gauge degrees of freedom on the single D4-brane that, in turn, can be effectively represented by a  chiral WZW model at level 1 on ${\cal C} $, ${\rm WZW}^{{\rm level} \, 1}_{{\frak g}^\vee_{{\rm aff}}}$, where ${\frak g}^\vee_{{\rm aff}}$ is the \emph{Langlands dual} of the affine $G$-algebra $\frak g_{{\rm aff}}$. This is consistent with our observation after (\ref{AGT-fluxbrane-dual-interval}) that  the symmetries of the 2d theory along ${\cal C}$ ought to be rooted in the Langlands dual Lie algebra $\frak g^\vee$ (and therefore $\frak g^\vee_{\rm aff}$). 

Now turn Omega-deformation back on. As indicated in (\ref{equivalent IIA system 2 - AGT}), as one traverses around a closed loop in ${\cal C}$, the $x_4$-$x_5$ plane in ${\mathbb R^{4}}\vert_{\epsilon_3; \, x_{4,5}} \subset {\mathbb R^{5}}\vert_{\epsilon_3; \, x_{4,5}}$ would be rotated by an angle of $\epsilon_3$ together with an $SU(2)_R$-symmetry rotation of the supersymmetric $SO(2N)$ gauge theory along $ {\mathbb R^{4}}\vert_{\epsilon_3; \, x_{4,5}}$. According to our discussion in $\S$5.1 which led us to (\ref{5d BPS}) and slightly beyond, we find that Omega-deformation in this instance would effect a ${\bf g}'$-automorphism of ${\cal M}_{SO(2N), m}$ as we traverse around a closed loop in ${\cal C}$, where ${\cal M}_{SO(2N), m}$ is the moduli space of $SO(2N)$-instantons on $\mathbb R^4$ with instanton number $m$; ${\bf g}' =  {\rm exp} \, \beta (\epsilon_3 J_3 + {{\vec a} \cdot {\vec T}'})$; $J_3$ is the rotation generator of the $x_4$-$x_5$ plane corrected with an appropriate amount of $SU(2)_R$-symmetry to commute with the D6/$\textrm{O6}^-$ worlvolume supercharges; ${\vec T}' = (T'_1 \dots, T'_{{\rm rank} \, SO(2N)})$ are the generators of the Cartan subgroup $T' \subset SO(2N)$;  and ${\vec a} = (a_1, \dots, a_{{\rm rank} \, SO(2N)})$ are the corresponding purely imaginary Coulomb moduli of the $SO(2N)$ gauge theory on $\mathbb R^4 \vert_{\epsilon_3; \, x_{4,5}}$. In fact, since ${\cal M}_{SO(2N), m}$ is also the space of self-dual connections of an $SO(2N)$-bundle on $\mathbb R^4$, and since these self-dual connections correspond to differential one-forms valued in the Lie algebra $\frak {so}(2N)$, Omega-deformation also means that there is a ${\bf g}'$-automorphism of the space of elements of $\frak {so}(2N)$ and thus $SO(2N)$,  as we traverse a closed loop in ${\cal C}$.

Note at this point that in the above, $\frak g^\vee_{\rm aff}  \simeq \frak {so}(2N)^{(n)}_{\rm aff}$, where $\frak {so}(2N)^{(n)}_{\rm aff}$ is a $\mathbb Z_n$-twisted affine $SO(2N)$-algebra. What this means is that ${\rm WZW}^{{\rm level} \, 1}_{{\frak g}^\vee_{{\rm aff}}}$ can be regarded as a (chiral half of a) $SO(2N)$ WZW model at level 1 that is $\mathbb Z_n$-twisted on ${\cal C}$. Since a $\cal G$ WZW model on $\Sigma$ is a bosonic sigma-model on $\Sigma$ with target the $\cal G$-manifold, according to the last paragraph,  it would mean that Omega-deformation would effect a ${\bf g}'$-automorphism  of the target space of ${\rm WZW}^{{\rm level} \, 1}_{{\frak g}^\vee_{{\rm aff}}}$ as we traverse a closed loop in ${\cal C}$, where ${\bf g}' \in U(1) \times T'$.  In turn, according to footnote~\ref{gauging worldsheet}, it would mean that in the presence of Omega-deformation, we would have to non-dynamically gauge ${\rm WZW}^{{\rm level} \, 1}_{{\frak g}^\vee_{{\rm aff}}}$ by $ U(1) \times T'$. 

That being said, notice also from (\ref{equivalent IIA system 2 - AGT}) that as one traverses around a closed loop in ${\cal C}$, the $x_6$-$x_7$ plane in ${\mathbb R^{3} / {\cal I}_3}\vert_{\epsilon_3; \, x_{6,7}}$ would be rotated by an angle of $\epsilon_3$ together with an $R$-symmetry rotation of the supersymmetric gauge theory living on the single D4-brane, i.e., Omega-deformation is also being turned on along the D4-brane. Now recall from our arguments leading up to (\ref{partially gauged CFT - CDG}) that because the gauge field on the D4-brane --  unlike the gauge field on the D6-branes -- is dynamical,  one has to reduce away in the I-brane system the WZW model associated with the D4-brane. Similarly,  the aforedescribed Omega-deformation along the D4-brane would act not to enlarge  but to \emph{reduce} the $ U(1) \times T'$ Omega-deformation factor in the previous paragraph by $R = U(1) \times \mathbb T$, where $U(1) \subset R$ is associated with the $\epsilon_3$-rotation of the $x_6$-$x_7$ plane in ${\mathbb R^{3} / {\cal I}_3}\vert_{\epsilon_3; \, x_{6,7}}$, and ${\mathbb T} \subset R$ is the Cartan of the gauge group on the D4-brane. In short, we would in fact have to non-dynamically gauge ${\rm WZW}^{{\rm level} \, 1}_{{\frak g}^\vee_{{\rm aff}}}$ not by $ U(1) \times T'$ but by ${\cal T} \subset T'$. 

At any rate, because $SO(2N)/T' \simeq SO(2N, \mathbb C) / B_+$, where $B_+$ is a Borel subgroup, it would mean that $SO(2N)/{\cal T} \simeq (SO(2N, \mathbb C) / B_+) \times (T' / \cal T)$. Also, $T' / \cal T$ is never bigger than the Cartan subgroup $C \subset B_+ = C \times N_+$, where $N_+$ is the subgroup of strictly upper triangular matrices which are nilpotent and traceless whose Lie algebra is $\frak n_+$. Altogether, this means that our gauged WZW model which corresponds to the coset model $SO(2N)/{\cal T}$, can also be studied as an $S$-gauged  $SO(2N, \mathbb C)$ WZW model which corresponds to the coset model  $SO(2N, \mathbb C) / S$, where $N_+ \subseteq S \subset B_+$. As physically consistent $\cal H$-gauged $\cal G$ WZW models are such that $\cal H$ is necessarily a connected subgroup of $\cal G$, it will mean that $S = N_+$. Therefore, what we ought to ultimately consider is an $N_+$-gauged $SO(2N, \mathbb C)$ WZW model.

Before we proceed any further, let us make a slight deviation to highlight an important point regarding the effective geometry of ${\cal C}$. As the simple roots of $N_+$ form a subset of the simple roots of $SO(2N,\mathbb C)$, the level of the affine $N_+$-algebra ought to be the equal to the level of the affine $SO(2N,\mathbb C)$-algebra~\cite{Ketov} which is 1.  However, it is clear from our discussion hitherto that the affine $N_+$-algebra, in particular its level, will depend nontrivially on the Omega-deformation parameters which may or may not take integral values; in other words, its level will \emph{not} be equal to 1. A resolution to this conundrum is as follows. A deviation of the level of the affine $N_+$-algebra from 1 would translate into a corresponding deviation of its central charge; since a central charge arises due to an introduction of a macroscopic scale in the 2d system which results from a curvature along ${\cal C}$~\cite{CFT text}, it would mean that Omega-deformation ought to deform  the \emph{a priori} flat ${\cal C}  = \Sigma_{n,t}$ into a curved Riemann surface with the same topology -- that is, a Riemann sphere with two punctures -- such that the anomalous deviation in the central charge and thus level, can be consistently ``absorbed'' in the process (see also footnote~\ref{deform C}). Thus, we effectively have $ {\cal C} =  {\bf S}^2 / \{ 0, \infty \}$, so $\cal C$ can be viewed as an ${\bf S}^1_n$ fibration of $\mathbb I_t$ whose fiber has zero radius at the two end points $z = 0$ and $z = \infty$, where `$z$' is a holomorphic coordinate on $\cal C$.

Coming back to our main discussion, it is clear that in the schematic notation of $\S$3.2,  our $N_+$-gauged $SO(2N, \mathbb C)$ WZW model can be expressed as the partially gauged chiral CFT
\be
\frak {so}(2N)^{(n)}_{\rm aff, 1} / {\frak{n}_+}^{(n)}_{{\rm aff}, p}
\label{AGT-CDG-chiral CFT}
\ee
on $\cal C$, where the level $p$ would, according to our discussions hitherto, necessarily depend on the Omega-deformation parameters $\epsilon'_1 = \beta \epsilon_1$ and $\epsilon'_2 = \beta \epsilon_2$. ($p$, being a purely real number, should not depend on the purely imaginary parameter $\vec a' = \beta \vec a$).

In sum, the sought-after spacetime BPS states ought to be given by the states of the partially gauged chiral CFT in (\ref{AGT-CDG-chiral CFT}), and via $\S$B and~\cite[eqn.~(6.67)]{review}, we find that this chiral CFT realizes  ${\cal W}({\frak g}^\vee_{\rm aff})$ -- a  $\mathbb Z_n$-twisted version of the affine $\cal W$-algebra ${\cal W}(\widehat {\frak {so}(2N)})$ obtained from $\frak {so}(2N)_{\rm aff}$ via a quantum Drinfeld-Sokolov reduction. In other words,  the states of the chiral CFT would be furnished by a Verma module  $\widehat{{\cal W}}({\frak g}^\vee_{\rm aff})$ over ${\cal W}({\frak g}^\vee_{\rm aff})$, and the Hilbert space ${\cal H}^{\Omega'}_{\rm BPS}$ of spacetime BPS states on the RHS of (\ref{AGT-M-duality-CDG}) can be expressed as
\be
{\cal H}^{\Omega'}_{\rm BPS}  = \widehat{{\cal W}}({\frak g}^\vee_{\rm aff}).
\label{AGT-CDG-H=W}
\ee

\bigskip\noindent{\it A Pure AGT Correspondence for the $C$--$D$--$G$ Groups}

Clearly, the physical duality of the compactifications in (\ref{AGT-M-duality-CDG}) will mean that ${\cal H}^\Omega_{\rm BPS}$ in (\ref{BPS-AGT-CDG}) is equivalent to ${\cal H}^{\Omega'}_{\rm BPS}$ in (\ref{AGT-CDG-H=W}), i.e.,
\be
 \boxed{\bigoplus_{m} ~{\rm IH}^\ast_{U(1)^2 \times T} \, {\cal U}({\cal M}_{G, m}) = \widehat{{\cal W}}({\frak g}^\vee_{\rm aff})}
 \label{AGT-duality-D}
\ee
Thus, we have a 4d-2d duality relation in the sense of  (\ref{GL-relation D}), (\ref{GL-relation C}) and (\ref{GL-relation G}). 

According to footnote~\ref{central charge SO(2N)} and (\ref{cc-simply-laced}) -- bearing in mind that (i) $\textrm{dim} \,\frak {so}(2N) = 2N^2 -N$;  (ii) ${\rm rank} \, \frak {so}(2N) = N$; and (iii) $h^\vee_{\frak {so}(2N)} = 2N-2$ -- the central charge of  ${\cal W}({\frak g}^\vee_{\rm aff})$ is
\be
 c_{D}  = N -   (2N-2) (2N^2 -N) \left( \alpha_+ + \alpha_- \right)^2,
\label{cc-D}
\ee
where $\alpha_+ \alpha_- = -1$; $\alpha_+ = 1/ \sqrt{k'+ 2N-2}$; and $k' \in \mathbb R$ is the effective level of the underlying affine Lie algebra $\frak {so}(2N)_{\rm aff}$.  Note at this point that (\ref{AGT-CDG-chiral CFT}) means that we can also write $c_D = c(\frak {so}(2N)^{(n)}_{\rm aff, 1}) - c({\frak{n}_+}^{(n)}_{{\rm aff}, p})$, and since the central charge $c(\frak {so}(2N)^{(n)}_{\rm aff, 1}) = N$, according to (\ref{cc-D}), we can also write $c({\frak{n}_+}^{(n)}_{{\rm aff}, p}) =  (2N-2) (2N^2 -N) \left( \alpha_+ + \alpha_- \right)^2$. 

As $p$ will depend on $\epsilon'_1 = \beta \epsilon_1$ and $\epsilon'_2 = \beta \epsilon_2$, so would $\alpha_{\pm}$. Because $\alpha_+ \alpha_- = -1$, it would mean that we can write $\alpha_+ = if(\epsilon'_1, \epsilon'_2)$ and $\alpha_- = i f^{-1}(\epsilon'_1, \epsilon'_2)$, where $f(\epsilon'_1, \epsilon'_2)$ is some possibly complex function. Because (\ref{AGT-M-duality-CDG}) is symmetric under the exchange $\epsilon'_1 \leftrightarrow \epsilon'_2$, so must $c_D$; in particular, we ought to have $f(\epsilon'_1, \epsilon'_2)+  f^{-1}(\epsilon'_1, \epsilon'_2) = f(\epsilon'_2, \epsilon'_1)+  f^{-1}(\epsilon'_2, \epsilon'_1)$.  Because $\alpha_+ = 1/ \sqrt{k'+ 2N-2} = if(\epsilon'_1, \epsilon'_2)$ would go from positive real to negative purely complex as we vary $k'$, it would mean that $f(\epsilon'_1, \epsilon'_2)$ must also go from negative purely complex to negative real as we vary the $\epsilon'_i$'s. Because we have a geometrical ${\bf g}'' = {\rm exp} [(\epsilon'_1 + \epsilon'_2)J_3] = {\rm exp} [(\lambda \epsilon'_1 + \lambda \epsilon'_2) \lambda^{-1}J_3]$ automorphism associated with the Omega-deformation in (\ref{equivalent IIA system 2 - AGT}), and since $c_D$ is only a function of $\epsilon'_{1,2}$ and not of $J_3$, we ought to have $c_D(\epsilon'_1, \epsilon'_2) = c_D(\lambda\epsilon'_1, \lambda\epsilon'_2)$; in other words, $\alpha_+ + \alpha_-$ ought to be invariant under $\epsilon'_i \to \lambda \epsilon'_i$, where $\lambda$ is some real constant. Altogether therefore, it would mean that we can write $\alpha_+ =  - i \sqrt {\epsilon'_1 / \epsilon'_2} =  - i \sqrt {\epsilon_1 / \epsilon_2}$ and $\alpha_- =  - i \sqrt {\epsilon'_2 / \epsilon'_1} =  - i \sqrt {\epsilon_2 / \epsilon_1}$; in turn, (\ref{cc-D}) would be given by
\be
\label{c-D-epsilon}
\boxed{c_{D, \epsilon_{1,2} }  = N  + (2N-2) (2N^2 -N) {(\epsilon_1 + \epsilon_2)^2 \over {\epsilon_1\epsilon_2}}}
\ee
where in addition, we would have
\be
\label{k'-D}
\boxed{k'  = - 2N + 2 - b^{-2} \qquad {\rm and} \qquad b = \sqrt {\epsilon_1 / \epsilon_2}}
\ee
so that one can also write
\be
c_{D, \epsilon_{1,2} } = c(\frak {so}(2N)^{(n)}_{\rm aff, 1}) + c(\Omega_{\epsilon_{1,2}}),
\ee
where
\be
 c(\Omega_{\epsilon_{1,2}}) =  h^\vee_{\frak {so}(2N)} {\rm dim} \, \frak {so}(2N) \left(b + {1 \over b}\right)^2
\label{cc-D-Omega}
\ee
can be regarded as an Omega-deformation-induced central charge. (Notice that there is no $\beta$-dependence in the formulas (\ref{c-D-epsilon})--(\ref{cc-D-Omega}); this is consistent with the fact that these formulas are supposed to be globally-defined on $\cal C$ independent of the varying radius of ${\bf S}^1_n \subset \cal C$.)

The Verma module $\widehat{{\cal W}}({\frak g}^\vee_{\rm aff})$ is generated by the application of creation operators $W^{(s_i)}_{m < 0}$ on its $\mathbb Z_n$-twisted highest  weight state $| \Delta \rangle$, where the $W^{(s_i)}_{m < 0}$'s are the negative modes of the spin-$s_i$ fields $W^{(s_i)}(z)$ on $\cal C$ which span ${\cal W}({\frak g}^\vee_{\rm aff})$, and $m \in \mathbb Z/n$. On the other hand, $| \Delta \rangle$ is annihilated by the annihilation operators $W^{(s_i)}_{m > 0}$, where the $W^{(s_i)}_{m > 0}$'s are the positive modes of the $W^{(s_i)}(z)$ fields on $\cal C$ which also span ${\cal W}({\frak g}^\vee_{\rm aff})$. Nonetheless, we have $W^{(s_i)}_{ 0} | \Delta \rangle = \Delta^{(s_i)} | \Delta \rangle$, where the $\Delta^{(s_i)}$'s are Weyl-invariant polynomials in ${\bf a} = {\bf J}_0 + (\alpha_+ + \alpha_-) \rho$; ${\bf J}_0 = (J^1_0, \dots, J_0^{{\rm rank} \, \frak {so}(2N)})$ are the zeroth modes of the ${\rm rank} \, \frak {so}(2N)$ untwisted scalar bosonic fields in the free-field realization of ${\cal W}({\frak g}^\vee_{\rm aff})$; and $\rho$ is the Weyl vector of $\frak {so}(2N)$~\cite{review}. For example, $W^{(2)}_{ 0} | \Delta \rangle = L_0 | \Delta \rangle = \Delta^{(2)} | \Delta \rangle$, where $L_0$ is the zeroth mode of the stress tensor $T(z) = W^{(2)}(z)$, and $\Delta^{(2)} = \left({\bf a}^2 - (\alpha_+ + \alpha_-)^2 \rho^2 \right) /2 = {\bf a}^2/2  + (2N-2)(2N^2 - N)(\epsilon_1 + \epsilon_2)^2/24 \epsilon_1 \epsilon_2$. (See~\cite[eqn.~(6.18)]{review}, and note that just like the quantities in (\ref{c-D-epsilon})--(\ref{cc-D-Omega}), $\Delta^{(2)}$ should be $\beta$-independent, as is the case.) 

Recall at this point that $L_0$ generates translations along the ${\bf S}^1_n$ fiber in $\cal C$, and since the presence of Omega-deformation means that there is a rotation of an $\mathbb R^4$ space and the gauge field over it as we go around the ${\bf S}^1_n$ (c.f.~our earlier discussion on a ${\bf g}'$-automorphism), $L_0$ should be related to the rotation parameters $(\epsilon_1, \epsilon_2, \vec a)$. Indeed, we saw in the last paragraph that $L_0$ has eigenvalues which depend on $(\epsilon_1, \epsilon_2, \bf a)$, and since $\bf a$, like $\vec a$, is a vector whose number of components even coincides with that of $\vec a$ when $G = SO(2N)$, we can naturally identify $ \bf a$ with $-i\vec a$, where a factor of $-i$ is needed because $\vec a$ is purely imaginary while $\bf a$ is purely real. That said, because of (\ref{AGT-duality-D}), it would mean that the symmetries of $\Delta^{(2)}$ ought to be compatible with the symmetries of the partition function $Z_{\rm BPS} (\epsilon_1, \epsilon_2, \vec a, \beta)$ of ${\cal H}^{\Omega}_{\rm BPS}$ in (\ref{5d BPS}); in particular, since $Z_{\rm BPS} (\epsilon_1, \epsilon_2, \vec a, \beta)$ is invariant under the simultaneous rescalings $(\beta, \epsilon_1, \epsilon_2, \vec a) \to ( \zeta^{-1} \beta, \zeta \epsilon_1, \zeta \epsilon_2, \zeta \vec a)$, where $\zeta$ is some real constant, the $\beta$-independent $\Delta^{(2)}$ must be  invariant under the simultaneous rescalings $(\epsilon_1, \epsilon_2, \vec a) \to (\zeta \epsilon_1, \zeta \epsilon_2, \zeta \vec a)$. Furthermore, because (\ref{AGT-M-duality-CDG}) is symmetric under the exchange $\epsilon_1 \leftrightarrow \epsilon_2$, so must $\Delta^{(2)}$. In sum, we ought to have ${\bf a} \sim -i  \vec a / \sqrt {\epsilon_1\epsilon_2}$, whence we can write
\be
\label{L_0-D}
\boxed{W^{(2)}_{ 0} | \Delta \rangle = \Delta^{(2)} | \Delta \rangle}
\ee 
where
\be
\label{Delta(2)-D}
\boxed{\Delta^{(2)} = {(2N-2)(2N^2 - N) \over 24}{(\epsilon_1 + \epsilon_2)^2 \over \epsilon_1 \epsilon_2} - { \gamma'  {\vec a}^2 \over  \epsilon_1\epsilon_2}}
\ee
for some real constant $\gamma'$.
 
In the limit that $\beta \to 0$, it is well-known~\cite{abcd} that
 \be
 \label{abcd BPS relation-D}
 Z_{\rm BPS} (\epsilon_1, \epsilon_2, \vec a, \beta) =  \sum_m Z_{{\rm BPS}, m} (\epsilon_1, \epsilon_2, \vec a, \beta)
 \ee 
 of (\ref{5d BPS}) behaves such that $Z_{{\rm BPS}, m} (\epsilon_1, \epsilon_2, \vec a, \beta \to 0)  \sim \beta^{-2mh^\vee_{\frak g}} \, Z^{\rm 4d}_{{\rm BPS}, m} (\epsilon_1, \epsilon_2, \vec a)$, whence the Nekrasov instanton partition function $Z_{\rm inst}  (\Lambda, \epsilon_1, \epsilon_2, \vec a)= \sum_m \Lambda^{2mh^\vee_{\frak g}} \, Z^{\rm 4d}_{{\rm BPS}, m} (\epsilon_1, \epsilon_2, \vec a) $ can be written as 
\be
Z_{\rm inst}(\Lambda, \epsilon_1, \epsilon_2, \vec a) =  \sum_m \Lambda^{2mh^\vee_{\frak g}} \, Z'_{{\rm BPS}, m} (\epsilon_1, \epsilon_2, \vec a, \beta \to 0), 
\label{Z_inst-5d-D}
\ee
where $Z'_{{\rm BPS}, m} = l_m \beta^{2mh^\vee_{\frak g}} Z_{{\rm BPS}, m}$; $l_m$ is some constant; and $\Lambda$ can be interpreted as the inverse of the observed scale of the $\mathbb R^4\vert_{\epsilon_1, \epsilon_2}$ space on the LHS of (\ref{AGT-M-duality-CDG}).

Note at this point that equivariant localization~\cite{Atiyah-Bott} implies that ${\rm IH}^\ast_{U(1)^2 \times T} \, {\cal U}({\cal M}_{G, m})$ must be endowed with an orthogonal basis $\{| {{\vec p}_m} \rangle \}$  that is in one-to-one correspondence with the fixed-point set $\{ {{\vec p}_m} \}$ of the $U(1)^2 \times T$-action on ${\cal U}({\cal M}_{G, m})$. Thus, since $Z'_{{\rm BPS}, m}$ is a weighted count of the states in ${\cal H}^\Omega_{{\rm BPS}, m} = {\rm IH}^\ast_{U(1)^2 \times T} \, {\cal U}({\cal M}_{G, m})$, it would mean that one can write
\be
Z'_{{\rm BPS}, m} (\epsilon_1, \epsilon_2, \vec a, \beta \to 0) =  \sum_{{{\vec p}_m}} \,  l^2_{{{\vec p}_m}} (\epsilon_1, \epsilon_2, \vec a)\langle {{\vec p}_m} | {{\vec p}_m} \rangle,
\label{rhs of Z_inst-5d-D}
\ee
where $l_{{{\vec p}_m}} (\epsilon_1, \epsilon_2, \vec a) \in \mathbb R$, and the dependence on $\epsilon_1$, $\epsilon_2$ and $\vec a$ arises because the energy level of each state (given by the eigenvalue of the $L_0$ operator which generates translation along ${\bf S}^1_n \subset \Sigma_{n,t}$ in (\ref{AGT-M-duality-CDG}) whence there is an Omega-deformation twist of the theory along the orthogonal spaces indicated therein) ought to depend on these Omega-deformation parameters.

Notice that (\ref{rhs of Z_inst-5d-D}) also means that
\be
Z'_{{\rm BPS}, m} (\epsilon_1, \epsilon_2, \vec a, \beta \to 0) =   \langle \Psi_m | \Psi_m \rangle,
\label{Z'-D}
\ee
where
\be
|\Psi_m \rangle = \bigoplus_{{{\vec p}_m}} \,  l_{{{\vec p}_m}}    | {{\vec p}_m} \rangle.
\ee 
Here, the state $| \Psi_m \rangle \in {\rm IH}^\ast_{U(1)^2 \times T} \, {\cal U}({\cal M}_{G, m})$, and $ \langle \cdot | \cdot \rangle$ is a Poincar\'e pairing in the sense of~\cite[$\S$2.6]{J-function}.

Now consider the state
\be
| \Psi \rangle = \bigoplus_m  \Lambda^{m h^\vee_{\frak g}} | \Psi_m \rangle.
\ee
By substituting (\ref{Z'-D}) in the RHS of (\ref{Z_inst-5d-D}), and by noting that $\langle \Psi_m | \Psi_n \rangle = \delta_{mn}$, one can immediately see that
\be
Z_{\rm inst} (\Lambda, \epsilon_1, \epsilon_2, \vec a) = \langle \Psi |  \Psi \rangle,
\label{Psi | Psi, D}
\ee
where $| \Psi \rangle \in \bigoplus_m {\rm IH}^\ast_{U(1)^2 \times T} \, {\cal U}({\cal M}_{G, m})$.  In turn, the duality relation (\ref{AGT-duality-D}) would mean that 
\be
\boxed{| \Psi \rangle = | q, \Delta \rangle}
\label{psi = delta, D}
\ee
whence  
\be
\label{q | q, D}
\boxed{Z_{\rm inst} (\Lambda, \epsilon_1, \epsilon_2, \vec a) = \langle q, \Delta   | q, \Delta \rangle}
\ee
where $| q, \Delta \rangle \in \widehat{{\cal W}}({\frak g}^\vee_{\rm aff})$. (The meaning of the label `$q$' will be clear shortly.) Since the RHS of (\ref{q | q, D}) is defined at $\beta \to 0$ (see the RHS of (\ref{Z_inst-5d-D})), and since we have in $\cal C$ a \emph{common} boundary condition at $z = 0$ and $z=\infty$,  $| q, \Delta \rangle$ and $\langle q, \Delta |$ ought to be a state and its dual associated with the puncture at $z = 0$ and $z=\infty$, respectively (as $z = 0, \infty$ are the points in $\cal C$ where the ${\bf S}^1_n$ fiber   has zero radius). This is depicted in fig.~2.  

\begin{figure}
  \centering
    \includegraphics[width=0.4\textwidth]{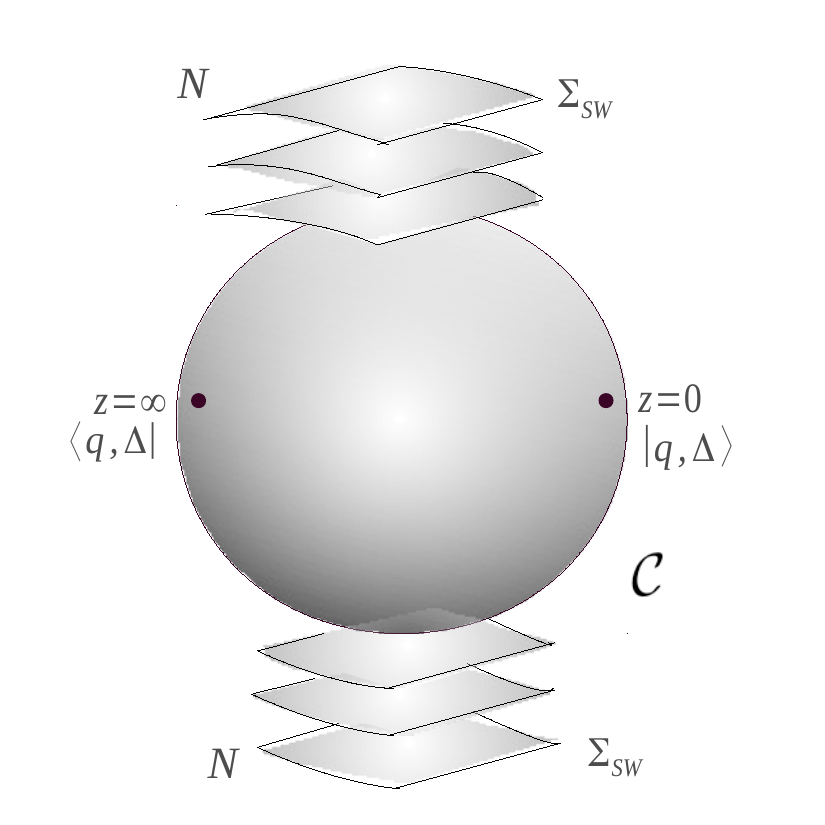}
  \caption{$\cal C$ and its $2N$-fold cover $\Sigma_{SW}$ with the states $\langle q, \Delta |$ and $| q, \Delta \rangle$ at $z=0$ and $\infty$}
\end{figure}

At any rate, note that if we only have $N$ D6-branes and 1 D4-brane wrapping $\cal C$ in (\ref{equivalent IIA system 2 - AGT}), we would (as explained in the last subsection) just have an $N \times 1 = N$-fold cover of $\cal C$. In the presence of  the O6$^-$-plane however, there will be a mirror image of this configuration on the ``opposite side'' whence this cover is doubled, i.e., in (\ref{equivalent IIA system 2 - AGT}), we effectively have a $2(N \times 1) = 2N$-fold cover $\Sigma_{SW}$ of $\cal C$. This is also depicted in fig.~2. Incidentally, $\Sigma_{SW}$ is also the Seiberg-Witten curve which underlies $Z_{\rm inst} (\Lambda, \epsilon_1, \epsilon_2, \vec a)$! Moreover, it is by now well-established (see~\cite{Lerche} and references therein) that for $n=1$, i.e., $G = SO(2N)$, $\Sigma_{SW}$ can be described in terms of the algebraic relation 
\be
\Sigma_{SW}: \lambda^{2N} + \phi_2(z) \lambda^{2N-2} + \dots + \phi_{2N-2}(z) \lambda^2 +  \phi^2_N(z) = 0, 
\ee
where $\lambda = y  dz /z$ (for some complex variable $y$) is a section of $T^\ast \cal C$; the $\phi_s(z)$'s are $(s,0)$-holomorphic differentials on $\cal C$ given by
 \be
 \label{phi(z)-D}
\phi_j (z)= u_j \left({dz \over z}\right)^j \quad {\rm and} \quad \phi_{2N-2}(z) =\left( z + u_{2N-2} + {\Lambda^{2N-2} \over z}\right)\left({dz \over z}\right)^{2N-2}, 
\ee
where $j = 2, 4, \dots, 2N-4, N$; while for weights $\lambda_1, \dots, \lambda_{2N}$ of the $2N$-dimensional representation of $SO(2N)$, and for $s=2, 4,  \dots, 2N -2, N$,
\be
\label{u's-D}
u_s = (-1)^{s+1} \sum_{k_1 \neq \dots \neq k_s} e_{\lambda_{k_1}} e_{\lambda_{k_2}} \dots e_{\lambda_{k_s}} (\vec a) \quad {\rm and} \quad e_{\lambda_{r}}  = \vec a \cdot \lambda_r.  
\ee
This is consistent with our results established in $\S$B that for $G = SO(2N)$, we have, on $\cal C$, the following $(s_i, 0)$-holomorphic differentials 
\be
\label{W's-D}
W^{(s_i)}(z) = \left(\sum_{l \in \mathbb Z} {W^{(s_i)}_l \over z^l} \right) \left( dz \over z \right)^{s_i}, \quad {\rm where} \quad s_i = e_i + 1 = 2, 4, \dots, 2N-2, N,
\ee
whence we can naturally identify, up to some constant factor,  $\phi_s (z)$ with  $W^{(s)}(z)$. (In fact, a $U(1)$ $R$-symmetry of the 4d theory along $\mathbb R^4\vert_{\epsilon_1, \epsilon_2}$ on the LHS of (\ref{AGT-M-duality-CDG}) which underlies $Z_{\rm inst} (\Lambda, \epsilon_1, \epsilon_2, \vec a)$ and $\Sigma_{SW}$, can be identified with the rotational symmetry of ${\bf S}^1_n$; the duality relation (\ref{AGT-M-duality-CDG}) then means that the corresponding $U(1)$ $R$-charge of the $\phi_s(z)$  operators that define  $\Sigma_{SW}$, ought to match, up to a constant, the conformal dimension of the $W^{(s)}(z)$ operators on $\cal C$, which is indeed the case.)

At $z = 0$ where we have the state $|q, \Delta \rangle$, we find, after comparing (\ref{phi(z)-D}) with (\ref{W's-D}), that 
\be
\label{W-2-D}
\boxed{W^{(s)}_{l \geq 2} \, |q, \Delta \rangle = 0, \quad {\rm for} \quad s = 2,4, \dots, 2N-2, N}
\ee
$W^{(s)}_0|q, \Delta \rangle \sim u_s|q, \Delta \rangle$ and $W^{(2N-2)}_1 |q, \Delta \rangle = q |q, \Delta \rangle  \sim \Lambda^{2N-2} |q, \Delta \rangle$. To determine the exact form of the relation involving $W^{(s)}_0$, note that as in our derivation of (\ref{Delta(2)-D}), i.e., the eigenvalue $\Delta^{(2)}$ of $W^{(2)}_0$,  the eigenvalues of $W^{(s)}_0$ must be invariant under the simultaneous rescalings $(\epsilon_1, \epsilon_2, \vec a) \to (\zeta \epsilon_1, \zeta \epsilon_2, \zeta \vec a)$ and the exchange $\epsilon_1 \leftrightarrow \epsilon_2$; since (\ref{u's-D}) tells us that $u_s$ is of order $s$ in $\vec a$, it must be that
\be
\label{Ws0-D}
\boxed{W^{(s)}_0|q, \Delta \rangle  = { u_s \over (\epsilon_1 \epsilon_2)^{s / 2}} \,  |q, \Delta \rangle, \quad {\rm for} \quad s = 2, 4, \dots, 2N-2,N} 
\ee
To determine the exact form of the relation involving $W^{(2N-2)}_1$, recall that since the underlying worldvolume theory of the $N$ M5-branes on the LHS of (\ref{AGT-M-duality-CDG}) is scale-invariant, it would mean that in addition to possessing the symmetries of $Z_{\rm BPS} (\epsilon_1, \epsilon_2, \vec a, \beta)$ in (\ref{5d BPS}), the $W^{(N)}_1$-eigenvalue $q \sim \Lambda^{2N-2}$ ought to also be invariant under the rescaling $(\Lambda, \beta) \to (\zeta^{-1} \Lambda, \zeta \beta)$; thus, as the rescaling $(\beta, \epsilon_1, \epsilon_2, \vec a) \to ( \zeta \beta, \zeta^{-1} \epsilon_1, \zeta^{-1} \epsilon_2, \zeta^{-1} \vec a)$ is a symmetry of $Z_{\rm BPS} (\epsilon_1, \epsilon_2, \vec a, \beta)$, the $\beta$- and $\vec a$-independent $q$ must be  invariant under the rescaling $(\Lambda, \epsilon_1, \epsilon_2) \to (\zeta^{-1} \Lambda, \zeta^{-1} \epsilon_1, \zeta^{-1} \epsilon_2)$. Furthermore, because (\ref{AGT-M-duality-CDG}) is symmetric under the exchange $\epsilon_1 \leftrightarrow \epsilon_2$, so must $q$. In sum,  it must be that
\be
\label{WN1-D}
\boxed{W^{(2N-2)}_1 |q, \Delta \rangle = q |q, \Delta \rangle, \quad q = {\Lambda^{2N-2} \over (\epsilon_1\epsilon_2)^{N-1} }}
\ee
Recall here that the $W^{(s_i)}_l$'s generate ${\cal W}({\frak {so}(2N)}^\vee_{\rm aff})$, and that on $\widehat{{\cal W}}({\frak {so}(2N)}^\vee_{\rm aff})$,  the  $W^{(s_i)}_{l < 0}$'s and $W^{(s_i)}_{l > 0}$'s act as creation and annihilation operators, respectively; in particular, $W^{(N)}_1$ is an annihilation operator, so (\ref{WN1-D}) means that $ |q, \Delta \rangle$ is actually a\emph{ coherent state}, i.e., an eigenstate of an annihilation operator.

What about when $n=2$ whence we have $G = USp(2N-2)$? According to~\cite{abcdefg}, instead of (\ref{phi(z)-D}), we now have
\be
\label{twisted phi(z)-D}
\phi_s (z)= u_s \left({dz \over z}\right)^s,  \quad {\tilde \phi}_j (z)= 0, \quad {\tilde \phi}_{2N-2}(z) = \left(z^{1/2} + {\Lambda^{2N-2} \over z^{1/2}}\right)\left({dz \over z}\right)^{2N-2},
\ee 
where the $\tilde \phi_s(z)$'s are also $(s,0)$-holomorphic differentials on $\cal C$ with modes in $\mathbb Z$ and $\mathbb Z + 1/2$. This is again consistent with our results established in $\S$B and after (\ref{AGT-CDG-chiral CFT}) that for $n=2$, we have, on $\cal C$, the following $(s_i, 0)$-holomorphic differentials 
\be
\label{twisted W's-D}
W^{(s_i)}(z) = \left(\sum_{l \in \mathbb Z} {W^{(s_i)}_l \over z^l} \right) \left( dz \over z \right)^{s_i} \qquad {\rm and} \qquad {\tilde W}^{(s_i)}(z) = \left(\sum_{l \in \mathbb Z} {{\tilde W}^{(s_i)}_{l + 1/2} \over z^{l+1/2}} \right) \left( dz \over z \right)^{s_i},
 \ee
where $s_i = 2, 4 \dots, 2N-2, N$, whence we can naturally identify, up to some constant factor,  $\phi_s (z)$ with  $W^{(s)}(z)$ and  $\tilde\phi_s (z)$ with  $\tilde W^{(s)}(z)$.

At $z = 0$ where the state $|q, \Delta \rangle$ is, we find, after comparing (\ref{twisted phi(z)-D}) with (\ref{twisted W's-D}), that instead of (\ref{W-2-D}), we have
\be
\label{W-1-D}
\boxed{W^{(s)}_{l \geq 1} \, |q, \Delta \rangle = 0, \quad {\rm for} \quad s = 2, 4,  \dots, 2N-2, N}
\ee
We also have (\ref{Ws0-D}), and
\be
\label{twisted W-2-D}
\boxed{\tilde W^{(s)}_{l \geq 3/2} \, |q, \Delta \rangle = 0, \quad {\rm for} \quad s = 2,4,  \dots, 2N-2,N}
\ee
and instead of (\ref{WN1-D}), we have $\tilde W^{(2N-2)}_{1/2} |q, \Delta \rangle = q |q, \Delta \rangle  \sim \Lambda^{2N-2} |q, \Delta \rangle$. By employing the same reasoning used to derive (\ref{WN1-D}), we find that  
\be
\label{tildeWN1-D}
\boxed{\tilde W^{(2N-2)}_{1/2} |q, \Delta \rangle = q |q, \Delta \rangle, \quad q = {\Lambda^{2N-2} \over (\epsilon_1\epsilon_2)^{N -1} }}
\ee
Recall here that the $W^{(s_i)}_l$'s and $\tilde W^{(s_i)}_{l + 1/2}$'s generate ${\cal W}({\frak {so}(2N)}^{(2)}_{\rm aff}) = {\cal W}({\frak {usp}(2N-2)}^\vee_{\rm aff})$, and that on $\widehat{{\cal W}}({\frak {usp}(2N-2)}^\vee_{\rm aff})$,  the  $\{W^{(s_i)}_{l < 0}, \tilde W^{(s_i)}_{l < 0}\}$ and $\{W^{(s_i)}_{l > 0}, \tilde W^{(s_i)}_{l > 0}\}$ act as creation and annihilation operators, respectively; in particular, $\tilde W^{(2N-2)}_{1/2}$ is an annihilation operator, so (\ref{tildeWN1-D}) means that $ |q, \Delta \rangle$ is again a\emph{ coherent state}.

What about when $n=3$ (with $N=4$) whence we have $G = G_2$? According to~\cite{abcdefg}, instead of (\ref{phi(z)-D}), we now have
\be
\label{twisted phi(z)-DD}
\phi_s (z)= u_s \left({dz \over z}\right)^s,  \quad {\tilde \phi}_j (z)= 0, \quad {\tilde \phi}_{6}(z) = \left(z^{1/3} + {\Lambda^{6} \over z^{1/3}}\right)\left({dz \over z}\right)^{6},
\ee 
where $s=2,4, 6$; $j = 2,4$; and the $\tilde \phi_s(z)$'s are $(s,0)$-holomorphic differentials on $\cal C$ with modes in $\mathbb Z$, $\mathbb Z + 1/3$ and $\mathbb Z + 2/3$. This is again consistent with our results established in $\S$B and after (\ref{AGT-CDG-chiral CFT}) that for $n=3$ (with $N=4$), we have, on $\cal C$, the following $(s_i, 0)$-holomorphic differentials 
\be
\label{twisted W's-DD}
W^{(s_i)}(z) = \left(\sum_{l \in \mathbb Z} {W^{(s_i)}_l \over z^l} \right) \left( dz \over z \right)^{s_i}, \quad {\tilde W}^{(s_i)}(z) = \left(\sum_{k=1}^2 \sum_{l \in \mathbb Z} {{\tilde W}^{(s_i)}_{l + k/3} \over z^{l+k/3}} \right) \left( dz \over z \right)^{s_i},
 \ee
where $s_i = 2, 4, 6$, whence  we can naturally identify, up to some constant factor,  $\phi_s (z)$ with  $W^{(s)}(z)$ and  $\tilde\phi_s (z)$ with  $\tilde W^{(s)}(z)$.

At $z = 0$ where the state $|q, \Delta \rangle$ is, we find, after comparing (\ref{twisted phi(z)-DD}) with (\ref{twisted W's-DD}), that instead of (\ref{W-2-D}), we have
\be
\label{W-1-DD}
\boxed{W^{(s)}_{l \geq 1} \, |q, \Delta \rangle = 0, \quad {\rm for} \quad s = 2, 4, 6}
\ee
We also have (\ref{Ws0-D}) (where $N=4$), and
\be
\label{twisted W-2-DD}
\boxed{\tilde W^{(s)}_{l \geq 2/3} \, |q, \Delta \rangle = 0, \quad {\rm for} \quad s = 2,4,6}
\ee
and instead of (\ref{WN1-D}), we have $\tilde W^{(6)}_{1/3} |q, \Delta \rangle = q |q, \Delta \rangle  \sim \Lambda^{6} |q, \Delta \rangle$. By employing the same reasoning used to derive (\ref{WN1-D}), we find that  
\be
\label{tildeWN1-DD}
\boxed{\tilde W^{(6)}_{1/3} |q, \Delta \rangle = q |q, \Delta \rangle, \quad q = {\Lambda^{6} \over (\epsilon_1\epsilon_2)^{3} }}
\ee
Recall here that the $W^{(s_i)}_l$'s, $\tilde W^{(s_i)}_{l + 1/3}$'s and $\tilde W^{(s_i)}_{l + 2/3}$'s generate ${\cal W}({\frak {so}(2N)}^{(3)}_{\rm aff}) = {\cal W}({{\frak g}}^\vee_{2 \, \rm aff})$, and that on $\widehat{{\cal W}}({\frak g}^\vee_{2 \, \rm aff})$,  the  $\{W^{(s_i)}_{l < 0}, \tilde W^{(s_i)}_{l < 0}\}$ and $\{W^{(s_i)}_{l > 0}, \tilde W^{(s_i)}_{l > 0}\}$ act as creation and annihilation operators, respectively; in particular, $\tilde W^{(6)}_{1/3}$ is an annihilation operator, so (\ref{tildeWN1-DD}) means that $ |q, \Delta \rangle$ is also a\emph{ coherent state}.

Thus, in arriving at the above boxed relations (i) (\ref{AGT-duality-D}), (\ref{c-D-epsilon}), (\ref{k'-D}), (\ref{L_0-D}), (\ref{Delta(2)-D}), (\ref{psi = delta, D}), (\ref{q | q, D}), (\ref{W-2-D}), (\ref{Ws0-D}), (\ref{WN1-D}), (ii) (\ref{AGT-duality-D}), (\ref{c-D-epsilon}), (\ref{k'-D}), (\ref{L_0-D}), (\ref{Delta(2)-D}), (\ref{psi = delta, D}), (\ref{q | q, D}), (\ref{Ws0-D}), (\ref{W-1-D}), (\ref{twisted W-2-D}), (\ref{tildeWN1-D}) and (iii) (\ref{AGT-duality-D}), (\ref{c-D-epsilon}), (\ref{k'-D}), (\ref{L_0-D}), (\ref{Delta(2)-D}), (\ref{psi = delta, D}), (\ref{q | q, D}), (\ref{Ws0-D}), (\ref{W-1-DD}), (\ref{twisted W-2-DD}), (\ref{tildeWN1-DD}), we have just furnished a fundamental physical derivation of the pure AGT correspondence for the (i) $D_{N}$, (ii) $C_{N-1}$ and (iii) $G_2$ groups!

\newsection{Generalizations of the Pure AGT Correspondence and the Case with Matter}

\newsubsection{A ``Ramified'' Generalization of the Pure AGT Correspondence}

Let us now derive, purely physically, a ``ramified'' generalization of the pure AGT correspondence for the $A$, $B$, $C$, $D$ and $G$ groups. To this end, recall from (\ref{M-theory defect}) and (\ref{M-theory defect dual}) that the 4d worldvolume defect is equivalent to a geometric background of the underlying M-theory compactification, just like $\mathbb R^4/\mathbb Z_k$ and $TN_N^{R \to 0}$; the same can be said about the 4d worldvolume defect in (\ref{OM-theory defect}) and (\ref{OM-theory 8 defect}) -- it is equivalent to a geometric background of the underlying M-theory compactification, just like $\mathbb R^4 / \mathbb Z_k$ and $SN^{R \to 0}_N$.  Recall also that our discussion in $\S$5.1 about turning on Omega-deformation is independent of the geometric background of the underlying M-theory/string compactification. Altogether therefore, in the presence of the 4d worldvolume defect, our arguments hitherto which led us to (\ref{AGT-M-duality-AB}) and (\ref{AGT-M-duality-CDG}) would mean that in place of them, we ought to have the following \emph{physically dual} compactifications 
\be
\underbrace{\mathbb R^4\vert_{\epsilon_1, \epsilon_2}  \times \Sigma_{n,t}}_{\textrm{$N$ M5-branes + 4d defect}}\times \mathbb R^{5}\vert_{\epsilon_3; \,  x_{6,7}}  \quad \Longleftrightarrow  \quad   {\mathbb R^{5}}\vert_{\epsilon_3; \, x_{4,5}} \times \underbrace{{\cal C}  \times TN_N^{R\to 0}\vert_{\epsilon_3; \, x_{6,7}}}_{\textrm{$1$ M5-branes + 4d defect}},
\label{AGT-M-duality-AB-defect}
\ee
and 
\be
\underbrace{\mathbb R^4\vert_{\epsilon_1, \epsilon_2}  \times \Sigma_{n,t}}_{\textrm{$N$ M5-branes + OM5-plane + 4d defect}}\times \mathbb R^{5}\vert_{\epsilon_3; \,  x_{6,7}}  \quad \Longleftrightarrow  \quad   {\mathbb R^{5}}\vert_{\epsilon_3; \, x_{4,5}} \times \underbrace{{\cal C}  \times SN_N^{R\to 0}\vert_{\epsilon_3; \, x_{6,7}}}_{\textrm{$1$ M5-branes + 4d defect}},
\label{AGT-M-duality-CDG-defect}
\ee
respectively, where we have a common half-BPS boundary condition at the tips of $\mathbb I_t \subset \Sigma_{n,t} = {\bf S}^1_n \times \mathbb I_t$; the radius of ${\bf S}^1_n$ is $\beta$; $\mathbb I_t \ll \beta$; $\cal C$ is \emph{a priori} the same as $\Sigma_{n,t} $; the 4d worldvolume defect on the LHS of (\ref{AGT-M-duality-AB-defect}) and (\ref{AGT-M-duality-CDG-defect}) wraps $\Sigma_{n,t}$ and the $z$-plane in $\mathbb R^4\vert_{\epsilon_1, \epsilon_2} \simeq \mathbb C_z\vert_{\epsilon_1} \times \mathbb C_w\vert_{\epsilon_2}$; the\emph{ dual} 4d worldvolume defect on the RHS of (\ref{AGT-M-duality-AB-defect}) and (\ref{AGT-M-duality-CDG-defect}) wraps $\cal C$ and the $x_8$-$x_9$ directions in $TN_N^{R\to 0}\vert_{\epsilon_3; \, x_{6,7}}$ and $SN_N^{R\to 0}\vert_{\epsilon_3; \, x_{6,7}}$, and here, the $x_9$-direction is spanned by the ${\bf S}^1$-fiber of  $TN_N^{R\to 0}\vert_{\epsilon_3; \, x_{6,7}}$ and $SN_N^{R\to 0}\vert_{\epsilon_3; \, x_{6,7}}$, while the $x_6$-$x_7$-$x_8$-directions are spanned by their $\mathbb R^3\vert_{\epsilon_3; \, x_{6,7}}$ base. As usual, there is a $\mathbb Z_n$-outer-automorphism of ${\mathbb R^4}\vert_{\epsilon_1, \epsilon_2}$, $ TN_N^{R\to 0}\vert_{\epsilon_3; \, x_{6,7}}$  and $ SN_N^{R\to 0}\vert_{\epsilon_3; \, x_{6,7}}$ as we go around the ${\bf S}^1_n$ circle and identify the circle under an order $n$ translation, and the $\epsilon_i$'s are parameters of the Omega-deformation along the indicated planes described in detail in $\S$5.1.

\bigskip\noindent{\it The Spectrum of Spacetime BPS States on the LHS of (\ref{AGT-M-duality-AB-defect}) and (\ref{AGT-M-duality-CDG-defect})}

Let us now determine the spectrum of spacetime BPS states on the LHS of (i) (\ref{AGT-M-duality-AB-defect}) and (ii) (\ref{AGT-M-duality-CDG-defect}) that define a ``ramified'' generalization of the partition function in (\ref{5d BPS}). In the absence of Omega-deformation whence $\epsilon_i = 0$, according to our discussions in $\S$5.1 and $\S$4.3, the spacetime BPS states would be captured by the topological sector of the $\cN = (4,4)$ sigma-model on  $\Sigma_{n,t}$ with target the moduli space ${\cal M}_{G, \mathbb L}$ of ``ramified'' $G$-instantons on $\mathbb R^4$, where (i) for $n=1$ or 2 (with even $N$), $G = SU(N)$ or $SO(N+1)$; (ii) for $n=1$, 2 and 3 (with $N=4$), $G = SO(2N)$, $USp(2N-2)$ and $G_2$; while $\mathbb L \subset G$ is a Levi subgroup which characterizes the 4d worldvolume defect (as explained in $\S$2.3 and $\S$4.3). However, in the presence of Omega-deformation, our discussion immediately after (\ref{5d BPS}) now means that as one traverses a closed loop in $\Sigma_{n,t}$, there would be a  $\bf g$-automorphism of ${\cal M}_{G, \mathbb L}$, where ${\bf g} \in U(1) \times U(1) \times T$, and $T \subset G$ is the Cartan subgroup. Consequently, the spacetime BPS states of interest would, in the presence of Omega-deformation, be captured by the topological sector of a non-dynamically ${\bf g}$-gauged version of the aforementioned sigma-model (see footnote~\ref{gauging worldsheet}).  Hence, according to~\cite{mine-equivariant} and our arguments in $\S$4.3 which led us to (i)   (\ref{BPS-M-ram-A}) and (\ref{BPS-M-ram-B}), (ii) (\ref{BPS-M-ram-D}), (\ref{BPS-M-ram-C}) and (\ref{BPS-M-ram-G}), we can express the Hilbert space ${\cal H}^{\Omega}_{\rm BPS}$ of spacetime BPS states on the LHS of (i) (\ref{AGT-M-duality-AB-defect}) and (ii)  (\ref{AGT-M-duality-CDG-defect}) as
\be
{\cal H}^\Omega_{\rm BPS} = \bigoplus_{a'} {\cal H}^\Omega_{{\rm BPS}, a'}  =  \bigoplus_{a'} ~{\rm IH}^\ast_{U(1)^2 \times T} \, {\cal U}({\cal M}_{G, \mathbb L, a'}), 
\label{BPS-AGT-AB-defect}
\ee
where ${\rm IH}^\ast_{U(1)^2 \times T} \, {\cal U}({\cal M}_{G, a'})$ is the $\mathbb Z_n$-invariant (in the sense of (i) (\ref{HBPS-eff}) and (ii) (\ref{HBPS-eff-USp(2N-2)}) and (\ref{HBPS-eff-G_2}), when (i) $n=2$ and (ii) $n = 2$ and 3) $U(1)^2 \times T$-equivariant intersection cohomology of the Uhlenbeck compactification ${\cal U}({\cal M}_{G, \mathbb L, a'})$ of the (singular) moduli space ${\cal M}_{G, \mathbb L, a'}$ of ``ramified'' $G$-instantons on $\mathbb R^4$ with ``ramified'' instanton number $a'$. Here, the positive number $a' = a + {\rm Tr} \, \alpha\frak m$, where $a$ is the ordinary instanton number; Tr is a quadratic form on $\frak g$; $\alpha \in \frak t$ is the holonomy parameter that is the commutant of $\mathbb L$; $\frak t$ is the Lie algebra of $T$; ${\frak m} \in \Lambda_{\rm cochar}$ is the ``magnetic charge''; and $ \Lambda_{\rm cochar}$ is the cocharacter lattice of $G$.

\bigskip\noindent{\it The Spectrum of Spacetime BPS States on the RHS of (\ref{AGT-M-duality-AB-defect}) and (\ref{AGT-M-duality-CDG-defect})}

Let us next ascertain the corresponding spectrum of spacetime BPS states on the RHS of (i) (\ref{AGT-M-duality-AB-defect}) and (ii) (\ref{AGT-M-duality-CDG-defect}). Bearing in mind footnote~\ref{junya's twist} which tells us that the underlying worldvolume theory of the single M5-brane (plus 4d worldvolume defect) is conformal along (i) $TN_N^{R\to 0}\vert_{\epsilon_3; \, x_{6,7}}$ and (ii) $SN_N^{R\to 0}\vert_{\epsilon_3; \, x_{6,7}}$  in (i) (\ref{AGT-M-duality-AB-defect}) and (ii) (\ref{AGT-M-duality-CDG-defect}), by repeating our arguments in $\S$3.1 and $\S$3.2 which led us beyond (i) (\ref{equivalent IIA system 1}) and (ii) (\ref{equivalent IIA system 2}), and from our discussion surrounding (\ref{metric}), we find that the spacetime BPS states would be furnished by the ``ramified'' I-brane theory in
\be
{\rm (i)} \qquad \textrm{IIA}: \quad \underbrace{ {\mathbb R}^5\vert_{\epsilon_3; x_{4,5}} \times {\cal C} \times {\mathbb R}^3\vert_{\epsilon_3; x_{6,7}}}_{\textrm{I-brane on ${\cal C} = N \textrm{D6} \cap 1\textrm{D4}\cap{\rm 3d \, defect}$}},
\label{equivalent IIA system 1 - AGT-AB-defect}
\ee
and
\be
{\rm (ii)} \qquad \textrm{IIA}: \quad \underbrace{ {\mathbb R}^5\vert_{\epsilon_3; x_{4,5}} \times {\cal C} \times {\mathbb R}^3/{\cal I}_3\vert_{\epsilon_3; x_{6,7}}}_{\textrm{I-brane on ${\cal C} = N \textrm{D6} \cap 1\textrm{D4}\cap{\rm 3d \, defect}$}}.
\label{equivalent IIA system 1 - AGT-CDG-defect}
\ee
Here, we have a stack of $N$ coincident D6-branes whose worldvolume is given by ${\mathbb R}^5\vert_{\epsilon_3; x_{4,5}} \times {\cal C}$; a single D4-brane whose worldvolume is given by (i) ${\cal C} \times \mathbb R^3\vert_{\epsilon_3; x_{6,7}}$ and (ii) ${\cal C} \times \mathbb R^3/{\cal I}_3\vert_{\epsilon_3; x_{6,7}}$; and a 3d worldvolume defect which wraps ${\cal C}$ and the $x_8$-direction in (i) $\mathbb R^3\vert_{\epsilon_2; x_{6,7}} = \mathbb R \times \mathbb R^2\vert_{\epsilon_3}$ and (ii) ${\mathbb R}^3/{\cal I}_3\vert_{\epsilon_3; x_{6,7}} = (\mathbb R \times \mathbb R^2\vert_{\epsilon_3})/ {\cal I}_3$.

If the 4d worldvolume defect is absent or trivial whence $\mathbb L = G$, our arguments that took us from (i) (\ref{equivalent IIA system 1 - AGT}) to (\ref{AGT-AB-chiral CFT}) and (ii) (\ref{equivalent IIA system 2 - AGT}) to (\ref{AGT-CDG-chiral CFT}),  would mean that the spacetime BPS states ought to be furnished by the states of a partially gauged chiral CFT on $\cal C$ which, in the schematic notation of $\S$3.1 and $\S$3.2,  can   be expressed as\footnote{To arrive at the following expressions, we recall that $\frak {sl}(N)^{(1)}_{\rm aff} \simeq \frak {sl}(N)^\vee_{\rm aff}$, $\frak {sl}(N)^{(2)}_{\rm aff} \simeq \frak {so}(N+1)^\vee_{\rm aff}$ (for even $N$), $\frak {so}(2N)^{(1)}_{\rm aff} \simeq \frak {so}(2N)^\vee_{\rm aff}$, $\frak {so}(2N)^{(2)}_{\rm aff} \simeq \frak {usp}(2N-2)^\vee_{\rm aff}$ and $\frak {so}(2N)^{(3)}_{\rm aff} \simeq \frak g^\vee_{2 \, \rm aff}$ (with $N=4$).\label{affine iso}}
\be
\hspace{-2.0cm}{\rm (i)} \, \, \frak {g}^\vee_{\mathbb C \, \rm aff, 1} / {\frak{n}}^{\vee}_{+ \, {\rm aff}, p_1},  \quad  \textrm{$\frak {g}^\vee_{\mathbb C \, \rm aff} = {\frak {sl}(N)}^\vee_{\rm aff}$, ${\frak{so}(N+1)}^\vee_{\rm aff}$ (with even $N$) if $n=1$, 2,}
\label{AGT-AB-chiral CFT-defect}
\ee
 and
\be
{\rm (ii)} \, \, \frak {g}^\vee_{\mathbb C \, \rm aff, 1}  / {\frak{n}}^\vee_{+ \, {\rm aff}, p_2}, \quad \textrm{$\frak {g}^\vee_{\mathbb C \, \rm aff} = {\frak {so}(2N)}^\vee_{\rm aff}$, ${\frak{usp}(2N-2)}^\vee_{\rm aff}$, $\frak g^\vee_{2 \, \rm aff}$ (with $N = 4$) if $n=1$, 2, 3.}
\label{AGT-CDG-chiral CFT-defect}
\ee
Here, $\cal C$ is effectively ${\bf S}^2 / \{0, \infty\}$, i.e., it can be regarded as an ${\bf S}^1_n$ fibration of $\mathbb I_t$ whose fiber has zero radius at the two end points $z = 0$ and $z = \infty$; `$z$' is a holomorphic coordinate on $\cal C$; $\frak n^\vee_{+ \, \rm aff} \subset \frak g^\vee_{\mathbb C \, \rm aff}$ is a \emph{Langlands dual} affine Lie subalgebra whose associated nilpotent Lie algebra consists of strictly upper-traingular matrices; and the level $p_i$ necessarily depends on the relevant Omega-deformation parameters $\epsilon'_1 = \beta \epsilon_1$ and $\epsilon'_2 = \beta \epsilon_2$, although $p_i$, being a purely real number, should not depend on the purely imaginary parameter $\vec a' = \beta \vec a$.

However, if the 4d worldvolume defect is nontrivial whence $\mathbb L \neq G$, then, our arguments which led us to (i) (\ref{partially gauged CFT - A - ram}) and (\ref{partially gauged CFT - B - ram}) and (ii) (\ref{partially gauged CFT - D - ram}), (\ref{partially gauged CFT - C - ram}) and (\ref{partially gauged CFT - G - ram}), would mean that in place of (\ref{AGT-AB-chiral CFT-defect}) and (\ref{AGT-CDG-chiral CFT-defect}), we ought to have
\be
\hspace{-0.4cm}{\rm (i)} \, \, {\frak {g}^\vee_{\mathbb C \, {\rm aff}, r} \over {\frak{n}}^{\vee}_{+ \, {\rm aff}, p'_1} \otimes [\frak {g}^\vee_{\mathbb C \, {\rm aff}, r} / \frak {p}^\vee_{{\rm aff}, r}]}, \, \, \,  \textrm{$\frak {g}^\vee_{\mathbb C \, \rm aff} = {\frak {sl}(N)}^\vee_{\rm aff}$, ${\frak{so}(N+1)}^\vee_{\rm aff}$ (with even $N$) if $n=1$, 2,} 
\label{AGT-AB-chiral CFT-defect-modded}
\ee
 and
\be
\hspace{-0.5cm}{\rm (ii)} \, \,  {\frak {g}^\vee_{\mathbb C \, {\rm aff}, q} \over {\frak{n}}^{\vee}_{+ \, {\rm aff}, p'_2} \otimes [\frak {g}^\vee_{\mathbb C \, {\rm aff}, q} / \frak {p}^\vee_{{\rm aff}, q}]}, \, \, \, \textrm{$\frak {g}^\vee_{\mathbb C \, \rm aff} = {\frak {so}(2N)}^\vee_{\rm aff}$, ${\frak{usp}(2N-2)}^\vee_{\rm aff}$, $\frak g^\vee_{2 \, \rm aff}$ (with $N = 4$) if $n=1$, 2, 3,}
\label{AGT-CDG-chiral CFT-defect-modded}
\ee
where $\frak p^\vee_{\rm aff} \subset \frak g^\vee_{\mathbb C \, \rm aff}$ is a (semi-lower triangular) parabolic  \emph{Langlands dual} affine Lie subalgebra that is associated with $\mathbb L$; the second factor in the denominator is due to the dual 4d worldvolume defect; and the levels $\{p'_i, r, q \} \in \mathbb R$.\footnote{We have, for convenience, replaced the levels $\{1, p_1\}$ and $\{1, p_2 \}$ in (\ref{AGT-AB-chiral CFT-defect}) and (\ref{AGT-CDG-chiral CFT-defect}) with the levels  $\{r, p'_1\}$ and $\{q, p'_2\}$ in (\ref{AGT-AB-chiral CFT-defect-modded}) and (\ref{AGT-CDG-chiral CFT-defect-modded}), keeping the overall central charge of the underlying partially gauged chiral CFT the same.}

Note that one can also regard the entire factor in the denominator of (\ref{AGT-AB-chiral CFT-defect-modded}) and (\ref{AGT-CDG-chiral CFT-defect-modded}) as being due to an \emph{Omega-deformed dual 4d worldvolume defect} which (i) effects a pure Omega-deformation that brings us back to (\ref{AGT-AB-chiral CFT-defect}) and (\ref{AGT-CDG-chiral CFT-defect}), respectively, when the  defect is trivial, i.e., when $\mathbb L = G$ whence $\frak p^\vee = \frak g^\vee_\mathbb C$; (ii) effects a trivial Omega-deformation -- so that the entire factor in the denominator of (\ref{AGT-AB-chiral CFT-defect-modded}) and (\ref{AGT-CDG-chiral CFT-defect-modded}) is equal to identity -- when the defect is full, i.e., when $\mathbb L = T$ (as we shall explain later).

\bigskip\noindent{\it A ``Ramified'' Generalization of the Pure AGT Correspondence for the $A$ Groups}

Let us now focus on (\ref{AGT-AB-chiral CFT-defect-modded}) with $n=1$ whence $\frak {g}^\vee_{\mathbb C \, \rm aff} = {\frak {sl}(N)}^\vee_{\rm aff}$. Note that the factor in the denominator of (\ref{AGT-AB-chiral CFT-defect-modded}) means that we are gauging the ${\frak {sl}(N)}^\vee_{\rm aff}$ WZW model on $\cal C$ by a subgroup $\cal S$ whose Lie algebra is $\frak s = \frak{n}^{\vee}_+ \oplus [{\frak {sl}(N)}^\vee  \ominus \frak {p}^\vee]$. Since the second nilpotent factor $ [{\frak {sl}(N)}^\vee  \ominus \frak {p}^\vee]$ is due to the dual 4d worldvolume defect which, in turn, is characterized by dual nilpotent orbits $O_{e^\vee}$ (c.f.~$\S$4.3), according to $\S$2.3, $\S$4.3, and the fact that $\frak{n}^{\vee}_+ \simeq \frak{n}_+$ for simply-laced Lie algebras, we can also write   
\be
\label{cal S}
{\cal S} = N_+ \times {\cal  P}_{[N]^t} / {\cal P}_{[n_I]^t},
\ee
where $N_+ \subset SL(N,\mathbb C)$ is the nilpotent subgroup of strictly upper triangular matrices, and ${\cal  P}_{[m]^t} \subset SL(N,\mathbb C)$ is a semi-lower triangular parabolic subgroup defined by the partition $[m]^t$ (see $\S$2.3) that is a transpose of the partition $[m]$ of $N$ (in the sense of a Young diagram defined in the British convention). Since $ {\cal  P}_{[N]^t} =  {\cal  P}_{[1, \dots, 1]} = B_-$, where $B_- \subset SL(N,\mathbb C)$ is a lower-triangular Borel subgroup, (\ref{cal S}) would mean that $\cal S$ is a nilpotent subgroup of strictly upper-triangular matrices, i.e., we can also write 
\be
\label{a-plus}
\frak s = \frak a_+,
\ee
where $\frak a_+ \subset {\frak {sl}(N)}$ is a nilpotent Lie subalgebra of strictly upper triangular matrices. 

Thus, in this case, we find (bearing in mind the isomorphism relations in footnote~\ref{affine iso}) that the sought-after spacetime BPS states ought to be given by the states of the partially gauged chiral CFT 
\be
\label{coset-AGT-A}
{\frak {sl}(N)}_{{\rm aff}, r} / \frak a_{+ \, {\rm aff}, p_A},
\ee
where the levels $p_A$ and $r$ may not be the same, as the central charge of $ \frak a_{+ \, {\rm aff}, p_A}$, like that of ${\frak{n}}^{\vee}_{+ \, {\rm aff}, p_1} \simeq {\frak{n}}_{+ \, {\rm aff}, p_1}$ in (\ref{AGT-AB-chiral CFT-defect}), must also contribute to an anomalous shift in the overall central charge which can then be ``absorbed'' by the curvature of $\cal C$, as explained in $\S$5.2. This partially gauged chiral CFT, like the one in (\ref{AGT-AB-chiral CFT}), can be realized as a gauged $SL(N,\mathbb C)$ WZW model, although the Lie algebra of the gauge group is now $\frak a_+$ instead of $\frak n_+$.  

Note that $\frak a_+$ is such that in an appropriate basis of $\frak {sl}(N)$, one can always find an element $\delta$ of the Cartan subalgebra of $\frak {sl}(N)$ whereby
\be
\label{positive eigenspace}
[\delta, x] = l x
\ee
for some $x \in \frak a_+$ and positive  \emph{integer} $l$. Take for example $N=3$ and $[n_I] = [2,1]$; let $E_{ij}$ denote an $N \times N$ matrix whose $(i,j)$ component is one while the rest are zero; then, from (\ref{cal S}), we have $x = \alpha_1 E_{13} + \alpha_2E_{23}$, where the $\alpha_i$'s are real constants, and as explicitly verified in~\cite{Tjin}, $x$ indeed satisfies (\ref{positive eigenspace}). As another example, one can take $N=4$ and $[n_I] = [2,1,1]$, $[2,2]$ or $[3,1]$; again, one can, for each case, compute $x$ using (\ref{cal S}), and as explicitly verified in~\cite{Tjin}, it will always satisfy (\ref{positive eigenspace}). 

Hence, if we were to repeat the computation in Appendix B with gauge group $\cal S$ instead of $N_+$, we would physically realize the general BRST algorithm in~\cite{Tjin-Boer}. What this means is that the chiral CFT would realize ${\cal W}({\frak {su}}(N)_{\rm aff}, \rho_{\cal A})$ -- an untwisted affine $\cal W$-algebra obtained from ${\frak {sl}}(N)_{\rm aff}$ via a quantum Drinfeld-Sokolov reduction that is associated with the embedding $\rho_{\cal A}: \frak {sl}(2) \to \frak {sl}(N)$ (which, through the Jacobson-Morozov theorem, is determined by $\frak a_+$ and therefore, the partition $[n_I]$ which characterizes the underlying 4d worldvolume defect).  In other words,  the states of the chiral CFT would be furnished by a Verma module  $\widehat{{\cal W}}({\frak {su}}(N)_{\rm aff}, \rho_{\cal A})$ over ${\cal W}({\frak {su}}(N)_{\rm aff}, \rho_{\cal A})$, and the Hilbert space ${\cal H}^{\Omega'}_{\rm BPS}$ of spacetime BPS states on the RHS of (\ref{AGT-M-duality-AB-defect}) when $n=1$, can be expressed as
\be
{\cal H}^{\Omega'}_{\rm BPS}  = \widehat{{\cal W}}({\frak {su}}(N)_{\rm aff}, \rho_{\cal A}).
\label{AGT-AB-H=W-defect}
\ee

Clearly, the physical duality of the compactifications in (\ref{AGT-M-duality-AB-defect}) will mean that ${\cal H}^\Omega_{\rm BPS}$ in (\ref{BPS-AGT-AB-defect}) (when $n=1$) is equivalent to ${\cal H}^{\Omega'}_{\rm BPS}$ in (\ref{AGT-AB-H=W-defect}), i.e.,
\be
 \boxed{\bigoplus_{a'} ~{\rm IH}^\ast_{U(1)^2 \times T} \, {\cal U}({\cal M}_{SU(N), \mathbb L, a'}) = \widehat{{\cal W}}({\frak {su}}(N)_{\rm aff}, \rho_{\cal A})}
 \label{AGT-duality-A-defect}
\ee
Thus, we have a ``ramified'' generalization of the duality relation (\ref{AGT-duality-A}) for $G = SU(N)$. 

Accordingly, $c_A$ in (\ref{cc-A}) ought to be replaced by~\cite{Tjin-Boer} 
\be
 c_{\cal A}  = {\rm dim} \, {\frak {sl}}(N)_0  - {1 \over 2} {\rm dim} \, {\frak {sl}}(N)_{1 \over 2} - 12 \left | \alpha_+ \rho   + \alpha_- t_0\right |^2. 
\label{cc-defect-general}
\ee
Here, ${\frak {sl}}(N)_j = \{ x \in {\frak {sl}}(N) \, \vert \, [t_0, x] = jx \}$; $t_0 = \rho_{\cal A}(\sigma^3)$, where $\sigma^3 \in {\frak {sl}}(2)$ is a Cartan element; $\rho$ is the Weyl vector of ${\frak {sl}}(N)$; $\alpha_+ = 1 / \sqrt{k_{\cal A} + N}$, where $\alpha_+\alpha_- = -1$; and $k_{\cal A} \in \mathbb R$ is the effective level of the underlying affine Lie algebra ${\frak {sl}}(N)_{\rm aff}$.   

If we have a trivial 4d worldvolume defect whence $\mathbb L = SU(N)$ so $[n_I] = [N]$,  from (\ref{cal S}), we find  that $\frak a_+ = \frak n_+$ whence $\rho_{\cal A}$ would be principal; in this  case, $t_0 = {\rho}^\vee = \rho$, ${\rm dim} \, {\frak {sl}}(N)_0  - {1 \over 2} {\rm dim} \, {\frak {sl}}(N)_{1 \over 2} = {\rm rank} \, \frak {sl}(N) = N-1$, so $c_{\cal A}$ coincides with (\ref{cc-simply-laced}) (for the $A$ groups).  As such, when the defect is trivial (i.e.~absent), ${\cal W}({\frak {su}}(N)_{\rm aff}, \rho_{\cal A}) = {\cal W}({\frak {su}}(N)_{\rm aff})$ with central charge $c_{\cal A} = c_A$, consistent with our ``unramified'' results in $\S$5.2. As further explained in $\S$5.2, $c_{\cal A}$ would also depend on the Omega-deformation parameters $\epsilon_{1,2}$ through $\alpha_+ = -i \sqrt{\epsilon_1 / \epsilon_2}$ and $\alpha_- = -i \sqrt{\epsilon_2 / \epsilon_1}$. 

Since Omega-deformation is independent of the choice of 4d worldvolume defect, the manner in which $c_{\cal A}$ depends on $\epsilon_{1,2}$ would not change as we vary $\rho_{\cal A}$ away from being principal, i.e., we have $\alpha_+ = -i \sqrt{\epsilon_1 / \epsilon_2}$ and $\alpha_- = -i \sqrt{\epsilon_2 / \epsilon_1}$ in (\ref{cc-defect-general}) for \emph{all} $\rho_{\cal A}$. Therefore, we can also write $c_{\cal A}$ as 
\be
\boxed{ c_{{\cal A}, \epsilon_{1,2}}  = {\rm dim} \, {\frak {sl}}(N)_0  - {1 \over 2} {\rm dim} \, {\frak {sl}}(N)_{1 \over 2} + 12 \left | \sqrt{\epsilon_1 \over \epsilon_2} \, \rho   + \sqrt{\epsilon_2 \over \epsilon_1} \, t_0\right |^2} 
\label{cc-defect-omega-deformed}
\ee
Notice that consistent with the LHS of (\ref{AGT-M-duality-AB-defect}), $c_{{\cal A}, \epsilon_{1,2}}$ is also asymmetric under the exchange $\epsilon_1 \leftrightarrow \epsilon_2$ whenever we have a nontrivial defect. In addition, we also have
\be
\label{k'-defect}
\boxed{k_{\cal A}  = - N - {\epsilon_2 \over \epsilon_1}}
\ee

Note at this point that a rigorous definition of a ``ramified'' generalization of the Nekrasov instanton partition function  in~\cite[$\S$6.6]{AGT-math} means that we can actually ``ramify'' the arguments which took us from (\ref{abcd BPS relation}) to (\ref{q | q}). As such, in the presence of a nontrivial 4d worldvolume defect whose nature is encoded in $\rho_{\cal A}$, we can, via (\ref{AGT-duality-A-defect}), write the ``ramified'' Nekrasov instanton partition function as
\be
\label{q | q, defect}
\boxed{Z_{\rm inst} (SU(N), \epsilon_1, \epsilon_2, \vec a, \mathbb L) = \langle \rho_{\cal A}, \Delta   | \rho_{\cal A}, \Delta \rangle}
\ee
where
\be
\label{coherent-A}
\boxed{| \rho_{\cal A}, \Delta \rangle = \bigoplus_{a'}  A^{a'}  | \Psi_{a', \mathbb L_{\cal A}} \rangle}
\ee
Here, $| \rho_{\cal A}, \Delta \rangle \in \widehat{{\cal W}}({\frak {su}}(N)_{\rm aff}, \rho_{\cal A})$; $A^{a'}$ is some real number; $ | \Psi_{a', \mathbb L_{ \cal A}} \rangle \in {\rm IH}^\ast_{U(1)^2 \times T} \, {\cal U}({\cal M}_{SU(N), \mathbb L, a'})$ is also a state in $ \widehat{{\cal W}}({\frak {su}}(N)_{\rm aff}, \rho_{\cal A})$ with energy level $n_{a'}$ determined by the ``ramified'' instanton number $a'$ (as one recalls that  $n_{a'}$ is a constant shift of the eigenvalue $a'$ of the $L_0$ operator which generates translations along the ${\bf S}^1_n$ circle in (\ref{AGT-M-duality-AB-defect})); and $ \langle \cdot | \cdot \rangle$ is a Poincar\'e pairing in the sense of~\cite[$\S$2.6]{J-function}. The label $\Delta$ just means that $\widehat{{\cal W}}({\frak {su}}(N)_{\rm aff}, \rho_{\cal A})$ is generated by the application of creation operators (furnished by the negative-mode elements of ${{\cal W}}({\frak {su}}(N)_{\rm aff}, \rho_{\cal A})$) on the highest weight state $| \Delta \rangle$.  

As in the ``unramified'' case, since the RHS of (\ref{q | q, defect}) is defined in the limit that the ${\bf S}^1_n$ fiber in $\cal C$ has \emph{zero} radius, and since we have in $\cal C$ a \emph{common} boundary condition at $z = 0$ and $z=\infty$ (where the radius of the  ${\bf S}^1_n$ fiber is zero),  $| \rho_{\cal A}, \Delta \rangle$ and $\langle \rho_{\cal A}, \Delta |$ ought to be a state and its dual associated with the puncture at $z = 0$ and $z=\infty$, respectively. Furthermore, as the RHS of (\ref{coherent-A}) is a sum over states of all possible energy levels, it would mean that $| \rho_{\cal A}, \Delta \rangle$ is actually a \emph{coherent state}. 

Thus, in arriving at the boxed relations (\ref{AGT-duality-A-defect}), (\ref{cc-defect-omega-deformed}), (\ref{k'-defect}), (\ref{q | q, defect}) and (\ref{coherent-A}), we have just furnished a fundamental physical derivation of a ``ramified'' pure AGT correspondence for the $A_{N-1}$ groups! (Given a specific $\mathbb L$ and hence $[n_I]$, the ``ramified'' version of the relations (\ref{W-2})--(\ref{WN1}) can be straightforwardly obtained, albeit rather tediously, via the computational technique introduced in~\cite[$\S$3.3]{Kanno-Tachikawa}.)

\bigskip\noindent{\it A ``Ramified'' Generalization of the Pure AGT Correspondence for the $B$ Groups}

Let us now focus on (\ref{AGT-AB-chiral CFT-defect-modded}) with $n=2$ and even $N$ whence $\frak {g}^\vee_{\mathbb C \, \rm aff} = {\frak {so}(N+1)}^\vee_{\rm aff}$. Recall at this point from footnote~\ref{affine iso} that ${\frak {so}(N+1)}^\vee_{\rm aff} = {\frak {sl}(N)}^{(2)}_{\rm aff} = {\frak {sl}(N)}^{\vee \, (2)}_{\rm aff}$, which means that our proceeding analysis would be exactly the same as that for the $A$ groups above, except that the affine Lie algebras involved are now $\mathbb Z_2$-twisted. As such, the Hilbert space ${\cal H}^{\Omega'}_{\rm BPS}$ of spacetime BPS states on the RHS of (\ref{AGT-M-duality-AB-defect}) when $n=2$ with even $N$, can be expressed as
\be
{\cal H}^{\Omega'}_{\rm BPS}  = \widehat{{\cal W}}({\frak {so}}(N+1)^\vee_{\rm aff}, \rho_{\cal A}), 
\label{AGT-AB-H=W-defect-B}
\ee
where $\widehat{{\cal W}}({\frak {so}}(N+1)^\vee_{\rm aff}, \rho_{\cal A})$ is a Verma module over ${\cal W}({\frak {so}}(N+1)^\vee_{\rm aff}, \rho_{\cal A})$  -- a $\mathbb Z_2$-twisted version of the affine $\cal W$-algebra ${\cal W}({\frak {su}}(N)_{\rm aff}, \rho_{\cal A})$  obtained from ${\frak {sl}}(N)_{\rm aff}$ via a quantum Drinfeld-Sokolov reduction that is associated with the embedding $\rho_{\cal A}: \frak {sl}(2) \to \frak {sl}(N)$ (which, as explained above, encodes the nature of the underlying 4d worldvolume defect).

Clearly, the physical duality of the compactifications in (\ref{AGT-M-duality-AB-defect}) will mean that ${\cal H}^\Omega_{\rm BPS}$ in (\ref{BPS-AGT-AB-defect}) (when $n=2$ with even $N$) is equivalent to ${\cal H}^{\Omega'}_{\rm BPS}$ in (\ref{AGT-AB-H=W-defect-B}), i.e.,
\be
 \boxed{\bigoplus_{a'} ~{\rm IH}^\ast_{U(1)^2 \times T} \, {\cal U}({\cal M}_{SO(N+1), \mathbb L, a'}) = \widehat{{\cal W}}({\frak {so}}(N+1)^\vee_{\rm aff}, \rho_{\cal A})}
 \label{AGT-duality-B-defect}
\ee
where the equivariant intersection cohomology is $\mathbb Z_2$-invariant in the sense explained below (\ref{BPS-AGT-AB-defect}). Thus, we have a ``ramified'' generalization of the duality relation (\ref{AGT-duality-A}) for $G = SO(N+1)$. 

According to footnote~\ref{central charge}, the central charge of a twisted $\cal W$-algebra (obtained as a coset theory of twisted affine Lie algebras, such as in our case) would be the same as its untwisted version. As such, the central charge $c_{\cal B}$ in this case would be the same as $c_{\cal A}$, i.e., 
\be
\boxed{ c_{{\cal B}, \epsilon_{1,2}}  = {\rm dim} \, {\frak {sl}}(N)_0  - {1 \over 2} {\rm dim} \, {\frak {sl}}(N)_{1 \over 2} + 12 \left | \sqrt{\epsilon_1 \over \epsilon_2} \, \rho   + \sqrt{\epsilon_2 \over \epsilon_1} \, t_0\right |^2} 
\label{cc-defect-omega-deformed-B}
\ee
Likewise, the level of the underlying $\mathbb Z_2$-twisted affine Lie algebra ${\frak {so}(N+1)}^\vee_{\rm aff}$ is
\be
\label{k'-defect-B}
\boxed{k_{\cal B}  = - N - {\epsilon_2 \over \epsilon_1}}
\ee
and 
\be
\label{q | q, defect-B}
\boxed{Z_{\rm inst} (SO(N+1), \epsilon_1, \epsilon_2, \vec a, \mathbb L) = \langle \rho_{\cal A}, \Delta_2   | \rho_{\cal A}, \Delta_2 \rangle}
\ee
where
\be
\label{coherent-B}
\boxed{| \rho_{\cal A}, \Delta_2 \rangle = \bigoplus_{a'}  B^{a'}  | \Psi_{a', \mathbb L_{\cal B}} \rangle}
\ee
Here, $| \rho_{\cal A}, \Delta_2 \rangle \in \widehat{{\cal W}}({\frak {so}}(N+1)^\vee_{\rm aff}, \rho_{\cal A})$; ${B}^{a'}$ is some real number; $ | \Psi_{a', \mathbb L_{\cal B}} \rangle \in {\rm IH}^\ast_{U(1)^2 \times T} \, {\cal U}({\cal M}_{SO(N+1), \mathbb L, a'})$ is also a state in $ \widehat{{\cal W}}({\frak {so}}(N+1)^\vee_{\rm aff}, \rho_{\cal A})$ with energy level $n_{a'}$ determined by the ``ramified'' instanton number $a'$; and the label $\Delta_2$ just means that $\widehat{{\cal W}}({\frak {so}}(N+1)^\vee_{\rm aff}, \rho_{\cal A})$ is generated by the application of creation operators (furnished by the negative-mode elements of ${{\cal W}}({\frak {so}}(N+1)^\vee_{\rm aff}, \rho_{\cal A})$) on the $\mathbb Z_2$-twisted highest weight state $| \Delta \rangle$.  

As in the $n=1$ case,  $| \rho_{\cal A}, \Delta_2 \rangle$ and $\langle \rho_{\cal A}, \Delta_2 |$ ought to be a state and its dual associated with the puncture at $z = 0$ and $z=\infty$ on $\cal C$, respectively. Furthermore, as the RHS of (\ref{coherent-B}) is a sum over states of all possible energy levels, it would mean that $| \rho_{\cal A}, \Delta_2 \rangle$ is actually a \emph{coherent state}. 

Thus, in arriving at the boxed relations (\ref{AGT-duality-B-defect}), (\ref{cc-defect-omega-deformed-B}), (\ref{k'-defect-B}), (\ref{q | q, defect-B}) and (\ref{coherent-B}), we have just furnished a fundamental physical derivation of a ``ramified'' pure AGT correspondence for the $B_{N/2}$ groups! (The ``ramified'' version of the relations (\ref{W-2})--(\ref{WN1}) can be obtained via a $\mathbb Z_2$-twisted generalization of the computational technique introduced in~\cite[$\S$3.3]{Kanno-Tachikawa}.)

 \bigskip\noindent{\it A ``Ramified'' Generalization of the Pure AGT Correspondence for the $D$ Groups}

Let us now focus on (\ref{AGT-CDG-chiral CFT-defect-modded}) with $n=1$ whence $\frak {g}^\vee_{\mathbb C \, \rm aff} = {\frak {so}(2N)}^\vee_{\rm aff}$. Note that the factor in the denominator of (\ref{AGT-CDG-chiral CFT-defect-modded}) means that we are gauging the ${\frak {so}(2N)}^\vee_{\rm aff}$ WZW model on $\cal C$ by a subgroup $\cal S$ whose Lie algebra is $\frak s = \frak{n}^{\vee}_+ \oplus [{\frak {so}(2N)}^\vee  \ominus \frak {p}^\vee]$. Notice that the second factor $ [{\frak {so}(2N)}^\vee  \ominus \frak {p}^\vee]$ is spanned by strictly upper triangular matrices which are thus nilpotent like the matrices that span the first factor ${\frak n}^\vee_+$; this just reflects the fact that the second factor is due to the dual 4d worldvolume defect which is in turn characterized by dual nilpotent orbits $O_{e^\vee}$ (c.f.~$\S$4.3). As such, we find that $\cal S$ would be a nilpotent subgroup of strictly upper-triangular matrices, i.e., we can also write 
\be
\label{d-plus}
\frak s = \frak d^\vee_+,
\ee
where $\frak d^\vee_+ \subset \frak {so}(2N)^\vee$ is a nilpotent Lie subalgebra of strictly upper triangular matrices. 

Thus, in this case, we see (bearing in mind the isomorphism relations in footnote~\ref{affine iso}) that the sought-after spacetime BPS states ought to be given by the states of the partially gauged chiral CFT 
\be
\label{partial-gauged CFT-defect-D}
{\frak {so}(2N)}_{{\rm aff}, q} / \frak d_{+ \, {\rm aff}, p_D},
\ee
where the levels $p_D$ and $q$ may not be the same, as the central charge of $ \frak d_{+ \, {\rm aff}, p_D}$, like that of ${\frak{n}}^{\vee}_{+ \, {\rm aff}, p_2} \simeq {\frak{n}}_{+ \, {\rm aff}, p_2}$ in (\ref{AGT-CDG-chiral CFT-defect}), must also contribute to an anomalous shift in the overall central charge which can then be ``absorbed'' by the curvature of $\cal C$, as explained in $\S$5.3. This partially gauged chiral CFT, like the one in (\ref{AGT-CDG-chiral CFT}), can be realized as a gauged $SO(2N,\mathbb C)$ WZW model, although the Lie algebra of the gauge group ${\cal S}$ is now $\frak d_+$ instead of $\frak n_+$.  

In an appropriate basis of $\frak {so}(2N)$, one can always find an element $H$ of the Cartan subalgebra of $\frak {so}(2N)$ such that
\be
\label{positive eigenspace-D}
[H, x] = k x
\ee
for some $x \in \frak d_+$ and positive  \emph{integer} $k$.\footnote{Note that this claim is only true if we restrict ourselves to the subset of 4d worldvolume defects whereby there exists a nilpotent element $e = \rho_{\cal D} (\sigma^3) \in \frak d_+$ (where $\sigma^3 \in \frak {sl}(2)$) such that the embedding $\rho_{\cal D}: \frak {sl}(2) \to \frak {so}(2N)$ realizes the conditions for an $H$-compatible halving as spelt out in~\cite[Appendix C, after eqn.~(C.10)]{ref for gauged WZW} (for the $D$ series). For simplicity and brevity of discussion, we shall henceforth assume our 4d worldvolume defects to be such.\label{special integral defects}} Hence, if we were to repeat the computation in Appendix B with gauge group ${\cal S}$ instead of $N_+$, we would physically realize the general BRST algorithm in~\cite{Tjin-Boer}. What this means is that the chiral CFT would realize ${\cal W}({\frak {s0}}(2N)_{\rm aff}, \rho_{\cal D})$ -- an untwisted affine $\cal W$-algebra obtained from ${\frak {so}}(2N)_{\rm aff}$ via a quantum Drinfeld-Sokolov reduction that is associated with the embedding $\rho_{\cal D}: \frak {sl}(2) \to \frak {so}(2N)$ (which, through the Jacobson-Morozov theorem, is determined by $\frak d_+$ that is in turn determined by the underlying 4d worldvolume defect).  In other words,  the states of the chiral CFT would be furnished by a Verma module  $\widehat{{\cal W}}({\frak {so}}(2N)_{\rm aff}, \rho_{\cal D})$ over ${\cal W}({\frak {so}}(2N)_{\rm aff}, \rho_{\cal D})$, and the Hilbert space ${\cal H}^{\Omega'}_{\rm BPS}$ of spacetime BPS states on the RHS of (\ref{AGT-M-duality-CDG-defect}) when $n=1$, can be expressed as
\be
{\cal H}^{\Omega'}_{\rm BPS}  = \widehat{{\cal W}}({\frak {so}}(2N)_{\rm aff}, \rho_{\cal D}).
\label{AGT-CDG-H=W-defect}
\ee

Clearly, the physical duality of the compactifications in (\ref{AGT-M-duality-CDG-defect}) will mean that ${\cal H}^\Omega_{\rm BPS}$ in (\ref{BPS-AGT-AB-defect}) (when $n=1$) is equivalent to ${\cal H}^{\Omega'}_{\rm BPS}$ in (\ref{AGT-CDG-H=W-defect}), i.e.,
\be
 \boxed{\bigoplus_{a'} ~{\rm IH}^\ast_{U(1)^2 \times T} \, {\cal U}({\cal M}_{SO(2N), \mathbb L, a'}) = \widehat{{\cal W}}({\frak {so}}(2N)_{\rm aff}, \rho_{\cal D})}
 \label{AGT-duality-D-defect}
\ee
Thus, we have a ``ramified'' generalization of the duality relation (\ref{AGT-duality-D}) for $G = SO(2N)$. 

Accordingly, $c_D$ in (\ref{cc-D}) ought to be replaced by~\cite{Tjin-Boer} 
\be
 c_{\cal D}  = {\rm dim} \, {\frak {so}}(2N)_0  - {1 \over 2} {\rm dim} \, {\frak {so}}(2N)_{1 \over 2} - 12 \left | \alpha_+ \rho   + \alpha_- t_0\right |^2. 
\label{cc-defect-general-D}
\ee
Here, ${\frak {so}}(2N)_j = \{ x \in {\frak {so}}(2N) \, \vert \, [t_0, x] = jx \}$; $t_0 = \rho_{\cal D}(\sigma^3)$, where $\sigma^3 \in {\frak {sl}}(2)$ is a Cartan element; $\rho$ is the Weyl vector of ${\frak {so}}(2N)$; $\alpha_+ = 1 / \sqrt{k_{\cal D} + 2N-2}$, where $\alpha_+\alpha_- = -1$; and $k_{\cal D} \in \mathbb R$ is the effective level of the underlying affine Lie algebra ${\frak {so}}(2N)_{\rm aff}$.

If we have a trivial 4d worldvolume defect whence $\mathbb L = SO(2N)$ so ${\frak p}^\vee = \frak {so}(2N)^\vee$ (and the second factor in $\frak s$ is trivial), we have $\frak d_+ = \frak n_+$ whence $\rho_{\cal D}$ would be principal; in this  case, $t_0 = {\rho}^\vee = \rho$, ${\rm dim} \, {\frak {so}}(2N)_0  - {1 \over 2} {\rm dim} \, {\frak {so}}(2N)_{1 \over 2} = {\rm rank} \, \frak {so}(2N) = N$, so $c_{\cal D}$ coincides with (\ref{cc-simply-laced}) (for the $D$ groups).  As such, when the defect is trivial (i.e.~absent), ${\cal W}({\frak {so}}(2N)_{\rm aff}, \rho_{\cal D}) = {\cal W}({\frak {so}}(2N)_{\rm aff})$ with central charge $c_{\cal D} = c_D$, consistent with our ``unramified'' results in $\S$5.3. As further explained in $\S$5.3, $c_{\cal D}$ would also depend on the Omega-deformation parameters $\epsilon_{1,2}$ through $\alpha_+ = -i \sqrt{\epsilon_1 / \epsilon_2}$ and $\alpha_- = -i \sqrt{\epsilon_2 / \epsilon_1}$. 

Since Omega-deformation is independent of the choice of 4d worldvolume defect, the manner in which $c_{\cal D}$ depends on $\epsilon_{1,2}$ would not change as we vary $\rho_{\cal D}$ away from being principal, i.e., we have $\alpha_+ = -i \sqrt{\epsilon_1 / \epsilon_2}$ and $\alpha_- = -i \sqrt{\epsilon_2 / \epsilon_1}$ in (\ref{cc-defect-general-D}) for \emph{all }$\rho_{\cal D}$. Therefore, we can also write $c_{\cal D}$ as 
\be
\boxed{ c_{{\cal D}, \epsilon_{1,2}}  = {\rm dim} \, {\frak {so}}(2N)_0  - {1 \over 2} {\rm dim} \, {\frak {so}}(2N)_{1 \over 2} + 12 \left | \sqrt{\epsilon_1 \over \epsilon_2} \, \rho   + \sqrt{\epsilon_2 \over \epsilon_1} \, t_0\right |^2} 
\label{cc-defect-omega-deformed-D}
\ee
Notice that consistent with the LHS of (\ref{AGT-M-duality-CDG-defect}), $c_{{\cal D}, \epsilon_{1,2}}$ is also asymmetric under the exchange $\epsilon_1 \leftrightarrow \epsilon_2$ whenever we have a nontrivial defect. In addition, we also have
\be
\label{k'-defect-D}
\boxed{k_{\cal D}  = - 2N + 2 - {\epsilon_2 \over \epsilon_1}}
\ee

Note at this point that a rigorous definition of a ``ramified'' generalization of the Nekrasov instanton partition function  in~\cite[$\S$6.6]{AGT-math} means that we can actually ``ramify'' the arguments which took us from (\ref{abcd BPS relation-D}) to (\ref{q | q, D}). As such, in the presence of a nontrivial 4d worldvolume defect whose nature is encoded in $\rho_{\cal D}$, we can, via (\ref{AGT-duality-D-defect}), write the ``ramified'' Nekrasov instanton partition function as
\be
\label{q | q, defect-D}
\boxed{Z_{\rm inst} (SO(2N), \epsilon_1, \epsilon_2, \vec a, \mathbb L) = \langle \rho_{\cal D}, \Delta   | \rho_{\cal D}, \Delta \rangle}
\ee
where
\be
\label{coherent-D}
\boxed{| \rho_{\cal D}, \Delta \rangle = \bigoplus_{a'}  D^{a'}  | \Psi_{a', \mathbb L_{\cal D}} \rangle}
\ee
Here, $| \rho_{\cal D}, \Delta \rangle \in \widehat{{\cal W}}(\frak {so}(2N)_{\rm aff}, \rho_{\cal D})$; $D^{a'}$ is some real number; $ | \Psi_{a', \mathbb L_{ \cal D}} \rangle \in {\rm IH}^\ast_{U(1)^2 \times T} \, {\cal U}({\cal M}_{SO(2N), \mathbb L, a'})$ is also a state in $ \widehat{{\cal W}}({\frak {so}}(2N)_{\rm aff}, \rho_{\cal D})$ with energy level $n_{a'}$ determined by the ``ramified'' instanton number $a'$ (as one recalls that  $n_{a'}$ is a constant shift of the eigenvalue $a'$ of the $L_0$ operator which generates translations along the ${\bf S}^1_n$ circle in (\ref{AGT-M-duality-CDG-defect})); and $ \langle \cdot | \cdot \rangle$ is a Poincar\'e pairing in the sense of~\cite[$\S$2.6]{J-function}. The label $\Delta$ just means that $\widehat{{\cal W}}({\frak {so}}(2N)_{\rm aff}, \rho_{\cal D})$ is generated by the application of creation operators (furnished by the negative-mode elements of ${{\cal W}}({\frak {so}}(2N)_{\rm aff}, \rho_{\cal D})$) on the highest weight state $| \Delta \rangle$.  

As in the ``unramified'' case, since the RHS of (\ref{q | q, defect-D}) is defined in the limit that the ${\bf S}^1_n$ fiber in $\cal C$ has \emph{zero} radius, and since we have in $\cal C$ a \emph{common} boundary condition at $z = 0$ and $z=\infty$ (where the radius of the  ${\bf S}^1_n$ fiber is zero),  $| \rho_{\cal D}, \Delta \rangle$ and $\langle \rho_{\cal D}, \Delta |$ ought to be a state and its dual associated with the puncture at $z = 0$ and $z=\infty$, respectively. Furthermore, as the RHS of (\ref{coherent-D}) is a sum over states of all possible energy levels, it would mean that $| \rho_{\cal D}, \Delta \rangle$ is actually a \emph{coherent state}. 

Thus, in arriving at the boxed relations (\ref{AGT-duality-D-defect}), (\ref{cc-defect-omega-deformed-D}), (\ref{k'-defect-D}), (\ref{q | q, defect-D}) and (\ref{coherent-D}), we have just furnished a fundamental physical derivation of a ``ramified'' pure AGT correspondence for the $D_{N}$ groups! (The ``ramified'' version of the relations (\ref{W-2})--(\ref{WN1}) can be obtained via an $SO(2N)$ generalization of the computational technique introduced in~\cite[$\S$3.3]{Kanno-Tachikawa}.)

\bigskip\noindent{\it A ``Ramified'' Generalization of the Pure AGT Correspondence for the $C$--$G$ Groups}

Let us now focus on (\ref{AGT-CDG-chiral CFT-defect-modded}) for $n=2$ and 3 (with $N=4$) whence $\frak {g}^\vee_{\mathbb C \, \rm aff} = {\frak {usp}(2N-2)}^\vee_{\rm aff}$ and ${\frak g}^\vee_{2 \, {\rm aff}}$, respectively. Recall at this point from footnote~\ref{affine iso} that ${\frak {usp}(2N-2)}^\vee_{\rm aff} = {\frak {so}(2N)}^{(2)}_{\rm aff} = {\frak {so}(2N)}^{\vee \, (2)}_{\rm aff}$ and ${\frak g}^\vee_{2 \, {\rm aff}} =  {\frak {so}(2N)}^{(3)}_{\rm aff} = {\frak {so}(2N)}^{\vee \, (3)}_{\rm aff}$ (where $N=4$), which means that our proceeding analysis would be exactly the same as that for the $D$ groups above, except that the affine Lie algebras involved are now $\mathbb Z_2$- and $\mathbb Z_3$-twisted, accordingly. As such, the Hilbert space ${\cal H}^{\Omega'}_{\rm BPS}$ of spacetime BPS states on the RHS of (\ref{AGT-M-duality-CDG-defect}) can be expressed as
\be
{\cal H}^{\Omega'}_{\rm BPS}  = \widehat{{\cal W}}(\frak {g}^\vee_{\rm aff}, \rho_{\cal D}), 
\label{AGT-AB-H=W-defect-CG}
\ee
where $\widehat{{\cal W}}(\frak {g}^\vee_{\rm aff}, \rho_{\cal D})$ is a Verma module over ${\cal W}(\frak {g}^\vee_{\rm aff}, \rho_{\cal D})$  -- a $\mathbb Z_n$-twisted version of the affine $\cal W$-algebra ${\cal W}({\frak {so}}(2N)_{\rm aff}, \rho_{\cal D})$  obtained from ${\frak {so}}(2N)_{\rm aff}$ via a quantum Drinfeld-Sokolov reduction that is associated with the embedding $\rho_{\cal D}: \frak {sl}(2) \to \frak {so}(2N)$ (which, as explained above, encodes the nature of the underlying 4d worldvolume defect).

Clearly, the physical duality of the compactifications in (\ref{AGT-M-duality-CDG-defect}) will mean that ${\cal H}^\Omega_{\rm BPS}$ in (\ref{BPS-AGT-AB-defect}) (when $n=2$, and when $n=3$ with $N=4$) is equivalent to ${\cal H}^{\Omega'}_{\rm BPS}$ in (\ref{AGT-AB-H=W-defect-CG}) (when $n=2$, and when $n=3$ with $N=4$), i.e.,
\be
 \boxed{\bigoplus_{a'} ~{\rm IH}^\ast_{U(1)^2 \times T} \, {\cal U}({\cal M}_{G, \mathbb L, a'}) = \widehat{{\cal W}}(\frak {g}^\vee_{\rm aff}, \rho_{\cal D})}
 \label{AGT-duality-CG-defect}
\ee
where the equivariant intersection cohomology is $\mathbb Z_n$-invariant in the sense explained below (\ref{BPS-AGT-AB-defect}). Thus, we have a ``ramified'' generalization of the duality relation (\ref{AGT-duality-D}) for $G = USp(2N - 2)$ and $G_2$. 

According to footnote~\ref{central charge SO(2N)}, the central charge of a twisted $\cal W$-algebra (obtained as a coset theory of twisted affine Lie algebras, such as in our case) would be the same as its untwisted version. As such, the central charge $c_{G}$ in this case would be the same as $c_{\cal D}$, i.e., 
\be
\boxed{ c_{G, \epsilon_{1,2}}  = {\rm dim} \, {\frak {so}}(2N)_0  - {1 \over 2} {\rm dim} \, {\frak {so}}(2N)_{1 \over 2} + 12 \left | \sqrt{\epsilon_1 \over \epsilon_2} \, \rho   + \sqrt{\epsilon_2 \over \epsilon_1} \, t_0\right |^2} 
\label{cc-defect-omega-deformed-CG}
\ee
Likewise, the level of the underlying $\mathbb Z_n$-twisted affine Lie algebra ${\frak g}^\vee_{\rm aff}$ is
\be
\label{k'-defect-CG}
\boxed{k_{G}  = - 2N +2 - {\epsilon_2 \over \epsilon_1}}
\ee
and 
\be
\label{q | q, defect-CG}
\boxed{Z_{\rm inst} (G, \epsilon_1, \epsilon_2, \vec a, \mathbb L) = \langle \rho_{\cal D}, \Delta_n   | \rho_{\cal D}, \Delta_n \rangle}
\ee
where
\be
\label{coherent-CG}
\boxed{| \rho_{\cal D}, \Delta_n \rangle = \bigoplus_{a'}  G^{a'}  | \Psi_{a', \mathbb L_{G}} \rangle}
\ee
Here, $| \rho_{\cal D}, \Delta_n \rangle \in \widehat{{\cal W}}({\frak g}^\vee_{\rm aff}, \rho_{\cal D})$; ${G}^{a'}$ is some real number; $ | \Psi_{a', \mathbb L_{G}} \rangle \in {\rm IH}^\ast_{U(1)^2 \times T} \, {\cal U}({\cal M}_{G, \mathbb L, a'})$ is also a state in $ \widehat{{\cal W}}({\frak g}^\vee_{\rm aff}, \rho_{\cal D})$ with energy level $n_{a'}$ determined by the ``ramified'' instanton number $a'$; and the label $\Delta_n$ just means that $\widehat{{\cal W}}({\frak g}^\vee_{\rm aff}, \rho_{\cal D})$ is generated by the application of creation operators (furnished by the negative-mode elements of ${{\cal W}}({\frak g}^\vee_{\rm aff}, \rho_{\cal D})$) on the $\mathbb Z_n$-twisted highest weight state $| \Delta \rangle$.  

As in the $n=1$ case,  $| \rho_{\cal D}, \Delta_n \rangle$ and $\langle \rho_{\cal D}, \Delta_n |$ ought to be a state and its dual associated with the puncture at $z = 0$ and $z=\infty$ on $\cal C$, respectively. Furthermore, as the RHS of (\ref{coherent-CG}) is a sum over states of all possible energy levels, it would mean that $| \rho_{\cal D}, \Delta_n \rangle$ is actually a \emph{coherent state}. 

Thus, in arriving at the boxed relations (\ref{AGT-duality-CG-defect}), (\ref{cc-defect-omega-deformed-CG}), (\ref{k'-defect-CG}), (\ref{q | q, defect-CG}) and (\ref{coherent-CG}), we have just furnished a fundamental physical derivation of a ``ramified'' pure AGT correspondence for the $C_{N-1}$ and $G_2$ groups! (The ``ramified'' version of the relations (\ref{W-2})--(\ref{WN1}) can be obtained via a $\mathbb Z_n$-twisted, $SO(2N)$ generalization of the computational technique introduced in~\cite[$\S$3.3]{Kanno-Tachikawa}.)

\bigskip\noindent{\it  The ``Fully-Ramified'' Pure AGT Correspondence for the $A$--$B$ Groups}

Let us now specialize our above discussion to the case of a \emph{full} 4d worldvolume defect whence $\mathbb L = T$. For the $A$--$B$ groups, this means that $[n_I] = [1, \dots, 1]$ in (\ref{cal S}). As such, $\frak a_+$ in (\ref{a-plus}) is trivial.  Hence, the full defect ``undoes'' the quantum Drinfeld-Sokolov reduction, and in place of (\ref{coset-AGT-A}), we have 
\be
\label{coset-AGT-A-full}
{\frak {sl}(N)}^{(n)}_{{\rm aff}, k_{\cal AB}},
\ee
where $n=1$ and $2$ (with even $N$) for the $A_{N-1}$ and $B_{N/2}$ groups, respectively.  

Thus, in place of (\ref{AGT-duality-A-defect}) and (\ref{AGT-duality-B-defect}), we have 
\be
 \boxed{\bigoplus_{a'} ~{\rm IH}^\ast_{U(1)^2 \times T} \, {\cal U}({\cal M}_{G, T, a'}) = {\widehat {\frak g}}^\vee_{{\rm aff}, k_{\cal AB}}}
 \label{AGT-duality-AB-fulldefect}
\ee
where $G = SU(N)$ and $SO(N+1)$ (with even $N$), and  ${\widehat {\frak g}}^\vee_{{\rm aff}, k_{\cal AB}}$ is a Verma module over the Langlands dual affine Lie algebra ${\frak g}^\vee_{{\rm aff}, k_{\cal AB}}$ at level $k_{\cal AB}$. 

From (\ref{k'-defect}) and (\ref{k'-defect-B}), we get 
\be
\label{K-ab}
\boxed{k_{\cal AB}  = - N  - {\epsilon_2 \over \epsilon_1}}
\ee
In turn, the central charge is
\be
\boxed{ c_{{\cal AB}, \epsilon_{1,2}}  = {\epsilon_1 \over \epsilon_2} (N^3 - N) + N^2 -1} 
\label{cc-defect-omega-deformed-AB-full}
\ee

Recall at this point that if  the defect were to be trivial, (i) ${\frak g}^\vee_{{\rm aff}, k_{\cal AB}}$ would be replaced by ${\cal W}({\frak g}^\vee_{\rm aff}, \rho_{\cal A})$ with principal $\rho_{\cal A}$; (ii) the $\mathbb Z_n$-twisted highest weight state $| \vec j, \Delta \rangle \in {\widehat {\frak g}}^\vee_{{\rm aff}, k_{\cal AB}}$ would be replaced by the $\mathbb Z_n$-twisted highest weight state $| \Delta \rangle \in \widehat{{\cal W}}({\frak g}^\vee_{\rm aff}, \rho_{\cal A})$;  (iii) the zeroth modes ${\bf J} = (J^1_0, \dots, J_0^{N-1})$ of the $N-1$ untwisted scalar bosonic fields in the free-field realization of ${\cal W}({\frak g}^\vee_{\rm aff}, \rho_{\cal A})$ would be given by  ${\bf J} = {\bf a} + i (b + b^{-1}) \rho$, where $b = \sqrt {\epsilon_1 / \epsilon_2}$; (iv) the conformal dimension $\Delta^{(2)}$ of $| \Delta \rangle$ would be given by $\Delta^{(2)} = ({\bf a}^2 + (b + b^{-1})^2\rho^2) / 2$. Notice that since both $\bf J$ and $\Delta^{(2)}$ are $\rho_{\cal A}$-independent, we can expect them to take the same form at $\rho_{\cal A} = 0$, i.e., when we actually have a full defect. 

That being said, the ``unramified'' configuration  (\ref{AGT-M-duality-AB}) -- which underlies the above-stated expressions for $\bf J$ and $\Delta^{(2)}$ -- is symmetric under the exchange  $\epsilon_1 \leftrightarrow \epsilon_2$; on the other hand, the ``ramified'' configuration  (\ref{AGT-M-duality-AB-defect}) -- which underlies the story with the full defect -- is not; in other words, unlike the above-stated expressions for $\bf J$ and $\Delta^{(2)}$ for when the defect is trivial, the expressions for $\bf J$ and $\Delta^{(2)}$ for when the defect is full should \emph{not} be symmetric under the exchange $\epsilon_1 \leftrightarrow \epsilon_2$. Thus,  the expressions for $\bf J$ and $\Delta^{(2)}$ for when the defect is full should be given by the above-stated expressions for $\bf J$ and $\Delta^{(2)}$ \emph{less} the $b$- or $b^{-1}$-dependent term. In turn, this means that (i) the expression for the conformal dimension  $ \Delta^{(2)}_{\vec j}$ of $| \vec j, \Delta \rangle$ ought to be given by the above-stated expression for $\Delta^{(2)}$ less the $b$- or $b^{-1}$-dependent term; (ii)   the highest weight  ${\vec j} =  i b^{-1} {\bf J}'$ associated with $| \vec j, \Delta \rangle$ is such that the expression for ${\bf J}'$ ought to be given by the above-stated expression for $\bf J$ less the $b$- or $b^{-1}$-dependent term.\footnote{Note that according to~\cite[Appendix C]{Alday-Tachikawa}, the relation between $\vec j$ and ${\bf J}'$ is actually $\vec j = - b^{-1} {\bf J}'$; in other words, there is an extra factor of `$-i$' in our definition of the relation. The reason for our deviation is as follows. Recall that the vector ${\bf a}$ in $\S$5.2  is purely real in our conventions; this implies that $\epsilon_1$ and $\epsilon_2$ must be opposite  in sign  whence $b^{-1}$ is purely imaginary; thus, since $\vec j$ (like ${\bf J}'$) must also be purely real, one has to insert an extra factor of `$-i$' in the relation.}   

Therefore, as $ \Delta^{(2)}_{\vec j} = - \vec j \cdot (\vec j + 2\rho)/2b^{-2}$ by definition, a consistent solution would involve dropping the $b^{-1}$-dependent term in $\bf J$ and $\Delta^{(2)}$ such that $\vec j = i b^{-1} {\bf a} - \rho$ and $ \Delta^{(2)}_{\vec j} = ({\bf a}^2 + b^2\rho^2) / 2$. Since we can identify ${\bf a}$ with $- i \vec a / \sqrt{\epsilon_1 \epsilon_2}$ (see $\S$5.2),  we can write
\be
\label{j-weight-AB}
\boxed{\vec j = {\vec a \over \epsilon_1} - \rho}
\ee
and
\be
\label{j-dim-AB}
\boxed{L_0 | \vec j, \Delta \rangle = \Delta^{(2)}_{\vec j}  | \vec j, \Delta \rangle  \quad {\rm where} \quad \Delta^{(2)}_{\vec j} =  {\epsilon_1 \over 2\epsilon_2}\left[ \rho^2 - {{\vec a}^2 \over \epsilon^2_1}\right]}
\ee

Hence, in place of (\ref{q | q, defect}) and (\ref{q | q, defect-B}), and in place of (\ref{coherent-A}) and (\ref{coherent-B}), we have 
\be
\label{q | q, full-defect-AB}
\boxed{Z_{\rm inst} (G, \epsilon_1, \epsilon_2, \vec a, T) = \langle 0, \Delta_n   | 0, \Delta_n \rangle}
\ee
and
\be
\label{coherent-AB-full-defect}
\boxed{| 0, \Delta_n \rangle = \bigoplus_{a'}  {\cal G}^{a'}  | \Psi_{a', T_{\cal AB}} \rangle}
\ee
Here, $|0, \Delta_n \rangle \in {\widehat {\frak g}}^\vee_{{\rm aff}, k_{\cal AB}}$; ${\cal G}^{a'}$ is some real number; and $ | \Psi_{a', T_{\cal AB}} \rangle \in {\rm IH}^\ast_{U(1)^2 \times T} \, {\cal U}({\cal M}_{G, T, a'})$ is also a state in ${\widehat {\frak g}}^\vee_{{\rm aff}, k_{\cal AB}}$ with energy level $n_{a'}$ determined by the ``ramified'' instanton number $a'$. The label $\Delta_n$ just means that ${\widehat {\frak g}}^\vee_{{\rm aff}, k_{\cal AB}}$ is generated by the application of creation operators (furnished by the negative-mode elements of ${\frak g}^\vee_{{\rm aff}, k_{\cal AB}}$) on the $\mathbb Z_n$-twisted highest weight state $| {\vec j}, \Delta \rangle$.  

As  before,  $| 0, \Delta_n \rangle$ and $\langle 0, \Delta_n |$ ought to be a state and its dual associated with the puncture at $z = 0$ and $z=\infty$ in $\cal C$, respectively. Furthermore, as the RHS of (\ref{coherent-AB-full-defect}) is a sum over states of all possible energy levels, it would mean that $| 0, \Delta_n \rangle$ is actually a \emph{coherent state}. 

Thus, in arriving at the boxed relations (\ref{AGT-duality-AB-fulldefect}), (\ref{K-ab}), (\ref{cc-defect-omega-deformed-AB-full}),  (\ref{j-weight-AB}), (\ref{j-dim-AB}), (\ref{q | q, full-defect-AB}) and (\ref{coherent-AB-full-defect}), we have just furnished a fundamental physical derivation of a  ``\emph{fully-ramified}'' pure AGT correspondence for the $A_{N-1}$ and (for even $N$) the $B_{N/2}$ groups! (The ``ramified'' version of the relations (\ref{W-2})--(\ref{WN1}) can be obtained via a $\mathbb Z_n$-twisted generalization of the computational technique introduced in~\cite[$\S$3.3]{Kanno-Tachikawa}.)

\bigskip\noindent{\it  The ``Fully-Ramified'' Pure AGT Correspondence for the $C$--$D$--$G$ Groups}

Let us now turn our attention to the $C$--$D$--$G$ groups. Unlike $\frak a_+$ in (\ref{coset-AGT-A}) for the $A$--$B$ groups, we do not have, for the $C$--$D$--$G$ groups, an explicit description of $\frak d_+$ in  (\ref{partial-gauged CFT-defect-D}) in terms of some partition $[n_I]$ which describes $\mathbb L$, i.e., apart from the obvious case of a trivial 4d worldvolume defect where $\mathbb L = G$ whence $\frak d_+ = \frak n_+$, we cannot determine the exact form of $\frak d_+$ in all generality. Nevertheless, one can still deduce the exact form of $\frak d_+$ for when the 4d worldvolume defect is full. 

To this end, first note that by shifting the center-of-mass of the $N$ M5-branes + 4d defect system in (\ref{AGT-M-duality-CDG-defect}) away from the OM5-plane, the $SO(2N)$ gauge group which underlies the original $N$ M5-branes + OM5-plane + 4d defect system would reduce to an $SU(N)$ gauge group which underlies the now effective $N$ M5-branes + 4d defect system. Second,  note that shifting the center-of-mass of the $N$ M5-branes + 4d defect system in (\ref{AGT-M-duality-CDG-defect}) will not modify the intrinsic properties of distinguished defects, i.e., a full or trivial defect will remain as such, regardless. Third, note that the action of Omega-deformation, as effected by a background fluxbrane, is also independent of this shift in the center-of-mass of the $N$ M5-branes + 4d defect system. Altogether, this means that if we start with configuration  (\ref{AGT-M-duality-CDG-defect}) with a full defect and shift the center-of-mass of the $N$ M5-branes + 4d defect system away from the OM5-plane, we will end up with configuration (\ref{AGT-M-duality-AB-defect}) with a full defect, and vice-versa. Therefore, since our above ``fully-ramified'' analysis for the $A$--$B$ groups is also independent of the center-of-mass of the $N$ M5-branes + 4d defect system, we can conclude that the ``fully-ramified'' analysis for the $C$--$D$--$G$ groups ought to be the same, except that one has to replace $\frak sl(N)$ with $\frak {so}(2N)$ everywhere. In particular, instead of (\ref{coset-AGT-A-full}), we now have
\be
\label{coset-AGT-D-full}
{\frak {so}(2N)}^{(n)}_{{\rm aff}, k_{\cal CDG}},
\ee
where $n=1$, $2$ and $3$ (with $N = 4$) for the $D_N$, $C_{N-1}$ and $G_2$ groups, respectively. Hence, we deduce that $\frak d_+$ is actually trivial, like $\frak a_+$ was.  

Also, in place of (\ref{AGT-duality-AB-fulldefect}) is  
\be
 \boxed{\bigoplus_{a'} ~{\rm IH}^\ast_{U(1)^2 \times T} \, {\cal U}({\cal M}_{G, T, a'}) = {\widehat {\frak g}}^\vee_{{\rm aff}, k_{\cal CDG}}}
 \label{AGT-duality-CDG-fulldefect}
\ee
where $G = SO(2N)$, $USp(2N-2)$ and $G_2$, and  ${\widehat {\frak g}}^\vee_{{\rm aff}, k_{\cal CDG}}$ is a Verma module over the Langlands dual affine Lie algebra ${\frak g}^\vee_{{\rm aff}, k_{\cal CDG}}$ at level $k_{\cal CDG}$. 

From (\ref{k'-defect-D}) and (\ref{k'-defect-CG}), we get 
\be
\label{K-cdg}
\boxed{k_{\cal CDG}  = - 2N +2  - {\epsilon_2 \over \epsilon_1}}
\ee
In turn, the central charge is
\be
\boxed{ c_{{\cal CDG}, \epsilon_{1,2}}  = {\epsilon_1 \over \epsilon_2} (4N^3 - 6N^2 + 2N) + 2N^2 -N} 
\label{cc-defect-omega-deformed-CDG-full}
\ee

By the same arguments which led us to (\ref{j-weight-AB}) and (\ref{j-dim-AB}), we can also state the following. The highest weight ${\vec l} $ associated with the highest weight state $| \vec l, \Delta \rangle \in {\widehat {\frak g}}^\vee_{{\rm aff}, k_{\cal CDG}}$, can be written as
\be
\label{j-weight-CDG}
\boxed{\vec l = {\vec a \over \epsilon_1} - \rho}
\ee
and moreover, 
\be
\label{j-dim-CDG}
\boxed{L_0 | \vec l, \Delta \rangle = \Delta^{(2)}_{\vec l}  | \vec l, \Delta \rangle  \quad {\rm where} \quad \Delta^{(2)}_{\vec l} =  {\epsilon_1 \over 2\epsilon_2}\left[ \rho^2 - {{\vec a}^2 \over \epsilon^2_1}\right]}
\ee

Hence, in place of (\ref{q | q, defect-D}) and (\ref{q | q, defect-CG}), and in place of (\ref{coherent-D}) and (\ref{coherent-CG}), we have 
\be
\label{q | q, full-defect-CDG}
\boxed{Z_{\rm inst} (G, \epsilon_1, \epsilon_2, \vec a, T) = \langle 0, \Delta_n   | 0, \Delta_n \rangle}
\ee
and
\be
\label{coherent-CDG-full-defect}
\boxed{| 0, \Delta_n \rangle = \bigoplus_{a'}  {\mathscr G}^{a'}  | \Psi_{a', T_{\cal CDG}} \rangle}
\ee
Here, $|0, \Delta_n \rangle \in {\widehat {\frak g}}^\vee_{{\rm aff}, k_{\cal CDG}}$; ${\mathscr G}^{a'}$ is some real number; and $ | \Psi_{a', T_{\cal CDG}} \rangle \in {\rm IH}^\ast_{U(1)^2 \times T} \, {\cal U}({\cal M}_{G, T, a'})$ is also a state in ${\widehat {\frak g}}^\vee_{{\rm aff}, k_{\cal CDG}}$ with energy level $n_{a'}$ determined by the ``ramified'' instanton number $a'$. The label $\Delta_n$ just means that ${\widehat {\frak g}}^\vee_{{\rm aff}, k_{\cal CDG}}$ is generated by the application of creation operators (furnished by the negative-mode elements of ${\frak g}^\vee_{{\rm aff}, k_{\cal CDG}}$) on the $\mathbb Z_n$-twisted highest weight state $| {\vec l}, \Delta \rangle$.  

Again,  $| 0, \Delta_n \rangle$ and $\langle 0, \Delta_n |$ ought to be a state and its dual associated with the puncture at $z = 0$ and $z=\infty$ in $\cal C$, respectively. Furthermore, as the RHS of (\ref{coherent-CDG-full-defect}) is a sum over states of all possible energy levels, it would mean that $| 0, \Delta_n \rangle$ is actually a \emph{coherent state}. 

Thus, in arriving at the boxed relations (\ref{AGT-duality-CDG-fulldefect}), (\ref{K-cdg}), (\ref{cc-defect-omega-deformed-CDG-full}),  (\ref{j-weight-CDG}), (\ref{j-dim-CDG}), (\ref{q | q, full-defect-CDG}) and (\ref{coherent-CDG-full-defect}), we have just furnished a fundamental physical derivation of a  ``\emph{fully-ramified}'' pure AGT correspondence for the $D_{N}$, $C_{N-1}$ and $G_2$ groups! (The ``ramified'' version of the relations (\ref{W-2})--(\ref{WN1}) can be obtained via a $\mathbb Z_n$-twisted, $SO(2N)$ generalization of the computational technique introduced in~\cite[$\S$3.3]{Kanno-Tachikawa}.)

\newsubsection{An ALE Generalization of the Pure AGT Correspondence}

Let us now derive, purely physically, an ALE generalization of the pure AGT correspondence for the $A$, $B$, $C$, $D$ and $G$ groups. For brevity, we shall consider only the fully-resolved ALE space of $A$-type with $k$ centers, $\widetilde{\mathbb R^4 / \mathbb Z_k}$.   

\bigskip\noindent{\it An ALE Generalization of the Pure AGT Correspondence for the $A$--$B$ Groups} 

To this end, first note that according to our analysis in $\S$4.1, replacing $\mathbb  R^4$ on the LHS of (\ref{AGT-M-duality-AB}) with $\widetilde{\mathbb R^4 / \mathbb Z_k}$ would mean that we have to replace (\ref{AGT-AB-chiral CFT}) with the Omega-deformed version of (\ref{effective CFT-fully-resolved-A}). Bearing in mind the relation (\ref{coset}), and the fact that Omega-deformation effectively acts only on the $\mathbb Z_n$-twisted affine CFT associated with the stack of $N$ D6-branes in (\ref{equivalent IIA system Witten}) (whence our result at $k=1$ would indeed be the same as that found in $\S$5.2), we find that (\ref{AGT-AB-chiral CFT}) has to be replaced by
\be
{\frak{su}(k)^{(n)}_{{\rm aff}, N} \over{[\frak{u}(1)^{(n)}_{{\rm aff}, N}}]^{k-1}}  \otimes {{\frak{sl}(N)^{(n)}_{{\rm aff}, k}} \over {\frak{n}_+}^{(n)}_{{\rm aff}, l}}.
\label{coset CFT-AGT-ALE}
\ee
Here, the level $l \in \mathbb R$, and $\frak n_+$ is a nilpotent subalgebra of strictly upper triangular matrices. 

Recall at this point that the central charge due to the Omega-deformation factor $ {\frak{n}_+}^{(n)}_{{\rm aff}, l}$ in (\ref{coset CFT-AGT-ALE}), is given by (\ref{cc-A-Omega}) when we have a \emph{single} D4-brane intersecting the $N$ D6-branes, i.e., when $k=1$, as shown in (\ref{equivalent IIA system 1 - AGT}). Recall also that this central charge is proportional to the curvature of $\cal C$ induced by Omega-deformation; thus, when we have $k$ D4-branes intersecting the $N$ D6-branes whence the curvature of $\cal C$ would  be  ``diluted'' over $k$ D4-branes, we ought to divide its value by $k$. This means that the second factor in (\ref{coset CFT-AGT-ALE}) ought to obey the following (conformal) equivalence of coset realizations:
\be
\label{coset equivalence AGT-ALE}
{{\frak{sl}(N)^{(n)}_{{\rm aff}, k}} \over {\frak{n}_+}^{(n)}_{{\rm aff}, p}} = {{\frak{su}(N)^{(n)}_{{\rm aff}, k}} \over{[\frak{u}(1)^{(n)}_{{\rm aff}, k}}]^{N-1}} \otimes {[\frak{u}(1)^{(n)}_{{\rm aff}, k}}]^{N-1}_{\rm Toda},
\ee
where the subscript ``Toda'' indicates that the affine CFT is realized by a $\mathbb Z_n$-twisted Toda field theory with ${\rm rank} \, \frak {su}(N)$ scalar fields and central charge
\be
\label{c-toda-AB}
c_{\rm Toda} (\epsilon_1, \epsilon_2) = {\rm rank} \, \frak {su}(N) +  {h^\vee_{\frak {su}(N)} {\rm dim} \, \frak {su}(N) \over k} \left(b + {1 \over b}\right)^2.
\ee
Here, $h^\vee_{\frak {su}(N)}$ is the dual Coxeter number of $\frak {su}(N)$, and $b = \sqrt{\epsilon_1 / \epsilon_2}$.

Therefore, via (\ref{coset equivalence AGT-ALE}), we can also express (\ref{coset CFT-AGT-ALE}) as
\be
\left[{\frak{su}(k)^{(n)}_{{\rm aff}, N} \over{[\frak{u}(1)^{(n)}_{{\rm aff}, N}}]^{k-1}} \right] \otimes \left[{{\frak{su}(N)^{(n)}_{{\rm aff}, k}} \over{[\frak{u}(1)^{(n)}_{{\rm aff}, k}}]^{N-1}} \otimes {[\frak{u}(1)^{(n)}_{{\rm aff}, k}}]^{N-1}_{\rm Toda} \right].
\label{coset CFT-AGT-AB-ALE-paratoda}
\ee
The first factor in the above product is a $\mathbb Z_n$-twisted parafermionic coset theory of $SU(k)$ at level $N$, and from (\ref{c-toda-AB}), one can see that the second factor  is a $\mathbb Z_n$-twisted version of the $k^{\rm th}$ paratoda theory of $SU(N)$ described in~\cite[$\S$2]{warner}. In light of the isomorphism relations in footnote~\ref{affine iso}, we can also write the affine algebras associated with (\ref{coset CFT-AGT-AB-ALE-paratoda}) as
\be
\label{coset-para-AB}
{\cal G}^\vee_{{\rm para}, N} \otimes {\cal W}_k({\frak g}^\vee_{\rm aff}).
\ee
Here, ${\cal G}^\vee_{{\rm para},N}$ is the parafermionic coset of the \emph{Langlands dual} affine Lie algebra ${\cal G}^\vee_{\rm aff}$ at level $N$, where ${\cal G}_{\rm aff} = {\frak {su}(k)}_{\rm aff}$ or ${\frak {so}(k+1)}_{\rm aff}$  when $n=1$ or 2 (with even $k$), and ${\cal W}_k({\frak g}^\vee_{\rm aff})$ is the $k$-th para-$\cal W$-algebra derived from the \emph{Langlands dual} affine Lie algebra ${\frak g}^\vee_{\rm aff}$, where  $\frak g_{\rm aff} = \frak {su}(N)_{\rm aff}$ or $\frak {so}(N+1)_{\rm aff}$  when $n=1$ or $2$ (with even $N$).

Hence, in place of (\ref{AGT-duality-A}), we have 
\be
 \boxed{\bigoplus_{m, w_2} ~{\rm IH}^\ast_{U(1)^2 \times T} \, {\cal U}({\cal M}^{w_2}_{G, m}(\widetilde{\mathbb R^4 / \mathbb Z_k})) = \widehat{{\cal G}}^\vee_{{\rm para}, N} \otimes \widehat{{\cal W}_k}({\frak g}^\vee_{\rm aff})}
 \label{AGT-duality-AB-ALE}
\ee
Here, ${\cal U}({\cal M}^{w_2}_{G, m}(\widetilde{\mathbb R^4 / \mathbb Z_k}))$ is the Uhlenbeck compactification of the moduli space  ${\cal M}^{w_2}_{G, m}(\widetilde{\mathbb R^4 / \mathbb Z_k})$ of $G$-instantons of instanton number $m$ on $\widetilde{\mathbb R^4 / \mathbb Z_k}$ of class $w_2 \in H^2(\widetilde{\mathbb R^4 / \mathbb Z_k}, \pi_1(G))$; $G = SU(N)$ or $SO(N+1)$ when $n= 1$ or $2$ (with even $N$); and $\widehat{{\cal G}}^\vee_{{\rm para}, N}$ and $\widehat{{\cal W}_k}({\frak g}^\vee_{\rm aff})$ are Verma modules over ${\cal G}^\vee_{{\rm para}, N}$ and  ${\cal W}_k({\frak g}^\vee_{\rm aff})$, respectively.

From (\ref{coset CFT-AGT-AB-ALE-paratoda}), (\ref{c-toda-AB}), and footnote~\ref{central charge}, we find that the central charge of the affine algebra which underlies the RHS of (\ref{AGT-duality-AB-ALE}) is
\be
\label{cc-AGT-ALE-AB}
\boxed{c^k_{{A}, \epsilon_1, \epsilon_2} =  k(N  - 1)  +  {(N^3 -N) \over k} {(\epsilon_1 + \epsilon_2)^2 \over \epsilon_1 \epsilon_2}}
\ee
When $k=1$, $c^{k}_{{A}, \epsilon_1, \epsilon_2}$ indeed reduces to (\ref{c-A-epsilon}), as expected.

Since we can straightforwardly generalize from $\mathbb R^4$ to $\widetilde{\mathbb R^4 / \mathbb Z_k}$ the arguments which took us from (\ref{abcd BPS relation}) to (\ref{q | q}), via (\ref{AGT-duality-AB-ALE}), we can write the $\widetilde{\mathbb R^4 / \mathbb Z_k}$ Nekrasov instanton partition function as
\be
\label{q | q, ALE-AB}
\boxed{Z_{\rm inst} (G, \epsilon_1, \epsilon_2, \vec a, k) = \langle k, \Delta   | k, \Delta \rangle}
\ee
where
\be
\label{coherent-ALE-AB}
\boxed{| k, \Delta \rangle = \bigoplus_{m}  A^{m}  | \Psi_{m, k} \rangle}
\ee
Here, $| k, \Delta \rangle \in \widehat{{\cal G}}^\vee_{{\rm para}, N} \otimes \widehat{{\cal W}_k}({\frak g}^\vee_{\rm aff})$; $A^{m}$ is some real number; $ | \Psi_{m, k} \rangle \in \bigoplus_{w_2}  {\rm IH}^\ast_{U(1)^2 \times T} \, {\cal U}({\cal M}^{w_2}_{G, m}(\widetilde{\mathbb R^4 / \mathbb Z_k}))$ is also a state in $\widehat{{\cal G}}^\vee_{{\rm para}, N} \otimes \widehat{{\cal W}_k}({\frak g}^\vee_{\rm aff})$ with energy level $n_{m}$ determined by the instanton number $m$ (as one recalls that  $n_{m}$ is a constant shift of the eigenvalue $m$ of the $L_0$ operator which generates translations along the ${\bf S}^1_n$ circle in (\ref{AGT-M-duality-AB})); and $ \langle \cdot | \cdot \rangle$ is a Poincar\'e pairing in the sense of~\cite[$\S$2.6]{J-function}. The label $\Delta$ just means that $\widehat{{\cal G}}^\vee_{{\rm para}, N} \otimes \widehat{{\cal W}_k}({\frak g}^\vee_{\rm aff})$ is generated by the application of creation operators on the highest weight state $| \Delta \rangle$.  

As in the $\mathbb R^4$ case, since the RHS of (\ref{q | q, ALE-AB}) is defined in the limit that the ${\bf S}^1_n$ fiber in $\cal C$ has \emph{zero} radius, and since we have in $\cal C$ a \emph{common} boundary condition at $z = 0$ and $z=\infty$ (where the radius of the  ${\bf S}^1_n$ fiber is zero),  $| k, \Delta \rangle$ and $\langle k, \Delta |$ ought to be a state and its dual associated with the puncture at $z = 0$ and $z=\infty$, respectively. Furthermore, as the RHS of (\ref{coherent-ALE-AB}) is a sum over states of all possible energy levels, it would mean that $| k, \Delta \rangle$ is actually a \emph{coherent state}. 

Thus, in arriving at the boxed relations (\ref{AGT-duality-AB-ALE}), (\ref{cc-AGT-ALE-AB}),  (\ref{q | q, ALE-AB}) and (\ref{coherent-ALE-AB}), we have just furnished a fundamental physical derivation of an  $\widetilde{\mathbb R^4 / \mathbb Z_k}$ pure AGT correspondence for the $A_{N-1}$ and (for even $N$) the $B_{N/2}$ groups! 
 
\bigskip\noindent{\it An ALE Generalization of the Pure AGT Correspondence for the $C$--$D$--$G$ Groups} 

Similarly, if we replace $\mathbb  R^4$ on the LHS of (\ref{AGT-M-duality-CDG}) with $\widetilde{\mathbb R^4 / \mathbb Z_k}$, it would mean that we have to replace (\ref{AGT-CDG-chiral CFT}) with an Omega-deformed $D$-type version of (\ref{effective CFT-fully-resolved-A}). 

A $D$-type version of  (\ref{effective CFT-fully-resolved-A}) would be given by an $\widetilde{\mathbb R^4 / \mathbb Z_k}$ generalization of (\ref{partially gauged CFT - CDG}). Such a generalization would be furnished by the I-brane system in (\ref{equivalent IIA system 2}), but now with $k$ \emph{non-coincident} D4-branes; in other words, the original $SO(k)$ gauge symmetry associated with the stack of D4-branes would now be reduced to a $U(1)^{k/2}$ gauge symmetry. In turn, according to our explanations in $\S$3.2, since the gauge groups on the D4- and D6-branes must be of the \emph{same} type, it would mean that we ought to associate a $U(1)^N$ gauge symmetry with the D6-branes, i.e., the D6-branes would be pulled apart and away from the O$6^-$-plane by the 6-4 strings as the D4-branes become non-coincident. Thus, via the arguments which led us to  (\ref{effective CFT-fully-resolved-A}), and the fact that only the $U(1)^{k/2}$ gauge symmetry associated with the non-coincident D4-branes is dynamical, we find that a $D$-type version of  (\ref{effective CFT-fully-resolved-A}) would be given by
\be
\label{effective CFT-fully-resolved-D}
{[\frak{u}(1)^{(n)}_{{\rm aff}, 2N}]^{k/2} \over [\frak{u}(1)^{(n)}_{{\rm aff}, 2N}]^{k/2} } \otimes {\frak{so}(k)^{(n)}_{{\rm aff}, 2N} \over{[\frak{u}(1)^{(n)}_{{\rm aff}, 2N}}]^{k/2}} \otimes \left[{\frak{so}(2N)^{(n)}_{{\rm aff}, k} \over [\frak{u}(1)^{(n)}_{{\rm aff}, k}]^N} \otimes [\frak{u}(1)^{(n)}_{{\rm aff}, k}]^N \right].
\ee

Bearing in mind the fact that Omega-deformation effectively acts only on the $\mathbb Z_n$-twisted affine CFT associated with the set of $N$ D6-branes (whence our result at $k=1$ would indeed be the same as that found in $\S$5.3), we find that an Omega-deformed version of (\ref{effective CFT-fully-resolved-D}) would be given by
\be
{[\frak{u}(1)^{(n)}_{{\rm aff}, 2N}]^{k/2} \over [\frak{u}(1)^{(n)}_{{\rm aff}, 2N}]^{k/2} } \otimes {\frak{so}(k)^{(n)}_{{\rm aff}, 2N} \over{[\frak{u}(1)^{(n)}_{{\rm aff}, 2N}}]^{k/2}} \otimes \left[{\frak{so}(2N)^{(n)}_{{\rm aff}, k} \over {\frak{n}_+}^{(n)}_{{\rm aff}, q} \otimes [\frak{u}(1)^{(n)}_{{\rm aff}, k}]^N} \otimes [\frak{u}(1)^{(n)}_{{\rm aff}, k}]^N \right],
\label{coset CFT-AGT-ALE-D}
\ee
where the level $q \in \mathbb R$, and $\frak n_+$ is a nilpotent subalgebra of strictly upper triangular matrices. Therefore, we have to replace (\ref{AGT-CDG-chiral CFT}) with (\ref{coset CFT-AGT-ALE-D}).

Recall at this point that the central charge due to the Omega-deformation factor $ {\frak{n}_+}^{(n)}_{{\rm aff}, q}$ in (\ref{coset CFT-AGT-ALE-D}), is given by (\ref{cc-D-Omega}) when we have a \emph{single} D4-brane intersecting the $N$ D6-branes, i.e., when $k=1$, as shown in (\ref{equivalent IIA system 2 - AGT}). Recall also that this central charge is proportional to the curvature of $\cal C$ induced by Omega-deformation; thus, when we have $k$ D4-branes intersecting the $N$ D6-branes whence the curvature of $\cal C$ would  be  ``diluted'' over $k$ D4-branes, we ought to divide its value by $k$. This means that the last factor in (\ref{coset CFT-AGT-ALE-D}) ought to obey the following (conformal) equivalence of coset realizations:
\be
\label{coset equivalence AGT-ALE-D}
{\frak{so}(2N)^{(n)}_{{\rm aff}, k} \over {\frak{n}_+}^{(n)}_{{\rm aff}, q} \otimes [\frak{u}(1)^{(n)}_{{\rm aff}, k}]^N} \otimes [\frak{u}(1)^{(n)}_{{\rm aff}, k}]^N = {\frak{so}(2N)^{(n)}_{{\rm aff}, k} \over [\frak{u}(1)^{(n)}_{{\rm aff}, k}]^N} \otimes [\frak{u}(1)^{(n)}_{{\rm aff}, k}]^N_{\rm Toda},
\ee
where the subscript ``Toda'' indicates that the affine CFT is realized by a $\mathbb Z_n$-twisted Toda field theory with ${\rm rank} \, \frak {so}(2N)$ scalar fields and central charge
\be
\label{c-toda-CDG}
c_{\rm Toda} (\epsilon_1, \epsilon_2) = {\rm rank} \, \frak {so}(2N) +  {h^\vee_{\frak {so}(2N)} {\rm dim} \, \frak {so}(2N) \over k} \left(b + {1 \over b}\right)^2.
\ee
Here, $h^\vee_{\frak {so}(2N)}$ is the dual Coxeter number of $\frak {so}(2N)$, and $b = \sqrt{\epsilon_1 / \epsilon_2}$.

Therefore, via (\ref{coset equivalence AGT-ALE-D}), we can also express (\ref{coset CFT-AGT-ALE-D}) as
\be
\left[{\frak{so}(k)^{(n)}_{{\rm aff}, 2N} \over{[\frak{u}(1)^{(n)}_{{\rm aff}, 2N}}]^{k/2}} \right] \otimes \left[{\frak{so}(2N)^{(n)}_{{\rm aff}, k} \over [\frak{u}(1)^{(n)}_{{\rm aff}, k}]^N} \otimes [\frak{u}(1)^{(n)}_{{\rm aff}, k}]^N_{\rm Toda} \right].
\label{coset CFT-AGT-CDG-ALE-paratoda}
\ee
The first factor in the above product is a $\mathbb Z_n$-twisted \emph{generalized} parafermionic coset theory of $SO(k)$ at level $2N$, and from (\ref{c-toda-CDG}), one can see that the second factor  is a $\mathbb Z_n$-twisted version of the $k^{\rm th}$ paratoda theory of $SO(2N)$ described in~\cite[$\S$2]{warner}. In light of the isomorphism relations in footnote~\ref{affine iso}, we can also write the affine algebras associated with (\ref{coset CFT-AGT-CDG-ALE-paratoda}) as
\be
\label{coset-para-CDG}
{\mathscr G}^\vee_{{\rm para}, 2N} \otimes {\cal W}_k({\frak g}^\vee_{\rm aff}).
\ee
Here, ${\mathscr G}^\vee_{{\rm para}, 2N}$ is the generalized parafermionic coset of the \emph{Langlands dual} affine Lie algebra ${\mathscr G}^\vee_{\rm aff}$ at level $2N$, where ${\mathscr G}_{\rm aff} = {\frak {so}(k)}_{\rm aff}$, ${\frak {usp}(k-2)}_{\rm aff}$ or  ${\frak g}_{2 \, {\rm aff}}$ when $n=1$, 2 or 3 (with $k = 8$), and ${\cal W}_k({\frak g}^\vee_{\rm aff})$ is the $k$-th para-$\cal W$-algebra derived from the \emph{Langlands dual} affine Lie algebra ${\frak g}^\vee_{\rm aff}$, where  $\frak g_{\rm aff} = \frak {so}(2N)_{\rm aff}$, $\frak {usp}(2N-2)_{\rm aff}$ or ${\frak g}_{2 \, {\rm aff}}$  when $n=1$, $2$ or 3 (with $N = 4$).

Hence, in place of (\ref{AGT-duality-D}), we have 
\be
 \boxed{\bigoplus_{m, w_2} ~{\rm IH}^\ast_{U(1)^2 \times T} \, {\cal U}({\cal M}^{w_2}_{G, m}(\widetilde{\mathbb R^4 / \mathbb Z_k})) = \widehat{{\mathscr G}}^\vee_{{\rm para}, 2N} \otimes \widehat{{\cal W}_k}({\frak g}^\vee_{\rm aff})}
 \label{AGT-duality-CDG-ALE}
\ee
Here, ${\cal U}({\cal M}^{w_2}_{G, m}(\widetilde{\mathbb R^4 / \mathbb Z_k}))$ is the Uhlenbeck compactification of the moduli space  ${\mathscr M}^{w_2}_{G, m}(\widetilde{\mathbb R^4 / \mathbb Z_k})$ of $G$-instantons of instanton number $m$ on $\widetilde{\mathbb R^4 / \mathbb Z_k}$ of class $w_2 \in H^2(\widetilde{\mathbb R^4 / \mathbb Z_k}, \pi_1(G))$; $G = SO(2N)$, $USp(2N-2)$ or $G_2$ when $n= 1$, $2$ or $3$ (with $N=4$); and $\widehat{{\mathscr G}}^\vee_{{\rm para}, 2N}$ and $\widehat{{\cal W}_k}({\frak g}^\vee_{\rm aff})$ are Verma modules over ${\mathscr G}^\vee_{{\rm para}, 2N}$ and  ${\cal W}_k({\frak g}^\vee_{\rm aff})$, respectively.

From (\ref{coset CFT-AGT-CDG-ALE-paratoda}), (\ref{c-toda-CDG}), and footnote~\ref{central charge SO(2N)}, it is clear that the central charge of the affine algebra which underlies the RHS of (\ref{AGT-duality-CDG-ALE}) is $c^k_{{D}, \epsilon_1, \epsilon_2} =  kN - c([\frak{u}(1)^{(n)}_{{\rm aff}, 2N}]^{k/2}) + {(2N-2) (2N^2 -N) } \newline {(\epsilon_1 + \epsilon_2)^2} / k {\epsilon_1\epsilon_2}$,  where $c(\dots)$ is the central charge of the indicated affine algebra. That said, when $k=1$, $c^{k}_{{D}, \epsilon_1, \epsilon_2}$ should reduce to (\ref{c-D-epsilon}); this implies that $c([\frak{u}(1)^{(n)}_{{\rm aff}, 2N}]^{1/2}) = 0$; in turn, this means that $c([\frak{u}(1)^{(n)}_{{\rm aff}, 2N}]^{k/2}) = k \times c([\frak{u}(1)^{(n)}_{{\rm aff}, 2N}]^{1/2}) = 0$. Thus, the central charge of the affine algebra which underlies the RHS of (\ref{AGT-duality-CDG-ALE}) is actually
 \be
\label{cc-AGT-ALE-CDG}
\boxed{c^k_{{D}, \epsilon_1, \epsilon_2} =  kN  + {(2N-2) (2N^2 -N) \over k} {(\epsilon_1 + \epsilon_2)^2 \over {\epsilon_1\epsilon_2}}}
\ee
 
Since we can straightforwardly generalize from $\mathbb R^4$ to $\widetilde{\mathbb R^4 / \mathbb Z_k}$ the arguments which took us from (\ref{abcd BPS relation-D}) to (\ref{q | q, D}), via (\ref{AGT-duality-CDG-ALE}), we can write the $\widetilde{\mathbb R^4 / \mathbb Z_k}$ Nekrasov instanton partition function as
\be
\label{q | q, ALE-CDG}
\boxed{Z_{\rm inst} (G, \epsilon_1, \epsilon_2, \vec a, k) = \langle k, \Delta   | k, \Delta \rangle}
\ee
where
\be
\label{coherent-ALE-CDG}
\boxed{| k, \Delta \rangle = \bigoplus_{m}  D^{m}  | \Psi_{m, k} \rangle}
\ee
Here, $| k, \Delta \rangle \in \widehat{{\mathscr G}}^\vee_{{\rm para}, 2N} \otimes \widehat{{\cal W}_k}({\frak g}^\vee_{\rm aff})$; $D^{m}$ is some real number; $ | \Psi_{m, k} \rangle \in \bigoplus_{w_2}  {\rm IH}^\ast_{U(1)^2 \times T} \, {\cal U}({\cal M}^{w_2}_{G, m}(\widetilde{\mathbb R^4 / \mathbb Z_k}))$ is also a state in $\widehat{{\mathscr G}}^\vee_{{\rm para}, 2N} \otimes \widehat{{\cal W}_k}({\frak g}^\vee_{\rm aff})$ with energy level $n_{m}$ determined by the instanton number $m$ (as one recalls that  $n_{m}$ is a constant shift of the eigenvalue $m$ of the $L_0$ operator which generates translations along the ${\bf S}^1_n$ circle in (\ref{AGT-M-duality-CDG})); and $ \langle \cdot | \cdot \rangle$ is a Poincar\'e pairing in the sense of~\cite[$\S$2.6]{J-function}. The label $\Delta$ just means that $\widehat{{\mathscr G}}^\vee_{{\rm para}, 2N} \otimes \widehat{{\cal W}_k}({\frak g}^\vee_{\rm aff})$ is generated by the application of creation operators on the highest weight state $| \Delta \rangle$.  

As in the $\mathbb R^4$ case, since the RHS of (\ref{q | q, ALE-CDG}) is defined in the limit that the ${\bf S}^1_n$ fiber in $\cal C$ has \emph{zero} radius, and since we have in $\cal C$ a \emph{common} boundary condition at $z = 0$ and $z=\infty$ (where the radius of the  ${\bf S}^1_n$ fiber is zero),  $| k, \Delta \rangle$ and $\langle k, \Delta |$ ought to be a state and its dual associated with the puncture at $z = 0$ and $z=\infty$, respectively. Furthermore, as the RHS of (\ref{coherent-ALE-CDG}) is a sum over states of all possible energy levels, it would mean that $| k, \Delta \rangle$ is actually a \emph{coherent state}. 

Thus, in arriving at the boxed relations (\ref{AGT-duality-CDG-ALE}), (\ref{cc-AGT-ALE-CDG}),  (\ref{q | q, ALE-CDG}) and (\ref{coherent-ALE-CDG}), we have just furnished a fundamental physical derivation of an  $\widetilde{\mathbb R^4 / \mathbb Z_k}$ pure AGT correspondence for the $D_{N}$, $C_{N-1}$ and the $G_2$ groups!

\newsubsection{The AGT Correspondence with Matter}

Let us now extend our derivation of the pure AGT correspondence in $\S$5 to include matter. For concreteness, we shall restrict ourselves to the $A$-type superconformal quiver gauge theories described by Gaiotto in~\cite{N=2}. 

From Gaiotto's construction in~\cite{N=2}, it is clear that in order to obtain the corresponding Nekrasov instanton partition function with matter of mass $\bf m$ and Coulomb moduli $\vec {\bf a}$, i.e., $Z_{\rm inst} (G, \epsilon_1, \epsilon_2, \vec {\bf a}, \bf m)$, we would need to insert, along $\mathbb R^4\vert_{\epsilon_1, \epsilon_2}$ on the LHS of (\ref{AGT-M-duality-AB}), 4d worldvolume defects of the type studied in~\cite{TD}.  These defects are  characterized by Young diagrams.

We considered such 4d worldvolume defects in our derivation of a ``ramified'' pure AGT correspondence  in $\S$6.1. There, the original defect spanned a complex plane in $\mathbb R^4\vert_{\epsilon_1, \epsilon_2}$ and wrapped $\Sigma_{n,t} = {\bf S}^1_n \times \mathbb I_t$, whence we could appeal to the chain of dualities described in $\S$2.3 to write down the duality relation (\ref{AGT-M-duality-AB-defect}).  However, if the original defect were to span the entire  $\mathbb R^4\vert_{\epsilon_1, \epsilon_2}$, as in the case at hand, it is no longer clear how one can appeal to duality arguments of the kind furnished in $\S$2.3 to arrive at a duality relation like (\ref{AGT-M-duality-AB-defect}). 

Nevertheless, recall from $\S$5.1--$\S$5.2 that we actually have a pair of M9-branes at the ends of $\Sigma_{n,t}$ and ${\cal C} = {{\bf S}^2} / \{0, \infty\}$   in (\ref{AGT-M-duality-AB}), whereby the M9-branes at the ends of $\cal C$ at $z = \{0,\infty\}$ have a nine-dimensional worldvolume which spans the directions transverse to $\cal C$ (as the ${\bf S}^1_n$-fiber of $\cal C$ that their underlying ten-dimensional worldvolumes wrap has zero radius at those points). Also, according to Gaiotto's generalization~\cite{irr} of the analysis in~\cite{N=2} to asymptotically-free theories, and our discussions leading up to (\ref{q | q}), one ought to associate to each puncture at $z = \{0,\infty\}$ in $\cal C$, a 4d worldvolume defect (which would underlie the coherent state in (\ref{q | q})).  Last but not least, note that the 4d worldvolume defects that Gaiotto had considered, can be realized by intersecting M-branes in the M-theoretic picture of his story~\cite{GMN-WKB}. Altogether therefore, this means that instead of inserting 4d worldvolume defects along $\mathbb R^4\vert_{\epsilon_1, \epsilon_2}$ on the LHS of (\ref{AGT-M-duality-AB}), one can also obtain $Z_{\rm inst} (G, \epsilon_1, \epsilon_2, \vec {\bf a}, \bf m)$ by inserting appropriate M9-branes which intersect the M5-branes along  ${\bf S}^1_n \times \mathbb R^4\vert_{\epsilon_1, \epsilon_2}$ whilst taking the radius $\beta$ of ${\bf S}^1_n$ to zero (recall this from $\S$5.2). In the limit that $\beta \to 0$, on the dual side, we would have\emph{ instantonic} M9-branes which sit  at specific points in $\cal C$ --  the ``time'' degree of freedom of the M9-branes along the ${\bf S}^1_n$-fiber in $\cal C$ is effectively reduced to a point as $\beta \to 0$. 

In the pure case with Nekrasov instanton partition function  $Z_{\rm inst} (G, \epsilon_1, \epsilon_2, \vec a)$, the M9-branes in the original compactification in the limit $\beta \to 0$ and the corresponding CFT on $\cal C$ in the dual compactification that are behind our derivation of the pure AGT correspondence in $\S$5.2, are depicted in fig.~3. In fig.~3, the vertical planes represent the spatial part of the M9-branes; $X^9\vert_{\epsilon_i} = \mathbb R^4\vert_{\epsilon_1, \epsilon_2} \times \mathbb R^5\vert_{\epsilon_3; x_{6,7}}$, where four of the spatial directions of the M5-branes are along $\mathbb R^4\vert_{\epsilon_1, \epsilon_2} \subset X^9\vert_{\epsilon_i}$; $\epsilon_3 = \epsilon_1 + \epsilon_2$; $l$ and $\vec a$ are the instanton number and Coulomb moduli of the underlying 4d gauge theory along $\mathbb R^4\vert_{\epsilon_1, \epsilon_2}$; $V_{q, \Delta}$ and $V^\ast_{q, \Delta}$ is a vertex operator and its dual with higher order poles that represent the coherent state $|q, \Delta \rangle$ and its dual $\langle q, \Delta |$ in (\ref{q | q}) of the CFT on $\cal C$; and the two points on $\cal C$ where the vertex operators are located are also where the two instantonic M9-branes which are dual to the two original M9-branes, sit. Note that each of the two planes in fig.~3 along which $l \in \mathbb Z_+$, can be thought of as a coalescence of the planes in fig.~4 along which $l$ takes its minimum value $l_{\rm min}$; this is consistent with the fact~\cite{irr} that the class of punctures in fig.~3 arise from a coalescence of the class of punctures  in fig.~4.  

\begin{figure}
  \centering
    \includegraphics[width=0.8\textwidth]{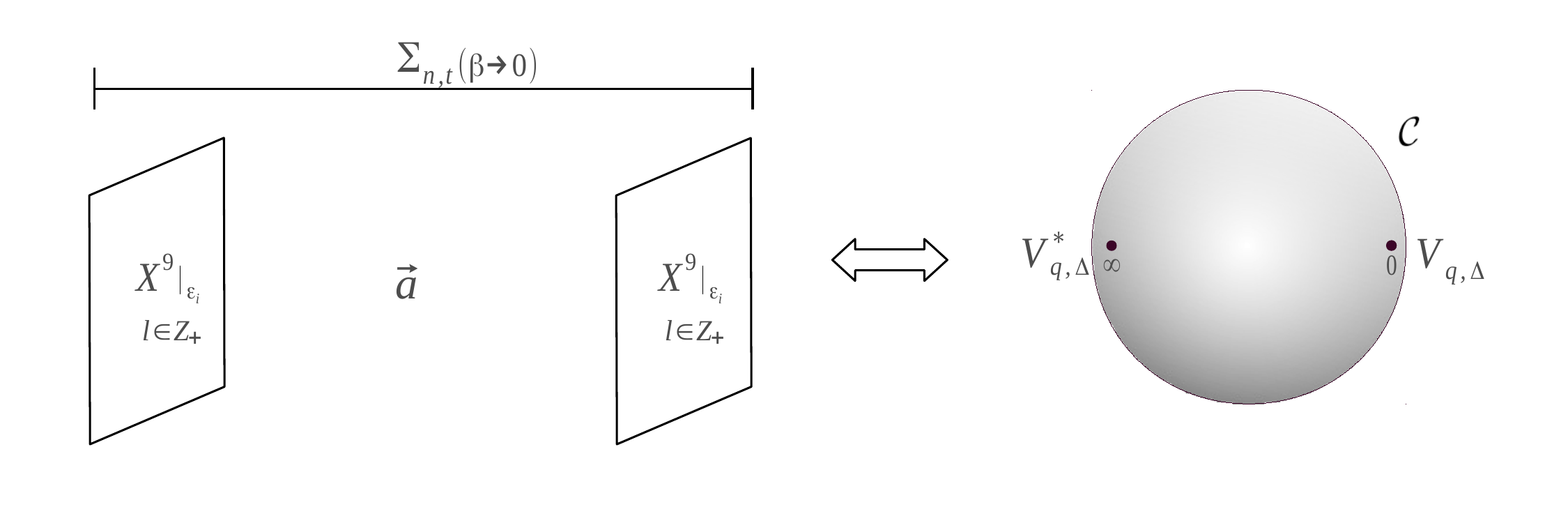}
  \caption{A pair of M9-branes in the original compactification in the limit $\beta \to 0$ and the corresponding CFT on $\cal C$ in the dual compactification that are behind our derivation of the pure AGT correspondence in $\S$5.2.}
\end{figure}

In the case with matter with Nekrasov instanton partition function $Z_{\rm inst} (G, \epsilon_1, \epsilon_2, \vec {\bf a}, \bf m)$, the AGT correspondence can be constructed out of the sphere with a small hole and the cylinder depicted in fig.~4. In fig.~4, $l_{\rm min}$ and $\vec a_i$ are the minimal instanton number and the Coulomb moduli of the underlying 4d gauge theory along $\mathbb R^4\vert_{\epsilon_1, \epsilon_2}$; $V^Q_{\vec \alpha}$ is a vertex operator associated with the unshaded plane that represents the state $|V^Q_{\vec \alpha} \rangle$ of the CFT on $\cal C$ whose conformal weight depends on $Q = (\epsilon_1 +   \epsilon_2) /  \sqrt{\epsilon_1 \epsilon_2}$ and $\vec \alpha$ (or its relevant mass substitute, as we shall explain below); $V_{\vec a_{i}, \vec a_{i+1}}$ is a vertex operator which is associated with the shaded plane that transforms the theory with parameter $\vec a_i$ to the adjacent theory with parameter $\vec a_{i+1}$;  and the second correspondence, which is actually a conformal equivalence, arises because we are dealing with a CFT on $\cal C$. As the 4d gauge theory along $\mathbb R^4\vert_{\epsilon_1, \epsilon_2}$ in the original compactification in fig.~4b (like the one in fig.~3) is asymptotically-free, the observed scale of the eleven-dimensional spacetime $\Sigma_{n,t} \times \mathbb R^4\vert_{\epsilon_1, \epsilon_2} \times \mathbb R^5\vert_{\epsilon_3; x_{6,7}}$ would be inversely proportional to $g^2$, its gauge coupling squared; in turn, this means that the length of the cylinder on the dual side ought to be proportional to $1/g^2$, as indicated. As $l = l_{\rm min}$ along the planes,  the 4d-2d correspondence between $l$ and the conformal weight of CFT states on $\cal C$ that we have derived hitherto, means that the vertex operators in fig.~4 are all \emph{primary} operators. Last but not least, note that the CFT on $\cal C$ with $\cal W$-algebra symmetry can be thought of as a conformal Toda field theory with background charge $Q$; with an appropriate metric on $\cal C$, one can localize $Q$ to the poles~\cite{Ketov}; in other words, one can regard $Q$ to be zero at the point where $V_{\vec a_i, \vec a_{i+1}}$ is inserted (which explains the absence of the superscript `$Q$').

\begin{figure}
  \centering
    \includegraphics[width=1.0\textwidth]{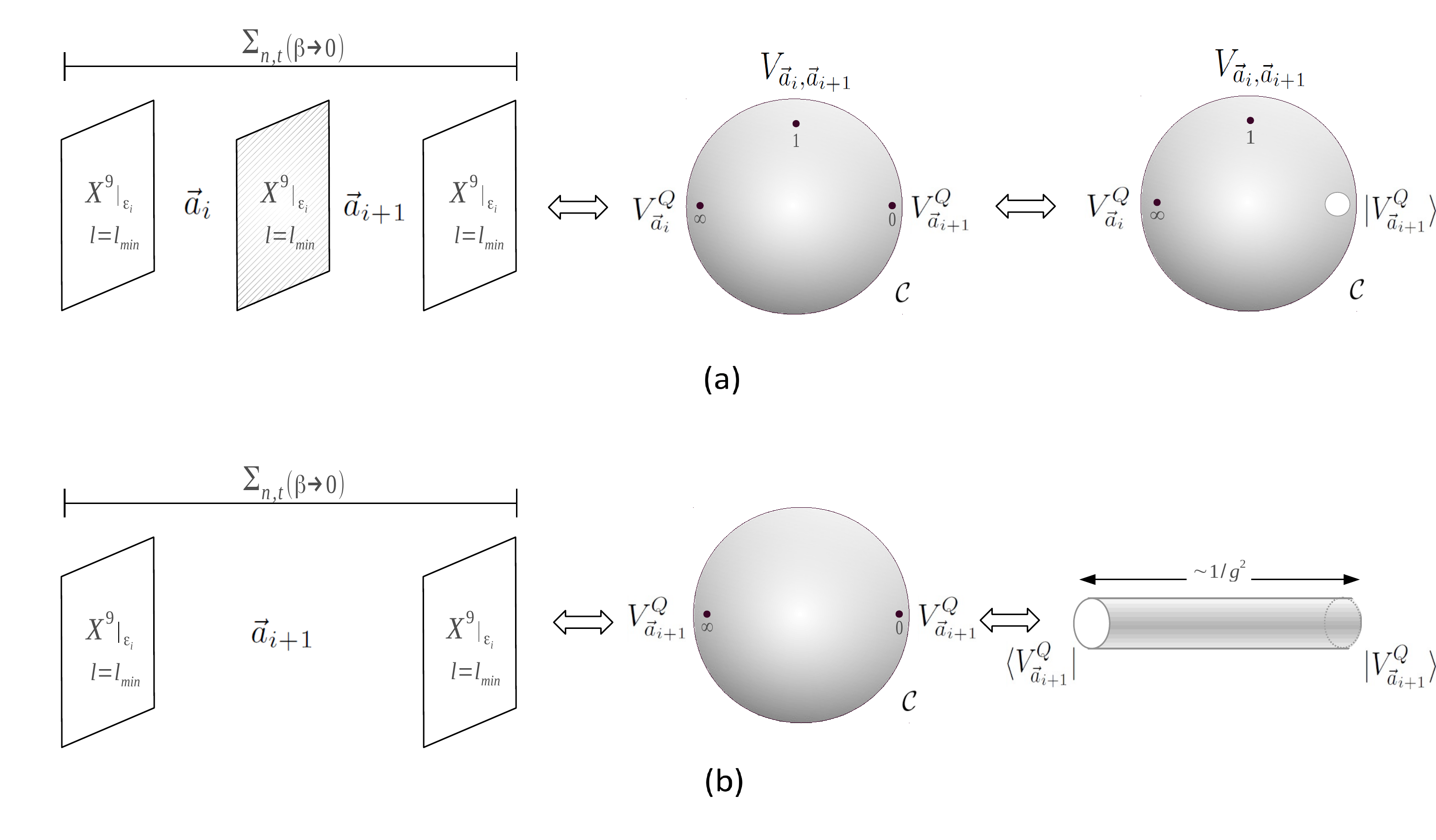}
  \caption{Building blocks of our derivation of the AGT correspondence with matter. (a) a sphere with vertex operators $V^Q_{\vec a_{i}}$ and $V_{\vec a_i, \vec a_{i+1}}$ at $z=\infty$ and $1$, respectively, and a small hole at $z = 0$ with corresponding boundary state $|V^Q_{\vec a_{i+1}} \rangle$; (b) a cylinder of length $\sim 1 / g^2$, with boundary states $\langle V^Q_{\vec a_{i+1}}|$ and $|V^Q_{\vec a_{i+1}} \rangle$.}
\end{figure}

\bigskip\noindent{\it The AGT Correspondence for a Conformal Linear Quiver of $n$ $SU(N)$ Gauge Groups}

Let us now consider an illuminating example of a conformal linear quiver of $n$ $SU(N)$ gauge groups, where $N > 2$. The linear quiver diagram (in the formulation of~\cite{N=2}) and the various steps that lead us to the overall Riemann surface $\Sigma$ on which our 2d CFT lives, are depicted in fig.~5. In fig.~5, the circles and boxes denote the gauge and flavor symmetry groups, respectively; in the second step, we strip away the circles and boxes, use a filled and circled dot to indicate the two different types of external legs corresponding to different flavor symmetry groups, and represent the gauge group with a bounded line; in the third step, we depict the correspondence with Riemann surfaces (in accordance with~\cite{N=2}), where $g_r$ is the gauge coupling associated with the $SU(N)_r$ gauge group; and in the final step, we glue together the individual Riemann surfaces to form the overall Riemann surface $\Sigma$ on which our 2d  CFT lives. 

\begin{figure}
  \centering
    \includegraphics[width=1.0\textwidth]{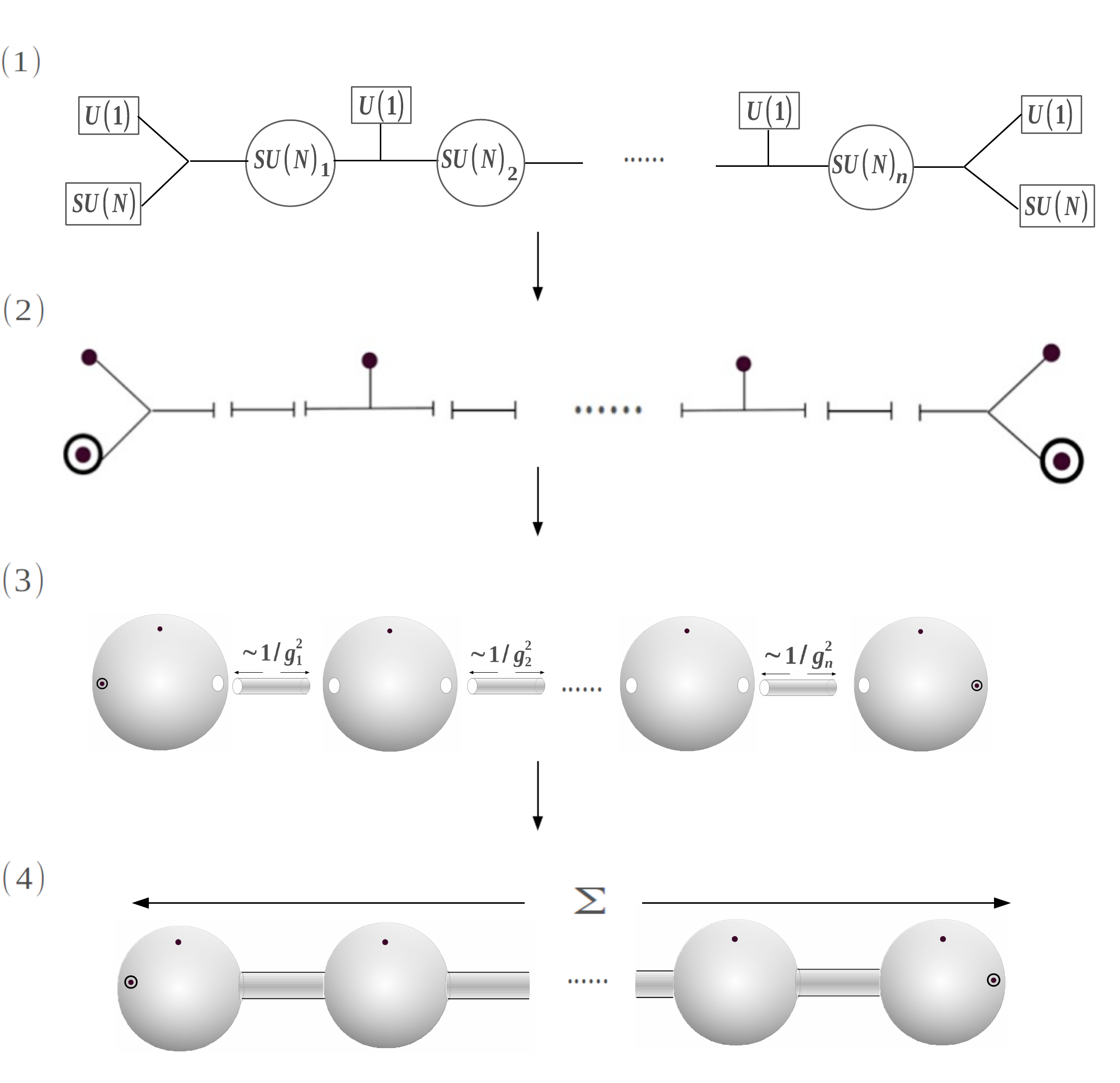}
  \caption{The linear quiver diagram and the various steps that lead us to the overall Riemann surface $\Sigma$ on which our 2d CFT lives.}
\end{figure}

Comparing the individual Riemann surfaces in fig.~5 with the building blocks in fig.~4, whilst noting that one can, for our present purpose, replace the primary operator $V^Q_{\vec a_i}$ in fig.~4(a) with a small hole and the state $\langle V^Q_{\vec a_i}|$, it is clear that we have to glue the eleven-dimensional theories in fig.~4 along the unshaded boundary planes in order to obtain $Z_{\rm inst} (G, \epsilon_1, \epsilon_2, \vec {\bf a}, \bf m)$ on the 4d gauge theory side. On the 2d CFT side, notice that the overall Riemann surface $\Sigma$ is actually conformally equivalent to a sphere; in other words, the efffective Riemann surface ${\cal C}_{\rm eff}$ on which our 2d CFT lives, is ${\bf S}^2$. Thus, the effective correspondence in this case which replaces fig.~3 in the pure case, would be as depicted in fig.~6. In fig.~6, the $m_k$'s are the mass parameters associated with the flavor groups; $\Phi_{\vec \alpha, m_{i}, \vec \beta}$ is an operator representing the shaded plane which transforms the theory with parameter $\vec \alpha$ to the adjacent theory with parameter $\vec \beta$; the subscript `$\vec j_p$' is the highest weight that defines the primary operators $V^Q_{\vec j_p}$ and $V_{\vec j_p}$; $q_r = e^{2 \pi i \tau_r}$, where $\tau_r = 4 \pi i / g^2_r + \theta_r / 2 \pi$; and the points where the $V_{\vec j_p}$'s are inserted are $z = 1, q_1, q_1q_2, \dots, q_1q_2\dots q_n$ because the insertion point $z = 1$ is propagated along the tube of length $\sim 1 / g_1^2$ in fig.~5 to the insertion point $z = 1 \cdot q_1 = q_1$, the insertion point $z = q_1$ is propagated along the tube of length $\sim 1 / g_2^2$ in fig.~5 to the insertion point $z = q_1q_2$, and so on.  Note also that in order to arrive at fig.~6, we have chosen the normalization $\langle V^Q_{\vec a_{k}}|V^Q_{\vec a_{k}} \rangle = 1$ so that pairs of unshaded boundary planes associated with the same $\vec a_{k}$ in fig.~4, upon being glued together, become identity planes which therefore effectively disappear.

\begin{figure}
  \centering
    \includegraphics[width=1.0\textwidth]{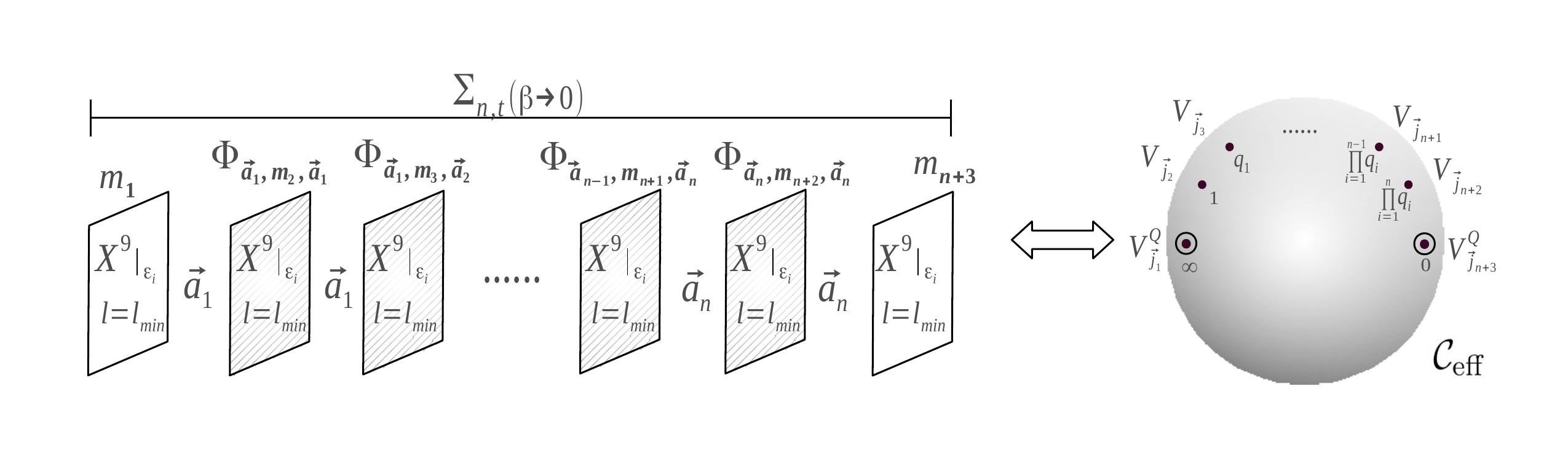}
  \caption{The effective correspondence when the 4d theory is a conformal linear quiver with $n$ $SU(N)$ gauge groups.}
\end{figure}

Clearly, the operator $\Phi_{\vec \alpha, m_{i}, \vec \beta}$ also transforms the space (\ref{BPS-AGT-AB}) of BPS states of the theory with parameter $\vec \alpha$ to that of the adjacent theory with parameter $ \vec \beta$. Thus, we can also describe $\Phi$ as the following map:
\be
\label{Phi-map}
\Phi_{\vec \alpha, m_{i}, \vec \beta}: {\cal H}_{\vec \alpha} \to {\cal H}_{\vec \beta}, \quad {\rm where} \quad {\cal H}_{\vec a_p} = \bigoplus_{l} ~{\rm IH}^\ast_{U(1)^2 \times T} \, {\cal U}({\cal M}_{SU(N), l}) \otimes \mathbb C(\epsilon_1, \epsilon_2, \vec a_p)
\ee
is the space of BPS states of the theory with parameter $\vec a_p$.

At any rate, in the case of a linear quiver of $n$ $SU(N)$ gauge groups, the expression (\ref{Z_inst-5d}) for the Nekrasov instanton partition function ought to be replaced by
\be
Z^{\rm lin}_{\rm inst}({\bf q}, \epsilon_1, \epsilon_2, {\vec {\bf a}}, {\bf m}) =  \sum_{l_1, l_2 \dots, l_n} q^{l_1}_1 q^{l_2}_2 \cdots q^{l_n}_n \, Z^{\rm lin}_{{\rm BPS}, l_1, l_2, \dots, l_n} (\epsilon_1, \epsilon_2, \vec {\bf a}, {\bf m}, \beta \to 0), 
\ee 
where $l_i$ is the instanton number associated with the $SU(N)_i$ gauge group, and $Z^{\rm lin}_{{\rm BPS}, l_1, l_2, \dots, l_n}$ is the partition function of the aforementioned BPS states associated with the left diagram in fig.~6. This partition function can be viewed as a sum over BPS states that propagate from the rightmost to the leftmost end of the diagram which undergo transformations of the kind described in (\ref{Phi-map}) due to the presence of the shaded planes; in other words, one can also write 
\be
\label{correlation of Phi's}
\hspace{-0.2cm}Z^{\rm lin}_{\rm inst}({\bf q}, \epsilon_1, \epsilon_2, {\vec {\bf a}}, {\bf m}) = {}_{m_1}\langle \emptyset \vert  \Phi_{\vec a_1, m_{2}, \vec a_{1}} \, q^{{\bf l}_1}_1 \, \Phi_{\vec a_1, m_{3}, \vec a_{2}} \, q^{{\bf l}_2}_2 \cdots  \Phi_{\vec a_{n-1}, m_{n+1}, \vec a_{n}} \, q^{{\bf l}_n}_n \,\Phi_{\vec a_{n}, m_{n+2}, \vec a_{n}} \vert \emptyset \rangle_{m_{n+3}},
\ee   
where ${}_{m_1}\langle \emptyset \vert$ and $ \vert \emptyset \rangle_{m_{n+3}}$ are the minimum energy BPS states at the leftmost and rightmost end of the diagram that are associated with $m_1$ and $m_{n+3}$, respectively, while ${\bf l}_i$ is an instanton number operator whose eigenvalue is the instanton number $l_i$ associated with the BPS states.

Note at this point that the duality relation in (\ref{AGT-duality-A}), the self-Langlands-duality of simply-laced affine Lie algebras, the discussion following (\ref{AGT-duality-A}), and the map (\ref{Phi-map}), also mean that 
 \be
\Phi_{\vec \alpha, m_{i}, \vec \beta}: {\cal V}_{{\bf j}(\vec \alpha)} \to {\cal V}_{{\bf j}(\vec \beta)}, 
\ee
where $ {\cal V}_{{\bf j}(\vec a_p)}$ is the Verma module over the ${\cal W}$-algebra ${\cal W}({\frak {su}}(N)_{\rm aff})$ of central charge 
\be
\label{cc-linear quiver}
\boxed{c = (N-1)  + (N^3 - N) {(\epsilon_1 + \epsilon_2)^2 \over \epsilon_1 \epsilon_2}}
\ee  
and highest weight  
\be 
\label{highest weight linear quiver}
{\bf j}(\vec a_p) = {-i\vec a_p \over {\sqrt{\epsilon_1 \epsilon_2}}}  + i Q \vec\rho,
\ee 
with $\vec \rho$ being the Weyl vector of $\frak {su}(N)$. Consequently, $\Phi$ can also be interpreted as a primary vertex operator $V$ acting on $\cal V$; this underlies the correspondence between $\Phi_{\vec \alpha, m_{i}, \vec \beta}$ and $V_{\vec j_{i}}$ in fig.~6.  Similarly, the duality relation in  (\ref{AGT-duality-A}), and the discussion following it, underlie the correspondence between ${}_{m_1}\langle \emptyset \vert$ and $ \vert \emptyset \rangle_{m_{n+3}}$ and $\langle V^Q_{\vec j_1}\vert$ and $\vert V^Q_{\vec j_{n+3}}\rangle$ in fig.~6. 

Hence, the correspondence depicted in fig.~6, and our explanations in the last three paragraphs, mean that we can write 
\be
\label{AGT-lin}
\hspace{-0.0cm}\boxed{Z^{\rm lin}_{\rm inst}({\bf q}, \epsilon_1, \epsilon_2, {\vec {\bf a}}, {\bf m}) = Z^{\rm lin}({\bf q}, \epsilon_i, {\bf m}) \cdot \left\langle V^Q_{\vec j_1} (\infty) V_{\vec j_2} (1) V_{\vec j_3} (q_1) \dots  V_{\vec j_{n+2}} (q_1q_2 \dots q_{n}) V^Q_{\vec j_{n+3}} (0) \right\rangle_{{\bf S}^2}}
\ee
The independence of the factor $Z^{\rm lin}$ on $\vec {\bf a}$ is because the $\vec a_p$'s have already been ``contracted'' in the correlation function: see the RHS of (\ref{correlation of Phi's}).   

According to (\ref{highest weight linear quiver}), the fact that the $a_p$'s and the $m_k$'s have the same dimension, and the fact that $V^Q_{j_1}$ and $V^Q_{j_{n+3}}$ ought to depend on $m_1$ and $m_{n+3}$, respectively, one can conclude that 
\be
\label{vecj-1}
\boxed{\vec j_1 = {-i{\vec m}_1  \over {\sqrt{\epsilon_1 \epsilon_2}}}  + {i \vec\rho \, (\epsilon_1 + \epsilon_2)  \over \sqrt{\epsilon_1 \epsilon_2}} \quad {\rm and} \quad \vec j_{n+3} = {-i \vec m_{n+3} \over {\sqrt{\epsilon_1 \epsilon_2}}}  + {i\vec\rho \, (\epsilon_1 + \epsilon_2)  \over \sqrt{\epsilon_1 \epsilon_2}}}
\ee
where the $N-1$ component vectors $\vec m_1$ and $\vec m_{n+3}$ depend on $m_1$ and $m_{n+3}$. 

Similarly, one can conclude, after recalling that $Q$ vanishes where the $ V_{\vec j_u}$ operators are inserted, that  
\be
\label{vecj-2}
\boxed{\vec j_u = {-i{\vec m}_u  \over {\sqrt{\epsilon_1 \epsilon_2}}} \quad {\rm for} \quad u = 2, 3, \dots, n+2}
\ee
where the $N-1$ component vector $\vec m_i$ depends on $m_i$. 

\begin{figure}
  \centering
    \includegraphics[width=0.4\textwidth]{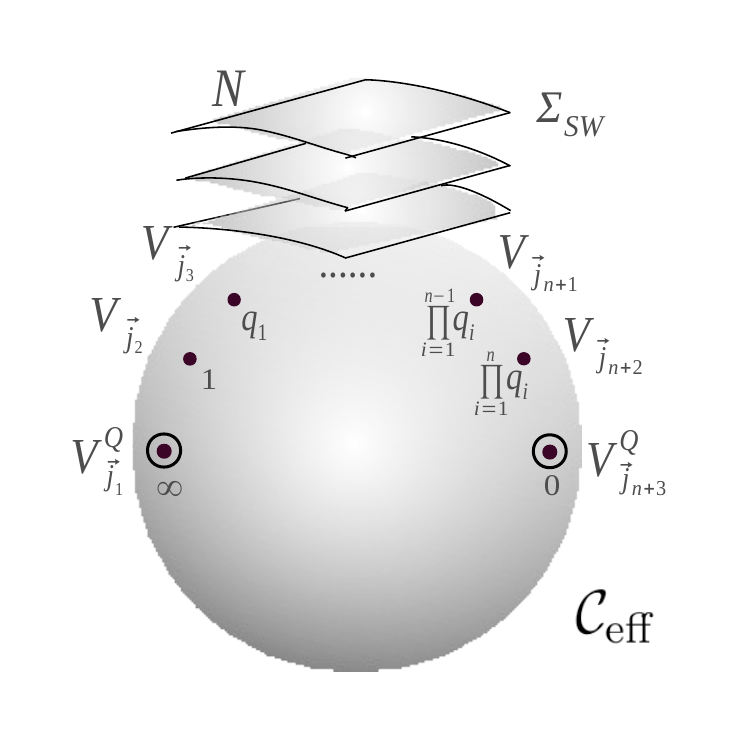}
  \caption{${\cal C}_{\rm eff}$ and its $N$-fold cover $\Sigma_{SW}$ with primary operators inserted at the $n+3$ punctures $z= \infty, 1, q_1, q_1q_2, \dots, q_1q_2 \dots q_n, 0$.}
\end{figure}

Last but not least, note that like in $\S$5.2, we effectively have $N$ D6-branes and $1$ D4-brane wrapping ${\cal C}_{\rm eff}$ (which one can see by ``gluing'' the configuration in (\ref{equivalent IIA system 1 - AGT}) according to our description above), i.e., we effectively have an $N \times 1 = N$-fold cover $\Sigma_{SW}$ of ${\cal C}_{\rm eff}$. This is depicted in fig.~7. Incidentally, $\Sigma_{SW}$ is also the Seiberg-Witten curve which underlies  $Z^{\rm lin}_{\rm inst}({\bf q}, \epsilon_1, \epsilon_2, {\vec {\bf a}}, {\bf m})$! In fact, $\Sigma_{SW}$ can be described in terms of the algebraic relation~\cite{N=2} 
\be
\Sigma_{SW}: \lambda^N + \sum_{k=2}^{N} \lambda^{N-k} \phi_k(z) = 0,
\ee 
where $\lambda = y  dz /z$ (for some complex variable $y$) is a section of $T^\ast {\cal C}_{\rm eff}$, and the $\phi_k(z)$'s are $(k,0)$-holomorphic differentials on ${\cal C}_{\rm eff}$  with poles at the punctures $z = \infty, 1, q_1, q_1q_2, \dots, \newline q_1q_2 \dots q_n$ that are determined by the matter content of the 4d theory. In particular, near the puncture $z = z_s$, we have 
\be
\phi_2(z) \sim {u^{(2)}_s dz^2 \over (z - z_s)^2}, 
\ee
and from the correspondence between $\phi_2(z)$ and the holomorphic stress tensor $W^{(2)}(z)$ (established in $\S$5.2, which thus also applies here), we have 
\be
W^{(2)} (z) {\mathbb V}_{\vec j_s} (z_s) \sim {u^{(2)}_s  \over (z - z_s)^2}  {\mathbb V}_{\vec j_s} (z_s),
\ee
where ${\mathbb V}_{\vec j_s} (z_s)$ can be $V^Q_{\vec j_s} (z_s)$ or $V_{\vec j_s} (z_s)$.  In other words, the conformal dimension of the primary operator $ {\mathbb V}_{\vec j_s} (z_s)$ is equal to $u^{(2)}_s$, i.e., we have
\be
\label{weight-lin}
\boxed{ {{\vec j_s}^2 \over 2} -  {  \vec j_s \cdot i \vec\rho \, (\epsilon_1 + \epsilon_2) \over \sqrt{\epsilon_1 \epsilon_2}}  = u^{(2)}_s, \quad {\rm where} \quad s = 1, 2, \dots, n+3}
\ee
from which we can ascertain the explicit form of the mass vectors $\vec m_s$ in (\ref{vecj-1}) and (\ref{vecj-2}). 

Thus, in arriving at the boxed relations (\ref{cc-linear quiver}), (\ref{AGT-lin}), (\ref{vecj-1}), (\ref{vecj-2}) and (\ref{weight-lin}), we have just derived the AGT correspondence for a conformal linear quiver of $n$ $SU(N)$ gauge groups!

\bigskip\noindent{\it The AGT Correspondence for a Conformal Necklace Quiver of $n$ $SU(N)$ Gauge Groups}

Let us now consider another illuminating example of a conformal necklace quiver of $n$ $SU(N)$ gauge groups, where $N > 2$. The necklace quiver diagram (in the formulation of~\cite{N=2}) and the various steps that lead us to the overall Riemann surface $\Sigma$ on which our 2d CFT lives, are depicted in fig.~8. In fig.~8, the circles and boxes denote the gauge and flavor symmetry groups, respectively; in the second step, we strip away the circles and boxes, use a circled dot to indicate the external leg corresponding to the flavor symmetry group, and represent the gauge group with a bounded curve; in the third step, we depict the correspondence with Riemann surfaces (in accordance with~\cite{N=2}), where $g_r$ is the gauge coupling associated with the $SU(N)_r$ gauge group; and in the final step, we glue together the individual Riemann surfaces to form the overall Riemann surface $\Sigma$ on which our 2d  CFT lives. 

\begin{figure}
  \centering
    \includegraphics[width=1.0\textwidth]{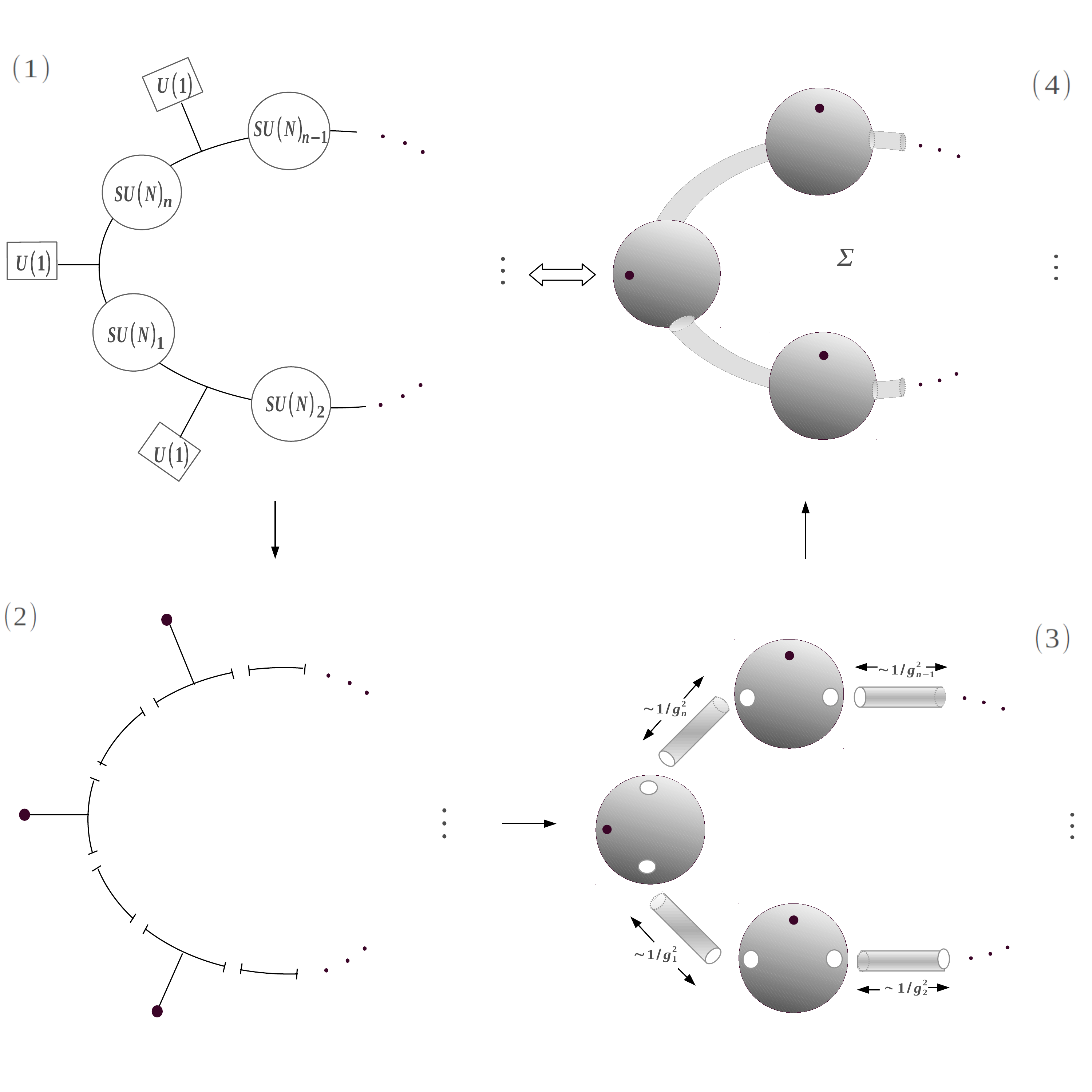}
  \caption{The necklace quiver diagram and the various steps that lead us to the overall Riemann surface $\Sigma$ on which our 2d CFT lives.}
\end{figure}

Comparing the individual Riemann surfaces in fig.~8 with the building blocks in fig.~4, whilst noting that one can, for our present purpose, replace the primary operator $V^Q_{\vec a_i}$ in fig.~4(a) with a small hole and the state $\langle V^Q_{\vec a_i}|$, it is clear that we have to glue in a loop the eleven-dimensional theories in fig.~4 along the unshaded boundary planes in order to obtain $Z_{\rm inst} (G, \epsilon_1, \epsilon_2, \vec {\bf a}, \bf m)$ on the 4d gauge theory side. On the 2d CFT side, notice that the overall Riemann surface $\Sigma$ is actually conformally equivalent to a torus; in other words, the efffective Riemann surface ${\cal C}_{\rm eff}$ on which our 2d CFT lives, is ${\bf T}^2$. Thus, the effective correspondence in this case which replaces fig.~3 in the pure case, would be as depicted in fig.~9. In fig.~9, the $m_k$'s are the mass parameters associated with the flavor groups; $\Phi_{\vec a_i, m_{i}, \vec a_{i+1}}$ is an operator representing the shaded plane which transforms the theory with parameter $\vec a_i$ to the adjacent theory with parameter $\vec a_{i+1}$; the subscript `$\vec j_p$' is the highest weight that defines the primary operators $V^Q_{\vec j_p}$ and $V_{\vec j_p}$; $q_r = e^{2 \pi i \tau_r}$, where $\tau_r = 4 \pi i / g^2_r + \theta_r / 2 \pi$; and the points where the $V_{\vec j_p}$'s are inserted are $z = 1, q_1, q_1q_2, \dots, q_1q_2\dots q_{n-1}$ because the insertion point $z = 1$ is propagated along the tube of length $\sim 1 / g_1^2$ in fig.~8 to the insertion point $z = 1 \cdot q_1 = q_1$, the insertion point $z = q_1$ is propagated along the tube of length $\sim 1 / g_2^2$ in fig.~8 to the insertion point $z = q_1q_2$, and so on.  Note also that in order to arrive at fig.~9, we have chosen the normalization $\langle V^Q_{\vec a_{k}}|V^Q_{\vec a_{k}} \rangle = 1$ so that pairs of unshaded boundary planes associated with the same $\vec a_{k}$ in fig.~4, upon being glued together, become identity planes which therefore effectively disappear.

\begin{figure}
  \centering
    \includegraphics[width=1.0\textwidth]{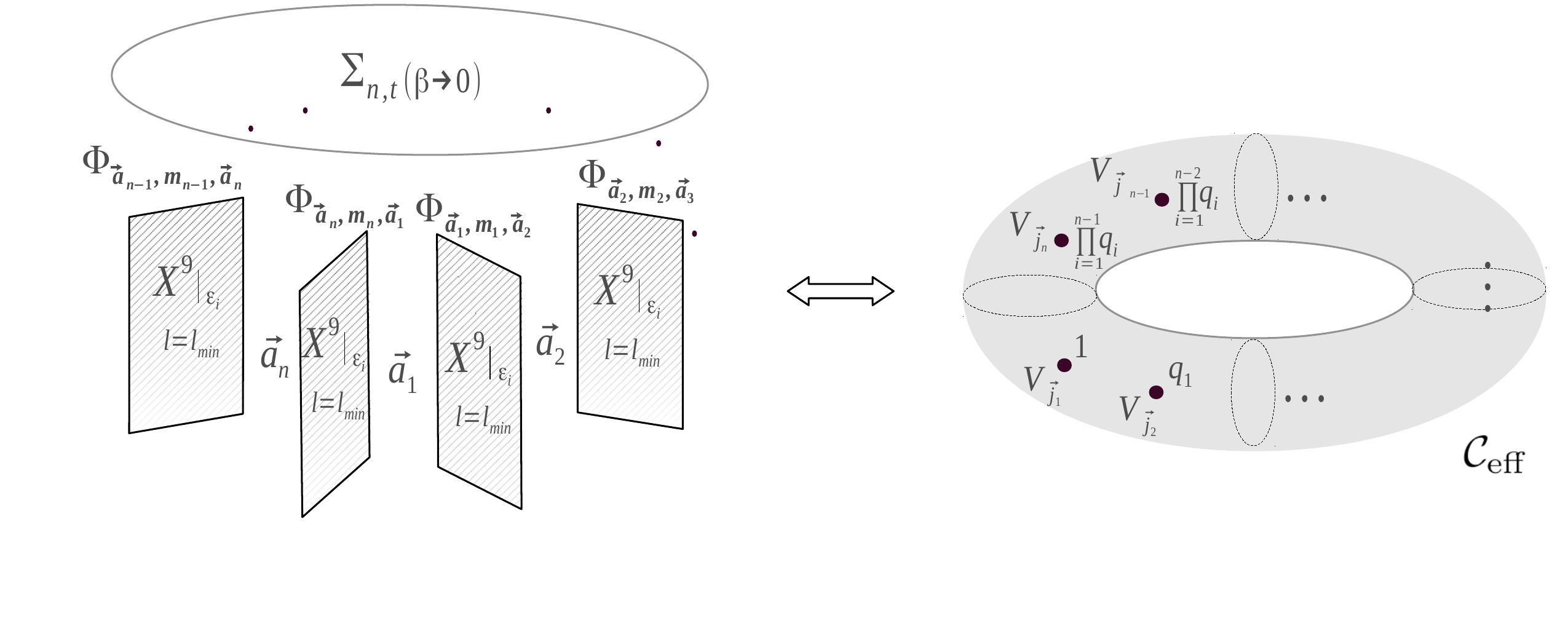}
  \caption{The effective correspondence when the 4d theory is a conformal necklace quiver with $n$ $SU(N)$ gauge groups.}
\end{figure}

Clearly, the operator $\Phi_{\vec a_i, m_{i}, \vec a_{i+1}}$ also transforms the space (\ref{BPS-AGT-AB}) of BPS states of the theory with parameter $\vec a_i$ to that of the adjacent theory with parameter $ \vec a_{i+1}$. Thus, we can also describe $\Phi$ as the following map:
\be
\label{Phi-map-neck}
\Phi_{\vec a_i, m_{i}, \vec a_{i+1}}: {\cal H}_{\vec a_i} \to {\cal H}_{\vec a_{i+1}}, \quad {\rm where} \quad {\cal H}_{\vec a_p} = \bigoplus_{l} ~{\rm IH}^\ast_{U(1)^2 \times T} \, {\cal U}({\cal M}_{SU(N), l}) \otimes \mathbb C(\epsilon_1, \epsilon_2, \vec a_p)
\ee
is the space of BPS states of the theory with parameter $\vec a_p$.

At any rate, in the case of a necklace quiver of $n$ $SU(N)$ gauge groups, the expression (\ref{Z_inst-5d}) for the Nekrasov instanton partition function ought to be replaced by
\be
Z^{\rm neck}_{\rm inst}({\bf q}, \epsilon_1, \epsilon_2, {\vec {\bf a}}, {\bf m}) =  \sum_{l_1, l_2 \dots, l_n} q^{l_1}_1 q^{l_2}_2 \cdots q^{l_n}_n \, Z^{\rm neck}_{{\rm BPS}, l_1, l_2, \dots, l_n} (\epsilon_1, \epsilon_2, \vec {\bf a}, {\bf m}, \beta \to 0), 
\ee 
where $l_i$ is the instanton number associated with the $SU(N)_i$ gauge group, and $Z^{\rm neck}_{{\rm BPS}, l_1, l_2, \dots, l_n}$ is the partition function of the aforementioned BPS states associated with the left diagram in fig.~9. This partition function can be viewed as a sum over BPS states that propagate \emph{around} the diagram which undergo transformations of the kind described in (\ref{Phi-map-neck}) due to the presence of the shaded planes; in other words, one can also write 
\be
\label{correlation of Phi's-neck}
\hspace{-0.2cm}Z^{\rm neck}_{\rm inst}({\bf q}, \epsilon_1, \epsilon_2, {\vec {\bf a}}, {\bf m}) = {\rm Tr}_{{\cal H}_{\vec a_1}} \, (q^{{\bf l}_1}_1 q^{{\bf l}_2}_2 \dots q_n^{{\bf l}_n}) \,  \Phi_{\vec a_1, m_{1}, \vec a_{2}} \,  \Phi_{\vec a_2, m_{2}, \vec a_{3}}  \cdots \Phi_{\vec a_{n-1}, m_{n-1}, \vec a_{n}} \,\Phi_{\vec a_{n}, m_{n}, \vec a_{1}},
\ee   
where ${\bf l}_i$ is an instanton number operator whose eigenvalue is the instanton number $l_i$ associated with the $i$th BPS states.

Note at this point that the duality relation in (\ref{AGT-duality-A}), the self-Langlands-duality of simply-laced affine Lie algebras, the discussion following (\ref{AGT-duality-A}), and the map (\ref{Phi-map-neck}), also mean that 
 \be
\Phi_{\vec a_i, m_{i}, \vec a_{i+1}}: {\cal V}_{{\bf j}(\vec a_i)} \to {\cal V}_{{\bf j}(\vec a_{i+1})}, 
\ee
where $ {\cal V}_{{\bf j}(\vec a_p)}$ is the Verma module over the ${\cal W}$-algebra ${\cal W}({\frak {su}}(N)_{\rm aff})$ of central charge 
\be
\label{cc-neck quiver}
\boxed{c = (N-1)  + (N^3 - N) {(\epsilon_1 + \epsilon_2)^2 \over \epsilon_1 \epsilon_2}}
\ee  
and highest weight  
\be 
\label{highest weight neck quiver}
{\bf j}(\vec a_p) = {-i\vec a_p \over {\sqrt{\epsilon_1 \epsilon_2}}}  + i Q \vec\rho,
\ee 
with $\vec \rho$ being the Weyl vector of $\frak {su}(N)$. Consequently, $\Phi$ can also be interpreted as a primary vertex operator $V$ acting on $\cal V$; this underlies the correspondence between $\Phi_{\vec a_{i-1}, m_{i-1}, \vec a_{i}}$ and $V_{\vec j_{i}}$ in fig.~9.  

Hence, the correspondence depicted in fig.~9, and our explanations in the last three paragraphs, mean that we can write 
\be
\label{AGT-neck}
\hspace{-0.0cm}\boxed{Z^{\rm neck}_{\rm inst}({\bf q}, \epsilon_1, \epsilon_2, {\vec {\bf a}}, {\bf m}) = Z^{\rm neck}({\bf q}, \epsilon_i, {\bf m}) \cdot \left\langle V_{\vec j_1} (1) V_{\vec j_2} (q_1) V_{\vec j_3} (q_1q_2) \dots  V_{\vec j_{n}} (q_1q_2 \dots q_{n-1})  \right\rangle_{{\bf T}^2}}
\ee
The independence of the factor $Z^{\rm neck}$ on $\vec {\bf a}$ is because the $\vec a_p$'s have already been ``contracted'' in the correlation function: see the RHS of (\ref{correlation of Phi's-neck}).   

According to (\ref{highest weight neck quiver}), the fact that the $a_p$'s and the $m_s$'s have the same dimension, the fact that $V_{j_s}$ ought to depend on $m_{s-1}$, and recalling that $Q$ vanishes where $ V_{\vec j_s}$ is inserted, one can conclude that 
\be
\label{vecj-neck}
\boxed{\vec j_s = {-i{\vec m}_{s-1}  \over {\sqrt{\epsilon_1 \epsilon_2}}} \quad {\rm for} \quad s = 1, 2, \dots, n}
\ee
where $\vec m_0 = \vec m_n$, and the $N-1$ component vector $\vec m_k$ depends on $m_k$.

\begin{figure}
  \centering
    \includegraphics[width=0.5\textwidth]{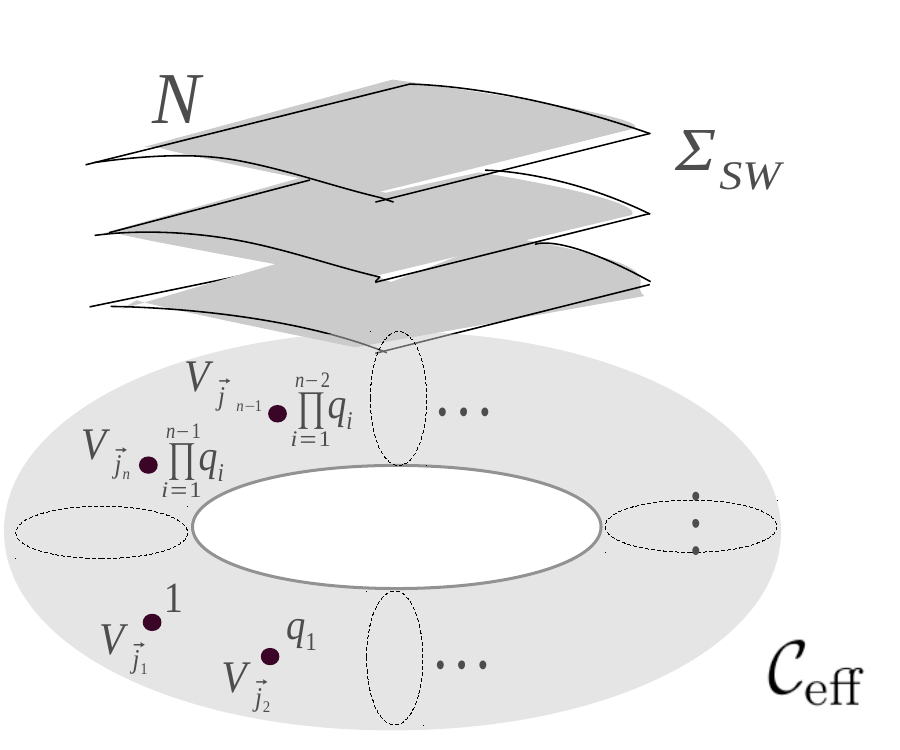}
  \caption{${\cal C}_{\rm eff}$ and its $N$-fold cover $\Sigma_{SW}$ with primary operators inserted at the $n$ punctures $z= 1, q_1, q_1q_2, \dots, q_1q_2 \dots q_{n-1}$.}
\end{figure}

Last but not least, note that like in $\S$5.2, we effectively have $N$ D6-branes and $1$ D4-brane wrapping ${\cal C}_{\rm eff}$ (which one can see by ``gluing'' the configuration in (\ref{equivalent IIA system 1 - AGT}) according to our description above), i.e., we effectively have an $N \times 1 = N$-fold cover $\Sigma_{SW}$ of ${\cal C}_{\rm eff}$. This is depicted in fig.~10. Incidentally, $\Sigma_{SW}$ is also the Seiberg-Witten curve which underlies  $Z^{\rm neck}_{\rm inst}({\bf q}, \epsilon_1, \epsilon_2, {\vec {\bf a}}, {\bf m})$! In fact,  $\Sigma_{SW}$ can be described in terms of the algebraic relation~\cite{N=2} 
\be
\label{SW curve - neck}
\Sigma_{SW}: \lambda^N + \sum_{k=2}^{N} \lambda^{N-k} \phi_k(z) = 0,
\ee 
where $\lambda = y  dz /z$ (for some complex variable $y$) is a section of $T^\ast {\cal C}_{\rm eff}$, and the $\phi_k(z)$'s are $(k,0)$-holomorphic differentials on ${\cal C}_{\rm eff}$  with poles at the punctures $z = 1, q_1, q_1q_2, \dots, \newline q_1q_2 \dots q_{n-1}$ that are determined by the matter content of the 4d theory. In particular, near the puncture $z = z_s$, we have 
\be
\phi_2(z) \sim {u^{(2)}_s dz^2 \over (z - z_s)^2}, 
\ee
and from the correspondence between $\phi_2(z)$ and the holomorphic stress tensor $W^{(2)}(z)$ (established in $\S$5.2, which thus also applies here), we have 
\be
W^{(2)} (z) {V}_{\vec j_s} (z_s) \sim {u^{(2)}_s  \over (z - z_s)^2}  {V}_{\vec j_s} (z_s).
\ee
In other words, the conformal dimension of the primary operator $ {V}_{\vec j_s} (z_s)$ is equal to $u^{(2)}_s$, i.e., we have
\be
\label{weight-neck}
\boxed{ {{\vec j_s}^2 \over 2} -  {  \vec j_s \cdot i \vec\rho \, (\epsilon_1 + \epsilon_2) \over \sqrt{\epsilon_1 \epsilon_2}}  = u^{(2)}_s, \quad {\rm where} \quad s = 1, 2, \dots, n}
\ee
from which we can ascertain the explicit form of the mass vectors $\vec m_s$ in (\ref{vecj-neck}). 

Thus, in arriving at the boxed relations (\ref{cc-neck quiver}), (\ref{AGT-neck}), (\ref{vecj-neck}) and (\ref{weight-neck}), we have just derived the AGT correspondence for a conformal necklace quiver of $n$ $SU(N)$ gauge groups! 

\part{\Large Integrable Systems}

\newsection{The AGT Correspondence, Chiral Fermions, Integrable Systems, and the ``Ramified'' Geometric Langlands Correspondence for Curves}

\newsubsection{The AGT Correspondence with Matter and Chiral Fermions}

Let us consider the topological string limit $\epsilon_1 + \epsilon_2 = \epsilon_3 =  0$ in our derivation of the AGT correspondence with matter in $\S$6.3. For brevity, and to make contact with results by Nekrasov-Okounkov in~\cite{NO}, we shall consider only the conformal necklace quiver of $n$ $SU(N)$ gauge groups. (The analysis for the conformal linear quiver with $n$ $SU(N)$ gauge groups is similar.) 

In the topological string limit $\epsilon_1 = - \epsilon_2 = \hbar$, Omega-deformation on the RHS of (\ref{AGT-M-duality-AB}) effectively vanishes. According to our discussions in $\S$5, the partially gauged chiral CFT behind (\ref{AGT-AB-chiral CFT}) would then be \emph{ungauged}. Consequently, the $\cal W$-algebra ${\cal W}(\frak {su}(N)_{\rm aff})$ that appears in $\S$6.3, ought to be replaced throughout by the affine Lie algebra $\frak {su}(N)_{\rm aff, 1}$ of level 1. This means that instead of (\ref{correlation of Phi's-neck}), the Nekrasov instanton partition function would now be given by
 \be
\label{correlation of Phi's-neck-ts limit}
Z^{\rm neck}_{\rm inst}({\bf q}, \hbar, {\vec {\bf a}}, {\bf m}) = {\rm Tr}_{{\cal H}_{\vec a_1}} \, (q^{{\bf l}_1}_1 q^{{\bf l}_2}_2 \dots q^{{\bf l}_n}_n) \,  \Phi_{\vec a_1, m_{1}, \vec a_{2}} \,  \Phi_{\vec a_2, m_{2}, \vec a_{3}}  \cdots \Phi_{\vec a_{n-1}, m_{n-1}, \vec a_{n}} \,\Phi_{\vec a_{n}, m_{n}, \vec a_{1}},
\ee   
where 
\be
 \label{Phi-map-ts limit}
\Phi_{\vec a_i, m_{i}, \vec a_{i+1}}: {\cal V}_{{\bf j}(\vec a_i)} \to {\cal V}_{{\bf j}(\vec a_{i+1})},
\ee
with $ {\cal V}_{{\bf j}(\vec a_p)}$ being the Verma module over ${\frak {su}}(N)_{\rm aff, 1}$ of central charge 
\be
\label{cc-neck quiver-ts limit}
\boxed{c = N-1}
\ee  
and highest weight  
\be 
\label{highest weight neck quiver-ts limit}
\boxed{{\bf j}(\vec a_p) = -{\vec a_p \over \hbar}}
\ee 
$\vec a_p$ is the Coulomb moduli of the $p^{\rm th}$ $SU(N)$ gauge group;  $q_r = e^{2 \pi i \tau_r}$, with $\tau_r$ being the complexified gauge coupling of the $r^{\rm th}$ $SU(N)$ gauge group; ${\bf l}_i$ is the instanton number operator of the $i$th $SU(N)$ gauge group; and $m_s$ is the mass of the $s^{\rm th}$ bifundamental matter.

From (\ref{Phi-map-ts limit}), we see that $\Phi$ can be interpreted as a primary vertex operator $V$ acting on $\cal V$. Furthermore, recall that (i) we have $N$ chiral fermions which live on ${\cal C}_{\rm eff}$ in fig.~10, that realize ${\frak {su}}(N)_{\rm aff, 1}$; (ii) the duality of the compactifications in (\ref{AGT-dual pair-A}) means that the instanton number $l_i$ of the gauge theory corresponds to the energy level $L_{0,i}$ (of the module with highest weight  ${\vec j}_i$) of the chiral CFT on ${\cal C}_{\rm eff}$. In other words, we can also write 
\be
\label{AGT-neck-ts limit}
\boxed{Z^{\rm neck}_{\rm inst}({\bf q}, \hbar, {\vec {\bf a}}, {\bf m}) = {\rm Tr}_{{\cal H}^{(N)}_{\vec a \over \hbar}} \, \, (q^{L_{0, 1}}_1 q^{L_{0,2}}_2 \dots q^{L_{0,n}}_n) \, \, V_{\vec j_1} (1) \,  V_{\vec j_2} (q_1)  \, V_{\vec j_3} (q_1q_2) \dots  V_{\vec j_{n}} (q_1q_2 \dots q_{n-1})}
\ee
where ${\cal H}^{(N)}_{\vec a \over \hbar}$ is the Fock space of $N$ chiral fermions defined by the highest weight ${\vec a / \hbar} = {\bf j}(\vec a_1)$ of ${\frak {su}}(N)_{\rm aff, 1}$; $V_{\vec j_s}(z_s)$ is a primary operator inserted at $z = z_s$ in ${\cal C}_{\rm eff} = {\bf T}^2$, that is associated with the highest weight $\vec j_s$; and from (\ref{vecj-neck}),  
\be
\label{vecj-neck-ts limit}
\boxed{\vec j_s = -  {{\vec m}_{s-1}  \over \hbar} \quad {\rm for} \quad s = 1, 2, \dots, n}
\ee
where $\vec m_0 = \vec m_n$, and the $N-1$ component vector $\vec m_k$ depends on $m_k$.

For $n=1$, the 4d quiver gauge theory reduces to an $SU(N)$ theory with a massive adjoint hypermultiplet, or the ${\cal N} = 2^\ast$ theory. In this case, our above results coincide with those by Nekrasov-Okounkov in~\cite[$\S$6.3]{NO}. Hence, the boxed relations (\ref{cc-neck quiver-ts limit}), (\ref{highest weight neck quiver-ts limit}), (\ref{AGT-neck-ts limit}) and (\ref{vecj-neck-ts limit}) serve as a bifundamental quiver generalization of the results in \emph{loc.~cit.}.

\newsubsection{The Nekrasov-Okounkov Conjecture and the Tau-Function of Toda Lattice Hierarchy}

We shall now derive a conjecture by Nekrasov-Okounkov~\cite{NO}, and elucidate the connection between the Nekrasov instanton partition function and the tau-function of Toda lattice hierarchy.   

 \bigskip\noindent{\it The Nekrasov-Okounkov Conjecture}
 
To this end, let us consider the topological string limit $\epsilon_1 + \epsilon_2 = \epsilon_3 =  0$ in our derivation of the pure AGT correspondence for $G$ in $\S$5. In this limit, Omega-deformation on the RHS of (\ref{AGT-M-duality-AB}) and (\ref{AGT-M-duality-CDG}) effectively vanishes. According to our discussions in $\S$5, (i) $\cal C$ in the I-brane configurations (\ref{equivalent IIA system 1 - AGT}) and (\ref{equivalent IIA system 2 - AGT}) would become flat again, i.e., $\cal C$ would return to becoming the finite cylinder $\Sigma_{n,t} = {\bf S}^1_n \times \mathbb I_t$; (ii) the partially gauged chiral CFT behind (\ref{AGT-AB-chiral CFT}) and (\ref{AGT-CDG-chiral CFT}) would be \emph{ungauged}. This means that instead of (\ref{q | q}) and (\ref{q | q, D}), we would now have
\be
\label{q | q - hbar}
\boxed{Z_{\rm inst} (G, \Lambda, \hbar, \vec a)  =  \langle {u}_{\hbar}  | \Lambda^{2 n h^\vee L_0} | {u}_{\hbar} \rangle}
\ee
Here, $\Lambda$ is the scale; $\hbar = \epsilon_1 = - \epsilon_2$; $\vec a$ is the Coulomb moduli of the underlying 4d pure $G$ theory; $| u_{\hbar} \rangle \in {\widehat {\frak g}^\vee_{{\rm aff}, 1}}$, where  $\widehat {\frak g}^\vee_{{\rm aff}, 1}$ is the integrable highest weight module over the \emph{Langlands dual} affine Lie algebra $\frak g_{{\rm aff}, 1}^\vee$ of level 1 and central charge $c_{G,\hbar}$; $| u_{\hbar} \rangle$ is a \emph{coherent} state generated from the primary state $| \Delta_\hbar \rangle$ of conformal dimension $\Delta_\hbar$; $h^\vee$ is the dual Coxeter number of the Lie algebra $\frak g$; $n = 1$ for $G = SU(N)$ and $SO(2N)$; $n = 2$ for $G = SO(N+1)$ (with even $N$) and $USp(2n-2)$; $n =3$ for $G = G_2$; and $L_0$ is the generator of time translations along $\Sigma_{n,t}$ which propagates the state $| {u}_{\hbar} \rangle$ at one end by a distance $\sim 1 / g^2 \sim {\rm ln}  \, \Lambda^{2 n h^\vee}$ to the other end whence it is annihilated by the state $\langle {u}_{\hbar} |$,  where $g$ is the underlying gauge coupling.\footnote{As the 4d gauge theories along $\mathbb R^4\vert_{\epsilon_1, \epsilon_2}$ in the original compactifications (\ref{AGT-M-duality-AB}) and (\ref{AGT-M-duality-CDG}) are, in this case, asymptotically-free, the observed scale of the eleven-dimensional spacetime $ \mathbb R^4\vert_{\epsilon_1, \epsilon_2} \times \Sigma_{n,t} \times \mathbb R^5$ ought to be inversely proportional to $g^2$; in particular, this means that the length of $\Sigma_{n,t}$ ought to be proportional to $1/g^2$.} 

For $G = SU(N)$ and $SO(N+1)$ (with even $N$), we have, from (\ref{c-A-epsilon}) and (\ref{Delta(2)}), 
\be
\label{c-A-epsilon-hbar}
\boxed{c_{G, \hbar}  = N-1  
\quad
{\rm and} \quad
\Delta_\hbar = {  { \gamma  {\vec a}^2 \over  \hbar^2}}}
\ee
where $\gamma$ is some real constant.

For $G = SO(2N)$, $USp(2N-2)$ and $G_2$ (with $N = 4$), we have, from (\ref{c-D-epsilon}) and (\ref{Delta(2)-D}), 
\be
\label{c-D-epsilon-hbar}
\boxed{c_{G, \hbar}  = N  
\quad
{\rm and}
\quad
\Delta_\hbar = {  { \gamma'  {\vec a}^2 \over  \hbar^2}}}
\ee
where $\gamma'$ is some real constant.

Note that in arriving at the boxed relation (\ref{q | q - hbar}), and its accompanying boxed relations (\ref{c-A-epsilon-hbar}) and (\ref{c-D-epsilon-hbar}), we have just derived the Nekrasov-Okounkov conjecture in~\cite [$\S$5.4]{NO}! (Strictly speaking, the Nekrasov-Okounkov conjecture holds for the full dual partition function $Z_D$. Nevertheless, since (i) $Z_D$ is just a linear sum of the full partition function at different values of $\vec a$ but with the same underlying highest weight $\vec j$ when the complex parameter $\xi$ in eqn.~(5.1) of \emph{loc.~cit.} is set to zero, where $\vec a^2 \sim \vec j^2$; (ii) the conjecture also holds at $\xi = 0$; (iii)  the perturbative part of the full partition function is just some constant at each different value of $\vec a$; the conjecture also holds for a linear sum of $Z_{\rm inst}$ with the same underlying highest weight $\vec j$. This last statement is what our aforementioned results imply.)

 \bigskip\noindent{\it The $G = SU(N)$ Case and the Tau-Function of Toda Lattice Hierarchy}

Let us now focus on the $G = SU(N)$ case where we necessarily have $n=1$ such that there is no twist of the chiral CFT on $\Sigma_{n,t}$ as we go around ${\bf S}^1_n$. According to our discussions in $\S$3 and $\S$5, we would have $N$ untwisted chiral fermions on $\Sigma_{n,t}$ which realize $ {{\frak su}(N)}_{{\rm aff}, 1}$. As such, by comparing the RHS of (\ref{q | q - hbar}) with the RHS of \cite[eqn.~(5.24)]{NO},  bearing in mind that \cite[eqn.~(5.24)]{NO} can also be written as \cite[eqn.~(5.25)]{NO},  we find that we can also express the Nekrasov instanton partition function as
\be
\label{q | q - hbar - chiral fermions}
Z_{\rm inst} (SU(N), \Lambda, \hbar, \vec a)  =  \langle p  | e^{{\cal J}_{1} \over \hbar} \, {\Lambda}^{2 N L_0} \, e^{{\cal J}_{-1} \over \hbar} | p \rangle,
\ee 
where $| p \rangle$ is a vacuum state in a standard fermionic Fock space $\cal H$ whose energy level is $p^2 /2$; $1 \leq p \leq N$; and ${\cal J}_{\mp 1}$ are creation and annihilation operators in $\cal H$, respectively, which are  constructed out of the chiral fermions.  

According to~\cite{Harnad}, the tau-function of Toda lattice hierarchy (in the  fermionic Fock space formulation) is given by 
\be
\tau_{p, \cal G} = \langle p  | e^{{\cal J}_{1} \over \hbar} \, {\cal G} \, e^{{\cal J}_{-1} \over \hbar} | p \rangle,
\ee 
where $| p \rangle \in  {\cal H}_0$, $\langle p  | \in {\cal H}_\infty$, and ${\cal G}: {\cal H}_0 \to {\cal H}_\infty$. This is just the RHS of (\ref{q | q - hbar - chiral fermions}) when ${\cal G} = {\Lambda}^{2 N L_0}$. Therefore, we have 
\be
\label{NN-conjecture}
\boxed{Z_{\rm inst} (SU(N), \Lambda, \hbar, \vec a)  = \tau_{p, {\Lambda}^{2 N L_0}}}
\ee
and since $\Sigma_{n,t}$ is conformally equivalent to a Riemann sphere with two disks deleted at $z= 0, \infty$, this is just Nekrasov's conjecture in~\cite[eqn.~(4.11)]{NN}!

\newsubsection{The ``Fully-Ramified'' Nekrasov Instanton Partition Function and Quantum Affine Toda Systems}

Let us consider the Nekrasov-Shatashvili  limit $\epsilon_2 = 0$ in our derivation of the ``fully-ramified'' pure AGT correspondence for $G$ in $\S$6.1. From (\ref{K-ab}) and (\ref{K-cdg}), we find that in place of (\ref{q | q, full-defect-AB}) and (\ref{q | q, full-defect-CDG}), the ``fully-ramified'' Nekrasov instanton partition function is now
\be
\label{q | q, full defect-NS limit}
Z_{\rm inst} (G, \epsilon_1, 0, \vec a, T) = \langle \epsilon_1, \Delta_n   | \epsilon_1, \Delta_n \rangle,
\ee
where $\vec a$ is the Coulomb moduli of the underlying pure $G$ theory on $\mathbb R^4\vert_{\epsilon_1, \epsilon_2 = 0}$; $T \subset G$ is the maximal torus that $G$ reduces to along $\mathbb R^2\vert_{\epsilon_1} \subset \mathbb R^4\vert_{\epsilon_1, \epsilon_2 = 0}$; $|\epsilon_1, \Delta_n \rangle \in {\widehat {\frak g}}^\vee_{{\rm aff}, {\rm crit}}$, and ${\widehat {\frak g}}^\vee_{{\rm aff}, {\rm crit}}$ is the integrable module of the \emph{Langlands dual} affine Lie algebra ${\frak g}^\vee_{{\rm aff}, {\rm crit}}$ at the \emph{critical level}.

Via the (twisted) dimension-one currents $J_{a_i}$ that generate ${\frak g}^\vee_{{\rm aff}, {\rm crit}}$ on ${\cal C} = {\bf S}^2$, one can define the (twisted) Segal-Sugawara operators  
\be 
S^{(s_i)}(z) = (k+h^{\vee}) T^{(s_i)}(z), \quad s_i = e_i +1, \quad i = 1, 2, \dots, {\rm rank}(\frak g),
\label{act 2}
\ee 
where the Lie algebra $\frak g = {\frak {sl}}(N)$ if $G = SU(N)$ and $SO(N+1)$ (with even $N$); $\frak g = {\frak {so}}(2N)$ if $G = SO(2N)$, $USp(2N-2)$ and $G_2$; $h^\vee$ is the dual Coxeter number of $\frak g$; the $e_i$'s are the exponents of $\frak g$; and the spin-$s_i$ ($s_i \geq 2$) operators $T^{(s_i)}$ are just higher spin generalizations of the holomorphic stress tensor $T^{(2)}$. Note that we can also write
\be S^{(s_i)}(z) = :{
d}^{a_1 a_2 a_3 \dots a_{s_i}}(k) (J_{a_1} J_{a_2}\dots
J_{a_{s_i}})  (z):,
\label{S^{(s_i)}(z)} 
\ee 
where ${d}^{a_1 a_2 a_3 \dots a_{s_i}}(k)$ is a completely symmetric traceless $\frak g$-invariant tensor of rank $s_i$ (which depends on the variable $k$); in other words, the ${\rm rank}(\frak g)$ number of $S^{(s_i)}$'s are Casimir operators.

 From (\ref{act 2}), one can see that the $S^{(s_i)}$'s generate in their OPE's with all other operators of
the quantum CFT on $\cal C$,  $(k+h^{\vee})$ times the field transformations generated by the $T^{(s_i)}$'s. Therefore, at the critical level $k=-h^\vee$, the $S^{(s_i)}$'s  generate \emph{no} field transformations at all: their OPE's amongst themselves, and with all other field operators, are regular. Hence, on any correlation function of operators, the $S^{(s_i)}$'s effectively act as $c$-numbers. In particular, this means that the RHS of (\ref{q | q, full defect-NS limit}) -- which can be interpreted as a two-point correlation function of coherent state operators -- is a simultaneous eigenfunction of the commuting $S^{(s_i)}$'s. As the $S^{(s_i)}$'s generically act as  order-$s_i$ differential operators in their action on a correlation function of primary state operators (see for example~\cite[$\S$15.7]{CFT text}), and since a coherent state can be obtained by applying creation operators on a primary state, i.e., a coherent state operator can be derived from a primary state operator, our discussion hitherto would mean that $Z_{\rm inst} (G, \epsilon_1, 0, \vec a, T)$ ought to be a simultaneous eigenfunction of ${\rm rank}(\frak g)$ commuting differential operators $D_1, D_2, \dots D_{\rm rank (\frak g)}$ derived from the Casimirs of the \emph{Langlands dual} of an affine $G$-algebra. Furthermore, these commuting $D_l$'s should also define a completely integrable system with spectral curve the (twisted) Seiberg-Witten curve $\Sigma_{SW}$ in fig.~1 or 2 when $G = \{ SU(N), SO(N+1) \}$ or $\{ SO(2N), USp(2N-2), G_2 \}$, respectively.    

Indeed, note that the coherent state $| \epsilon_1, \Delta_n \rangle$ is also known as a Whittaker vector in representation theory whence its norm on the RHS of (\ref{q | q, full defect-NS limit}) is a Whittaker function associated with ${\frak g}^\vee_{{\rm aff}, {\rm crit}}$; in turn, according to~\cite[$\S$2]{etingof}, this would mean that  $Z_{\rm inst} (G, \epsilon_1, 0, \vec a, T)$ must be a simultaneous eigenfunction  of ${\rm rank}(\frak g)$ quantum Toda Hamiltonians ${\cal D}^{(1)}_{\rm Toda}, {\cal D}^{(2)}_{\rm Toda}, \dots, {\cal D}^{(\rm rank ({\frak g}))}_{\rm Toda}$ that are associated with the \emph{Langlands dual} of an affine $G$-algebra, i.e., 
\be
\label{toda IS}
\boxed{{\cal D}^{(l)}_{\rm Toda} \cdot Z_{\rm inst} (G, \epsilon_1, 0, \vec a, T) = {\cal E}^{(l)}_{\rm Toda}  \, Z_{\rm inst} (G, \epsilon_1, 0, \vec a, T) }
\ee
where the ${\cal D}^{(l)}_{\rm Toda}$'s are Casimir differential operators; the ${\cal E}^{(l)}_{\rm Toda}$'s are complex eigenvalues; and $l = 1, 2, \dots, {\rm rank}(\frak g)$.  Furthermore, it has also been shown in~\cite{MW} that the spectral curve of the quantum affine Toda system defined by (\ref{toda IS}), is just $\Sigma_{SW}$.

Note that our result is also consistent with~\cite[Corollary~3.7(2)]{J-function}, where ${\cal Z}^{\rm aff}_{G, B}$ and $\epsilon$ in \emph{loc.~cit.} correspond respectively to $Z_{\rm inst} (G, \epsilon_1, 0, \vec a, T)$  and $\epsilon_2$.

\newsubsection{The ``Fully-Ramified'' Nekrasov Instanton Partition Function, Hitchin Systems, and the ``Ramified'' Geometric Langlands Correspondence for Curves}

We shall now elucidate the relation between the ``fully-ramified'' Nekrasov instanton partition function and Hitchin systems, and connect our story to the ``ramified'' geometric Langlands correspondence for curves. 

\bigskip\noindent{\it The Relation to Hitchin Systems}

To this end, first note that a pure $G$ theory can also be interpreted as the $m \to \infty$, $q = e^{2 \pi i \tau} \to 0$ limit of a $G$ theory with an adjoint hypermultiplet of mass $m$ and complexified gauge coupling $\tau$, where $m e^{  i \pi  \tau /  h^\vee}$ remains fixed with $h^\vee$ being the dual Coxeter number of $G$.  Second, note that in this limit, the Hamiltonian and spectral curve of the (twisted) elliptic Calogero-Moser system associated with the aforementioned $G$ theory with adjoint matter, reduce to the Hamiltonian and spectral curve of the quantum Toda system associated with the Langlands dual of an affine $G$-algebra~\cite{Phong}. These two points and (\ref{toda IS}) therefore imply that 
\be
\label{CM IS}
\boxed{{\cal D}^{(l)}_{\rm CM} \cdot Z_{\rm inst} (G, q, \epsilon_1, 0, \vec a, m, T) = {\cal E}^{(l)}_{\rm CM}  \, Z_{\rm inst} (G, q, \epsilon_1, 0, \vec a, m, T)}
\ee    
where $Z_{\rm inst} (G, q, \epsilon_1, 0, \vec a, m, T)$ is the ``fully-ramified'' Nekrasov instanton partition function in the Nekrasov-Shatashvili  limit $\epsilon_2 = 0$ of the $G$ theory on $\mathbb R^4\vert_{\epsilon_1, \epsilon_2 = 0}$ with an adjoint hypermultiplet of mass $m$, i.e., for the ${\cal N} = 2^\ast$ theory on $\mathbb R^4\vert_{\epsilon_1, \epsilon_2 = 0}$; $\vec a$ are its Coulomb moduli; $T \subset G$ is the maximal torus that $G$ reduces to along $\mathbb R^2\vert_{\epsilon_1} \subset \mathbb R^4\vert_{\epsilon_1, \epsilon_2 = 0}$; the ${\cal D}^{(l)}_{\rm CM}$'s are differential operators that correspond to the quantum Hamiltonians of the (twisted) elliptic Calogero-Moser system for $G$; the ${\cal E}^{(l)}_{\rm CM}$'s are complex eigenvalues; and $l = 1, 2, \dots, {\rm rank}(G)$. In the case where $G = SU(2)$, the eigenvalue ${\cal E}^{(1)}_{\rm CM}$ has been determined explicitly in~\cite[$\S$4.3]{Alday-Tachikawa}. 

Let us henceforth assume that $G = SU(N)$. The untwisted elliptic Calogero-Moser system in this case is also known to be equivalent to the Hitchin system on a torus with one puncture over which the singularity of the associated Higgs field is regular~\cite{Donagi}. Thus, from (\ref{CM IS}), we can write
\be
\label{Hitchin IS}
\boxed{{\cal D}^{(l)}_{\rm H} \cdot Z_{\rm inst} (SU(N), q, \epsilon_1, 0, \vec a, m, T) = {\cal E}^{(l)}_{\rm H}  \, Z_{\rm inst} (SU(N), q, \epsilon_1, 0, \vec a, m, T) }
\ee    
where the ${\cal D}^{(l)}_{\rm H}$'s are the quantum Hitchin Hamiltonians for $SU(N)$, and the ${\cal E}^{(l)}_{\rm H}$'s are complex eigenvalues. In other words, $Z_{\rm inst} (SU(N), q, \epsilon_1, 0, \vec a, m, T)$ \emph{is also a simultaneous eigenfunction of the quantum Hitchin Hamiltonians for $SU(N)$}. This confirms the conjecture by Alday-Tachikawa in~\cite{Alday-Tachikawa}, at least for a genus one Riemann surface with a single puncture. 
  
 \bigskip\noindent{\it The Connection to the ``Tamely-Ramified'' Geometric Langlands Correspondence for a Single-Punctured Elliptic Curve}
  
Notice that (\ref{Hitchin IS}) means that $Z_{\rm inst} (SU(N), q, \epsilon_1, 0, \vec a, m, T)$  is a $\cal D$-module; moreover, this $\cal D$-module is associated with the Hitchin system for $SU(N)$ on a genus one complex curve with a  ``tamely-ramified'' point. Hence, according to~\cite{rubtsov}, $Z_{\rm inst} (SU(N), q, \epsilon_1, 0, \vec a, m, T)$ \emph{is a $\cal D$-module in the ``tamely-ramified'' geometric Langlands correspondence for $SU(N)$ at genus one with a single puncture}!  
  
One can also obtain this result as follows. Firstly, from the LHS of (\ref{AGT-M-duality-AB-defect}) and fig.~9, we find that we have, along the six-dimensional worldvolume of the $N$ M5-branes  in the original compactification that underlie $Z_{\rm inst} (SU(N), q, \epsilon_1, 0, \vec a, m, T)$, the configuration   
\be
\label{overlap-1}
  \rlap{$\overbrace{\phantom{{\mathbb R}^2\vert_{\epsilon_2 = 0} \times {\mathbb R}^2\vert_{\epsilon_1}}}^{\textrm{part of M9-plane}}$} {\mathbb R}^2\vert_{\epsilon_2 = 0}
\times \underbrace{{\mathbb R}^2\vert_{\epsilon_1} \times {\bf T}^2_{{\rm marked}, \beta \to 0}}_{\textrm {4d defect}},
\ee
where ${\bf T}^2_{{\rm marked}, \beta \to 0}$ is a torus with a marked point over which the (spatial part of the) M9-plane sits, and $\beta$ is the radius of one of its two circles. 

Note that footnote~\ref{junya's twist} means that the precise metrics on ${\mathbb R}^2\vert_{\epsilon_2 = 0}$ and ${\mathbb R}^2\vert_{\epsilon_1}$ are not essential in our forthcoming analysis; in particular, one can place on ${\mathbb R}^2\vert_{\epsilon_2 = 0} = \bR^2$ a ``cigar-like'' metric 
\be
ds^2 = dr^2 + f(r) d\theta, \quad 0 \leq r < \infty, \quad 0 \leq \theta \leq 2 \pi, 
\ee
with $f(r) \sim r^2$ for $r \to 0$ and $f(r) \to \rho^2$ for $r \to \infty$. Obviously, $\rho$ is the asymptotic radius of the circle ${\bf S}^1$ parameterized by $\theta$. We can assume that $f(r)$ is identically equal to $\rho$ for sufficiently large $r$, say $r \geq r_0$. 

Likewise, one can also place on ${\mathbb R}^2\vert_{\epsilon_1}$ the above ``cigar-like'' metric, where the rotation of the plane associated with a nonzero $\epsilon_1$ would therefore correspond to a rotation of the  circle ${\tilde {\bf S}}^1$ of the cigar. Moreover, since the rotation of the plane confines the physical excitations close to the origin, i.e., close to the tip of the cigar, we can conveniently consider the truncated cigar with length $r \leq R$, where $R >> \rho, r_0$. 

Thus, if we denote $\bR^2\vert_{\epsilon_2 =0}$ and  $\bR^2\vert_{\epsilon_1}$ endowed with the above ``cigar-like'' metrics as $D$ and $D_{R, \epsilon_1}$, respectively, we can also express (\ref{overlap-1}) as 
\be
\label{overlap-2}
  \rlap{$\overbrace{\phantom{D \times D_{R, \epsilon_1}}}^{\textrm{part of M9-plane}}$} D
\times \underbrace{D_{R, \epsilon_1} \times {\bf T}^2_{{\rm marked}, \beta \to 0}}_{\textrm {4d defect}}.
\ee

Secondly, notice that we can regard $D \times D_{R, \epsilon_1}$ as a nontrivial ${\bf S}^1 \times {\tilde {\bf S}}^1$ fibration of $\bR_+ \times \mathbb I$, where $\bR_+$ and $\mathbb I$ are a half-line and an interval, respectively. However, macroscopically at low-energies whence the curvature of the cigar tips is not observable, $D \times D_{R, \epsilon_1}$ is effectively a \emph{trivial} ${\bf S}^1 \times {\tilde {\bf S}}^1$ fibration of $\bR_+ \times \mathbb I$. Therefore, where the minimal energy limit of the M5-brane worldvolume theory is concerned, we can simply take (\ref{overlap-2}) to be
\be
\label{overlap-3}
  \rlap{$\overbrace{\phantom{{\bf S}^1 \times \bR_+ \times {\tilde {\bf S}}^1 \times \mathbb I}}^{\textrm{part of M9-plane}}$} {\bf S}^1 \times \bR_+
\times \underbrace{{\tilde {\bf S}}^1 \times \mathbb I
 \times {\bf T}^2_{{\rm marked}, \beta \to 0}}_{\textrm {4d defect}}.
\ee    

Thirdly, according to $\S$6.3 and~\cite{XN-regular, NW}, the compactification  at minimal energy of the M5-brane worldvolume theory on ${\bf S}^1 \times {\tilde {\bf S}}^1 \times {\bf T}^2_{{\rm marked}, \beta \to 0}$ in (\ref{overlap-3}), would result in an ${\cal N} = (4,4)$ sigma-model on $\Sigma = \bR_+ \times \mathbb I$ with target space ${\cal M}_H$,  where ${\cal M}_H$ is the ``tamely-ramified''  Hitchin fibration associated with $SU(N)$ and the single-punctured Riemann surface ${\bf T}^2_{\rm punc}$. As $\Sigma$ is a worldsheet of an \emph{open} string which starts propagating at time $t=0$, there must be branes at the two ends of $\mathbb I$. Since the sigma-model is an $A$-model, these branes must necessarily be $A$-branes.

Fourthly, as argued in~\cite{NW},  the distinguished $A$-brane at the tip of $ D_{R, \epsilon}$ is a space-filling canonical coisotropic brane ${\cal B}_{cc}$ of type $(A, B, A)$ with respect to the hyperk\"ahler structure of ${\cal M}_H$. Hence, on one end of $\mathbb I$, we have the brane ${\cal B}_{cc}$. 

What about the brane ${\cal B}_L$ at the far end of $D_{R, \epsilon}$? According to~\cite{GW}, $({\cal B}_{cc}, {\cal B}_{cc})$ strings correspond to holomorphic differential operators that act on some line bundle over ${\cal M}_H$. Since our $({\cal B}_{cc}, {\cal B}_{L})$ strings ought to encode the minimal energy spectrum of the underlying M5-brane worldvolume theory, i.e., the space of $({\cal B}_{cc}, {\cal B}_{L})$ strings ought to furnish a space of states and \emph{not} a space of differential operators, ${\cal B}_{L}$ \emph{cannot} be  ${\cal B}_{cc}$. As such, being an $A$-brane that is not space-filling, ${\cal B}_L$ can only be middle-dimensional Lagrangian; in other words, ${\cal B}_L$ can either wrap the fiber or base of ${\cal M}_H$.  

As in the last subsection, we have along ${\cal C}_{\rm eff} = {\bf T}^2_{\rm punc}$ in the dual compactification given by the RHS of a generalization (to include a 4d worldvolume defect) of fig.~9 with $n=1$, the spin-$s_i$ Casimir operators $S^{(s_i)}$ whose action on any correlation function of operators  is a multiplication by a $c$-number. Also, the correspondence in the generalization of fig.~9 tells us that the $S^{(k)}$'s -- like the $W^{(k)}$'s in $\S$5.2 -- can be identified with the $(k, 0)$-holomorphic differentials $\phi_k$ (with an order-$k$ pole at $z=1$) that define the underlying SW curve which is an $N$-fold cover of ${\bf T}^2_{\rm punc}$. Since the $\phi_k$'s are the commuting Hamiltonians $H_k$ that define ${\cal M}_H$~\cite{GMN-WKB}, the last two statements mean that we effectively have the condition $H_k - h_k = 0$, where the $h_k$'s are complex constants. This condition defines the fiber $\bf F$ of ${\cal M}_H$~\cite{GW}. Thus, as $H_k \sim {\rm {Tr}} \, \Phi^k$, where the 4d scalar field $\Phi$ of the ``fully-ramified'' ${\cal N } = 2^\ast$ $SU(N)$ theory on ${\bf S}^1 \times {\tilde {\bf S}}^1 \times \Sigma$  survives as a sigma-model scalar field on $\Sigma$ after compactifying on ${\bf S}^1 \times {\tilde {\bf S}}^1$, from the concluding remarks in the last paragraph, one can deduce  that ${\cal B}_L = \bf F$. Hence, on the other end of $\mathbb I$, we have the brane $\bf F$. 

From the above four points, one can conclude that the minimal energy spectrum of the M5-brane worldvolume theory which is captured by $Z_{\rm inst} (SU(N), q, \epsilon_1, 0, \vec a, m, T)$, is furnished by the space of $({\cal B}_{cc}, \bf F)$ strings. In turn, since the space of $({\cal B}_{cc}, \bf F)$ strings in ${\cal M}_H$ also furnishes a $\cal D$-module~\cite{GW}, we find that $Z_{\rm inst} (SU(N), q, \epsilon_1, 0, \vec a, m, T)$ is also a $\cal D$-module in the ``tamely-ramified'' geometric Langlands correspondence for $SU(N)$ on ${\bf T}^2_{\rm punc}$. 

 \bigskip\noindent{\it The Connection to the  ``Tamely-Ramified'' Geometric Langlands Correspondence for a Multi-Punctured Elliptic Curve}

One could also replace ${\bf T}^2_{{\rm marked}, \beta \to 0}$ and the single M9-plane in (\ref{overlap-1}) with an $n$-marked torus ${\bf T}^2_{{\rm marked}, n, \beta \to 0}$ and $n$ M9-planes, respectively, and repeat the above analysis. Everything follows verbatim, where the only change is the target of the A-model which is now ${\cal M}_{H, n}$ -- the ``tamely-ramified''  Hitchin fibration associated with $SU(N)$ and the $n$-punctured Riemann surface ${\bf T}^2_{{\rm punc}, n}$. 

Therefore, if $Z^{\rm neck}_{\rm inst} (SU(N), {\bf q}, \epsilon_1, 0, \vec {\bf a}, {\bf m}, T)$ is the ``fully-ramified'' Nekrasov instanton partition function in the Nekrasov-Shatashvili limit $\epsilon_2 = 0$ of a conformal necklace quiver theory of $n$ $SU(N)$ gauge groups, then \emph{$Z^{\rm neck}_{\rm inst} (SU(N),  {\bf q}, \epsilon_1, 0, \vec {\bf a}, {\bf m}, T)$ is also a $\cal D$-module in the ``tamely-ramified'' geometric Langlands correspondence for $SU(N)$ on ${\bf T}^2_{{\rm punc}, n}$, where $n  \geq 1$}! 
 
 In turn, according to~\cite{rubtsov}, \emph{ $Z^{\rm neck}_{\rm inst} (SU(N), {\bf q}, \epsilon_1, 0, \vec {\bf a}, {\bf m}, T)$ is also a simultaneous eigenfunction of the quantum Hitchin Hamiltonians for $SU(N)$}. This confirms the conjecture by Alday-Tachikawa in~\cite{Alday-Tachikawa}, at least for a genus one Riemann surface with $n  \geq 1$ punctures.

 \bigskip\noindent{\it The Connection to the ``Tamely-Ramified'' Geometric Langlands Correspondence for a Multi-Punctured Rational Curve}

Let us now turn our attention to  $Z^{\rm lin}_{\rm inst} (SU(N), {\bf q}, \epsilon_1, 0, \vec {\bf a}, {\bf m}, T)$ -- the ``fully-ramified'' Nekrasov instanton partition function in the Nekrasov-Shatashvili limit $\epsilon_2 = 0$ of a conformal linear quiver theory of $n$ $SU(N)$ gauge groups. From the LHS of (\ref{AGT-M-duality-AB-defect}) and fig.~6, we find that we have, along the six-dimensional worldvolume of the $N$ M5-branes  in the original compactification that underlie  $Z^{\rm lin}_{\rm inst} (SU(N), {\bf q}, \epsilon_1, 0, \vec {\bf a}, {\bf m}, T)$, the configuration   
\be
\label{overlap-1-rational}
  \rlap{$\overbrace{\phantom{{\mathbb R}^2\vert_{\epsilon_2 = 0} \times {\mathbb R}^2\vert_{\epsilon_1}}}^{\textrm{part of M9-planes}}$} {\mathbb R}^2\vert_{\epsilon_2 = 0}
\times \underbrace{{\mathbb R}^2\vert_{\epsilon_1} \times {C}_{n+3, \beta \to 0}}_{\textrm {4d defect}},
\ee
where ${C}_{n+3, \beta \to 0}$ is a short cylinder with $n+3$ marked points over which the (spatial part of the) $n+3$ number of M9-planes sit, and $\beta$ is the radius of its circle. 

By repeating our arguments which took us from (\ref{overlap-1}) to (\ref{overlap-3}), we find that where the minimal energy limit of the M5-brane worldvolume theory is concerned, we can simply take (\ref{overlap-1-rational}) to be
\be
\label{overlap-3-rational}
  \rlap{$\overbrace{\phantom{{\bf S}^1 \times \bR_+ \times {\tilde {\bf S}}^1 \times \mathbb I}}^{\textrm{part of M9-planes}}$} {\bf S}^1 \times \bR_+
\times \underbrace{{\tilde {\bf S}}^1 \times \mathbb I
 \times {C}_{n+3, \beta \to 0}}_{\textrm {4d defect}}.
\ee    

According to $\S$6.3 and~\cite{XN-regular, NW}, the compactification  at minimal energy of the M5-brane worldvolume theory on ${\bf S}^1 \times {\tilde {\bf S}}^1 \times {C}_{n+3, \beta \to 0}$ in (\ref{overlap-3-rational}), would result in an ${\cal N} = (4,4)$ sigma-model on $\Sigma = \bR_+ \times \mathbb I$ with target space ${\mathscr M}_{H, n+3}$,  where ${\mathscr M}_{H, n+3}$ is the ``tamely-ramified''  Hitchin fibration associated with $SU(N)$ and the $n+3$-punctured Riemann sphere ${\bf S}^2_{{\rm punc}, n+3}$. 

As before, the sigma-model is an $A$-model whence the branes at the ends of $\mathbb I$ must be $A$-branes; in particular, the sigma-model describes a $({\cal B}_{cc}, {\mathscr B}_L)$ string, where ${\mathscr B}_L$ is a Lagrangian brane that can either wrap the fiber or base of ${\mathscr M}_{H, n+3}$.  So which does  ${\mathscr B}_L$ wrap?  

As in the last subsection, we have along ${\cal C}_{\rm eff} = {\bf S}^2_{{\rm punc}, n+3}$ in the dual compactification given by the RHS of a generalization (to include a 4d worldvolume defect) of fig.~6, the spin-$s_i$ Casimir operators $S^{(s_i)}$ whose action on any correlation function of operators  is a multiplication by a $c$-number. Also, the correspondence in the generalization of fig.~6 tells us that the $S^{(k)}$'s -- like the $W^{(k)}$'s in $\S$5.2 -- can be identified with the $(k, 0)$-holomorphic differentials $\phi_k$ (which have an order-$k$ pole at each puncture) that define the underlying SW curve which is an $N$-fold cover of ${\bf S}^2_{{\rm punc}, n+3}$. Since the $\phi_k$'s are the commuting Hamiltonians $H_k$ that define ${\mathscr M}_{H, n+3}$~\cite{GMN-WKB}, the last two statements mean that we effectively have the condition $H_k - h_k = 0$, where the $h_k$'s are complex constants. This condition defines the fiber $\bf F$ of ${\mathscr M}_{H, n+3}$~\cite{GW}. Thus, as $H_k \sim {\rm {Tr}} \, \Phi^k$, where the 4d scalar field $\Phi$ of the ``fully-ramified'' conformal linear quiver theory on ${\bf S}^1 \times {\tilde {\bf S}}^1 \times \Sigma$  survives as a sigma-model scalar field on $\Sigma$ after compactifying on ${\bf S}^1 \times {\tilde {\bf S}}^1$, from the concluding remarks in the last paragraph, one can deduce  that ${\mathscr B}_L = \bf F$. 

Thus, one can conclude that the minimal energy spectrum of the M5-brane worldvolume theory which is captured by $Z^{\rm lin}_{\rm inst} (SU(N), {\bf q}, \epsilon_1, 0, \vec a, {\bf m}, T)$, is furnished by the space of $({\cal B}_{cc}, \bf F)$ strings. In turn, since the space of $({\cal B}_{cc}, \bf F)$ strings in ${\mathscr M}_{H, n+3}$ also furnishes a $\cal D$-module~\cite{GW}, we find that \emph{$Z^{\rm lin}_{\rm inst} (SU(N), {\bf q}, \epsilon_1, 0, \vec a, {\bf m}, T)$ is also a $\cal D$-module in the ``tamely-ramified'' geometric Langlands correspondence for $SU(N)$ on ${\bf S}^2_{{\rm punc}, n+3}$, where $n  \geq 1$}!

In turn, according to~\cite{rubtsov}, \emph{ $Z^{\rm lin}_{\rm inst} (SU(N), {\bf q}, \epsilon_1, 0, \vec {\bf a}, {\bf m}, T)$ is also a simultaneous eigenfunction of the quantum Hitchin Hamiltonians for $SU(N)$}. This confirms the conjecture by Alday-Tachikawa in~\cite{Alday-Tachikawa}, at least for a genus zero Riemann surface with $n+3$ punctures, where $n  \geq 1$.

\part{\Large Appendix}

\appendix

\newsection{The Multi-Taub-NUT Space, Sen's Four-Manifold and String/M-Theory}

\newsubsection{The Geometry of Multi-Taub-NUT Space}

The multi-Taub-NUT space, which we will here denote broadly as $TN_k$, is a hyperk\"ahler four-manifold that can be regarded as a nontrivial singular ${\bf S}^1$ fibration of ${\mathbb R}^3$. It has the metric~\cite{BBS}
\be
ds^2_{TN_k} = {1\over U(\vec{r})}(d\alpha + \chi)^2 + U(\vec{r})d\vec{r}^2,
\label{TN}
\ee
where $\alpha$ is a compact periodic coordinate, and ${\vec r} = (r^1, r^2, r^3)$ is a three-vector in $\mathbb R^3$. The function $U(\vec{r})$ and the 1-form $\chi$ are defined by
\be
U(\vec{r}) = 1 + {R\over 2}\sum_{a=1}^k {1 \over{|\vec{r} - \vec {r}_a}|}, \qquad\qquad  d\chi = *_3~dU,
\label{TN function}
\ee
where $*_3$ is Poincar\'e duality in three-dimensions. Smoothness requirements of the metric (\ref{TN}) dictate that $\alpha$ must have period $2 \pi R$. Hence, the actual radius of the circle fiber is given by~\cite{BBS}
\be
{\widetilde R}(\vec{r}) = U(\vec{r}) ^{-1/2} R.
\label{radius}
\ee

Notice from (\ref{TN function}) and (\ref{radius}) that the circle fiber shrinks to zero size at the $k$ points $\vec{r}_1, \vec{r}_2, \dots, \vec{r}_k$ in $\mathbb R^3$. That is, there are $k-1$ line segments that connect each pair of neighboring points, and over each of these $k-1$ line segments, there is a circle fibration which degenerates at the end points. In other words, $TN_k$ is generically a perfectly smooth four-manifold with $k-1$ homologically independent two-spheres given by the circle fibrations of the line segments.

Notice also from (\ref{TN function}) and (\ref{radius}) that at infinity, i.e. $\vec{r} \to \infty$, we have ${\widetilde R}(\infty) = R$. Consequently, one can see from (\ref{TN}) that the geometry of $TN_k$ at infinity approximates $\mathbb R^3 \times {{\bf S}^1}$, where ${\bf S}^1$ has a fixed radius of $R$. However, the ${\bf S}^1$ factor is actually nontrivially fibered over the ${\bf S}^2 $ submanifold of $\mathbb R^3 \cong {\bf S}^2 \times \mathbb R$ at infinity, where the fibration can be viewed as a monopole bundle of charge (or first Chern-class) $k$, i.e.,
\be
\int_{{\bf S}^2} d\chi = 2\pi k.
\label{monopole charge}
\ee
This point will be important in our below discussion of $TN_k$ as an M-theory background and its interpretation as D6-branes in the corresponding type IIA theory.

Last but not least, note that as we ``decompactify" the asymptotic radius of the circle by letting $R \to \infty$, the geometry of $TN_k$ will be that of a $\it{resolved}$ ALE space of type $A_{k-1}$; the intersection matrix of the two-spheres just gives the Cartan matrix of the $A_{k-1}$ Lie algebra. In order to obtain a $\it{singular}$ ALE space of type $A_{k-1}$ such as $\mathbb R^4 / \mathbb Z_k$, one just needs to bring together all the $k$ points $\vec{r}_1, \vec{r}_2, \dots, \vec{r}_k$ to the origin in $\mathbb R^3$, such that the $k-1$ homologically independent two-spheres all collapse to result in an $A_{k-1}$ singularity at $0$. This has an interpretation in terms of enhanced gauge symmetries in the context of string/M-theory as we will explain below.

\newsubsection{The Multi-Taub-NUT Space in a IIA/M-Theory Correspondence}

The $k$ Kaluza-Klein monopoles solution in M-theory can be described by the metric
\be
ds^2 = -dt^2 + \sum_{m=1}^{6} dy^m dy^m + ds^2_{TN_k},
\label{monopole solution}
\ee
where the $y^m$'s denote the space-like worldvolume coordinates on the six-dimensional solitons in type IIA that are represented by the above solution in M-theory. In order to ascertain what these solitons are, let us take the ``eleventh circle'' to be the circle fiber of $TN_k$. Then, a D0-brane in type IIA can be interpreted as a Kaluza-Klein excitation along the ``eleventh circle''. The D0-brane is electrically charged under the gauge field $C_\mu = g_{\mu 10}$ after a Kaluza-Klein reduction. Therefore, its magnetic dual, the D6-brane, must be magnetically charged under the same gauge field. Since a Kaluza-Klein monopole must correspond to a magnetically charged soliton, we find that the six-dimensional space with coordinates $y^m$ ought to be filled by D6-branes after a type IIA compactification of M-theory along the circle fiber of $TN_k$.

That one has $k$ D6-branes is consistent with  the fact that the circle fibration of $TN_k$ is actually a monopole bundle of charge $k$ at infinity via (\ref{monopole charge}). Note also that the $\vec{r}_a$'s can be interpreted as the location of the Kaluza-Klein monopoles in $\mathbb R^3 \in TN_k$. This means that the $k$ D6-branes will be localized at the $k$ points $\vec{r}_1, \vec{r}_2, \dots, \vec{r}_k$ in $\mathbb R^3 \in TN_k$. Therefore, as one brings the $k$ points together towards $0$, all $k$ D6-branes will coincide and the worldvolume theory will possess an enhanced non-abelian $U(k)$ gauge symmetry. Hence, upon a compactification along the circle fiber of M-theory on a $TN_k$ that has an $A_{k-1}$ singularity at its origin, one will obtain an equivalent description in terms of a stack of $k$ $\it{coincident}$ D6-branes that span the directions transverse to $TN_k$ in type IIA string theory. One can also understand this enhancement of gauge symmetries as follows~\cite{sen}. Starting with a non-singular $TN_k$ manifold, there are M2-branes which wrap the $k-1$ two spheres in $TN_k$. Upon compactification along the circle fiber, these M2-branes become open strings in type IIA which connect between neighboring D6-branes which are non-coincident. As we bring all the $\vec{r}_a$'s together, the $k-1$ two-spheres in $TN_k$ collapse, and we have an enhanced gauge symmetry in M-theory due to extra massless gauge fields that originate from the M2-branes which now have zero-volume, in the transverse spacetime directions. In the equivalent IIA picture, this corresponds to the open strings becoming massless as the $k$ D6-branes become coincident, which consequently results in an enhanced non-abelian gauge symmetry in the transverse spacetime directions along the worldvolume of the D6-branes.

Another relevant point would be the following. In order for the tension of a soliton described by the monopole solution (\ref{monopole solution}) to agree with the tension of a D6-brane in type IIA string theory, one has to set $R = g^A_s l_s$, where $g^A_s$ is the IIA string coupling and $l_s$ is the string length scale~\cite{BBS}. In particular, a compactification of M-theory along the circle fiber of $TN_k$ where the asymptotic radius $R$ is either large or small, will result in an equivalent IIA theory that is either strongly- or weakly-coupled, respectively.

\newsubsection{The Multi-Taub-NUT Space, NS5-Branes and T-Duality}

Consider the following ten-dimensional background in type IIA or IIB string theory:
\be
ds^2 = -dt^2 + \sum_{l=1}^{5} dy^l dy^l + ds^2_{TN_k}.
\label{type II background}
\ee
Notice that the metric (\ref{TN}) enjoys a $U(1)$ isometry which acts to shift the value of $\alpha$. Consequently, this allows for the application of T-duality transformations to the above background solution. In doing so, one will obtain the following T-dual solution~\cite{CJ, Tong}:
\be
ds^2 = -dt^2 + \sum_{l=1}^{5} dy^l dy^l + V(\vec{x})(d\theta^2 + d\vec{r}^2),
\label{type II background-T-dual}
\ee
where $\theta$ is a compact coordinate of period $2\pi$ which parameterizes the dual ${\bf S}^1$, and
\be
V(\vec{x}) = {1\over R^2} + {1\over 2} \sum_{a=1}^{k} {1\over |\vec{x} - \vec{x}_a|},
\label{V}
\ee
where $\vec{x} = (\theta, \vec{r})$ is taken to mean a position in a full $\mathbb R^4$. From (\ref{V}) and (\ref{type II background-T-dual}), we see that the asymptotic radius of the dual circle is indeed given by $1/R$ as expected under T-duality.

Note that the solution given by (\ref{type II background-T-dual}) consists of $k$ objects which are pointlike in the $\mathbb R^4$, and which are also magnetic sources of the NS-NS potential $B_{\mu \nu}$~\cite{CJ}. In fact, they just correspond to $k$ NS5-branes which span the space with coordinates $y^l$, that are also arranged in a circle on $\theta$ and localized on the rest of $\mathbb R^4$ according to the centers $\vec{x}_a, a = 1, 2, \dots, k$. Reversing the above arguments, we conclude that one can do a T-duality along any circle that is $\it{transverse}$ to a stack of $k$ $\it{coincident}$ NS5-branes in type IIA(IIB) string theory, and obtain a dual background with no NS5-branes but with a $TN_k$ manifold that has an $A_{k-1}$ $\it{singularity}$ at the origin in type IIB(IIA) string theory. In addition, notice that the asymptotic radius $R$ of the dual, singular $TN_k$ background must tend to zero if the radius ${V(\vec{x})}^{1/2}$ of the circle transverse to the NS5-branes is to be infinitely large at any point $\vec{r} \in \mathbb R^3$.

Last but not least, note that in going from (\ref{type II background}) to (\ref{type II background-T-dual}) under  T-duality transformations, only components of the solution transverse to the NS5-brane worldvolume get modified. In other words, the components of the solution along the worldvolume directions have no structure and are therefore trivial. Consequently, an application of T-duality along any worldvolume direction will map us back to the same NS5-brane solution given by (\ref{type II background-T-dual}). (This is to be contrasted with a $D_p$-brane, where T-duality along a direction parallel or transverse to its worldvolume will result in a $D_{p-1}$ or $D_{p+1}$-brane, respectively.)

\newsubsection{The Geometry of Sen's Four-Manifold}

Consider the following four-manifold characterized by a nontrivial ${\bf S}^1$ fibration of $\mathbb R^3$ with metric~\cite{sen}
\be
ds^2 = {1 \over W(\vec{r})} (d\alpha + \chi)^2 + W(\vec{r}) d\vec{r}^2,
\label{sen's manifold}
\ee
modded out by the transformation
\be
(\vec{r} \rightarrow -\vec{r}, \qquad \alpha \rightarrow -\alpha),
\label{reflection}
\ee
where $\alpha$ is a compact periodic coordinate of the ${\bf S}^1$ fiber, and $\vec{r} = (r^1, r^2, r^3)$ is a three-vector in $\mathbb R^3$. The function $W(\vec{r})$ and the 1-form $\chi$ are defined by
\be
W(\vec{r}) = 1 - {2R \over {|\vec{r}|} }  + {R\over 2}\sum_{a=1}^k \left( {1\over |\vec{r} - \vec{r}_a|} + {1\over |\vec{r} + \vec{r}_a|} \right), \qquad d\chi = *_3~dW,
\label{SN function}
\ee
where $*_3$ is Poincar\'e duality in three-dimensions, and where the asymptotic radius of the circle fiber is $R$ (before the identification in (\ref{reflection})).

Note that the metric is invariant under the reflection (\ref{reflection}); $W(\vec{r})$ is invariant under $(\vec{r} \rightarrow -\vec{r})$ and $\chi$ changes sign under the reflection. However, the metric is singular at $\vec{r} =0$. This singularity can be removed by replacing the metric near $\vec{r} =0$ by the Atiyah-Hitchin metric~\cite{sen18}, which is completely nonsingular after we perform the reflection (\ref{reflection}). We shall here denote this effectively smooth hyperk\"ahler four-manifold  broadly as Sen's four-manifold or $SN_k$.

In the region where $\vec{r} \to \infty$, we see from (\ref{SN function}) that $W(\vec{r}) \to 1$. Hence, from (\ref{sen's manifold}) and (\ref{reflection}), we find that $SN_k$ approximates $(\mathbb R^3 \times {{\bf S}^1})/{\cal I}_4$ far away from the origin at infinity, where ${\cal I}_4$ denotes an independent action on the two factors $\mathbb R^3$ and ${\bf S}^1$ that is defined in (\ref{reflection}).  As mentioned earlier, the ${\bf S}^1$ factor has a fixed radius of $R$.

At the $k$ points $\vec{r}_1, \dots, \vec{r}_k$ in $SN_k$, the circle fiber shrinks to zero size, as one can see from (\ref{SN function}) and (\ref{sen's manifold}). Consequently, the circle fibrations of the line segments connecting each of these neighboring points will result in a set of $k-1$ two-spheres. Because the reflection (\ref{reflection}) is a symmetry of the space, there is an identification $\vec{r}_a \sim - \vec{r}_a$. As such, there will be additional two spheres coming from the extra circle fibrations of the line segments that connect the points $\vec{r}_i$ and $-\vec{r}_{i+1}$. In short, the homologically independent two-spheres will define an intersection matrix that is the Cartan matrix of a $D_k$ Lie algebra~\cite{sen}. If we let all the $\vec{r}_a$'s approach the origin, the areas of all the two-spheres vanish, and we obtain a $D_k$ singularity. As we shall explain below, this observation is consistent with the fact that such an $SN_k$ background in string/M-theory would lead to an enhanced $SO(2k)$ gauge symmetry.

\newsubsection{Sen's Four-Manifold in a IIA/M-Theory Correspondence}

Consider the following eleven-dimensional background in M-theory:
\be
ds^2 = -dt^2 + \sum_{m=1}^{6} dy^m dy^m + ds^2_{SN_k},
\label{orientifold solution}
\ee
where the $y^m$'s denote the space-like worldvolume coordinates on the six-dimensional solitons in type IIA that are represented by the above solution in M-theory. In order to ascertain what these solitons are, first note that near $\vec{r}=0$, the metric of $SN_k$ agrees with the Atiyah-Hitchin or AH space. It is known that upon a type IIA compactifcation of M-theory along the circle fiber of such an AH space, one would get an orientifold six-plane~\cite{sen17}. Second, note that near the point $\vec{r} = \vec{r}_a$ or its image $-\vec{r}_a$ (under ${\cal I}_4$) for $1\leq a \leq k$, the metric agrees with the one near a Kaluza-Klein monopole. Moreover, far away from the origin at infinity, the metric looks like the multi-Taub-NUT space at infinity albeit identified under the action of ${\cal I}_4$. In all, this means that (\ref{orientifold solution}) represents an M-theory background which upon compactification along the circle fiber, gives us $k$ D6-branes and an $\textrm{O}6^{-}$-plane in type IIA string theory which span the directions transverse to $SN_k$ given by the coordinates $y^m$.\footnote{As emphasized in \cite{sen} itself, the M-theory background given by (\ref{orientifold solution}) is only an approximate solution to the exact one describing the D6-branes and  $\textrm{O}6^{-}$-plane in type IIA string theory. However, it differs from the exact solution by terms that vanish exponentially as we move away from the origin. Since our discussion in the main text will only involve an analysis of $SN_k$ near the boundary at infinity, this deviation from the exact solution will not affect us.}

Note also that the $\vec{r}_a$'s can be interpreted as the location of the Kaluza-Klein monopoles in $SN_k$. This means that the $k$ D6-branes will be localized at the $k$ points $\vec{r}_1, \vec{r}_2, \dots, \vec{r}_k$ in $SN_k$. Therefore, as one brings the $k$ points together towards $0$, all $k$ D6-branes will coincide on top of the  $\textrm{O}6^{-}$-plane and the worldvolume theory will possess an enhanced non-abelian $SO(2k)$ gauge symmetry.\footnote{One has an $SO(2k)$ gauge symmetry because of the presence of an  $\textrm{O}6^{-}$-plane, i.e., the orientifold six-plane that is associated with a worldsheet parity operator whose eigenvalue is $-1$.} Hence, upon a compactification along the circle fiber of M-theory on an $SN_k$ that has a $D_k$ singularity at its origin, one will obtain an equivalent description in terms of a stack of $k$ $\it{coincident}$ D6-branes on top of an  $\textrm{O}6^{-}$-plane that span the directions transverse to $SN_k$ in type IIA string theory. One can also understand this enhancement of gauge symmetries from the perspective of M2-branes wrapping the two-spheres in $SN_k$ and open strings in type IIA connecting between the D6-branes~\cite{sen}. Since the discussion is analogous to the one before on $TN_k$, we shall skip it.

Once again, in order for the tension of a soliton described by the monopole solution (\ref{orientifold solution}) to agree with the tension of a D6-brane in type IIA string theory, one must have $R \sim g^A_s l_s$. Therefore, a compactification of M-theory along the circle fiber of $SN_k$ where the asymptotic radius $R$ is either large or small, will result in an equivalent IIA theory that is either strongly- or weakly-coupled, respectively.

\newsubsection{Sen's Four-Manifold, NS5-branes/ON5-planes and T-Duality}

Consider the following ten-dimensional background in either type IIA or IIB string theory:
\be
ds^2 = -dt^2 + \sum_{l=1}^{5} dy^l dy^l + ds^2_{SN_k}.
\label{type II background Sen}
\ee
Notice that the metric (\ref{sen's manifold}), just like the metric (\ref{TN}), enjoys a $U(1)$ isometry which acts to shift the value of $\alpha$. Consequently, this allows for the application of T-duality transformations to the above background solution, just like in the multi-Taub-NUT example.  Far away from the origin,\footnote{As mentioned earlier, our analysis in the main text will only involve the physics of the background near infinity. As such, it suffices to just discuss what happens away from the origin.} the T-dual background will therefore look like
\be
ds^2 = -dt^2 + \sum_{l=1}^{5} dy^l dy^l + Y(\vec{r})(d\theta^2 + d\vec{r}^2),
\label{type II background-T-dual Sen}
\ee
where $\theta$ is a compact coordinate of period $2\pi$ which parameterizes the dual ${\bf S}^1$, and
\be
Y(\vec{x}) = {1\over R^2} - {2 \over {|\vec{x}|}}  + {1\over 2}\sum_{a=1}^k \left( {1\over |\vec{x} - \vec{x}_a|} + {1\over |\vec{x} + \vec{x}_a|} \right),
\label{Y}
\ee
where $\vec{x} = (\theta, \vec{r})$ is taken to mean a position in a full $\mathbb R^4$. From (\ref{Y}) and (\ref{type II background-T-dual Sen}), we see that the asymptotic radius of the dual circle is indeed given by $1/R$ as expected under T-duality.

Note that the solution given by (\ref{type II background-T-dual Sen}) consists of $k$ objects which are pointlike in the $\mathbb R^4$, and which are also magnetic sources of the NS-NS potential $B_{\mu \nu}$~\cite{CJ}. In fact, they just correspond to $k$ NS5-branes which span the space with coordinates $y^l$, that are localized on the $\mathbb R^4$ according to the centers $\pm\vec{x}_a, a = 1, 2, \dots, k$. The reason why we ended up with a dual background that appears to have $k$ instead of $k$ NS5-branes is because the background represented by (\ref{sen's manifold})--(\ref{SN function}), and therefore the type II background (\ref{type II background Sen}), incorporates a reflection in the spatial directions transverse to the NS5-branes, which, effectively doubles the number of NS5-branes present. This means that the T-dual solution (\ref{type II background-T-dual Sen}) really corresponds to a background which only has $k$ dynamical NS5-branes and an ON$5^-$-plane, whereby the `-' superscript just indicates that its presence will result in an orthogonal gauge symmetry in the worldvolume theory as required, while the `N' just denotes that it can only be associated with NS5-branes~\cite{hanany}. Reversing the above arguments, we conclude that one can do a T-duality along any circle that is $\it{transverse}$ to a stack of $k$ $\it{coincident}$ NS5-branes on top of an ON$5^-$-plane in type IIA(IIB) string theory, and obtain a dual background with no NS5-branes and no ON$5^-$-plane but with an $SN_k$ manifold that has a $D_k$ $\it{singularity}$ at the origin in type (IIB)(IIA) string theory. In addition, notice that the asymptotic radius $R$ of the dual, singular $SN_k$ background must tend to zero if the radius ${Y(\vec{x})}^{1/2}$ of the circle transverse to the NS5-branes is to be infinitely large over any point $\vec{r} \in \mathbb R^3$.

Finally, note that in going from (\ref{type II background Sen}) to (\ref{type II background-T-dual Sen}) under  T-duality transformations, only components of the solution transverse to the NS5-brane/ON$5^-$-plane worldvolume get modified. In other words, the components of the solution along the worldvolume directions have no structure and are therefore trivial. Consequently, an application of T-duality along any worldvolume direction will map us back to the same NS5-brane/ON$5^-$-plane solution given by (\ref{type II background-T-dual Sen}).

\newsection{A Gauged WZW Model and Affine $\cal W$-Algebras from a Quantum Drinfeld-Sokolov Reduction}

\medskip\noindent{\it An $N_+$-Gauged $\cal G$ WZW Model}

The action of the most general WZW model can be written as
\be S_{\textrm{WZ}}(g) = { k ' \over {4 \pi}} \int_{\Sigma} d^2z \  \textrm{Tr}
(\partial_{z} g^{-1} \partial_{\bar z} g)  + { i k ' \over {24
\pi}} \int_{B; \partial B = \Sigma} d^3 x \ \textrm{Tr} (g^{-1} d
g)^3, \label{WZW action}
 \ee
where $k'$ is the level, the worldsheet $\Sigma$ is a closed Riemann surface, and  $g$ is a worldsheet scalar field
valued in any connected Lie group  $\cal G$ that is also periodic along one of the worldsheet
directions with period $2 \pi$. The trace $\textrm{Tr}$ is the
usual matrix trace in the defining representation of  $\cal G$ whose Lie algebra is $\frak g$.

A non-dynamically gauged version of (\ref{WZW action}) can be written as
\begin{eqnarray}
S_{\textrm{gauged}} (g, A_z, A_{\bar z}) & = & S_{\textrm{WZ}} (g)  + {k' \over {2\pi}}\int_{\Sigma} d^2z \  \textrm{Tr}  [ A_z (\partial_{\bar z} g g^{-1} + {\bar M}) -  A_{\bar z}(g^{-1} \partial_z g + {M})   \nonumber \\
&& \hspace{3.0cm}  + A_z g A_{\bar z} g^{-1} - A_z A_{\bar z}],
\label{gauged WZW action}
\end{eqnarray}
where the worldsheet one-form gauge field $A= A_z dz + A_{\bar z}
d{\bar z}$ is valued in $\frak h$, the Lie algebra of a subgroup
 $\cal H$ of  $\cal G$. Notice that $S_{\textrm{gauged}} (g, A_z, A_{\bar z})$
differs slightly from the standard form of a gauged WZW model
commonly found in the physical literature -- additional $\bar M$
and $M$ constant matrices have been incorporated in the
$\partial_{\bar z} g g^{-1}$ and $g^{-1} \partial_z g $ terms of
the standard action, so that one can later use them to derive the
correct form of the holomorphic stress tensor without reference to
a coset formalism. Setting $\bar M$ and $M$ to the zero matrices
simply takes us back to the standard action for the gauged WZW
model. As required, $S_{\textrm{gauged}} (g, A_z, A_{\bar z})$ is
invariant under the standard (chiral) local gauge transformations
\be g \to hgh^{-1}; \ \ \ A_z \to
\partial_z h \cdot h^{-1} + h A_z h^{-1}; \ \ \ A_{\bar z} \to
\partial_{\bar z} h \cdot h^{-1} + h A_{\bar z} h^{-1},
\label{gauge tx} \ee where $h = e^{\lambda (z, \bar z)} \in {\cal H}$ for
any $\lambda (z, \bar z) \in {\frak h}$.\footnote{A similar model
has been considered in \cite{ref for gauged WZW}. However, the
action in that context is instead invariant under a
$\it{non}$-$\it{chiral}$ local gauge transformation. Moreover, it
does not contain the $A_zA_{\bar z}$ term present in a standard
gauged WZW model.} The invariance of (\ref{gauged WZW action})
under the gauge transformations in (\ref{gauge tx}) can be
verified as follows. 

Firstly, note that the $\bar
M(M)$-independent terms make up the usual Lagrangian for the
standard gauged WZW action, which is certainly invariant under the
gauge transformations of (\ref{gauge tx}). Next, note that under
an infinitesimal gauge transformation $h \simeq 1+ \lambda$, the
terms $\textrm{Tr} (A_{z} \ \bar M)$ and $\textrm{Tr} (A_{\bar z}
\ M)$ change as
\begin{eqnarray} \label{variation Tr 1}\delta \textrm{Tr} (A_{z}
\ \bar M) & = & \textrm{Tr} (\partial_{z} \lambda \
\bar M) - \textrm{Tr} (\bar M \ [\lambda, A_z]),  \\
\delta \textrm{Tr} (A_{\bar z} \ M) & = & \textrm{Tr}
(\partial_{\bar z} \lambda \ M) - \textrm{Tr} (M \ [\lambda,
A_{\bar z}]). \label{variation Tr} \end{eqnarray} Since we really want to
consider the case where  $\cal H$ is a nilpotent subgroup of  a \emph{complex} Lie group $\cal G$, $\lambda$ and $A$ will be valued in the Lie algebra of
a maximally $\it{solvable}$ (nilpotent) subgroup of  $\cal G$. As such, the second
term on the RHS of (\ref{variation Tr 1}) and (\ref{variation
Tr}) will be zero \cite{ref for gauged WZW}. What  remains are
total divergence terms that will vanish upon integration on
$\Sigma$ because it has no boundaries. Therefore,
unless  $\cal H$ is a nilpotent subgroup of  $\cal G$ (or any other solvable
subgroup of  $\cal G$), one cannot incorporate $\bar M$ and $M$ in the
action and still maintain the requisite gauge invariance. This
explains why generalizations of gauged WZW models with these
constant matrices $\bar M$ and $M$ have not appeared much in the
physical literature. Nevertheless, this generalization can be
considered in our case. As we shall see shortly, this
generalization will allow us to obtain the correct form of the
holomorphic stress tensor of the $\cal H$-gauged $\cal G$ WZW model
without any explicit reference to a coset formalism.

The classical equations of motion that follow from the field
variations in (\ref{gauge tx}) are
\begin{eqnarray}
\label{1}
\delta A_z & : & D_{\bar z} g g^{-1}|_h = -M_+, \\
\label{2}
\delta A_{\bar z} & : & g^{-1} D_z g |_h = -M-, \\
\label{3}
\delta g & : & D_{\bar z} (g^{-1} D_z g ) = F_{z \bar z}, \\
\label{4} \delta g & : & D_{z} (D_{\bar z} g g^{-1}) = F_{\bar z
z},
\end{eqnarray}
where $F_{z \bar z} = \partial_z A_{\bar z} - \partial_{\bar z}
A_z + [A_z, A_{\bar z}]$ and $F_{\bar z z} = \partial_{\bar z} A_z
- \partial_{z} A_{\bar z} + [A_{\bar z}, A_{z}]$ are the
non-vanishing components of the field strength, and the covariant
derivatives are given by $D_z =
\partial_z + [A_z, \ ]$ and $D_{\bar z} = \partial_{\bar z} +
[A_{\bar z}, \ ]$. By imposing the condition of (\ref{2}) in
(\ref{3}), and by imposing the condition of (\ref{1}) in
(\ref{4}), since $M_{\pm}$ are constant matrices, we find that we
have the zero curvature condition $F_{z \bar z} = F_{\bar z z} =
0$ as expected of a non-dynamically gauged WZW model. This means
that $A_z$ and $A_{\bar z}$ are trivial on-shell. One is then free
to use the gauge invariance to set $A_z$ and/or $A_{\bar z}$ to a
constant such as zero. In setting $A_z = A_{\bar z} =0$ in
(\ref{3}) and (\ref{4}), noting that $F_{z \bar z}= F_{\bar z z} =
0$, we have the relations \be
\partial_{\bar z} (g^{-1} \partial_z g ) = 0 \qquad \textrm{and} \qquad \partial_{z} (\partial_{\bar z} g g^{-1}) =
0. \label{con} \ee In other words, we have a $\frak g$-valued,
holomorphic conserved current $J(z) = g^{-1} \partial_z g$, and a
$\frak g$-valued antiholomorphic conserved current $\bar J(\bar z)
= \partial_{\bar z} g g^{-1}$, both of which are dimension one and
generate affine symmetries on $\Sigma$. The action in (\ref{gauged
WZW action}) can thus be written as \begin{eqnarray} \label{gauged
WZW action simplified}
S_{\textrm{gauged}} (g, A_z, A_{\bar z}) & = & S_{\textrm{WZ}} (g)  + {k' \over {2\pi}}\int_{\Sigma} d^2z \  \textrm{Tr}  [ A_z ({\bar J}(\bar z) + \bar M) -  A_{\bar z}(J(z) + M)   \nonumber \\
&& \hspace{3.0cm}  + A_z g A_{\bar z} g^{-1} - A_z A_{\bar z}].
\end{eqnarray}

In fact, (\ref{gauged WZW action simplified}) can be further simplified as follows.
Firstly, since  $\cal G$ is a connected group, its Lie algebra $\frak g$ will have a
Cartan decomposition ${\frak g} = {\frak n}_- \oplus {\frak c}
\oplus {\frak n}_+$, where $\frak c$ is the Cartan subalgebra, and
${\frak n}_{\pm}$ are the nilpotent subalgebras of strictly upper
and lower triangular matrices, respectively. The Borel subalgebras will
then be given by ${\frak b}_{\pm} = {\frak c} \oplus {\frak
n}_{\pm}$, and they correspond to the Borel subgroups $B_{\pm} = C \times N_\pm$.
With
respect to this decomposition of the Lie algebra $\frak g$, we can
write $ J(z) = \sum_{a=1}^{\textrm{dim} \, {{\frak n}_-}} J^a_- (z)
t^{-}_a + \sum_{a=1}^{\textrm{dim} \, {{\frak c}}} J^a_c(z) t^{c}_a +
\sum_{a=1}^{\textrm{dim} \, {{\frak n}_+}} J^a_+(z) t^{+}_a$, and $
{\bar J}(\bar z) = \sum_{a=1}^{\textrm{dim} \, {{\frak n}_-}} {\bar
J}^a_- (\bar z) t^{-}_a + \sum_{a=1}^{\textrm{dim} \, {{\frak c}}}
{\bar J}^a_c(\bar z) t^{c}_a + \sum_{a=1}^{\textrm{dim} \, {{\frak
n}_+}} {\bar J}^a_+(\bar z) t^{+}_a$, where $t^{-}_a \in {\frak
n}_-$, $t^{c}_a \in {\frak c}$, and $t^{+}_a \in {\frak n}_+$. One
can also write $M = \sum_{a=1}^{\textrm{dim} \, {{\frak n}_-}} M^a_-
t^{-}_a + \sum_{a=1}^{\textrm{dim} \, {{\frak c}}} M^a_c t^{c}_a  +
\sum_{a=1}^{\textrm{dim} \, {{\frak n}_+}} M^a_+ t^{+}_a$, and $\bar M
= \sum_{a=1}^{\textrm{dim} \, {{\frak n}_-}} {\bar M}^a_- t^{-}_a +
\sum_{a=1}^{\textrm{dim} \, {{\frak c}}} {\bar M}^a_c t^{c}_a  +
\sum_{a=1}^{\textrm{dim} \, {{\frak n}_+}} {\bar M}^a_+ t^{+}_a$,
where $M^a_{\pm ; c}$(${\bar M}^a_{\pm ; c}$) are arbitrary number
constants. 

Secondly, note that the gauged WZW model would be anomalous unless we have $\textrm{Tr}_{L} (tt') -
\textrm{Tr}_{R}(t't) = 0$, where $t,t' \in {\frak h}$, and the traces $\textrm{Tr}_{L}$ and $\textrm{Tr}_{R}$ are taken over some $L$ and $R$ representation of $\cal G$~\cite{holo WZW}. Since in this paper, $\frak h = {\frak n}_+$ is spanned by traceless square matrices whose products with one another remain traceless, this anomaly-cancellation condition would indeed be satisfied. Thus, let us proceed to write $A_{z} = \sum_{a
=1}^{\textrm{dim} \, {\frak n}_+} {\tilde A}_{z}^a t^+_a$, and
$A_{\bar z} =   \sum_{a =1}^{\textrm{dim} \, {\frak n}_+} {\tilde
A}_{\bar z}^a t^+_a$.

Finally, note that since $\textrm{Tr}
(t^{\alpha}_a t^{\beta}_b) = \delta_{a,b}\delta^{\alpha, \beta}$,
the trace of the second term on the RHS of (\ref{gauged WZW
action simplified}) will be nonvanishing only  for components of
$J(z)$($\bar J(\bar z)$) and $M$($\bar M$) that are associated with
their expansion in ${\frak n}_+$. 

Let us write ${J}^+(z) =
\sum_{a=1}^{\textrm{dim} \, {\frak n}_+} {J}^a_+ (z) t^{+}_a$ and
${M}^+ = \sum_{a=1}^{\textrm{dim} \, {{\frak n}_+}} {M}^a_+ t^{+}_a$.
Let us also write ${\bar J}^+(\bar z) =
\sum_{a=1}^{\textrm{dim} \, {\frak n}_+} {\bar J}^a_+ (\bar z)
t^{+}_a$ and ${\bar M}^+ = \sum_{a=1}^{\textrm{dim} \, {{\frak n}_+}}
{\bar M}^a_+ t^{+}_a$. Then, from the above three points, one can express the action of an
$N_+$-gauged $\cal G$ WZW model as
\begin{eqnarray} S_{\textrm{$N_+$-gauged}} (g, A_{z}, A_{\bar z}, J^+,
{\bar J}^+)& = & S_{\textrm{WZ}} (g) - {k' \over {2\pi}} \int_{\Sigma} d^2z \
\sum_{l=1}^{\textrm{dim} \, {{\frak n}_+}} \left[ {\tilde A}^l_{\bar z}( J^l_+(z) + {M}^l_+) -
{\tilde A}^l_{z}( \bar J^l_+(\bar z) + {\bar
M}^l_+) \right]\nonumber \\
&& \hspace{4.5cm} - \textrm {Tr} [ {A}_z g {A}_{\bar z} g^{-1} -
{A}_z {A}_{\bar z}].
\label{B-gauged WZW action G}
\end{eqnarray}

Due to the $N_+$-gauge invariance of the theory, we must divide the
measure in any path integral computation by the volume of the
$N_+$-gauge symmetry. That is, the partition function has to take
the form \be Z_{G} = \int_{\Sigma} { {[g^{-1}dg, d{\tilde
A}^l_{z}, d{\tilde A}^l_{\bar z}]} \over {(\textrm{gauge
volume})}} \ \textrm{exp} \left(i S_{G}(g, A_z, A_{\bar z},
J^+, \bar J^+) \right). \ee One must now fix this gauge invariance
to eliminate the non-unique degrees of freedom. One can do this by
employing the BRST formalism which requires the introduction of
Faddev-Popov ghost fields. In order to obtain the
$\it{holomorphic}$ BRST transformations of the fields, one simply
replaces the infinitesimal position-dependent parameters
${\epsilon}^l$ of $h = \textrm{exp}( -
\sum_{l=1}^{\textrm{dim} \, {\frak n}_+} \epsilon^l t^+_m) \in N_+$ in the
corresponding $\textrm{\it left-sector}$ of the gauge
transformations in (\ref{gauge tx}) with the ghost fields $c^l$,
which then gives us \be \delta_{\textrm{BRST}}(g) = -c^l t^+_l g,
\quad \delta_{\textrm{BRST}}(\tilde A^l_{\bar z}) = - D_{\bar z}
c^l, \quad \delta_{\textrm{BRST}}(\textrm{others}) =0. \label{BRST
tx SL(N)} \ee The components of the ghost field $c(z) =
\sum_{l=1}^{\textrm{dim} \, {\frak n}_+} c^l (z) t^+_l$ and those of
its anti-ghost partner $b(z) = \sum_{l=1}^{\textrm{dim} \, {\frak
n}_+} b^l (z) t^+_l$ will transform as \be \delta_{\textrm{BRST}}
(c^l) = - {1\over 2}f_{mk}^l c^mc^k, \quad
\delta_{\textrm{BRST}}(b^l) = {\tilde B}^l, \quad
\delta_{\textrm{BRST}} {(\tilde B^l)} = 0, \ee where the
$f^l_{mk}$'s are the structure constants of  ${\frak n}_+$. Also, the ${\tilde B}^l$'s are the
Nakanishi-Lautrup auxiliary fields that are the BRST transforms of
the $b^l$'s. They also serve as Lagrange multipliers to impose the
gauge-fixing conditions.

In order to obtain the $\it{antiholomorphic}$ BRST transformations
of the fields, one employs the same recipe with the corresponding
$\textrm{\it right-sector}$ of the gauge transformations in
(\ref{gauge tx}), with the infinitesimal position-dependent gauge
parameter now replaced by the ghost field $\bar c^l$, which then
gives us \be \bar \delta_{\textrm{BRST}}(g) = \bar c^l t^+_l g,
\quad \bar \delta_{\textrm{BRST}}({\tilde A}^l_{z}) = - D_{z} \bar
c^l, \quad \bar\delta_{\textrm{BRST}}(\textrm{others})
=0.\label{BRST tx 1 SL(N)}\ee The components of the ghost field
${\bar c}(\bar z) = \sum_{l=1}^{\textrm{dim} \, {\frak n}_+} {\bar
c}^l (\bar z) t^+_l$ and those of its anti-ghost partner $\bar
b(\bar z) = \sum_{l=1}^{\textrm{dim} \, {\frak n}_+} {\bar b}^l (\bar
z) t^+_l$ will transform as \be \bar\delta_{\textrm{BRST}} (\bar
c^l) = - {1\over 2}f_{mk}^l {\bar c}^m {\bar c}^k, \quad
\bar\delta_{\textrm{BRST}}(\bar b^l) = {\tilde {\bar B}^l}, \quad
\bar\delta_{\textrm{BRST}} {(\tilde {\bar B}^l)} = 0. \ee In the
above, the $\tilde {\bar B}^l$'s are the Nakanishi-Lautrup
auxiliary fields that are the antiholomorphic BRST transforms of
the $\bar b^l$ fields. They also serve as Lagrange multipliers to
impose the gauge-fixing conditions.

Since the BRST transformations in (\ref{BRST tx SL(N)}) and
(\ref{BRST tx 1 SL(N)}) are just infinitesimal versions of the
gauge transformations in (\ref{gauge tx}), $S_{{\textrm{$N_+$-gauged}}} (g, A_z,
A_{\bar z}, J^+, \bar J^+)$ will be invariant under them. An important point to note is that in addition to $(\delta_{\textrm{BRST}} + \bar\delta_{\textrm{BRST}}) \cdot
(\delta_{\textrm{BRST}} + \bar\delta_{\textrm{BRST}}) = 0$, the
holomorphic and antiholomorphic BRST-variations are also
separately nilpotent, i.e., $\delta^2_{\textrm{BRST}} = 0$ and
$\bar\delta^2_{\textrm{BRST}}=0$, whence
$\delta_{\textrm{BRST}}\cdot \bar\delta_{\textrm{BRST}}= - \bar
\delta_{\textrm{BRST}} \cdot \delta_{\textrm{BRST}}$. This means
that the BRST-cohomology of the $N_+$-gauged $\cal G$ WZW   model
can be decomposed into $\it{independent}$ holomorphic and
antiholomorphic sectors that are just complex conjugate of each
other, and that it can be computed via a spectral sequence,
whereby the first two complexes will be furnished by its
holomorphic and antiholomorphic BRST-cohomologies, respectively.
Since we will only be interested in the \emph{chiral half}
of the $N_+$-gauged $\cal G$ WZW   model, we shall henceforth focus on the $\it{holomorphic}$
BRST-cohomology of the $N_+$-gauged $\cal G$ WZW   model.

By the usual recipe of the BRST formalism, one can fix the gauge
by adding to the BRST-invariant action $S_{\textrm{$N_+$-gauged}} (g, A_z,
A_{\bar z}, J^+, \bar J^+)$, a BRST-exact term. Since the BRST
transformation by $(\delta_{\textrm{BRST}} + \bar
\delta_{\textrm{BRST}})$  is nilpotent, the new total action will
still be BRST-invariant as required. The choice of the BRST-exact
operator will then define the gauge-fixing conditions. A
consistent choice of the BRST-exact operator that will give us the
requisite action for the ghost and anti-ghost fields is \be
S_{\textrm{$N_+$-gauged}} (g, A_z, A_{\bar z}, J^+, \bar J^+) +
(\delta_{\textrm{BRST}} + \bar \delta_{\textrm{BRST}})
\left({k'\over 2\pi} \int_{\Sigma} d^2 z \
\sum_{l=1}^{\textrm{dim} \, {\frak n}_+} {\tilde A}^l_{\bar z} b^l +
{\tilde A}^l_{z} {\bar b}^l \right),\nonumber \ee where one will
indeed have the desired total action, which can be written as
\begin{eqnarray} \label{total desired action for SL(N)}
S_{\textrm{WZW}}(g)  -  {k' \over {2\pi}} \int_{\Sigma} d^2z \ \{
\sum_{l=1}^{\textrm{dim} \, {\frak n}_+} \left[ {\tilde A}^l_{\bar z}(
J^l_+(z) + {M}^l_+ - \tilde B^l) - {\tilde A}^l_{z}( \bar
J^l_+(\bar z) + {\bar M}^l_+ + \tilde {\bar B}^l)
\right]  \nonumber \\
 - \textrm {Tr} [ {A}_z g {A}_{\bar z} g^{-1} - {A}_z {A}_{\bar
z}] \} +  {k'\over {2\pi}}\int_{\Sigma} d^2z \
\sum_{l=1}^{\textrm{dim} \, {\frak n}_+} \left ( c^l
D_{\bar z} b^l + + \bar c^l D_{z} \bar b^l \right). \nonumber \\
\end{eqnarray} From the equations of motion by varying the ${\tilde
B}^l$'s, we have the conditions $\tilde A^l_{\bar z} = 0$ for
$l=1,\dots, {\textrm{dim} \, {\frak n}_+}$. From the equations of
motion by varying the $\tilde {\bar B}^l$'s, we also have the
conditions $\tilde A^l_{z} = 0$ for $l=1,\dots,{\textrm{dim} \, {\frak
n}_+}$. Thus, the partition function of the $N_+$-gauged $\cal G$ WZW   model
can also be expressed as
 \be
Z_{G} = \int [ g^{-1} dg, db, dc, d\bar b, d\bar c] \
\textrm{exp} \left ( iS_{\textrm{WZW}}(g) + {i k'\over
{2\pi}}\int_{\Sigma} d^2z \ \textrm{Tr} (c \cdot \partial_{\bar z}
b) (z) + \textrm{Tr} (\bar c \cdot \partial_{z} \bar b) (\bar z)
\right), \label{Z_SL(N)} \ee where the $\it{holomorphic}$ BRST
variations of the fields which leave the effective action in
(\ref{Z_SL(N)}) $\it{invariant}$ are now given by \begin{eqnarray}
\delta_{\textrm{BRST}} (g) = -c^mt^+_m g, & \quad
\delta_{\textrm{BRST}} (c^l) = -{1\over 2} f^l_{mk}c^m c^k,
\quad \delta_{\textrm{BRST}}(b^l) =  J^l_+ + M^l_+ - f^l_{mk}b^mc^k,\nonumber \\
\hspace{-2.0cm}\quad \delta_{\textrm{BRST}}(\textrm{others}) = 0.
& \label{BRST variations SL(N)}
 \end{eqnarray}

The holomorphic BRST-charge generating the field variations in
(\ref{BRST variations SL(N)}) will then be given by \be
Q_{\textrm{BRST}} = \oint {dz \over {2 \pi i}} \left(
\sum_{l=1}^{\textrm{dim} \, {\frak n}_+} \ c^l (z) (J^l_+(z) + M^l_+)
-{1\over 2} \sum_{l=1}^{\textrm{dim} \, {\frak n}_+} f^l_{mk}b^mc^lc^k
(z) \right). \label{Q_BRST,WZW SL(N)} \ee The free-field action of
the left-moving ghost fields in (\ref{Z_SL(N)}) implies that we
have the usual OPE's of ($\textrm{dim} \, {\frak n}_+$) free $bc$
systems. From these free $bc$ OPE's, one can verify that
$Q_{\textrm{BRST}}$ in (\ref{Q_BRST,WZW SL(N)}) will indeed
generate the field variations in (\ref{BRST variations SL(N)}).

Though we did not make this obvious in our discussion hitherto, by integrating out the $\tilde A^l_{\bar z}$'s in
(\ref{B-gauged WZW action G}), and using
the above conditions $\tilde A^l_{z} =0$ for $l=1,\dots,
\textrm{dim} \, {\frak n}_+$, we find that we actually have the
relations $(J^l_+(z) + M^l_+) =0$ for $l=1,\dots,
\textrm{dim} \, {\frak n}_+$. These relations -- involving the currents
associated with the Lie algebra ${\frak n}_+$ of the group $N_+$ that
we are modding out from  $\cal G$ -- will lead us directly to the correct form
of the holomorphic stress tensor for the gauged WZW model without
reference to a coset formalism, as we shall now see.

In the holomorphic BRST-cohomology are $\textrm{dim} \, {\frak g}$ currents $J^a(z)$ (spanned by the $J^l_+(z)$'s, $J^l_-(z)$'s and the $J^l_c(z)'s$)  that generate an affine  $\cal G$
OPE-algebra at level $k'$. As such, one can construct a
holomorphic stress tensor using the Sugawara formalism as \be
T_{\cal G}(z) = {: d_{ab}(J^a J^b) (z):\over {k'+ h^{\vee}}}, \ee where $d_{ab}$ is the Cartan-Killing metric on $\frak g$.  
However, as shown above, one will have the conditions $J^l_+ =
-M^l_+$ for $l=1,2,\dots, \textrm{dim} \, {\frak n}_+$. In order for
the conformal dimensions of the $J^l_+$'s to be compatible with these
conditions, one must define a modified holomorphic stress tensor:
\be T_{\textrm{modified}}(z) = T_{\cal G}(z) + \vec{l} \cdot
\partial\vec{J_c}(z),\ee where $\vec{J_c}(z)$ is a
$\textrm{rank} \, \frak g$-dimensional vector with components being the $J^l_c$
currents associated with the Cartan subalgebra $\frak c$, and
$\vec{l}$ is a sum of simple positive roots of $\frak g$. In
order for the above conditions involving the $J^l_+$'s to be compatible with
the fact that $Q_{\textrm{BRST}}$ generating the holomorphic
variations of the fields must be a scalar of dimension zero, the
$\textrm{dim} \, {\frak n}_+$ left-moving ghost systems
$(b^l, c^l)$ must have conformal dimensions $(h^l, 1-h^l)$ for
$l=1,2,\dots, \textrm{dim} \, {\frak n}_+$, where $h^l$ is the conformal
dimension of the corresponding $J^l_+$ current  under
$T_{\textrm{modified}}(z)$. In sum, by
including the holomorphic stress tensor contribution from the action of the free left-moving ghost fields, we can write the total
holomorphic stress tensor of the $N_+$-gauged $\cal G$ WZW   model
as \be T_{\textrm{$N_+$-gauged}}(z)  = {: d_{ab}(J^a J^b) (z):\over {k'+
h^{\vee}}} + \sum_{a=1}^{\textrm{dim}{\frak c}}
\partial_z J^a_c (z) -  \sum_{l
\in \vartriangle_+} \left [ h^l b^{l}\partial_z c^{l}(z)
+ (h^l - 1)(\partial_z b^{l} c^{l})(z) \right].\label{T_total}\ee Here,
$\vartriangle_+$ is the set of positive roots of $\frak g$, 
and for $\alpha \in \vartriangle_+$, we have $(\rho^{\vee}, \alpha) =1$ if and only if $\alpha$ is a simple root of $\frak g$, where $\rho^{\vee}$ is the ``dual Weyl vector'' of $\frak g$.

\bigskip\noindent{\it The Holomorphic BRST-Cohomology and Affine ${\cal W}$-Algebras from a Quantum Drinfeld-Sokolov Reduction}

We shall now show that the holomorphic BRST-cohomology of the $N_+$-gauged $\cal G$ WZW model will be spanned by local operators whose Laurent modes generate ${\cal W}(\widehat{\frak g})$ -- an affine ${\cal W}$-algebra  obtained from $\widehat {\frak g}$ via a quantum Drinfeld-Sokolov reduction, where $\widehat {\frak g}$ is the affine version of $\frak g$ at level $k'$. To this end, let us first review the purely algebraic approach to generating
${\cal W}(\widehat{\frak g})$ via a quantum
Drinfeld-Sokolov (DS) reduction~\cite{DS-reduction}.

The quantum DS-reduction scheme can be summarized as
the following steps~\cite{review}. Firstly, one starts with a triple
$(\widehat{\frak g}, \widehat{\frak g}', \chi)$, where
$\widehat{\frak g}'$ is an affine subalgebra of $\widehat{\frak
g}$, and $\chi$ is a 1-dimensional representation of
$\widehat{\frak g}'$. Next, one imposes the first class
constraints $ g \sim \chi(g)$ , $\forall g \in {\widehat{\frak
g}'}$, via a BRST procedure. The cohomology of the BRST operator
$Q$ on the set of normal-ordered expressions in currents, ghosts
and their derivatives, is what is called the Hecke algebra
$H^i_Q(\widehat{\frak g}, \widehat{\frak g}', \chi)$ of the triple
$(\widehat{\frak g}, \widehat{\frak g}', \chi)$. For generic
values of $k'$, the Hecke algebra vanishes for $i \neq 0$; the
existing zeroth cohomology $H^0_Q (\widehat{\frak g},
\widehat{\frak g}', \chi)$ is just spanned by a set of local
operators associated with the triple $(\widehat{\frak g},
\widehat{\frak g}', \chi)$, whose Laurent modes generate a closed
affine $\cal W$-algebra. We shall denote the affine $\cal W$-algebra associated
with this set of operators  as ${\cal W}_{DS}[\widehat{\frak g},
\widehat{\frak g}', \chi]$. Note that ${\cal
W}_{DS}[\widehat{\frak g}, \widehat{\frak g}', \chi]$ is
just ${\cal W}(\widehat{\frak g})$. Let us
be more explicit about how one can go about defining ${\cal
W}_{DS}[\widehat{\frak g}, \widehat {\frak g}', \chi]$ and
therefore ${\cal W}(\widehat{\frak g})$,  now that we
have sketched the general idea behind the DS-reduction scheme.
f
In order for ${\cal W}_{DS} [\widehat{\frak g}, \widehat{\frak
g}', \chi]$ to be an affine ${\cal W}$-algebra, one has to choose
a suitable triple $(\widehat{\frak g}, \widehat{\frak g}',\chi)$. A
suitable triple can be obtained by considering a principal
$\frak{sl}(2)$ embedding in $\frak g$. Let us now describe this
embedding. Suppose we have an $\frak {sl}(2)$ subalgebra $\{t_3,
t_+, t_-\}$ of $\frak g$. The adjoint representation of $\frak g$
decomposes into $\frak{sl}(2)$ representations of spin $j_k$, where $k =
1,\dots, s$, for example. Then, one may write the $\widehat {\frak
g}$ current $J(z) = \sum^{\textrm{dim} \, {\frak g}}_a J^a (z) t_a$ as
\be J(z) = \sum^s_{k=1} \sum^{j_k}_{m= -j_k}  J^{k,m} (z) t_{k,m},
\ee where $t_{k,m}$ corresponds to the generator of spin $j_k$ and
isospin $m$ under the $\frak{sl}(2)$ subalgebra. In particular, we
have the correspondences $t_{1,1} = t_+$, $t_{1,0} = t_3$, and
$t_{1,-1} = t_-$. The $\frak {sl}(2)$ subalgebra ${t_3, t_+, t_-}$
can be characterized by a ``dual Weyl vector'' $\rho^{\vee}$, i.e., as mentioned above, for $\alpha \in \vartriangle_+$, where $\vartriangle_+$ is
the set of positive roots of $\frak g$, we have $(\rho^{\vee},
\alpha) =1$ if and only if $\alpha$ is a simple root of $\frak g$.
The $\frak{sl}(2)$ root $\hat \alpha$ is given by $\hat \alpha =
\rho/(\rho, \rho)$, and $t_3 = \rho \cdot {\frak c}$, where $\frak
c$ is the Cartan sublagebra of $\frak g$.

Take $\widehat {\frak g}'$ to be the affine Lie subalgebra
$\widehat {\frak n}_+$ generated by all $J^{k,m}(z)$, where $m > 0$.
Denoting the currents corresponding to positive roots $\alpha$ by
$J^{\alpha}(z)$, and choosing $t_{1,1} = \sum_i e^{\alpha_i}$, one
can then impose the condition (which realizes the required
first-class constraint $g \sim \chi(g)$) \be
\chi_{DS}(J^{\alpha}(z)) = 1 \ (\textrm{for simple roots} \
\alpha_i, ), \quad \chi(J^{\alpha}(z)) =0 \  (\textrm{otherwise}).
\label{constraint} \ee

Next, we introduce pairs of ghost fields $(b^{\alpha}(z),
c_{\alpha}(z))$, one for every positive root $\alpha \in
\vartriangle_+$. By definition, they obey the OPE $b_{\alpha}(z)
c_{\beta} (z') \sim \delta_{\alpha \beta} / (z-z')$, where the
$\alpha, \beta$ (and $\gamma$) indices run over the basis of
$\frak n_+$. The BRST operator that is consistent with
(\ref{constraint}) will then be given by $Q = Q_0 + Q_1$, where
\be Q_0 = \oint {dz \over {2\pi i}} \ \left (J^{\alpha}(z)
c_{\alpha}(z) - {1\over 2}f^{\alpha \beta}_{\gamma} (b^{\gamma}
c_{\alpha}c_{\beta})(z) \right) \label{Q_0} \ee is the standard
differential associated with $\widehat{\frak n}_+$, $f^{\alpha
\beta}_{\gamma}$ are the structure constants of ${\frak n}_+$, and
\be Q_1 =   -\oint {dz \over {2\pi i}}\ \chi_{DS}(J^{\alpha}(z))
c_{\alpha}(z). \label{Q_1} \ee They satisfy \be Q^2 = Q^2_0 =
Q^2_1 = \{Q_0, Q_1\} = 0. \label{Q relations} \ee The resulting
$Q$-cohomology is just the Hecke algebra $H^0_Q (\widehat{\frak
g}, \widehat{\frak g}', \chi)$, which is spanned by a set of local
operators whose Laurent modes generate ${\cal
W}_{DS}[\widehat{\frak g}, \widehat{\frak g}', \chi] = {\cal
W}(\widehat{\frak g})$. Note that (\ref{Q relations}) implies
that one can compute the Hecke algebra via a spectral sequence of
a double complex with differentials being $Q_0$ and $Q_1$
accordingly.

The variation of the various fields under the action of $Q$ can
also be computed using the OPE's of the affine algebra
$\widehat{\frak g}$, the OPE's of the ghost fields, and the
explicit forms of $Q_0$ and $Q_1$ in (\ref{Q_0}) and (\ref{Q_1})
above, and they are given by
\begin{eqnarray}
\label{field variations for DS-1}
\delta c_{\alpha}(z) & = & -\frac{1}{2}f^{\beta \gamma}_{\alpha}(
c_{\beta}c_{\gamma}) (z),  \\
\label{field variations for DS-2} \delta b^{\alpha}(z) & = &
J^{\alpha} (z) - \chi_{DS} (J^{\alpha}(z)) - f^{\alpha
\beta}_{\gamma} (b^{\gamma} c_{\beta})(z).
\end{eqnarray}

Note also that ${\cal W}_{DS}[\widehat{\frak g}, \widehat{\frak
g}', \chi]$ and thus ${\cal W}(\widehat{\frak g})$, will at
least contain the Virasoro algebra. The explicit form of the
stress tensor whose Laurent modes will generate the Virasoro
algebra is (after omitting the normal-ordering symbol)\be
T_{DS}(z) = {{d_{ab}J^a(z)J^b(z)} \over {(k'+h^{\vee})}} +
\sum^{\textrm{dim}\frak c}_{c=1} \partial_z J^c(z) + \sum_{\alpha
\in \vartriangle_+} ((\rho^{\vee}, \alpha) -1)b^{\alpha}\partial_z
c_{\alpha}(z) + (\rho^{\vee}, \alpha)(\partial_z b^{\alpha}
c_{\alpha})(z), \label{stress tensor of DS-reduction}\ee where the
$J^c(z)$'s are just the affine currents that are valued in the
Cartan subalgebra $\frak c$ of the Lie algebra $\frak g$. Note
that with respect to $T_{DS}(z)$, the conformal dimensions of the
pair $(b^{\alpha}(z), c_{\alpha}(z))$ will be given by
$(1-(\rho^{\vee}, \alpha), (\rho^{\vee}, \alpha))$. The central
charge of this Virasoro subalgebra and therefore that of ${\cal
W}(\widehat{\frak g})$, is 
\be c =  l - 12|\alpha_+ \rho + \alpha_-\rho^{\vee}|^2,
\label{central charge of DS-reduction}
\ee
where $l = \textrm{rank}\, \frak g$; $\alpha_+ \alpha_- = -1$; $\alpha_+ = 1/ \sqrt{k'+h^{\vee}}$; and $h^\vee$ is the dual Coxeter number of $\frak g$. 

In the case of a simply-laced Lie algebra $\frak g$, we have $\frak g = {\frak g}^\vee$ and $\rho = \rho^{\vee}$, where ${\frak g}^\vee$ is the Langlands dual Lie algebra. From the Freudenthal-de Vries strange formula
$
{\rho^2 / {2 h^{\vee}}} = {\textrm{dim} \,\frak g / 24},
$
one can then simplify $c$ to
\be
 c_{ADE}  = l - h^\vee  \textrm{dim} \,\frak g \left( \alpha_+ + \alpha_- \right)^2.
\label{cc-simply-laced}
\ee

At any rate, notice that in the $N_+$-gauged $\cal G$ WZW   model, for any $J^l_+$ which has $h^l\neq 0$, the corresponding $M^l_+$  must be set to zero for consistency. This means from our above discussion, that one can identify $M^l_+$ with $-\chi_{DS}(J^l_+(z))$. With this identification, one can see that the field variations in (\ref{BRST variations SL(N)}) agree with the field variations in (\ref{field variations for DS-1}) and (\ref{field variations for DS-2}). In addition, we find that $Q_{\textrm{BRST}}$ in (\ref{Q_BRST,WZW SL(N)}) also coincides with $Q = Q_0 + Q_1$, where $Q_0$ and $Q_1$ are given in (\ref{Q_0}) and (\ref{Q_1}), respectively. Moreover, $T_{\textrm{$N_+$-gauged}}(z)$ of (\ref{T_total}) is just $T_{DS}(z)$ of (\ref{stress tensor of DS-reduction}). Hence, we see that the holomorphic BRST-cohomology of the $N_+$-gauged $\cal G$ WZW   model physically realizes, in all generality, the purely algebraic DS-reduction scheme of generating the Hecke algebra.

We can summarize the results in this appendix as follows. Let us label the local operators of the Hecke algebra as $W^{(s_i)}(z)$, where $i=1, 2, \dots, \textrm{rank}\, \frak g$; $s_i = e_i +1$, the $e_i$'s being the exponents of $\frak g$; and $W^{(s_i)}(z)$ are higher spin-$s_i$ analogs of $T_{\textrm{$N_+$-gauged}}(z)$, where $W^{(2)}(z) = T_{\textrm{$N_+$-gauged}}(z)$. Then, we find that the holomorphic BRST-cohomology of the $N_+$-gauged $\cal G$ WZW   model will be spanned by local operators $W^{(s_i)}(z)$ whose Laurent modes generate ${\cal W}(\widehat{\frak g})$ with central charge (\ref{central charge of DS-reduction}), or in the simply-laced case, (\ref{cc-simply-laced}). Hence, just as a chiral half of a regular $\cal G$ WZW model realizes $\widehat {\frak g}$ whence its spectrum would be generated (modulo null states) by a Verma module over $\widehat {\frak g}$, a chiral half of an $N_+$-gauged $\cal G$ WZW model realizes  ${\cal W}(\widehat{\frak g})$ whence its spectrum would be generated (modulo null states) by a Verma module over ${\cal W}(\widehat{\frak g})$.

\end{document}